\documentclass[10pt]{amsart}

\usepackage[a4paper,top=3cm,bottom=3cm,left=2cm,right=2cm]{geometry}
\usepackage[T1]{fontenc} 
\usepackage[utf8]{inputenc}
\usepackage[english]{babel}

\usepackage{amssymb}
\usepackage{amsthm}
\usepackage{amsmath}
\usepackage{amsfonts}
\usepackage{amscd}
\usepackage{bbm}
\usepackage{enumerate}
\usepackage{times}
\usepackage[mathscr]{eucal}
\usepackage{indentfirst}
\usepackage{verbatim}
\usepackage{lipsum}
\usepackage{mathrsfs}
\usepackage{bm}
\usepackage{dsfont}
\usepackage{latexsym}
\usepackage{yhmath}
\usepackage{braket}
\usepackage{hyperref}
\usepackage{accents}

\usepackage{subfig}
\usepackage{tabularx}

\usepackage[normalem]{ulem}

\usepackage[svgnames]{xcolor}
\hypersetup{
	colorlinks = true,
	citecolor = teal, 
	linkcolor=red,
	urlcolor=blue}

\usepackage[style=numeric,backend=bibtex,natbib=true,maxnames=5,giveninits=true]{biblatex}

\usepackage{orcidlink}

\usepackage{caption}
\usepackage{float}

\allowdisplaybreaks
\makeindex

\newtheorem{theorem}{Theorem}[section]
\newtheorem{claim}[theorem]{Claim}
\newtheorem{lemma}[theorem]{Lemma}
\newtheorem{proposition}[theorem]{Proposition}
\newtheorem{definition}[theorem]{Definition}

\newtheorem{remark}[theorem]{Remark}
\newtheorem{corollary}[theorem]{Corollary}
\newtheorem{assumption}[theorem]{Assumption}

\numberwithin{equation}{section}
\allowdisplaybreaks

\newcommand{\nc}{\normalcolor}

\newcommand{\dif}{\mathrm{d}}
\newcommand{\E}{\mathbb{E}}
\newcommand{\R}{\mathbb{R}}
\newcommand{\C}{\mathbb{C}}

\newcommand{\N}{\mathbf{N}}
\newcommand{\supp}{\mathrm{supp}}
\newcommand{\ii}{\mathrm{i}}
\newcommand{\ee}{\mathrm{e}}
\newcommand{\rd}{\mathrm{d}}
\newcommand{\dd}{\mathrm{d}}
\newcommand{\del}{\partial}
\newcommand{\delab}{\partial}

\newcommand{\vecL}{R}

\DeclareMathOperator{\dist}{dist}
\DeclareMathOperator{\sign}{sign}
\DeclareMathOperator{\spec}{spec}

\newcommand{\other}[1]{\widetilde{#1}}
\newcommand{\Brwn}{\mathfrak{B}}
\newcommand{\Tr}{\mathrm{Tr}}
\newcommand{\G}{\mathcal{G}}

\newcommand{\norm}[1]{\left\lVert#1\right\rVert}
\newcommand{\vertiii}[1]{{\left\vert\kern-0.3ex\left\vert\kern-0.3ex\left\vert #1 
		\right\vert\kern-0.3ex\right\vert\kern-0.3ex\right\vert}}

	\DeclareFontFamily{U}{mathx}{}
	\DeclareFontShape{U}{mathx}{m}{n}{<-> mathx10}{}
	\DeclareSymbolFont{mathx}{U}{mathx}{m}{n}
	\DeclareMathAccent{\widehat}{0}{mathx}{"70}
	\DeclareMathAccent{\widecheck}{0}{mathx}{"71}

\newcommand{\tin}{t_{\mathrm{init}}}
\newcommand{\tfin}{t_{\mathrm{final}}}

\newcommand{\etaf}{\eta_{\mathfrak{f}}}
\newcommand{\kapd}{\varkappa}

\newcommand{\abvD}{\mathbb{D}^\mathrm{abv}}
\newcommand{\globD}{\mathbb{D}^{\mathrm{glob}}}

\newcommand{\bddD}{\mathbb{D}^\mathrm{bdd}}
\newcommand{\regD}{\mathbb{D}^{\mathrm{pert}}}

\newcommand{\smallD}{\mathbb{D}^{\mathrm{small}}}

\newcommand{\smallsymbol}[1]{\scalebox{0.65}{#1}}
\newcommand{\dift}{{\text{\smallsymbol{$\Delta$}} t}}
\newcommand{\indset}[1]{[\! [#1] \!]}

\newcommand{\Mdom}{\mathfrak{M}_\beta}
\newcommand{\tempR}{\widehat{R}}

\newcommand\regreg[1]{%
	{
		\mathop{\kern0pt #1}\limits^{
			\vbox to-1.85ex{
				\kern-2ex 
				\hbox to 0pt{\hss\normalfont\kern.25em \sbullet{}\kern-.2em \sbullet{}\hss}%
				\vss 
			}%
		}%
	}%
}

\newcommand\ringring[1]{%
	{%
		\mathop{\kern0pt #1}\limits^{%
			\vbox to-1.85ex{
				\kern-2ex %
				\hbox to 0pt{\hss\normalfont\kern.25em \r{} \kern-.35em \r{} \hss}%
				\vss 
			}%
		}%
	}%
}

\makeatletter
\newcommand{\sbullet}{%
	\hbox{\fontfamily{lmr}\fontsize{.6\dimexpr(\f@size pt)}{0}\selectfont\textbullet}}
\DeclareRobustCommand{\reg}[1]{\accentset{\sbullet}{#1}}
\makeatother
\newcommand{\arb}{\nu} 

\newcommand{\remove}[1]{{comment}}

\title{Anomalous rate of eigenstate thermalisation at singularities of the density of states}

\date{\today}

\DeclareFieldFormat*{title}{#1}
\AtEveryBibitem{\clearfield{month}}
\AtEveryBibitem{\clearfield{number}}
\AtEveryBibitem{\clearfield{publisher}}
\AtEveryBibitem{\clearfield{url}}
\DeclareFieldFormat*{eprint}{
	\hfil\penalty90\hfilneg\space \ifhyperref{
		\href{https://arxiv.org/abs/#1}{arXiv\addcolon#1} 
	}{
		arXiv\addcolon\nolinkurl{#1}
	} 
}
\DeclareFieldFormat*{doi}{
	\hfil\penalty90\hfilneg\space \ifhyperref{
		\href{https://doi.org/#1}{DOI\addcolon\addnbspace#1}
	}{
		DOI\addcolon\addnbspace\nolinkurl{#1}
	} 
}
\begin{document}
	\begin{minipage}{0.85\textwidth}	\vspace{0.5cm}
	\end{minipage}
	\begin{center}
		\large\bf Anomalous rate of eigenstate thermalisation at singularities of the density of states
	\end{center}
	\vspace{0.75cm}
	
	\renewcommand{\thefootnote}{\fnsymbol{footnote}}

	\noindent
	\mbox{}%
	\hfill%
	\begin{minipage}{0.25\textwidth}
		\centering
		{L\'aszl\'o Erd\H{o}s}\footnotemark[1]~\orcidlink{0000-0001-5366-9603}\\
		\footnotesize{\textit{lerdos@ist.ac.at}}
	\end{minipage}
	\hfill%
	\begin{minipage}{0.25\textwidth}
		\centering
		{Joscha Henheik}\footnotemark[2]~\orcidlink{0000-0003-1106-327X}\\
		\footnotesize{\textit{joscha.henheik@unige.ch}}
	\end{minipage}
	\hfill%
	\begin{minipage}{0.25\textwidth}
		\centering
		{Volodymyr Riabov}\footnotemark[1]~\orcidlink{0009-0007-4989-7524}\\
		\footnotesize{\textit{vriabov@ist.ac.at}}
	\end{minipage}
	\hfill%
	\mbox{}%
	\footnotetext[1]{Institute of Science and Technology Austria, Am Campus 1, 3400 Klosterneuburg, Austria.  Supported by the ERC Advanced Grant "RMTBeyond" No.~101020331.
	}
		\footnotetext[2]{University of Geneva, Rue du Conseil-Général 7-9, 1205 Geneva, Switzerland.  Supported by the ERC Advanced Grant "RMTBeyond" No.~101020331 and the ERC Consolidator Grant "ProbQuant" (jointly with the Swiss State Secretariat for Education, Research
			and Innovation).
	}
	
	\renewcommand*{\thefootnote}{\arabic{footnote}}
	\vspace{0.25cm}
	
	\begin{center}
		\begin{minipage}{0.91\textwidth}\footnotesize{ {\bf Abstract.}} 
			We prove the Eigenstate Thermalisation Hypothesis (ETH), also known as Quantum Unique Ergodicity (QUE), for large $N\times N$ mean-field random matrices with general correlation structure. We identify the microcanonical ensemble and establish the optimal fluctuation scale of eigenvector overlaps around it.
			Our results invalidate the inverse-density scaling predicted by Feingold and Peres \cite{FeinPeres} (and incorporated into Srednicki's ansatz \cite{Srednicki}) in the physics literature of quantum chaos, based upon popular semiclassical theory, and uncover the genuine mechanism which relies on multi-resolvent local laws. 
			Although fluctuations are expected to increase as the density of states vanishes, and indeed scale as  $N^{-1/2}$ in the special cusp regime, rather than $N^{-1}$ in the bulk, we find, unexpectedly, that the same $N^{-1}$ rate persists at regular spectral edges.
			Hence, generically, in the absence of cusps, the entire eigenbasis fluctuates on the same scale as a Haar unitary. 
			This anomaly stems from delicate cancellations in the solution of the underlying matrix Dyson equation, which form the core of our analysis.
		\end{minipage}
	\end{center}
	
	\vspace{3mm}

	{\small
		\footnotesize{\noindent\textit{Date}: \today}\\
		\footnotesize{\noindent\textit{Keywords and phrases}:  Eigenstate Thermalization Hypothesis, cusp singularity, spectral edge, local law, correlated random matrix
	}
	
	\vspace{3mm}
	
	\thispagestyle{headings} 
	\normalsize

	\begingroup
	\setcounter{tocdepth}{1}
	\hypersetup{linkcolor=black}
	\tableofcontents
	\setcounter{tocdepth}{100}
	\endgroup

	\newpage
	\section{Introduction}		
	The long time evolution of sufficiently chaotic classical dynamical systems  tends to thermal equilibrium
	states that can be effectively described by statistical mechanics using only a few parameters given by
	conservation laws. For the simplest {\it ergodic} systems, where energy is the only conserved
	quantity, the thermal state is unique on each energy shell and it is given by the equipartition principle.
	The exact analogue of this phenomenon fails for quantum systems; quantum evolution of an isolated system
	permanently retains some information about the initial state. Nevertheless, 
	some principles of equilibrium statistical mechanics can still be applied to chaotic quantum systems
	if  the classical concept of ergodicity is replaced with 
	the {\it Eigenstate Thermalisation Hypothesis (ETH)}, also known as {\it Quantum Unique Ergodicity}.
	Crudely speaking, ETH asserts that the eigenfunctions ${\bm u}_i$ of a sufficiently 
	chaotic quantum Hamilton operator are uniformly distributed in the phase space
	in the sense that quantum expectations $\langle {\bm u}_i, A{\bm u}_i\rangle$ 
	(also called {\it eigenvector overlaps}) of physical observables $A$
	converge to the microcanonical average of $A$ as the size of the system increases.

	ETH is now considered to be one of the principal signatures of quantum chaos.  
	Another such signature is the presence of universal Wigner–Dyson local eigenvalue statistics, as asserted by the celebrated Bohigas-Giannoni-Schmit (BGS) conjecture \cite{BoGiSch}, motivated by the quantization of the Sinai billiard. 
	Although the original paper \cite{BoGiSch} focused on eigenvalues, the modern physics literature attributes the fundamental connection between ETH and spectral universality to the BGS conjecture, see, e.g., \cite{magan2024two}.
	
	On the mathematical side, the first rigorous results related to ETH were Shnirelman’s quantum ergodicity theorem \cite{Shnirelman}
	and its extensions \cite{ColinDeVerdiere, Zelditch, RudnickSarnak}, concerning the eigenfunctions
	of the Laplace--Beltrami operator on compact manifolds with ergodic geodesic flow and establishing equidistribution for a density-one subsequence of eigenfunctions.

	In the physics literature, the quantitative study of eigenvector overlaps goes back to Feingold and Peres \cite{FeinPeres}.
	Based upon semiclassical path theory, they predicted that, after subtracting the appropriate energy-dependent microcanonical average, the typical size of  $\langle {\bm u}_i, A{\bm u}_i\rangle$ is inversely proportional to the square root of the unnormalized local eigenvalue density at the corresponding energy $\lambda_i$.
	Later, Deutsch \cite{deutsch} proposed \emph{eigenstate thermalisation} as a mechanism for the emergence of statistical mechanics in isolated quantum systems. 
	Subsequently, Srednicki~\cite{Srednicki} formulated a heuristic ansatz for the fluctuations of the overlaps,
	\begin{equation} \label{eq:Srednicki}
		\langle {\bm u}_i, A{\bm u}_j\rangle = \delta_{ij} a(E) + \ee^{-S(E)/2} f(E, \lambda_i - \lambda_j) r_{ij},
	\end{equation}
	where $E := \tfrac{1}{2}(\lambda_i + \lambda_j)$, $S(E)$ is the thermodynamic entropy, $a$ and $f$ are smooth functions of their arguments, depending on the observable $A$, and $r_{ij}$ are random variables with zero mean and unit variance.
	Since $\ee^{S(E)}$ is proportional to the unnormalized density of states, Srednicki’s ansatz incorporates the inverse-density scaling predicted by Feingold~and~Peres.
	The scaling of the variance by the inverse density of states has since become a standard feature of the physics literature on ETH \cite{d2016quantum, wang2025eigenstate}. 
	More recently, the refined ETH ansatz \cite{reimann2021refining, dabelow2022thermalization} and its extensions to higher-order correlations of eigenvector overlaps have been systematically used to understand various thermalisation and dynamical  properties of quantum systems \cite{foini2019eigenstate, pappalardi2022eigenstate,foini2025out, wang2026eigenstate}.
	In these formulations, too, the leading fluctuation scale is proportional to the same inverse-density factor.
	ETH has been extensively investigated numerically \cite{brenes2021out,pappalardi2024microcanonical, mondaini2017eigenstate, noh2021eigenstate}. 
	Albeit these numerical studies explicitly exclude the regions near the spectral edges, the ansatz is expected to hold at the edges of the (many-body) spectrum \cite[Appendix~A]{kim2014testing}.
	
	\bigskip
	 
	Random matrix theory (RMT) is a natural testbed for ETH. In fact, it is currently the only setting in which the fluctuations in the ETH ansatz can be mathematically rigorously analyzed (for more physical models such 
	as non-integrable spin chains only numerical evidence is available, e.g. \cite{brenes2021out,pappalardi2024microcanonical}). 
	For the simplest case, where the eigenvectors ${\bm u}_i$ are the normalized eigenfunctions of a standard $N\times N$ 
	Wigner matrix\footnote{Wigner matrices are defined by having centred i.i.d. entries up to Hermitian symmetry.}
	$H$, i.e. $H{\bm u}_i=\lambda_i{\bm u}_i$, we have~\cite{cipolloni2023eigenstate}
	\begin{equation}\label{fullWignerETH}
		\max_{j,k} \big| \langle {\bm u}_j, A {\bm u}_k\rangle - \langle A \rangle \delta_{jk}\big| \lesssim N^{-1/2+\epsilon}
		\big\langle | A-  \langle A \rangle|^2\big\rangle^{1/2},  
	\end{equation}
	with very high probability, for any fixed $\epsilon>0$,
	uniformly for any  deterministic matrix  $A\in \C^{N\times N}$. Here $\langle A\rangle: = \frac{1}{N}
	\Tr A$ is the normalized trace. This estimate is optimal, both in terms
	of the $N$-power (up to the $N^\epsilon$) and in the Hilbert-Schmidt norm of the traceless part of $A$. In fact,
	$N^{1/2}[ \langle {\bm u}_j, A {\bm u}_k\rangle - \langle A \rangle \delta_{jk}] $ is Gaussian
	for each pair of bulk indices\footnote{Bulk index means
		that the corresponding eigenvalues $\lambda_j, \lambda_k$
		are separated away from the spectral edges of the Wigner semicircle law.} $j,k$
	with a variance proportional to  $\langle | A-  \langle A \rangle|^2\rangle$, see~\cite{A2}
	for diagonal overlaps and~\cite{benigni2024fluctuations} for the general case.
	Prior to~\cite{cipolloni2023eigenstate}, the first proof of ETH for Wigner matrices\footnote{We remark that for Gaussian ensembles,
		i.e. when $H$ is a GOE/GUE matrix, the
		ETH is trivial since the eigenvectors are exactly Haar distributed.} was given in~\cite{ETHpaper} 
	with the optimal $N^{-1/2+\epsilon}$ convergence
	rate but with the cruder operator norm $\|A\|$,
	\begin{equation}\label{AWignerETH}
		\big| \langle {\bm u}_j, A {\bm u}_k\rangle - \langle A \rangle \delta_{jk}\big| \lesssim N^{-1/2+\epsilon}\|A\|.
	\end{equation}
	The optimal dependence on $A$ in~\eqref{fullWignerETH} for bulk indices was first established for projections \cite{BenigniLopatto2103.12013, 2303.11142}, subsequently for general $A$ in~\cite{A2}. We also mention \cite{sugimoto2023eigenstate} for ETH in a translation-invariant Wigner matrix model. 
	
	Beyond Wigner matrices, ETH in the form~\eqref{AWignerETH}
	was proven for a restricted class of generalized Wigner matrices \cite{GenWigETH}
	and certain \emph{deformed  ensembles}, i.e. $H=D+W$, where $D$ is deterministic 
	and $W$ has centred  i.i.d.~entries \cite{iid, equipart}. In the latter case
	the thermal equilibrium  $\langle A\rangle$
	is modified to $\langle A \Im M (\lambda_j)\rangle/\langle \Im M (\lambda_j)\rangle$, 
	where $M$ is the solution of the matrix Dyson equation (MDE), see~\eqref{MDE} below,
	reflecting that the deformation changes the structure of the microcanonical ensemble, in particular it becomes energy 
	dependent.
	Dropping identical distribution while retaining independence, 
	ETH  in the bulk   was established with the optimal rate 
	with very high probability for {\it Wigner-type} matrices
	\cite{WigTypeETH} and even for a large class of \emph{random band matrices} \cite{erdHos2025zigzag}.
	We mention that, as a necessary step towards Wigner-Dyson universality,
	ETH for special observables and in a weaker form\footnote{These results were proven only in expectation instead
		of very high probability and the rate was far from optimal.}
	had previously been proved  for certain Gaussian random band matrices
	in the bulk \cite{yau2025delocalization}  and at the edge  \cite{yang2025delocalizationedge}, as well as in higher dimensions \cite{dubova2025delocalization2, dubova2025delocalization3}
and for other closely related random block models \cite{yang2025delocalizationRBSO, truong2025localization, fan2025localization}. 
	
	\bigskip
	
	In the present paper, we establish ETH with the optimal fluctuation scale for general mean-field random matrices, including models with correlated entries. 
	Our result holds in all spectral regimes
	and it is optimal not only in the bulk, but also at the spectral edges and the possible cusp singularities\footnote{
	In what is known as \emph{shape analysis}, it was proven in~\cite{ajanki2019quadratic, AEK2020} 
		that, under very general conditions on the matrix ensemble,
		the density $\rho$ can have only two types of singularities;  in addition to
		the {\it regular edges} where $\rho$ behaves as a square root, it may exhibit cubic-root {\it cusps}.
		Moreover, there is a one-parameter family of intermediate {\it almost cusp} regimes describing
		the universal shape of the density as  two intervals of the support of $\rho$ merge into one through a cusp.} of
	the (normalized) {\it self-consistent density} $\rho$ of the eigenvalues.
	In this respect, it complements the comprehensive results on local spectral universality\footnote{
		For general 
		correlated matrices the Wigner-Dyson bulk universality
		was proven in~\cite{slowcorr}, the Tracy-Widom edge universality in~\cite{edgelocallaw}
		and finally the Pearcey cusp universality in~\cite{cuspuniv}; for earlier results in more restricted setups,
		see references therein.
		} proven for correlated ensembles.
	We show that the fluctuation scale of the eigenvector overlaps is not a function of the local density alone; instead, it depends on a more intricate stability structure encoded in the solution of the matrix Dyson equation, which we identify fully explicitly.  
	
	Since $N\rho \sim \ee^{S}$ in the notations of Srednicki's ansatz~\eqref{eq:Srednicki}, our result disproves the simple inverse-density prediction of Feingold and Peres in the random-matrix setting.
	In particular, we identify two distinct regimes of vanishing density $\rho$ in which the overlap fluctuations occur on  different scales, neither matching the inverse-density scaling.  
	More precisely, $\langle{\bm u}_i, A{\bm u}_i\rangle$ fluctuates exactly on the same $N^{-1/2}$ scale at the edge as in the bulk,
	while at the cusp\footnote{
	We remark that near the edge $\rho \sim N^{-1/3}$ and near the cusp $\rho \sim N^{-1/4}$.} the typical fluctuation scale is $N^{-1/4}$, except for a  
	codimension-one subspace of exceptional observables.
	We also identify the correct scale in every intermediate regime.

	\bigskip 
	
	Why is this result surprising? 
	As we will explain  later, in the small density regime
	the ETH problem in general has \emph{three} critical 
	modes. These are the unstable eigenvectors of three closely related \emph{stability operators},
	each of which has an isolated small eigenvalue and the criticality of the mode is due to the smallness of its eigenvalue. 
	Typically, the most unstable mode yields the leading deterministic term, the microcanonical ensemble;
	this mode is always critical since the corresponding eigenvalue is small in any spectral regime, including the bulk.
	The other two modes are expected to influence any excessive fluctuation. Since they are
	critical only in the small density regime, they do not affect the fluctuation in the bulk, which therefore
	lives on the standard $N^{-1/2}$ scale as if the eigenbasis were a Haar unitary.  
	However, both critical modes predict an excessive fluctuation much 
	larger than $N^{-1/2}$, whenever $\rho\ll1$, that apparently contradicts our result at the edge, but not  at the cusp.
	
	At this point we encounter two unexpected but closely related {\it anomalies} that eventually resolve the contradiction.
	The first anomaly is that the three critical directions are (essentially) coplanar, i.e. there are effectively only \emph{two}, rather than three, critical modes. The other anomaly is that despite the presence of the second small eigenvalue,
	the fluctuation surprisingly avoids the corresponding second critical mode in the edge regime but not in the cusp regime.
	However, the second mode still \emph{influences} the choice of the microcanonical ensemble; the correct choice
	is a specific  linear combination of the two most critical modes.
	
	Finally, we note that the ubiquity of the  $N^{-1/2}$ fluctuation scale may not sound surprising in light of~\eqref{fullWignerETH}, 
	but the Wigner matrices have many extra symmetries that make the analysis of the bulk and the edge regimes
	basically identical. In particular all three unstable modes coincide; each of them is the identity matrix.
	While  our analysis reveals that the $N^{-1/2}$ fluctuation scale remains universal  for general 
	ensembles in the bulk and at the edge, the mechanism  is  very different from that for the Wigner ensembles. The subtlety is indicated by  a fluctuation  of order $N^{-1/4}$ at the cusp 
	which agrees with neither the Feingold–Peres prediction nor the universal $N^{-1/2}$ fluctuation in the bulk and at the edge.
	Moreover, we also conclude that in the cusp regime the components of the 
	eigenvectors have non-negligible correlations that are inconsistent with the general belief that
	eigenvectors, possibly with some natural weight factor, are 
	asymptotically Haar distributed.

	\subsection{Main results, informally} \label{intro1}
	To explain our main results, we first recall that the resolvent $G(z):=(H-z)^{-1}$, $z\in \C\setminus\R$,
	of an $N\times N$ Hermitian mean-field random matrix  $H$ tends to become deterministic in the large $N$
	limit. Its deterministic approximation, denoted by $M(z)\in \C^{N\times N}$, is the unique solution to the 
	Dyson equation
	\begin{equation}\label{MDE}
		-M(z)^{-1} =  z-D + \mathcal{S}\bigl[M(z)\bigr], \qquad (\Im z) \bigl(\Im M(z)\bigr) >0,
	\end{equation}
	where $D:= \E H$ and  $\mathcal{S}$ is the \emph{self-energy operator} (see~\eqref{eq:S_def} later)
	encompassing all second order correlations
	of the  matrix elements of $H$. The pair $(D, \mathcal{S})$ constitutes the basic data of our random
	matrix ensemble, in particular they determine the self-consistent density of eigenvalues as 
	$\rho(E): = \frac{1}{\pi}\langle \Im M(E+\ii 0)\rangle$. 
	{\it Mean field models} are characterized by the property that $\mathcal{S}$ is comparable
	with the normalized trace, up to constant factors;
	$\mathcal{S}[R] \sim \langle R \rangle$ for any positive matrix $R\ge 0$.
	For example,  for deformed Wigner matrices $\mathcal{S}[\cdot] = \langle\cdot \rangle$.
	The {\it (single-resolvent) local law} is a 
	concentration result that   asserts that $G(z)$ is close to $M(z)$ with very high probability
	if tested against a deterministic matrix $A$ as long as $|\Im z|\gg N^{-1}$:
	\begin{equation}\label{1GLL}
		\big\langle (G(z)-M(z))A \rangle \lesssim (N |\Im z|)^{-1}\|A\|_{\mathrm{hs}}, \qquad  
		\|A\|_{\mathrm{hs}}:=\langle AA^*\rangle^{1/2}.
	\end{equation}
	Single-resolvent local laws provide a key technical input for many fundamental results in RMT, such as eigenvalue {\it rigidity}, eigenfunction {\it delocalisation}
	and \emph{universality of local spectral statistics} as demonstrated in~\cite{cuspuniv}
	for the most general correlated model, preceded by numerous earlier results in 
	more restrictive settings. 
	
	It is well known since~\cite{ETHpaper} that ETH requires a more sophisticated local law
	than~\eqref{1GLL}. First, we need to understand an object of the form
	$\langle \Im G(z) A \Im G(z) A^*\rangle$, hence a 
	{\it two-resolvent} local law  is needed with the complication that the corresponding deterministic
	approximation is {\it not} the naively guessed $\langle \Im M(z) A \Im M(z) A^*\rangle$.
	Second, we need to exploit that for a certain codimension-one subspace of observables (called {\it regular}),
	the fluctuations of $\langle G(z)A\rangle$ and $\langle G(z_1)AG(z_2)A\rangle$ are considerably smaller
	than for general $A$’s. For Wigner matrices, regular observables are exactly the  traceless ones, $\langle A\rangle=0$,
	but for more general models the concept of regularity depends on the spectral parameters of the neighboring 
	resolvents $G$. For example, when $z=z_1=\bar z_2$, 
	regularity is given by the condition $\langle A \Im M(z)\rangle=0$,
	where $ \Im M(z)$ arises as the (approximate) unstable direction of the \emph{stability operator}
	$\mathcal{B}(z):=1- M(z)\mathcal{S}[\cdot] M(\bar z)$.
	It is useful to introduce the notation $\mathring{A} : = A- \langle A \Im M (z)\rangle/\langle \Im M (z)\rangle$
	for the ($z$-dependent) regularization of any $A$ and a similar concept applies for general $z_1, z_2$.

	The following estimate, based upon the spectral theorem, is
	the basic step to connecting ETH with a two-resolvent chain:
	\begin{equation}
		\Big|\langle {\bm u}_j, A {\bm u}_k\rangle -\delta_{jk}  \frac{\langle A \Im M(z_1)\rangle}{\langle \Im M (z_1)\rangle}
		\Big|^2  = \big|\langle {\bm u}_j, \mathring{A}  {\bm u}_k\rangle \big|^2 \lesssim
		N\eta_1\eta_2 \langle \Im G(z_1) \mathring{A} \Im G(z_2) \mathring{A}^*\rangle.
		\label{main}\end{equation}
	Here  the spectral parameters $z_1=\gamma_j+\ii \eta_1$, $z_2=\gamma_k+\ii \eta_2$ 
	are given by the deterministic approximations
	of the eigenvalues  $\lambda_j, \lambda_k$, i.e. by the quantiles $\gamma_j, \gamma_k$
	of the density $\rho$,  with  small imaginary parts $\eta_1, \eta_2$ comparable with the local eigenvalue spacing.
	We will prove that $\Im G(z_1) \mathring{A} \Im G(z_2)$
	concentrates around its deterministic approximation, denoted by $ \widehat{M}(z_1, \mathring{A}, z_2)$
	and computed explicitly from the solution $M$ of~\eqref{MDE}, 
	and use this approximation to estimate the right-hand side of~\eqref{main}. We have the following sharp upper bound,  
	\begin{equation}\label{Var}
		N\eta_1\eta_2 \langle \widehat{M}(z_1, \mathring{A}, z_2) \mathring{A}^*\rangle 
		\lesssim \frac{1}{N}  \Big[ \frac{| \langle {R}_{jk}  \mathring{A}
			\rangle|^2}{|\beta_{jk}|}+ \|  \mathring{A}\|_{\mathrm{hs}}^2\Big],
	\end{equation}
	which completes the estimate in~\eqref{main}.  Here
	$\beta_{jk} $ is the smallest eigenvalue of the stability operator $\mathcal{B}_{jk}:=1- M(z_1)\mathcal{S}[\cdot] M(z_2)$
	and the matrix ${R}_{jk}$, defined explicitly later with $\| {R}_{jk}\|_{\mathrm{hs}}\lesssim 1$,
	expresses the \emph{discrepancy} between the unstable direction of 
	$1- M(z_1)\mathcal{S}[\cdot] M(z_2)$ and that of $1- M(z_1)\mathcal{S}[\cdot] M(\bar z_2)$.
	We note that in all previously studied situations  
	the first term  in the r.h.s. of~\eqref{Var} was smaller or at most comparable with 
	$\|  \mathring{A}\|_{\mathrm{hs}}^2$, giving rise to an order $\|  \mathring{A}\|_{\mathrm{hs}}^2/N$ 
	estimate on the variance of the overlaps.
	This is because either only  the bulk regime was considered, where $\beta_{jk}\sim 1$, or Wigner matrices 
	were studied, where ${R}_{jk}=0$ since the  unstable directions of the two stability operators coincide.
	However, for generic models we have $\| {R}_{jk}\|_{\mathrm{hs}}\sim 1$ and $\beta_{jk}$ becomes small
	in the regime where the density is small.
	
	To be more specific,  
	for simplicity, we now assume  that $j=k$, i.e. we consider only the diagonal overlaps. 
	Then $\beta\sim \rho\sim N^{-1/3}$ at a regular edge and $\beta\sim\rho^2 \sim N^{-1/2}$
	at the exact cusp. So, given the optimality of~\eqref{Var}, the Feingold-Peres prediction of order $1/(N\rho)$ 
	for the variance of $\langle {\bm u}_j, A {\bm u}_j\rangle$ seems to be correct at least for generic ensembles,
	even though their paper was much more heuristic and it did not use  stability operators.
	While their prediction is wrong at the cusp
	and at the edges of  the Wigner ensemble,  from a physics perspective, they 
	may be viewed as two different types of non-generic cases. Cusps in the self-consistent density profile exist
	only for a codimension-one  manifold in the space of data $(D, \mathcal{S})$.
	Wigner matrices are even more special, although their key property
	for this analysis, ${R}_{jk}=0$, is also a codimension-one  condition.

	So far, the picture suggested by~\eqref{Var} is consistent with earlier ETH proofs in more special cases.
	The main instability in the problem, the very small eigenvalue of the 
	stability operator $1- M(z)\mathcal{S}[\cdot] M(\bar z)$ is neutralized by  
	the proper choice of the regularization, $A\to \mathring{A}$, since  $\Im M$  is the corresponding 
	eigenvector\footnote{Up to 
		an irrelevant error of order $\eta$.}.
	This shift identifies 
	the expected value of the diagonal overlap. The instability
	of the other operator $1- M(z)\mathcal{S}[\cdot] M(z)$ is less severe, 
	but it is still present in the regime where the density is small, so naively it is natural 
	to expect that eventually its smallest eigenvalue, $\beta_{jk}$, governs the fluctuation
	of $\langle {\bm u}_j, A {\bm u}_k\rangle$.

	The surprising novelty of our work is  to observe and 
	prove that this natural heuristic is wrong at the regular edges:
	the true variance of  $\langle {\bm u}_j, A {\bm u}_j\rangle$ is of order $1/N$, the same as in the bulk,
	uniformly throughout the spectrum away from the possible cusps. In particular, the Feingold-Peres
	prediction is \emph{wrong in all regimes}\footnote{Note that~\cite{cipolloni2023eigenstate}  already revealed that Feingold-Peres is wrong at the edges of  Wigner matrices, but its mechanism 
		relied on the non-generic property that $M(z)$ is constant times the identity matrix.} where the density is small, $\rho\ll1$.
	While both estimates \eqref{main}--\eqref{Var} are essentially optimal, they miss the key point: the best deterministic approximation to $\langle {\bm u}_j, A {\bm u}_j\rangle$, the microcanonical ensemble,
	is not given solely by $\Im M(z)$:  it has a nontrivial correction term. 
	This correction  is small but essential since 
	the size of~\eqref{main},  correctly predicted by~\eqref{Var}, comes from the slightly wrong 
	choice of the microcanonical ensemble.
	
	To find the correction in the regime where $\rho\ll1$, 
	we solve the quadratic variational problem
	\begin{equation}\label{ymin}
		\min_{y\in \C} \langle \widehat{M}(z, A-yI , z) (A-yI)^*\rangle 
	\end{equation}
	which serves as the deterministic approximation for the right-hand side of~\eqref{main}
	with an adjusted shift in the definition of $\mathring{A}$.
	We find that the unique minimum attained at $y= y_{\rm min}$ 
	is \emph{much smaller} than the value   with the previous choice 
	$y=  \langle A \Im M (z)\rangle/\langle \Im M (z)\rangle$. 
	We give an explicit formula for $y_{\rm min}$ that identifies the correct
	microcanonical ensemble, in fact it turns out that $y_{\rm min}$ is essentially 
	$\langle A \Im M (\Re z)\rangle/\langle \Im M (\Re z)\rangle$, i.e. 
	the same $\Im M$ as before but evaluated \emph{exactly} on the real axis rather than at distance $\eta$ from it. 
	   However, setting $\eta_1=\eta_2=0$ directly
	in~\eqref{main} is not allowed since $\langle \Im G(z_1) \mathring{A} \Im G(z_2) \mathring{A}^*\rangle$
	would blow up.

	After replacing $\mathring{A}$ by the improved regularization $\reg{A} := A- y_{\rm min} I$, 
	 $$\ring{A} \mapsto \reg{A}$$
	at the end of a very involved analysis,  we 
	obtain that the large $|\beta|^{-1}$ factor on right-hand side of~\eqref{Var} improves to 
	$\alpha^{-2}:=(\Delta^{2/3}+|\beta|)^{-1}$ in the vicinity of a boundary point of $\mbox{supp}(\rho)$
	where $\Delta$ is 
	the size of the nearest gap.
	In particular, near the \emph{regular edges}, we have $\Delta\sim 1$ and thus the large factor $|\beta|^{-1}$
	in~\eqref{Var} is changed to $1/\alpha^2\sim 1$, while at the exact  cusp ($\Delta=0$, $\rho=0$),
	and in the \emph{almost-cusp} regime with a non-vanishing minimum ($\Delta=0$, $0\ne \rho\ll 1$)
	the previous estimate~\eqref{Var} with $|\beta|^{-1}$  is optimal.
	Of course this large  factor is effective only if $\reg{A}$ is not specifically orthogonal to the discrepancy direction ${R}$;
	in a subspace of codimension one in the space of all observables, the variance remains of order $1/N$,
	also at the cusp.
	
	In summary, we prove that 
	\begin{equation}\label{newETH}
		\Big| \langle {\bm u}_j, A {\bm u}_j\rangle - \frac{\langle A \Im M(\gamma_j)\rangle}{\langle \Im M(\gamma_j)\rangle}
		\Big|^2 \lesssim \frac{1}{N}  \Big[ \frac{| \langle {R}_{jj}  \reg{A}
			\rangle|^2}{\alpha_{jj}^2}+ \|  \reg{A}\|_{\mathrm{hs}}^2\Big],
	\end{equation}
	i.e. the leading (deterministic)
	behavior of the diagonal overlap $\langle {\bm u}_j, A {\bm u}_j\rangle$, the microcanonical state, 
	is given by  $\Im M(\gamma_j)$,  while 
	the leading fluctuation is proportional to $\langle  {R}_{jj}  \reg{A}\rangle$ where $\reg{A}$ is
	the improved regularization
	of $A$.  Hence, 
	for the purpose of describing the overlap 
	for any observable $A$,  we have a \emph{three-term} decomposition of $A$: the components of $A$
	in the direction $\Im M(\gamma_j)$ and $  {R}_{jj}^*$ play a special role; for observables
	orthogonal to these two directions, $\langle {\bm u}_j, A {\bm u}_j\rangle$ is of order $N^{-1/2}$.
	Although we focus here on diagonal overlaps, $j=k$, for simplicity, we prove very similar results for any overlaps.

	Focusing only on the exact edge and cusp regimes for simplicity of the presentation,
	our results   
	can  be  compactly summarized in  Table~\ref{table}. 
	
		\begin{table}[htb] 
		\centering
		\begin{tabular}{l | c c c}
			$\mathrm{Var}[\langle {\bm u}_j, A{\bm u}_j\rangle]$ &   Bulk & Edge & Cusp   \\ \hline 
			Wigner ensemble &  $\frac{1}{N}$ &  $\frac{1}{N}$  & n.a. \\
			Feingold-Peres \cite{FeinPeres} prediction:  $(N\rho)^{-1}$ &  $\frac{1}{N}$ &  $\frac{1}{N^{2/3}}$  &  $\frac{1}{N^{3/4}}$ \\ 
			General ensemble, estimate from \eqref{main}--\eqref{Var}: $(N|\beta|)^{-1}$ &  $\frac{1}{N}$ &  $\frac{1}{N^{2/3}}$  &  $\frac{1}{N^{1/2}}$ \\ 
			General ensemble, our optimal result:  $(N\alpha^2)^{-1}$ 
			&  $\frac{1}{N}$ &  $\frac{1}{N}$  &  $\frac{1}{N^{1/2}}$ \\ 
		\end{tabular}
		\caption{Summary of fluctuation sizes for diagonal overlaps in different spectral regimes. For Wigner matrices, no cusp is present. Feingold and Peres \cite{FeinPeres} predicted a $1/(N \rho)$ behavior. The bounds in \eqref{main}--\eqref{Var} suggest the estimate by $1/(N |\beta|)$ for general ensembles. However, as we show, the optimal bound is $1/(N \alpha^2)$ with $\alpha^2 = \Delta^{2/3}+|\beta|$.}
		\label{table}
	\end{table}

	\medskip
	
	The key mechanism for the improvement of the $1/|\beta|$ factor in~\eqref{Var} to $1/\alpha^2$
	after the correct $y_{\rm min}$ shift
	is that,  near the edge, both stability
	operators, $1- M(z)\mathcal{S}[\cdot] M(\bar z)$ and $1- M(z)\mathcal{S}[\cdot] M(z)$
	have a small eigenvalue (in fact, their sizes are comparable exactly at the edge) and  the
	corresponding eigenvectors are close but not exactly the same in general. 
	It is therefore natural that   the proper 
	concept of regularisation of $A$ contains  not only $\Im M(z)$, but a carefully chosen linear combination of 
	\emph{both} unstable eigenvectors which amounts to a small but essential correction to $\Im M(z)$.
	It is not too surprising that the corrected direction is essentially $\Im M(\Re z)$. But it is very 
	surprising that this small correction substantially reduces the value of the minimum in~\eqref{ymin},
	effectively neutralizing the instability effect of the small $\beta$.
	This subtle mechanism was not visible in earlier works 
	because the correction term is negligible 
	both in the bulk regime and for Wigner matrices for two different reasons as mentioned above.

	We admit  that we do not have an independent heuristics for the surprising $1/N$ variance at the regular edge.
	The mechanism outlined above and carefully 
	developed in the current paper gives a systematic way to compute the true fluctuation via 
	solving the minimization problem~\eqref{ymin},
	but it does not hint \emph{why}  the true minimum of~\eqref{ymin} produces a factor $1/\alpha^2$ that
	is  \emph{much smaller} than
	$1/|\beta|$ in~\eqref{Var}. It is the outcome of a long and very delicate analysis  of these two stability operators that exploits
	several hidden cancellation effects. It seems there is no short way even to guess the answer without
	going through this analysis which forms the core of our work.

	\subsection{Some consequences and an example}
	The ubiquitous $1/N$ variance throughout the entire spectrum for generic mean-field models indicates a
	certain type of \emph{universality}:
	the entire eigenbasis $\{ {\bm u}_i\}_{i=1}^N$ asymptotically behaves as the columns of a Haar unitary matrix
	as far as the \emph{fluctuations} of overlaps $\langle {\bm u}_j, A {\bm u}_j\rangle$ are concerned
	(the deterministic part, given by $\Im M$ to leading order, is naturally model dependent).
	In this paper we identify only their correct fluctuation scale, but they are expected
	to follow a Gaussian distribution as proved for Wigner matrices using eigenvector moment flow \cite{bourgade2017eigenvector}, see also the very recent  direct analysis of the eigenvector Dyson Brownian motion \cite{benigni2026quantitative}
	and references therein.

	For illustration, consider a
	\emph{deformed Wigner matrix}, $H_0=D + W \in \C^{N \times N}$ with $N$ even,
	where $W$ is a Wigner matrix  and $D$ is a diagonal deterministic matrix with $N/2$ ones in the upper part and 
	$N/2$ minus ones  in the lower part of the diagonal, $D=\mathrm{diag}(1,\ldots 1, -1, \ldots, -1)$. 
	We choose this example because it contains all spectral regimes, in particular it
	features a cusp at the origin to highlight its special properties.
	For simplicity, take $A$ to be the orthogonal projector onto the first $N/2$ coordinates, $A=\mathrm{diag}(1,\ldots 1, 0, \ldots, 0)$, i.e.
	\begin{equation}\label{ex}
		\langle  {\bm u}_i, A  {\bm u}_i\rangle = \sum_{a\le N/2} |{\bm u}_i(a)|^2.
	\end{equation}
	
	\begin{figure}[h]
		\centering
		\includegraphics[width=.795\textwidth]{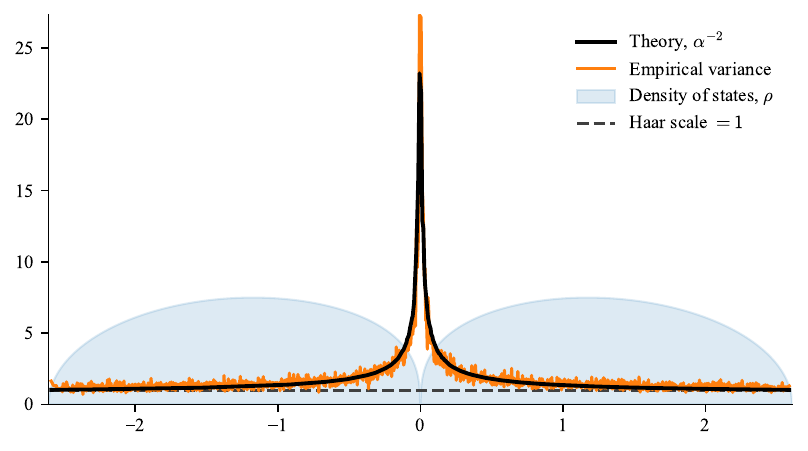}
		\caption{
			Comparison between the empirical variance of the normalized diagonal eigenvector overlaps
			\(\sqrt{N}\langle \bm u_j, A \bm u_j\rangle\) (orange), plotted against the corresponding quantiles
			\(\gamma_j\) of the self-consistent density \(\rho\), and the theoretical prediction \(\alpha^{-2}\) (black).
			The data is obtained from \(100\) realizations of a GUE of size \(N=2000\) with an additive deformation  $D$. The observable $A$ and the deformation $D$ are chosen as in the discussion above.
			Superimposed is the blue shaded region indicating the density of states~\(\rho(E)\). The dotted grey line shows the standard ETH variance scale of order $N^{-1}$.
		}
		\label{fig:overlaps}
	\end{figure}

	Figure~\ref{fig:overlaps} shows the empirical variance of  the overlaps  
	$\langle  {\bm u}_i, A  {\bm u}_i\rangle$ across the spectrum. 
	A simple calculation shows that  $\langle A\Im M(E)\rangle / \langle \Im M(E)\rangle$ 
	is larger for positive energies, $E>0$,
	than for negative energies, by a factor of order unity, but the fluctuation remains the same
	throughout the bulk and the edges, while it becomes much bigger near the cusp.
	Note that all eigenvectors, including the ones near the cusp,
	are fully delocalized~\cite[Corollary 2.10]{cuspuniv}, i.e. 
	the typical size of   $|{\bm u}_i(a)|^2$ for each entry of ${\bm u}_i$ is of order $N^{-1}$. 
	The variance of their “half-sum”~\eqref{ex} being of order $N^{-1}$ in the bulk and at the edge is compatible with the expected approximate independence of the different entries of ${\bm u}_i$, as if ${\bm u}_i$ were Haar distributed,
	but the large $N^{-1/2}$ variance for cusp eigenvectors must come from
	strong correlations among different entries of ${\bm u}_i$. Therefore the cusp eigenvectors 
	have very different internal structure, in particular they are far from being Haar distributed.

	We note that in the recent physics literature by Pappalardi~et~al.~(see, e.g., \cite{pappalardi2022eigenstate, vallini2025refinements} and references therein)
	a very strong form of ETH was assumed as a starting ansatz for detailed analysis of chaotic quantum systems.
	Albeit the presence of cusps is non-generic, its anomalous ETH behavior indicates that
	the ansatz used in these works may have limitations.
	
	\subsection{Main ideas and novelties in the proofs}
	
	Given the relation~\eqref{main}, we need an optimal two-resolvent local law for 
	$ \langle \Im G(z_1) \reg{A} \Im G(z_2) \reg{A}^*\rangle$ with the redefined 
	concept of regularization $\reg{A} $ of the observable $A$. Similarly to the single $G$ case,
	local laws  are concentration results for (normalized) traces of products 
	$G_{[1,k]} : = G_1 A_1 G_2 A_2 \ldots A_{k-1}G_k$
	of resolvents $G_i =G(z_i)$
	and deterministic observables $A_i$ in between. Schematically, they take the form
	\begin{equation}\label{LL}
		\Big| \langle \big( G_{[1,k]}   - M_{[1,k]}\big) A_k \rangle\Big|\lesssim  
		\frac{ | \langle M_{[1,k]} A_k \rangle |}{\sqrt{N\ell}},
		\qquad \ell : = \min_i |\rho(z_i)\Im z_i|,
	\end{equation}
	with very high probability up to a tolerance factor $N^\epsilon$.
	Here $M_{[1,k]}$ is the  deterministic approximation for $G_{[1,k]}$ that can be computed explicitly
	via a recursion on $k$ from the solution of the MDE~\eqref{MDE}. A similar recursion was presented in
	\cite[Definition 4.1]{iid} and especially in \cite[Sec. 11]{erdHos2025zigzag}, which can be directly generalized to
	the current setup.  For example, for $k=2$ we have
	\begin{equation}\label{M12}
		M_{[1,2]} = \mathcal{B}_{12}^{-1}[ M_1 A M_2], \qquad \mathcal{B}_{12}:=1- M_1\mathcal{S}[\cdot] M_2, \qquad M_i=M(z_i).
	\end{equation}
	Note that as long as $|\Im z_i|$ is much bigger than the local eigenvalue spacing, hence $N\ell\gg 1$, the error
	term is typically negligible\footnote{For atypical
		observables $A_k$ that are (almost) orthogonal to $M_{[1,k]}^*$ 
		the leading term $\langle M_{[1,k]} A_k \rangle$ is unusually small or may even vanish.
		The fluctuation of $G_{[1,k]}A_k$ does not follow this special pattern, 
		so  $\langle M_{[1,k]} A_k \rangle$ in the right-hand side of~\eqref{LL} is not always
		the correct way to measure the fluctuation. The correct choice is a Hilbert-Schmidt type norm,
		namely  the square root of the deterministic $M$-term
		of a symmetrized chain of length $2k$ that comes from a certain quadratic variation  of $G_{[1,k]}A_k$.
	}
	compared with the leading term $\langle M_{[1,k]} A_k \rangle$. Thus the structure of the local 
	laws is primarily governed by the corresponding $M$-terms. In this paper only $k=1, 2$ chains are considered
	(some longer chains arising along the proof are reduced by using Ward identities and various trace inequalities),
	but  chains of arbitrary length $k$ can also be  accurately analysed, see~\cite{multiG} for Wigner matrices and~\cite{erdHos2025zigzag} for
	band matrices. In fact, \cite{multiG} introduced, for the first time, the concept of
	an entire \emph{hierarchy} of local laws that needs to be controlled simultaneously, combined 
	with a truncation procedure, analogously to the BBGKY hierarchy in 
	many-body physics.
	Our main local law result (Theorem~\ref{thm:locallawreg}) is for the special two-resolvent chain 
	$ \langle \Im G(z_1) A_1 \Im G(z_2) A_2\rangle$ with regular observables $\reg{A}_i=A_i$.
	However, along the way, we also prove two new optimal single resolvent local laws for $\langle (G-M)A\rangle$
	with regular $A$ as well as for  $\Im G - \Im M$. These additional results 
	 naturally complement our previous single-resolvent local law
	for $G-M$ without regularity in~\cite{cuspuniv}.
	
	We use the recent  
	\emph{zigzag} strategy that is currently the most effective method  to prove any local law: 
	It consists of an iteration of two complementary stochastic flows in the space of random matrices; starting
	from a global law that is proven separately,  the \emph{zig step}
	adds a small Gaussian component  and simultaneously moves the spectral parameter $z$
	closer to the real axis along a specific ODE, the \emph{characteristic flow}. In turn, the \emph{zag step}
	removes this Gaussian component while leaving the spectral parameter intact. This procedure is repeated
	finitely many times until $\Im z$ reaches the smallest possible scale (local eigenvalue spacing).\footnote{The first version of the characteristic flow appeared in a paper by Pastur~\cite{pastur},
		who related   the complex Burgers equation to the evolution of the resolvent along the
		Ornstein-Uhlenbeck flow. The idea was later revived by Lee and Schnelli in~\cite{lee2015edge} and then by
		von Soosten and Warzel in~\cite{Soosten}.
		The full power of the characteristic flow method has gradually been realized in the context of Dyson Brownian motion
		in  \cite{huang2019rigidity, adhikari2020dyson, adhikari2023local, aggarwal2024edge},
		and later for matrix models, e.g., in  \cite{bourgade2021extreme,
			landon2022almost, landon2024single}. Its  combination with a GFT argument---that is, the full zigzag strategy---first appeared in \cite{mesoCLT}. 
		The full version, systematically controlling the fluctuations 
		of resolvent chains of \emph{arbitrary} length, was introduced in~\cite{cipolloni2023eigenstate, OTOC}, while the latter also introduced the term \emph{zigzag}.
		Since then, the full power of the zigzag method was successfully exploited in several models and regimes that were previously inaccessible 
		due to instabilities~\cite{cipolloni2023universality, cipolloni2024eigenvector, WigTypeETH, cuspuniv}.}
	Apart from the global law, the other key input of the zigzag strategy is a good bound (called $M$-bounds) on the 
	deterministic approximations on various resolvent chains.	The improved $M$-bounds with the new concept of regularization are the fundamental novelties of the paper.
	
	The most important $M$-term is $\langle \widehat{M}(z_1, \reg{A}, z_2) \reg{A}\rangle$,
	corresponding to $\langle \Im G(z_1) \reg{A} \Im G(z_2) \reg{A}\rangle$, and our result 
	(Theorem~\ref{prop:M2_bounds})  shows that
	\begin{equation}\label{Mhat}
		\langle \widehat{M}(z_1, A_1, z_2) A_2\rangle \lesssim |\rho_1\rho_2|  \vertiii{A_1}_{z_1, z_2} \vertiii{A_2}_{z_1, z_2},
		\qquad \rho_i := \rho(z_i),
	\end{equation} 
	holds for regular observables,
	where the norm is defined by
	\begin{equation}\label{triple}
		\vertiii{A}_{z_1, z_2} : = \frac{ |\langle R A \rangle|}{\alpha} + \norm{A}_{\mathrm{hs}}.
	\end{equation}
	Here $\alpha =\alpha(z_1, z_2)$ is the improved stability factor and $R=R(z_1, z_2)$ 
	is the special direction that comes from the proper concept of regularity of $A$, as explained in 
	Section~\ref{intro1} in the special $z_1=z_2$ case needed for diagonal overlaps.
	
	The estimate~\eqref{Mhat} is very subtle as it simultaneously encompasses three different effects, all of which are necessary to obtain our optimal ETH bound:
	\begin{itemize}
		\item[(i)] The factor $\rho_1\rho_2$ to account for the presence of  the \emph{imaginary parts} $\Im G_1$, $\Im G_2$
		of the resolvents instead of $G_1, G_2$,
		\item[(ii)] The new concept of regularization that results in the improved factor $1/\alpha^2$ instead of $1/|\beta|$
		in~\eqref{Var},
		\item[(iii)] The identification of $R$ that determines the special direction for $A$ where the $1/N$ fluctuation holds  
		even at the cusp.
	\end{itemize}
	In previous works, simultaneous improvements from the presence of $\Im G$ and regularity (items (i) and (ii))
	were obtained only in the much simpler setting of standard Wigner matrices~\cite{cipolloni2023eigenstate},
	where the critical eigenvectors of the two  stability operators coincide, hence automatically $R=0$,
	moreover all formulas for the deterministic $M_{[1,k]}$-quantities are scalar and explicit. 
	Neither the potentially dangerous  effect of $\beta$ and its improvement to $\alpha^2$, nor the delicate choice of $R$ 
	was present.  
	
	Extracting  the correct
	$\rho_1\rho_2$ factor  (item (i)) requires a delicate four-fold cancellation among the four $M_{[1,2]}$ terms
	from~\eqref{M12}, arising as the $M$-terms after writing out both $\Im G(z) = (2\ii)^{-1}( G(z)-G(\bar z))$. Note that none
	of these $M_{[1,2]}$ terms is small, and only their linear combination produces the small quadratic factor in $\rho$.
	While the \emph{shape analysis} of the single $M$ from~\cite{AEK2020} exhibited the effect that $\Im M \sim \rho$
	in the perturbative $\rho\ll 1$ regime, a naive extension of this perturbation result to $M_{[1,2]}$
	yields only a linear $\rho$ factor and not the optimal quadratic one.

	Finding the correct regularization (item (ii)) is even more delicate. Just focusing on the
	diagonal case, $z_1=z_2$, we explained in Section~\ref{intro1}
	that the singular directions of the two stability operators $\mathcal{B}_{z,\bar z} = 1- M\mathcal{S}[\cdot] M^*$ 
	and $\mathcal{B}_{z, z} = 1- M\mathcal{S}[\cdot] M$ play a critical role. The first observation is
	that there is a third stability operator, namely $\mathcal{B}_{\bar z,\bar z}$, whose singular direction may 
	also affect the argument. We find, however, that these three singular directions are essentially coplanar
	(up to a tolerable $O(\rho^2)$ error). This is based upon a subtle  algebraic identity which 
	can be viewed as the  generalization of the  explicit formula
	$$
	M(\bar z)^2+ M(z)^2 = 2|M(z)|^2 + O(\rho^2)
	$$
	in the deformed Wigner case,  $\mathcal{S}=\langle \cdot\rangle$, where
	$M(z_1) M(z_2)$ is the singular  direction of $\mathcal{B}_{z_1, z_2}$ with very high precision.
	Such explicit expression for the singular directions is not available in the general case
	and we do not see any intuitive or conceptual reason why these three directions remain 
	coplanar in general, but they do, as our analysis shows.
	
	The second observation concerning item (ii) is that choosing the regularization very carefully 
	allows us to substantially improve on the effect of the $1/|\beta|$-instability. This is a delicate
	choice \emph{within} the two-dimensional subspace spanned by the two singular directions
	and requires another  singular perturbation argument since these two directions are collinear up to order $O(\rho)$.
	As explained in Section~\ref{intro1}, the outcome is unexpected and still lacks intuitive explanation.
	
	Finally, item (iii) is the byproduct of the correct definition of regularization, but the dependence of $R$ on 
	the spectral parameters is also delicate: it turns out to behave more regularly than naive perturbation theory
	suggests and this is essential to identify the correct microcanonical ensemble as $\Im M(\Re z)\approx \Im M(\gamma_j)$ in 
	our  ETH result~\eqref{newETH}.

	To extract all these three effects and thus prove~\eqref{Mhat} we use a \emph{dynamical} approach.
	The main observation is that  the gist of  the zig-part
	of the zigzag strategy, the cancellation effect along the characteristic flow, is very effective 
	at the level of the deterministic $M_{[1,k]}$-quantities as well. For spectral parameters far
	away from $\mbox{supp}(\rho)$ the main singularities are harmless (all $\mathcal{B}$ operators
	have bounded inverse) and even item (i), extracting the $\rho_1\rho_2$ effect is much easier.
	In particular, following the evolution of $\widehat{M}$ along the characteristic flow as $z$ approaches the real axis
	very accurately unfolds its singular behavior  without overestimating it.
	We do not know how to get this level of precision via a static argument, i.e., to prove~\eqref{Mhat}
	directly without embedding it into a flow.  We note that this powerful
	 dynamical proof of $M$-bounds, even for long chains,
	first appeared in the context of random band matrices~\cite{erdHos2025zigzag}, but the setup there was more concrete
	and the analysis was in the bulk, neither edges nor cusps were present.
	To handle these delicate singular regimes, we need a very precise shape analysis for the $M$-terms
	connected with \emph{two} spectral variables or \emph{two} resolvents
	 like $M_{[1,2]}$, $\widehat{M}$ and $\mathcal{B}_{z_1, z_2}$. 
	Shape analysis was previously developed in~\cite{ajanki2019quadratic, AEK2020}  only for $M(z)$ corresponding to a
	\emph{single} resolvent.

	Finally, we mention an important idea concerning the zigzag strategy that we heavily use
	to streamline the proofs.
	Traditionally, both static and dynamic approaches
	for proving local laws relied on a fairly involved  hierarchy of inequalities or differential equations
	involving averaged and isotropic versions of resolvent chains of different length (see, e.g, \cite{cipolloni2024eigenvector, nonHermdecay}). 
	In the context of small randomness, the very recent work \cite{smallrand} introduced
	the idea that a specific two-resolvent chain, namely $\langle \Im G A\Im G A^*\rangle$ with regular $A$,
	should be the  central quantity in the analysis. Longer resolvent chains can be reduced to shorter ones
	by Schwarz  or other trace inequalities, $\Im G$’s naturally reappear through the Ward identity $GG^*= \Im G/\Im z$.
	This greatly simplifies the analysis since essentially only one quantity has to be monitored along the flow
	and there is no hierarchy or longer chains $G_{[1,k]}$ with $k\ge 3$. However, in general, this simplification also has some trade-offs. First, it does not give optimal error estimates 
	in local laws for longer chains. Second, it is optimized for chains with $\Im G$’s and it does not
	directly give bounds for general chains that contain some $G$’s as well. It is possible to
	use integral representations to express $G$’s in terms of $\Im G$ (see  Section~\ref{sec:other}
	and, e.g., \cite[Lemma 4.6]{cipolloni2023eigenstate} and \cite[Lemma 4.11]{WigTypeETH}), 
	but they are genuinely non-local in the spectral parameters, hence
	the concept of regularity must also be adjusted in these integral representations.
	Based upon our  detailed analysis leading to the correct concept of regularization, the necessary control on 
	the so-called \emph{re-regularization} procedure (see, e.g., \cite[Lemma 3.3]{iid}) is available,
	but the implementation is still quite technical, especially in the edge and cusp regime. 
	Fortunately, these complications are not relevant for the main results of the present paper, i.e. 
	for the local law for $\langle \Im G A\Im G A^*\rangle$, needed for ETH, as well as
	for the extensions of the single-resolvent local law.

	\medskip
	
	We briefly summarize the structure of the paper.
	In the next Section~\ref{sec:main} we precisely define our model, give the conditions
	on the correlation structure,  state our main Theorem~\ref{thm:main} on the eigenstate thermalisation hypothesis (ETH)
	and interpret the anomalous fluctuation  rates in all spectral regimes. 
	In Section~\ref{sec:LL} we define the concept of regularity and  
	formulate the  two-resolvent local laws:  the main Theorem~\ref{thm:locallawreg} for regular observables 
	in the critical perturbative spectral domain, which directly implies ETH, 
	and the auxiliary Theorem~\ref{thm:locallawnonreg}  for general observables away from criticality.
	The size of the fluctuation in the local laws is governed by a new observable norm, $\vertiii{\cdot}_{z, w}$
	(Definition~\ref{def:triple}) which we also use to express the optimal bounds on the deterministic 
	$M$-terms of two-resolvent chains (Theorem~\ref{prop:M2_bounds}).  	
	We complement our main two-resolvent local law by a single-resolvent local law with regular observables 
	(Theorem~\ref{thm:singleG}) and mention several possible extensions.
	Section~\ref{sec:Mboundsproof} proves the new and very delicate optimal $M$-bounds with several 
	necessary ingredients deferred to Sections~\ref{sec:M_flow}, \ref{sec:pert}, \ref{sec:stab} and Appendix~\ref{app:technicals}, 
	as well as the demonstration of the optimality in Appendix~\ref{app:Moptimal}.
	Armed with the $M$-bounds, Section~\ref{sec:zigzag} sets up the zigzag strategy and proves the main local law,
	Theorem~\ref{thm:locallawreg},
	while Sections~\ref{sec:zig_proof} and~\ref{sec:zagproof} contain the zig and zag steps, respectively.
	The initial step of the zigzag method, the global law, is proven in Section~\ref{sec:glob}.  
	The other local laws are proven in Sections~\ref{sec:proofnonreg}, \ref{sec:proof1G}, and
	Appendix~\ref{app:outside}.

	\subsection*{Summary of frequent notations}	
	We denote the complex upper half-plane by $\mathbb{H}$, that is, $\mathbb{H} := \{z \in \mathbb{C} : \Im z > 0\}$, and its closure by $\overline{\mathbb{H}} := \mathbb{H}\cup\mathbb{R}$.
		We use the notation $[N]$ to represent the index set $\{1,\dots, N\}$. The letters $a$, $b$, $j$, and $k$ are used to denote integer indices, while $\alpha$ (with various subscripts) denotes elements of $[N]^2$. All unrestricted summations of the form $\sum_a$ and $\sum_{\alpha}$ are understood to run over $a \in [N]$ and $\alpha \in [N]^2$, respectively. 
	
	We denote vectors in $\mathbb{C}^{N\times N}$ using boldface letters, e.g., $\bm x$. The scalar product on $\mathbb{C}^N$ is defined by $\langle \bm x, \bm y\rangle := \sum_{j=1}^N \overline{x_j}y_j$, and the corresponding Euclidean norm is denoted by $\norm{\bm x} := \langle \bm x, \bm x\rangle^{1/2}$. Matrices are denoted by capital letters and the Hermitian real and imaginary parts of a matrix $X \in \mathbb{C}^{N\times N}$ are denoted by
	\begin{equation}
		\Re X := \tfrac{1}{2}(X + X^*), \qquad \Im X := \tfrac{1}{2\ii} (X-X^*).
	\end{equation} 
	Unless explicitly stated otherwise, all matrices we consider are $N\times N$. We denote the identity super-operator on $\mathbb{C}^{N\times N}$ by $\mathrm{Id} \equiv \mathrm{Id}_N$, that is
	\begin{equation}
		\mathrm{Id}[X] \equiv \mathrm{Id}_N[X]  := X, \qquad X \in \mathbb{C}^{N\times N}.
	\end{equation}
	
For any $a\in [N]$ and vectors $\bm x$ and $\bm y$, we use the following notation:
	\begin{equation*}
		A_{\bm x \bm y} := \langle \bm x, A\bm y\rangle, \quad A_{\bm x a} := \langle \bm x, A \bm{e}_a\rangle, \quad A_{a \bm y} := \langle\bm{e}_a , A \bm y\rangle,
	\end{equation*}
	where $\bm{e}_a$ is the standard $a$-th basis vector of $\mathbb{C}^N$.	For a matrix $X \in \mathbb{C}^{N\times N}$, we denote its normalized trace by $\langle X\rangle := N^{-1}\Tr [X]$, its operator norm by $\norm{X}$, and its (normalized) Hilbert-Schmidt norm by $\norm{X}_{\mathrm{hs}}$, and the maximal size of its entry by $\norm{X}_{\max}$ that is
	\begin{equation}
		\norm{X} := \sup_{\norm{\bm x} = 1} \norm{X\bm x}, \quad	\norm{X}_{\mathrm{hs}} := \langle XX^* \rangle^{1/2}, \quad \norm{X}_{\max} := \max_{ij} |X_{ij}|. 
	\end{equation}
	For a (linear) super-operator $\mathcal{F} : \mathbb{C}^{N\times N} \to \mathbb{C}^{N\times N}$, we denote its norms induced by the operator, Hilbert-Schmidt, and maximum norms on the domain and codomain by $\norm{\mathcal{F}}$, $\norm{\mathcal{F}}_{\mathrm{hs} \to \norm{\cdot}}$, $\norm{\mathcal{F}}_{\max \to \norm{\cdot}}$, and $\norm{\mathcal{F}}_{\mathrm{hs} \to \mathrm{hs}}$, respectively.

		We use $c$ and $C$ to denote unspecified, positive constants -- small and large, respectively -- that are independent of $N$ and may change from line to line. 	For two positive quantities $\mathcal{X}$ and $\mathcal{Y}$, we write $\mathcal{X} \lesssim \mathcal{Y}$ if there exists a constant $C > 0$ that depends only on the \emph{model parameters} in Assumptions~{\ref{ass:boundedexp}--\ref{ass:Mbdd}} and on the radius $C_{\mathrm{bdd}}$ in \eqref{eq:bdd_def} (unless explicitly stated otherwise), such that $\mathcal{X} \le C \mathcal{Y}$. We use the notation $\mathcal{X} \sim \mathcal{Y}$ if both $\mathcal{X} \lesssim \mathcal{Y}$ and $\mathcal{Y} \lesssim \mathcal{X}$ hold. For an arbitrary quantity $\mathcal{X}$ and a positive quantity $\mathcal{Y}$, we use the notation $\mathcal{X} = \mathcal{O}(\mathcal{Y})$ to indicate that $|\mathcal{X}|\lesssim \mathcal{Y}$.
	
		Let $\Omega := \{\Omega^{(N)}(u) \, \lvert\, N\in\mathbb{N},\, u\in\mathcal{U}^{(N)} \}$  be a family of events depending on $N$ and possibly on a parameter $u$ that varies over some parameter set $\mathcal{U}^{(N)}$. We say that $\Omega$ holds \textit{with very high probability} (w.v.h.p.) uniformly in $u \in \mathcal{U}^{(N)}$ if, for any $D > 0$,
		\begin{equation*}
			\sup\limits_{u \in \mathcal{U}^{(N)}} \mathbb{P}\bigl[\Omega^{(N)}(u)\bigr] \ge 1 - N^{-D},
		\end{equation*}
		for any $N \ge N_0(D)$. We often discard the explicit dependence of $\Omega^{(N)}$ and $\mathcal{U}^{(N)}$ on $N$, and simply refer to $\Omega$ as a very-high-probability event. A bound is said to hold w.v.h.p. if it holds on a very-high-probability event.
		
		Finally, we introduce the common notion of \emph{stochastic domination} (see, e.g., \cite{semicirclegeneral}): For two families
		\begin{equation*}
			X = \left(X^{(N)}(u) \mid N \in \N, u \in U^{(N)}\right) \quad \text{and} \quad Y = \left(Y^{(N)}(u) \mid N \in \N, u \in U^{(N)}\right)
		\end{equation*}
		of non-negative random variables indexed by $N$, and possibly an additional parameter $u$
		from a parameter space $U^{(N)}$, we say that $X$ is stochastically dominated by $Y$, if for all $\epsilon, D >0$ we have 
		\begin{equation*}
			\sup_{u \in U^{(N)}} \mathbf{P} \left[X^{(N)}(u) > N^\epsilon Y^{(N)}(u)\right] \le N^{-D}
		\end{equation*}
		for large enough $N \ge N_0(\epsilon, D)$. In this case we write $X \prec Y$. If for some complex family of random variables we have $\vert X \vert \prec Y$, we also write $X = O_\prec(Y)$.

	\section{Main results}\label{sec:main}

	\subsection{Assumptions}
	We consider $N\times N$ random matrices $H \equiv H^{(N)}$ in the real-symmetric or complex-Hermitian symmetry class with expectation matrix
	\begin{equation} \label{eq:D_def}
		D \equiv D^{(N)} := \E H^{(N)} . 
	\end{equation}
		
	\begin{assumption}[Bounded expectation] \label{ass:boundedexp}
		There exists a constant $C_D > 0$ such that $\Vert D \Vert \le C_D$, uniformly in $N$. 
	\end{assumption}
	
	  The covariance structure of the matrix elements of $H$ is described by two very closely related
	super-operators, denoted by $\mathcal{S}$ and $\Sigma$, 
	both acting on deterministic $N\times N$ matrices. Although they contain exactly the same information, they are used differently in the proof, so it is convenient to introduce separate notation.  \nc

	The \emph{self-energy super-operator} $\mathcal{S} \equiv \mathcal{S}_H$ corresponding to the random matrix $H$ is defined via its action on deterministic $N\times N$ matrices as follows:
	\begin{equation} \label{eq:S_def}
		\mathcal{S}[X] \equiv \mathcal{S}_H[X] := \E\bigl[(H - \E H)\, X \, (H- \E H)\bigr], \qquad X \in \mathbb{C}^{N\times N}. 
 	\end{equation}
 	In the standard basis of $\mathbb{C}^{N\times N}$, the entries of the super-operator $\mathcal{S}$ are given by
 	\begin{equation}
 		\mathcal{S}_{ab, ij} = \mathrm{Cov}\bigl[H_{ai}, H_{jb}\bigr] = N^{-1}\kappa(ai, jb), \qquad i,j,a,b \in \indset{N},
 	\end{equation}
 	  where $\kappa(ai, jb)$ denotes the second order cumulant (covariance) of the rescaled 
 	matrix elements $\sqrt{N} H_{ai}, \sqrt{N}H_{jb}$. \nc
 	\begin{assumption} [Self-Energy Norms] \label{ass:S_norms}
 		We assume that there exists a positive $N$-independent constant $C_2$, such that 
 		\begin{equation} \label{eq:kappa2_norms}
 			\begin{split}
 				\vertiii{\mathcal{S}}_2  &:= \sup_{\norm{X}_{\mathrm{hs}} = \norm{Y}_{\mathrm{hs}} = 1} \Bigl\lvert N \sum_{ai, jb} |\mathcal{S}_{ab, ij}|\, X_{ai} Y_{jb}\Bigr\rvert  \le C_2,\\
 				\vertiii{\mathcal{S}}_2^\mathrm{iso} &:= \inf_{\mathcal{S} = \mathcal{S}_{\mathrm{c}} + \mathcal{S}_{\mathrm{d}}} \bigl(\vertiii{\mathcal{S}_\mathrm{c}}_\mathrm{c}  +  \vertiii{\mathcal{S}_\mathrm{d}}_\mathrm{d}\bigr) \le C_2,
 			\end{split}
 		\end{equation}
 		where the norms $\vertiii{\cdot}_\mathrm{c}$ and $\vertiii{\cdot}_\mathrm{d}$ are defined as 
 		\begin{equation} \label{eq:cross_dir_norm_defs}
 			\begin{split}
 				\vertiii{\mathcal{S}_{\mathrm{c}}}_\mathrm{c} &:=  
 				\sup\limits_{\norm{\bm x} = 1} \sup\limits_{\norm{\bm u} = \norm{\bm v} = 1} \Bigl\lvert N\sum_{ij} \overline{u}_i v_j \Bigl(\sum_b \Bigl\lvert \sum_a x_a (\mathcal{S}_{\mathrm{c}})_{ab, ij}\Bigr\rvert^2\Bigr)^{1/2} \Bigr\rvert \,, \\
 				\vertiii{\mathcal{S}_{\mathrm{d}}}_\mathrm{d} &:= 
 				\sup\limits_{\norm{\bm x} = 1} \sup\limits_{\norm{\bm u} = \norm{\bm v} = 1} \Bigl\lvert N\sum_{ib} \overline{u}_i v_b \Bigl(\sum_j \Bigl\lvert \sum_a x_a (\mathcal{S}_{\mathrm{d}})_{ab, ij} \Bigr\rvert^2\Bigr)^{1/2} \Bigr\rvert \,. 
 			\end{split}
 		\end{equation}
 	\end{assumption}
 	The subscripts $\mathrm{c}, \mathrm{d}$ above refer to the (non-unique) decomposition
 	of the self-energy operator into its \emph{cross} and \emph{direct}
 	parts (see \cite[Remark 2.8]{slowcorr}). In the independent case (i.e.\emph{Wigner-type matrices}), this decomposition admits a natural realization,
 	$$
 	\mathcal{S}_{ab, ij} = S_{ai} \delta_{ab}\delta_{ij} + T_{ai}  \delta_{aj}\delta_{bi}, \qquad  
 	S_{ai} : = \E \big| H_{ai} -\E H_{ai}\big|^2, \qquad T_{ai} : = \E \big( H_{ai} -\E H_{ai}\big)^2,
 	$$
 	corresponding to the two natural covariances: between $H_{ai}$ and $H_{ia}= \overline{H}_{ai}$,
 	and between $H_{ai}$ with itself. It is immediate that \eqref{eq:kappa2_norms} holds true
 	if the  matrix of variances $S$ satisfies the usual \emph{mean-field} condition, $\max_{ai} S_{ai}\le C_2/N$.
 	The correlated model considered here extends the independent setting by allowing nontrivial correlations between $H_{ai}$ and $H_{jb}$ even when $|a-j|+ |i-b|$ or $|a-b|+|i-j|$ 
 	is nonzero. The boundedness conditions in Assumption~\ref{ass:S_norms} enforce a suitable decay of correlations, which preserves the mean-field character of the model and, in particular, ensures that the spectrum remains uniformly bounded in $N$. 
 	A primary example is given by polynomially decaying correlation structures; see \cite[Example 2.6]{slowcorr}.

	Complementary to the self-energy operator $\mathcal{S}$, defined in \eqref{eq:S_def}, we define the   \emph{covariance tensor} \nc $\Sigma \equiv \Sigma_H$ via its action on deterministic $N\times N$ matrices, 
	\begin{equation} \label{eq:CovTensor}
		\Sigma[X] \equiv \Sigma_H[X] := \E\bigl[ \Tr [(H - \E H)\, X] (H- \E H)\bigr], \qquad X \in \mathbb{C}^{N\times N}. 
	\end{equation}
	The entries of $\Sigma$ in the standard basis of $\mathbb{C}^{N\times N}$ are given by 
	\begin{equation} \label{eq:SigmaSentries}
		\Sigma_{ab, ij}  = \mathrm{Cov}\bigl[ H_{ji},  H_{ab}\bigr] = N^{-1}\kappa(ji, ab) = \mathcal{S}_{jb, ia}. 
	\end{equation}
	  As an example, for complex-Hermitian Wigner matrices
	satisfying the standard additional condition $\E H_{ab}^2=0$, we have 
	$$ \mathcal{S}[X] =\langle X\rangle,
	\quad \Sigma[X] = X,
	$$
	while for real symmetric Wigner matrices 
	$$\mathcal{S}[X] =\langle X\rangle +N^{-1}X^\mathfrak{t}, \quad
	\Sigma[X] = X+X^\mathfrak{t},
	$$
	where $X^\mathfrak{t}$ denotes the transpose of $X$. 
	
	In addition to the upper bounds on the covariances, we impose suitable lower bounds to guarantee that the model exhibits sufficient randomness. The most convenient such assumption is the following \emph{fullness} condition. \nc
	
	\begin{assumption}[Fullness] \label{ass:full}
		We assume that there exists a constant $c_{\rm full} > 0$ such that the covariance tensor $\Sigma$ satisfies
		\begin{equation} \label{eq:fullness}
			N\,  \Tr\bigl[X\Sigma[X]\bigr]    \ge c_{\rm full} \, \Tr [X^2],
		\end{equation}
		for any deterministic matrix $X = X^*\in \mathbb{C}^{N\times N}$ if $H$ is complex Hermitian, and $X = X^*\in \mathbb{R}^{N\times N}$ if $H$ is real symmetric. 	 
	\end{assumption}
	
	Some parts of the analysis require a somewhat weaker condition, which can be conveniently 
	expressed in terms of $\mathcal{S}$. \nc
	We say that a self-energy operator $\mathcal{S}$ enjoys the \textit{flatness estimates} with constants $c, C > 0$ if and only if
	\begin{equation} \label{eq:flatness}
		c \langle R \rangle \le \mathcal{S}[R] \le C \langle R \rangle, \qquad R \ge 0,
	\end{equation}
	where $R$ is in the same symmetry class as $H$, and the inequalities are understood in the sense of quadratic forms. 
	Note that the first bound in \eqref{eq:kappa2_norms} and \eqref{eq:fullness} imply that the $\mathcal{S}$ satisfies the \textit{flatness} estimates with $c = c_{\rm full}$ and $C = C_2$.
		
	Let $M(z)$ be the unique solution \cite{helton2007operator} to the matrix Dyson equation (MDE) with data-pair $(D,\mathcal{S})$, defined in \eqref{eq:D_def} and \eqref{eq:S_def}, 
	\begin{equation} \label{eq:MDE}
		- M(z)^{-1} = z - D + \mathcal{S}\bigl[M(z)\bigr], \qquad (\Im z) \bigl(\Im M(z)\bigr) >0, \quad z\in \mathbb{C}\backslash\mathbb{R}. 
	\end{equation}
	  The solution $M$ is known as the \emph{self-consistent Green function}, since it captures the
	leading deterministic behavior of the resolvent of $H$. \nc
	
	  The corresponding \textit{self-consistent density of states} (scDOS) $\rho$ is a probability density function on $\mathbb{R}$, recovered from $\langle M\rangle$ via the Stieltjes inversion formula, 
	\begin{equation} \label{eq:scDOS}
		\rho(E) := \lim_{\eta\to+0} \pi^{-1}\bigl\langle \Im M(E+\ii\eta) \bigr\rangle.
	\end{equation}
	For data-pairs $(D,\mathcal{S})$ satisfying Assumption~\ref{ass:boundedexp} and the flatness estimates \eqref{eq:flatness}, the density $\rho(E)$ is bounded and compactly supported, as shown in \cite{AEK2020} (see Lemma~\ref{lemma:M_structure} below). \nc 
		
	For a pair of positive constants $c, C$, we define the set of admissible energies  $\mathcal{I}_{c,C} \subset \mathbb{R}$ as 
	\begin{equation} \label{eq:adm_E}
		\mathcal{I}_{c, C} := \Bigl\{ E \in \mathbb{R} \, : \, \norm{M(z)} \le C (1 + |z|)^{-1} \quad \text{for all} \quad z\in \mathbb{C} \quad\text{with} \quad \Re z \in [E - c, E+ c]  \Bigr\}.
	\end{equation}
	\begin{assumption} [Bounded Self-Consistent Green Function] \label{ass:Mbdd}
		We assume that there exists a pair of positive constants $c_M, C_M$ such that the set of admissible energies $\mathcal{I} \equiv \mathcal{I}_{c_M, C_M}$ is non-empty. 
	\end{assumption}
	  Assumption~\ref{ass:Mbdd} imposes a bound on $M$, which does not follow
	from the previously assumed bounds on $D$ and $\mathcal{S}$. For verifiable sufficient conditions, see \cite[Section 9]{AEK2020}. 
	It was shown in \cite{ajanki2019quadratic} that, in neighborhoods of admissible energies, the behavior of $\rho$ is governed by universal \emph{shape functions}; see also \cite[Theorem~7.1]{AEK2020}. 
	In particular, at the boundary of its support, $\rho$ vanishes like a square root. We refer to this regime as the \emph{spectral edge}. Near its zeroes in the interior of the support, $\rho$ exhibits cubic-root behavior, and such points are called \emph{cusps}. These are the only two possible types of singularities of $\rho$ near the admissible energies.

	As the last assumption on our random matrix $H$, we suppose certain higher-order correlation structure on  the entries $(w_{ab})_{a,b\in\indset{N}}$ of   $$W := H - \E H \,,$$
	formulated through the norms of their (normalized) joint \emph{cumulants}\footnote{Let $\bm{w} = (w_1, ..., w_k)$ be a random vector. Recall that its joint cumulants, $\kappa_{\bm m}$ with $\bm m \in (\mathbb{N}\cup\{0\})^k$, are traditionally given as the coefficients of the $\log$-characteristic function
		\begin{equation*}
			\log \E \ee^{\ii \bm{w} \cdot \bm t} = \sum_{\bm{m}} \kappa_{\bm m}  \frac{(\ii \bm t)^{\bm m}}{\bm m!} \,. 
		\end{equation*}
		For $\bm w = (\sqrt{N}w_{\alpha_1}, ... , \sqrt{N}w_{\alpha_k})$ we use the notation $\kappa(\alpha_1, ..., \alpha_k) \equiv \kappa(\sqrt{N}w_{\alpha_1}, ... , \sqrt{N}w_{\alpha_k}) := \kappa_{(1,...,1)}$ %
		and note that, by construction, $\kappa(\alpha_1, ... , \alpha_k)$ is invariant under permutations of its arguments. For example, for $k=2$, $\kappa(\alpha_1, \alpha_2) = N \E [w_{\alpha_1} w_{\alpha_2}]$.
	},
	\begin{equation} \label{eq:cumulants}
		\kappa(\alpha_1, ... , \alpha_k) \equiv \kappa(\sqrt{N}w_{\alpha_1}, ... , \sqrt{N}w_{\alpha_k})\,. 
	\end{equation}
	In the sequel, a double index $\alpha_i \in \indset{N}^2$ is represented by two single indices $a_i, b_i \in \indset{N}$, identifying $\alpha_i \equiv (a_i, b_i)$. For brevity, we often use the notation $a_ib_i = (a_i, b_i)$. 

We formulate Assumption \ref{ass:cumulants} only in the real symmetric case. For complex Hermitian matrices, we require the cumulant norms introduced below to be bounded for all choices of real and imaginary parts in each of the arguments of a cumulant, i.e.~for $\kappa(\alpha_1^{\mathfrak{X}_1}, ... ,  \alpha_k^{\mathfrak{X}_k}) = \kappa(\sqrt{N} \mathfrak{X}_1 w_{\alpha_1}, ... , \sqrt{N} \mathfrak{X}_k w_{\alpha_k})$ and all choices of $\mathfrak{X}_i \in \{\Re , \Im \}$ (see~\cite[Appendix~C]{slowcorr}).

	\begin{assumption}[Higher-order correlation structure] \label{ass:cumulants}
	The matrix entries $(w_\alpha)_\alpha$ of $W$ satisfy the following. 
	\begin{itemize}
		\item[(i)] The {cumulants} $\kappa(\alpha_1, ... , \alpha_k)$ have bounded  %
		matrix norms (viewed as an $N^2 \times N^2$ matrix), 
		i.e.~for all $k \ge 2$ there exists a constant $C_k > 0$ such that\footnote{ We remark that the constants $C_k$ in the bounds \eqref{eq:summcum}--\eqref{eq:kappa_3_av_norm} could also be replaced by $C_{k, \arb} N^\arb$ for any $\arb > 0$, where $C_{k,\arb}$ is a positive constant. All our proofs hold under this more general condition, but we omit it for simplicity.}
		\begin{equation} \label{eq:summcum}
			\vertiii{\kappa}_k := \bigg\Vert \sum_{\alpha_1, ... , \alpha_{k-2}} |\kappa(\alpha_1, ... , \alpha_{k-2}, *, *) | \bigg\Vert \le C_k \,. 
		\end{equation}
		Moreover, we assume that
		\begin{equation} \label{eq:kappa_3_av_norm}
			\vertiii{\kappa}_3^\mathrm{av} :=  N^{-3/2} \sup\limits_{\substack{X,Y,Z\in\C^{N\times N} : \\ \Vert X \Vert, \Vert Y \Vert \le 1\,, \ \Vert Z \Vert_{\rm hs} \le 1}}  \ \sum\limits_{a_1b_1}\sum\limits_{a_2b_2}\sum\limits_{a_3b_3} \bigl\lvert \kappa(a_1b_1,a_2b_2,a_3b_3) \bigr\rvert |X_{b_1a_2}|\, |Y_{b_2a_3}|\, |Z_{b_3a_1}| \le C_3\,. 
		\end{equation}
		
		\item[(ii)] There exists a positive $\mu > 0$, such that for every $\alpha \in \indset{N}^2$ there exists an index set $\mathcal{N}(\alpha) \subset \indset{N}^2$ of cardinality $|\mathcal{N}(\alpha)| \le N^{1/2 - \mu}$ with the property that $w_\alpha $ is independent of  $w_\beta$ for all $\beta \notin \mathcal{N}(\alpha)$. That is, every $w_\alpha$ is correlated with at most $N^{1/2-\mu}$ other matrix elements and independent of the rest.
	\end{itemize}
\end{assumption}

Note that in the case of random matrices with independent entries (\emph{Wigner-type} ensembles), Assumption~\ref{ass:cumulants} simplifies to an upper bound
$$
	\max_{\alpha\in\indset{N}^2}\E \bigl[\lvert \sqrt{N} w_{\alpha} \rvert^k\bigr] \le C_k, \qquad k \ge 2.  
$$ 
Furthermore, for Wigner-type matrices, Assumption~\ref{ass:full} is equivalent to the lower flatness bound in \eqref{eq:flatness}.

We refer to the   $N$-independent constants in Assumptions~\ref{ass:boundedexp}--\ref{ass:cumulants} as \textit{model parameters}.   
In the sequel, all implicit constants may depend on these model parameters; their dependencies are explicit, but we do not track them.

	\subsection{Eigenstate Thermalisation Hypothesis for correlated random matrices}  	Before stating the main result, we introduce the necessary notation.
	Recall that  $\rho$ denotes the self-consistent density of states corresponding to~$H$, see~\eqref{eq:scDOS}. For any $i\in \indset{N}$, let $\gamma_i \in \supp\,\rho$ denote the $i/N$-quantile $\rho$, that is, 
	\begin{equation} \label{eq:quant_def}
		\gamma_i := \inf \biggl\{ x \in \supp\,\rho \,:\, \int_{-\infty}^x \rho(y)\mathrm{d}y = \frac{i}{N} \biggr\}.
	\end{equation}
	
	We define the local fluctuation scale $\etaf(E)$ around a fixed energy $E \in \supp\,\rho$ implicitly via\footnote{
		The local fluctuation scale $\etaf(E)$ can equivalently be defined, up to multiplicative constants of order one, by
		$$
		\int_{|x| \le \etaf(E)} \rho(E+x)\mathrm{d}x = \frac{1}{N}, \qquad E\in \supp\,\rho. 
		$$
	}
	\begin{equation} \label{eq:etaf_def}
		\rho\bigl(E+ \ii\etaf(E)\bigr) \etaf(E) = N^{-1},
	\end{equation}
	where $\rho(z) := \pi^{-1} \langle \Im M(z) \rangle$ for $z \in \mathbb{C}\backslash\mathbb{R}$ denotes the harmonic extension of $\rho(x)$, defined in \eqref{eq:scDOS}. Note that under this convention $\rho(z)$ is negative for $z$ in the lower half-plane. 
	
	For all $z \in \mathbb{H}$, let $\sigma(z)$ denote the real-valued function\footnote{  
		In the regime when the self-consistent density of states $\rho(z)$ is small, the magnitude of $\sigma(z)$ measures the proximity of $z$ to a near-cusp singularity. In particular, $|\sigma(z)| \sim 1$ near a regular-edge point (adjacent to a large gap in the support of $\rho$), while $|\sigma(z)| \approx 0$ near a sharp cubic-root cusp.  
		}
	\begin{equation} \label{eq:sigma_def} 
		\sigma(z) := \biggl\langle \sign \Re \mathscr{U}(z) \bigl(\rho(z)^{-1}\Im \mathscr{U}(z)\bigr)^3 \biggr\rangle, \quad \mathscr{U}(z) := \frac{(\Im M(z))^{-1/2} (\Re M(z)) (\Im M(z))^{-1/2} + \ii I }{\bigl\lvert (\Im M(z))^{-1/2} (\Re M(z)) (\Im M(z))^{-1/2} + \ii I \bigr\rvert}.
	\end{equation} 
	
	Next, we define the matrix $R \in \mathbb{C}^{N\times N}$, which generates the anomalous fluctuation mode. Let $M'(z) := \tfrac{\dd}{\dd z}M(z)$ denote the derivative of the solution $M$. Then, for all $z \in \mathbb{H}$, we define
	\begin{equation} \label{eq:R_def_p2}
	M^\perp(z) :=  M'(z) - \frac{\bigl\langle M'(z) \Im M(z) \bigr\rangle}{\norm{\Im M(z)}_\mathrm{hs}^2} \Im M(z).
	\end{equation}	 
	While $M^\perp(z)$ is defined only for $z\in\mathbb{H}$, the corresponding orthogonal projector $\mathcal{R}_z : \mathbb{C}^{N\times N} \to \mathbb{C}^{N\times N}$, given by
	\begin{equation} \label{eq:R_def}
		\mathcal{R}_z[\,\cdot\,] := \frac{\big\langle \big(M^\perp(z)\big)^* (\,\cdot\,)\big\rangle}{\norm{M^\perp(z)}_\mathrm{hs}^2}M^\perp(z), \qquad \text{if}\,\,  M^\perp(z) \neq 0, \quad \text{and} \quad \mathcal{R}_z := 0 \,\,\text{otherwise}
	\end{equation} 
	admits an extension $\mathcal{R}_E$ to the set of admissible energies $E \in \mathcal{I} \subset \R$ (recall Assumption~\ref{ass:Mbdd}), which we construct in Lemma~\ref{lemma:R_extend}. Finally, for $E \in \mathcal{I}$ and $A \in \C^{N \times N}$, we denote 
			\begin{equation} \label{eq:Acirc_def}
		\reg{A}^{E, E} :=  A - \Bigl\langle A \frac{\Im M(E)}{\pi \rho(E)}\Bigr\rangle I,
	\end{equation} 
	and for $E_1, E_2 \in \mathcal{I}$ we set\footnote{ 
		Verbatim, the function $\sigma(z)$ defined in \eqref{eq:sigma_def}, is H\"older regular only in the regime $\rho(z) \le \rho_*$, for some sufficiently small $\rho_*\sim 1$ (see Lemma \ref{lemma:MDE_props} below), and thus admits a continuous extension $\sigma(E)$ only  to energies $E$ satisfying $\rho(E) \le \rho_*$. 
		Nevertheless, we may replace $\sigma(E)$ by $\other{\sigma}(E) := \sigma(E) \chi(\rho(E))$, where $\chi(E) \in C^\infty$ is a cutoff function satisfying $\chi(x) = 1$ for $0 \le x \le \tfrac{1}{2}\rho_* $ and $\chi(x) = 0$ for $x \ge \rho_*$. 
		Then $\other{\sigma}(E)$ is $1/3$-H\"older and satisfies $\rho(E) + |\other{\sigma}(E)| \sim \rho(E) + |\sigma(E)|$ on all of $\mathcal{I}$. 
		Hence, replacing $\sigma$ with $\other{\sigma}$ in \eqref{eq:ETH_alpha_jk} yields a quantity equivalent to $\other{\alpha}$ up to $\mathcal{O}(1)$ factors. We therefore ignore this inessential distinction between $\sigma$ and $\other{\sigma}$ to simplify the presentation. 
	}
	\begin{equation} \label{eq:ETH_alpha_jk}
		\other{\alpha}(E_1, E_2) := \rho(E_1) + |\sigma(E_1)| + \rho(E_2) + |\sigma(E_2)| + |E_1 - E_2|^{1/3} \,. 
	\end{equation}
	Equipped with these notations, we are ready to formulate our main result. Its proof is formally concluded in Section~\ref{subsec:ETHProof} below. 
	\begin{theorem}[Eigenstate Thermalisation Hypothesis for correlated random matrices] \label{thm:main} 
		Let $H$ be a correlated random matrix satisfying Assumptions~\ref{ass:boundedexp}--\ref{ass:cumulants} with $\mathcal{I} = \R$ in Assumption \ref{ass:Mbdd}.\footnote{See Remark \ref{rmk:Iextend} below for an extension to general $\mathcal{I} \neq \R$.} Let $(\bm u_i)_{i=1}^N$ denote the orthogonal set of eigenvectors of $H$, corresponding to the ordered eigenvalues $\lambda_1 \le \lambda_2\le \dots \le \lambda_N$.
		Then, using the notations \eqref{eq:Acirc_def}--\eqref{eq:ETH_alpha_jk}, for any deterministic matrix $A \in \mathbb{C}^{N\times N}$ and any pair of indices $j,k \in \indset{N}$ we have
		\begin{equation} \label{eq:ETH}
			\biggl\lvert \bigl\langle \bm u_j, A \bm u_k \bigr\rangle - \delta_{jk}\Bigl\langle A\frac{ \Im M(\gamma_j) }{\pi \rho(\gamma_j)}\Bigr\rangle \biggr\rvert^2
			\prec 			\frac{1}{N}\min_{E \in \{\gamma_j, \gamma_k\}}
\Biggl(\frac{ \bigl\lVert \mathcal{R}_{E}\bigl[\reg{A}^{E, E}\bigr]  \bigr\rVert_\mathrm{hs}}{ \other{\alpha}(\gamma_j, \gamma_k) + N^{-1/4}} + \lVert \reg{A}^{E,E}\rVert_{\mathrm{hs}}\Biggr)^2 \,. 
		\end{equation}
	\end{theorem}
	In the following remark, we extend Theorem \ref{thm:main} to the case $\mathcal{I} \neq \R$. 
	\begin{remark}[The case $\mathcal{I} \neq \R$] \label{rmk:Iextend}
In case that $\mathcal{I} \neq \R$, Theorem \ref{thm:main} remains valid for energies inside the admissible set $\mathcal{I}$: Consider the map
\begin{equation} \label{eq:jkdef}
\mathrm{ind} : \R \to  \indset{N} \cup \{0\} \,, \qquad E \mapsto \left\lceil N\int_{-\infty}^{E} \rho(x) \dd x \right\rceil \,. 
\end{equation}
Then, ETH as in \eqref{eq:ETH} remains true for those indices $j, k \in \indset{N}$ such that $\mathrm{ind}^{-1}(\{j\}) \cap \mathcal{I} \neq \emptyset$ and $\mathrm{ind}^{-1}(\{k\}) \cap \mathcal{I} \neq \emptyset$, 
upon replacing $\gamma_j$ and  $\gamma_k$ by some $E_1 \in \mathrm{ind}^{-1}(\{j\}) \cap \mathcal{I}$ and $E_2 \in \mathrm{ind}^{-1}(\{k\}) \cap \mathcal{I}$, respectively, with the convention that if $j=k$, we pick $E_1 = E_2$.  We prove this extension of Theorem \ref{thm:main} in Section \ref{subsec:ETHProof}. 
	\end{remark}
	
\subsection{Anomalous rates of ETH and other comments}
	We now explain the implications of \eqref{eq:ETH} and suppose, for simplicity, that the admissible energies exhaust all of $\R$, i.e.~$\mathcal{I} = \R$. We distinguish two regimes, characterized by whether the quantiles $\gamma_j, \gamma_k$ are far away (less involved regime) or close (more involved regime). Take now any $E \in \{\gamma_j, \gamma_k\}$. 

	If $|\gamma_j - \gamma_k| > \delta_*$ for some $\delta_* \sim 1$, then we recover the usual size of overlap fluctuations, which agree with Haar-distributed random vectors, 
	\begin{equation}
		\Bigl\lvert \bigl\langle \bm u_j, A \bm u_k \bigr\rangle \Bigr\rvert^2
		\prec  \frac{\lVert A\rVert_{\mathrm{hs}}^2}{N}, \qquad |\gamma_j - \gamma_k| > \delta_*. 
	\end{equation}
	This follows since 
	\begin{equation} \label{eq:HSestimate}
	\lVert \mathcal{R}_{E}[\reg{A}^{E,E}] \rVert_\mathrm{hs} \lesssim \Vert \reg{A}^{E,E} \Vert_{\rm hs} \lesssim \Vert A \Vert_{\rm hs}
	\end{equation}
 by \eqref{eq:R_def}, the fact that $\Im M(E) \sim \rho(E)I$ (see \eqref{eq:imM_bound} below), and $\other{\alpha}(\gamma_j, \gamma_k) \gtrsim 1$. 
	
	In the complementary regime $|\gamma_j - \gamma_k| \le \delta_*$, the bound \eqref{eq:ETH} naturally induces the following (non-orthogonal) three-term decomposition
	\begin{equation} \label{eq:A_decomp}
		A = \Bigl\langle A \frac{\Im M(E)}{\pi \rho(E)}\Bigr\rangle\, I +  \mathcal{R}_{E}[\reg{A}^{E,E}] + \regreg{A}^{E, E},
	\end{equation}
	where  the matrix $\regreg{A}^{E, E}$ is defined as
	\begin{equation} \label{eq:ringring_def}
		\regreg{A}^{E, E} := \reg{A}^{E, E}-  \mathcal{R}_{E}[\reg{A}^{E,E}] \quad \text{with} \quad \reg{A}^{E,E} \quad \text{given by \eqref{eq:Acirc_def}} \,. 
	\end{equation}
	Overlaps with the first identity component in \eqref{eq:A_decomp} are trivially deterministic, $\langle \bm u_j, I \bm u_k \rangle =\delta_{jk}$, yielding the leading term of $\langle \bm u_j, A \bm u_k \rangle$ in \eqref{eq:ETH}. 
	
	The second component, $\mathcal{R}_{E}[\reg{A}^{E,E}]$, carries the anomalous fluctuation mode (when it is present),
	\begin{equation} \label{eq:anomalousmode}
		\Bigl\lvert \bigl\langle \bm u_j, \mathcal{R}_{E}[\reg{A}^{E,E}] \bm u_k \bigr\rangle \Bigr\rvert^2 \prec \frac{\lVert \mathcal{R}_E[\reg{A}^{E,E}] \rVert_{\rm hs}^2}{N\other{\alpha}(\gamma_j, \gamma_k)^2 + N^{1/2}},
	\end{equation}
	where we used \eqref{eq:ETH} for the observable $ \mathcal{R}_{E}[\reg{A}^{E,E}]$ instead of $A$ together with  $\other{\alpha}(\gamma_j, \gamma_k) \lesssim 1$,  and $\langle \mathcal{R}_{E}[\reg{A}^{E,E}] \Im M(E) \rangle = 0$ by \eqref{eq:R_def_p2}--\eqref{eq:R_def}. 
	
	Finally, the remaining third component on the right-hand side of \eqref{eq:A_decomp}, always exhibits the usual Haar-like behavior,
	\begin{equation} \label{eq:secondmode}
		\Bigl\lvert \bigl\langle \bm u_j, \regreg{A}^{E, E} \bm u_k \bigr\rangle \Bigr\rvert^2
		=\Bigl\lvert \bigl\langle \bm u_j, A \bm u_k \bigr\rangle - \delta_{jk} \Bigl\langle A \frac{\Im M(E)}{\pi \rho(E)}\Bigr\rangle - \bigl\langle \bm u_j, \mathcal{R}_{E}[\reg{A}^{E,E}] \bm u_k \bigr\rangle   \Bigr\rvert^2
		\prec \frac{\lVert \regreg{A}^{E, E}\rVert_\mathrm{hs}^2 }{N},
	\end{equation}
	where we used \eqref{eq:ETH} with $ \regreg{A}^{E, E}$ instead of $A$ as well as $\mathcal{R}_E[\regreg{A}^{E,E}] = 0$ by \eqref{eq:ringring_def}. Note that, using \eqref{eq:HSestimate}, the Hilbert-Schmidt norms on the right-hand side~of \eqref{eq:anomalousmode} and \eqref{eq:secondmode} can be replaced by $\Vert A \Vert_{\rm hs}^2$, in particular, all three terms in the decomposition \eqref{eq:A_decomp} are controlled in Hilbert-Schmidt norm by $\Vert A \Vert_{\rm hs}$. 
	
	Next, we decode the amplitude $N\other{\alpha}^2+N^{1/2}$ of the anomalous fluctuation mode in the three principal spectral regimes (see Figure~\ref{fig:shapes}). 
	Throughout this explanation we focus on highlighting the size of the anomalous fluctuation mode, that is, of \eqref{eq:anomalousmode}. 
	In all computations below we use the explicit formula  \eqref{eq:ETH_alpha_jk} as well as the shape analysis for the self-consistent density of states $\rho$ from \cite{AEK2020} (see also \eqref{eq:rho_comp} below).
	We refer the reader to Figure~\ref{fig:alphas} for an illustration of $\other{\alpha}(E,E)^{-2}$ in the relevant near-cusp regimes. \nc  
	
	\begin{figure}[h]
		\begin{minipage}{0.32\textwidth}
			\centering
			\includegraphics[width=\textwidth]{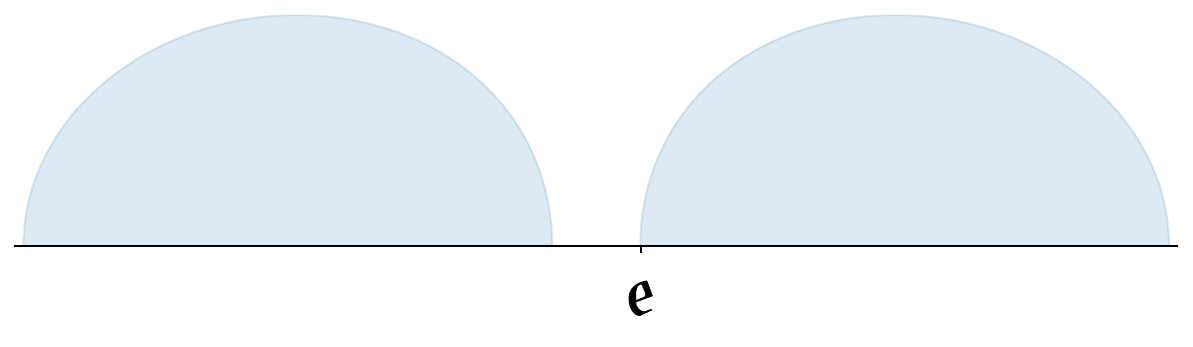}\\
			(a) $\lambda > 1$
		\end{minipage}
		\hfill
		\begin{minipage}{0.32\textwidth}
			\centering
			\includegraphics[width=\textwidth]{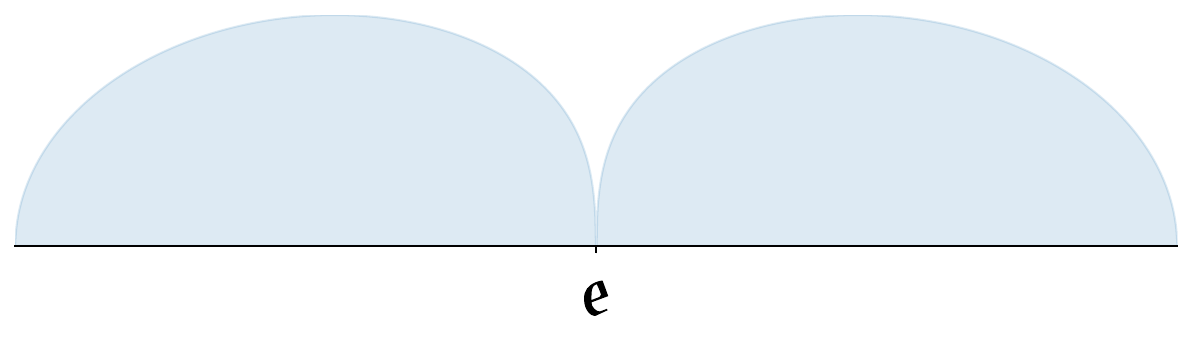}\\
			(b) $\lambda = 1$
		\end{minipage}
		\hfill
		\begin{minipage}{0.32\textwidth}
			\centering
			\includegraphics[width=\textwidth]{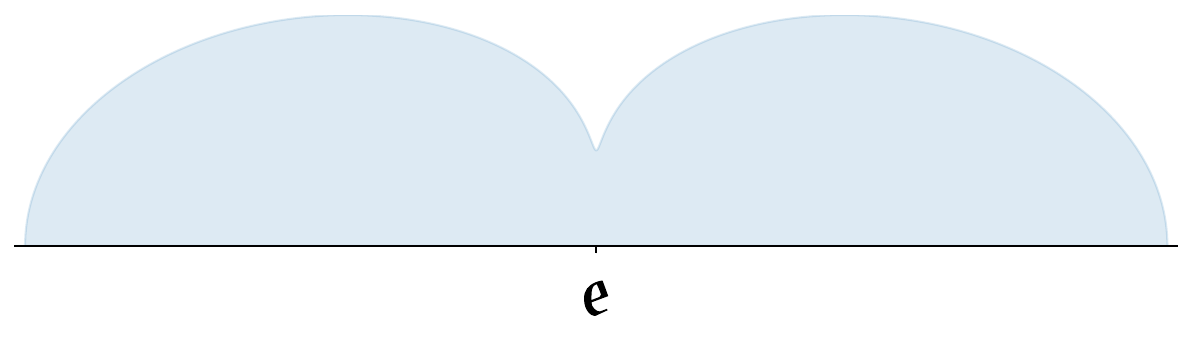}\\
			(c) $\lambda < 1$
		\end{minipage}
		\caption{ 
			Various spectral regimes realized using the self-consistent density of states $\rho$ of a deformed Wigner matrix $W + \lambda D$ at different values of $\lambda \ge 0$ around the reference energy $\mathfrak{e}$: for $\lambda > 1 + c$, the two end-points of the gap around the origin on panel (a) exhibit square-root behavior of a \textit{regular edge}, regime~(ii);
			for $\lambda = 0$ on panel (b), the singularity at the origin is a sharp cubic-root \textit{cusp}, regime~(iii);
			if $\lambda < 1-c$, the origin is absorbed by the \textit{bulk} spectrum, regime~(i); The transitionary \textit{small gap} (iv) and \textit{small local minimum} (v) regimes correspond to panels~(a) and~(c) respectively, with~$|1-\lambda| \ll 1$.}
		\label{fig:shapes}
	\end{figure}

	\begin{corollary}[Anomalous Fluctuations in Specific Spectral Regimes] \label{rem:regimes}
		Let $\mathfrak{e}$ be a reference energy in the spectrum and assume that $\gamma_j, \gamma_k$ satisfy  $|\gamma_j - \mathfrak{e}| \wedge |\gamma_k - \mathfrak{e}| \le c_*$ for some sufficiently small $c_* \sim 1$.
		We assume that $\mathcal{R}_{\gamma_j} \neq 0$, and abbreviate 
		$$
		\other{\alpha}_{jk} \equiv \other{\alpha}(\gamma_j, \gamma_k) + N^{-1/4}, \qquad R_j \in \mathrm{Range} \bigl(\mathcal{R}_{\gamma_j}\bigr), \quad \text{with}\quad \norm{R_j}_\mathrm{hs} = 1. 
		$$ 
		Then, in the "pure" spectral regimes, the size of the anomalous fluctuation mode is determined by the following asymptotics.
		\begin{itemize}
			\item[(i)]
			\textbf{Bulk regime}.
			Assume that $\mathfrak{e}$ is in the bulk of the spectrum, that is, $\rho(\mathfrak{e}) \sim 1$.  
			Then, 
			\begin{equation}
			N\other{\alpha}_{jk}^2 \sim N.
			\end{equation}
			Consequently, the anomalous mode is suppressed and we recover ETH in its standard form,
			\begin{equation} \label{eq:ETHstandard}
				\biggl\lvert \bigl\langle \bm u_j, A \bm u_k \bigr\rangle - \delta_{jk}\Bigl\langle A\frac{ \Im M(\gamma_j) }{\pi \rho(\gamma_j)}\Bigr\rangle \biggr\rvert^2
				\prec  \frac{1}{N} \min_{E \in \{\gamma_j, \gamma_k\}}\lVert \reg{A}^{E, E}\rVert_{\mathrm{hs}}^2. 
			\end{equation}
			
			\item[(ii)]
			\textbf{Regular edge regime}. 
			Assume that  $\mathfrak{e}$ is a \emph{regular edge} of $\rho$, that is, $\mathfrak{e} \in \supp\,\rho$, $\rho(\mathfrak{e}) = 0$ and $|\sigma(\mathfrak{e})| \sim 1$ (or equivalently, $\mathfrak{e}$ is an end-point of a support interval of $\rho$ adjacent to a gap of size $\Delta \gtrsim 1$). 
			Then, close to the edge, i.e., for $\gamma_j, \gamma_k$ satisfy $|\gamma_j - \mathfrak{e}| + |\gamma_k - \mathfrak{e}| \lesssim N^{-2/3}$, we have
			\begin{equation}
				N\other{\alpha}_{jk}^2 \sim N.
			\end{equation}
			Thus the anomalous fluctuation mode is dominated by the overall Haar-like fluctuations and \eqref{eq:ETHstandard} remains valid. 
			
			\item[(iii)]
			\textbf{Physical cusp regime}.
			Assume that  $\mathfrak{e}$ is a \emph{physical cusp} of $\rho$, that is, a local minimum of $\rho\rvert_{\supp\,\rho}$ satisfying  $\rho(\mathfrak{e}) + |\sigma(\mathfrak{e})| \lesssim N^{-1/4}$ (or equivalently, $\mathfrak{e}$ is either a positive local minimum with $\rho(\mathfrak{e}) \lesssim N^{-1/4}$ or an end-point of a support interval adjacent to a gap of size $\Delta \lesssim N^{-3/4}$). 
			Then, close to the cusp, i.e., for $|\gamma_j - \mathfrak{e}| + |\gamma_k - \mathfrak{e}| \lesssim N^{-3/4}$, we have 
			\begin{equation} \label{eq:ETH_cusp}
				N\other{\alpha}_{jk}^2 \sim N^{1/2}. 
			\end{equation} 
			In particular, the diagonal overlap $\langle \bm u_j, R_j \bm u_j \rangle$ fluctuates on the scale $N^{-1/4}$.
			\end{itemize}
			The following two cases interpolate between the regimes above, both in the sense of the local behavior of $\rho$ near $\mathfrak{e}$, and in the sense  of the location of $\gamma_j, \gamma_k$ in the spectrum.
			\begin{itemize}
			\item[(iv)] \textbf{Small local minimum} \textnormal{(cusp--bulk transition)}. 
			Assume that $\mathfrak{e}$ is a small positive local minimum of $\rho$.  Then, 
			\begin{equation} \label{eq:cusp_lmin_long}
				N\other{\alpha}_{jk}^2 \sim  N \bigl(\rho(\mathfrak{e}) + N^{-1/4}\bigr)^2 + N|\gamma_j - \mathfrak{e}|^{2/3} + N|\gamma_k - \mathfrak{e}|^{2/3}. 
			\end{equation}
			 
 			On the one hand, close to the local minimum, i.e., for $|\gamma_j - \mathfrak{e}| \lesssim \rho(\mathfrak{e})^3 + N^{-3/4}$,   
			the  diagonal overlap~$\langle \bm u_j, R_j \bm u_j \rangle$ fluctuates on the scale $N^{-1/2}\bigl(\rho(\mathfrak{e}) + N^{-1/4}\bigr)^{-1}$. 
			In particular, taking $\rho(\mathfrak{e}) \sim 1$, we recover the bulk regime~(i), and for $\rho(\mathfrak{e}) \lesssim N^{-1/4}$, we recover the physical cusp regime~(iii). 
		 
		 	On the other hand, away from the minimum, i.e., for  $\rho(\mathfrak{e})^3 + N^{-3/4} \lesssim |\gamma_j - \mathfrak{e}| \wedge |\gamma_k - \mathfrak{e}|$, we have  
		 	\begin{equation}\label{away}
				N\other{\alpha}_{jk}^2 \sim  N |\gamma_j - \mathfrak{e}|^{2/3} + N|\gamma_k - \mathfrak{e}|^{2/3}. 
			\end{equation}
			In particular, if $|\gamma_j - \mathfrak{e}| + |\gamma_k - \mathfrak{e}| \sim 1$, we recover the bulk regime from case~(i).

			\item[(v)]
			\textbf{Small gap regime} \textnormal{(cusp--edge and edge--bulk transitions)}. 
			Assume that $\mathfrak{e}$ is an end-point of a support interval of $\rho$ adjacent to a small gap of size $0 < \Delta \lesssim 1$. 
			Then,
			\begin{equation}
				N\other{\alpha}_{jk}^2 \sim N\bigl(\Delta + N^{-3/4}\bigr)^{2/3} + N|\gamma_j - \mathfrak{e}|^{2/3} + N|\gamma_k - \mathfrak{e}|^{2/3}.  
			\end{equation}
			
			On the one hand, near the edge, i.e., for $|\gamma_j - \mathfrak{e}| \lesssim  \Delta + N^{-3/4}$, the  diagonal overlap~$\langle \bm u_j, R_j \bm u_j \rangle$ fluctuates on the scale  $N^{-1/2}(\Delta + N^{-3/4})^{-1/3}$. 
			In particular, taking $\Delta \sim 1$, we recover the regular edge regime~(ii), and taking $\Delta \lesssim N^{-3/4}$, we recover the  physical cusp regime from case~(iii) above. 
			
			On the other hand, away from the edge, i.e. for $\Delta + N^{-3/4} \lesssim |\gamma_j - \mathfrak{e}| \wedge |\gamma_k - \mathfrak{e}|$, we have 
			\begin{equation}
				N\other{\alpha}_{jk}^2 \sim  N |\gamma_j - \mathfrak{e}|^{2/3} + N|\gamma_k - \mathfrak{e}|^{2/3},
			\end{equation}
			recovering the same fluctuations as in \eqref{away} above.
		\end{itemize}
	\end{corollary}

	\begin{figure}[h]
		\begin{minipage}{0.32\textwidth}
			\centering
			\includegraphics[width=\textwidth]{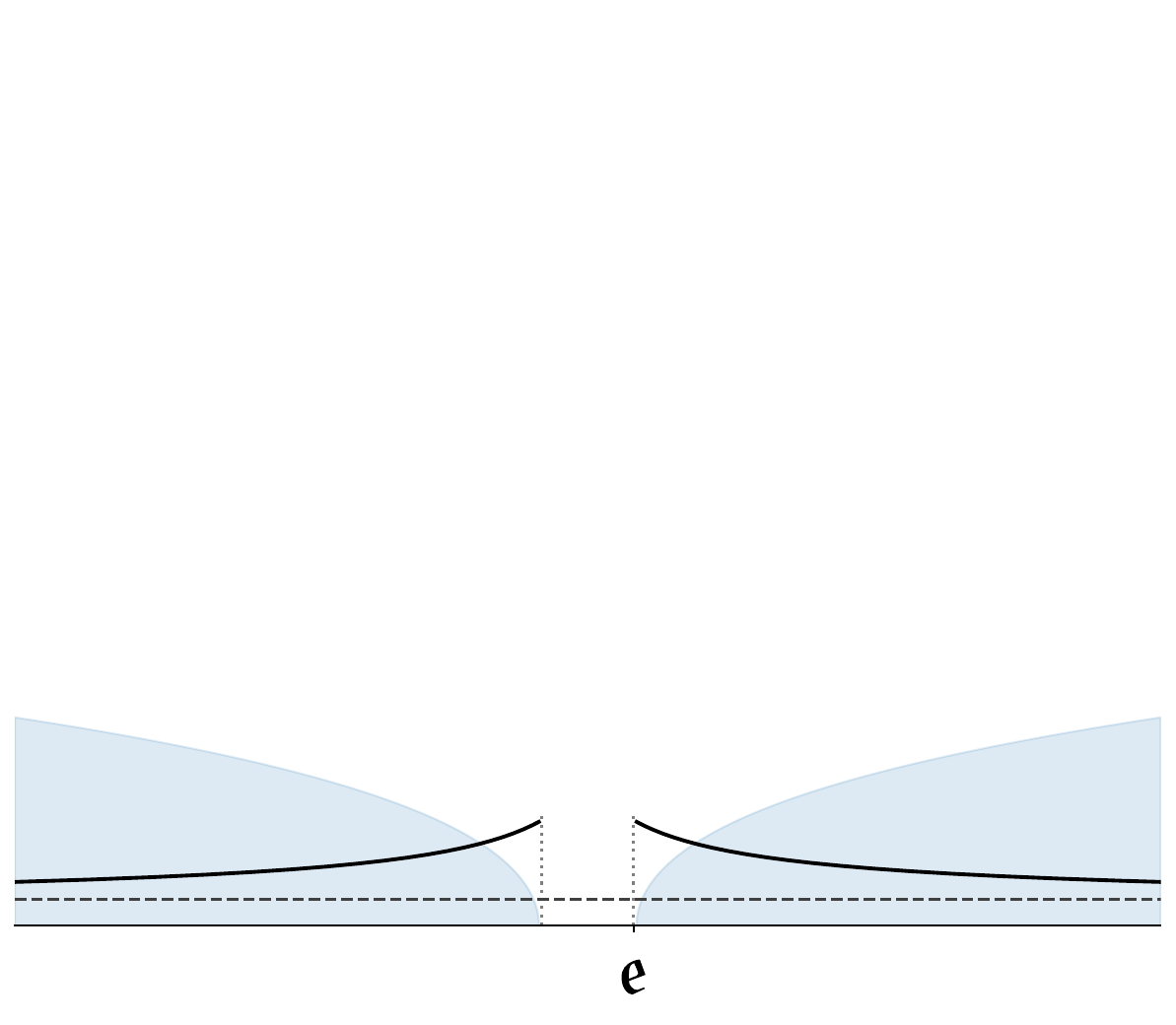}\\
			(a)  $\Delta(\mathfrak{e}) > 0, \quad  \rho(\mathfrak{e}) = 0$
		\end{minipage}
		\hfill
		\begin{minipage}{0.32\textwidth}
			\centering
			\includegraphics[width=\textwidth]{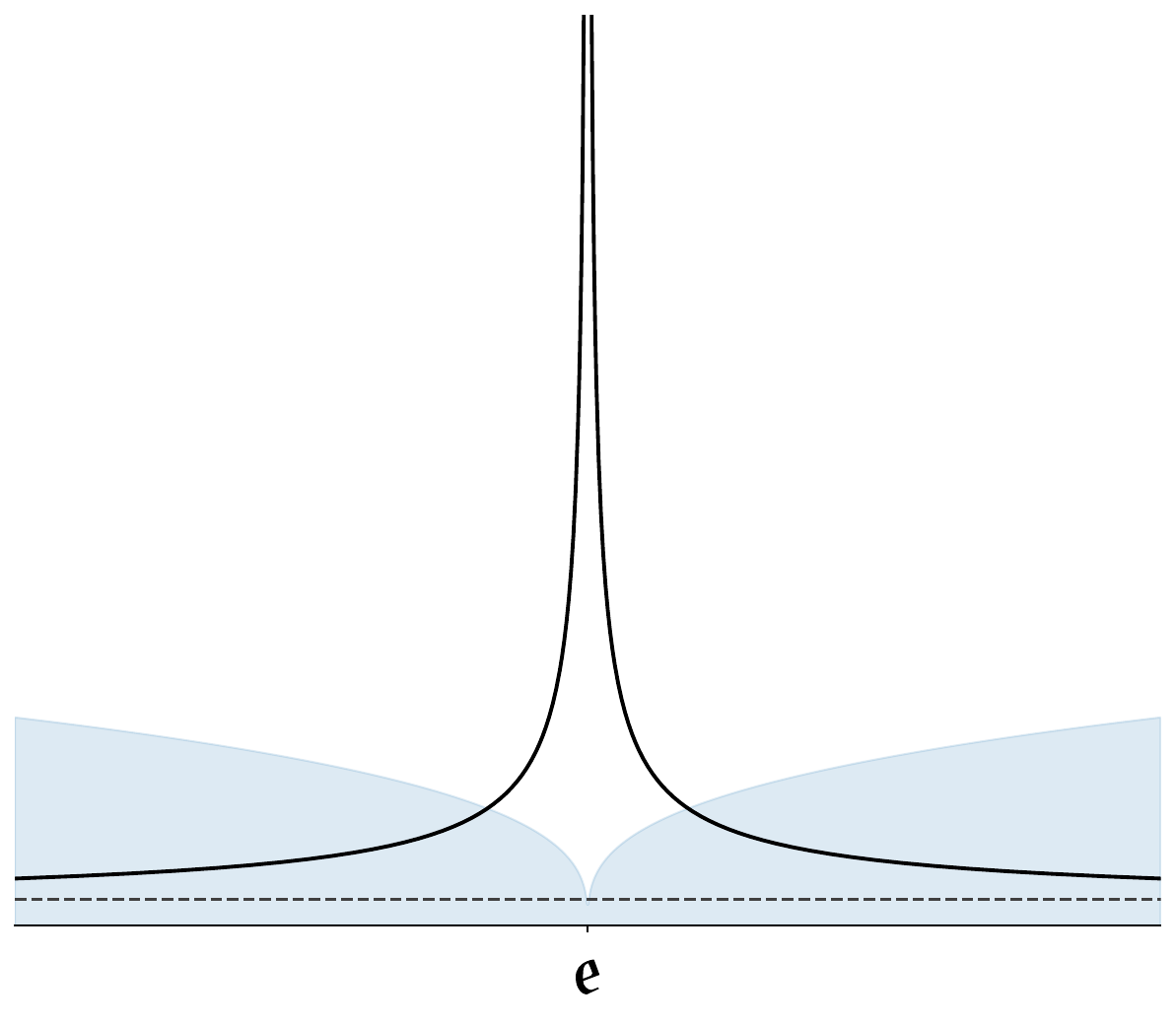}\\
			(b) $\Delta(\mathfrak{e}) =0, \quad  \rho(\mathfrak{e}) = 0$
		\end{minipage}
		\hfill
		\begin{minipage}{0.32\textwidth}
			\centering
			\includegraphics[width=\textwidth]{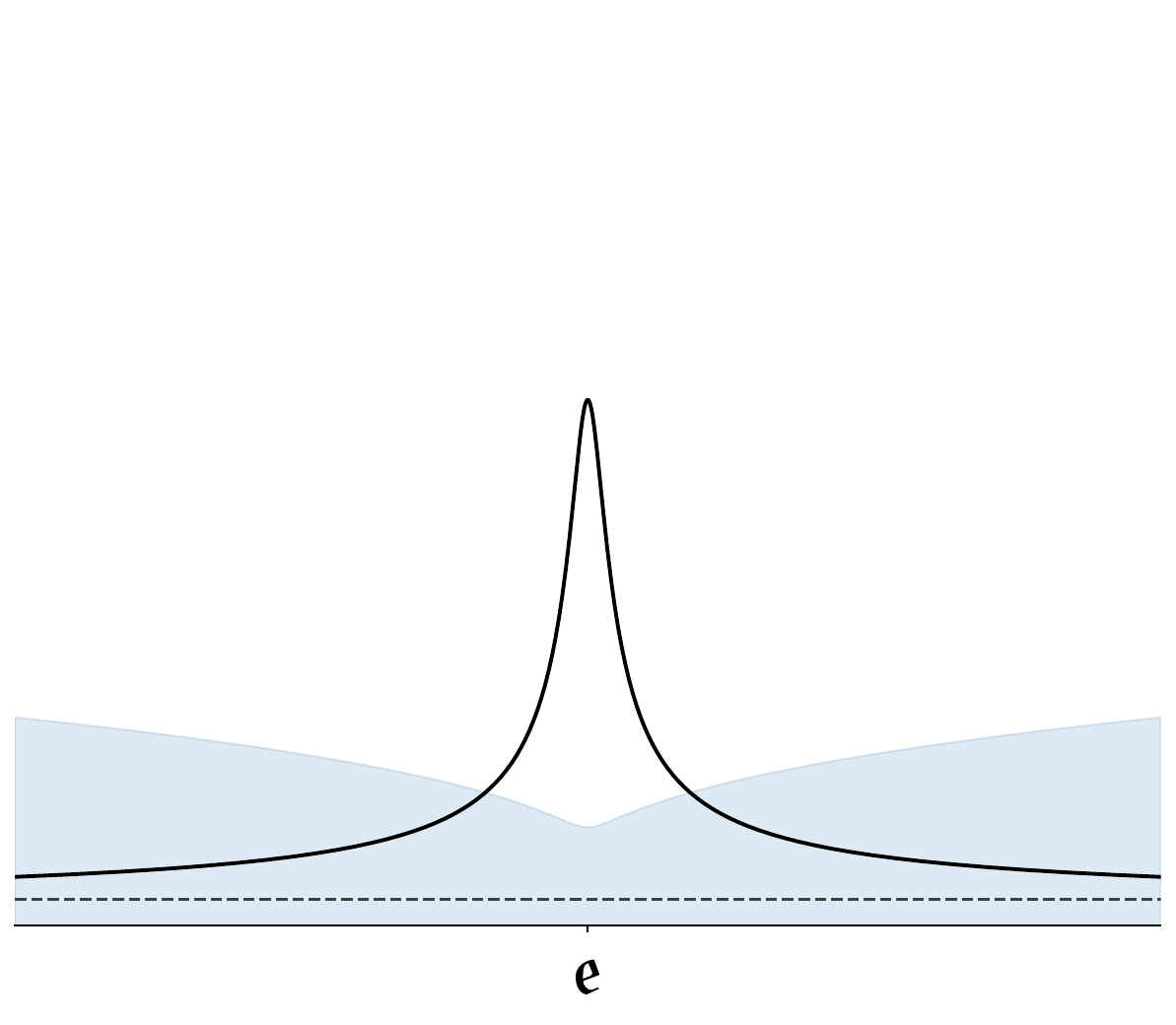}\\
			(c) $\Delta(\mathfrak{e}) = 0, \quad  \rho(\mathfrak{e}) > 0$
		\end{minipage}
		\caption{ 
			Depicted are the profiles of the anomalous fluctuation amplitude $\other{\alpha}(E,E)^{-2}$ 
			as a function of the energy $E$ in the three near-cusp spectral regimes: around a \textit{small gap} in the support of $\rho$ on panel (a), in the vicinity of a sharp \textit{cusp} on panel (b), and near a \textit{small positive local minimum} of $\rho$ on panel (c). 
			Superimposed is the blue shaded region indicating the density of states~\(\rho(E)\). 
			The horizontal dashed grey lines mark the standard ETH fluctuation scale.
			The vertical dotted lines in panel (a) indicate the end-points of the gap in the support of $\rho$.} 
		\label{fig:alphas}
	\end{figure}
	
	\section{Multi-Resolvent Local Laws: Proof of Theorem \ref{thm:main}}\label{sec:LL}
	The goal of this section is to give the proof of our main result, Theorem \ref{thm:main},  based on a two-resolvent local law formulated in Theorems \ref{thm:locallawreg}--\ref{thm:locallawnonreg} below. We start by collecting various preliminaries in Section \ref{subsec:prelim}, including bounds on the deterministic approximation $M$ in 
	Theorem \ref{prop:M2_bounds}. Afterwards, in Section \ref{subsec:LL}, we formulate the just mentioned local law and discuss various extensions in Section \ref{subsec:interpret}. Finally, in Section \ref{subsec:ETHProof}, we conclude the proof of Theorem \ref{thm:main}. The rest of this paper, starting from Section \ref{sec:M_flow}, is devoted to establishing 
	Theorem \ref{prop:M2_bounds} and the local laws in Theorems \ref{thm:locallawreg}--\ref{thm:locallawnonreg}. 
	
	\subsection{Preliminaries} \label{subsec:prelim} We begin by collecting a few preliminaries. 
	
	\subsubsection{Stability operator}
	The key object governing the size of both the deterministic approximation to the two-resolvent chain $G(z)AG(w)$ and its fluctuations is the \textit{two-body stability operator}, defined by 
	\begin{equation} \label{eq:stab_def}
		\mathcal{B}_{z, w}[\,\cdot\,] := \mathrm{Id}[\,\cdot\,] - M(z)\mathcal{S}[\,\cdot\,]M(w), \qquad z,w \in \C \setminus \R \,, 
	\end{equation}
	where $\mathcal{S}$ is the self-energy defined in \eqref{eq:S_def}.  
	In the sequel we show that the operator $\mathcal{B}_{z, w}$ has at most one small eigenvalue (see Lemma~\ref{lemma:stab_bound}), while the remainder of its spectrum is separated from zero by a disk of small, $N$-independent radius.
	
	Next, for any $z,w \in \mathbb{C}\backslash\mathbb{R}$, we introduce the \textit{two-body stability factor} $\other{\beta}(z,w)$\footnote{
		Note that the definition of $\other{\beta}(z,w)$ is not symmetric in its arguments: in general, $\other{\beta}(z,w) \neq \other{\beta}(w,z)$.
		Nevertheless, we show in the sequel that these two quantities are comparable up to multiplicative constants of order one (see \eqref{eq:beta_sym}).
	}, defined as
	\begin{equation} \label{eq:betaf_def}
		\other{\beta}(z, w) :=  1\wedge \begin{cases}
			\rho(z)^{-1}\Im z  + \rho(z)^2 + |\rho(z) \sigma(z)|   + \bigl(|\rho(z)| + |\sigma(z)|\bigr)^{1/2} |z-w|^{1/2} + |z-w|^{2/3}, \quad &(\Im z)(\Im w) > 0,\\
			\min\Bigl\{ |\rho(z)|^{-1} |z-w|, \, \other{\beta}(z, \overline{w})\Bigr\}, \quad &(\Im z)(\Im w) < 0.
		\end{cases}
	\end{equation}
	Here, we recall that the real-valued function $\sigma(z)$ for $z\in\mathbb{H}$ is defined in \eqref{eq:sigma_def},
	and we extend $\sigma(z)$ into the lower half-plane via $\sigma(\overline{z}) := \sigma(z)$. 
	In the sequel, we show that $\other{\beta}(z,w)$ controls the size of the smallest eigenvalue and, consequently, controls the norm of the inverse operator $\mathcal{B}_{z, w}^{-1}$, see Lemma~\ref{lemma:stab_bound} below.

	We now provide an informal explanation of the terms appearing in the definition of $\other{\beta}(z,w)$ in \eqref{eq:betaf_def}. In the sequel we show that, in the relevant regime of $\other{\beta}$ being small, it is behaves like $\other{\beta}(z,w) \sim |z-w| / |\langle M(z) \rangle - \langle M(w)\rangle|$, see \eqref{eq:beta_assymp}.  For $z, w$ in the same half-plane, the first three summands, $\rho(z)^{-1}\Im z + \rho(z)^2 + |\rho(z)\sigma(z)|$, correspond to the one-body stability factor, governing the linear stability of the MDE \eqref{eq:MDE} at the point $z$; the second term, $(|\rho(z)| + |\sigma(z)|)^{1/2} |z-w|^{1/2}$ dominates in the vicinity of a regular edge (where $|\sigma(z)|\sim 1$), and arises from the square-root behavior of $\rho$ in this regime; finally, the last term $|z-w|^{2/3}$ is dominant in the near-cusp regimes (where $|\sigma(z)| \approx 0$), and corresponds to the cubic-root behavior of $\rho$. 	
	We remark that although the definition of $\other{\beta}(z,w)$ is not symmetrical, this asymmetry plays no role due to~\eqref{eq:beta_sym} below.

	Throughout the rest of this paper, we fix a (large) $N$-independent constant $C_{\mathrm{bdd}} > 0$, and carry out the analysis in a large rectangle of size\footnote{Our results remain valid outside of this region, after adjusting the control quantities by an appropriate power of the multiplicative factor $1 + |z_j|$ and its inverse (see, e.g., \cite{cuspuniv}, where such factors were carefully recorded). For brevity and since local laws in the large $|z_j|$ regime are largely uninteresting, we do not pursue this extension here. } $C_{\mathrm{bdd}}$ centered at the origin. We regard  $C_{\mathrm{bdd}}$ as an additional model parameter in the sequel. Further, we restrict our analysis to the domain where the solution $M(z)$ is uniformly bounded, $\bddD \subset \mathbb{C}$,  defined as 
	\begin{equation} \label{eq:bdd_def}
		 \bddD := \bigl\{ E+\ii \eta \in \mathbb{C} \,:\, \dist(E, \mathcal{I}) \le  \tfrac{1}{2}c_M, \quad |E| \le C_{\mathrm{bdd}}, \quad 0 < |\eta| \le C_{\mathrm{bdd}}  \bigr\}, 
	\end{equation}
	where $\mathcal{I} \equiv \mathcal{I}_{c_M, C_M}$ is the set of admissible energies, defined in \eqref{eq:adm_E}. By Assumption \ref{ass:Mbdd}, $M$ is bounded in a $\tfrac{1}{2}c_M$-vicinity of $\bddD$;  the additional security distance is necessary to perform shape analysis for $z \in\bddD$. 
	
	We introduce the quantity $\other{\alpha}$, which controls the size of the anomalous fluctuation mode, and appears in the local law error estimates below (see Theorem~\ref{thm:locallawreg}). It extends $\other{\alpha}(E_1, E_2)$ from \eqref{eq:ETH_alpha_jk} away from the real line (see Lemma \ref{lemma:alphabeta}~(iii) below) and is defined as 
	\begin{equation} \label{eq:alpha_defin}
	 	\other{\alpha}(z,w) := |\sigma(z)| + \other{\beta}(z,w)^{1/2}, \qquad z,w \in \mathbb{H} \,. 
	\end{equation}
	\begin{lemma}[Properties of the Two-Body Quantities] \label{lemma:alphabeta}
		The quantities $\other{\beta}(z,w)$ and $\other{\alpha}(z,w)$, defined in \eqref{eq:betaf_def} and \eqref{eq:alpha_defin}, respectively, satisfy the following properties:
		\begin{itemize}
			\item[(i)] The quantities $\other{\beta}$ and $\other{\alpha}$ are symmetric (up to multiplicative constants of order unity), 
			\begin{equation} \label{eq:beta_sym}
				\other{\beta}(\bar z, \bar w) = \other{\beta}(z,w) \sim \other{\beta}(w,z), \qquad z,w \in \bddD,
			\end{equation}
			\begin{equation} \label{eq:alpha_sym}
				 \other{\alpha}(z, w) \sim \other{\alpha}(w, z) \lesssim 1, \qquad z,w \in \bddD\cap\mathbb{H}. 
			\end{equation}
			\item[(ii)] For $z \in \bddD\cap\mathbb{H}$ satisfying $\Re z \in \supp\,\rho$, we have the comparison
			\begin{equation} \label{eq:alp_beta_in_supp}
				\other{\beta}(z,z) \sim \rho(z)\other{\alpha}(z,z). 
			\end{equation}
			\item[(iii)]  For all $E_1, E_2 \in \mathcal{I}$, the quantity $\other{\alpha}(z_1, z_2)$ with $z_j := E_j +\ii\eta_j$ and $0 < \eta_j \lesssim 1$ for $j \in \{1,2\}$, satisfies
			\begin{equation} \label{eq:alp_eta_comp}
				\other{\alpha}(z_1, z_2) \sim \other{\alpha}(E_1, E_2) + \eta_1^{1/3} + \eta_2^{1/3}, 
			\end{equation} 	
			where $\other{\alpha}(E_1, E_2)$ is defined \eqref{eq:ETH_alpha_jk}. 
			In particular,
			\begin{equation} \label{eq:alpha_realline}
				\other{\alpha}(z_1, z_2) \gtrsim \other{\alpha}(E_1, E_2) + N^{-1/4},
			\end{equation}
			provided $\eta_j \ge \etaf(E_j)$ for $j \in \{1,2\}$, where $\etaf(E)$ is the local fluctuation scale defined in \eqref{eq:etaf_def}. 
		\end{itemize}
	\end{lemma}
	We prove Lemma~\ref{lemma:alphabeta} in Appendix~\ref{app:beta_prop}.

	\nc
 	\begin{lemma} [Two-Body Stability] \label{lemma:stab_bound}
 		Let $z, w$ be a pair of spectral parameters in $\bddD$. Then, the \textit{two-body stability operator} $\mathcal{B}_{z, w}$, defined in \eqref{eq:stab_def} satisfies the bound
 		\begin{equation} \label{eq:stab_beta_bound}
 			\norm{\mathcal{B}_{z, w}^{-1}}_{\mathrm{hs}\to\mathrm{hs}} + \norm{\mathcal{B}_{z, w}^{-1}} \lesssim \other{\beta}(z, w)^{-1},
 		\end{equation}
 		where $\other{\beta}(z, w)$ is defined in \eqref{eq:betaf_def}. 
 		
 		Furthermore, there exist   (small) \nc positive thresholds $\beta_*, \varepsilon_* \sim 1$, such that, if $z,w \in \bddD$ satisfy $\other{\beta}(z,w) \le \beta_*$, then 
 		\begin{equation}\label{eq:stab_well_struc}
 			\norm{(\zeta\,\mathrm{Id} - \mathcal{B}_{z, w})^{-1}}_{\mathrm{hs}\to\mathrm{hs}} + \norm{(\zeta\,\mathrm{Id} - \mathcal{B}_{z, w})^{-1}} \lesssim 1, 
 		\end{equation}
 		for all $\zeta \in\mathbb{C}$ satisfying $|\zeta| \ge \varepsilon_*$ and $|1 - \zeta| \ge 1 - 2\varepsilon_*$;
 		moreover  the disk of radius $\varepsilon_*$ around the origin contains 
 		a single eigenvalue with algebraic multiplicity one, that is,  for the corresponding spectral projection we have 
 		\begin{equation} \label{eq:Pi_rank1}
 			\mathrm{rank}\,\Pi_{z,w} = 1, \qquad \Pi_{z,w} := \frac{1}{2\pi\ii} \oint_{|\zeta| = \varepsilon_*} (\zeta\, \mathrm{Id}- \mathcal{B}_{z, w})^{-1} \mathrm{d}\zeta. 
 		\end{equation}
 		The small isolated eigenvalue $\beta_{z,w} = \Tr[\mathcal{B}_{z,w}\Pi_{z,w}]$ of $\mathcal{B}_{z,w}$ satisfies 
 		\begin{equation} \label{eq:beta_conj}
 			\beta_{z,w} = \overline{\beta_{\bar w, \bar z}}, 
 		\end{equation}
 		\begin{equation} \label{eq:beta_assymp}
 			|\beta_{z,w}| \sim \biggl\lvert \frac{z-w}{\bigl\langle M(z) - M(w) \bigr\rangle} \biggr\rvert \sim \other{\beta}(z,w) \,. 
 		\end{equation} 
 	\end{lemma}
 	We prove Lemma~\ref{lemma:stab_bound} in Section~\ref{sec:stab}.   
	The \textit{one-body} stability operators, corresponding to the special cases $z = w$ and $z = \overline{w}$ in the most general $C^*$-algebra setup were studied in \cite[Section 5]{AEK2020}, see Lemma~\ref{lemma:stab1} below. 
	A complete analysis of the two-body stability operator for Wigner-type matrices with a diagonal deformation $D$ was carried out in \cite{riabov2025linear}. 
 	 
\subsubsection{Regularity}	The central concept of our proof is the notion of \emph{regular observables}, which are defined as follows: 
For a pair of spectral parameters $z,w \in \mathbb{C}\backslash\mathbb{R}$ and an arbitrary deterministic $B \in \C^{N \times N}$, the deterministic approximation $M(z_1,B, z_2)$ to the resolvent chain $G(z) B G(w)$ is given by
 	\begin{equation} \label{eq:M2_def}
 		M(z, B, w) := \mathcal{B}_{z, w}^{-1} \bigl[M(z) B \, M(w)\bigr],
 	\end{equation}
 	where the \textit{two-body stability operator}  $\mathcal{B}_{z, w}$ is  defined in \eqref{eq:stab_def}. Moreover, we denote the deterministic approximation to the chain $\Im G(z) B  \Im G(w)$ by $\widehat{M}(z, B, w)$, that is
 	\begin{equation} \label{eq:ImM2_def}
 		\widehat{M}(z, B, w) := \frac{1}{4}\Bigl(M(\overline{z}, B, w) + M(z, B, \overline{w}) - M(z, B, w) - M(\overline{z}, B, \overline{w})\Bigr). 
 	\end{equation}

 	Equipped with this notation, we now define the regular observables, which we consistently denote by $A$ throughout this section, while the letter $B$ represents generic observables.
 	
 	\begin{definition}[Regular observables] \label{def:reg}
	Let  $z, w \in \C \setminus \R$. 
Then, we say that a deterministic matrix $A\in\mathbb{C}^{N\times N}$ is \emph{regular} with respect to $(z, w)$, or simply $(z, w)$-regular,  if and only if
\begin{equation} \label{eq:reg_def}
	\Big\langle \widehat{M}(z, I, w) A\Big\rangle = 0.
\end{equation}
Moreover, for an arbitrary $A \in \C^{N \times N}$, we denote its \emph{$(z, w)$-regularization} by
\begin{equation} \label{eq:regA}
\reg{A} = \reg{A}^{z, w} := A - \Upsilon(z, A, w) I \,, \quad \text{with} \quad \Upsilon(z, A, w) := \frac{\big\langle \widehat{M}(z, I, w) A\big\rangle}{\big\langle \widehat{M}(z, I, w) \big\rangle} \,. 
\end{equation}
In particular, $A$ is $(z, w)$-regular, if and only if $A = \reg{A}^{z, w}$. 
 	\end{definition}
 	It easily follows from Definition \ref{def:reg}, that the notion of regularity is invariant under interchanging $z$ and $w$, and taking complex conjugates in one or both arguments. Additionally, since $\widehat{M}(z, I,w)$ is self-adjoint, $A$ is $(z, w)$-regular, if and only if $A^*$ is $(z,w)$-regular:
	\begin{equation}\label{regsymm}
	 \reg{A}^{z,w} = \reg{A}^{w,z},   \quad  \reg{A}^{z,w} =\reg{A}^{z^{(*)}, w^{(*)}}, \quad (\reg{A})^* = (A^*)^\bullet.
	\end{equation}
	
	The main motivation for this definition is that  the shift $\Upsilon$ is the solution of the minimization problem	
		\begin{equation}\label{yminzw}
		\min_{y\in \C} \bigl \langle \widehat{M}(z, A-yI , w) (A-yI)^* \bigr\rangle.
	\end{equation}
	\nc
	 As already explained in the introduction, 
 	we note that this concept of regularity differs from the one used in \cite{iid, equipart, cipolloni2024eigenvector} for deformed Wigner matrices and \cite{WigTypeETH} for Wigner-type matrices. 
 	It is not only more symmetric,  but,  most importantly, it more accurately compensates
 	for the instability of $\mathcal{B}_{z, w}$.  For Wigner matrices, however, 
 	all regularity concepts are the same  and they all amount to $A$ being traceless, $\langle A\rangle=0$,
 	independently of $z, w$. 
 	We remark that for $z = w = E + \ii \eta$ with $\eta \to +0$, our Definition~\ref{def:reg} of regularity is consistent with $\langle \Im M(E) A \rangle = 0$, which was used to define regularity in  
	\cite{iid, equipart, cipolloni2024eigenvector, WigTypeETH},  see Footnote \ref{ftn:regularities} below).

Regular observables are designed to counteract the divergence coming from the small eigenvalue of the two-body stability operator $\mathcal{B}_{z, w}$ by having a sufficiently small projection onto the corresponding rank-one eigenspace. Hence, considering $\reg{A}$ instead of $A$ is really important only on the sub-domain in $\bddD\times\bddD$ where the stability factor with the arguments in the opposite half-planes is sufficiently small. More precisely, for any constant $c > 0$, we define 
 	\begin{equation} \label{eq:regD_def}
 		\regD_c := \bigl\{ (z,w) \in \bddD\times\bddD \, : \, \max\{\other{\beta}(z^+, w^-), \other{\beta}(w^+, z^-)  \}\le c \bigr\} \,. 
 	\end{equation}
 	where, for a spectral parameter $z\in\mathbb{C}$, we use the notation
 	\begin{equation}
 		z^\pm := \Re z \pm \ii|\Im z|. 
 	\end{equation}
 	In essence, the  domain $\regD_c \subset \mathbb{C}^2$ consists of pairs of spectral parameters $(z,w)$ that are both close to $\supp\,\rho \subset \mathbb{R}$, defined in \eqref{eq:scDOS}, and close to each other.
 	In the sequel, we often use \eqref{eq:regD_def} with $c = \beta_*$, where $\beta_* \sim 1$ is the threshold from Lemma \ref{lemma:stab_bound}, and hence we abbreviate 
 	\begin{equation} \label{eq:regD_beta}
 		\regD \equiv \regD_{\beta_*}~.
 	\end{equation}

 \subsubsection{Bounds on the deterministic approximation}	
In this section, we present the optimal bounds on the deterministic approximations $M(z, A, w)$ and  $\widehat{M}(z, A, w)$ for regular matrices.

 	 	To this end, we now introduce a convenient norm on the space of $N\times N$ matrices: For all $z,w \in \mathbb{H}$, 
 	 	 we introduce the pair of deterministic matrices $V(z,w)$ and $U(z,w)$, given by  
 	 	\begin{equation} \label{eq:V2_def}
 	 		V(z, w) := \frac{M(z) - M(w)^*}{z - \overline{w}}, \qquad U(z, w) := \frac{M(z) - M(w)}{z-w},
 	 	\end{equation}
 	 	and the associated projected version of $U(z,w)$, $U^{\perp V}(z,w)$, given by
 	 		\begin{equation} \label{eq:UperpV}
 	 		U^{\perp V}(z,w) :=  U(z,w) - \frac{\bigl\langle V(z,w)^* U(z,w)  \bigr\rangle}{\norm{V(z,w)}_\mathrm{hs}^2} V(z,w).
 	 	\end{equation}		
 	 	 Let $\vecL(z,w)$ denote the following convenient rescaling of $U^{\perp V}(z,w)$, defined in \eqref{eq:UperpV},  
 	 	\begin{equation} \label{eq:L_def}
 	 		\vecL(z, w) := \frac{\other{\beta}(z,w)}{\rho(w)} U^{\perp V}(z,w), \qquad  z, w \in  \mathbb{H}.
 	 	\end{equation} 
The factor $\rho(w)$ is the natural normalization since $U(z,w)/\langle U(z,w)\rangle$ is an order $\rho(w)$ perturbation of $V(z,w)/\langle V(z,w) \rangle$, as we show in the sequel. 
		The factor $\other{\beta}(z, w)$ is introduced to ensure useful regularity properties as a function of $z$ and $w$ (see~\eqref{eq:L_reg}).  
		
As was explained in the introduction, the anomalous $\other{\alpha}^{-1}$-large fluctuation in the local law and the ETH is generated by the discrepancy between the unstable direction of the opposite and same half-plane stability operators, $\mathcal{B}_{z, \bar{w}}$ and $\mathcal{B}_{z,w}$, respectively. The matrices $V$ and $U$ are natural approximations for these unstable directions of $\mathcal{B}_{z, \bar{w}}$ and $\mathcal{B}_{z,w}$, respectively, and the vector $R$ then measures the aforementioned discrepancy between them. 
 	\begin{definition} [Observable Norm]\label{def:triple}
 		Let  $(z, w) \in \regD$, defined in \eqref{eq:regD_beta}. 
		Then, we define the (quasi-)norm $\vertiii{\cdot}_{z, w}$ on the space of matrices $B \in \mathbb{C}^{N\times N}$ by
 		\begin{equation} \label{eq:tripplenorm}
 			 \vertiii{B}_{z, w} := \frac{ \bigl\lvert \bigl\langle \vecL(z^+, w^+) \reg{B}^{z, w}  \bigr\rangle \bigr\rvert}{\other{\alpha}(z^+, w^+)} + \norm{\reg{B}^{z, w}  }_{\mathrm{hs}}\,, \qquad \reg{B}^{z, w} := B - \Bigl\langle B \frac{\widehat{M}(z, I, w)}{\langle \widehat{M}(z, I, w)\rangle } \Bigr\rangle \,I,
 		\end{equation} 
 		where the quantity $\other{\alpha}(z^+, w^+)$ \nc and the matrix $\vecL$ are defined in \eqref{eq:alpha_defin} and \eqref{eq:L_def}, respectively. 
 		\end{definition}
		We remark that in the definition of $U^{\perp V}$ in \eqref{eq:UperpV} (and hence in $\vecL$ from \eqref{eq:L_def}) we chose the spectral parameters to be in the upper half-plane for concreteness. The alternative definition with spectral parameters in the lower half-plane would be equally valid. 
 		
			The matrix $\vecL$ and observable norm  $\vertiii{\cdot}$ defined above satisfy the following properties.
			\begin{lemma} [Properties of $\vecL$ and  $\vertiii{\cdot}$] \label{lemma:tripple_prop}
				Let $(z, w) \in \regD$, defined in \eqref{eq:regD_beta}. 
				Then, the matrix ${\vecL}(z, w)$ satisfies 
				\begin{equation} \label{eq:L_bound}
					\bigl\lVert {\vecL}(z^+, w^+)\bigr\rVert_\mathrm{hs} \lesssim 1. 
				\end{equation}
				
				 The quasi-norm $\vertiii{\cdot}_{z, w}$ vanishes identically on the span of the identity matrix $I \in \mathbb{C}^{N\times N}$, that is,
				\begin{equation} \label{eq:tripl_I=0}
					\vertiii{B + \zeta\, I }_{z, w} = \vertiii{B}_{z, w}, \qquad \zeta \in \mathbb{C}, \qquad B \in \mathbb{C}^{N\times N}. 
				\end{equation} 
				Furthermore, the norm $\vertiii{\cdot}_{z, w}$ satisfies the following symmetries (up to multiplicative constants of order unity)
				\begin{equation} \label{eq:symmetries}
					\vertiii{B^*}_{w, z} \sim \vertiii{B}_{z, w} \sim \vertiii{B}_{w, z},
				\end{equation}
				for all $B \in \mathbb{C}^{N\times N}$. 	
			\end{lemma} 
			We defer the proof of Lemma~\ref{lemma:tripple_prop} to Section~\ref{sec:beta_reg}. 
 	Equipped with the norm $\vertiii{\cdot}_{z, w}$, the bounds on the two-body deterministic approximations can be expressed as follows. 	
 	\begin{theorem} [$M$-Bounds] \label{prop:M2_bounds} For $z_1, z_2 \in \C\setminus \R$, denote $\eta_j := |\Im z_j|$, $\varkappa_j := \dist(z_j, \supp\, \rho)$, and $\rho_j := \rho(z_j)$, all for $j \in \indset{2}$. Then, we have the following bounds on the deterministic approximation: 
 		\begin{itemize}
\item[(i)] \textnormal{[Perturbative spectral domain]}   		Let $(z_1, z_2) \in \regD$, defined in \eqref{eq:regD_beta}. Then, for any $(z_1, z_2)$-regular observables $A_1, A_2 \in \mathbb{C}^{N\times N}$ we have 
 		\begin{equation} \label{eq:ImM2_bounds}
	\Bigl\lvert \bigl\langle \widehat{M}(z_1, A_1, z_2) A_2 \bigr\rangle \Bigr\rvert \lesssim |\rho_1\rho_2|\, \vertiii{A_1}_{z_1, z_2} \, \vertiii{A_2}_{z_2, z_1}~,
\end{equation}
and  
\begin{equation} \label{eq:M2_aux_bound}
	\bigl\lvert \bigl\langle M(z_1, A_1, z_2) \bigr\rangle \bigr\rvert \lesssim \vertiii{A_1}_{z_1, z_2}~.
\end{equation}
We also note that $\langle \widehat{M}(z_1, A_1, z_2) \rangle = 0$ by definition of regularity. 
 		
\item[(ii)] \textnormal{[Non-perturbative spectral domain]}   	Let $(z_1, z_2) \in (\bddD)^2 \setminus \regD$. Then, for any observables $B_1, B_2 \in \mathbb{C}^{N\times N}$ we have 
 		\begin{equation} \label{eq:ImM2_nobeta}
 			\Bigl\lvert \bigl\langle \widehat{M}(z_1, B_1, z_2) B_2 \bigr\rangle \Bigr\rvert \lesssim |\rho_1\rho_2|\,  \norm{B_1}_{\mathrm{hs}}\norm{B_2}_{\mathrm{hs}}~,
\end{equation}
as well as 
\begin{equation} \label{eq:M2_nobeta}
 			\Bigl\lvert \bigl\langle M(z_1, B, z_2) \bigr\rangle \Bigr\rvert \lesssim  \norm{B}_{\mathrm{hs}}~,
\end{equation}

\item[(iii)] Furthermore, for all $z_1, z_2 \in \bddD$, the quantity $\langle \widehat{M}(z_1, I, z_2) \rangle$ satisfies
\begin{equation} \label{eq:Mhat_comp}
	\big| \langle \widehat{M}(z_1, I, z_2) \rangle  \big| \sim |\rho_1 \rho_2| \Bigg[\frac{ |\rho_1|^{-1}\eta_1 +  |\rho_2|^{-1}\eta_2 }{|z_1 - z_2|^2 + \varkappa_1^2 + \varkappa_2^2}\Bigg]~.
\end{equation}
 		\end{itemize}
 	\end{theorem}
 	We prove Theorem~\ref{prop:M2_bounds} in Section~\ref{sec:Mboundsproof}.  We point out that the bound provided in \eqref{eq:ImM2_bounds} in Theorem \ref{prop:M2_bounds}~(i) is \emph{optimal} and refer to Appendix \ref{app:Moptimal} for a proof of this fact.

 	\subsection{Local laws for one and two resolvents} \label{subsec:LL}
 	Equipped with the definitions and notation spelled out in the previous sections, we present the local laws that we prove in this paper. 
 	
 	The first theorem below concerns two resolvent local laws with spectral parameters in the \emph{perturbative} spectral domain, i.e.~in which one has a small eigenvalue of the two-body stability operator. 
 	\begin{theorem}[Two resolvent local laws in the perturbative spectral domain] \label{thm:locallawreg} Let $H$ be a $N \times N$ correlated random matrix satisfying Assumptions \ref{ass:boundedexp}--\ref{ass:cumulants}. Fix $\varepsilon> 0$ and consider spectral parameters $(z_1, z_2) \in \regD$ (recall \eqref{eq:regD_def} and \eqref{eq:regD_beta}) and denote $\eta_j := |\Im z_j|$,  $\varkappa_j := \dist(z_j, \supp \rho)$, and $\rho_j := \rho(z_j)$, all for $j \in \indset{2}$. Assume that $\ell := \min_j  |\rho_j| \eta_j  \ge N^{-1+\varepsilon}$ and let $A_1, A_2 \in \C^{N \times N}$ be $(z_1, z_2)$-regular deterministic observables. Then, denoting $G_j := G(z_j) = (H- z_j)^{-1}$, the following local laws hold. 
 		\begin{subequations}
 		\begin{itemize}
	\item[(a)] For two $\Im G$'s and two regular observables we have
	\begin{equation} \label{eq:ImG2_ll}
		\biggl\lvert \Bigl\langle \bigl(\Im G_1  {A}_1 \Im G_2 - \widehat{M}(z_1, {A}_1, z_2)\bigr)  {A}_2  \Bigr\rangle \biggr\rvert \prec \frac{|\rho_1\rho_2|}{\sqrt{N\ell}} \, \vertiii{A_1}_{z_1, z_2} \vertiii{A_2}_{z_1, z_2} \,. 
	\end{equation}
		\item[(b)] For two $\Im G$'s and one regular observable we have 
	\begin{equation} \label{eq:ImG2_ll1A}
		\biggl\lvert \Bigl\langle \Im G_1  {A}_1 \Im G_2 - \widehat{M}(z_1, {A}_1, z_2)  \Bigr\rangle \biggr\rvert \prec \frac{|\rho_1\rho_2|}{N\ell} \, \Bigg[\frac{ |\rho_1|^{-1}\eta_1 +  |\rho_2|^{-1}\eta_2 }{|z_1 - z_2|^2 + \varkappa_1^2 + \varkappa_2^2}\Bigg]^{1/2} \vertiii{A_1}_{z_1, z_2}  \,. 
	\end{equation}
			\item[(c)] For two $\Im G$'s and no regular observable we have
	\begin{equation} \label{eq:ImG2_ll0A}
		\biggl\lvert \Bigl\langle \Im G_1   \Im G_2 - \widehat{M}(z_1, I, z_2)  \Bigr\rangle \biggr\rvert \prec \frac{|\rho_1\rho_2|}{N\ell} \, \Bigg[\frac{ |\rho_1|^{-1}\eta_1 +  |\rho_2|^{-1}\eta_2 }{|z_1 - z_2|^2 + \varkappa_1^2 + \varkappa_2^2}\Bigg]   \,.  
	\end{equation}
				\item[(d)] Finally, for two $G$'s and one regular observable we have (analogously for $1 \leftrightarrow 2$)
\begin{equation} \label{eq:GG_ll}
	\biggl\lvert \bigl\langle G_1  {A}_1 G_2 - M(z_1, {A}_1, z_2)  \bigr\rangle \biggr\rvert \prec \sqrt{\frac{|\rho_1 \rho_2|}{N^2\ell \eta_1 \eta_2}}\, \vertiii{A_1}_{z_1, z_2} \,. 
\end{equation}
\end{itemize}
 		\end{subequations}
 	\end{theorem}
 	The proof of Theorem \ref{thm:locallawreg} is given in Section \ref{sec:zigzag}. We point out that the main objective in Section \ref{sec:zigzag} is to prove \eqref{eq:ImG2_ll}. The other three local laws in Theorem \ref{thm:locallawreg} concern auxiliary objects arising within the \emph{zigzag strategy}, a recursive tandem of the characteristic flow method (\emph{zig}) and a Green function comparison (GFT) argument (\emph{zag}), which we follow to obtain Theorem \ref{thm:locallawreg}. We point out that the local law \eqref{eq:ImG2_ll0A} without regular observables could, in principle, be obtained completely independently of the other three ones, conducting an independent zigzag proof. In the actual proof in Section~\ref{sec:zigzag} (and the subsequent Sections \ref{sec:zig_proof}--\ref{sec:zagproof}), however, we follow all four quantities in Theorem \ref{thm:locallawreg} simultaneously, for brevity of the presentation.

 	Next, we turn to two resolvent local laws in the \emph{non-perturbative} regime, in which there is no small eigenvalue of the two-body stability operator.

 	\begin{theorem}[Two resolvent local laws in the non-perturbative spectral domain] \label{thm:locallawnonreg}
 		Let $H$ be a $N \times N$ correlated random matrix satisfying Assumptions \ref{ass:boundedexp}--\ref{ass:cumulants}. Fix $\varepsilon > 0$ and consider spectral parameters $(z_1, z_2) \in (\bddD \times \bddD) \setminus \regD$ (recall \eqref{eq:regD_def}) and denote $\eta_j := |\Im z_j|$,  $\varkappa_j := \dist(z_j, \supp \rho)$, and $\rho_j := \rho(z_j)$, all for $j \in \indset{2}$. Assume that $\ell := \min_j  |\rho_j| \eta_j  \ge N^{-1+\varepsilon}$ and let $B_1, B_2 \in \C^{N \times N}$ be arbitrary deterministic observables.  Then, denoting $G_j := G(z_j) = (H- z_j)^{-1}$, the following local laws hold. 
 		 		\begin{subequations}
 			\begin{itemize}
 				\item[(a)] For two $\Im G$'s and two general observables we have
 		\begin{equation} \label{eq:ImM2_ll_nobeta}
	\biggl\lvert \Bigl\langle \bigl(\Im G_1  B_1 \Im G_2 - \widehat{M}(z_1, B_1, z_2)\bigr)  B_2  \Bigr\rangle \biggr\rvert \prec \frac{|\rho_1\rho_2|}{\sqrt{N\ell}} \, \norm{B_1}_{\mathrm{hs}} \norm{B_2}_{\mathrm{hs}} ,
\end{equation}
 				\item[(b)] For two $\Im G$'s and one general observable we have 
 				\begin{equation} \label{eq:ImG2_ll1Anonreg}
 					\biggl\lvert \Bigl\langle \Im G_1  B_1 \Im G_2 - \widehat{M}(z_1, B_1, z_2)  \Bigr\rangle \biggr\rvert \prec \frac{|\rho_1\rho_2|}{N\ell} \, \Big[ |\rho_1|^{-1}\eta_1 +  |\rho_2|^{-1}\eta_2 \Big]^{1/2} \norm{B_1}_{\rm hs}  \,. 
 				\end{equation}
 				\item[(c)] For two $\Im G$'s and no regular observable we have
 				\begin{equation} \label{eq:ImG2_ll0Anonreg}
 					\biggl\lvert \Bigl\langle \Im G_1   \Im G_2 - \widehat{M}(z_1, I, z_2)  \Bigr\rangle \biggr\rvert \prec \frac{|\rho_1\rho_2|}{N\ell} \, \Big[ |\rho_1|^{-1}\eta_1 +  |\rho_2|^{-1}\eta_2 \Big]   \,.  
 				\end{equation}
 				\item[(d)] Finally, for two $G$'s and one general observable we have 
 		\begin{equation} \label{eq:GG_llnobeta}
	\biggl\lvert \bigl\langle G_1  B_1 G_2 - M(z_1, B_1, z_2)  \bigr\rangle \biggr\rvert \prec \sqrt{\frac{|\rho_1 \rho_2|}{N^2\ell \eta_1 \eta_2 }}\, \norm{B_1}_{\mathrm{hs}}.
\end{equation}
 			\end{itemize}
 		\end{subequations}
 	\end{theorem}
 	
 	 The proof of Theorem \ref{thm:locallawnonreg}, also based on the zigzag strategy, is given in Section~\ref{sec:proofnonreg}.  	A fundamental input of all our zigzag arguments are the corresponding $M$-bounds from 
	 Theorem \ref{prop:M2_bounds}, which, as mentioned above, are proven by a dynamical method as well (see Section \ref{sec:Mboundsproof}).

	 Finally, in the  following Theorem \ref{thm:singleG}, 
we state a single resolvent local law with regular observable. A similar local law 
with a general observable was already proven in~\cite[Theorem 2.8]{cuspuniv}, taking the form 
\begin{equation} \label{eq:generalLL}
	\big\lvert \bigl\langle \big(G(z)-M(z)\big)  B  \bigr\rangle \big\rvert \prec \frac{1}{N\eta}\,\norm{B}_{\rm hs},
\end{equation}
for $z \in \bddD$ with $N \eta |\rho(z)| \ge N^\epsilon$. In Theorem \ref{thm:singleG}, regularity of the observable improves 
the general bound \eqref{eq:generalLL}. It is a relatively simple consequence of the optimal  two-resolvent local laws
with regular observable; the proof is given in Section \ref{sec:proof1G}.
 	
 	\begin{theorem}[Single resolvent local law] \label{thm:singleG}
Let $z \in \bddD$ be a spectral parameter satisfying $\ell := |\rho(z)| \eta \ge N^{-1+\varepsilon}$ with $\eta := |\Im z|$,  $|z| \lesssim C$, and $\other{\beta}(z, \overline{z}) \le \beta_*$. Then, for any $(z,z)$-regular observable $A$ we have 
\begin{equation} \label{eq:Gll}
	\big\lvert \bigl\langle (G-M)  A  \bigr\rangle \big\rvert \prec \frac{|\rho|^{1/2}}{N\eta^{1/2}}\,\vertiii{A}_{z,z},
\end{equation}
where $G \equiv G(z)$, $\rho \equiv \rho(z)$, and $M \equiv M(z)$. 
 	\end{theorem}
We did not state a separate local law for $\Im G$ since
replacing $G$ with $\Im G$ does not give any  improvement in~\eqref{eq:Gll} 
as long as $\Re z$ is in the support of $\rho$	(unlike
for 2G local laws, cf. \eqref{eq:ImG2_ll1A}, \eqref{eq:GG_ll}). The situation outside of the support
is discussed in Section~\ref{subsub:outside}.

 	Finally, we mention that in Theorems \ref{thm:locallawreg}--\ref{thm:locallawnonreg} we only recorded those local laws needed in the proof of \eqref{eq:ImG2_ll} and \eqref{eq:ImM2_ll_nobeta} and thus our main result, Theorem \ref{thm:main}. Our method, however, easily yields local laws for $\langle \Im GA G \rangle$ and $\langle G \Im G  \rangle$ as well (cf.~\eqref{eq:newout}--\eqref{eq:newout2}).

\subsection{Interpretations, extensions and improvements} \label{subsec:interpret}
We have several comments on Theorems~\ref{thm:locallawreg}--\ref{thm:singleG}: 

\subsubsection{Effect of regularity}
Regularity of observables strongly influences the local law bounds: In case of a single resolvent, we find from \eqref{eq:Gll} and \cite[Theorem 2.8]{cuspuniv} that, uniformly in spectral parameters\footnote{For simplicity of the presentation, in this entire section, we assume that the set of admissible energies exhausts the entire real line, i.e.~$\mathcal{I} = \R$. } $z \in \mathbb{H}$ satisfying $\ell := \rho \eta \ge N^{-1+\epsilon}$ (where we denoted $\eta := \Im z$ and $\rho := \rho(z)$) and $|z| \le C$, we have
	\begin{equation} \label{eq:general1G}
\big| \langle (G-M)B \rangle\big| \prec \frac{\rho^{1/2}}{N \eta^{1/2}} \left( \frac{1}{\ell^{1/2}} \left| \frac{\langle  \widehat{M} B \rangle}{\langle  \widehat{M} \rangle} \right|  + \frac{\Vert \mathcal{R}[\reg{B}]\Vert_{\rm hs}}{\other{\alpha}} + \Vert \regreg{B} \Vert_{\rm hs}\right)
	\end{equation}
	for a general observable $B \in \C^{N \times N}$, for which we define 
	\begin{equation}
\reg{B} := B - \frac{\langle \widehat{M} B \rangle}{\langle  \widehat{M} \rangle} I \quad \text{and} \quad \regreg{B} := \reg{B} - \mathcal{R}[\reg{B}] \,. 
	\end{equation}
	Here, we used the abbreviations $G \equiv G(z)$, $\widehat{M} = \widehat{M}(z, I,z)$, $\other{\alpha} \equiv \other{\alpha}(z,z)$, and $\mathcal{R} = \mathcal{R}_z$ (recall \eqref{eq:R_def}). 
	
	Note that \emph{(first order) regularity} of an observable, i.e.~if $B = \reg{B}$, improves the local law bound by a factor $\sqrt{\ell}/\other{\alpha} \lesssim 1$ since the first term on the right-hand side of~\eqref{eq:general1G} is absent 
	as $\langle  \widehat{M} B \rangle=0$. This generalizes the \emph{$\sqrt{\eta}$-rule} obtained in the bulk for Wigner matrices in \cite{multiG, A2}, in which case $\ell \sim \eta$ and $\other{\alpha} \sim 1$. Moreover, in case of \emph{second order regularity}, i.e.~if $B = \reg{B} = \regreg{B}$, the local law further improves by a factor $\other{\alpha} \lesssim 1$ since
	both the first and second terms in the right-hand side of~\eqref{eq:general1G} are absent, as 
	$\langle  \widehat{M} B \rangle= 0 $ and $\mathcal{R}[\reg{B}]=  0 $. 
	In other words, any observable $B$ may be decomposed as
	$$
	B  = \frac{\langle \widehat{M} B \rangle}{\langle  \widehat{M} \rangle} I + 
\mathcal{R}[\reg{B}]  +  \regreg{B} \,. 
	$$
	The first term, proportional to the identity, corresponds to the largest fluctuation mode on the right-hand side~of \eqref{eq:general1G}; the second term, $\mathcal{R}[\reg{B}]$, corresponds to the second largest fluctuation mode on the right-hand side~of \eqref{eq:general1G}; finally, the last term corresponds to the smallest fluctuation mode on the right-hand side~of \eqref{eq:general1G}. We point out that, removing both the first and second term, corresponds to essentially removing both bad directions of the stability operators $\mathcal{B}_{z,\overline{z}}$ and $\mathcal{B}_{z,z}$. The last term, $\regreg{B}$, lies in the complement where the fluctuation in the local law is insensitive to the instabilities of the problem.

\subsubsection{Outside of the support} \label{subsub:outside}If the real part of some of the spectral parameters are away from the support of the scDOS, i.e. $\Re z_i \not\in \supp \rho$, then the local laws in Theorems \ref{thm:locallawreg}--\ref{thm:singleG} can be improved. These  improved laws have exactly the same form as before just the parameter $\ell$ is redefined; the new parameter $\widetilde{\ell}$
coincides with $\ell$ in the support $\rho$, otherwise $\widetilde{\ell}> \ell$.
We now state the local laws in a unified form, deferring a (sketch of) their proofs to Appendix \ref{app:outside}. 

In case of one resolvent, for $z \in \mathbb{H}$ define 
$$\varkappa := \dist(z, \supp \rho) \quad \text{and} \quad\widetilde{\ell} := \frac{\rho \varkappa^2}{\eta}.$$
 Then the local law \eqref{eq:general1G} holds true verbatim upon replacing $\ell \to \widetilde{\ell}$. Moreover, the imaginary part of the resolvent satisfies
	\begin{equation} \label{eq:generalImG}
	\big| \langle (\Im G-\Im M)B \rangle\big| \prec \frac{\rho}{N \widetilde{\ell}^{1/2}} \left( \frac{1}{\widetilde{\ell}^{1/2}} \left| \frac{\langle  \widehat{M} B \rangle}{\langle \widehat{M} \rangle} \right|  +  \, \frac{\Vert \mathcal{R}[\reg{B}] \Vert_{\rm hs} }{\other{\alpha}}+ \Vert \regreg{B} \Vert_{\rm hs}\right) \,. 
\end{equation}
In particular, comparing \eqref{eq:general1G} and \eqref{eq:generalImG}, we notice that 
taking the imaginary part improves the bound by a factor  $\sqrt{\ell/\widetilde{\ell}} = \sqrt{\rho \eta/\widetilde{\ell}} \le 1$ outside of the spectrum. 

We conclude the discussion of the case with one resolvent by specializing the observable to the rank-one operator $B = N \ket{\bm y} \bra{\bm x}$, to obtain the isotropic local law
\begin{equation}
	\big| (G - M)_{\bm x \bm y}\big| \prec \sqrt{ \frac{\rho}{ N\eta } } \, \Vert \bm x \Vert \, \Vert \bm y \Vert \qquad \text{and} \qquad
	\big| (\Im G - \Im M)_{\bm x \bm y}\big| \prec 
	\frac{\rho}{\sqrt{N \widetilde{\ell}}} \, \Vert \bm x \Vert \, \Vert \bm y \Vert \,. 
\end{equation}

In case of two resolvents with spectral parameters $z_1, z_2 \in \C$ and denoting $$\varkappa_i := \dist(z_i, \supp \rho) \quad \text{for} \quad i \in \indset{2} \quad \text{and} \quad\widetilde{\ell} := \min_{i \in \indset{2}} \big( |\rho_i| \varkappa_i^2/\eta_i\big),$$
 the local laws in \eqref{eq:ImG2_ll}--\eqref{eq:GG_ll} and \eqref{eq:ImM2_ll_nobeta}--\eqref{eq:GG_llnobeta} hold true when replacing $\ell \to \widetilde{\ell}$.

\subsubsection{Other two-resolvent local laws}\label{sec:other} As mentioned above, in Theorems \ref{thm:locallawreg}--\ref{thm:locallawnonreg}, we provided those two-resolvent local laws that are needed for proving our main result on ETH, Theorem \ref{thm:main}. We also mentioned that our proof easily yields the local laws for $\langle \Im GAG \rangle$ and $\langle G \Im G \rangle$. However, local laws for $\langle \Im GAGA \rangle$ and $\langle GAGA\rangle$, effectively taking into account two regular observables, require a more sophisticated approach: In fact, with the aid of the integral representation (see \cite[Lemma~4.11]{WigTypeETH})
\begin{equation} \label{eq:cone}
G(z) = \frac{1}{\pi} \int_\R \frac{\Im G(\psi(x))}{x - \psi^{-1}(z)} \dd x
\end{equation}
where $\psi$ parametrizes the boundary of a suitable infinite cone in $\C$ containing $z$, we can obtain the above mentioned versions of \eqref{eq:ImG2_ll} where one or two of the $\Im G$'s are replaced by a $G$ (and consequentially the corresponding $\rho$ on the right-hand side~of the bound is replaced by one).  This procedure requires to \emph{re-regularize} the observables, i.e.~adjust the notion of regularity to the changing spectral parameters within the integral representation \eqref{eq:cone}. Such a procedure has frequently been applied in the bulk of the spectrum in various contexts, see, e.g., \cite{iid, equipart, WigTypeETH, cipolloni2024eigenvector}. 
The re-regularization procedure is possible owing to the regularity properties of the matrices $\langle  V(z,w) \rangle^{-1} V(z,w)$ from \eqref{eq:V2_def} and $\vecL(z,w)$ from \eqref{eq:L_def}, which are established in Section~\ref{sec:beta_reg}, see also the proof of \eqref{eq:Exp_err} of Lemma~\ref{lemma:re-reg} in Section~\ref{sec:re-reg}.

So far we focused on \emph{averaged} local laws, i.e. where (normalized) traces of chains of resolvents and
observables were considered. \emph{Isotropic} local  laws, i.e.
matrix elements of such chains, e.g. $\langle {\bm x}, (GAG) {\bm y}\rangle$
with deterministic vectrors ${\bm x, \bm y}$, 
can  be considered directly by choosing a special observable in the averaged chain. 
For example, by choosing, say, $A_2$, to be a regularized version of the rank-one operator $N \ket{\bm y} \bra{\bm x}$ in \eqref{eq:ImG2_ll}, one can immediately deduce isotropic two-resolvent local laws. 

\subsubsection{General multi-resolvent local laws}
Although we focus on two-resolvent local laws here, our general proof method could easily handle general resolvent chains of arbitrary length in any spectral regime, including edges and cusps. We do not state them in this paper but explain the main ingredients of their analysis.  
We remark that the deterministic approximations $M_{[j,k]} \equiv M_{[j,k]}(z_j, B_j, z_{j+1}, B_{j+1}, \dots, B_{k-1}, z_k)$ to the long resolvent chains are defined via the recursive relation (see \cite[Definition 4.1]{iid})
\begin{equation}
	M_{[1,k]} = \mathcal{B}^{-1}_{z_1, z_k}\biggl[M_1 B_1 M_{[2,k]} + \sum_{j=2}^{k-1}M_1\mathcal{S}\bigl[M_{[1,j]}\bigr]M_{[j,k]}\biggr], \qquad k \ge 2,
\end{equation}
where we abbreviate $M_j := M(z_j)$.  
Similarly to Theorem~\ref{thm:locallawreg}, the basic idea is to first analyze strictly alternating (average) chains of $\Im G$'s and \emph{regular} observables as the fundamental building block, since this enables us to deduce results for general products of resolvents and deterministic matrices. In fact, general deterministic matrices $B$ can be decomposed as $B = c I + \reg{B}$, where $\reg {B}$ is regular, and higher powers of $\Im G$, resulting from the $cI$ part, can be linearized with the aid of the resolvent identity (or a suitable integral representation such as \cite[Lemma 5.1]{iid}). Moreover,  one can employ the integral identity \eqref{eq:cone} to deduce local laws for $G$ from those for $\Im G$. The main advantage of focusing on $\Im G$ instead of $G$ is that the critical effect of the 
smallness of the  density can be tracked only via $\Im G$ that is much more localized in the spectrum than $G$.

Finally, our approach avoiding isotropic objects in the zag step, which significantly streamlines the argument compared to previous papers such as \cite{cipolloni2023eigenstate, WigTypeETH, cipolloni2024eigenvector}, can easily be extended to longer resolvent chains as well.

\subsection{Two-resolvent bound: Proof of Theorem \ref{thm:main}} \label{subsec:ETHProof}
In this subsection we derive Theorem~\ref{thm:main} from the following two-resolvent bound (see, e.g., \cite[Lemma~3.2]{ETHpaper} or \cite[Section~2.2]{cipolloni2023eigenstate}), which is an immediate consequences of the $M$-bounds,
Theorem~\ref{prop:M2_bounds},
and the two-resolvent local laws,  Theorems~\ref{thm:locallawreg}--\ref{thm:locallawnonreg}: 	Let $E_1, E_2 \in \mathcal{I}\cap\supp\,\rho$ be a pair of admissible energies. Then, for any small enough fixed $\varepsilon > 0$ and $z_j := E_j +\ii N^\epsilon \etaf(E_j)$ for $j \in \{1,2\}$ it holds that
\begin{equation} \label{eq:mainLLbound}  
	\Bigl\langle \Im G(z_1) \reg{A}^{z_1, z_2} \Im G(z_2) \bigl(\reg{A}^{z_1, z_2}\bigr)^* \Bigr\rangle \lesssim \begin{cases}
\rho(z_1)\rho(z_2) \vertiii{\reg{A}^{z_1, z_2}}_{z_1, z_2}^2 \quad &(z_1, z_2) \in \regD \\
\rho(z_1)\rho(z_2) \norm{\reg{A}^{z_1, z_2}}_{\rm hs}^2 \quad &(z_1, z_2) \in (\bddD)^2 \setminus \regD
	\end{cases},
\end{equation}
 with very high probability, for any $A \in \mathbb{C}^{N\times N}$.

In the following,  we immediately explain how to obtain the version in Remark \ref{rmk:Iextend} with a general $\mathcal{I} \subset \R$. As shown in \cite[Corollary~2.11 (b)]{cuspuniv}, we have the following form of eigenvalue rigidity: For any\footnote{The slight enlargement of $\mathcal{I}$ is immediate by redefining $c_M \to c_M/2$ in the definition of $\mathcal{I}$.} $\widetilde{E}_1, \widetilde{E}_2 \in (\mathcal{I} + [-c_M/2 , c_M/2]) \cap \supp \rho$ and defining $\tilde{j} = \mathrm{ind}(\widetilde{E}_1),\tilde{k} = \mathrm{ind}(\widetilde{E}_2)$ as in \eqref{eq:jkdef}, it holds that
\begin{equation} \label{eq:rigidity}
	|\lambda_{\tilde{j}} - \widetilde{E}_1| \prec \etaf(\widetilde{E}_1) \qquad \text{and} \qquad |\lambda_{\tilde{k}} - \widetilde{E}_2| \prec \etaf(\widetilde{E}_2) \,. 
\end{equation}

As a first step, we now control the typical size of the eigenvector overlaps by a two-resolvent chains with an observable $A$ regularized with respect to $(z_1, z_2)$ in the sense of \eqref{eq:regA} at positive imaginary parts $\Im z_1$ and $\Im z_2$. Fix $j,k \in \indset{N}$ such that $\mathrm{ind}^{-1}(\{j\}) \cap \mathcal{I} \neq \emptyset$ and $\mathrm{ind}^{-1}(\{k\}) \cap \mathcal{I} \neq \emptyset$ and take, as in \eqref{eq:jkdef},  $E_1 \in \mathrm{ind}^{-1}(\{j\}) \cap \mathcal{I}$ as well as $E_2 \in \mathrm{ind}^{-1}(\{k\}) \cap \mathcal{I} $ with $E_1 = E_2$ if $j=k$. 

In the following presentation, we focus on the situation in which $z_i = E_i + \ii N^\epsilon \etaf(E_i)$, $i \in \indset{2}$, are such that $(z_1, z_2) \in \regD$, i.e.~the first case in \eqref{eq:mainLLbound}; the other case is briefly discussed below. 
Using \eqref{eq:rigidity}, given an arbitrary $\epsilon > 0$,
we have, by spectral decomposition,  additionally using \eqref{eq:mainLLbound}, 
\begin{equation} \label{eq:ETHbound}
	\begin{split}
		|\langle \bm u_j, \reg{A}^{z_1, z_2} \bm u_k \rangle|^2 &\prec  \etaf(E_1) \etaf(E_2) \sum_{\tilde{j}, \tilde{k}} \frac{\etaf(E_1) }{(\lambda_{\tilde{j}} - E_1)^2 + \etaf(E_1)^2} \frac{\etaf(E_2) }{(\lambda_{\tilde{k}} - E_2)^2 + \etaf(E_2)^2} |\langle \bm u_{\tilde{j}}, \reg{A}^{z_1, z_2} \bm u_{\tilde{k}} \rangle|^2 \\
		&\lesssim N^{1+2\epsilon} \etaf(E_1) \etaf(E_2) \big\langle \Im G(z_1) \reg{A}^{z_1, z_2} \Im G(z_2) (\reg{A}^{z_1, z_2})^* \big\rangle \\
		&\lesssim N^{1+2\epsilon} \etaf(E_1) \etaf(E_2) \rho(z_1)\rho(z_2) \vertiii{\reg{A}^{z_1, z_2}}_{z_1, z_2}^2 \lesssim \frac{N^{4\epsilon}}{N} \vertiii{\reg{A}^{z_1, z_2}}_{z_1, z_2}^2 \,. 
	\end{split}
\end{equation}
Here we used  monotonicity  of $\eta \mapsto \eta^2 /(x^2 + \eta^2)$ in the second step. In the ultimate step, we used that $\rho(z_i) \lesssim N^{\epsilon} \rho(E_i + \ii \etaf(E_i))$ by monotonicity  of $\eta \mapsto 1 /(x^2 + \eta^2)$, additionally using the Stieltjes representation of $\rho(z)$.

Subsequently, we re-express the bound on the overlap in \eqref{eq:ETHbound} given by \eqref{eq:mainLLbound} in terms of the quantities appearing on the right-hand side of \eqref{eq:ETH}, and account for the discrepancy between the left-hand sides of \eqref{eq:ETHbound} and \eqref{eq:ETH}. To this end, we use the following technical lemma, proved in Section~\ref{sec:beta_reg}.
\begin{lemma} [Bounds on re-regularization errors] \label{lemma:re-reg}
	For all $E \in \mathcal{I}\cap\supp\,\rho$ and $0 < \eta \lesssim 1$, the re-regularization error between regularity with respect to $(z,z)$ for $z:=E+\ii\eta$ and $(E,E)$ admits the bound
	\begin{equation} \label{eq:Exp_err}
		\biggl\lvert \Bigl\langle \frac{\Im M(E)}{\pi \rho(E)} \reg{A}^{z,z} \Bigr\rangle \biggr\rvert =  	\biggl\lvert \Bigl\langle \frac{\Im M(E)}{\pi \rho(E)} A \Bigr\rangle - \Bigl\langle \frac{\widehat{M}(z, I,z) }{\langle \widehat{M}(z,I,z) \rangle} A \Bigr\rangle  \biggr\rvert\lesssim    
		\frac{\eta}{ \other{\alpha}(z,z)} \vertiii{\reg{A}^{z,z}}_{z,z} ~,
	\end{equation}
	for all  $A\in\mathbb{C}^{N\times N}$, where $\reg{A}^{z,z}$ is as defined in \eqref{eq:regA}.
	
	Furthermore, let $(z_1, z_2) := (E_1 + \ii\eta_1, E_2 + \ii\eta_2) \in \regD \cap (\mathbb{H}^2)$. Assume that $\eta_j \ge \etaf(E_j)$ for $j \in \{1,2\}$, where $\etaf$ is defined in \eqref{eq:etaf_def}. Then, the norm $\vertiii{\cdot}_{z_1, z_2}$ satisfies
	\begin{equation} \label{eq:re-triple}
		\vertiii{\reg{A}^{z_1,z_2}}_{z_1, z_2} \lesssim \min_{E \in \{E_1, E_2\}}\left(\frac{\bigl\lVert  \mathcal{R}_{E} \bigl[\reg{A}^{E, E}\bigr]   \bigr\rVert_\mathrm{hs}}{\other{\alpha}(E_1, E_2) + N^{-1/4}} + \bigl\lVert \reg{A}^{E, E} \bigr\rVert_\mathrm{hs}\right), \qquad A \in \mathbb{C}^{N\times N}, 
	\end{equation} 
	where we recall the definition of $\reg{A}^{E, E}$ from \eqref{eq:Acirc_def}.
	\footnote{\label{ftn:regularities}
	We point out that, for all $E \in \mathcal{I}\cap\supp\,\rho$ and $A \in \mathbb{C}^{N\times N}$, 
	\begin{equation} \label{eq:convergence}
		\lim_{\eta \to +0} \Bigl(\reg{A}^{z,z}\bigr\rvert_{z=E+\ii\eta}\Bigr) = A - \Bigl\langle A \frac{\Im M(E)}{\pi \rho(E)} \Bigr\rangle\, I  = \reg{A}^{E,E},
	\end{equation}
	where $\reg{A}^{z,z}$ defined in \eqref{eq:regA}, while $\reg{A}^{E,E}$ is given by \eqref{eq:Acirc_def}. 
	In particular, Definition \ref{def:reg} of regularity with $z_1 = z_2 = E+\ii\eta$ is consistent with \eqref{eq:Acirc_def} in the limit $\eta \to +0$. 
The convergence \eqref{eq:convergence} is a consequence of the bound \eqref{eq:Exp_err}, together with \eqref{eq:alp_eta_comp}, which implies $\other{\alpha}(z,z) \gtrsim (\Im z)^{1/3}$.  
 }
\end{lemma}

First, we note that, by $\langle \bm u_j, \bm u_k \rangle = \delta_{jk}$, 
\begin{equation} \label{eq:ETHbound2}
	\begin{split}
		\left|\langle \bm u_j, A \bm u_k \rangle - \delta_{jk}\Bigl\langle \frac{\Im M(E_1)}{\pi \rho(E_1)} A\Bigr\rangle \right|^2 &= 		\left|\langle \bm u_j, \reg{A}^{z_1, z_2} \bm u_k \rangle - \delta_{jk}\Bigl\langle \frac{\Im M(E_1)}{\pi \rho(E_1)} \reg{A}^{z_1, z_1}\Bigr\rangle \right|^2 \\
		&\lesssim \left|\langle \bm u_j, \reg{A}^{z_1, z_2} \bm u_k \rangle \right|^2 + \delta_{jk} \left| \Bigl\langle \frac{\Im M(E_1)}{\pi \rho(E_1)} \reg{A}^{z_1, z_1}\Bigr\rangle \right|^2   \,.
	\end{split}
\end{equation}
By \eqref{eq:Exp_err} and \eqref{eq:alp_beta_in_supp}, together with simple monotonicity estimates, the second term on the right-hand side~admits the bound
\begin{equation} \label{eq:ETHbound3}
	\begin{split}
\delta_{jk} \left| \Bigl\langle \frac{\Im M(E_1)}{\pi \rho(E_1)} \reg{A}^{z_1, z_1}\Bigr\rangle \right|^2 &\lesssim \delta_{jk} N^{2\epsilon} \left(\frac{\etaf(E_1)}{\other{\alpha}(z_1, z_1)} \vertiii{\reg{A}^{z_1, z_1}}_{z_1, z_1}\right)^2 \\
&\lesssim \delta_{jk} N^{4\epsilon} \left(\frac{\etaf(E_1) \rho(E_1 + \ii \etaf)}{\other{\beta}(z_1, z_1)} \vertiii{\reg{A}^{z_1, z_1}}_{z_1, z_1}\right)^2 \lesssim   \delta_{jk} \frac{N^{4\epsilon}}{N} \vertiii{\reg{A}^{z_1, z_1}}_{z_1, z_1}^2
	\end{split} 
\end{equation}
where in the last step we additionally used that $\other{\beta}(z_1, z_1) \gtrsim N^{-1/2}$, as a consequence of $\eta_1 \ge \etaf(E_1)$. 
 
In the regime $(z_1, z_2) \in (\bddD)^2 \setminus \regD$, we note that necessarily $j \neq k$ and hence \eqref{eq:ETHbound3} is irrelevant.
 All of the other above steps naturally hold when replacing $\vertiii{\,\cdot\,}$ by $\Vert \, \cdot \, \Vert_{\rm hs}$, in particular using the second estimate in \eqref{eq:mainLLbound} instead of the first. 

Therefore, combining \eqref{eq:ETHbound2} with \eqref{eq:ETHbound} (for the first term on the right-hand side~of \eqref{eq:ETHbound2}), \eqref{eq:ETHbound3} (for the second term on the right-hand side~of \eqref{eq:ETHbound2}) and \eqref{eq:re-triple} (to estimate the $\vertiii{\, \cdot \,}$ norm), additionally using arbitrariness of $\epsilon > 0$, we conclude
\begin{equation}
		\left|\langle \bm u_j, A \bm u_k \rangle - \delta_{jk}\Bigl\langle \frac{\Im M(E_1)}{\pi \rho(E_1)} A\Bigr\rangle \right|^2 \prec \frac{1}{N} \min_{i \in \indset{2}}\biggl(\frac{\Vert \mathcal{R}_{E_i}[\reg{A}^{E_i, E_i}]\Vert_{\rm hs}}{\other{\alpha}(E_1, E_2) + N^{-1/4}} + \bigl\lVert \reg{A}^{E_i, E_i} \bigr\rVert_\mathrm{hs}\biggr)^2 \,. 
\end{equation}

Finally, if $\mathcal{I} = \R$, the rigidity bounds in \eqref{eq:rigidity} can be simplified to $|\lambda_j - \gamma_j| \prec \etaf(\gamma_j)$ and $|\lambda_k - \gamma_k| \prec \etaf(\gamma_k)$ for all indices $j,k \in \indset{N}$ (additionally using \cite[Theorem 2.9]{cuspuniv}). This concludes the proof of Theorem \ref{thm:main} and its extension in Remark~\ref{rmk:Iextend}. \qed 

\subsection{Regularity and pre-regularity}\label{sec:prereg}

As explained in the introduction, the new concept of regularity (or regularization) of an observable $A$
 in Definition~\ref{def:reg},  based
upon $\widehat{M}(z, I, w)$ and motivated by minimization problem~\eqref{yminzw}
is  more accurate than the analogous concept in previous papers relying on the (approximate) singular direction 
of the stability operator $\mathcal{B}_{z^+, w^-}$. Regularity now defines the ultimate direction 
that minimizes~\eqref{ymin}  exactly, it determines the correct microcanonical state
and the optimal fluctuation size for the overlap (Theorem~\ref{thm:main}) and the local laws (see \eqref{eq:general1G} and
Theorem~\ref{thm:locallawreg}). However the  singular direction of $\mathcal{B}_{z^+, w^-}$,
that can very well be approximated by $M(z^+)-M(w^-)$, 
still plays a major role in our proofs, so we introduce the corresponding regularity concept that we call 
\emph{pre-regular}\footnote{
	We point out that in prior works \cite{iid, equipart, WigTypeETH}, the concept of \emph{regularity} was based upon the unstable direction of the opposite-half-plane stability operator $\mathcal{B}_{z^+,w^-}$, and is essentially identical to what we call \emph{pre-regularity} in the present work.
}: 
 	\begin{definition}[Observable Pre-Regularity] \label{def:prereg}
 		Let  $(z, w) \in \regD$, where $\regD$ is defined in \eqref{eq:regD_beta}.   
 		Then, we say that a deterministic matrix $A\in\mathbb{C}^{N\times N}$ is \emph{pre-regular}
		 with respect to $(z, w)$, or simply $(z, w)$-pre-regular,  if and only if
 		\begin{equation} \label{eq:prereg_def}
 			\Bigl\langle \bigl(M(z^+)-M(w^-)\bigr) A\Bigr\rangle = 0.
 		\end{equation}
	For a general matrix $A\in\mathbb{C}^{N\times N}$ we define its \emph{$(z, w)$-pre-regularization}
		as
\begin{equation}\label{eq:preregularization}
   \ring{A}^{z,w}:  = A -  \Bigl\langle A \frac{M(z^+) - M(w^-)}{\langle M(z^+) - M(w^-)\rangle } \Bigr\rangle \, I. 
\end{equation} 
Note that the denominator does not vanish by~\eqref{eq:beta_assymp} and~\eqref{eq:betaf_def}
  as long as $(z, w) \in \regD$.  Clearly, $A$ is $(z, w)$-pre-regular if and only if $A=  \ring{A}^{z,w}$. \nc
 	\end{definition}  
	Note that for given spectral parameters  $(z, w) \in \regD$ and for any matrix $A$, we have 
	defined two related concepts;
	regularization\footnote{
	In fact, regularity can be defined for any pair $(z, w) \in \C\setminus\R$, while pre-regularity requires
	restriction to the perturbative regime to ensure $\langle M(z^+) - M(w^-)\rangle\ne0$.}, $\reg{A}^{z,w}$ (Definition~\ref{def:reg}), and pre-regularization,  $\ring{A}^{z,w}$. It is straightforward to check that they 
	satisfy the relations
$(\reg{A})^\circ =\ring{A}$,   $(\ring{A})^\bullet = \reg{A}$.
While pre-regularity represents the non-optimal choice (compare \eqref{Var} and \eqref{newETH}) and it might be considered obsolete, it is important in the proofs. In fact, as its name suggests, the main analysis involving regular
observables goes through their pre-regularization as a first step because pre-regularity has several important
properties that regularity does not enjoy. 
It is defined directly through $M(z)$ which simplifies its analysis. But most importantly, pre-regularity is preserved
along the zig-flow, see~\eqref{eq:preregpreserved} later. The trade-off is that,  
unlike regularization~\eqref{regsymm}, pre-regularization $\ring{A}^{z,w}$
is sensitive to the order of spectral parameters, i.e. $\ring{A}^{z,w}\ne \ring{A}^{w,z}$ in general.

\nc
\begin{remark}[On choosing the regularization in ETH] \label{rmk:ETHreg}
We mention that for all $(z,w)\in\regD$, the norm $\vertiii{\, \cdot \, }_{z,w}$ could have alternatively  been defined as 
\begin{equation} \label{eq:tripalternative}
	\vertiii{B}'_{z, w} := \frac{ \bigl\lvert \bigl\langle \vecL(z^+, w^+) \ring{B}^{z, w}  \bigr\rangle \bigr\rvert}{\other{\alpha}(z^+, w^+)} + \bigl\lVert\ring{B}^{z, w}  \bigr\rVert_{\mathrm{hs}}\,, 
\end{equation} 
i.e. using the pre-regularization of $B$ instead of its regularization.  \nc
Indeed, it would not affect the estimates in Proposition \ref{prop:M2_bounds} and Theorem \ref{thm:locallawreg}, since, as we show in Claim~\ref{claim:Ups} below, 
\begin{equation}
	\vertiii{B}'_{z, w} \sim  \vertiii{B}_{z, w} \,. 
\end{equation}

In fact, by inspecting the proof of Claim~\ref{claim:Ups} in the later Section~\ref{sec:techM_bounds}, it is straightforward to see that replacing $\reg{B}^{z,w}$ on the right-hand side of \eqref{eq:tripplenorm} by $B - \langle B \other{V}(z,w) \rangle$ for any vector $\other{V}(z,w)$ satisfying 
\begin{equation}
	\bigl\langle \other{V}(z,w) \bigr\rangle = 1, \qquad \biggl\lVert \other{V}(z,w) - \frac{\widehat{M}(z,I,w)}{\langle \widehat{M}(z,I,w) \rangle} \biggr\rVert_\mathrm{hs} \lesssim \other{\alpha}(z,w),
\end{equation}
would yield a norm equivalent to $\vertiii{\cdot}_{z,w}$. Other possible choices of $\other{V}(z,w)$ include:
\begin{itemize}
	\item[(i)] $(\pi \rho(E))^{-1}\Im M(E)$, for~$E \in \{\Re z, \Re w \}$, as we have used on the right-hand side of \eqref{eq:ETH}, (see also \eqref{eq:Exp_err}).
	
	\item[(ii)] any accumulation point of the map $z,w \mapsto \langle \widehat{M}(z,I,w) \rangle^{-1} \widehat{M}(z,I,w)$ as $\Im z, \Im w \to +0$.
	Note that, in general, this map does not have a well-defined boundary value on the set of admissible energies $\mathcal{I}\times \mathcal{I}$, unless $z = w$.
\end{itemize}
A similar freedom (up to errors of order $\other{\alpha}(z,w)$) is available in choosing the direction of the anomalous fluctuation mode~$R(z,w)$. In particular, this allows us to pass to the one-body projector $\mathcal{R}_E$ on the right-hand side of \eqref{eq:ETH} even for $j\ne k$, which is conceptually the simplest choice to present our ETH result. \nc 

\end{remark}
 	Several specialized quantities associated with the two-body stability operator, depending on an ordered pair of spectral parameters $z,w\in \mathbb{C}$, which are introduced in Section~\ref{sec:LL} above, are often used in the forthcoming proofs.
		For the reader’s convenience, we collect them in Table~\ref{glossary} at the end of the paper, accompanied by brief explanations to provide intuition and context as well as precise references to their definitions.

 	\section{Dynamical \texorpdfstring{$M$}{M}-bounds: Proof of Theorem \ref{prop:M2_bounds}} \label{sec:Mboundsproof} 
 	In this section we present the analysis of the deterministic approximations $M(z_1, A, z_2)$ to the two-resolvent chains, which constitutes the main technical novelty of the present work.
 	In particular, we prove the key optimal estimate \eqref{eq:ImM2_bounds} for the deterministic quantity $\langle\widehat{M}(z_1, A_1, z_2)A_2\rangle$, which approximates $\langle\Im G(z_1) A_1 \Im G (z_2)A_2\rangle$.
 	The latter quantity plays a central role in the probabilistic analysis carried out in Sections~\ref{sec:zig_proof}--\ref{sec:zagproof}, where it is used to control fluctuations of resolvent chains.
 	In this sense, the estimate \eqref{eq:ImM2_bounds} provides the crucial deterministic input for the zigzag strategy. 
 	
 	In concert with the probabilistic analysis in the sequel, we prove the deterministic estimates in 
	Theorem~\ref{prop:M2_bounds} using the dynamical characteristic flow approach. We begin by preparing the necessary notation and collecting preliminary results about the flow. \nc 
 	
 	\subsection{Stability along the Characteristic Flow}
 	Let $(D,\mathcal{S})$ be the target MDE data-pair satisfying the Assumptions~\ref{ass:boundedexp}--\ref{ass:S_norms}, and let the covariance tensor $\Sigma$, corresponding to $\mathcal{S}$ via (c.f., \eqref{eq:SigmaSentries})
 	\begin{equation} \label{eq:Sigma_ent_def}
 		\Sigma_{ab, ij} :=   \mathcal{S}_{jb, ia}, \qquad a,b,i,j \in \indset{N},
 	\end{equation}
 	satisfy Assumption~\ref{ass:full} with some $c_\mathrm{full} > 0$. Let $T > 0 $ be a terminal time, and let $(D_t, \mathcal{S}_t)$ be the solution to the ordinary differential equations 
 	\begin{equation} \label{eq:datapair_flow}
 		\mathrm{d} D_t = -\tfrac{1}{2}D_t \mathrm{d}t, \qquad 	\mathrm{d} \mathcal{S}_t[\,\cdot\,] = \bigl(-\mathcal{S}_t[\,\cdot\,] + \langle \,\cdot \, \rangle\bigr) \mathrm{d}t, \qquad t \in [0, T],
 	\end{equation}
 	satisfying the final conditions $D_T = D$ and $\mathcal{S}_T = \mathcal{S}$. 
 	In the sequel, we denote the solution to the  MDE \eqref{eq:MDE} with data $(D_t, \mathcal{S}_t)$ by $M_t(z)$, and at the terminal time $t=T$ we write $M_T = M$, where $M$ solves the MDE~\eqref{eq:MDE} with data $(D, \mathcal{S})$. 
 	The key properties of the data-pair flow \eqref{eq:datapair_flow} and the associated solution $M_t$ are summarized in the following lemma. 
 	Its proof is deferred to Section~\ref{sec:flow_aux}.
 	
 	\begin{lemma} [Characteristic Flow Properties] \label{lemma:flow}
 		There exists a terminal time $T \sim 1$, such that the following properties hold:
 		\begin{itemize}
 			\item[(i)] For all $0 \le t \le T$, we have the bound
 			\begin{equation} \label{eq:termT_bound}
 				1 + (c_\mathrm{full} - 1)\ee^{T-t} \ge \frac{c_\mathrm{full}}{2}\wedge1, \qquad t \in [0, T]. 
 			\end{equation}
 			In particular, for any matrix $R \ge 0$, the self-energy operator $\mathcal{S}_t$, solving \eqref{eq:datapair_flow}, satisfies
 			\begin{equation} \label{eq:S_t_flatness}
 				\mathcal{S}_t[R] \sim \langle R \rangle, \qquad R\ge 0,
 			\end{equation}
 			in the sense of quadratic forms, uniformly in $t \in [0,T]$. 
 			
 			\item[(ii)] Let $M_t(z)$ denote the solution to the MDE \eqref{eq:MDE} with the time-dependent data-pair $(D_t, \mathcal{S}_t)$ solving \eqref{eq:datapair_flow}. For any $z \in \mathbb{C}\backslash\mathbb{R}$, let $z_t \equiv z_{t,T}$ denote the trajectory of the characteristic flow equation
 			\begin{equation} \label{eq:char_flow}
 				\mathrm{d} z_t = -\bigl(\tfrac{1}{2}z_t + \langle M_t(z_t) \rangle\bigr) \mathrm{d}t, \qquad t\in[0,T],  \qquad z_T = z. 
 			\end{equation}
 			Then, along the flow \eqref{eq:char_flow}, the matrix $M_{t}(z_{t})$ satisfies the ordinary differential equation
 			\begin{equation} \label{eq:dMt}
 				\mathrm{d} M_{t}(z_{t}) = \tfrac{1}{2} M_{t}(z_{t}) \mathrm{d}t, \qquad t \in[0,T], 
				\qquad M_T(z_T) =M(z). \nc
 			\end{equation}
 			Let $\rho_t(x) := \pi^{-1}\lim\limits_{\eta \to +0} \langle \Im M_t(x+\ii\eta) \rangle$ denote the self-consistent density of states corresponding to MDE \eqref{eq:MDE} with data-pair $(D_t, \mathcal{S}_t)$.  Then, the quantity $\kapd_t := \dist(z_t, \supp\,\rho_t)$ satisfies 
 			\begin{equation} \label{eq:kapd_rhoeta_t}
 				\rho_t(z_t)^{-1} \Im z_t \gtrsim 1\quad   \Longleftrightarrow \quad \kapd_t \gtrsim 1,
 			\end{equation}
 			where $\rho_t(z_t) := \pi^{-1}\langle \Im M_t(z_t) \rangle$.
 			
 			\item[(iii)] Let $z_1$, $z_2 \in \bddD$. 
 			Then,  the trajectories $z_{j,t}$, solving \eqref{eq:char_flow} with final condition $z_{j,T} = z_j$ for $j \in \{1,2\}$, satisfy
 			\begin{equation} \label{eq:z_diff_comp}
 				\Bigl\lvert \frac{ z_{1,t} - z_{2,t} }{ z_1 - z_2 }\Bigr\rvert \sim 1 + (T-t) \bigl\lvert u(z_1, z_2) \bigr\rvert, \qquad u(z_1, z_2) := \frac{m(z_1) - m(z_2)}{z_1 - z_2}, \qquad t \in [0,T],
 			\end{equation}
 			where $m(z) := \langle M(z) \rangle$.  \nc
 		\end{itemize}
 	\end{lemma}
 	Here and in the sequel, we parametrize the trajectories of the characteristic flow \eqref{eq:char_flow} by prescribing their terminal values at a fixed time $T$. 
 	We denote such trajectories by, e.g., $z_t, z_{j,t}$ or $w_t$, and formulate all estimates for the relevant quantities in terms of their terminal values, i.e., $z = z_T, z_j = z_{j,T}$ or $w = w_T$.

 	The main novel estimate in Lemma~\ref{lemma:flow} is the comparison \eqref{eq:z_diff_comp}, which quantifies the relative motion of two characteristics of \eqref{eq:char_flow}. 
 	In particular, it implies that as $t$ decreases from $T$ to $0$, the separation between trajectories $z_{1,t}$ and $z_{2,t}$ is non-decreasing, up to multiplicative constants of order one.
 	Lemma~\ref{lemma:flow} also has the following immediate consequences.
 	
 	Let $z_t$ be a trajectory of \eqref{eq:char_flow} satisfying $z_T = z$. \nc Note that solving the equation \eqref{eq:dMt} for $M_t(z_t)$ explicitly yields 
 	\begin{equation} \label{eq:M_t}
 		M_t(z_t) = \ee^{(t-T)/2} M(z), \qquad t\in[0,T].
 	\end{equation} 
 	Similarly, explicitly solving \eqref{eq:char_flow} and using \eqref{eq:M_t}, we obtain  
 	\begin{equation} \label{eq:z_t}
 		z_t = \ee^{(T-t)/2} z + (\ee^{(T-t)/2} - \ee^{(t-T)/2}) \bigl\langle M(z) \bigr\rangle, \qquad t\in[0,T].
 	\end{equation}
 	Therefore, since $T \lesssim 1$, we deduce that $M_t(z_t)$ and $\rho_t(z_t) = \pi^{-1}\langle \Im M_t(z_t)\rangle$ satisfy
 	\begin{equation} \label{eq:Mt_bound}
 		\norm{M_t(z_t)} \sim  \norm{M(z)}, \qquad t \in [0, T],
 	\end{equation}
 	\begin{equation} \label{eq:rho_t}
 		\rho_t(z_t) \sim \rho(z), \qquad t \in [0, T].
 	\end{equation}
 	Hence, taking the imaginary part of \eqref{eq:z_t}, we conclude that
 	\begin{equation} \label{eq:eta_t}
 		\Im z_t \sim \Im z + (T-t)\,\rho(z).
 	\end{equation}
 	The definition of $\mathcal{I}$ in \eqref{eq:adm_E} together with \eqref{eq:Mt_bound} implies the bound
 	\begin{equation} \label{eq:M_bound}
 		\norm{M_t(z_t)} \lesssim 1, \qquad z \in \bddD.
 	\end{equation}
 	
 	 To analyze $\widehat{M}(z_1, A, z_2)$ dynamically, we embed it into the characteristic flow \eqref{eq:char_flow}.
 	To this end, we define the time-dependent two-body stability operator \nc 
 	\begin{equation} \label{eq:stab_t_def}
 		\mathcal{B}_{t,z_1, z_2}[\,\cdot\,] := \mathrm{Id}[\,\cdot\,] - M_{1,t} \mathcal{S}_t[\,\cdot\,] M_{2,t}, \qquad t \in [0,T], 
 	\end{equation}
 	where $M_{j,t} := M_t(z_{j,t})$, the trajectories $z_{j,t}$ solve \eqref{eq:char_flow} with final conditions $z_{j,T}=z_j$, \nc and $\mathcal{S}_t$ evolves according to \eqref{eq:datapair_flow}. 
 	
 	 Our first intermediate result is a bound on the inverse of this time-dependent stability operator. Its proof is deferred to Section~\ref{sec:stab}. \nc  	
 	\begin{lemma}[Stability along the Flow] \label{lemma:stab_t}
 		For any $z_1, z_2 \in \bddD$ and all $t \in [0,T]$, the stability operator $\mathcal{B}_{t,z_1, z_2}$ defined in \eqref{eq:stab_t_def} satisfies the bound
 		\begin{equation} \label{eq:stab_t_bound}
 			\norm{\mathcal{B}_{t,z_1, z_2}^{-1}}_{\mathrm{hs}\to\mathrm{hs}} + \norm{\mathcal{B}_{t,z_1, z_2}^{-1}} \lesssim \other{\beta}_t(z_1, z_2)^{-1}, \qquad t \in [0,T],
 		\end{equation}
 		where  the function  $\other{\beta}_t(z,w)$ is defined as 
 		\begin{equation} \label{eq:betaf_t_def}
 			\other{\beta}_t(z,w) \equiv \other{\beta}_{t,T}(z,w) := \other{\beta}(z,w) + (T-t). 
 		\end{equation}
 	\end{lemma}
 	 
 	Note that the quantity $\other{\beta}_t(z, w)$ is defined by simply adding  $T-t$ to $\other{\beta}(z,w)$, where $z=z_T, w=w_T$ are the spectral parameters at the terminal time of the flow, instead of replacing $z$, $w$, $\rho$, and $\sigma$ by their time-dependent counterparts in the definition~\eqref{eq:betaf_def} of $\other{\beta}(z,w)$. 
 	This formulation is chosen for technical convenience, and the alternative, fully time-dependent definition would lead to an equivalent quantity, up to multiplicative constants of order one. 
	Indeed, it is straightforward to verify that for all $z,w \in \bddD\cap\mathbb{H}$ with $|z|+|w| \lesssim 1$, we have
 	\begin{equation} \label{eq:beta_tvsbeta+t}
 		\rho_t(z_t)^{-1} \Im z_t + \rho_t(z_t)^2 + \rho_t(z_t)\bigl\lvert \sigma_t(z_t) \bigr\rvert + \bigl(\rho_t(z_t)+\bigl\lvert \sigma_t(z_t) \bigr\rvert\bigr)^{1/2}|z_t - w_t|^{1/2} + |z_t - w_t|^{2/3} \sim \other{\beta}_t(z,w).
 	\end{equation}
 	The lower bound on the left-hand side of \eqref{eq:beta_tvsbeta+t} follows directly from \eqref{eq:stab_t_bound} and \eqref{eq:beta_assymp}.  
 	For the  upper bound, we observe that
 	\begin{equation}
 		\rho_t(z_t) \sim \rho(z), \qquad 
 		\rho_t(z_t)^{-1}\Im z_t \sim \rho(z)^{-1}\Im z + (T-t), \qquad 
 		\sigma_t(z_t) = \sigma(z), 
 	\end{equation}
 	 where we used  \eqref{eq:rho_t} and \eqref{eq:eta_t} for the first two comparisons, and \eqref{eq:sigma_def} together with \eqref{eq:M_t} for the last relation. 
 	Hence, it remains to control the terms involving $|z_t-w_t|$. Using \eqref{eq:z_diff_comp} followed by the Cauchy–-Schwarz inequality yields
 	\begin{equation}
 		\bigl(\rho_t(z_t)+\bigl\lvert \sigma_t(z_t) \bigr\rvert\bigr)^{1/2}|z_t - w_t|^{1/2} \lesssim \bigl(\rho(z)+ \lvert \sigma(z)  \rvert\bigr)^{1/2}|z - w|^{1/2} + (T-t)^{1/2}\other{\beta}(z,w)^{1/2} \lesssim \other{\beta}_t(z,w),
 	\end{equation}
 	where we used the bound $|u(z,w)| \lesssim \other{\beta}(z,w)^{-1}$ by \eqref{eq:u_one}--\eqref{eq:u_sim} below and $\other{\beta}(z,w) \gtrsim (\rho(z)+ \lvert \sigma(z)  \rvert )^{1/2}|z - w|^{1/2}$ by \eqref{eq:betaf_def}. Similarly, since $\other{\beta}(z,w) \gtrsim |z-w|^{2/3}$ by \eqref{eq:betaf_def}, we obtain the remaining 
 	\begin{equation}
 		|z_t - w_t|^{2/3} \lesssim |z-w|^{2/3} + (T-t)^{2/3} \other{\beta}^{1/3} \lesssim \other{\beta}_t(z,w). 
 	\end{equation} 
 	An analogous comparison holds for $\other{\beta}_t(z,\bar w)$ with the arguments in the opposite half-planes. Indeed, for $z,w \in \bddD\cap\mathbb{H}$,
 	\begin{equation}
 		\rho_t(z_t)^{-1} |z_t - \overline{w}_t| \lesssim \rho(z)^{-1} |z-t| + (T-t), \quad \text{provided} \quad \rho(z)^{-1} |z-\overline{w}| \lesssim \other{\beta}(z,w).
 	\end{equation}
 	
 	Thus, a posteriori, the definition \eqref{eq:stab_t_bound} sharply captures the essential behavior of the inverse stability operator $\mathcal{B}_{t,z,w}^{-1}$: its norm improves   inverse-linearly in $T-t$ as $t$ decreases from $T$ to $0$.  
 	This reflects a guiding principle of our proof: we separate the simpler time dependence of the relevant quantities from their more intricate dependence on the spectral parameters, the latter being encoded through their values at the terminal time.

 	\nc
 	\subsection{$M$-Bounds along the Flow} 
 	Using \eqref{eq:stab_t_def}, we define the time-dependent version of the quantity $M(z_1, B, z_2)$. 
 	For a pair of spectral parameter $z_1, z_2$, and for any observable $B \in \mathbb{C}^{N\times N}$,
	we define $M_t(z_{1}, B, z_2)$ as
 	\begin{equation} \label{eq:Mt_def}
 		M_t(z_{1}, B, z_2) := \mathcal{B}_{t,z_1,z_2}^{-1}\bigl[M_{1,t} B\, M_{2,t}\bigr], \qquad t \in [0,T],
 	\end{equation}
 	and denote
 	\begin{equation} \label{eq:ImM2t_def}
 		\widehat{M}_t(z_{1}, B, z_2) := \frac{M_t(z_1,B, \overline{z}_2)  + M_t(\overline{z}_1, B, z_2) -M_t(z_1, B, z_2) - M_t(\overline{z}_1,B, \overline{z}_2)}{4}, \qquad t \in [0,T].
 	\end{equation}
 	
 	We note that the size of the time-dependent $M$-term defined \eqref{eq:Mt_def} depends strongly on the observable $B$. More precisely, there exists a codimension-one subspace of observables, related to the unstable direction of the super-operator $\mathcal{B}_{t,z_1,\bar z_2}$, for which $M_t(z_{1}, B, z_2)$ becomes much smaller. 
	Although   $\mathcal{B}_{t,z_1,\bar z_2}$ itself and its unstable direction are genuinely time-dependent, 
	a very good approximation to the corresponding unstable eigenvector can be chosen in a \textit{time-independent} way.  
	 This is exactly the concept of pre-regularization from Definition~\ref{def:prereg}: if  $(z_1, z_2) \in \regD$
	 and $A$ is $(z_1, z_2)$-pre-regular, i.e. $\langle (M(z_1^+)-M(z_2^-)) A\rangle = 0$, then, for all $t\in [0,T]$,
 	\begin{equation}\label{eq:preregpreserved}
 		\Bigl\langle \bigl(M_t(z_{1,t}^+)-M_t(z_{2,t}^-)\bigr) A\Bigr\rangle = 0, 
 	\end{equation}
 	since the matrix $M_t(z_{1,t}^+)-M_t(z_{2,t}^-)$ depends on time only via multiplication by a scalar factor $\ee^{(t-T)/2}$  (see \eqref{eq:M_t}), and hence its direction is invariant in time~$t\in[0,T]$. \nc
 	
 	In the present section, we work with pre-regular observables $A$ as basic objects.  In particular, we adhere to the convention that from now on the letter $A$ denotes pre-regular observables unless explicitly stated otherwise.
	To relate our results to the regular (in the sense of Definition~\ref{def:reg}) observables at the terminal time $t=T$, we now introduce the concept of time-dependent regularization along the characteristic flow. 
 	For all $(z_1, z_2) \in \bddD$, all times  $t\in[0,t]$, and $B \in \mathbb{C}^{N\times N}$, we define the time-dependent correction coefficient $\Upsilon_t(z_1, B, z_2)$ as 
 	\begin{equation} \label{eq:Ups_def}
 		 \Upsilon_t(z_1, B, z_2) := \frac{\bigl\langle \widehat{M}_t(z_1, I, z_2) B \bigr\rangle}{\bigl\langle \widehat{M}_t(z_1, I, z_2) \bigr\rangle},
 	\end{equation}
 	where $\widehat{M}_t(z_1, I, z_2)$ is defined in \eqref{eq:ImM2t_def}. 
 	Then, we define the time-dependent regularization of $B$ by 
 	\begin{equation} \label{eq:regB_t_def}
 		\reg{B}_t \equiv \reg{B}^{t,z_1, z_2} := B - \Upsilon_t(z_1, B, z_2)\,I, \qquad t\in[0,T]. 
 	\end{equation}
	Note that at time $t=T$, we have $\reg{B}^{T,z_1, z_2} = \reg{B}^{z_1, z_2}$. 
	
	In the sequel, we frequently compute the correction to regularity $\Upsilon_t(z_1,A,z_2)$ and the regularization $\reg{A}_t^{z_1,z_2}$ for observables $A$ that are already pre-regular with respect to the same pair of spectral parameters $(z_1,z_2)$. Accordingly, we suppress the dependence on $(z_1,z_2)$ in the notation unless a different pair of spectral parameters is used, in which case we state it explicitly. Thus, whenever $A$ is pre-regular with respect to $(z_1,z_2)$ in the sense of Definition~\ref{def:prereg}, we abbreviate
	\begin{equation}\label{eq:regA_def}
		\Upsilon_t(A):=\Upsilon_t(z_1,A,z_2), \qquad
		\reg{A}_t:=\reg{A}^{t,z_1,z_2}.
	\end{equation}
	\nc Moreover, it is straightforward to check that the $(z_1, z_2)$-pre-regular observable $A$ can be recovered from its regularized version via
	\begin{equation} \label{eq:un-reg}
		A = \reg{A}_t - \Bigl\langle \reg{A}_t \frac{M(z_1^+) - M(z_2^-)}{\langle M(z_1^+) - M(z_2^-)\rangle } \Bigr\rangle \, I. 
	\end{equation}
 	Finally, in the sequel, we abbreviate $\reg{A}_{j,t} := \reg{(A_j)}_t$.
 	
 	We also introduce the time-dependent version of the norm $\vertiii{\cdot}_{z_1, z_2}$ from \eqref{eq:tripplenorm}. 	
 	For all $(z_1, z_2) \in \regD$ and 
 	matrices $B \in \mathbb{C}^{N\times N}$, we define 
 	\begin{equation} \label{eq:size_t_def}
 		\vertiii{B}_{t,z_1, z_2} := \frac{\bigl\lvert \bigl\langle {\vecL}(z_1^+, z_2^+) \reg{B}^{t,z_1,z_2}  \bigr\rangle \bigr\rvert}{ \other{\alpha}_t(z_1^+, z_2^{+})  } + \norm{\reg{B}^{t,z_1,z_2}}_{\mathrm{hs}}, \qquad t\in[0,T],
 	\end{equation}
 	where the matrix ${\vecL}(z_1, z_2)$ is defined in \eqref{eq:L_def}, and the quantity $\other{\alpha}_t(z,w)$ is given by
 	\begin{equation} \label{eq:alp_t_def}
 		\other{\alpha}_t(z, w) := \lvert \sigma(z)\rvert + \other{\beta}_t(z,w)^{1/2}, \qquad z, w \in \mathbb{H}, \qquad t\in[0,T]. 
 	\end{equation}
 	Here $\sigma$ is defined in \eqref{eq:sigma_def} and $\other{\beta}_t$ is the stability factor from \eqref{eq:betaf_t_def}.
 	Note that, by \eqref{eq:alpha_sym} and \eqref{eq:betaf_t_def}, 
 	\begin{equation}
 		\other{\alpha}_t(z, w) \sim \other{\alpha}_t(w, z), \qquad z,w \in \bddD\cap\mathbb{H}.
 	\end{equation}  
 	Moreover, $\other{\alpha}_t(z, w)$ is decreasing in time $t \in [0,T]$, and the norm in~\eqref{eq:size_t_def} depends on $t$ only through the scalar quantity $\other{\beta}_t$;  	in particular, the matrix ${\vecL}(z_1^+, z_2^+)$ depends solely on the spectral parameters at the terminal time $T$. 
 		This simplification is possible because, applying the formulas~\eqref{eq:R_def}--\eqref{eq:R_def_p2} for $\widehat{R}(z,w)$ with time dependent $M_t$, $z_t$, and $w_t$, one finds that the direction of $R(z,w)$ is in fact independent of time by~\eqref{eq:M_t}.
 		Furthermore, at the terminal time $t=T$, $\vertiii{A}_{T,z_1, z_2} = \vertiii{A}_{z_1, z_2}$, where the latter time-independent norm is defined in \eqref{eq:tripplenorm}.

 	We now state the main result of this section, which provides bounds on $M_t(z_{1}, \reg{A}_t , z_2)$ along the characteristic flow~\eqref{eq:char_flow}. The time-independent Theorem~\ref{prop:M2_bounds} is then obtained as a straightforward corollary by evaluating the estimates below at the final time $t=T$.  
 	\begin{proposition} \label{prop:Mt_bound}
 		Let $z_1, z_2 \in \regD$, defined in \eqref{eq:regD_beta}. Then, for any observables $A_1$ and $A_2 \in \mathbb{C}^{N\times N}$, pre-regular with respect to $(z_1, z_2)$ and $(z_2, z_1)$, respectively, in the sense of Definition~\ref{def:prereg}, the quantity $\langle \widehat{M}_t(z_1, \reg{A}_{1,t}, z_2)\reg{A}_{2,t}\rangle$ satisfies the bound 
 		\begin{equation} \label{eq:imMt_bound}
 			\Bigl\lvert \bigl\langle \widehat{M}_t(z_1, \reg{A}_{1,t}, z_2) \reg{A}_{2,t}\bigr\rangle \Bigr\rvert \lesssim |\rho(z_1)\rho(z_2)|\, \vertiii{A_1}_{t,z_1, z_2} \vertiii{A_2}_{t,z_2,z_1}~, 
 			\qquad t\in[0,T].
 		\end{equation}
	Furthermore, the auxiliary quantity $\langle M_t(z_1, \reg{A}_{1,t}, z_2) \rangle$ satisfies, for all $t \in [0,T]$, 
		\begin{equation} \label{eq:GImG_bounds}
			\begin{split}
				\Bigl\lvert \bigl\langle M_t(z_1, \reg{A}_{1,t}, z_2) - M_t(\overline{z}_1, \reg{A}_{1,t}, z_2) \bigr\rangle \Bigr\rvert &\lesssim \frac{|\rho(z_1)|}{\other{\alpha}_t(z_1^+, z_2^+)}\, \vertiii{A_1}_{t,z_1, z_2},\\
				\Bigl\lvert \bigl\langle M_t(z_1, \reg{A}_{1,t}, z_2) - M_t(z_1, \reg{A}_{1,t}, \overline{z}_2) \bigr\rangle \Bigr\rvert &\lesssim \frac{|\rho(z_2)|}{\other{\alpha}_t(z_1^+, z_2^+)}\, \vertiii{A_1}_{t,z_1, z_2},
			\end{split}
		\end{equation}
 		\begin{equation} \label{eq:M2aux_t}
 			\bigl\lvert \bigl\langle M_t(z_1, \reg{A}_{1,t}, z_2) \bigr\rangle \bigr\rvert \lesssim  \vertiii{A_1}_{t,z_1, z_2}.
 		\end{equation} 
 	\end{proposition}
 	We remark that the bounds~\eqref{eq:imMt_bound}--\eqref{eq:M2aux_t} are optimal for general observables $A_j$ and spectral parameters $z_j$, which can be seen, for example, by an explicit computation in the case of deformed Wigner matrices  (see Appendix~\ref{app:Moptimal}). \nc
	
	Furthermore, we record the bound on the correction to regularity, and a closely related equivalent definition of~$\vertiii{\cdot}_{t,z_1,z_2}$, the proof of which we defer to the end of the present section.
	\begin{claim} \label{claim:Ups}
		Let $z_1, z_2 \in \regD\cap(\mathbb{H}^2)$, and let $A \in \mathbb{C}^{N\times N}$ be a $(z_1, z_2)$-pre-regular observable in the sense of Definition~\ref{def:prereg}. Then, the time-dependent correction to regularity $\Upsilon_t(z_1, A, z_2)$, defined in \eqref{eq:Ups_def}, admits the bound 
		\begin{equation} \label{eq:Ups_bound}
			\bigl\lvert \Upsilon_t(z_1, A, z_2) \bigr\rvert \lesssim \other{\beta}_t(z_1, \overline{z}_2) \vertiii{A}_{t,z_1, z_2}, \qquad t\in[0,T],
		\end{equation}
		where we recall the definition of $\other{\beta}_t$ from \eqref{eq:betaf_t_def}.
		 
		Moreover, the norm $\vertiii{A}_{t,z_1, z_2}$, defined in \eqref{eq:size_t_def}, satisfies the following comparison 
		\begin{equation} \label{eq:trip_equiv}
			\vertiii{A}_{t,z_1, z_2} = \frac{\bigl\lvert \bigl\langle {\vecL}(z_1, z_2) \reg{A}_t \bigr\rangle \bigr\rvert}{ \other{\alpha}_t(z_1, z_2)  } + \bigl\lVert\reg{A}_t\bigr\rVert_{\mathrm{hs}}  \sim \frac{\bigl\lvert \bigl\langle {\vecL}(z_1, z_2) A \bigr\rangle \bigr\rvert}{ \other{\alpha}_t(z_1, z_2)  } + \norm{A}_{\mathrm{hs}}, \qquad t\in[0,T].
		\end{equation}
	\end{claim}

	\subsubsection{Proof Ingredients}		
  	Before presenting the proof of Proposition~\ref{prop:Mt_bound}, we collect the requisite ingredients. 
 	In the regime where the stability operators $\mathcal{B}_{t,w_1, w_2}^{-1}$ are bounded by an order-one constant for any choice of $w_j \in \{z_j, \overline{z}_j\}$, the bounds in Proposition~\ref{prop:Mt_bound} follow directly from the following simple claim, which we prove in Appendix~\ref{app:stab_rho_cancel}.
 	\begin{claim}  \label{claim:stab_rho_cancel}
 		Let $z_1, z_2 \in \bddD\cap\mathbb{H}$, and let $t \in [0,T]$ be a fixed time. 
 		Assume that the stability operators $\mathcal{B}_{t, w_1, w_2}$,  defined as in \eqref{eq:stab_t_def}, satisfy the bound
 		\begin{equation} \label{eq:stab_bound_assume}
 			\norm{\mathcal{B}_{t,w_1, w_2}^{-1}}_{\mathrm{hs}\to \mathrm{hs}} \lesssim 1, \qquad w_j \in \{z_j, \overline{z}_j\}, \quad j\in \{1,2\},
 		\end{equation}
 		for some positive $N$-independent constant $C \ge 1$, where $\overline{z}_{j,t}$ is the complex conjugate of the trajectory $z_{j,t}$. 
 		Let  $\mathcal{M}_{w_1, w_2}$ denote the super-operators mapping $A \mapsto M_t(w_1, A, w_2)$, that is,
 		\begin{equation} \label{eq:M2op_def}
 			\mathcal{M}_{w_1, w_2}[\,\cdot\,] \equiv \mathcal{M}_{t,w_1, w_2}[\,\cdot\,] := \mathcal{B}_{t,w_1, w_2}^{-1}\bigl[M_t(w_{1,t})\, \mathrm{Id}[\,\cdot\,]\, M_t(w_{2,t})\bigr], \qquad w_j \in \{z_j, \overline{z}_j\}, \quad j\in \{1,2\}.
 		\end{equation}
 		Then, for all $w_1 \in \{z_1, \overline{z}_1\}$ and $w_2 \in \{z_2, \overline{z}_2\}$, the differences of $\mathcal{M}_{w_1, w_2}$ satisfy the bounds
 		\begin{equation} \label{eq:M2op_rho_cancel}
 			\norm{\mathcal{M}_{z_1, w_2} - \mathcal{M}_{\overline{z}_1, w_2} }_{\mathrm{hs}\to \mathrm{hs}} \lesssim \rho(z_1) , \quad \norm{\mathcal{M}_{w_1, z_2} - \mathcal{M}_{w_1, \overline{z}_2} }_{\mathrm{hs}\to \mathrm{hs}} \lesssim \rho (z_2), 
 		\end{equation}
 		\begin{equation} \label{eq:M2op_rhorho_cancel}
 			\norm{\mathcal{M}_{z_1, \overline{z}_2} + \mathcal{M}_{\overline{z}_1, z_2}  -\mathcal{M}_{z_1, z_2}  - \mathcal{M}_{\overline{z}_1, \overline{z}_2}  }_{\mathrm{hs}\to \mathrm{hs}} \lesssim \rho(z_1) \rho(z_2).
 		\end{equation}
 	\end{claim} 
 	In particular, Lemma~\ref{lemma:stab_t} together with  Claim~\ref{claim:stab_rho_cancel} imply that the bounds \eqref{eq:imMt_bound}--\eqref{eq:M2aux_t} hold at time $t = 0$. 
 	Hence, to complete the proof of Proposition~\ref{prop:Mt_bound}, we need to control the change of  $M_t(z_1, A, z_2)$ along the flow \eqref{eq:char_flow}. 
 	To this end, we differentiate the definition \eqref{eq:stab_t_def} in time, and use \eqref{eq:datapair_flow} together with \eqref{eq:dMt}, to obtain 
 	\begin{equation}
 		\frac{\mathrm{d}}{\mathrm{d}t}\mathcal{B}_{t,z_1, z_2}[\,\cdot\,] := - M_{1,t} \langle \,\cdot\,\rangle M_{2,t}, \qquad t \in [0,T].
 	\end{equation}
 	Therefore, we deduce that 
 	\begin{equation} \label{eq:dstab}
 		\begin{split}
 			\frac{\mathrm{d}}{\mathrm{d}t}\mathcal{B}_{t,z_1, z_2}^{-1}[\,\cdot\,] &= \mathcal{B}_{t,z_1, z_2}^{-1}[M_{1,t}M_{2,t}] \bigl\langle \mathcal{B}_{t,z_1, z_2}^{-1}[\,\cdot\,] \bigr\rangle =  \Bigl\langle M_{1,t}^{-1}\frac{M_{1,t} - M_{2,t}}{z_{1,t} - z_{2,t}}M_{2,t}^{-1} \,\mathrm{Id}[\,\cdot\,] \Bigr\rangle\, \frac{M_{1,t} - M_{2,t}}{z_{1,t} - z_{2,t}} \\
 			&= \frac{(z_{1} - z_{2})^2}{(z_{1,t} - z_{2,t})^2} \Bigl\langle M_{1}^{-1}\frac{M_{1} - M_{2}}{z_{1} - z_{2}}M_{2}^{-1} \,\mathrm{Id}[\,\cdot\,] \Bigr\rangle\, \frac{M_{1} - M_{2}}{z_{1} - z_{2}} ,
 		\end{split}
 	\end{equation}
 	where $M_j := M(z_j)$, we used \eqref{eq:M_t} and  the identities
 	\begin{equation} \label{eq:B12_identity}
 		\mathcal{B}_{t,z_1, z_2}^{-1}\bigl[M_{1,t}M_{2,t}\bigr] = \frac{M_{1,t} - M_{2,t}}{z_{1,t} - z_{2,t}}, \qquad  (\mathcal{B}_{t, z_1, z_2}^{-1})^\mathfrak{t}\bigl[I\bigr] = M_{1,t}^{-1}\frac{M_{1,t} - M_{2,t}}{z_{1,t} - z_{2,t}}M_{2,t}^{-1},
 	\end{equation}
 	obtained by subtracting two copies of the MDE \eqref{eq:MDE} with   spectral parameters $z_{1,t}$, $z_{2,t}$ and \nc time-dependent data $(D_t, \mathcal{S}_t)$.  	
 	Differentiating \eqref{eq:Mt_def} in time, using \eqref{eq:dMt} and \eqref{eq:dstab}, we conclude that, for all $t\in[0,T]$,
 	\begin{equation} \label{eq:dM2}
 		\frac{\mathrm{d}}{\mathrm{d}t} M_t(z_1, A, z_2) = M_t(z_1, A, z_2) +  \ee^{t-T}\frac{(z_{1} - z_{2})^2}{(z_{1,t} - z_{2,t})^2} \Bigl\langle \frac{M(z_{1}) - M(z_{2})}{z_{1} - z_{2}}\, A\Bigr\rangle\, \frac{M(z_{1}) - M(z_{2})}{z_{1} - z_{2}},
 	\end{equation} 	
	where we recall that $z_{j}=z_{j,T}$ and $M(z_j)=M_T(z_{j,T})$.
 	Here and in the sequel, all divided-difference quantities, such as $\frac{M_{1} - M_{2}}{z_{1} - z_{2}}$, are understood in the limiting sense in case $z_1 = z_2$. 
 	Applying Duhamel's principle to the equation \eqref{eq:dM2}, we obtain, for all $t \in [0,T]$, 
 	\begin{equation} \label{eq:M2_duhamel}
 		M_t(z_1, A, z_2) = \ee^{t} M_0(z_1, A, z_2) +  \Biggl[\ee^{t-T}\int_{0}^t \frac{(z_1 - z_2)^2}{(z_{1,s} - z_{2,s})^2} \mathrm{d}s\Biggr] \, \Bigl\langle \frac{M(z_{1}) - M(z_2)}{z_1 - z_2}\, A\Bigr\rangle\, \frac{M(z_{1}) - M(z_2)}{z_1 - z_2}.
 	\end{equation}	
 	In the sequel, we use the \eqref{eq:M2_duhamel} to control the terms $M_t(z_1, A, z_2)$ with pre-regular observables $A$. 
 	
 	To obtain the optimal bounds on $M_t(z_1, \reg{A}_t, z_2)$ with $\reg{A}_t$ regular in the sense of \eqref{eq:regA_def}, we need to capture an intricate cancellation between $M_t(z_1, A, z_2)$ and a term involving the corresponding correction to regularity $\Upsilon_t(z_1, A, z_2)$, defined in~\eqref{eq:Ups_def}. 
 	This cancellation gives rise to the factor $\other{\alpha}_t(z_1, z_2)^{-1}$ on the right-hand side of \eqref{eq:size_t_def}, and its presence can be viewed as a consequence of $\langle \widehat{M}_t(z_1, I, z_2) \rangle$ appearing in the denominator of $\Upsilon_t$, together with the following two statements, which we prove in Appendix~\ref{app:shape}.   

 	\begin{lemma} \label{lemma:m_hat}
 		Let $z_1, z_2 \in \bddD\cap \mathbb{H}$. Then,   $\widehat{M}_t(z_1, I, z_2)$, defined in \eqref{eq:ImM2t_def} is flat in the sense that 
 		\begin{equation} \label{eq:Mhat_flat}
 			\norm{\widehat{M}_t(z_1, I, z_2)} \sim \bigl\langle \widehat{M}_t(z_1, I, z_2) \bigr\rangle, \qquad t\in [0,T],  
 		\end{equation}
 		\nc and satisfies the comparison 
 		\begin{equation} \label{eq:ImGImG_comp}
 			\bigl\langle \widehat{M}_t(z_1, I, z_2) \bigr\rangle \sim \frac{\rho(z_{1})\Im z_{2,t} + \rho(z_2)\Im z_{1,t}}{|z_{1,t} - z_{2,t}|^2 + \kapd_{1,t}^2 + \kapd_{2,t}^2 }, \qquad t\in [0,T] ,
 		\end{equation}
 		for all $t\in [0,T]$, where we recall that $z_{j,t}$ solves \eqref{eq:char_flow}, and $\kapd_{j,t} := \kapd_t(z_{j,t}) = \dist(z_{j,t}, \supp\,\rho_t)$. 
 	\end{lemma}
 	
 	\begin{claim} \label{claim:alpha_kappa}
 		For all $z_1, z_2 \in \bddD\cap \mathbb{H}$, the quantity $\other{\alpha}(z_1, z_2)$, defined in \eqref{eq:alp_t_def}, satisfies
 		\begin{equation} \label{eq:alpha_comp}
 			\other{\alpha}_t(z_1, z_2)^2 \sim \frac{ \other{\beta}_t(z_1, z_2)^2\other{\beta}_t(z_1, \overline{z}_2)^2}{|z_{1,t} - z_{2,t}|^2 + \kapd_{1,t}^2 + \kapd_{2,t}^2 } , \qquad t\in [0,T].  
 		\end{equation}	
 	\end{claim}
 	
 	In particular, it follows form \eqref{eq:ImGImG_comp} and \eqref{eq:alpha_comp} that 
 	\begin{equation} \label{eq:m_hat_alpha}
 		\bigl\langle \widehat{M}_t(z_1, I, z_2) \bigr\rangle 
 		\sim \frac{\rho(z_{1})\Im z_{2,t} + \rho(z_2)\Im z_{1,t}}{ \other{\beta}_t(z_1, z_2)^2\other{\beta}_t(z_1, \overline{z}_2)^2} \, \other{\alpha}_t(z_1, z_2)^2  , \qquad t\in [0,T]. 
 	\end{equation}
 	
 	The quantity $\other{\alpha}(z_1, z_2)$ plays another, equally important role: it gives a natural bound on the local Lipschitz factor of the divided difference quantity $ \frac{z_1 - z_2}{\langle M_1 - M_2 \rangle }$
	 viewed as a function\footnote{
 		Since $M(z)$ solves the MDE \eqref{eq:MDE}, $z = D - \mathcal{S}[M] -M^{-1}$ is a function of $M$.
	} of $M_1$ and $M_2$.
 	\begin{lemma} \label{lemma:beta_reg}
 		Let $z_1, z_2 \in \bddD$ be a pair of spectral parameters satisfying $\min\{\other{\beta}(z_1^+, z_2^+), \other{\beta}(z_2^+, z_1^+)\} \le \beta_*$, where $\beta_*$ is the constant from Lemma~\ref{lemma:stab_bound}. Then, for all $w\in\{z_1, \overline{z}_1, \overline{z}_2\}$ we have the bounds
 		\begin{equation} \label{eq:u_reg}
 			\begin{split}
 				\biggl\lvert  \frac{z_{1,t} - z_{2,t}}{\langle M_t(z_{1,t}) - M_t(z_{2,t}) \rangle} - \frac{z_{1,t} - w_t}{\langle M_t(z_{1,t}) - M_t(w_t) \rangle} \biggr\rvert
 				 \lesssim \other{\alpha}(z_1^+, z_2^+)\, \bigl\lvert \bigl\langle M(w) - M(z_2) \bigr\rangle \bigr\rvert, \qquad t \in [0,T]
 			\end{split}
 		\end{equation}  		
 		 where $\alpha(z_1^+, z_2^+) := \alpha_T(z_1^+, z_2^+)$. 
 	\end{lemma} 	 	
 	We prove Lemma~\ref{lemma:beta_reg} in Section~\ref{sec:quadratic}.
 	In the sequel, we use Lemma~\ref{lemma:beta_reg} in the form of the following simple corollary: For all $z_1, z_2 \in \bddD\cap\mathbb{H}$  and $t\in [0,T]$,  
 	\begin{equation} \label{eq:beta_reg}
 		\begin{split}
 			\biggl\lvert \Im\biggl[\frac{z_{1,t} - z_{2,t}}{\langle M_{1,t} - M_{2,t} \rangle} \biggr] \biggr\rvert 
 			&\lesssim \other{\alpha}(z_1, z_2)\, \bigl(\rho(z_1) + \rho(z_2)\bigr), \qquad 
 			\biggl\lvert \frac{\Im z_{1,t}}{\pi\rho_t(z_{1,t})} - \frac{\Im z_{2,t}}{\pi\rho_t(z_{2,t})} \biggr\rvert
 			 \lesssim \other{\alpha}(z_1, z_2)\,\bigl\lvert \langle M_1 - M_2 \rangle \bigr\rvert, 
 		\end{split}
 	\end{equation}
 	which follows from the bound \eqref{eq:u_reg} and its copy with the roles of $z_1$ and $z_2$ interchanged by triangle inequality. Here we used the symmetry of $\other{\alpha}_t(z_1, z_2)$ by \eqref{eq:alpha_sym}.

 	Furthermore, in Section~\ref{sec:Gamma_proof}, we show that the divided difference matrices $\frac{M_{1} - M_{2}}{z_{1} - z_{2}}$ satisfy the following key second-order cancellations.
 	\begin{lemma} \label{lemma:Gamma}  
 		Let $z_1, z_2 \in \bddD\cap\mathbb{H}$ be a pair of spectral parameters satisfying $\max\{\other{\beta}(z_1, z_2), \other{\beta}(z_2, z_1)\} \le \beta_*$. Then, 
 		\begin{equation} \label{eq:Gamma_cancel}  
 			\biggl\lVert \frac{1}{\rho_2} \, \frac{M_1 - M_2}{\langle M_1 - M_2 \rangle}
 			+ \frac{1}{\rho_1} \, \frac{M_1^* - M_2^*}{\langle M_1^* - M_2^* \rangle}  
 			- \frac{\rho_1 + \rho_2}{\rho_1\rho_2}  \, \frac{M_1 - M_2^*}{\langle M_1 - M_2^* \rangle} \biggr\rVert_{\mathrm{hs}} \lesssim 
 			\bigl\lvert \langle M_1 - M_2^* \rangle \bigr\rvert,
 		\end{equation}  
 		\begin{equation} \label{eq:Gamma_sizes}
 			\biggl\lVert \frac{M_1^* - M_2^*}{\langle M_1^* - M_2^* \rangle} - \frac{M_1 - M_2^*}{\langle M_1 - M_2^* \rangle} \biggr\rVert_{\mathrm{hs}} \lesssim \rho_1 , \qquad \biggl\lVert \frac{M_1 - M_2}{\langle M_1 - M_2 \rangle} - \frac{M_1 - M_2^*}{\langle M_1 - M_2^* \rangle} \biggr\rVert_{\mathrm{hs}} \lesssim \rho_2, 
 		\end{equation}
 		where $M_j := M(z_j)$ and $\rho_j := \rho(z_j)$.
 		Furthermore, for all $t\in[0,T]$, we have the bounds
 		\begin{equation} \label{eq:1/u-4fold}
 			\biggl\lvert \Re\biggl[\frac{z_{1,t} - z_{2,t}}{\langle M_{1,t} - M_{2,t} \rangle} - \frac{z_{1,t} - \overline{z}_{2,t}}{\langle M_{1,t} - M_{2,t}^* \rangle}\biggr] \biggr\rvert \lesssim \rho_1 \rho_2. 
 		\end{equation}
 		\begin{equation} \label{eq:1/u-rho_eta}
 			\biggl\lvert 2\Re\biggl[\frac{z_{1,t} - z_{2,t}}{\langle M_{1,t} - M_{2,t} \rangle}\biggr] - \frac{\Im z_{1,t}}{\rho_t(z_{1,t})} - \frac{\Im z_{2,t}}{\rho_t(z_{2,t})} \biggr\rvert \lesssim \bigl\lvert \langle M_1 - M_2^* \rangle \bigr\rvert^2. 
 		\end{equation}
 	\end{lemma}
 	
 	We also employ the following technical claim, proved in Appendix~\ref{app:beta_prop}, which asserts that a large difference between the values of $\other{\beta}(z, \overline{w})$ and $\other{\beta}(z, w)$ can only arise when densities $\rho(z)$ and $\rho(w)$ are order unity. 
 	\begin{claim} \label{claim:stab_diff_rho}
 		There exists a threshold $c_* \sim 1$, satisfying $c_* < \tfrac{1}{2}\beta_*$~,  such that for all $z, w \in \bddD\cap\mathbb{H}$, if the stability factors satisfy the chain of inequalities
 		\begin{equation} \label{eq:stab_gap}
 			\max\bigl\{\other{\beta}(z, \overline{w}), \other{\beta}(\overline{w}, z)\bigr\}  \le c_* < \tfrac{1}{2}\beta_* \le  \max\bigl\{\other{\beta}(z, w), \other{\beta}(w, z)\bigr\},
 		\end{equation}
 		then
 		\begin{equation} \label{eq:rho_large}
 			\rho(z) \gtrsim 1, \qquad \rho(w) \gtrsim 1.
 		\end{equation} 
 	\end{claim} 
 	
 	\subsubsection{Proof of $M$-Bounds along the Flow} 	
 	Equipped with Claims~\ref{claim:stab_rho_cancel} and~\ref{claim:stab_diff_rho}, and Lemmas~\ref{lemma:m_hat}--\ref{lemma:Gamma}, we are ready to prove Proposition~\ref{prop:Mt_bound}.
 	
 	\begin{proof} [Proof of Proposition~\ref{prop:Mt_bound}]
 		We proceed by considering three cases   depending on the proximity of the spectral parameters $z_1, z_2$ to the singularities of $\rho$ and on their mutual distance $|z_1 - z_2|$,   encoded in $\other{\beta}(z_1, z_2)$ and $\other{\beta}(z_1, \overline{z}_2)$. 
 		The most delicate regime is Case 1 below, in which both $z_1$ and $z_2$ are close to the same singularity.
 		In Case 2, both $z_1, z_2$ lie in the bulk of the spectrum and are close to each other, so that the stability operator with spectral parameters in opposite half-planes still possesses a small eigenvalue. 
 		Finally, Case 3 covers the remaining and simplest situation, in which the inverses of all relevant stability operators are uniformly bounded.  \nc 		
 		
 		\smallskip
 		\textbf{Case 1}. First, we complete the proof under the assumption
 		\begin{equation} \label{eq:2M_bound_case1}
 			\max\bigl\{\other{\beta}(z_1^+, z_2^+), \, \other{\beta}(z_2^+, z_1^+) \bigr\} \le \beta_*. 
 		\end{equation}
 		 
 		We begin by proving \eqref{eq:imMt_bound}. Without loss of generality, we can assume that $z_1, z_2 \in \mathbb{H}$.  		We define
 		\begin{equation} \label{eq:hatm_def}
 			\widehat{m}_t \equiv \widehat{m}_t(z_1,A_1, z_2, A_2) := \Bigl\langle \widehat{M}_t(z_1, \reg{A}_{1,t}, z_2)\reg{A}_{2,t} \Bigr\rangle ,
 		\end{equation}
 		It follows from the linearity of $A \mapsto \widehat{M}(z_1, A, z_2)$ that 
 		\begin{equation} \label{eq:hatm_1}
 			\widehat{m}_t = \bigl\langle \widehat{M}_t(z_1, A_1 , z_2) A_2\bigr\rangle - \frac{\bigl\langle \widehat{M}_t(z_1, I, z_2) A_1 \bigr\rangle\bigl\langle \widehat{M}_t(z_1, I, z_2) A_2 \bigr\rangle}{\bigl\langle \widehat{M}_t(z_1, I, z_2) \bigr\rangle}. 
 		\end{equation}
 		Here we used the identities  
 		\begin{equation} \label{eq:M12_dual}
 			\bigl\langle M_t(z_1, B, z_2) \bigr\rangle = \bigl\langle M_t(z_1, I, z_2) B \bigr\rangle, \qquad   M_t(z_1, I, z_2) =  M_t(z_2, I, z_1), \qquad t \in [0,T], \qquad B \in \mathbb{C}^{N\times N}, 
 		\end{equation}
 		which follow from \eqref{eq:Mt_def} and \eqref{eq:B12_identity}.  
 		First, we develop each of the two terms in \eqref{eq:hatm_1} into the form which will subsequently allow us to identify a cancellation between them. 
 		
 		Plugging \eqref{eq:M2_duhamel} into \eqref{eq:ImM2t_def}, we deduce that the first term on the right-hand side of \eqref{eq:hatm_1} satisfies
 		\begin{equation} \label{eq:hatM_AA_part}
  				\bigl\langle \widehat{M}_t(z_1, A_1 , z_2) A_2\bigr\rangle = -  \frac{u_t}{4} \frac{\langle U A_1 \rangle}{\langle U \rangle} \frac{\langle U A_2 \rangle}{\langle U \rangle}
  				- \frac{\overline{u}_t}{4} \frac{\langle U^* A_1 \rangle}{\langle U^* \rangle} \frac{\langle U^* A_2 \rangle}{\langle U^* \rangle} + \ee^t \mathfrak{m}_0, 
 		\end{equation} 
 		where the quantity $\mathfrak{m}_0$ is given by 
 		\begin{equation} \label{eq:m0_def}
 			\mathfrak{m}_0 := \bigl\langle \widehat{M}_0(z_1, A_1 , z_2) A_2\bigr\rangle  +  \frac{u_0}{4}\frac{\langle U A_1 \rangle}{\langle U \rangle} \frac{\langle U A_2 \rangle}{\langle U \rangle} + \frac{\overline{u}_0}{4}\frac{\langle U^* A_1 \rangle}{\langle U^* \rangle} \frac{\langle U^* A_2 \rangle}{\langle U^* \rangle}. 
 		\end{equation}
 		Here we introduce the notation   
 		\begin{equation} \label{eq:U_def}
 			U \equiv U_T, \qquad U_t :=  \frac{M_t(z_{1,t}) - M_t(z_{2,t})}{ z_{1,t} - z_{2,t} },  \qquad u_t :=  \langle U_t \rangle, \qquad  t \in [0,T],
 		\end{equation}
 		and used \eqref{eq:M_t}--\eqref{eq:z_t} to express the scalar integral on the right-hand side of \eqref{eq:M2_duhamel} as
 		\begin{equation} \label{eq:scal_int_1} 
 			\ee^{t-T}\int_{0}^t \frac{(z_1 - z_2)^2}{(z_{1,s} - z_{2,s})^2} \mathrm{d}s = \frac{u_t - \ee^tu_0}{(u_T)^2}. 
 		\end{equation}
 		In the sequel we show that $\mathfrak{m}_0$ admits the bound  
 		\begin{equation} \label{eq:m0_bound}
 			 \lvert \mathfrak{m}_0  \rvert \lesssim \rho_1 \rho_2 \norm{A_1}_\mathrm{hs}\norm{A_2}_\mathrm{hs},
 		\end{equation}
 		where we abbreviate $\rho_j := \rho(z_j)$ and recall that $\rho_t(z_j) \sim \rho_j$ for all $t\in[0,T]$. We defer the proof of \eqref{eq:m0_bound} until after analyzing the more singular contributions coming from the other terms in \eqref{eq:hatm_1} and \eqref{eq:hatM_AA_part}.
 		
 		Next, we  express the second term on the right-hand side of \eqref{eq:hatm_1}. To this end, we introduce the following notation complementary to \eqref{eq:U_def}, 
 		\begin{equation} \label{eq:Vt_def}
 			V \equiv V_T, \qquad V_t :=   \frac{M_t(z_{1,t}) - M_t^*(z_{2,t})}{ z_{1,t} - \overline{z}_{2,t} },  \qquad v_t :=  \langle V_t \rangle, \qquad  t \in [0,T].   
 		\end{equation}
		Note that by  \eqref{eq:Mt_def} and \eqref{eq:B12_identity}, we have   
$$ 
         U_t= M_t(z_1, I, z_2), \qquad V_t= M_t(z_1, I, \overline{z}_2), \qquad U_t^*= M_t(\overline{z}_1, I, \overline{z}_2),
         \qquad V_t^*= M_t(\overline{z}_1, I, z_2).
$$
 		Then, it follows from   \eqref{eq:ImM2t_def} 
		that 
 		\begin{equation} \label{eq:hatM_eq}
 			\widehat{M}_t(z_1, I , z_2) = \frac{1}{4}\bigl(V_t + V_t^* - U_t - U_t^*\bigr), \qquad \bigl\langle \widehat{M}_t(z_1, I , z_2) \bigr\rangle = \frac{1}{2}\Re\bigl[v_t-u_t\bigr], \qquad t\in[0,T]. 
 		\end{equation}
 		Recall that, by Definition~\ref{def:reg}, we have
 		\begin{equation} \label{eq:As_reg}
 			\langle A_1 V\rangle = \langle A_2 V^* \rangle = 0.
 		\end{equation} 
 		To account for the regularity of the observables when testing $\widehat{M}_t(z_1, I , z_2)$ against $A_j$, we eliminate the vector $V^*$ for $A_1$ and $V$ for $A_2$ from \eqref{eq:hatM_eq} using the trivial vanishing of the linear combination
 		\begin{equation} \label{eq:UVlin_dep}
 			(z_{1,t} - z_{2,t}) U_t + (\overline{z}_{1,t} - \overline{z}_{2,t}) U_t^*- (z_{1,t} - \overline{z}_{2,t}) V_t - (\overline{z}_{1,t} - z_{2,t}) V_t^* = 0. 
 		\end{equation}
 		Combining \eqref{eq:hatM_eq}, \eqref{eq:As_reg} and \eqref{eq:UVlin_dep}, we conclude that
 		\begin{equation} \label{eq:hatMA}
 			\bigl\langle \widehat{M}_t(z_1, I , z_2) A_1 \bigr\rangle = \frac{\ii}{2}  \Bigl\langle\frac{   U_t  \eta_{1,t} +    U_t^* \eta_{2,t}}{\overline{z}_{1,t} - z_{2,t}}A_1\Bigr\rangle, \qquad
 			\bigl\langle \widehat{M}_t(z_1, I , z_2) A_2 \bigr\rangle = \frac{-\ii}{2}  \Bigl\langle\frac{   U_t  \eta_{2,t} + U_t^* \eta_{1,t} }{z_{1,t}-\overline{z}_{2,t}}A_2\Bigr\rangle,
 		\end{equation}
 		where we recall that $\eta_{j,t} := \Im z_{j,t}$. 
 		
 		From  \eqref{eq:hatm_1}--\eqref{eq:hatM_AA_part}  and \eqref{eq:hatMA}, we conclude that, up to the  $\mathfrak{m}_0$ term bounded by \eqref{eq:m0_bound}, $\widehat{m}_t$ is given by a
		bilinear form on $A_1$, $A_2$ of rank at most two generated by $\{u_t^{-1} U_t, (u_t^{-1} U_t)^*\}$. 
		To accurately estimate the size of $\widehat{m}_t$, we perform a change of basis. Observe that the bound \eqref{eq:Gamma_cancel} of Lemma~\ref{lemma:Gamma} implies that the linear combination 
 		\begin{equation} \label{eq:Zdef}
 			Z \equiv Z(z_1, z_2) := \frac{1}{2\ii\rho_2}\frac{U}{ \langle U \rangle} + \frac{1}{2\ii\rho_1}\frac{U^*}{ \langle U^* \rangle}
 		\end{equation}
 		is of order $|m_1 - \overline{m}_2|$ when tested against $A_1$ and the same holds for $Z^*$ when tested against $A_2$ by conjugation.
 		More precisely, it follows from \eqref{eq:Gamma_cancel} and \eqref{eq:As_reg}
 		\begin{equation} \label{eq:ZA_bounds}
 			\bigl\lvert \langle Z A_1 \rangle \bigr\rvert \lesssim |m_1 - \overline{m}_2| \norm{A_1}_\mathrm{hs}, \qquad \bigl\lvert \langle Z^* A_2 \rangle \bigr\rvert  \lesssim |m_1 - \overline{m}_2| \norm{A_2}_\mathrm{hs}.	 
 		\end{equation}
 		We consider the complementary matrix $Y$, given by
 		\begin{equation} \label{eq:Ydef}
 			Y \equiv Y(z_1, z_2) :=  \frac{1}{2\ii\rho_2}\frac{U}{\langle U \rangle } - \frac{1}{2\ii\rho_1}\frac{U^*}{ \langle U^* \rangle}.
 		\end{equation}
 		
		Estimates \eqref{eq:beta_assymp}  and \eqref{eq:z_diff_comp} imply  that 
 		\begin{equation} \label{eq:uv_comp}
 			|u_t| \sim \other{\beta}(z_1, z_2)^{-1}, \qquad |v_t| \sim \other{\beta}(z_1, \overline{z}_2)^{-1}, \qquad t\in[0,T].
 		\end{equation}
 		Projecting $\langle U \rangle^{-1} U$ onto ${\vecL}(z_1, z_2)$, defined in \eqref{eq:L_def},
		using  \eqref{eq:Gamma_sizes} and \eqref{eq:As_reg}, together with \eqref{eq:ZA_bounds} and \eqref{eq:uv_comp}, we deduce from \eqref{eq:Zdef} and \eqref{eq:Ydef} that
 		\begin{equation} \label{eq:YA_bounds}
 			\bigl\lvert \langle Y A_1 \rangle \bigr\rvert \lesssim \bigl\lvert \bigl\langle {\vecL}(z_1, z_2) A_1 \bigr\rangle \bigr\rvert + |m_1 - \overline{m}_2| \norm{A_1}_\mathrm{hs}, \qquad  \bigl\lvert \langle Y^* A_2 \rangle \bigr\rvert  \lesssim \bigl\lvert \bigl\langle {\vecL}(z_2, z_1) A_2 \bigr\rangle \bigr\rvert + |m_1 - \overline{m}_2| \norm{A_2}_\mathrm{hs}.  
 		\end{equation} 		
 		Observe that the matrices $u_t^{-1} U_t$ and $v_t^{-1} V_t$ are time-independent by \eqref{eq:M_t}, hence, 
 		\begin{equation} \label{eq:UtoYZ}
 			U_t = \ii \rho_2 u_t \, (Y + Z), \qquad U_t^* = \ii \rho_1 \overline{u}_t \,(-Y + Z).   
 		\end{equation}
 		
 		Performing the change of coordinates $\{u_t^{-1} U_t, (u_t^{-1} U_t)^*\}  \mapsto \{Y, Z\}$, we obtain
 		\begin{equation} \label{eq:hatmt_expr}
 			\widehat{m}_t = \mathfrak{b}_{11,t} \langle Y A_1 \rangle \langle Y^* A_2 \rangle + \mathfrak{b}_{12,t} \langle Y A_1 \rangle \langle Z^* A_2 \rangle + \mathfrak{b}_{21,t} \langle Z A_1 \rangle \langle Y^* A_2 \rangle + \mathfrak{b}_{22,t} \langle Z A_1 \rangle \langle Z^* A_2 \rangle  +  \ee^t\mathfrak{m}_0,
 		\end{equation} 
 		where the scalar coefficients $(\mathfrak{b}_{ij,t})_{i,j=1}^2$ are given by 
 		\begin{equation} \label{eq:coeffb_def}
 			\begin{split}
 				\mathfrak{b}_{11,t} &:= 
 			 \frac{\rho_1\rho_2 \Re u_t}{2}
 			- \frac{ ( \rho_2    \eta_{1,t} + \rho_1  \eta_{2,t})^2|u_t|^2 - 4    \rho_1 \rho_2    \eta_{1,t} \eta_{2,t}  (\Re u_t)^2   }{2|z_{1,t}-\overline{z}_{2,t}|^2   \Re[v_t-u_t]},   \\
 			\mathfrak{b}_{12,t} = \overline{\mathfrak{b}}_{21,t} &:=  
 			  \frac{\ii \rho_1\rho_2   \Im u_t}{2}  
 			+  \frac{ \bigl((\rho_2    \eta_{1,t})^2 - (\rho_1  \eta_{2,t})^2\bigr)|u_t|^2 + 2 \ii \rho_1 \rho_2    \eta_{1,t} \eta_{2,t} \Im \bigl[u_t^2\bigr] }{2|z_{1,t}-\overline{z}_{2,t}|^2  \Re[v_t-u_t]},   \\
 			\mathfrak{b}_{22,t} &:=  
 			-  \frac{\rho_1\rho_2     \Re u_t}{2}
 			-  \frac{ ( \rho_2    \eta_{1,t} + \rho_1  \eta_{2,t})^2|u_t|^2  - 4    \rho_1 \rho_2    \eta_{1,t} \eta_{2,t}  (\Im u_t)^2  }{2|z_{1,t}-\overline{z}_{2,t}|^2  \Re[v_t-u_t]}.
 			\end{split}
 		\end{equation}

 		In the sequel, we show that the coefficients $\mathfrak{b}_{ij,t}$ of the bilinear form in \eqref{eq:hatmt_expr} admit the bounds
 		\begin{equation} \label{eq:b_bounds}
 			\bigl\lvert \mathfrak{b}_{11,t} \bigr\rvert \lesssim \frac{\rho_1 \rho_2}{\other{\alpha}_t(z_1, z_2)^2}, \qquad 
 			\bigl\lvert \mathfrak{b}_{12,t} \bigr\rvert \lesssim \frac{\rho_1 \rho_2}{|m_{1} - \overline{m}_{2}| \,\other{\alpha}_t(z_1, z_2) }, \qquad 
 			\bigl\lvert \mathfrak{b}_{22,t} \bigr\rvert \lesssim  \frac{\rho_1 \rho_2}{|m_{1} - \overline{m}_{2}|^2}. 
 		\end{equation}
 		Observe that, together with  \eqref{eq:m0_bound}, the bounds \eqref{eq:b_bounds} are  sufficient to establish the desired estimate on $\widehat{m}_t$. Indeed, plugging  \eqref{eq:m0_bound} and \eqref{eq:b_bounds} into \eqref{eq:hatmt_expr}, we deduce that
 		\begin{equation}  \label{eq:m_hat_bound}
 			\bigl\lvert \widehat{m}_t \bigr\rvert 
 			\lesssim \rho_1 \rho_2 \biggl(\frac{\bigl\lvert\langle Y A_1\rangle  \bigr\rvert }{\other{\alpha}_t(z_1, z_2)} + \frac{\bigl\lvert\langle Z A_1\rangle  \bigr\rvert }{|m_{1} - \overline{m}_{2}|}\biggr)\biggl(\frac{\bigl\lvert\langle Y^* A_2\rangle  \bigr\rvert }{\other{\alpha}_t(z_1, z_2)} + \frac{\bigl\lvert\langle Z^* A_2\rangle  \bigr\rvert }{|m_{1} - \overline{m}_{2}|}\biggr)
 			\lesssim \rho_1 \rho_2 \vertiii{A_1}_{t,z_1,z_2} \vertiii{A_2}_{t,z_2,z_1}~. 
 		\end{equation}
 		Here we used \eqref{eq:trip_equiv}, \eqref{eq:size_t_def}, \eqref{eq:ZA_bounds}, \eqref{eq:YA_bounds}, and the bound 
 		\begin{equation} \label{eq:m-m_beta_bound}
 			|m_{1,t}-\overline{m}_{2,t}| \lesssim |m_{1,t}-m_{2,t}| + \rho(z_1)  \lesssim \other{\beta}(z_1, z_2)^{1/2} \lesssim  \other{\alpha}_t(z_1, z_2),
 		\end{equation} 
 		which follows from \eqref{eq:betaf_def}, \eqref{eq:beta_assymp}, and \eqref{eq:alp_t_def}.
 		
 		We now turn to establishing \eqref{eq:b_bounds}. 		
 		We begin with the most challenging bound on $\mathfrak{b}_{11,t}$. 
 		First, we collect the relevant algebraic identities that we use to simplify
		the expression for $\mathfrak{b}_{11,t}$ in the sequel.
 		For all $u,v \in \mathbb{C}$, we have the following trivial formula
 		\begin{equation} \label{eq:uv_id}
 			\frac{\Re[v-u]\Re[u]}{|u|^2|v|^2} =  \Re\Bigl[\frac{1}{v}\Bigr]\Re\Bigl[\frac{1}{u} - \frac{1}{v}\Bigr] + \frac{|u|^2}{|v|^2}\biggl(\Im\Bigl[\frac{1}{u}\Bigr]\biggr)^2 - \biggl(\Im\Bigl[\frac{1}{v}\Bigr]\biggr)^2.
 		\end{equation} 
 		The definitions \eqref{eq:U_def} and \eqref{eq:Vt_def} imply that 
 		\begin{equation} \label{eq:cyclic2} 
 			\lvert m_{1,t} -\overline{m}_{2,t} \rvert^2 \Re\Bigl[\frac{1}{v_t}\Bigr]  - \lvert m_{1,t} - m_{2,t} \rvert^2 \Re\Bigl[\frac{1}{u_t}\Bigr] = 2 \pi (\rho_{1,t} \eta_{2,t} + \rho_{2,t} \eta_{1,t}), 
 		\end{equation}
 		where we recall that $m_{j,t} := \langle M_{t}(z_{j,t})\rangle$.
 		
 		We proceed by developing $\mathfrak{b}_{11,t}$. Using the definitions of $u_t$ and $v_t$  from \eqref{eq:U_def} and \eqref{eq:Vt_def}, applying \eqref{eq:uv_id} with $u := u_t$, $v := v_t$, and collecting terms, we obtain 
 		\begin{equation} 
 			\begin{split}
 				\frac{4\mathfrak{b}_{11,t}\Re[v_t-u_t]}{\ee^{T-t} |u_t|^{2}|v_t|^{2}} =&~ \frac{2\rho_{1,t}\rho_{2,t}}{|m_{1,t}-\overline{m}_{2,t}|^2} \Bigl( |m_{1,t}-\overline{m}_{2,t}|^2 \Re\Bigl[\frac{1}{v_t}\Bigr]   -  |m_{1,t}-m_{2,t}|^2  \Re \Bigl[\frac{1}{u_t}\Bigr]   \Bigr)\Re\Bigl[\frac{1}{u_t} \Bigr] \\
 				&-2\frac{ ( \rho_{2,t}    \eta_{1,t} + \rho_{1,t}  \eta_{2,t})^2}{|m_{1,t}-\overline{m}_{2,t}|^2}\\
 				=&~\frac{\rho_{1,t} \eta_{2,t} + \rho_{2,t} \eta_{1,t} }{\pi|m_{1,t}-\overline{m}_{2,t}|^2} \biggl(4\pi^2 \rho_{1,t}\rho_{2,t}   \Re\Bigl[\frac{1}{u_t} \Bigr]  -   2\pi( \rho_{2,t}    \eta_{1,t} + \rho_{1,t}  \eta_{2,t}) \biggr) 
 				\\=&~ \frac{ \rho_{1,t} \eta_{2,t} + \rho_{2,t} \eta_{1,t}  }{\pi} \Re\Bigl[\frac{1}{u_t} - \frac{1}{v_t}\Bigr].
 			\end{split}
 		\end{equation}
 		Here, we used the identity \eqref{eq:cyclic2} twice (once in the second and once in the third step), together with the fact that 
 		\begin{equation}
 			|m_{1,t}-\overline{m}_{2,t}|^2 - |m_{1,t}-m_{2,t}|^2 = 4\pi^2 \rho_{1,t}\rho_{2,t}. 
 		\end{equation}   
 		Therefore, by \eqref{eq:hatM_eq}, we deduce the  expression
 		\begin{equation} \label{eq:b11_expr}
 			\mathfrak{b}_{11,t} = \frac{1}{8\pi}\Re\Bigl[\frac{1}{u_t} - \frac{1}{v_t}\Bigr]  \frac{\rho_{1} \eta_{2,t} + \rho_{2} \eta_{1,t}}{ \bigl\langle \widehat{M}_t(z_1, I , z_2) \bigr\rangle }  |u_t|^2|v_t|^2.
 		\end{equation}
 		
 		Hence, using the estimates \eqref{eq:ImGImG_comp}--\eqref{eq:alpha_comp}, and the fact that  \eqref{eq:1/u-4fold}, we readily conclude the desired bound on $\mathfrak{b}_{11,t}$ in \eqref{eq:b_bounds} from the expression~\eqref{eq:b11_expr} and the estimate \eqref{eq:uv_comp}. 
 		
 		Next, we estimate the coefficient $\mathfrak{b}_{22,t}$. It follows from \eqref{eq:U_def}, \eqref{eq:hatM_eq} ,  and \eqref{eq:coeffb_def}  that
 		\begin{equation}
 			\mathfrak{b}_{22,t} = -\mathfrak{b}_{11,t}  - \frac{ ( \rho_2    \eta_{1,t})^2 + (\rho_1  \eta_{2,t})^2 }{2|m_{1,t}-\overline{m}_{2,t}|^2   \bigl\langle \widehat{M}_t(z_1, I , z_2) \bigr\rangle} |u_t|^2 |v_t|^2.
 		\end{equation}  
 		Therefore, using \eqref{eq:ImGImG_comp}--\eqref{eq:alpha_comp}, \eqref{eq:m-m_beta_bound}, \eqref{eq:uv_comp}, and the bound on $\mathfrak{b}_{11,t}$ from \eqref{eq:b_bounds} established above, we conclude that
 		\begin{equation} \label{eq:b22_bound}
 			\bigl\lvert \mathfrak{b}_{22,t}  \bigr\rvert \lesssim \frac{\rho_1 \rho_2}{\other{\alpha}_t(z_1, z_2)^2} + \frac{\rho_1 \rho_2}{|m_1 - \overline{m}_2|^2} \frac{\rho_1^{-1}\eta_{1,t} + \rho_2^{-1}\eta_{2,t}}{\other{\alpha}_t(z_1, z_2)^2}   \lesssim
 			\frac{\rho_1 \rho_2}{|m_1 - \overline{m}_2|^2}
  		\end{equation}
 		Here, in the last step we used \eqref{eq:alp_t_def}, the fact that $\rho_{j,t}^{-1}\eta_{j,t} \lesssim \other{\beta}_t(z_1, z_2)$ by \eqref{eq:rho_t}--\eqref{eq:eta_t}, \eqref{eq:betaf_t_def}, and \eqref{eq:beta_sym}, as well as the bound \eqref{eq:m-m_beta_bound}.
 		
 		Finally, we estimate $\mathfrak{b}_{12,t}$. We bound the real an imaginary parts separately. Similarly to \eqref{eq:b22_bound},  we deduce that real part of $\mathfrak{b}_{12,t}$ satisfies 
 		\begin{equation} \label{eq:re_b12}
 			\bigl\lvert \Re \mathfrak{b}_{12,t} \bigr\rvert = \frac{ \bigl\lvert \rho_2    \eta_{1,t} - \rho_1  \eta_{2,t} \bigr\rvert (\rho_2    \eta_{1,t}   + \rho_1  \eta_{2,t})}{4|m_{1,t}-\overline{m}_{2,t}|^2  \bigl\lvert \bigl\langle \widehat{M}_t(z_1, I , z_2) \bigr\rangle \bigr\rvert}|u_t|^2|v_t|^2
 			\lesssim \frac{\rho_1 \rho_2}{|m_{1,t}-\overline{m}_{2,t}| \other{\alpha}_t(z_1, z_2)}. 
 		\end{equation}
 		where in the second step we  used \eqref{eq:uv_comp} to estimate $|u_t|^2|v_t|^2$ and
		the second bound in \eqref{eq:beta_reg} for $\rho_2    \eta_{1,t} - \rho_1  \eta_{2,t}$. To bound the imaginary part, we employ the definitions \eqref{eq:U_def} and \eqref{eq:Vt_def} together with the identity \eqref{eq:cyclic2} to express $ \Im \mathfrak{b}_{12,t}$ in the form  
 		\begin{equation}
 			\bigl\lvert \Im \mathfrak{b}_{12,t} \bigr\rvert = \frac{\pi \rho_1\rho_2 \bigl\lvert \Im[u_t^{-1}] \bigr\rvert}{2|m_{1,t}-\overline{m}_{2,t}|^2 }
 			\frac{ \rho_{1,t} \eta_{2,t} + \rho_{2,t} \eta_{1,t} }{\bigl\lvert \bigl\langle \widehat{M}_t(z_1, I , z_2) \bigr\rangle \bigr\rvert } |u_t|^2|v_t|^2. 
 		\end{equation}
 		Proceeding analogously to \eqref{eq:re_b12} above and using 
		  the first equation in  \eqref{eq:beta_reg} together with $\rho_1+\rho_2\lesssim \rho_{1,t}+\rho_{2,t}
		\le |m_{1,t} - \overline{m}_{2,t}|$
		  to bound $\Im [u_t^{-1}]$, we conclude the desired estimate on $\Im \mathfrak{b}_{12,t}$. 
 		Therefore, \eqref{eq:b_bounds} is established. 
 		
 		It remains to establish \eqref{eq:m0_bound}.
 		Since  $T\sim 1$,   \eqref{eq:rho_t} and \eqref{eq:eta_t} imply that  
 		\begin{equation} \label{eq:init_is_away}
 			\rho_0(z_{j,0})^{-1} 
 			\Im z_{j,0} \gtrsim 1, \qquad j \in \{1,2\}.
 		\end{equation}
 		Consequently, $\other{\beta}_0(z_1, \overline{z}_2) \sim 1$ by \eqref{eq:betaf_t_def} and $|u_0| \sim 1$ by \eqref{eq:uv_comp}.
 		Moreover, \eqref{eq:rho_t}, \eqref{eq:ImM2t_def}, Lemma~\ref{lemma:stab_t} and \eqref{eq:M2op_rhorho_cancel} from   
 		Claim~\ref{claim:stab_rho_cancel} imply that 
 		\begin{equation} \label{eq:init_ImM2bound}
 			\bigl\lvert \bigl\langle \widehat{M}_0(z_1, A_1 z_2) A_2 \bigr\rangle \bigr\rvert  
 			\lesssim \rho_1\rho_2\norm{A_1}_{\mathrm{hs}}\norm{A_2}_{\mathrm{hs}},
 		\end{equation} 
 		which gives the desired bound on the first term in the definition of $\mathfrak{m}_0$ in \eqref{eq:m0_def}. 
 		To estimate the remaining terms, we use \eqref{eq:Gamma_sizes}   together with~\eqref{eq:As_reg},  
		 obtaining    
 		\begin{equation} \label{eq:UA_bounds}
 			\biggl\lvert\frac{\bigl\langle U A_1\bigr\rangle}{\langle U \rangle}\biggr\rvert \lesssim \rho_2 \norm{A_1}_{\mathrm{hs}}, \quad
 			\biggl\lvert\frac{\bigl\langle U^* A_1\bigr\rangle}{\langle U^* \rangle}\biggr\rvert \lesssim \rho_1 \norm{A_1}_{\mathrm{hs}}, \quad
 			\biggl\lvert\frac{\bigl\langle U A_2\bigr\rangle}{\langle U \rangle}\biggr\rvert \lesssim \rho_1 \norm{A_2}_{\mathrm{hs}}, \quad
 			\biggl\lvert\frac{\bigl\langle U^* A_2\bigr\rangle}{\langle U^* \rangle}\biggr\rvert \lesssim \rho_2 \norm{A_2}_{\mathrm{hs}}.		
 		\end{equation}
 		Hence, \eqref{eq:m0_bound} follows by plugging \eqref{eq:init_ImM2bound}, \eqref{eq:UA_bounds}, and $|u_0| \sim 1$ into \eqref{eq:m0_def}. 
 		This concludes the proof of \eqref{eq:imMt_bound} under the assumption \eqref{eq:2M_bound_case1}.
 	 
 		Next, we prove \eqref{eq:GImG_bounds}.
 		We only present a detailed proof of the first estimate in \eqref{eq:GImG_bounds}, since the second bound is completely analogous. 
 		The definitions of $\reg{A}_{1,t}$ in \eqref{eq:regA_def} and $\Upsilon_{1,t}$ in \eqref{eq:Ups_def} imply that
 		\begin{equation} \label{eq:Ups_ort}
 			\langle \widehat{M}_t(z_1, I, z_2) \reg{A}_{1,t}\rangle = 0. 
 		\end{equation} 
 		Hence, without loss of generality, we can assume that $z_1, z_2 \in \mathbb{H}$. Indeed,  by \eqref{eq:M12_dual} and \eqref{eq:Ups_ort}, we have the identities 
 		\begin{equation}\label{eq:MMh}
 			\bigl\langle M_t(z_1, \reg{A}_{1,t}, z_2) - M_t(\overline{z}_1, \reg{A}_{1,t}, z_2) \bigr\rangle = \bigl\langle M_t(z_1, \reg{A}_{1,t}, \overline{z}_2) - M_t(\overline{z}_1, \reg{A}_{1,t}, \overline{z}_2) \bigr\rangle = (2\ii)^2 h_t,
 		\end{equation}
 		where $h_t$ is given by (recall the definitions \eqref{eq:U_def} and \eqref{eq:Vt_def})
 		\begin{equation} \label{eq:h_def}
 			h_t \equiv h_t(z_1, A_1, z_2) := (4\ii)^{-1}\bigl\langle (U_t  - U_t^* + V_t - V_t^*) \reg{A}_{1,t} \bigr\rangle.
 		\end{equation}  
 		Therefore, to establish \eqref{eq:GImG_bounds}, it suffices to show that
 		\begin{equation} \label{eq:h_bound}
 			\lvert h_t \rvert \lesssim \frac{\rho_1}{\other{\alpha}(z_1, z_2)}\vertiii{A_1}_{t,z_1,z_2}, \qquad t\in[0,T]. 
 		\end{equation} 
 		
 		We proceed to establish \eqref{eq:h_bound}. 
 		First, we observe that \eqref{eq:UVlin_dep} implies 
 		\begin{equation} \label{eq:ImV_eq}
			V_t-V_t^* = \frac{2\Re [z_{1,t} - z_{2,t}]V_t}{\overline{z}_{1,t} - z_{2,t}}  - \frac{2\Re \bigl[(z_{1,t} - z_{2,t}) U_t\bigr]}{\overline{z}_{1,t} - z_{2,t}}.  			
 		\end{equation}
 		Taking the matrix imaginary part of \eqref{eq:ImV_eq} and collecting like terms, we obtain the identity
 		\begin{equation} \label{eq:ImV_id}
 			V_t - V_t^* = 4\ii \frac{\Re [z_{1,t} - z_{2,t}]}{\eta_{1,t} + \eta_{2,t}} \widehat{M}_t(z_1, I, z_2) 
 			+ \frac{\eta_{1,t} - \eta_{2,t}}{\eta_{1,t} + \eta_{2,t}} (U_t-U_t^*),
 		\end{equation}
 		where we recall the expression for $\widehat{M}_t(z_1, I, z_2) $ from \eqref{eq:hatM_eq}. 
 		In particular, plugging \eqref{eq:hatM_eq}, \eqref{eq:hatMA} and \eqref{eq:ImV_id} into \eqref{eq:Ups_ort}, and using \eqref{eq:h_def}, we obtain
 		\begin{equation}
 			h_t 
 			=\frac{1}{2\ii}\frac{\eta_{1,t}}{\eta_{1,t} + \eta_{2,t}}\biggl( \bigl\langle (U_t  - U_t^*) A_1 \bigr\rangle +  \Bigl\langle\frac{   U_t  \eta_{1,t} +    U_t^* \eta_{2,t}}{\overline{z}_{1,t} - z_{2,t}}A_1\Bigr\rangle \frac{2\Im u_t}{\Re[v_t - u_t]}\biggr) 
 			= \mathfrak{h}_{1,t} \langle Y A_1 \rangle + \mathfrak{h}_{2,t} \langle Z A_1 \rangle,
 		\end{equation}
 		where the coefficients $(\mathfrak{h}_{j,t})_{j=1}^2$  are obtained by performing the change of coordinates $\{u_t^{-1} U_t, (u_t^{-1} U_t)^*\}  \mapsto \{Y, Z\}$, with the matrices $Z$ and $Y$ defined in \eqref{eq:Zdef} and \eqref{eq:Ydef}, respectively, and are given by
		 \begin{equation} \label{eq:h1_def}
		 	\mathfrak{h}_{1,t} =  \frac{\eta_{1,t}}{\eta_{1,t} + \eta_{2,t}}\biggl(\frac{\rho_2 u_t  + \rho_1 \overline{u}_t}{2}   
		 	+   \frac{   \rho_2  \eta_{1,t} u_t - \rho_1 \eta_{2,t}\overline{u}_t}{(\overline{z}_{1,t} - z_{2,t})\Re[v_t - u_t]}   \Im u_t \biggr),
		 \end{equation}
		 \begin{equation} \label{eq:h2_def}
		 	\mathfrak{h}_{2,t} =  \frac{\eta_{1,t}}{\eta_{1,t} + \eta_{2,t}}\biggl(  \frac{\rho_2 u_t   -  \rho_1 \overline{u}_t}{2}  +   \frac{    \rho_2   \eta_{1,t} u_t +     \rho_1  \eta_{2,t} \overline{u}_t}{(\overline{z}_{1,t} - z_{2,t})\Re[v_t - u_t]}  \Im u_t \biggr).
		 \end{equation}
		  	  
 		We now bound $\mathfrak{h}_{1,t}$.  		
 		Taking the real part of \eqref{eq:h1_def} and bringing both terms to a common denominator, we deduce that 
 		\begin{equation}
 				\Re \mathfrak{h}_{1,t} =
 				\pi \rho_1 \frac{\rho_{2,t}\eta_{1,t}|u_t|^2|v_t|^2}
 				{|m_{1,t} - \overline{m}_{2,t}|^2\Re[v_t - u_t]} \biggl(2\Re \Bigl[\frac{1}{u_t}\Bigr] - \frac{\eta_{1,t}}{\pi\rho_{1,t}} - \frac{\eta_{2,t}}{\pi\rho_{2,t}}\biggr).
 		\end{equation}
 		Here, we used $\rho_{j,t} = \ee^{(t-T)/2}\rho_j$ by \eqref{eq:M_t},  \eqref{eq:cyclic2}, and the identities 
 		\begin{equation} \label{eq:cyclic1}
 			(z_{1,t}-\overline{z}_{2,t})u_t = (m_{1,t} - m_{2,t}) + 2\ii \eta_{2,t} u_t, \qquad (z_{1,t}-\overline{z}_{2,t})\overline{u}_t = (\overline{m}_{1,t} - \overline{m}_{2,t}) + 2\ii \eta_{1,t} \overline{u}_t,
 		\end{equation}
 		\begin{equation}
 			\Re[m_{1,t}-m_{2,t}] \Im u_t = - |u_t|^2 (\eta_{1,t} - \eta_{2,t}) + \pi (\rho_{1,t} - \rho_{2,t}) \Re u_t,
 		\end{equation}
 		which follow immediately from \eqref{eq:U_def}. Using the same identities for the imaginary part of \eqref{eq:h1_def}, we deduce that
 		\begin{equation}
 			\Im \mathfrak{h}_{1,t} = 0. 
 		\end{equation}
 		Hence, it follows from \eqref{eq:ImGImG_comp}--\eqref{eq:alpha_comp}, \eqref{eq:uv_comp}, and the bound \eqref{eq:1/u-rho_eta} that 
 		\begin{equation} \label{eq:h1_bound}
 			\bigl\lvert \mathfrak{h}_{1,t} \bigr\rvert \lesssim \frac{\rho_1}{\other{\alpha}_t(z_1, z_2)^2}. 
 		\end{equation}
 		
 		Similarly, the  coefficient $\mathfrak{h}_{2,t}$ satisfies
 		\begin{equation}
 			\mathfrak{h}_{2,t} = \pi\rho_1 \frac{\rho_{2,t}\eta_{1,t}|u_t|^2|v_t|^2}{|m_{1,t} - \overline{m}_{2,t}|^2\Re[v_t - u_t]} \biggl(\frac{\eta_{2,t}}{\pi\rho_{2,t}} - \frac{\eta_{1,t}}{\pi\rho_{1,t}} - 2\ii\Im\Bigl[\frac{1}{u_t}\Bigr] \biggr). 
 		\end{equation} 
 		Hence, using \eqref{eq:ImGImG_comp}--\eqref{eq:alpha_comp}, \eqref{eq:beta_reg}, \eqref{eq:uv_comp} and \eqref{eq:m-m_beta_bound}, we conclude that 
 		\begin{equation} \label{eq:h2_bound}
 			\bigl\lvert \mathfrak{h}_{2,t}\bigr\rvert \lesssim \frac{\rho_1 }{\other{\alpha}_t(z_1, z_2) | m_1 - \overline{m}_2|}. 
 		\end{equation}
 		In particular, plugging the bounds \eqref{eq:h1_bound} and \eqref{eq:h2_bound} into \eqref{eq:h_def} and using the estimates \eqref{eq:ZA_bounds} and \eqref{eq:YA_bounds}, together with \eqref{eq:m-m_beta_bound}, we readily conclude the desired \eqref{eq:h_bound} under the assumption \eqref{eq:2M_bound_case1}. 
 
 		Finally, we establish \eqref{eq:M2aux_t}. Using \eqref{eq:Ups_ort}, \eqref{eq:h_bound} established above, together with the fact that $\other{\alpha}(z_1^+, z_2^+) \gtrsim \rho(z_1^+) + \rho(z_2^+)$ by \eqref{eq:alp_t_def} and \eqref{eq:betaf_def}, we deduce that 
 		\begin{equation} \label{eq:MreGreG}
 			\bigl\lvert \bigl\langle M_t(z_1, \reg{A}_{1,t}, z_2) \bigr\rangle \bigr\rvert \lesssim |\widecheck{h}_t| + \vertiii{A_1}_{t, z_1, z_2}, \qquad 
 			\widecheck{h}_t \equiv \widecheck{h}_t(z_1, A_1, z_2) := \biggl\langle\reg{A}_{1,t}  \Re \frac{M_t(z_{1,t}^+) - M_t(z_{2,t}^+)}{z_{1,t}^+ - z_{2,t}^+}  \biggr\rangle. 
 		\end{equation}
 		Hence, we can assume without loss of generality that $z_1, z_2 \in \mathbb{H}$. As in the proof of \eqref{eq:GImG_bounds} above, performing the change of coordinates $\{u_t^{-1} U_t, (u_t^{-1} U_t)^*\}  \mapsto \{Y, Z\}$ (recall \eqref{eq:Zdef} and \eqref{eq:Ydef}), we obtain
 		\begin{equation} \label{eq:hcheck_ZY}
 			\widecheck{h}_t = 
 			\frac{1}{2}\biggl( \bigl\langle (U_t + U_t^*) A_1 \bigr\rangle - \ii  \Bigl\langle\frac{   U_t  \eta_{1,t} +    U_t^* \eta_{2,t}}{\overline{z}_{1,t} - z_{2,t}}A_1\Bigr\rangle \frac{2\Re u_t}{\Re[v_t - u_t]}\biggr) 
 			= \widecheck{\mathfrak{h}}_{1,t} \langle Y A_1 \rangle  + \widecheck{\mathfrak{h}}_{2,t} \langle Z A_1 \rangle,
 		\end{equation}
 		where the coefficients $\widecheck{\mathfrak{h}}_{1,t}$  and $\widecheck{\mathfrak{h}}_{2,t}$ are given by
 		\begin{equation}
 			\widecheck{\mathfrak{h}}_{1,t} = \ii\frac{\rho_2 u_t - \rho_1 \overline{u}_t}{2} + \frac{   \rho_2 \eta_{1,t} u_t - \rho_1 \eta_{2,t}\overline{u}_t }{(\overline{z}_{1,t} - z_{2,t})\Re[v_t - u_t]}   \Re u_t , \qquad \widecheck{\mathfrak{h}}_{2,t} = \ii\frac{\rho_2 u_t + \rho_1 \overline{u}_t}{2} +\frac{\rho_2 \eta_{1,t} u_t + \rho_1 \eta_{2,t} \overline{u}_t }{(\overline{z}_{1,t} - z_{2,t})\Re[v_t - u_t]}  \Re u_t. 
 		\end{equation}  
 		Using the identities \eqref{eq:cyclic2}, \eqref{eq:cyclic1}, together with 
 		\begin{equation}
 			\Re[m_{1,t}-m_{2,t}] \Re u_t  = |u_t|^2 \Re[z_{1,t} - z_{2,t}] - \pi (\rho_{1,t} - \rho_{2,t}) \Im u_t,	
 		\end{equation}
 		we find that the coefficients $\widecheck{\mathfrak{h}}_{1,t}$  and $\widecheck{\mathfrak{h}}_{2,t}$ are given by
		\begin{equation}
 			  \widecheck{\mathfrak{h}}_{1,t} = \rho_1\rho_{2,t} |u_t|^2  \frac{\Re [z_{1,t} - z_{2,t}] \bigl(  \rho_{1,t}^{-1}\eta_{1,t}  - \rho_{2,t}^{-1}\eta_{2,t} \bigr) + 2\pi ( \eta_{1,t} + \eta_{2,t}) \Im \bigl[u_t^{-1}\bigr] }{|z_{1,t} - \overline{z}_{2,t}|^2 \Re\bigl[v_t-u_t\bigr]},
 		\end{equation}
 		\begin{equation}
 			\widecheck{\mathfrak{h}}_{2,t} = \rho_1\rho_{2,t}  |u_t|^2 \frac{\Re [z_{1,t} - z_{2,t}] \bigl(  \rho_{1,t}^{-1}\eta_{1,t} + \rho_{2,t}^{-1}\eta_{2,t} \bigr) + 2\pi \ii   (\eta_{1,t} + \eta_{2,t})\Re \bigl[u_t^{-1}\bigr] }{ |z_{1,t} -\overline{z}_{2,t}|^2 \Re\bigl[v_t-u_t\bigr]}.
 		\end{equation}
 		Hence, using \eqref{eq:beta_assymp},  \eqref{eq:m_hat_alpha}, \eqref{eq:beta_reg} to estimate $\Im u_t^{-1}$,  \eqref{eq:uv_comp},   we obtain
 		\begin{equation} \label{eq:hhat1}
 			\bigl\lvert \widecheck{\mathfrak{h}}_{1,t} \bigr\rvert \lesssim  \frac{ \other{\beta}(z_1, \overline{z}_2)}{ |m_{1,t} - \overline{m}_{2,t}| \other{\alpha}_t(z_1, z_2)^2} \frac{\bigl\lvert \rho_{1,t}^{-1}\eta_{1,t}  - \rho_{2,t}^{-1}\eta_{2,t} \bigr\rvert}{\rho_{1,t}^{-1}\eta_{1,t}  + \rho_{2,t}^{-1}\eta_{2,t}}
 			+  \frac{ 1}{ \other{\alpha}_t(z_1, z_2)},
 		\end{equation}
 		where we used $|m_{1,t} - \overline{m}_{2,t}|\gtrsim \rho_1 + \rho_2$,  $\rho_1\eta_{2,t} +  \eta_{1,t} \gtrsim (\eta_{2,t} + \rho_2\eta_{1,t})(\rho_1 \wedge \rho_2)$, $\other{\alpha}_T \lesssim \other{\alpha}_t$ by \eqref{eq:alp_t_def}.
 		To proceed, we observe that, on the one hand, it follows from the second bound in \eqref{eq:beta_reg} that 
 		\begin{equation}
 			\frac{\bigl\lvert \rho_{1,t}^{-1}\eta_{1,t}  - \rho_{2,t}^{-1}\eta_{2,t} \bigr\rvert }{\rho_{1,t}^{-1}\eta_{1,t} + \rho_{2,t}^{-1}\eta_{2,t}} 
 			\lesssim 1 \wedge \frac{\other{\alpha}_T(z_1, z_2)|m_1 - m_2|}{\rho_{1,t}^{-1}\eta_{1,t} + \rho_{2,t}^{-1}\eta_{2,t}} 
 			\lesssim \frac{\other{\alpha}_T(z_1, z_2)|m_1 - m_2|}{\rho_{1,t}^{-1}\eta_{1,t} + \rho_{2,t}^{-1}\eta_{2,t} + \other{\alpha}_T(z_1, z_2)|m_1 - m_2| }.
 		\end{equation}
 		On the other hand, \eqref{eq:uv_comp} and \eqref{eq:u_reg} with $z_2 := \overline{z}_2$ and $w := \overline{z}_1$ implies that 
 		\begin{equation} \label{eq:v_bound}
 			\other{\beta}(z_1, \overline{z}_2) \sim |v_t^{-1}| \lesssim \rho_{1,t}^{-1}\eta_{1,t} + \other{\alpha}_T(z_1, z_2)|m_1 - m_2|.
 		\end{equation}
 		Therefore, combining \eqref{eq:hhat1}--\eqref{eq:v_bound}, we deduce that
 		\begin{equation} \label{eq:check_h1_bound}
 			\bigl\lvert \widecheck{\mathfrak{h}}_{1,t}\bigr\rvert \lesssim \other{\alpha}_t(z_1, z_2)^{-1}.
 		\end{equation}
 		Similarly, using \eqref{eq:uv_comp}, a trivial bounds $|\Re u_t^{-1}| \lesssim |u_t|^{-1}$ and $\other{\beta}_t(z_1, \overline{z}_2) \lesssim \other{\beta}_t(z_1, z_2) \lesssim \other{\alpha}_t(z_1, z_2)^2$ by \eqref{eq:alp_t_def}, and \eqref{eq:m_hat_alpha}, we obtain
 		\begin{equation} \label{eq:check_h2_bound}
 			\bigl\lvert \widecheck{\mathfrak{h}}_{2,t}\bigr\rvert \lesssim
 			|m_1 -\overline{m}_2|^{-1}.
 		\end{equation}
 		Combining \eqref{eq:ZA_bounds}, \eqref{eq:YA_bounds}, \eqref{eq:MreGreG}--\eqref{eq:hcheck_ZY}, \eqref{eq:check_h1_bound}--\eqref{eq:check_h2_bound}, we conclude the desired bound \eqref{eq:M2aux_t} under the assumption~\eqref{eq:2M_bound_case1}.

 		\nc 
 		\smallskip
 		\nc \textbf{Case 2}. Next, we complete the proof under the assumption
 		\begin{equation} \label{eq:2M_bound_case2}
 			\max\bigl\{\other{\beta}(z_1^+, z_2^-), \other{\beta}(z_2^-, z_1^+)\bigr\}  \le c_* < \beta_* \le  \max\bigl\{\other{\beta}(z_1^+, z_2^+), \other{\beta}(z_2^+, z_1^+)\bigr\},
 		\end{equation}
 		where $c_* \sim 1$ is the threshold from Claim~\ref{claim:stab_diff_rho}. In particular, it follows from \eqref{eq:rho_large}  that (recall that $\rho(z_j) \lesssim 1$ by \eqref{eq:M_bound}) 
 		\begin{equation} \label{eq:case2_conseq}
 			\rho(z_1) \sim 1, \qquad \rho(z_2) \sim 1.
 		\end{equation} 
 		In particular, under the assumption \eqref{eq:2M_bound_case2}, there are no $\rho$-gains in the bounds \eqref{eq:imMt_bound} and \eqref{eq:GImG_bounds}. Moreover, in this regime $\vertiii{A_1}_{t,z_1, z_2}\sim \norm{A_1}_\mathrm{hs}$ and $\vertiii{A_2}_{t,z_2, z_1}\sim \norm{A_2}_\mathrm{hs}$, since $\other{\alpha}_t(z_1^+, z_2^+)^2 \gtrsim \other{\beta}_t(z_1^+, z_2^+) \sim 1$. 
 		 
 		We now prove \eqref{eq:imMt_bound}. Once again, without loss of generality, we assume that $z_1,z_2\in\mathbb{H}$.
 		Using \eqref{eq:U_def}, \eqref{eq:B12_identity}, and \eqref{eq:stab_beta_bound}, we obtain the norm bound 
 		\begin{equation} \label{eq:Ut_norm_bound}
 			\norm{U_t}\lesssim \other{\beta}_t(z_1, z_2)^{-1} \lesssim 1, \qquad t\in[0,T], 
 		\end{equation}
 		where in the second step we used \eqref{eq:case2_conseq} and \eqref{eq:betaf_def}. Hence, it follows from \eqref{eq:hatMA} that 
 		\begin{equation}
 			\bigl\lvert \bigl\langle \widehat{M}_t(z_1, I , z_2) A_1 \bigr\rangle \bigr\rvert \lesssim \frac{\eta_{1,t} + \eta_{2,t}}{|z_{1,t} - \overline{z}_{2,t}|}\norm{A_1}_\mathrm{hs}, \quad
 			\bigl\lvert \bigl\langle \widehat{M}_t(z_1, I , z_2) A_2 \bigr\rangle \bigr\rvert \lesssim \frac{\eta_{1,t} + \eta_{2,t}}{|z_{1,t} - \overline{z}_{2,t}|}\norm{A_2}_\mathrm{hs}, \qquad t \in [0,T]. 
 		\end{equation}
 		Furthermore, plugging \eqref{eq:case2_conseq} into \eqref{eq:m_hat_alpha}, we deduce that 
 		\begin{equation} \label{eq:bulk_Mhat}
 			 \bigl\langle \widehat{M}_t(z_1, I, z_2) \bigr\rangle   \sim \frac{\eta_{1,t} + \eta_{2,t}}{\other{\beta}_t(z_1, \overline{z}_2)^2}, \qquad t \in [0,T]. 
 		\end{equation}
 		Therefore, using the fact that $\other{\beta}(z_1, z_2) \sim 1$ by assumption \eqref{eq:2M_bound_case2} and \eqref{eq:beta_sym}, together with 
 		$$|z_{1,t} - \overline{z}_{2,t}| \sim |m_{1,t} - \overline{m}_{2,t}| \other{\beta}_t(z_1, \overline{z}_2) \sim \other{\beta}_t(z_1, \overline{z}_2)$$
 		by \eqref{eq:M_bound}, \eqref{eq:z_diff_comp}, \eqref{eq:beta_assymp} and \eqref{eq:case2_conseq}, we deduce that
 		\begin{equation} \label{eq:bulk_Ups}
 			\bigl\lvert \Upsilon_{1,t} \bigr\rvert \lesssim \other{\beta}_t(z_1, \overline{z}_2) \norm{A_1}_\mathrm{hs}, \qquad \bigl\lvert \Upsilon_{2,t} \bigr\rvert \lesssim \other{\beta}_t(z_1, \overline{z}_2) \norm{A_2}_\mathrm{hs}, \qquad t \in [0,T].  
 		\end{equation}
 		Therefore, it follows from \eqref{eq:hatm_1}, \eqref{eq:bulk_Mhat}, and \eqref{eq:bulk_Ups}, that 
 		\begin{equation} \label{eq:bulk_hatM_bound1}
 			\widehat{m}_t = \bigl\langle \widehat{M}_t(z_1, A_1 , z_2) A_2\bigr\rangle + \mathcal{O}(\eta_{1,t} + \eta_{2,t})\norm{A_1}_\mathrm{hs}\norm{A_2}_\mathrm{hs}. 
 		\end{equation}
 		Hence, the contribution of the $\Upsilon_{j,t}$-correction terms is negligible in this regime. 
 		
 		To estimate the remaining term $\bigl\langle \widehat{M}_t(z_1, A_1 , z_2) A_2\bigr\rangle$, we plug  \eqref{eq:M2_duhamel} into \eqref{eq:ImM2t_def}, obtaining
 		\begin{equation} \label{eq:bulk_MAA_part}
 			\bigl\langle \widehat{M}_t(z_1, A_1 , z_2) A_2\bigr\rangle = \ee^t \bigl\langle \widehat{M}_0(z_1, A_1 , z_2) A_2\bigr\rangle -  \frac{\ee^{t}-1}{4} \Bigl(\langle U_t A_1 \rangle\langle U_0 A_2 \rangle
 			+ \langle U^*_t A_1 \rangle \langle U^*_0 A_2 \rangle \Bigr),
 		\end{equation}  
 		where we used \eqref{eq:M_t}--\eqref{eq:z_t} to express the scalar integral on the right-hand side of \eqref{eq:M2_duhamel} as (c.f. \eqref{eq:scal_int_1})
 		\begin{equation} \label{eq:scal_int_2}
 			\ee^{t-T}\int_{0}^t \frac{(z_1 - z_2)^2}{(z_{1,s} - z_{2,s})^2} \mathrm{d}s =   (\ee^{t}-1)   \frac{\ee^{(t-T)/2}(z_1 - z_2)}{z_{1,t} - z_{2,t}}\frac{\ee^{-T/2}(z_1 - z_2)}{z_{1,0} - z_{2,0}}. 
 		\end{equation}
 		Estimating the first term on the right-hand side of \eqref{eq:bulk_MAA_part} using \eqref{eq:init_ImM2bound}, and the second term using \eqref{eq:Ut_norm_bound}, we conclude from \eqref{eq:bulk_hatM_bound1} that 
 		\begin{equation}
 			\bigl\lvert  \widehat{m}_t \bigr\rvert \lesssim \norm{A_1}_\mathrm{hs}\norm{A_2}_\mathrm{hs}.
 		\end{equation}
 		Since $\rho_j \sim 1$ by \eqref{eq:case2_conseq}, this concludes the proof of \eqref{eq:imMt_bound} under the assumptions \eqref{eq:2M_bound_case2}.

 		Next, we prove \eqref{eq:GImG_bounds}. 
 		As in Case 1 above, it suffices to estimate $h_t$ defined in \eqref{eq:h_def}. Using \eqref{eq:Ups_ort} and \eqref{eq:ImV_eq}, we write
 		\begin{equation}
 			h_t = \frac{\eta_{1,t}}{\eta_{1,t} + \eta_{2,t}} \bigl\langle (\Im U_t) A_1 \bigr\rangle - \Upsilon_{1,t}\frac{\eta_{1,t}}{\eta_{1,t} + \eta_{2,t}} \bigl\langle  \Im U_t    \bigr\rangle,
 		\end{equation}
 		and, plugging in the bounds \eqref{eq:Ut_norm_bound} and \eqref{eq:bulk_Ups}, we conclude that $h_t \lesssim \norm{A_1}_\mathrm{hs}$, which is sufficient owing to $\rho_j \gtrsim 1$ by \eqref{eq:case2_conseq}.
 		Hence, this concludes the proof of \eqref{eq:GImG_bounds} given \eqref{eq:2M_bound_case2}. 	
 		
 		The estimate \eqref{eq:M2aux_t} follows completely analogously.

 		\smallskip
 		\textbf{Case 3}. Finally, we complete the proof under the assumption
 		\begin{equation} \label{eq:2M_bound_case3}
 			c_* \le \max\bigl\{\other{\beta}(z_1^+, z_2^-), \other{\beta}(z_2^-, z_1^+)\bigr\} \le \beta_*,
 		\end{equation}
 		where $c_*$ is the threshold from Case 2 above. In particular, \eqref{eq:beta_sym} implies that $\min\bigl\{\other{\beta}(z_1, \overline{z}_2), \other{\beta}(\overline{z}_2, z_1)\bigr\} \gtrsim 1$. 
 		In this case, by \eqref{eq:betaf_def} and \eqref{eq:beta_sym}, we have $\min \{\other{\beta}(w_1, w_2), \other{\beta}(w_2, w_1)\} \gtrsim 1$ for all $w_j \in \{z_j, \overline{z}_j\}$ with $j\in\{1,2\}$. 
 		Therefore, \nc  Lemma~\ref{lemma:stab_t} implies that $\lVert \mathcal{B}_{t,w_1, w_2}^{-1} \rVert_{\mathrm{hs}\to\mathrm{hs}} \lesssim 1$ for all $w_j \in \{z_j, \overline{z}_j\}$ with $j\in\{1,2\}$ and $t\in[0,T]$. 
 		Hence, the bounds \eqref{eq:imMt_bound}--\eqref{eq:GImG_bounds} follow immediately from Claim~\ref{claim:stab_rho_cancel}   using the formulas~\eqref{eq:Mt_def}--\eqref{eq:ImM2t_def},   and the trivial bound 
 		\begin{equation} \label{eq:far_Ups_bound}
 			\bigl\lvert \Upsilon_{j,t}\bigr\rvert \lesssim \norm{A_j}_\mathrm{hs},
 		\end{equation}
 		which is a consequence of \eqref{eq:Mhat_flat} and \eqref{eq:Ups_def}.   
 		The bound \eqref{eq:M2aux_t} follows immediately from \eqref{eq:Mt_def}, Lemma~\ref{lemma:stab_t}, \eqref{eq:2M_bound_case3}, and~\eqref{eq:far_Ups_bound}. 
 		This concludes the proof of Proposition~\ref{prop:Mt_bound}.		
 	\end{proof}

 	We conclude this subsection by proving the time-independent Theorem~\ref{prop:M2_bounds}.
 	\begin{proof}[Proof of Theorem \ref{prop:M2_bounds}]
 		In the case $\max\bigl\{\other{\beta}(z_1, \overline{z}_2), \other{\beta}(\overline{z}_2, z_1)\bigr\} \le \beta_*$, the bounds \eqref{eq:ImM2_bounds}--\eqref{eq:M2_aux_bound} follow immediately from Proposition~\ref{prop:Mt_bound}, since 
 		\begin{equation}
 			M(z_1, A, z_2) = M_T(z_1, A, z_2).
 		\end{equation}
 		
 		On the other hand, if $\max\bigl\{\other{\beta}(z_1, \overline{z}_2), \other{\beta}(\overline{z}_2, z_1)\bigr\} \ge \beta_*$, then $\other{\beta}(w_1, w_2) \gtrsim 1$ for all $w_j \in \{z_j, \overline{z}_j\}$ by \eqref{eq:betaf_def} and~\eqref{eq:beta_sym}. 
 		Therefore, it follows from \eqref{eq:stab_beta_bound} that 	$\lVert\mathcal{B}_{w_1, w_2}^{-1}\rVert_{\mathrm{hs}\to\mathrm{hs}} \lesssim 1$ for all $w_j \in \{z_j, \overline{z}_j\}$. 
 		Together with \eqref{eq:M_bound} at time $t=T$, this implies \eqref{eq:M2_nobeta} immediately. The remaining bound \eqref{eq:ImM2_nobeta} is an immediate consequence of Claim~\ref{claim:stab_rho_cancel}. 
 		
 		Finally, the comparison \eqref{eq:Mhat_comp} follows from \eqref{eq:ImGImG_comp} at time $t = T$. 
 		This concludes the proof of Theorem~\ref{prop:M2_bounds}. 
 	\end{proof}

 	\subsubsection{Technical $M$-bounds. Proof of Claims~\ref{claim:Ups} and~\ref{claim:ImGG_Mbound}} \label{sec:techM_bounds}
 	
 	\begin{proof}[Proof of Claim~\ref{claim:Ups}]
 		Observe that by  \eqref{eq:bulk_Ups} and \eqref{eq:far_Ups_bound} from Cases 2 and 3, respectively, in the proof of Proposition~\ref{prop:Mt_bound} above that the desired bound \eqref{eq:Ups_bound} is already established in those cases. Therefore, it remains to prove \eqref{eq:Ups_bound} under the assumption \eqref{eq:2M_bound_case1} of Case 1. 
 		Moreover $\other{\alpha}_t(z_1, z_2) \gtrsim \other{\beta}_t(z_1, z_2)^{1/2} \gtrsim \other{\beta}_t(z_1, \overline{z}_2)^{1/2} $, by \eqref{eq:betaf_def} and \eqref{eq:alp_t_def}; Hence, by~\eqref{eq:size_t_def}, the second estimate in \eqref{eq:Ups_bound} follows from the first.
 		
 		To establish  \eqref{eq:Ups_bound}, we plug \eqref{eq:UtoYZ} into \eqref{eq:hatMA}, obtaining 
 		\begin{equation} \label{eq:UpsYZ}
 			\Upsilon_t(z_1, A, z_2) = \mathfrak{u}_{1,t} \langle YA \rangle + \mathfrak{u}_{2,t} \langle Z A \rangle,
 		\end{equation}
 		where the matrices $Y$ and $Z$ are defined in \eqref{eq:Ydef} and \eqref{eq:Zdef}, respectively, and the coefficients $\mathfrak{u}_{1,t}$, $\mathfrak{u}_{2,t}$ are given by 
 		\begin{equation}
 			\mathfrak{u}_{1,t} := \frac{ (\rho_1\eta_{2,t} - \rho_2\eta_{1,t})\Re u_t  - \ii  (\rho_1 \eta_{2,t} + \rho_2\eta_{1,t}) \Im u_t }{2(\overline{z}_{1,t} - z_{2,t}) \bigl\langle \widehat{M}(z_1, I, z_2)\bigr\rangle}, \qquad 
 			\mathfrak{u}_{2,t} := \frac{-(\rho_1 \eta_{2,t} + \rho_2 \eta_{1,t})\Re u_t + \ii (\rho_1 \eta_{2,t}  - \rho_2 \eta_{1,t}) \Im u_t}{2(\overline{z}_{1,t} - z_{2,t}) \bigl\langle \widehat{M}(z_1, I, z_2)\bigr\rangle},
 		\end{equation}
 		where we abbreviate $\rho_j := \rho(z_j)$ and denote $\eta_{j,t} := \Im z_{j,t}$ for $j\in\{1,2\}$. Recall that $u_t := \langle U_t\rangle$ from \eqref{eq:U_def}.  Similarly to the proof of Proposition~\ref{prop:Mt_bound} above, we now proceed to estimate $\mathfrak{u}_{1,t}$ and $\mathfrak{u}_{2,t}$.
 		
 		First, we deduce that $\mathfrak{u}_{1,t}$ admits the bound
 		\begin{equation} \label{eq:u1_bound}
			\bigl\lvert \mathfrak{u}_{1,t} \bigr\rvert \sim \frac{\other{\beta}(z_1,\overline{z}_2) }{\other{\alpha}(z_1, z_2)^2 \lvert \overline{m}_{1,t} - m_{2,t} \rvert}\biggl\lvert  \frac{\rho_1\eta_{2,t} - \rho_2\eta_{1,t}}{\rho_1 \eta_{2,t} + \rho_2\eta_{1,t}}\Re \bigl[u_t^{-1}\bigr] + \ii \Im \bigl[u_t^{-1}\bigr]\biggr\rvert
			\lesssim 
			\frac{\other{\beta}(z_1,\overline{z}_2) }{\other{\alpha}(z_1, z_2)},
 		\end{equation}
 		Here, in the first step we used the comparisons \eqref{eq:m_hat_alpha}, \eqref{eq:uv_comp} and \eqref{eq:rho_t}, while in the second step we used \eqref{eq:beta_reg} and \eqref{eq:1/u-rho_eta} together with \eqref{eq:m-m_beta_bound} and $\other{\alpha}_t(z_1, z_2) \gtrsim \other{\alpha}_T(z_1, z_2)$ by \eqref{eq:alp_t_def}. Estimating $\mathfrak{u}_{2,t}$, we obtain
 		\begin{equation} \label{eq:u2_bound}
 			\bigl\lvert \mathfrak{u}_{2,t} \bigr\rvert \sim \frac{\other{\beta}(z_1,\overline{z}_2) }{\other{\alpha}(z_1, z_2)^2 \lvert \overline{m}_{1,t} - m_{2,t} \rvert}\biggl\lvert  \Re \bigl[u_t^{-1}\bigr] + \ii \frac{\rho_1\eta_{2,t} - \rho_2\eta_{1,t}}{\rho_1 \eta_{2,t} + \rho_2\eta_{1,t}} \Im \bigl[u_t^{-1}\bigr]\biggr\rvert
 			\lesssim \frac{\other{\beta}(z_1,\overline{z}_2) }{\lvert \overline{m}_{1,t} - m_{2,t} \rvert}.
 		\end{equation}
 		Combining \eqref{eq:UpsYZ}--\eqref{eq:u2_bound} with \eqref{eq:ZA_bounds} and \eqref{eq:YA_bounds}, as well as \eqref{eq:m-m_beta_bound}, we deduce that
 		\begin{equation} \label{eq:ups_R_bound}
 			\bigl\lvert \Upsilon_t(z_1, A, z_2) \bigr\rvert \lesssim \other{\beta}_t(z_1, \overline{z}_2)\biggl(\frac{\big| \langle \vecL(z_1, z_2) A \bigr\rangle\big|}{\other{\alpha}(z_1, z_2)} + \norm{A}_\mathrm{hs} \biggr). 
 		\end{equation}
 		In particular, it follows from \eqref{eq:betaf_def}, \eqref{eq:betaf_t_def}, and \eqref{eq:alp_t_def}, that 
 		\begin{equation} \label{eq:ups_pre_bound}
 			\bigl\lvert \Upsilon_t(z_1, A, z_2) \bigr\rvert \lesssim \other{\alpha}_t(z_1, z_2) \norm{A}_\mathrm{hs} \lesssim \other{\alpha}_t(z_1, z_2) \bigl\lVert \reg{A}_t\bigr\rVert_\mathrm{hs}~, 
 		\end{equation}
 		where in the second step we used \eqref{eq:M_bound}, \eqref{eq:stab_t_bound}, \eqref{eq:un-reg},   \eqref{eq:B12_identity}, \eqref{eq:Vt_def},  and \eqref{eq:uv_comp},  to deduce that 
 		\begin{equation}
 			\norm{A}_\mathrm{hs} = \bigl\lVert \reg{A}_t - v_t^{-1} \langle V_t \reg{A}_t \rangle\bigr\rVert_\mathrm{hs} \lesssim \bigl(1 + |v_t|^{-1} \norm{V_t}_\mathrm{hs}\bigr) \bigl\lVert \reg{A}_t  \bigr\rVert_\mathrm{hs} \lesssim \bigl\lVert \reg{A}_t  \bigr\rVert_\mathrm{hs}~.
 		\end{equation}
 		Then the desired comparison \eqref{eq:trip_equiv} follows immediately from \eqref{eq:ups_pre_bound} and \eqref{eq:size_t_def}. 
 		
 		The desired \eqref{eq:Ups_bound} follows by plugging \eqref{eq:trip_equiv} into \eqref{eq:ups_R_bound}. This concludes the proof of Claim~\ref{claim:Ups}. 
 	\end{proof}
 	
 	\begin{claim} \label{claim:ImGG_Mbound}
 		Let $z_1, z_2 \in \regD\cap(\mathbb{H}^2)$. Then, the matrices $U_t(z_1, z_2)$ and $V_t(z_1, z_2)$, defined in \eqref{eq:U_def} and \eqref{eq:Vt_def}, respectively, satisfy
 		\begin{equation} \label{eq:ImGG_Mbound}
 			\bigl\lvert \bigl\langle U_t(z_1, z_2) - V_t(z_1, z_2) \bigr\rangle \bigr\rvert \lesssim \frac{\other{\alpha}_T(z_1, z_2)\rho(z_2)}{\other{\beta}_t(z_1, z_2) \other{\beta}_t(z_1,\overline{z}_2)}, \qquad 
 			\bigl\lvert \bigl\langle U_t(z_1, z_2)^* - V_t(z_1, z_2) \bigr\rangle \bigr\rvert \lesssim \frac{\other{\alpha}_T(z_1, z_2)\rho(z_1)}{\other{\beta}_t(z_1, z_2) \other{\beta}_t(z_1,\overline{z}_2)}.
 		\end{equation} 
 	\end{claim}
 	
 	\begin{proof} [Proof of Claim~\ref{claim:ImGG_Mbound}]
 		We only present the proof of the first bound in \eqref{eq:ImGG_Mbound}, since the other bound is completely analogous. 
 		Note that the bound \eqref{eq:ImGG_Mbound} follows from Claim~\ref{claim:stab_rho_cancel} trivially in the regime when $\other{\beta}(z_1, \overline{z}_2) \gtrsim 1$. 
 		Furthermore, if $z_1, z_2$ satisfy \eqref{eq:stab_gap}, the bound \eqref{eq:ImGG_Mbound} follows trivially from \eqref{eq:rho_large} and \eqref{eq:beta_assymp}. 
 		Therefore, it remains to prove \eqref{eq:ImGG_Mbound} under the assumption
 		\begin{equation}
 			\max\{\other{\beta}(z_1, z_2), \other{\beta}(z_2, z_1)\} \le \beta_*. 
 		\end{equation}
 		We abbreviate $u_t := \langle U_t(z_1, z_2)\rangle$ and $v_t := \langle V_t(z_1, z_2)\rangle$. Then,
 		\begin{equation}
 			|v_t - u_t| = |v_t u_t| \bigl\lvert u_t^{-1} - v_t^{-1} \bigr\rvert \lesssim \frac{\other{\alpha}_T(z_1, z_2)\rho(z_2)}{\other{\beta}_t(z_1, z_2) \other{\beta}_t(z_1,\overline{z}_2)},
 		\end{equation}
 		where we used \eqref{eq:uv_comp} to estimate $|v_t u_t|$ and \eqref{eq:u_reg} to bound $u_t^{-1} - v_t^{-1}$. 
 		This concludes the proof of Claim~\ref{claim:ImGG_Mbound}.
 	\end{proof}

 	\section{Zigzag strategy: Proof of the local laws in Theorem \ref{thm:locallawreg}} \label{sec:zigzag}
 To streamline the presentation, we will henceforth assume that the admissible energies $\mathcal{I}$ from Assumption \ref{ass:Mbdd} exhaust the entire real line, $\mathcal{I} = \R$. The straightforward modifications needed for the case of general $\mathcal{I}$ are briefly discussed in Remark~\ref{rmk:Igeneral}.  We also remind the reader of the convention from Section \ref{sec:Mboundsproof} that the symbol $A$ stands for pre-regular observables. 

In order to prove Theorem \ref{thm:locallawreg}, we follow the \emph{zigzag strategy}, a recursive tandem of the characteristic flow method (\emph{zig step}, see Proposition \ref{prop:zig}) and a Green function comparison argument (\emph{zag step}, see Proposition \ref{prop:zag}). To initiate the procedure, we prove a \emph{global law} at large spectral scales (see Proposition \ref{prop:global}).  We start by collecting several preliminaries and setting up the overarching inductive proof scheme in Section \ref{subsec:zigzagprelim} and afterwards, in Section \ref{subsec:zigzagconclude}, provide the proof of Theorem~\ref{thm:locallawreg} by combining Propositions \ref{prop:global}, \ref{prop:zig}, and \ref{prop:zag} in an inductive way.

 	\subsection{Preliminaries: Setting up the inductive zigzag procedure} \label{subsec:zigzagprelim} In this section, we set up the overall proof scheme of the zigzag strategy, closely following \cite[Section 3]{cuspuniv}, but presented here for completeness. 
 	
 	\subsubsection{Two random matrix flows}		Along the zigzag proof, we will use two distinct flows in the space of $N\times N$ random matrices: the \textit{zig-flow} (which is just the standard Ornstein–Uhlenbeck process):
 	\begin{equation} \label{eq:OU_flow}
 		\rd H_t = -\frac{1}{2}H_t \rd t + \frac{\rd \Brwn_t}{\sqrt{N}},  \qquad t\ge 0;
 	\end{equation}
 	and the \textit{zag-flow} (a suitably modified Ornstein–Uhlenbeck process), distinguished by the superscript $t$: 
 	\begin{equation} \label{eq:zag_flow}
 		\rd H^t = -\frac{1}{2} \bigl(H^t - \E H^t\bigr)  \dif t + \Sigma_{H^0}^{1/2}\bigl[\rd \Brwn_t\bigr],    \qquad t \ge 0. 
 	\end{equation}
 	Here, $\Sigma_{H^0}$ is the covariance tensor of $H^0$, defined according to \eqref{eq:CovTensor}.
 	For both evolutions, \eqref{eq:OU_flow} and \eqref{eq:zag_flow}, we denote the real symmetric or complex Hermitian Brownian motion (depending on the symmetry class of $H$)  by $\Brwn_t$.

 	We point out that along the zig-flow \eqref{eq:OU_flow}, the covariance tensor $\Sigma_t := \Sigma_{H_t}$, corresponding to $H_t$ via \eqref{eq:CovTensor}, evolves according to the ODE
 	\begin{equation} \label{eq:Sigma_flow}
 		\rd \Sigma_t = \bigl(-\Sigma_t + \Sigma_{\mathrm{G}}\bigr)\rd t,
 	\end{equation}
 	where $\Sigma_{\mathrm{G}}$ is the covariance tensor of a GOE/GUE matrix in the same symmetry class as $H$. More precisely, $\Sigma_{\mathrm{G}}[X] = N^{-1}X$ in the complex Hermitian case, and $\Sigma_{\mathrm{G}}[X] = N^{-1}(X+X^\mathfrak{t})$ in the real-symmetric case, where $X^\mathfrak{t}$ denotes the transpose of $X$ (see the discussion below \eqref{eq:SigmaSentries}). Contrary to that, along the zag-flow \eqref{eq:zag_flow}, the expectation and the covariance tensor of $H^t$ (and hence the self-energy $\mathcal{S}_{H^t}$) are preserved. Therefore, the deterministic approximation $M$ (and all two-resolvent versions defined in \eqref{eq:M2_def}--\eqref{eq:ImM2_def}) remain unchanged along the zag-dynamic.

 	For any $t \ge 0$,  we define the flow maps $\mathfrak{F}_{\mathrm{zig}}^t$ and $\mathfrak{F}_{\mathrm{zag}}^t$ on the space of probability distribution $\mathcal{P}(\mathbb{C}^{N\times N})$ by
 	\begin{equation} \label{eq:zig_operator}
 		\mathfrak{F}_{\mathrm{zig}}^t\bigl[H \bigr] := H_t, \quad \text{ where } H_t \text{ solves (\ref{eq:OU_flow}) with the initial condition } H_0 = H.
 	\end{equation}
 	\begin{equation} \label{eq:zag_operator}
 		\mathfrak{F}_{\mathrm{zag}}^t\bigl[ H  \bigr] := H^t, \quad \text{ where } H^t \text{ solves (\ref{eq:zag_flow}) with the initial condition } H^0 = H.
 	\end{equation}
 	The key relation between the flow maps $\mathfrak{F}_{\mathrm{zig}}^t$ and $\mathfrak{F}_{\mathrm{zag}}^t$ is captured by the following lemma, which we import from \cite[Lemma~3.4]{cuspuniv}. 
 	\begin{lemma} [Relation between the zig- and zag-flows] \label{lemma:OUsurj}
 		Let $H$ be a random matrix satisfying the fullness condition \eqref{eq:fullness} with a constant $0 < c < 1$ instead of $c_{\mathrm{full}}$. Then there exists a random matrix $\mathfrak{H}_{c,t}(H)$ such that
 		\begin{equation} \label{eq:init_cond_for_zig}
 			\mathfrak{F}_{\mathrm{zig}}^t\bigl[\mathfrak{H}_{c,t}(H)\bigr] \,\overset{\rd}{=}\, \mathfrak{F}_{\mathrm{zag}}^{s(t)}\bigl[ H  \bigr], \quad 0 \le t \le -\log(1-c),
 		\end{equation}
 		
 		where the function $s(t) \equiv s_c(t)$ is defined as 
 		\begin{equation} \label{eq:init_cond_reverse}
 			s(t) \equiv s_c(t) := \log c - \log\bigl(c - 1 + \ee^{-t}\bigr),
 		\end{equation}
 		and satisfies
 		\begin{equation} \label{eq:s(t)_est}
 			s(t) \le 2c^{-1} t,\quad 0 \le t \le c/2. 
 		\end{equation}
 	\end{lemma}
 	
 	\subsubsection{Time-dependent spectral domains}
 	The base of our inductive proof is laid by a \emph{global law}. That is, we show that the two-resolvent chains in Theorem \ref{thm:locallawreg} concentrate around their deterministic approximations for spectral parameters having (at least) an order one distance from the spectrum of $H$. Note that, as shown in \cite[Theorem 2.9]{cuspuniv}, with very high probability, all eigenvalues of $H$ are within a distance $N^{-2/3 + \epsilon}$ of the support of the scDOS $\rho$. Hence, as shown in Proposition \ref{prop:global} below, for any fixed $\epsilon, \kappa > 0$, the global law holds on the domain\footnote{Note that we are slightly abusing notation, since this $\kappa$ has nothing to with the cumulants defined in \eqref{eq:cumulants}. We believe, however, that this should not lead to confusion.}
 	\begin{equation} \label{eq:Dglobdef}
\globD \equiv \globD(\epsilon, \kappa):= \left\{ z \in \C \setminus \R : |\Re z| \vee  |\Im z| \le \epsilon^{-1}\,, \ |\rho(z)| \ge N^{-1/\epsilon}\,, \ \dist(z, \supp \rho) \ge \kappa \right\} \,. 
 	\end{equation}
 	
 	Recall that the goal of our zigzag proof is to propagate the global law in $\globD$ to a \emph{local law} valid in the \emph{above the scale} regime, where $N |\Im z| |\rho(z)| $ is large. More precisely, for $\epsilon > 0$, we define it as
\begin{equation} \label{eq:Dabvdef}
\abvD \equiv \abvD(\epsilon) := \left\{ z \in \C \setminus \R  : |\Re z| \vee  |\Im z|\le  \epsilon^{-1}, \, N |\Im z| |\rho(z)| \ge N^\epsilon  \right\} \,. 
\end{equation}
Note that studying local laws in this regime is natural, since $N |\Im z| |\rho(z)|$ is the typical number of eigenvalues in an interval of length $|\Im z|$ around the energy $\Re z$.

 To achieve the propagation from global to local laws, as in Section \ref{sec:Mboundsproof}, we consider the following time-dependent version of the MDE \eqref{eq:MDE}, 
 	\begin{equation} \label{eq:MDEt}
 		-M_t(z)^{-1} = z - D_t + \mathcal{S}_t\bigl[M_t(z)\bigr], \quad z \in \mathbb{C}\backslash\mathbb{R},\quad   (\Im z) \Im M_t(z) > 0,
 	\end{equation}
 	where the data-pair $(D_t, \mathcal{S}_t)$ is the unique solutions to the  data-pair flow \eqref{eq:datapair_flow} with the {\it terminal conditions} $D_T = D =\E H$ and $\mathcal{S}_T = \mathcal{S} = \E [ (H-D) (\,\cdot\,) (H-D) ]$, respectively.  We also denote 
	$$\rho_t(z) := \pi^{-1} \langle \Im M_t(z) \rangle.$$

 	In the following we will fix the terminal time $T$ to be given by
the one from Lemma \ref{lemma:flow} That is, the total running time of our zig-flows will be given by the interval of times $t \in [0,T]$. 	Next, again just as in Section~\ref{sec:Mboundsproof}, given $M_t(z)$, we consider the {\it characteristic ODE} for the time dependent spectral parameter $z_t\in \C \setminus \R$ 
 	\begin{equation} \label{eq:z_flow}
 		\rd z_t = -\frac{1}{2}z_t\rd t - \bigl\langle M_t(z_t) \bigr\rangle \rd t.
 	\end{equation}
 	Moreover, we also define the (inverse) flow map $\varphi_{s,t} : \C\setminus \R \to \C\setminus \R$, uniquely defined as
 	\begin{equation} \label{eq:flowmap}
 		\varphi_{s,t}(z_t) := z_s \quad \text{with $z_s$ solving \eqref{eq:z_flow}} \,. 
 	\end{equation}
 	
 	For constants $c', C' > 0$, we define the time dependent version of the \emph{above the scale} domain as
 	\begin{equation*}
 		\begin{split}
 			\abvD_t = \abvD_t(\epsilon, c', C') := \Big\{ z \in \C  \setminus \R : \ &|\Re z| \vee |\Im z|\le  \epsilon^{-1} \cdot \big(1+ C'(T - t)\big),  \, N |\Im z| |\rho_{t}(z)| \ge N^\epsilon , \\
 			&|(\Im z)\rho_t(z)^{-1} | \ge c' \cdot \big( N^{-1+\epsilon} + (T - t)\big) \Big\}
 		\end{split}
 	\end{equation*}
 	Then we have the following key lemma on the propagation from the global to the local scale. The proof is completely analogous to that of \cite[Lemma 3.5]{cuspuniv} and hence omitted. 
 	\begin{lemma}[Relations between time-dependent domains] \label{lem:Dtrel}
 		Fix a small $\epsilon > 0$. Then there exist constants $c, C \sim 1$ such that the time dependent domains $\big(\abvD_t(\epsilon, c, C)\big)_{t \in [0,1]}$ satisfy the inclusion relation
 		\begin{equation} \label{eq:domaininterpolation}
 			\varphi_{s,t}(\abvD_t(\epsilon, c, C)) \subset \abvD_s(\epsilon, c, C)
 		\end{equation}
 		for all times $0 \le s \le t \le T$. Moreover, there exists a constant $\kappa \sim 1$ such that, when replacing $\rho \to \rho_{0}$ in the definition of $\globD(\epsilon, \kappa)$ in \eqref{eq:Dglobdef} and $\rho \to \rho_{T}$ in the definition of $\abvD(\epsilon)$ in \eqref{eq:Dabvdef}, it holds that 
 		\begin{equation} \label{eq:initialandfinaldomains}
 			\abvD_0 (\epsilon, c, C) \subset \globD(\epsilon, \kappa) \quad \text{and} \quad \abvD(\epsilon) \subset \abvD_T(\epsilon, c, C) \,. 
 		\end{equation}
 	\end{lemma}
 	The parameters $\epsilon, c, C, \kappa$ of the domains are henceforth considered fixed and shall often be omitted.

 	\subsubsection{Final preparations: Two sequences of random matrices}
As mentioned above, our zigzag proof will be conducted inductively, by repeating a tandem of small zig and zag steps. 
While the zig step could have been done in one move, immediately going from the global scale to the 
smallest target scale, the zag step, that adjusts for the added Gaussian component along the zig step,
cannot be run too long when $\Im z$ is small. The running times of both flows will be reduced as $\Im z_t$ 
becomes smaller. The  flow $z_t$~\eqref{eq:z_flow} 
will be the same in each zig step, but the initial conditions of the
matrix flows will be adjusted to match the output of the previous steps. Note that $z_t$ and
the matrix flows are naturally defined forward in time, but we are given our targets, the final matrix $H$ and spectral parameter $z$, thus the initial conditions of the flows need to be chosen properly.  This is possible thanks to the 
fullness condition~\eqref{eq:fullness} on $H$. 

 In order to set up this procedure, we fix an arbitrarily small step size exponent $\delta > 0$ and let $K = K(\delta, \epsilon)$ be the smallest integer such that $N^{-K \delta} \le N^{-1+\epsilon}$, where $\epsilon > 0$ is the \emph{above-the-scale exponent} from the definition of $\abvD$ in \eqref{eq:Dabvdef}. We now define a sequence of times $\{t_k\}_{k=0}^K$ as
\begin{equation} \label{eq:tkdef}
	t_0 := 0\,, \quad t_k := T - N^{-k \delta} \,, \quad k \in \{1, ... , K-1\}\,, \quad t_K := T \,. 
\end{equation}
Let $\{\dift_k\}_{k=1}^K$ denote the consecutive difference sequence of the $t_k$'s, that is
\begin{equation} \label{eq:dtkdef}
	\dift_k := t_k - t_{k-1} \quad \text{for} \quad k \in \{1, ... , K\} \,. 
\end{equation}

	For this sequence of times, we now define two sequences of random matrices along the zig- and zag-flow: Let $\Sigma_t$ solve the equation \eqref{eq:Sigma_flow} with the terminal condition $\Sigma_{T=1} = \Sigma$, where $\Sigma$,
defined via \eqref{eq:CovTensor}, is the covariance tensor of the target matrix $H$, for which
we eventually prove the local laws in Theorem~\ref{thm:main}, and point out that for all times $0 \le t \le T$, the solution $\Sigma_t$ satisfies
\begin{equation} \label{eq:fullness_alongflow}
	\Sigma_t \ge \other{c}\,\Sigma_{\mathrm{G}}, \quad \other{c} := \frac{c_{\mathrm{full} }}{10} \wedge 1,
\end{equation}
where $c_{\mathrm{full}}$ is the constant in Assumption~\ref{ass:full}.
Then, given the target random matrix ensemble $H$, we construct two sequences of random matrices, $\{H_k\}_{k=0}^{K}$ and $\{H^k\}_{k=1}^K$ recursively by 
\begin{equation} \label{eq:H_k_defs}
	H_K := H, \quad H^k := \mathfrak{F}^{s(\dift_k)}_{\mathrm{zag}}\bigl[H_k\bigr], \quad H_{k-1} := \mathfrak{H}_{\other{c},\dift_k}\bigl(H_k\bigr), \quad k\in \{1,\dots, K\},
\end{equation}
where $s(t) := s_{\other{\alpha}}(t)$ and $\mathfrak{H}_{\other{c},\dift_k}$ are given by Lemma~\ref{lemma:OUsurj}, and $\other{c}$ is the constant in \eqref{eq:fullness_alongflow}. Thus the zig flow in the $k$-th step will be run for time 
$\Delta t_k = N^{-(k-1)\delta} - N^{-k\delta}\sim N^{-(k-1)\delta} $, 
while the corresponding zag step runs for a comparable time $s(\Delta t_k)$. We point out that, by a simple 
backward inductive argument starting at $k=K$, the covariance tensor of both $H_k$ and $H^k$ is given by $\Sigma_{t_k}$, hence by \eqref{eq:fullness_alongflow}, $H_{k-1}$ is well-defined. 
 	
 	\subsection{Zigzag strategy: Proof of Theorem \ref{thm:locallawreg}} \label{subsec:zigzagconclude}
  		The goal of this section is to give the proof of Theorem \ref{thm:locallawreg}~(a) in an inductive way
		using the zigzag strategy. Notationally, we follow the previous convention that 
		we consider  time-dependent two-body quantities like 
$M_t(z_{1}, A, z_{2})$, $\widehat{M}_t(z_{1}, A, z_{2})$ from~\eqref{eq:Mt_def}--\eqref{eq:ImM2t_def},
 as functions of the spectral parameters $z_i=z_{i, T}\in \abvD(\epsilon)$ at the terminal time 
of the zig-flow, but $M_{i,t} \equiv M_t(z_{i,t})$ (see below~\eqref{eq:stab_t_def}) 
appearing in these definitions are evaluated at $z_{i,t}$. In particular, the initial condition $z_{i,0}$ is 
in the global regime, $z_{i,0}\in \globD(\epsilon, \kappa)$ for some $\kappa>0$, see~\eqref{eq:initialandfinaldomains}.  \nc
		Given a target random matrix $H$, the induction base is given by the following proposition, whose proof is provided in Section \ref{sec:glob}.

 	\begin{proposition}[Global law] \label{prop:global} 
 	Fix $T, \epsilon>0$, take $z_{1}, z_{2} \in \abvD(\epsilon)$ and let $z_{1, 0}, z_{2,0}$ be the
corresponding initial conditions. 
Then we have that the resolvents
\begin{equation*}
	G_{i,0} := \big(H_{k=0} - z_{i,0}\big)^{-1} \quad i \in \indset{2}\,,
\end{equation*}
satisfy the global laws
\begin{equation} \label{eq:2G1Aglobal}
	\left| \big\langle G_{1,0}^{(*)} A_1 G_{2,0}^{(*)} \big\rangle - \big\langle {M}_0(z_{1}^{(*)}, A_1, z_{2}^{(*)}) \big\rangle \right| \prec \frac{1}{N} \Vert A_1 \Vert_{\rm hs} \,, 
\end{equation}
and
\begin{equation} \label{eq:2G2Aglobal}
	\left| \big\langle \Im G_{1,0} A_1 \Im G_{2,0} A_2 \big\rangle - \big\langle \widehat{M}_0(z_{1}, A_1, z_{2}) A_2 \big\rangle \right| \prec\frac{ |\rho_0(z_{1,0}) \rho_0(z_{2,0})|}{N}\left( \Vert A_1 \Vert_{\rm hs} \, \Vert A_2 \Vert \wedge \Vert A_1 \Vert \, \Vert A_2 \Vert_{\rm hs}\right)  \,, 
\end{equation}
all uniformly in any\footnote{To better align with the formulation of the zig and zag steps in Propositions \ref{prop:zig} and \ref{prop:zag} below, we here locally violate the convention from Section~\ref{sec:Mboundsproof} that the letter $A$ is reserved for pre-regular observables.} deterministic $A_1, A_2 \in \C^{N \times N}$. 
 	\end{proposition}
Note that we formulated the global law for the initial condition of the characteristic flow to align with the setup  of the 
zigzag procedure.  However, as our proof in Section~\ref{sec:glob} shows, the global law is valid for any data-pair $(D_0, \mathcal{S}_0)$ 
satisfying Assumption~\ref{ass:boundedexp} and the flatness estimates \eqref{eq:flatness}, for
any spectral parameters $z_{i,0} \in \globD(\epsilon, \kappa)$
 (with $\rho \to \rho_0$ in \eqref{eq:Dglobdef}) and some fixed $\kappa,\epsilon>0$. Therefore, the validity of the global law 
 is independent of any embedding into a flow (recall that $M_0(z_1, A, z_2)$ actually depends only on $z_{i,0}$). 
  \nc
 	
Throughout the rest of the argument, we fix two target spectral parameters $z_1, z_2 \in \abvD(\epsilon)$ satisfying
\begin{equation*}
\max\{ \widetilde{\beta}(z_1^+, z_2^-), \widetilde{\beta}(z_1^-, z_2^+) \} \le \beta_*\,, \quad \text{i.e.} \quad (z_1, z_2) \in \regD_{\beta_*} \equiv \regD
\end{equation*}
 	with $\beta_* \sim 1$ from Lemma \ref{lemma:stab_bound} (recall also \eqref{eq:regD_def}--\eqref{eq:regD_beta}). Moreover, we consider two observables $A_1, A_2 \in \C^{N \times N}$ which we assume to be $(z_1, z_2)$- and $(z_2, z_1)$-pre-regular, respectively (see Definition \ref{def:prereg} -- note that pre-regularity is a time-independent concept!), and recall the time-dependent version of the observable norm $\vertiii{\cdot}_{z_1, z_2}$ from \eqref{eq:size_t_def} as well as 
	time-dependent correction to the full regularization (cf.~Definition \ref{def:reg}) from \eqref{eq:Ups_def}.

 	In the following, besides  $M_t(z_{1}, A, z_{2})$ and  $\widehat{M}_t(z_{1}, A, z_{2})$ from~\eqref{eq:Mt_def}--\eqref{eq:ImM2t_def}  we also use the notations $\eta_{i,t}:=|\Im z_{i,t}|$, 
	\nc $\rho_{i,t}:=\pi^{-1}\langle \Im M_{i,t}\rangle$, and $\ell_t := \min_i (\eta_{i,t} |\rho_{i,t}|)$.  
Finally, we introduce the shorthand notation
 		\begin{equation} \label{eq:gamma_def}
\gamma_t = \gamma_t({z_1, z_2}) := \frac{|\rho_{1,t}|^{-1} \eta_{1,t} + |\rho_{2,t}|^{-1} \eta_{2,t}}{|z_{1,t} - z_{2,t}|^2 + \varkappa_{1,t}^2 + \varkappa_{2,t}^2}
 		\end{equation}
 		to cast the local laws in a more compact form, and note that $\gamma_t({z_1^{(*)}, z_2^{(*)}}) \sim  \gamma_t({z_2^{(*)}, z_1^{(*)}})$ where $z^{(*)}_J$ stands for $z_j$ or $\overline{z}_j$ independently for $j \in \{1,2\}$. 
		 Note that, similarly to $M_t(z_1, A, z_2)$, even though $\gamma_t(z_1, z_2)$ depends on the time-evolved spectral parameters $z_{i,t}$, we compartmentalize this dependence into an explicit time subscript $t$ and the spectral argument $z_i = z_{i, T}$, corresponding to the terminal points of (and uniquely determining) the trajectories $z_{i,t}$. 

	 	With these preparations, the zigzag induction is then conducted along the following two propositions, for which we recall that  $(z_1, z_2) \in \regD$ are fixed spectral parameters
		and  $A$ denotes \emph{pre}-regular observables.
 	\begin{proposition}[Zig step] \label{prop:zig}
	Fix $k \in \{1, ... , K\}$ and denote 
\begin{equation*}
	G_{i,t} := \big(\mathfrak{F}_{\rm zig}^{t-t_{k-1}}\big[H_{k-1}\big] - z_{i,t}\big)^{-1}\,, \quad t_{k-1} \le t \le t_k \,, \quad i \in \indset{2}\,. 
\end{equation*}
Assume that the $G_{i, t}$'s satisfy the following two resolvent local laws: 
\begin{equation} \label{eq:2G1Azig}
	\left| \big\langle G_{1,t}^{(*)} \reg{A}_{1,t} G_{2,t}^{(*)} \big\rangle - \big\langle {M}_t(z_{1}^{(*)}, \reg{A}_{1,t}, z_{2}^{(*)}) \big\rangle \right| \prec 	\sqrt{\frac{|\rho_{1,t} \rho_{2,t}|}{N^2 \ell_t  \eta_{1,t} \eta_{2,t}}} \vertiii{A_1}_{t,z_1, z_2},
\end{equation}
and similarly for interchanged indices $1 \leftrightarrow 2$, as well as
\begin{subequations}\label{eq:2G2Azig}
	\begin{equation} 
		\left| \big\langle \Im G_{1,t}  \Im G_{2,t}  \big\rangle - \big\langle \widehat{M}_t(z_{1}, I, z_{2}) \big\rangle \right| \prec \frac{|\rho_{1,t} \rho_{2,t}|}{N \ell_t} \gamma_{t} \,, 
	\end{equation}
		\begin{equation} 
		\left| \big\langle \Im G_{1,t} \reg{A}_{1,t} \Im G_{2,t}  \big\rangle - \big\langle \widehat{M}_t(z_{1}, \reg{A}_{1,t}, z_{2}) \big\rangle \right| \prec \frac{|\rho_{1,t} \rho_{2,t}|}{N \ell_t} \gamma^{1/2}_{t} \vertiii{A_1}_{t, z_1, z_2} \,, 
	\end{equation}
	and similarly for interchanged indices $1 \leftrightarrow 2$, and
\begin{equation} \label{524c}
	\left| \big\langle \Im G_{1,t} \reg{A}_{1,t} \Im G_{2,t} \reg{A}_{2,t} \big\rangle - \big\langle \widehat{M}_t(z_{1}, \reg{A}_{1,t}, z_{2}) \reg{A}_{2,t}\big\rangle \right| \prec \frac{|\rho_{1,t} \rho_{2,t}|}{\sqrt{N \ell_t}} \vertiii{A_1}_{t,z_1, z_2} \, \vertiii{A_2}_{t,z_2, z_1} \,, 
\end{equation}
\end{subequations} 
all uniformly in $(z_1, z_2)$-pre-regular matrices $A_1  \in \C^{N \times N}$ and $(z_2, z_1)$-pre-regular matrices $A_2  \in \C^{N \times N}$, at time $t= t_{k-1}$. According to \eqref{eq:regA_def} and
the convention below~\eqref{eq:un-reg}, here we abbreviated $\reg{A}_{1,t} \equiv \reg{(A_1)}^{t, z_{1}, z_{2}}$
		and $\reg{A}_{2,t} \equiv \reg{(A_2)}^{t, z_{2}, z_{1}}$. \nc
Then the $G_{i, t}$'s satisfy the two resolvent local laws \eqref{eq:2G1Azig}--\eqref{eq:2G2Azig} 	
uniformly in $(z_1, z_2)$-pre-regular matrices $A_1  \in \C^{N \times N}$ and $(z_2, z_1)$-pre-regular matrices $A_2  \in \C^{N \times N}$ uniformly in all times $t\in [ t_{k-1}, t_k]$. 
 	\end{proposition}
 	
 	\begin{proposition}[Zag step] \label{prop:zag}
	Fix $k \in \{1, ... , K\}$ and denote $s_k := s(\dift_k)$ (cf.~Lemma \ref{lemma:OUsurj}). Let $H_k$ be the random matrix defined in \eqref{eq:H_k_defs}, and denote 
\begin{equation*}
	G_{i}^s := \big(\mathfrak{F}_{\rm zag}^{s}\big[H_{k}\big] - z_{i,t_k}\big)^{-1}\,, \quad 0 \le s \le s_k \,, \quad i \in \indset{2}\,. 
\end{equation*}
Assume that the $G_{i}^s$'s satisfy the following two resolvent local laws:
\begin{equation} \label{eq:2G1Azag}
	\left| \big\langle (G_{1}^s)^{(*)} \reg{A}_{1,t_k} (G_{2}^s)^{(*)} \big\rangle - \big\langle {M}_{t_k}(z_{1}^{(*)}, \reg{A}_{1,t_k}, z_{2}^{(*)}) \big\rangle \right| \prec \sqrt{\frac{|\rho_{1,t_k} \rho_{2,t_k}|}{N^2 \ell_t  \eta_{1,t_k} \eta_{2,t_k}}} \vertiii{A_1}_{t_k,z_1, z_2}
\end{equation}
and similarly for interchanged indices $1 \leftrightarrow 2$, as well as
\begin{subequations}\label{eq:2G2Azag}
	\begin{equation} 
	\left| \big\langle \Im G_{1}^s  \Im G_{2}^s  \big\rangle - \big\langle \widehat{M}_{t_k}(z_{1},I, z_{2}) \big\rangle \right| \prec \frac{|\rho_{1,t_k} \rho_{2,t_k}|}{N \ell_{t_k}}\gamma_{t_k} \,, 
\end{equation}
\begin{equation} 
	\left| \big\langle \Im G_{1}^s \reg{A}_{1,t_k} \Im G_{2}^s \big\rangle - \big\langle \widehat{M}_{t_k}(z_{1}, \reg{A}_{1,t_k}, z_{2}) \big\rangle \right| \prec \frac{|\rho_{1,t_k} \rho_{2,t_k}|}{N \ell_{t_k}} \gamma_{t_k}^{1/2} \vertiii{A_1}_{t_k,z_1, z_2} \,, 
\end{equation}
and similarly for interchanged indices $1 \leftrightarrow 2$, and 
	\begin{equation} 
	\left| \big\langle \Im G_{1}^s \reg{A}_{1,t_k} \Im G_{2}^s \reg{A}_{2,t_k} \big\rangle - \big\langle 
	\widehat{M}_{t_k}(z_{1}, \reg{A}_{1,t_k}, z_{2}, \reg{A}_{2,t_k}) \big\rangle \right| \prec \frac{|\rho_{1,t_k} \rho_{2,t_k}|}{\sqrt{N \ell_{t_k}}} \vertiii{A_1}_{t_k,z_1, z_2} \, \vertiii{A_2}_{t_k,z_2, z_1}
\end{equation}
\end{subequations} 
all uniformly in $(z_1, z_2)$-pre-regular matrices $A_1  \in \C^{N \times N}$ and $(z_2, z_1)$-pre-regular matrices $A_2  \in \C^{N \times N}$, at time $s= s_k$. 

Then the $G_{i}^s$'s satisfy the two-resolvent local laws \eqref{eq:2G1Azag}--\eqref{eq:2G2Azag} 	uniformly in $(z_1, z_2)$-regular matrices $A_1  \in \C^{N \times N}$ and $(z_2, z_1)$-regular matrices $A_2  \in \C^{N \times N}$, and uniformly in all times $s\in [0, s_k]$. 
 	\end{proposition}
 	\nc
 	
 	We can now easily conclude the proof of Theorem \ref{thm:locallawreg}. 
 	
 	\begin{proof}[Proof of Theorem \ref{thm:locallawreg}]
First, by Lemma \ref{lem:Dtrel}, we have that $	\abvD_0 (\epsilon, c, C) \subset \globD(\epsilon, \kappa)$. Moreover, we have $$\frac{|\rho_{1,0} \rho_{2,0}|}{ \ell_0  \eta_{1,0} \eta_{2,0}} \wedge \ell_0^{-1}\wedge \gamma_0 \gtrsim 1 $$
for $z_1, z_2 \in \abvD(\epsilon)$ and use that $\Vert A \Vert_{\rm hs} \le \Vert A \Vert \le N^{1/2} \Vert A\Vert_{\rm hs}$ as well as the fact that the observable norm is lower bounded by the Hilbert-Schmidt norm. Hence, the assumptions of Proposition \ref{prop:zig} are satisfied and we can thus apply Propositions \ref{prop:zig} and \ref{prop:zag} in tandem $K$ times (using that pre-regularity is a time-independent concept), to deduce Theorem~\ref{thm:locallawreg}, where we additionally use \eqref{eq:tripl_I=0} to get that the norms $\vertiii{\cdots}$ of $A_i$ and $ \reg{A}_{i,t}$ agree.  
 	\end{proof}

 	\begin{remark}[On locality of Assumption \ref{ass:Mbdd}] \label{rmk:Igeneral}
In case that the set of admissible energies (see \eqref{eq:adm_E}) does not exhaust all of $\R$, i.e.~$\mathcal{I} \subsetneq \R$, our zigzag proof holds verbatim, apart from intersecting the domains $\globD, \abvD, \abvD_t$ used along the argument with suitable vertical strips in the complex plane containing the admissible energies; see \cite[Remark~3.8]{cuspuniv} for details. 
 	\end{remark}

\section{Characteristic flow: Proof of Proposition \ref{prop:zig}} \label{sec:zig_proof}

We conduct the proof in the complex Hermitian case, the obvious modifications in the real symmetric case\footnote{For a detailed treatment, we refer to \cite[Section 4]{cipolloni2023eigenstate}.} are left to the reader. Also, note that it suffices to prove the statement only for fixed $z$'s and $t$'s, since uniformity follows by a standard simple grid argument. Lastly, we recall the time-dependent $M$-bounds from Proposition \ref{prop:Mt_bound} and the notations $\other{\beta}_t$ and $\other{\alpha}_t$ from \eqref{eq:betaf_t_def} and \eqref{eq:alp_t_def}, respectively. 

\subsection{Preliminaries} 
We begin with preparing the proof by introducing a few simplifying notations and recording basic estimates. Afterwards, we define the stopping times, providing the basic structure for the rest of this section. 

\subsubsection{Preparing the proof}
In the following, we drop the induction index $k$ as it remains fixed and introduce the shorthand notations $t_{\rm init} := t_{k-1}$ and $t_{\rm fin} := t_{k}$. Moreover, for $t \in [t_{\rm init}, t_{\rm final}]$, we denote $G_{i,t}:=\big(\mathfrak{F}_{\rm zig}^{t - t_{\rm init}}[H_{k-1}] - z_{i,t}\big)^{-1}$ for $z_{i,t} := \varphi_{t, t_{\rm final}}(z_i)$ with $\varphi$ defined in \eqref{eq:flowmap}, and $M_{i,t}:= M_{t}(z_{i,t})$ as well as $\eta_{i,t} := |\Im z_{i,t}|$ and $\rho_{i,t} := \pi^{-1} \langle \Im M_{i,t} \rangle$. 
Additionally, for $i \in \indset{2}$, we will also employ the shorthand notations
\begin{equation} \label{eq:timenotations}
	\nu_{i,t} := \frac{\eta_{i,t}}{|\rho_{i,t}|} \,,   \quad \other{\nu}_{t} := \frac{\eta_{1,t} + \eta_{2,t}}{|\rho_{1,t}| + |\rho_{2,t}|}\,, \quad \other{\alpha}_t := \other{\alpha}_t(z_1^+, z_2^+)\,, \quad \other{\beta}_t := \other{\beta}_t(z_1^+, z_2^+)\,, \quad \other{\beta}_{*,t} := \other{\beta}_t(z_1^+, z_2^-),
\end{equation}
and note the following ordering relation, that easily follows by their respective definitions: 
\begin{equation} \label{eq:timerelations}
	\min_{i \in \indset{2}} \nu_{i,t} \lesssim \other{\nu}_t \lesssim \nu_{1,t} + \nu_{2,t} \lesssim \other{\beta}_{*,t} \lesssim \other{\beta}_{t} \lesssim \other{\alpha}_t^2 \,. 
\end{equation}
As shown in \eqref{eq:m_hat_alpha}, we also have that $\gamma_t$ from \eqref{eq:gamma_def} satisfies
	\begin{equation} \label{eq:gamma}
		\gamma_t \sim \frac{\nu_{1,t} + \nu_{2,t}}{\other{\beta}_{*,t}^2 \other{\beta}_{t}^2} \other{\alpha}_t^2.
\end{equation}

{Moreover, recall the definition of the time-dependent correction $\Upsilon$ to pre-regular observables from \eqref{eq:Ups_def}. 
	To estimate their contribution along the dynamics, we record that the time-dependent correction parameter $\Upsilon_{t}$ satisfies the following (approximate) ODE, whose proof is given in Appendix \ref{app:Upsderv}. 
	
	\begin{lemma}[Derivative of $\Upsilon$] \label{lem:Upsderv} 
		Let $A_1$ be $(z_1, z_2)$-pre-regular and $A_2$ be $(z_2, z_1)$-pre-regular. Then, for any $t \in [\tin, \tfin]$ we have that $\Upsilon_{t}(A_i)$ defined in \eqref{eq:Ups_def} satisfies 
		\begin{equation}
	\begin{split}
				\left|	\frac{\dif}{\dif t}{\Upsilon}_{t}(A_i)  - \langle M_t(z_1, I, z_2) \reg{A}_{i,t}\rangle \right| \lesssim \mathcal{E}_{t} \vertiii{A_i}_{t, z_1, z_2} \quad \text{for} \quad i \in \indset{2}\,, 
	\end{split}
		\end{equation}
		where we introduced the shorthand notation
		\begin{equation} \label{eq:Upserror}
			\mathcal{E}_{t} := \frac{|\rho_{1,t}| + |\rho_{2,t}|}{\other{\alpha}_t} 
			+ \frac{\other{\beta}_t \other{\beta}_{*,t}}{ \other{\nu}_t (\nu_{1,t} + \nu_{2,t})^{1/2}\other{\alpha}_t }\,. 
		\end{equation}
	\end{lemma}
}
We also record that the time-dependent observable norms \eqref{eq:size_t_def} behave monotonically in time:  
\begin{equation} \label{eq:tripmonotone} 
	\vertiii{A_1}_{t, z_1, z_2} \ge \vertiii{A_1}_{s, z_1, z_2} \quad \text{and} \quad \vertiii{A_2}_{t, z_2, z_1} \ge \vertiii{A_2}_{s, z_2, z_1} \quad  \text{for all} \quad s \le t \,. 
\end{equation}

Finally, in subsequent estimates in Sections \ref{subsec:tau12}--\ref{subsec:tau6} we will frequently use the following integral estimate for $s , t \in [\tin, \tfin]$ with $s \le t$. The proof of Lemma \ref{lem:intestgeneral} is a simple exercise left to the reader. 
\begin{lemma}[Integral estimates] \label{lem:intestgeneral}
	Fix $k \in \N$. Now take $x_i \in (0,\infty)$ and $p_i\in \R$ for $i \in \indset{k}$ such that $x_{j} < x_{j+1}$ for all $j \in \indset{k-1}$ and $\sum_{i=1}^j p_i \le  q$ for all $j \in \indset{k}$ and some $q \le - 1$. Then for any $0  t \lesssim 1$ it holds that
	\begin{equation}
		\int_{0}^{t} \prod_{i=1}^{k} \big(x_i + (t-s)\big)^{p_i} \dd s \lesssim \big(1 + \mathbf{1}(q = - 1) |\log x_1| \big) x_1 \prod_{i=1}^{k} x_i^{p_i}.
	\end{equation}
\end{lemma}
In the proof below, we will choose the $x$-parameters from Lemma \ref{lem:intestgeneral} among the quantities (note the square at $\other{\alpha}$!)
\begin{equation}
	\widetilde{\nu}_{t}\,, \quad \nu_{i,t}\,, \ i \in \indset{2}\,, \quad \other{\alpha}^2_t \,,\quad  \other{\beta}_t\,, \quad \other{\beta}_{*,t} \,. 
\end{equation}
Note that all these quantities move linearly in time, e.g. $ \widetilde{\nu}_{r} \sim \widetilde{\nu}_{t} +(t-r)$, etc., see \eqref{eq:rho_t}--\eqref{eq:eta_t}, \eqref{eq:betaf_t_def}, \eqref{eq:alp_t_def}.

Thus, we will have, in particular that $|\log x_1| \lesssim \log N$ in all applications. Moreover, as an important special case of the general Lemma \ref{lem:intestgeneral}, we have, for $i \in \indset{2}$, 
\begin{equation} \label{eq:intrule1}
	\int_{s}^t \frac{|\rho_{i,r}|}{\eta_{i,r}} \dif r  = 	\int_{s}^t \frac{1}{\nu_{i,r}} \dif r\lesssim \log N \quad \text{and} \quad \int_{s}^t \frac{|\rho_{i,r}|}{\eta_{i,r}^\alpha} \dif r = \int_{s}^t \frac{1}{\nu_{i,r}\eta_{i,r}^{\alpha-1}} \dif r \lesssim \frac{1}{\eta_{i,t}^{\alpha - 1}} \quad \text{for} \ \alpha > 1 \,. 
\end{equation}

Lastly, we record the following further integral estimates as a simple consequence of \eqref{eq:char_flow} and \eqref{eq:beta_assymp}. Lemma \ref{lem:prop} will serve as a bound on propagator terms when applying a stochastic Gronwall argument in our proof below. 
\begin{lemma}[Propagator bounds] \label{lem:prop} Given $r > 0$ and target spectral parameters $z_1, z_2 \in \C \setminus \R$, denote
	\begin{equation}
		f_r = f_r(z_1, z_2) := \Re \left[\frac{\langle M_{1,r} - M_{2,r}\rangle}{ z_{1,r} - z_{2,r}}\right] \vee 0\,. 
	\end{equation}
	Then we have that
	\begin{equation}
		f_r \lesssim \other{\beta}_r(z_1, z_2)^{-1} \wedge \min_{i \in \indset{2}}\frac{|\rho_{i,r}|}{\eta_{i,r}} \,, \qquad 	\Phi_{s,t} := \exp\left( \int_{s}^{t} \left[\frac{1}{2} + f_r\right] \dd r\right) \lesssim  \frac{\other{\beta}_s(z_1, z_2)}{\other{\beta}_{t}(z_1, z_2)} \wedge \min_{i \in \indset{2}} \frac{\eta_{i,s}}{\eta_{i,t}}.
	\end{equation}
\end{lemma}

\subsubsection{Stopping time definition} In the proof below, we will control all quantities in \eqref{eq:2G1Azig}--\eqref{eq:2G2Azig} together. To do so, we fix $\xi \le \epsilon/10$ with $\epsilon > 0$ from the definition of $\abvD$ and introduce the following stopping time: 
\begin{equation}
		\label{eq:stoptime}
		\begin{split}
			\tau := \min_{i=1}^6 \tau_i 
		\end{split}
\end{equation}
as the minimum of the following six stopping times
\begin{align}
	\label{eq:stop1}\tau_1 := \inf\Bigg\{t \in  \mathcal{J} : &\max_* \max_{B_1 \in \{\reg{A}_{1,t}, \reg{A}_{2,t}^*\}}\frac{\big|\langle G_{1,t}^{(*)}B_1G_{2,t}^{(*)}-M_t(z_1^{(*)}, B_1, z_2^{(*)})\rangle\big|}{\vertiii{B_1}_{t,z_1, z_2}}\ge N^{2\xi}\sqrt{\frac{|\rho_{1,t}\rho_{2,t}|}{N^2 \ell_t \eta_{1,t}\eta_{2,t}}} \Bigg\} \,, \\
	\label{eq:stop2}\tau_2 := \inf\Bigg\{t \in  \mathcal{J} : &\max_*\max_{B_2 \in \{\reg{A}_{2,t}, \reg{A}_{1,t}^*\}}\frac{\big|\langle G_{2,t}^{(*)}B_2G_{1,t}^{(*)}-M_t(z_2^{(*)}, B_2, z_1^{(*)})\rangle\big|}{\vertiii{B_2}_{t,z_2, z_1}}\ge N^{2\xi}\sqrt{\frac{|\rho_{1,t}\rho_{2,t}|}{N^2 \ell_t \eta_{1,t}\eta_{2,t}}}\Bigg\}\,,  \\
	\label{eq:stop3}	\tau_3 := \inf\Bigg\{t \in \mathcal{J}  : & \ \big|\langle \Im G_{1,t}\Im G_{2,t}-\widehat{M}_t(z_1, I, z_2)\rangle\big|\ge N^{2\xi}\frac{|\rho_{1,t}\rho_{2,t}|}{N \ell_t } \gamma_{t}\Bigg\} \,,  \\
	\label{eq:stop4}	\tau_4 := \inf\Bigg\{t \in  \mathcal{J} : &\max_{\substack{B_1 \in \{\reg{A}_{1,t}, \reg{A}_{2,t}^*\} }}\frac{\big| \big\langle \Im G_{1,t}B_1 \Im G_{2,t} - \widehat{M}_t(z_1, B_1, z_2)\big\rangle \big|}{\vertiii{B_1}_{t,z_1, z_2} }\ge N^{3\xi} \frac{|\rho_{1,t}\rho_{2,t}|}{N \ell_t} \gamma^{1/2}_{t}\Bigg\} \,,  \\
	\label{eq:stop5}		\tau_5 := \inf\Bigg\{t \in  \mathcal{J} : &\max_{B_2 \in \{\reg{A}_{2,t}, \reg{A}_{1,t}^*\}}\frac{\big| \big\langle \Im G_{1,t} \Im G_{2,t} B_2- \widehat{M}_t(z_1, I, z_2)B_2\big\rangle \big|}{\vertiii{B_2}_{t,z_2, z_1} }\ge N^{3\xi} \frac{|\rho_{1,t}\rho_{2,t}|}{N \ell_t} \gamma^{1/2}_{t}\Bigg\}  \,, \\
	\label{eq:stop6} \tau_6 := \inf\Bigg\{t \in  \mathcal{J} : &\max_{\substack{B_1 \in \{\reg{A}_{1,t}, \reg{A}_{2,t}^*\} \\ B_2 \in \{\reg{A}_{2,t}, \reg{A}_{1,t}^*\}}}\frac{\big| \big\langle \big(\Im G_{1,t}B_1 \Im G_{2,t} - \widehat{M}_t(z_1, B_1, z_2)\big)B_2\big\rangle \big|}{\vertiii{B_1}_{t,z_1, z_2} \vertiii{B_2}_{t,z_2, z_1}}\ge N^\xi \frac{|\rho_{1,t}\rho_{2,t}|}{\sqrt{N \ell_t}}\Bigg\} \,. 
\end{align}
Here we used the shorthand notation $\mathcal{J} := [\tin, \tfin]$ and $\max_*$  in \eqref{eq:stop1}--\eqref{eq:stop2} is understood to be the maximum over all four possible ways to take complex conjugates of the two spectral parameters. To ease the notation, in the following, we will not write the unifying $B$ for the observables but stick to the $A$-notation, i.e.~working with the choice $B_1 = \reg{A}_{1,t}$ and $B_2 = \reg{A}_{2,t}$, as used in \eqref{eq:GAG}, \eqref{eq:imGimG}, \eqref{eq:imGAimG}, and \eqref{eq:imGAimGA} below, keeping in mind the uniformity expressed in \eqref{eq:stop1}--\eqref{eq:stop2} and \eqref{eq:stop4}--\eqref{eq:stop6}. Similarly, by interchanging indices $1 \leftrightarrow 2$ in \eqref{eq:GAG} and/or changing one or both of the spectral parameters to their adjoint, \eqref{eq:GAG} below will cover all the sixteen cases summarized in the two conditions of \eqref{eq:stop1}--\eqref{eq:stop2}; similarly for \eqref{eq:imGAimG} and \eqref{eq:stop4}--\eqref{eq:stop5}. 

In the following sections, we will analyze the flow equations for the quantities in the definition of $\tau_1$ and $\tau_2$ (Section~\ref{subsec:tau12}), $\tau_3$ (Section~\ref{subsec:tau3}), $\tau_4$ and $\tau_5$ (Section \ref{subsec:tau45}), and $\tau_6$ (Section \ref{subsec:tau6}). Our goal is to show that, for times $t \le \tau$ the bounds on the quantities in \eqref{eq:stop1}--\eqref{eq:stop6}, are \emph{strictly} better than prescribed by the stopping time. This allows to conclude $\tau = \tfin$ and thus the proof of Proposition \ref{prop:zig}. 

\subsection{Estimates of $\langle GAG \rangle$ for $\tau_1$ and $\tau_2$} \label{subsec:tau12} 
By Ito's formula, we have the following expression for the derivative (see, e.g., \cite[Section 4]{cipolloni2023eigenstate}):
\begin{equation}
	\begin{split}
		\label{eq:GAG}
		\dd \langle G_{1,t}\reg{A}_{1,t}G_{2,t} & -M_t(z_1, \reg{A}_{1,t}, z_2)\rangle \\
		=&\frac{1}{\sqrt{N}}\sum_{ab=1}^N \partial_{ab}\langle G_{1,t}\reg{A}_{1,t}G_{2,t}\rangle (\dd \Brwn_t)_{ab}+\langle G_{1,t}\reg{A}_{1,t}G_{2,t}-M_t(z_1, \reg{A}_{1,t}, z_2)\rangle (1 + \langle M_t(z_1, I, z_2)\rangle) \dd t \\
		& \quad +\big(\langle M_t(z_1, \reg{A}_{1,t}, z_2)\rangle - \tfrac{\dd}{\dd t} \Upsilon_{t}(A_1)\big) \langle G_{1,t}G_{2,t}-M_t(z_1, I, z_2)\rangle \dd t \\
		&\quad+\langle G_{1,t}\reg{A}_{1,t}G_{2,t}-M_t(z_1, \reg{A}_{1,t}, z_2)\rangle \langle G_{1,t}G_{2,t}- M_t(z_1, I, z_2)\rangle \dd t  \\
		&\quad+\langle G_{1,t}-M_{1,t}\rangle \langle G_{1,t}^2\reg{A}_{1,t}G_{2,t}\rangle \dd t+\langle G_{2,t}-M_{2,t}\rangle \langle G_{1,t}\reg{A}_{1,t}G_{2,t}^2\rangle \dd t
	\end{split}
\end{equation}
where we used the expression \eqref{eq:dM2} for the time derivatives of the $M$-terms.

We now start with the estimate of the terms in the right-hand side of \eqref{eq:GAG}. In the remainder of the section we always assume $t \in [\tin, \tfin \wedge \tau]$, i.e., in particular, neglect the $\wedge \tau$ in the writing henceforth. We begin with estimating the quadratic variation of the stochastic term in \eqref{eq:GAG}, whose integrand admits the bound
\begin{equation*}
	\begin{split}
		&\frac{1}{N^2} \left[ \langle G_{1,s}\reg{A}_{1,s}G_{2,s} G_{1,s}G_{1,s}^*G_{2,s}^*\reg{A}_{1,s}^* G_{1,s}^*\rangle| + |\langle G_{2,s}^*\reg{A}_{1,s}^* G_{1,s}^*G_{2,s}^* G_{2,s} G_{1,s}\reg{A}_{1,s}G_{2,s} \rangle| \right] \dd s \\
		\le & \frac{1}{N^2\eta_{1,s}^2}|\langle \Im G_{1,s}\reg{A}_{1,s}G_{2,s}\Im G_{1,s}G_{2,s}^*\reg{A}_{1,s}^*\rangle| \dd s+  \frac{1}{N^2\eta_{2,s}^2} |\langle \Im G_{2,s}\reg{A}_{1,s}^*G_{1,s}^*\Im G_{2,s}G_{1,s}\reg{A}_{1,s}\rangle | \dd s \\
 \lesssim& \frac{1}{N^2\eta_{1,s}\eta_{2,s}}\left(\frac{1}{\eta_{1,s}^2}+\frac{1}{\eta_{2,s}^2}\right) |\langle \Im G_{1,s}\reg{A}_{1,s}\Im G_{2,s}\reg{A}_{1,s}^*\rangle| \dd s \,, 
	\end{split}
\end{equation*}
where in the first step, we employed a Ward identity, $G_{i,s}G_{i,s}^* = \Im G_{i,s}/\Im z_{i,s}$. In the second step, we used another Ward identity and used a norm bound $\Vert \Im G_{i,s} \Vert \le |\Im z_{i,s}|^{-1}$. By the BDG inequality (see Appendix B.6, Eq.~(18) in \cite{martingale}) and application of \eqref{eq:intrule1} and \eqref{eq:tripmonotone}, we thus have, with very high probability,
\begin{equation} \label{eq:BDG}
	\begin{split}
		&\sup_{u \in [\tin, t]}  \left|\sum_{a,b=1}^N\int_{\tin}^{u}\partial_{ab} \langle  G_{1,s}\reg{A}_{1,s}  G_{2,s} \rangle\frac{(\dd \Brwn_s)_{ab}}{\sqrt{N}}\right| \\[2mm]
		\le \, &N^\xi \left(\int_{\tin}^{t} \frac{1}{N^2\eta_{1,s}\eta_{2,u}}\left(\frac{1}{\eta_{1,s}^2}+\frac{1}{\eta_{2,s}^2}\right) | \langle \Im G_{1,s}\reg{A}_{1,s}\Im G_{2,s}\reg{A}_{1,s}^*\rangle| \,\dd s\right)^{1/2} \\
		\lesssim \, & N^\xi \left(\int_{\tin}^{t} \frac{|\rho_{1,s} \rho_{2,s}|}{N^2\eta_{1,s}\eta_{2,s}}\left(\frac{1}{\eta_{1,s}^2}+\frac{1}{\eta_{2,s}^2}\right) \vertiii{{A}_{1}}_{s, z_1, z_2}^2 \,\dd s\right)^{1/2} 
		\lesssim \, 	 N^\xi \sqrt{\frac{|\rho_{1,t}\rho_{2,t}|}{N^2\eta_{1,t}\eta_{2,t}\ell_t}} \vertiii{{A}_{1}}_{t, z_1, z_2} \,, 
	\end{split}
\end{equation}
where we used the definition of the stopping time $\tau_6$ in \eqref{eq:stop6} and \eqref{eq:imMt_bound}. 

Similarly, using the single resolvent local law from \eqref{eq:generalLL}, we estimate the first term in the last line of \eqref{eq:GAG} (the second term can be treated analogously)
\begin{equation} \label{eq:lastline}
	\begin{split}
		\left|\int_{\tin}^t \hspace{-2mm} \dd s \, \langle G_{1,s}-M_{1,s}\rangle \langle G_{1,s}^2\reg{A}_{1,s}G_{2,s}\rangle \right|&\le \int_{\tin}^t \frac{N^\xi}{N\eta_{1,s}^2\eta_{2,s}^{1/2}}|\langle \Im G_{1,s}\rangle|^{1/2}  |\langle \Im G_{1,s}\reg{A}_{1,s}\Im G_{2,s}\reg{A}_{1,s}^*\rangle|^{1/2}\,\dd s \\
		&\lesssim \int_{\tin}^t \hspace{-2mm}\dd s \, \frac{ N^{\xi} |\rho_{1,s}| \, |\rho_{2,s}|^{1/2} \vertiii{{A}_{1}}_{s,z_1, z_2}}{N\eta_{1,s}^2\eta_{2,s}^{1/2}}  \lesssim  N^{\xi} \sqrt{\frac{\rho_{1,t}\rho_{2,t}}{N^2\eta_{1,t}\eta_{2,t}\ell_t}} \vertiii{A_1}_{t,z_1, z_2}\,, 
	\end{split}
\end{equation}
with very high probability. Here, we additionally used a Schwarz inequality and a single resolvent bound $|\langle \Im G_{1,s} \rangle| \lesssim |\rho_{1,s}|$ as a consequence of the single resolvent law from \eqref{eq:generalLL} and $N \ell_t \gtrsim 1$. 

Next, we use Lemma \ref{lem:Upsderv} to control the deterministic term in the  third line of \eqref{eq:GAG}, to find, recalling \eqref{eq:Upserror}, that
\begin{equation} \label{eq:GAGUps}
	\big|\langle M_t(z_1, \reg{A}_{1,t}, z_2)\rangle - 
	\tfrac{\dd}{\dd t}{\Upsilon}_{t}(A_1)\big| \lesssim  \mathcal{E}_t \vertiii{A_1}_{t, z_1, z_2} \,. 
\end{equation}
Moreover, we have the following bound on the stochastic term in the third line of \eqref{eq:GAG}: 
\begin{equation} \label{eq:GGz-z}
	\begin{split}
		\big|\langle G_{1,t}G_{2,t}-M_t(z_1, I, z_2)\rangle\big|&\le\frac{N^{\xi/2}}{N(\min_{i \in \indset{2}} \varkappa_{i,t})} \frac{1}{|z_{1,t} - z_{2,t}| + \varkappa_{1,t} + \varkappa_{2,t}}  \\
		&\lesssim \frac{N^{\xi/2}}{N }  \left(\frac{1}{|\rho_{1,t}| \nu_{1,t}}  + 	\frac{1}{|\rho_{2,t}| \nu_{2,t}}\right)\frac{\other{\alpha}_t}{\other{\beta}_t \other{\beta}_{*,t}} \le  \frac{N^{\xi/2}}{N\eta_{1,t}\eta_{2,t}} \,, 
	\end{split}
\end{equation}
which follows by Cauchy's integral formula and the single resolvent local law from \cite[Theorem~2.8]{cuspuniv}.\footnote{\label{ftn:cauchy}The contour in Cauchy's integral formulas is taken to be the counterclockwise oriented boundary of the union of the two disks $\{ w \in \C : |w - z_{i,t}| \le \varkappa_{i,t}/2\}$ for $i \in \indset{2}$.}  (Alternatively this can be proven using the characteristic flow as well.) 

We now multiply the two estimates \eqref{eq:GAGUps} and \eqref{eq:GGz-z} and control the resulting terms separately. The contribution from the first term on the right-hand side~of \eqref{eq:GAGUps} (i.e.~the first term on the right-hand side~of \eqref{eq:Upserror}) can be bounded as
\begin{equation} \label{eq:mixed1} 
	\begin{split}
		\int_{\tin}^t \hspace{-2mm} \dd s \, \frac{|\rho_{1,s}| + |\rho_{2,s}|}{\other{\alpha}_s} \vertiii{A_1}_{s, z_1, z_2}\frac{N^{\xi/2}}{N\eta_{1,s}\eta_{2,s}}
		&\lesssim \, N^{\xi} \frac{1}{N \other{\alpha}_t}  \left(\frac{1}{\eta_{1,t}} + \frac{1}{\eta_{2,t}}\right)\vertiii{A_1}_{t,z_1, z_2} \\
		&\lesssim \,  N^{\xi} \sqrt{\frac{|\rho_{1,t}\rho_{2,t}|}{N^2\eta_{1,t}\eta_{2,t}\ell_t}} \vertiii{A_1}_{t,z_1, z_2},
	\end{split}
\end{equation}
where we used the integration rules \eqref{eq:intrule1} together with \eqref{eq:tripmonotone} and the fact that $\other{\alpha}_s \gtrsim \other{\alpha}_t$ for $s \le t$ as well as $\other{\alpha}_t \ge \max_{i \in \indset{2}} \big(\eta_{i,t}/|\rho_{i,t}|\big)^{1/2}$ (following from \eqref{eq:timerelations}).  Note that we could even afford using the (generally crude) second estimate in the second line of \eqref{eq:GGz-z}. 

Next, we estimate the contribution coming from the second term in the definition of $\mathcal{E}_{t}$ in \eqref{eq:Upserror}. Ignoring an overall factor $N^{\xi/2}/N$, we have
\begin{equation} \label{eq:Upsest2}
	\begin{split}
		&	\int_{\tin}^t \dd s \, \left(\frac{1}{|\rho_{1,s}| \nu_{1,s}}  + 	\frac{1}{|\rho_{2,s}| \nu_{2,s}}\right) \frac{1}{(\nu_{1,s} + \nu_{2,s})^{1/2} \other{\nu}_s}
		\lesssim 	\frac{|\rho_{1,t}|  + |\rho_{2,t}|}{(\nu_{1,t} + \nu_{2,t})^{1/2} \other{\nu}_t |\rho_{1,t} \rho_{2,t}|}	 \lesssim  \sqrt{\frac{|\rho_{1,t} \rho_{2,t}|}{\ell_t \eta_{1,t} \eta_{2,t}}},
	\end{split}
\end{equation}
where we dropped the observable norm, which is simply estimated by monotonicity \eqref{eq:tripmonotone}. Moreover, we used Lemma~\ref{lem:intestgeneral} together with \eqref{eq:timerelations} and also the fact that $|\rho_{i,s}| \sim |\rho_{i,t}|$.
Collecting the estimates \eqref{eq:mixed1} and \eqref{eq:Upsest2}, we thus have that 
\begin{equation} \label{eq:Upsterms}
	\int_{\tin}^t  \text{third line of \eqref{eq:GAG}} \,  \dd s \lesssim \,  N^{\xi} \sqrt{\frac{|\rho_{1,t}\rho_{2,t}|}{N^2\eta_{1,t}\eta_{2,t}\ell_t}} \vertiii{A_1}_{t,z_1, z_2} \,. 
\end{equation}

Next, by means of \eqref{eq:intrule1}--\eqref{eq:tripmonotone} again, we estimate the fourth line of \eqref{eq:GAG} by
\begin{equation} \label{eq:timesG-M}
	\int_{\tin}^t N^{2\xi}\sqrt{\frac{|\rho_{1,s}\rho_{2,s}|}{N^2 \ell_s \eta_{1,s}\eta_{2,s}}} \frac{N^\xi}{N\eta_{1,s}\eta_{2,s}}\,\vertiii{A_1}_{s,z_1, z_2}\dd s\lesssim \frac{N^{2\xi}}{N\ell_t}\cdot N^{2\xi}\sqrt{\frac{|\rho_{1,t}\rho_{2,t}|}{N^2 \ell_t \eta_{1,t}\eta_{2,t}} }\vertiii{A_1}_{t,z_1, z_2} \,, 
\end{equation}
additionally using that $\ell_t \lesssim \ell_s$ for $t \ge s$. 

Finally, we handle the second term in the second line of \eqref{eq:GAG}. To do so, we abbreviate $X_t:=\langle G_{1,t}A_1G_{2,t}-M_{t}(z_1, A_1, z_2)\rangle$ and find, by application of a \emph{stochastic} version of the standard \emph{Gronwall argument} (see \cite[Lemma 5.6]{cipolloni2023universality}) together with the previously established bounds \eqref{eq:BDG}, \eqref{eq:lastline}, \eqref{eq:Upsterms}, and \eqref{eq:timesG-M} that, uniformly for $t \in [\tin, \tfin \wedge \tau]$, 
\begin{equation} \label{eq:GAGfinal}
	\begin{split}
		|X_t|^2 &\lesssim  N^{3\xi} {\frac{|\rho_{1,t}\rho_{2,t}|}{N^2\eta_{1,t}\eta_{2,t}\ell_{t}}} \vertiii{A_1}_{t,z_1, z_2}^2  
		+ \int_{\tin}^t \, \dd s N^{3\xi} {\frac{|\rho_{1,s}\rho_{2,s}|}{N^2\eta_{1,s}\eta_{2,s}\ell_{s}}} \vertiii{A_1}_{s,z_1, z_2}^2 f_s \Phi_{s,t}^2 \\
		&\lesssim N^{7\xi/2} {\frac{|\rho_{1,t}\rho_{2,t}|}{N^2\eta_{1,t}\eta_{2,t}\ell_{t}}} \vertiii{A_1}_{t,z_1, z_2}^2 \,, 
	\end{split}
\end{equation}
recalling the $\Phi$-notation from Lemma \ref{lem:prop}. Here, in order to control the initial condition, we additionally used that, by monotonicity of $\ell_t$ and $\eta_{i,t}$ in time,
\begin{equation*}
{ \frac{|\rho_{1,s}\rho_{2,s}|}{N^2 \ell_{s} \eta_{1,s}\eta_{2,s}} } \vertiii{A_1}_{s,z_1, z_2}^2 \lesssim 	{ \frac{|\rho_{1,t}\rho_{2,t}|}{N^2 \ell_t \eta_{1,t}\eta_{2,t}} } \vertiii{A_1}_{t,z_1, z_2}^2  \quad \text{for any} \quad s \le t \,. 
\end{equation*}
Moreover, to go to the second line, we used that, by Lemma \ref{lem:prop}, 
\begin{equation}
	f_s \lesssim \sqrt{\frac{|\rho_{1,s} \rho_{2,s}|}{\eta_{1,s} \eta_{2,s}}}\qquad \text{and} \qquad \Phi_{s,t} \lesssim \sqrt{\frac{\eta_{1,s} \eta_{2,s}}{\eta_{1,t} \eta_{2,t}}},
\end{equation}
and employed the integral estimate from Lemma \ref{lem:intestgeneral}. 

To conclude, for any time $t \in [\tin, \tfin \wedge \tau]$ the bound \eqref{eq:GAGfinal} on $|X_t|$ is \emph{strictly better} than prescribed by the stopping times \eqref{eq:stop1}--\eqref{eq:stop2}.

\subsection{Estimates of $\langle \Im G \Im G\rangle$ for $\tau_3$} \label{subsec:tau3}
Similarly to \eqref{eq:GAG}, we have 
\begin{equation}
	\begin{split}
		\label{eq:imGimG}
		\dd \big\langle \Im G_{1,t} &\Im G_{2,t} - \widehat{M}_t(z_1, I , z_2)\big\rangle \\
		=&\sum_{a,b=1}^N\partial_{ab} \langle \Im G_{1,t}  \Im G_{2,t}  \rangle\frac{(\dd \Brwn_t)_{ab}}{\sqrt{N}}+\big\langle \Im G_{1,t}  \Im G_{2,t} - \widehat{M}_t(z_1, I , z_2)\big\rangle \dd t \\
		&\quad+\left(\frac{\langle \Im G_{1,t} -\Im M_{1,t}\rangle}{\Im z_{1,t}}+\frac{\langle \Im G_{2,t} -\Im M_{2,t}\rangle}{\Im z_{2,t}}\right)\langle \Im G_{1,t}  \Im G_{2,t} \rangle \dd t \\
		&\quad -\frac{1}{4}\left[\big( \langle G_{1,t} G_{2,t}\rangle^2 - \langle M_t(z_1, I , z_2) \rangle^2 \big) + \big( \langle G_{1,t}^* G_{2,t}^*\rangle^2 - \langle M_t(\overline{z}_1, I , \overline{z}_2) \rangle^2 \big) \right] \dd t\\
		&\quad +\frac{1}{4}\left[\big( \langle G_{1,t}^* G_{2,t}\rangle^2  - \langle M_t(\overline{z_1}, I , z_2) \rangle^2 \big) + \big( \langle G_{1,t} G_{2,t}^*\rangle^2  - \langle M_t({z_1}, I , \overline{z}_2) \rangle^2 \big)\right] \dd t\\
		&\quad+\langle G_{1,t}-M_{1,t}\rangle \langle \Im G_{1,t}  \Im G_{2,t} G_{1,t}\rangle\dd t +\langle G_{2,t}-M_{2,t}\rangle \langle \Im G_{2,t} \Im G_{1,t}  G_{2,t}\rangle \dd t\\
		&\quad+\langle G_{1,t}^*-M_{1,t}^*\rangle \langle \Im G_{1,t}  \Im G_{2,t}  G_{1,t}^*\rangle \dd t+\langle G_{2,t}^*-M_{2,t}^*\rangle \langle \Im G_{2,t} \Im G_{1,t}  G_{2,t}^*\rangle \dd t.
	\end{split}
\end{equation}

We again start with the estimate of the martingale term on the right-hand side~of \eqref{eq:imGimG}, whose (integrand of the) quadratic variation, by application of two Ward identities, admits the bound
\begin{equation} \label{eq:imgimgQV}
	\begin{split}
		\left(\frac{1}{N^2 \eta_{1,s}^2} + \frac{1}{N^2 \eta_{2,s}^2}\right) \langle |\Im G_{1,s} \Im G_{2,s}|^2\rangle \dd s\lesssim \left(\frac{|\rho_{1,s} \rho_{2,s}|}{N^2 \eta_{1,s}^2} + \frac{|\rho_{1,s} \rho_{2,s}|}{N^2 \eta_{2,s}^2}\right) \Vert \Im G_{1,s} \Im G_{2,s}\Vert  \gamma_{s} \dd s \\
		\lesssim \frac{N^{\xi/2}}{N^2}\left(\frac{|\rho_{2,s}|}{ |\rho_{1,s}|\nu_{1,s}^2} + \frac{|\rho_{1,s}| }{|\rho_{2,s}| \nu_{2,s}^2}\right)  \, (\nu_{1,s} + \nu_{2,s}) \, \left(\frac{|\rho_{2,s}| \nu_{2,s}}{|\rho_{1,s}| \nu_{1,s}} + \frac{|\rho_{1,s}| \nu_{1,s}}{|\rho_{2,s}| \nu_{2,s}}\right) \left(\frac{ \other{\alpha}_s^2}{\other{\beta}_s^2 \other{\beta}_{*,s}^2}\right)^2 \dd s
	\end{split}
\end{equation}
where we used that, by definition of the stopping time $\tau_3$, Theorem~\ref{prop:M2_bounds}, 
and \eqref{eq:gamma}, we have 
\begin{equation} \label{eq:imgimginput}
	|\langle \Im G_{1,s} \Im G_{2,s}\rangle| \lesssim |\rho_{1,s} \rho_{2,s}| \,  \gamma_{s} \sim  |\rho_{1,s} \rho_{2,s}| \frac{(\nu_{1,s} + \nu_{2,s}) \other{\alpha}_s^2}{\other{\beta}_s^2 \other{\beta}_{*,s}^2}\,. 
\end{equation}
Additionally, we employed the norm bound 
\begin{equation} \label{eq:imgimgnorm}
	\Vert \Im G_{1,s} \Im G_{2,s}\Vert \lesssim N^{\xi/2}\frac{\max_{i \in \indset{2}} \eta_{i,s}}{\min_{i \in \indset{2}} \eta_{i,s}} \frac{1}{|z_{1,s} - z_{2,s}|^2 + \varkappa_{1,s}^2 + \varkappa_{2,s}^2} \lesssim N^{\xi/2} \left(\frac{|\rho_{2,s}| \nu_{2,s}}{|\rho_{1,s}| \nu_{1,s}} + \frac{|\rho_{1,s}| \nu_{1,s}}{|\rho_{2,s}| \nu_{2,s}}\right) \frac{ \other{\alpha}_s^2}{\other{\beta}_s^2 \other{\beta}_{*,s}^2}
\end{equation}
that simply follows by spectral decomposition and the well known \emph{eigenvalue rigidity}; see \eqref{eq:rigidity}. 
Hence, using the BDG inequality, we have, with very high probability, 
\begin{equation} \label{eq:martimgimg}
	\begin{split}
		&\sup_{u \in [\tin, t]} \left| \sum_{a,b = 1}^N \int_{\tin}^u  \partial_{ab} \langle \Im G_{1,s} \Im G_{2,s} \rangle \frac{\big(\dd \Brwn_s\big)_{ab}}{\sqrt{N}}\right| \\
		\le &\frac{N^\xi}{N} \left( \int_{\tin}^t \dd s \,  \left(\frac{|\rho_{2,s}|^2 \nu_{2,s}}{ |\rho_{1,s}|^2\nu_{1,s}^3} + \frac{|\rho_{1,s}|^2 \nu_{1,s} }{|\rho_{2,s}|^2 \nu_{2,s}^3} + \frac{1}{\nu_{1,s} \nu_{2,s}}\right) (\nu_{1,s} + \nu_{2,s}) \left(\frac{ \other{\alpha}_s^2}{\other{\beta}_s^2 \other{\beta}_{*,s}^2}\right)^2\right)^{1/2}  \\
		\lesssim &\frac{N^{\xi}}{N}  \left(1 + \frac{|\rho_{2,t}| }{ |\rho_{1,t}|\nu_{1,t}} + \frac{|\rho_{1,t}|  }{|\rho_{2,t}| \nu_{2,t}}\right) \left(\frac{ \other{\alpha}_t^2}{\other{\beta}_t^2 \other{\beta}_{*,t}^2}\right) \lesssim N^{\xi} \frac{|\rho_{1,t} \rho_{2,t}|}{N \ell_t} \gamma_{t}
	\end{split}
\end{equation}
where we used Lemma \ref{lem:intestgeneral} together with \eqref{eq:timerelations} and \eqref{eq:gamma} together with 
\begin{equation} \label{eq:elltrick}
	\frac{1}{\nu_{1,t} + \nu_{2,t}}\lesssim  \frac{(\min_{i \in \indset{2}}|\rho_{i,t}|)^2}{\ell_t}  \lesssim  \frac{|\rho_{1,t}\rho_{2,t}|}{\ell_t} \,. 
\end{equation}

Next, we estimate the last two lines on the right-hand side~of \eqref{eq:imGimG}, which, with the aid of the single resolvent law \eqref{eq:generalLL}, can be bounded as
\begin{equation} \label{eq:lasttwo}
	\begin{split}
		\int_{\tin}^t  \, \big| \text{last two lines of \eqref{eq:imGimG}} \big| \, \dd s \lesssim \int_{\tin}^t \frac{N^\xi}{N (\min_{i} \eta_{i,s})^2} |\langle \Im G_{1,s} \Im G_{2,s}\rangle| \dd s \\
		\lesssim \int_{\tin}^{t} \dd s \, \frac{N^\xi}{N}\left(\frac{1}{ |\rho_{1,s}|^2\nu_{1,s}^2} + \frac{1}{|\rho_{2,s}|^2\nu_{2,s}^2}\right) \big(\nu_{1,s} + \nu_{2,s}\big)\left(\frac{ \other{\alpha}_s^2}{\other{\beta}_s^2 \other{\beta}_{*,s}^2}\right)\lesssim N^{\xi} \frac{|\rho_{1,t} \rho_{2,t}|}{N \ell_t} \gamma_{t}
	\end{split}
\end{equation}
similarly to \eqref{eq:martimgimg}, additionally using \eqref{eq:imgimginput}. The first step in  \eqref{eq:lasttwo} follows from the trace inequality $|\langle X Y \rangle| \le \langle X \rangle \norm{Y}$ for positive $X\in \C^{N \times N}$ and general $Y \in \C^{N \times N}$, together with a resolvent norm bound on $|\langle \Im G_{1,s} \Im G_{2,s} G_{i,s}^{(*)} \rangle| \lesssim \eta_{i,s}^{-1}|\langle \Im G_{1,s} \Im G_{2,s}\rangle|$. 

The third line of \eqref{eq:imGimG} admits the exact same bound:
\begin{equation}
	\int_{\tin}^t  \, \big| \text{third line of \eqref{eq:imGimG}} \big| \, \dd s \lesssim N^{5\xi/4} \frac{|\rho_{1,t} \rho_{2,t}|}{N \ell_t} \gamma_{t} \,. 
\end{equation}

We finally turn to estimating the fourth and fifth line of \eqref{eq:imGimG}, which we rewrite using the algebraic identity
\begin{equation} \label{eq:algebraicimgimg}
	\begin{split}
		&- \langle G_{1,t}G_{2,t}\rangle^2 - \langle G_{1,t}^*G^*_{2,t}\rangle^2 + \langle G_{1,t}G^*_{2,t}\rangle^2   + \langle G_{1,t}^*G_{2,t}\rangle^2  \\
		= \ &4 \big(  \langle \Im G_{1,t}\Im G_{2,t}\rangle\langle G^*_{2,t}G_{1,t}\rangle +  \langle G_{1,t}^*G_{2,t}\rangle\langle \Im G_{2,t}\Im G_{1,t}\rangle \\
		&+  \langle \Im G_{1,t}G_{2,t}\rangle\langle \Im G_{2,t}G_{1,t}\rangle  +  \langle G_{1,t}^*\Im G_{2,t}\rangle\langle G_{2,t}^*\Im G_{1,t}\rangle     \big)
	\end{split}
\end{equation}
which holds analogously on the level of the $M$-terms. 

We start estimating the terms in the third line of \eqref{eq:algebraicimgimg} (i.e.~those where the two $\Im$'s occur in different traces), and their corresponding $M$-terms, for which we add and subtract the corresponding $G \times M$-type terms schematically as
\begin{equation} \label{eq:addsubtract}
G G - M M = M (G -M) + (G - M) M + (G-M) (G-M)
\end{equation}
 and control the resulting contributions as follows: First, all $(G-M) \times (G-M)$-type terms can be bounded as (note that we do not use the presence of $\Im$'s, but the simple bound \eqref{eq:GGz-z})
\begin{equation} \label{eq:G-MG-M}
	\int_{\tin}^t \dd s \, \frac{N^{\xi}}{N^2 }  \left(\frac{1}{|\rho_{1,s}| \nu_{1,s}}  + 	\frac{1}{|\rho_{2,s}| \nu_{2,s}}\right)^2 \left(\frac{\other{\alpha}_s}{\other{\beta}_s \other{\beta}_{*,s}}\right)^2 \lesssim \frac{N^{ \xi}}{N \ell_t} \frac{|\rho_{1,t} \rho_{2,t}|}{N \ell_t} \gamma_{t},
\end{equation}
where, similarly to \eqref{eq:martimgimg}--\eqref{eq:lasttwo}, we again used Lemma \ref{lem:intestgeneral} together with \eqref{eq:timerelations}. 

Next, the $M \times (G-M)$-type terms admit the bound 
\begin{equation} \label{eq:rho/ell}
	\begin{split}
		\int_{\tin}^t \dd s \, N^\xi \frac{|\rho_{1,s}| + |\rho_{2,s}|}{N } &\left(\frac{1}{|\rho_{1,s}| \nu_{1,s}} + \frac{1}{|\rho_{2,s}| \nu_{2,s}}\right) \left(\frac{\other{\alpha}_s}{\other{\beta}_s \other{\beta}_{*,s}}\right)^2 \\
		\lesssim &\frac{N^{ \xi} }{N} \left( \frac{|\rho_{1,t}|}{|\rho_{2,t}|} + \frac{|\rho_{2,t}|}{|\rho_{1,t}|} \right) \frac{1}{\nu_{1,t} + \nu_{2,t}} \gamma_t \lesssim N^{ \xi} \frac{|\rho_{1,t}||\rho_{2,t}|}{N \ell_t} \gamma_{t} \,, 
	\end{split}
\end{equation}
where we used Claim~\ref{claim:ImGG_Mbound} and \eqref{eq:alp_t_def} to bound 
\begin{equation} \label{eq:ReGImGbound}
	\begin{split}
		|\langle M_s(z_1, I, z_2)  -  &M_s(\overline{z}_1, I, z_2) \rangle| + |\langle M_s(z_1, I, z_2) -  M_s(z_1, I, \overline{z}_2) \rangle| \\
		&\lesssim \frac{|\rho_{1,s}| + |\rho_{2,s}|}{|z_{1,s} - z_{2,s}| + \varkappa_{1,s} + \varkappa_{2,s}} \sim  (|\rho_{1,s}| + |\rho_{2,s}|)  \frac{\other{\alpha}_s}{\other{\beta}_s \other{\beta}_{*,s}} \,. 
	\end{split}
\end{equation}

We are thus left with the terms from the second line of \eqref{eq:algebraicimgimg}. Again, we add and subtract the corresponding $G \times M$-type terms. The resulting $(G-M)\times (G-M)$-contribution can be handled exactly as in \eqref{eq:G-MG-M}; in fact, the presence of two $\Im$'s in one trace is not needed in estimating them, the bound \eqref{eq:GGz-z} is sufficient. Then,  the $M \times (G-M)$-type terms which have both $\Im$'s in the $M$-term can be bounded as
\begin{equation}
	\int_{\tin}^t \dd s \, N^\xi\frac{1}{N \nu_{1,s} \nu_{2,s}} (\nu_{1,s} + \nu_{2,s}) \left(\frac{\other{\alpha}_s}{\other{\beta}_s \other{\beta}_{*,s}}\right)^2 \lesssim N^{\xi} \frac{|\rho_{1,t} \rho_{2,t}|}{N \ell_t} \gamma_{t} \,. 
\end{equation}

The final part we thus need to handle is the term 
\begin{equation}
	\begin{split}
		&\big(1 + \langle M_s(z_1, I, \overline{z}_2) \rangle + \langle M_s(\overline{z}_1, I, {z}_2) \rangle\big) \big(\langle \Im G_{1,s} \Im G_{2,s} \rangle - \langle \widehat{M}_s(z_1, I, z_2) \rangle\big) \,, 
	\end{split}
\end{equation}
for which we combined the second term in the second line of \eqref{eq:imGimG} with the leftover contribution from \eqref{eq:algebraicimgimg}. To do so, we abbreviate $X_t := \langle \Im G_{1,t} \Im G_{2,t} \rangle - \langle \widehat{M}_t(z_1, I, z_2) \rangle$ and argue  similarly to \eqref{eq:GAGfinal}: As a consequence of the stochastic Gronwall argument \cite[Lemma 5.6]{cipolloni2023universality} and collecting the above bounds, we have (recall the definition of $\Phi$ in Lemma \ref{lem:prop})
\begin{equation} \label{eq:GGfinal}
	\begin{split}
		\hspace{-1mm}	|X_t|^2 \lesssim  N^{3\xi} {\frac{|\rho_{1,t}\rho_{2,t}|^2}{N^2\ell_{t}^2}} \gamma_t^2
		+ \int_{\tin}^t \hspace{-2mm} \dd s \frac{N^{3\xi}}{N^2} \left(\frac{|\rho_{1,s}|^2}{|\rho_{2,s}|^2 \nu_{2,s}^2} + \frac{|\rho_{2,s}|^2}{|\rho_{1,s}|^2 \nu_{1,s}^2}\right) \gamma_s^2
		f_s \Phi_{s,t}^4 \lesssim N^{7\xi/2} {\frac{|\rho_{1,t}\rho_{2,t}|^2}{N^2\ell_{t}^2}} \gamma_t^2 \,, 
	\end{split}
\end{equation}
uniformly in $t \in [\tin, \tfin \wedge \tau]$. Here, in order to control the initial condition, we additionally used  the monotonicity estimate
\begin{equation} \label{eq:monotonicitynontriv}
	\frac{|\rho_{1,s}||\rho_{2,s}|}{N \ell_{s}} \gamma_{s} \lesssim  \frac{|\rho_{1,t}||\rho_{2,t}|}{N \ell_t} \gamma_{t} \qquad \text{for any} \quad t \ge s \,, 
\end{equation}
which follows since $|\rho_{i,t}| \sim |\rho_{i,s}|$ for all times $s,t$ and $\ell_t^{-1} \gamma_t \gtrsim \ell_s^{-1} \gamma_s$ for all $s \le t$ (as a consequence of \eqref{eq:timerelations} and \eqref{eq:gamma}). Moreover, in the second step, we used \eqref{eq:gamma} and the bounds  
\begin{equation} \label{eq:Groninput}
	|f_s| \lesssim \frac{1}{\nu_{1,s} + \nu_{2,s}} \,, \qquad \Phi_{s,t} \lesssim  \frac{\other{\beta}_{*,s}}{\other{\beta}_{*,t}}  + \frac{\other{\beta}_{s}}{\other{\beta}_{t}} 
\end{equation}
from Lemma \ref{lem:prop}, yielding, with the aid of Lemma \ref{lem:intestgeneral} and \eqref{eq:timerelations}, 
\begin{equation}
	\begin{split}
		\int_{\tin}^{t} \dd s \left(\frac{|\rho_{1,s}|^2}{|\rho_{2,s}|^2 \nu_{2,s}^2} + \frac{|\rho_{2,s}|^2}{|\rho_{1,s}|^2 \nu_{1,s}^2}\right) \big(\nu_{1,s} + \nu_{2,s}\big) \left(\frac{\other{\alpha}_s}{\other{\beta}_s \other{\beta}_{*,s}}\right)^4 \left[\left( \frac{\other{\beta}_{*,s}}{\other{\beta}_{*,t}} \right)^4 + \left( \frac{\other{\beta}_{s}}{\other{\beta}_{t}} \right)^4\right] \\
		\lesssim \left(\frac{|\rho_{1,t}|^2}{|\rho_{2,t}|^2 \nu_{2,t}} + \frac{|\rho_{2,s}|^2}{|\rho_{1,s}|^2 \nu_{1,t}}\right) \big(\nu_{1,t} + \nu_{2,t}\big) \left(\frac{\other{\alpha}_t}{\other{\beta}_t \other{\beta}_{*,t}}\right)^4 \lesssim {\frac{|\rho_{1,t}\rho_{2,t}|^2}{\ell_{t}^2}} \gamma_t^2 \,. 
	\end{split}
\end{equation}

Hence, to summarize, for any time $t \in [\tin, \tfin \wedge \tau]$ the bound \eqref{eq:GGfinal} on $|X_t|$ is \emph{strictly better} than prescribed by the stopping time \eqref{eq:stop3}. 

\subsection{Estimates of $\langle \Im GA \Im G\rangle$ for $\tau_4$ and $\tau_5$} \label{subsec:tau45}  As mentioned below \eqref{eq:stop1}--\eqref{eq:stop6}, we only discuss the case $\langle \Im G_{1,t} \reg{A}_{1,t} \Im G_{2,t} \rangle$, the other three cases covered by $\tau_4$ and $\tau_5$ are analogous and thus omitted. 
Then, similarly to \eqref{eq:GAG} and \eqref{eq:imGimG} we have
\begin{equation}
	\begin{split}
		\label{eq:imGAimG}
		\dd \big\langle \Im G_{1,t}\reg{A}_{1,t} &\Im G_{2,t} - \widehat{M}_t(z_1, \reg{A}_{1,t} , z_2)\big\rangle \\
		=&\sum_{a,b=1}^N\partial_{ab} \langle \Im G_{1,t}\reg{A}_{1,t}  \Im G_{2,t}  \rangle\frac{(\dd \Brwn_t)_{ab}}{\sqrt{N}}+\big\langle \Im G_{1,t}\reg{A}_{1,t}  \Im G_{2,t} - \widehat{M}_t(z_1, \reg{A}_{1,t} , z_2)\big\rangle \dd t \\
		&\quad+\left(\frac{\langle \Im G_{1,t} -\Im M_{1,t}\rangle}{\Im z_{1,t}}+\frac{\langle \Im G_{2,t} -\Im M_{2,t}\rangle}{\Im z_{2,t}}\right)\langle \Im G_{1,t} \reg{A}_{1,t} \Im G_{2,t} \rangle \dd t \\
		&\quad -\frac{1}{4}\big( \langle G_{1,t}\reg{A}_{1,t} G_{2,t}\rangle\langle G_{2,t}G_{1,t}\rangle  - \langle M_t(z_1, \reg{A}_{1,t} , z_2) \rangle \langle M_t(z_2, I, z_1)\rangle\big) \dd t\\
		&\quad +\frac{1}{4}\big( \langle G_{1,t}^*\reg{A}_{1,t} G_{2,t}\rangle\langle G_{2,t}G_{1,t}^*\rangle  - \langle M_t(\overline{z_1}, \reg{A}_{1,t} , z_2) \rangle \langle M_t(z_2,I, \overline{z_1})\rangle\big) \dd t\\
		& \quad + \frac{1}{4}\big( \langle G_{1,t}\reg{A}_{1,t} G^*_{2,t}\rangle\langle G^*_{2,t}G_{1,t}\rangle  - \langle M_t(z_1, \reg{A}_{1,t} , \overline{z_2}) \rangle \langle M_t(\overline{z_2}, I, z_1)\rangle\big) \dd t\\
		&\quad -\frac{1}{4}\big( \langle G_{1,t}^*\reg{A}_{1,t} G^*_{2,t}\rangle\langle G^*_{2,t}G_{1,t}^*\rangle  - \langle M_t(\overline{z_1}, \reg{A}_{1,t} , \overline{z_2}) \rangle \langle M_t(\overline{z_2}, I, \overline{z_1})\rangle\big) \dd t\\
		& \quad  - \tfrac{\dd}{\dd t}{\Upsilon}_{t}(A_1) \big( \langle \Im G_{1,t} \Im G_{2,t} \rangle - \langle \widehat{M}_t(z_1, I, z_2)  \rangle  \big) \dd t \\
		&\quad+\langle G_{1,t}-M_{1,t}\rangle \langle \Im G_{1,t} \reg{A}_{1,t} \Im G_{2,t}  G_{1,t}\rangle\dd t +\langle G_{2,t}-M_{2,t}\rangle \langle \Im G_{2,t} \Im G_{1,t} \reg{A}_{1,t}  G_{2,t}\rangle \dd t\\
		&\quad+\langle G_{1,t}^*-M_{1,t}^*\rangle \langle \Im G_{1,t} \reg{A}_{1,t} \Im G_{2,t}  G_{1,t}^*\rangle \dd t+\langle G_{2,t}^*-M_{2,t}^*\rangle \langle \Im G_{2,t} \Im G_{1,t} \reg{A}_{1,t}  G_{2,t}^*\rangle \dd t.
	\end{split}
\end{equation}
The estimates of the various terms proceed in a very similar way to the previous two sections, which is why we will be rather brief. 

First, the estimates \eqref{eq:imgimgQV}--\eqref{eq:elltrick} are replaced as follows: The integrand of the quadratic variation of the martingale term admits the bound
\begin{equation} \label{eq:imgAimgQV}
	\begin{split}
		&\left(\frac{1}{N^2 \eta_{1,s}^2} + \frac{1}{N^2 \eta_{2,s}^2}\right) \langle \Im G_{1,s} \reg{A}_{1,s} \Im G_{2,s} \Im G_{1,s} \Im G_{2,s} \reg{A}_{1,s}^*\rangle \dd s \\
		\lesssim &\left(\frac{|\rho_{1,s} \rho_{2,s}|}{N^2 \eta_{1,s}^2} + \frac{|\rho_{1,s} \rho_{2,s}|}{N^2 \eta_{2,s}^2}\right) \Vert \Im G_{1,s} \Im G_{2,s}\Vert  \vertiii{A_1}_{s, z_1, z_2}^2 \dd s \\
		\lesssim &  \frac{1}{N^2} \left(\frac{|\rho_{2,s}|}{ |\rho_{1,s}|\nu_{1,s}^2} + \frac{|\rho_{1,s}| }{|\rho_{2,s}| \nu_{2,s}^2}\right)  \,   \left(\frac{|\rho_{2,s}| \nu_{2,s}}{|\rho_{1,s}| \nu_{1,s}} + \frac{|\rho_{1,s}| \nu_{1,s}}{|\rho_{2,s}| \nu_{2,s}}\right) \frac{ \other{\alpha}_s^2}{\other{\beta}_s^2 \other{\beta}_{*,s}^2}\vertiii{A_1}_{s, z_1, z_2}^2 \, \dd s
	\end{split}
\end{equation}
and we hence get, analogously to \eqref{eq:martimgimg}, 
\begin{equation*}
	\begin{split}
		&\sup_{u \in [\tin, t]} \left| \sum_{a,b = 1}^N \int_{\tin}^u  \partial_{ab} \langle \Im G_{1,s} \reg{A}_{1,s}\Im G_{2,s} \rangle \frac{\big(\dd \Brwn_s\big)_{ab}}{\sqrt{N}}\right| \\
		\lesssim& \frac{N^\xi}{N}\left( \int_{\tin}^{t} \dd s \, \left(\frac{|\rho_{2,s}|^2 \nu_{2,s}}{ |\rho_{1,s}|^2\nu_{1,s}^3} + \frac{|\rho_{1,s}|^2 \nu_{1,s} }{|\rho_{2,s}|^2 \nu_{2,s}^3} + \frac{1}{\nu_{1,s} \nu_{2,s}}\right) \frac{ \other{\alpha}_s^2}{\other{\beta}_s^2 \other{\beta}_{*,s}^2} \right)^{1/2} \hspace{-1mm}\vertiii{A_1}_{t, z_1, z_2}
		\lesssim N^{\xi} \frac{|\rho_{1,t} \rho_{2,t}|}{N \ell_t} \gamma_{t}^{1/2} \vertiii{A_1}_{t, z_1, z_2}\,, 
	\end{split}
\end{equation*}
additionally using \eqref{eq:tripmonotone}. 

To handle the third and the last two lines in \eqref{eq:imGAimG}, we estimate
\begin{equation} \label{eq:normbound}
	\begin{split}
		|\langle \Im G_{1,t} \reg{A}_{1,t} \Im G_{2,t} \rangle| &\lesssim |\rho_{1,t} \rho_{2,t}| \gamma_{t}^{1/2} \vertiii{A_1}_{t, z_1, z_2} \\
		|\langle \Im G_{1,t} \reg{A}_{1,t} \Im G_{2,t}G_{i,t}^{(*)} \rangle| &\lesssim \eta_{i,t}^{-1} |\rho_{1,t} \rho_{2,t}| \gamma_{t}^{1/2} \vertiii{A_1}_{t, z_1, z_2} \,. 
	\end{split}
\end{equation}
This follows by means of simple Schwarz inequalities and the trace inequality $|\langle X Y \rangle| \le \langle X \rangle \norm{Y}$ for positive $X\in \C^{N \times N}$ and general $Y \in \C^{N \times N}$, additionally using the definition of the stopping times \eqref{eq:stop3} and \eqref{eq:stop6} in combination with the corresponding $M$-bounds (see Proposition \ref{prop:Mt_bound} and Lemma \ref{lemma:m_hat}). Then, analogously to the bound \eqref{eq:lasttwo}, the estimate
\begin{equation} \label{eq:Diff12}
	\int_{\tin}^t \dd s \, \left(\frac{|\rho_{2,s}|}{|\rho_{1,s}|\nu_{1,s}^2} + \frac{|\rho_{1,s}|}{ |\rho_{2,s}|\nu_{2,s}^2}\right) \big(\nu_{1,s} + \nu_{2,s}\big)^{1/2} \frac{ \other{\alpha}_s}{\other{\beta}_s \other{\beta}_{*,s}} \lesssim \frac{|\rho_{1,t} \rho_{2,t}|}{\ell_t} \gamma_{t}^{1/2}
\end{equation}
allows us to conclude 
\begin{equation}
	\int_{\tin}^{t}  |\text{third, ninth and tenth line of \eqref{eq:imGAimG}}| \dd s \lesssim N^{\xi} \frac{|\rho_{1,t} \rho_{2,t}|}{N \ell_t} \gamma_{t}^{1/2} \vertiii{A_1}_{t, z_1, z_2} \,. 
\end{equation}

We are hence left with lines four, five, six, seven and eight in \eqref{eq:imGAimG}. For lines four to seven, we again employ an identity analogous to \eqref{eq:algebraicimgimg} in order to restore $\Im G$'s: 
\begin{equation} \label{eq:algebraicimgaimg}
	\begin{split}
		&- \langle G_{1,t}\reg{A}_{1,t}G_{2,t}\rangle\langle G_{2,t}G_{1,t}\rangle  
		+ \langle G_{1,t}^*\reg{A}_{1,t}G_{2,t}\rangle\langle G_{2,t}G_{1,t}^*\rangle \\
		&+ \langle G_{1,t}\reg{A}_{1,t}G^*_{2,t}\rangle\langle G^*_{2,t}G_{1,t}\rangle  - \langle G_{1,t}^*\reg{A}_{1,t}G^*_{2,t}\rangle\langle G^*_{2,t}G_{1,t}^*\rangle \\
		= \ &4 \big(  \langle \Im G_{1,t}\reg{A}_{1,t}\Im G_{2,t}\rangle\langle G^*_{2,t}G_{1,t}\rangle +  \langle G_{1,t}^*\reg{A}_{1,t}G_{2,t}\rangle\langle \Im G_{2,t}\Im G_{1,t}\rangle \\
		&+  \langle \Im G_{1,t}\reg{A}_{1,t}G_{2,t}\rangle\langle \Im G_{2,t}G_{1,t}\rangle  +  \langle G_{1,t}^*\reg{A}_{1,t}\Im G_{2,t}\rangle\langle G_{2,t}^*\Im G_{1,t}\rangle     \big)
	\end{split}
\end{equation}
which similarly holds on the $M$-level. As above, we start estimating the terms in the fourth line of \eqref{eq:algebraicimgaimg} (i.e.~those where the two $\Im$'s occur in different traces), for which we add and subtract the corresponding $G \times M$-type terms and control the resulting contributions as follows: The $(G-M) \times (G-M)$-type terms are bounded as (cf.~\eqref{eq:G-MG-M})
\begin{equation} \label{eq:toobad}
	\begin{split}
		&\int_{\tin}^{t} \dd s \, N^{5 \xi/2}\frac{1}{N^2  (\ell_s \nu_{1,s} \nu_{2,s})^{1/2} } \left(\frac{1}{|\rho_{1,s}| \nu_{1,s}}  + 	\frac{1}{|\rho_{2,s}| \nu_{2,s}}\right)\frac{\other{\alpha}_t}{\other{\beta}_t \other{\beta}_{*,t}} \vertiii{A_1}_{s, z_1, z_2} \\
		&\lesssim \frac{N^{5\xi/2}}{N \ell_t} \frac{|\rho_{1,t} \rho_{2,t}|}{N \ell_t} \gamma_{t}^{1/2} \vertiii{A_1}_{t, z_1, z_2}
	\end{split}
\end{equation}
using monotonicity of $\ell_t$, Lemma \ref{lem:intestgeneral}, \eqref{eq:elltrick} and \eqref{eq:GGz-z} as well as the definition of $\tau_1$, $\tau_2$ in \eqref{eq:stop1}--\eqref{eq:stop2}.  

Next, we estimate the $M \times (G-M)$-type terms resulting from the fourth line of \eqref{eq:algebraicimgaimg}. Here, by means of \eqref{eq:ReGImGbound}, \eqref{eq:stop1}--\eqref{eq:stop2}, and \eqref{eq:GImG_bounds} and \eqref{eq:GGz-z} together with 
\begin{equation}
	\frac{1}{(\min_i \eta_{i,s}) \other{\alpha}_s} \lesssim \sqrt{\frac{|\rho_{1,s} \rho_{2,s}|}{\ell_s \eta_{1,s} \eta_{2,s}}} \lesssim  \frac{1}{\sqrt{\ell_s\nu_{1,s} \nu_{2,s}}} 
\end{equation}
and additionally using \eqref{eq:elltrick}, we have the estimate (again employing monotonicity of $\ell_t$)
\begin{equation} \label{eq:toobad2}
	\int_{\tin}^{t} \dd s \, N^{2 \xi }\frac{|\rho_{1,s}| + |\rho_{2,s}|}{N  (\ell_s \nu_{1,s} \nu_{2,s})^{1/2}}   \frac{ \other{\alpha}_s}{\other{\beta}_s \other{\beta}_{*,s}} \vertiii{A_1}_{s, z_1, z_2} \lesssim N^{5\xi /2}\frac{|\rho_{1,t} \rho_{2,t}|}{N \ell_t} \gamma_{t}^{1/2} \vertiii{A_1}_{t, z_1, z_2} \,. 
\end{equation}

We are then left with the third line of \eqref{eq:algebraicimgaimg} (together with the respective $M$-terms). Again, adding and subtracting the corresponding $G \times M$-type terms, we first estimate the $(G-M) \times (G-M)$-contribution of the first term in the third line of \eqref{eq:algebraicimgaimg} as 
\begin{equation}
	\int_{\tin}^{t} \hspace{-2mm}\dd s \, N^{5\xi/2}\frac{1}{N^2 \ell_s \nu_{1,s} \nu_{2,s}} \big(\nu_{1,s} + \nu_{2,s}\big)^{1/2} \frac{ \other{\alpha}_s}{\other{\beta}_s \other{\beta}_{*,s}}  \vertiii{A_1}_{s, z_1, z_2} \lesssim \frac{N^{5\xi/2}}{N \ell_t} \frac{|\rho_{1,t} \rho_{2,t}|}{N \ell_t} \gamma_{t}^{1/2} \vertiii{A_1}_{t, z_1, z_2}
\end{equation}
with the aid of the second bound in \eqref{eq:GGz-z} and the stopping times \eqref{eq:stop4}--\eqref{eq:stop5}, additionally using monotonicity of $\ell_t$ and \eqref{eq:tripmonotone}. For the $(G-M) \times (G-M)$-contribution of the second term in the third line of \eqref{eq:algebraicimgaimg} we have 
\begin{equation}\label{eq:G-MG-MimgAimg}
	\begin{split}
		\int_{\tin}^{t} \hspace{-2mm}\dd s \, N^{5\xi/2}\frac{|\rho_{1,s} \rho_{2,s}|}{N^2  (\nu_{1,s} \nu_{2,s})^{1/2}} \left(\frac{1}{|\rho_{1,s}|^2 \nu_{1,s}} + \frac{1}{|\rho_{2,s}|^2 \nu_{2,s}}\right)^{3/2} \big(\nu_{1,s} + \nu_{2,s}\big) \frac{ \other{\alpha}_s^2}{\other{\beta}_s^2 \other{\beta}_{*,s}^2} \vertiii{A_1}_{s, z_1, z_2} \\
		\lesssim \frac{N^{5\xi/2}}{N \ell_t} \frac{|\rho_{1,t} \rho_{2,t}|}{N \ell_t^{1/2}} \gamma_{t} \vertiii{A_1}_{t, z_1, z_2} \lesssim \frac{N^{5\xi/2}}{N \ell_t} \frac{|\rho_{1,t} \rho_{2,t}|}{N \ell_t} \gamma_{t}^{1/2} \vertiii{A_1}_{t, z_1, z_2}
	\end{split}
\end{equation}
where we used 
\begin{equation} \label{eq:Diffest}
	\gamma_{s} \sim \big(\nu_{1,s} + \nu_{2,s}\big)  \frac{ \other{\alpha}_s^2}{\other{\beta}_s^2 \other{\beta}_{*,s}^2} \lesssim \frac{ \other{\alpha}_s^2}{\other{\beta}_s^2} \frac{1}{\other{\beta}_{*,s}} \lesssim \frac{1}{\max_{i \in \indset{2}} |\rho_{i,s}|^2} \frac{1}{\max_{i \in \indset{2}}\nu_{i,s}} \lesssim \frac{1}{\ell_s}
\end{equation}
for a square root of $\gamma_t$. 

Finally, we need to handle the $M \times (G-M)$-contribution of the third line of \eqref{eq:algebraicimgaimg}: For the first term, we note that $\langle \widehat{M}_t(z_1, \reg{A}_{1,t}, z_2) \rangle = 0$ by definition of $\reg{A}_{1,t}$ and thus only the term with $\langle M_t(\overline{z_2}, I, z_1) \rangle$ contributes.  For the second term, the part with $\langle \widehat{M}_t(z_1, I, z_2) \rangle \langle G_{1,t}^* \reg{A}_{1,t} G_{2,t} - M_t(\overline{z}_{1}, \reg{A}_{1,t}, z_2) \rangle$, we employ \eqref{eq:Diffest} a square root of $\gamma_s$ inside of the integral, to find this contribution to be bounded by
\begin{equation}\label{eq:G-MG-MimgAimg2}
	\begin{split}
		\int_{\tin}^{t} \dd s \, N^{2\xi} &\frac{|\rho_{1,s} \rho_{2,s}|}{N  (\nu_{1,s} \nu_{2,s})^{1/2}} \left(\frac{1}{|\rho_{1,s}|^2 \nu_{1,s}} + \frac{1}{|\rho_{2,s}|^2 \nu_{2,s}}\right) \big(\nu_{1,s} + \nu_{2,s}\big)^{1/2} \frac{ \other{\alpha}_s}{\other{\beta}_s \other{\beta}_{*,s}} \vertiii{A_1}_{s, z_1, z_2} \\
		&\lesssim N^{2\xi} \frac{|\rho_{1,t} \rho_{2,t}|}{N \ell_t} \gamma_{t}^{1/2} \vertiii{A_1}_{t, z_1, z_2},
	\end{split}
\end{equation}
analogously to \eqref{eq:G-MG-MimgAimg}. Therefore, it remains to control the terms 
\begin{equation} \label{eq:lasttwoterms}
	\langle M_t(z_1, I, \overline{z}_2) \rangle\langle \Im G_{1,t} \reg{A}_{1,t} \Im G_{2,t} - \widehat{M}_t(z_1, \reg{A}_{1,t}, z_2) \rangle \quad \text{and} \quad \langle M_t(z_1, \reg{A}_{1,t}, \overline{z}_2) \rangle\langle \Im G_{1,t}  \Im G_{2,t} - \widehat{M}_t(z_1, I, z_2) \rangle \,. 
\end{equation} 

For the second term in \eqref{eq:lasttwoterms} we involve the so far untouched eighth line of \eqref{eq:imGAimG} containing the factor $\frac{\dd}{\dd t}{\Upsilon}_{t}(A_1)$ and argue similarly to \eqref{eq:GAGUps}, \eqref{eq:mixed1}, and \eqref{eq:Upsest2}. In fact, armed with \eqref{eq:GAGUps}, we estimate the contribution analogous to \eqref{eq:mixed1} by (ignoring an overall $N^{2\xi}/N$)
\begin{equation} \label{eq:Ups1}
	\begin{split}
		&\int_{\tin}^{t} \dd s \, (|\rho_{1,s}| + |\rho_{2,s}|) \left(\frac{1}{|\rho_{1,s}|^2 \nu_{1,s}} + \frac{1}{|\rho_{2,s}|^2 \nu_{2,s}}\right) \big(\nu_{1,s} + \nu_{2,s}\big) \frac{ \other{\alpha}_s}{\other{\beta}_s^2 \other{\beta}_{*,s}^2} \\
		\lesssim 	&\int_{\tin}^{t} \dd s \, (|\rho_{1,s}| + |\rho_{2,s}|)  \left(\frac{1}{|\rho_{1,s}|^2 \nu_{1,s}} + \frac{1}{|\rho_{2,s}|^2 \nu_{2,s}}\right)  \frac{ \other{\alpha}_s}{\other{\beta}_s^2 \other{\beta}_{*,s}} \\
		\lesssim &\frac{\big(|\rho_{1,t}| + |\rho_{2,t}|\big)^3}{|\rho_{1,t} \rho_{2,t}|^2} \frac{1}{\other{\beta}_t (\nu_{1,t}+ \nu_{2,t})^{1/2}} \gamma_{t}^{1/2} \lesssim \frac{\gamma_{t}^{1/2}}{\ell_t},
	\end{split}
\end{equation}
where we additionally ignored the observable norm (that is again just handled by monotonicity). Moreover, we employed \eqref{eq:timerelations}, the first estimate in \eqref{eq:elltrick} and
\begin{equation} \label{eq:lowerbeta}
	\other{\beta}_t \gtrsim \max_{i \in \indset{2}} \nu_{i,t} + \max_{i \in \indset{2}} |\rho_{i,t}|^2 \,. 
\end{equation}
Next, we consider the analog of \eqref{eq:Upsest2} (again ignoring an overall $N^{2\xi}/N$). Assuming, for concreteness, that $\nu_{1,t} \le \nu_{2,t}$ (the complementary case is handled analogously) we find it to be bounded as
\begin{equation} \label{eq:Ups3}
\begin{split}
	&	\int_{\tin}^t  \dd s \,    \frac{\other{\alpha}_s(\nu_{1,s} + \nu_{2,s})^{1/2}}{ \other{\nu}_s  \other{\beta}_{*,s} \other{\beta}_s} \left(\frac{1}{|\rho_{1,s}|^2 \nu_{1,s}} + \frac{1}{|\rho_{2,s}|^2 \nu_{2,s}}\right) \\
	\lesssim & \int_{\tin}^t  \dd s \,    \frac{\other{\alpha}_s}{   \other{\beta}_{*,s} \other{\beta}_s \other{\nu}_s (\nu_{1,s} + \nu_{2,s})^{1/2}} \left(\frac{\nu_{2,s}}{|\rho_{1,s}|^2 \nu_{1,s}} + \frac{1}{|\rho_{2,s}|^2 }\right) \\
	\lesssim 	& \frac{\other{\alpha}_t}{\other{\beta}_{t} \other{\beta}_{*, t} (\nu_{1,t} + \nu_{2,t})^{1/2}} \left(\frac{\nu_{2,t}}{|\rho_{1,t}|^2 \other{\nu}_t} + \frac{1}{|\rho_{2,t}|^2}\right) \lesssim \frac{\gamma_t^{1/2}}{\nu_{1,t} + \nu_{2,t}}  \left(\frac{\nu_{2,t}}{|\rho_{1,t}|^2 \other{\nu}_t} + \frac{1}{|\rho_{2,t}|^2}\right) \lesssim \frac{\gamma_{t}^{1/2}}{\ell_t} \,, 
\end{split}
\end{equation}
where we used \eqref{eq:timerelations}--\eqref{eq:gamma}. Hence, to summarize \eqref{eq:Ups1}--\eqref{eq:Ups3}, we have that the second term in \eqref{eq:lasttwoterms} together with the $\tfrac{\dd}{\dd t}{\Upsilon}_{t}(A_1)$ part in \eqref{eq:imGAimG} admits the bound
\begin{equation}
\int_{\tin}^{t} \dd s \, \left| \big(\langle M_s(z_1, \reg{A}_{1,s}, \overline{z}_2) \rangle - \tfrac{\dd}{\dd s}{\Upsilon}_{s}(A_1)\big)\langle \Im G_{1,s}  \Im G_{2,s} - \widehat{M}_s(z_1, I, z_2) \rangle\right| \lesssim N^{5 \xi/2} \frac{|\rho_{1,t} \rho_{2,t}|}{N \ell_t} \gamma_{t}^{1/2} \,. 
\end{equation}

Lastly, we handle the first term in \eqref{eq:lasttwoterms} by a Gronwall argument, similarly to \eqref{eq:GAGfinal} and \eqref{eq:GGfinal}, controlling the initial condition exactly as in \eqref{eq:monotonicitynontriv}. In this way, abbreviating now $X_t := \langle \Im G_{1,t} \reg{A}_{1,t} \Im G_{2,t} - \widehat{M}_t(z_1, \reg{A}_{1,t}, z_2) \rangle$ we thus get (cf.~\cite[Lemma 5.6]{cipolloni2023universality}), similarly to the previous sections
\begin{equation} \label{eq:GAGfinal2}
\begin{split}
	|X_t|^2 &\lesssim N^{5 \xi} \frac{|\rho_{1,t}||\rho_{2,t}|^2}{(N \ell_t)^2} \gamma_{t} \vertiii{A_1}_{t, z_1, z_2}^2 \hspace{-1mm}+ \int_{\tin}^t \hspace{-2mm}\dd s \, \frac{N^{5 \xi}}{N^2} \left(\frac{|\rho_{1,s}|^2}{|\rho_{2,s}|^2 \nu_{2,s}^2} + \frac{|\rho_{2,s}|^2}{|\rho_{1,s}|^2 \nu_{1,s}^2}\right) \gamma_{s} \vertiii{A_1}_{s, z_1, z_2} f_s \Phi_{s,t}^2 \\[1mm]
	&\lesssim N^{11 \xi/2} \frac{|\rho_{1,t}||\rho_{2,t}|^2}{(N \ell_t)^2} \gamma_{t} \vertiii{A_1}_{t, z_1, z_2}^2
\end{split}
\end{equation}
for any $t \in [\tin, \tfin \wedge \tau]$. In the second step, we used \eqref{eq:gamma} and the bounds  \eqref{eq:Groninput} yielding, with the aid of Lemma~\ref{lem:intestgeneral} and the relations \eqref{eq:timerelations}, 
\begin{equation*}
\begin{split}
	\int_{\tin}^{t} \dd s \left(\frac{|\rho_{1,s}|^2}{|\rho_{2,s}|^2 \nu_{2,s}^2} + \frac{|\rho_{2,s}|^2}{|\rho_{1,s}|^2 \nu_{1,s}^2}\right)  \left(\frac{\other{\alpha}_s}{\other{\beta}_s \other{\beta}_{*,s}}\right)^2 \left[ \left(\frac{\other{\beta}_{*,s}}{\other{\beta}_{*,t}}\right)^2  +  \left(\frac{\other{\beta}_{s}}{\other{\beta}_{t}} \right)^2 \right] \\
	\lesssim \left(\frac{|\rho_{1,t}|^2}{|\rho_{2,t}|^2 \nu_{2,t}} + \frac{|\rho_{2,s}|^2}{|\rho_{1,s}|^2 \nu_{1,t}}\right)  \left(\frac{\other{\alpha}_t}{\other{\beta}_t \other{\beta}_{*,t}}\right)^2 \lesssim {\frac{|\rho_{1,t}\rho_{2,t}|^2}{\ell_{t}^2}} \gamma_t \,. 
\end{split}
\end{equation*}

Hence, to summarize, for any time $t \in [\tin, \tfin \wedge \tau]$ the bound on $|X_t|$ is \emph{strictly better} than prescribed by the stopping times \eqref{eq:stop4}--\eqref{eq:stop5}.

\subsection{Estimates of $\langle \Im GA \Im GA\rangle$ for $\tau_6$} \label{subsec:tau6}  Finally, we are left with estimating $\langle \Im G A \Im G A \rangle$. Similarly to \eqref{eq:GAG}, \eqref{eq:imGimG}, and \eqref{eq:imGAimG} we have
\begin{equation}
\begin{split}
		\label{eq:imGAimGA}
	\dd \big\langle \big(\Im G_{1,t}\reg{A}_{1,t} &\Im G_{2,t} - \widehat{M}_t(z_1, \reg{A}_{1,t} , z_2)\big)\reg{A}_{2,t}\big\rangle \\
	=&\sum_{a,b=1}^N\partial_{ab} \langle \Im G_{1,t}\reg{A}_{1,t}  \Im G_{2,t} \reg{A}_{2,t} \rangle\frac{(\dd \Brwn_t)_{ab}}{\sqrt{N}}+\big\langle \big(\Im G_{1,t}\reg{A}_{1,t}  \Im G_{2,t} - \widehat{M}_t(z_1, \reg{A}_{1,t} , z_2)\big)\reg{A}_{2,t}\big\rangle \dd t \\
	&\quad+\left(\frac{\langle \Im G_{1,t} -\Im M_{1,t}\rangle}{\Im z_{1,t}}+\frac{\langle \Im G_{2,t} -\Im M_{2,t}\rangle}{\Im z_{2,t}}\right)\langle \Im G_{1,t} \reg{A}_{1,t} \Im G_{2,t} \reg{A}_{2,t}\rangle \dd t \\
	&\quad -\frac{1}{4}\big( \langle G_{1,t}\reg{A}_{1,t} G_{2,t}\rangle\langle G_{2,t}\reg{A}_{2,t}G_{1,t}\rangle  - \langle M_t(z_1, \reg{A}_{1,t} , z_2) \rangle \langle M_t(z_2, \reg{A}_{2,t}, z_1)\rangle\big) \dd t\\
	&\quad +\frac{1}{4}\big( \langle G_{1,t}^*\reg{A}_{1,t} G_{2,t}\rangle\langle G_{2,t}\reg{A}_{2,t}G_{1,t}^*\rangle  - \langle M_t(\overline{z_1}, \reg{A}_{1,t} , z_2) \rangle \langle M_t(z_2, \reg{A}_{2,t}, \overline{z_1})\rangle\big) \dd t\\
	& \quad + \frac{1}{4}\big( \langle G_{1,t}\reg{A}_{1,t} G^*_{2,t}\rangle\langle G^*_{2,t}\reg{A}_{2,t}G_{1,t}\rangle  - \langle M_t(z_1, \reg{A}_{1,t} , \overline{z_2}) \rangle \langle M_t(\overline{z_2}, \reg{A}_{2,t}, z_1)\rangle\big) \dd t\\
	&\quad -\frac{1}{4}\big( \langle G_{1,t}^*\reg{A}_{1,t} G^*_{2,t}\rangle\langle G^*_{2,t}\reg{A}_{2,t}G_{1,t}^*\rangle  - \langle M_t(\overline{z_1}, \reg{A}_{1,t} , \overline{z_2}) \rangle \langle M_t(\overline{z_2}, \reg{A}_{2,t}, \overline{z_1})\rangle\big) \dd t\\
	& \quad - \tfrac{\dd}{\dd t}{\Upsilon}_{t}(A_1) \big( \langle \Im G_{1,t} \Im G_{2,t} \reg{A}_{2,t} \rangle - \langle \widehat{M}_t(z_1, I, z_2) \reg{A}_{2,t} \rangle  \big) \dd t \\
	& \quad - \tfrac{\dd}{\dd t}{\Upsilon}_{t}(A_2) \big( \langle \Im G_{1,t} \Im G_{2,t} \reg{A}_{1,t} \rangle - \langle \widehat{M}_t(z_1, I, z_2) \reg{A}_{1,t} \rangle  \big)\dd t\\ 
	&\quad+\langle G_{1,t}-M_{1,t}\rangle \langle \Im G_{1,t} \reg{A}_{1,t} \Im G_{2,t} \reg{A}_{2,t} G_{1,t}\rangle\dd t +\langle G_{2,t}-M_{2,t}\rangle \langle \Im G_{2,t} \reg{A}_{2,t}\Im G_{1,t} \reg{A}_{1,t}  G_{2,t}\rangle \dd t\\
	&\quad+\langle G_{1,t}^*-M_{1,t}^*\rangle \langle \Im G_{1,t} \reg{A}_{1,t} \Im G_{2,t} \reg{A}_{2,t} G_{1,t}^*\rangle \dd t+\langle G_{2,t}^*-M_{2,t}^*\rangle \langle \Im G_{2,t} \reg{A}_{2,t}\Im G_{1,t} \reg{A}_{1,t}  G_{2,t}^*\rangle \dd t.
\end{split}
\end{equation}

As in the previous sections, we begin by estimating the martingale term in the second line of \eqref{eq:imGAimGA}. By using again the BDG inequality, as in \eqref{eq:BDG}, we have
\begin{equation} \label{eq:QVimgaimga}
\begin{split}
	&\sup_{u \in [\tin , t]}\left|\sum_{a,b=1}^N\int_{\tin}^u\partial_{ab} \langle \Im G_{1,s}\reg{A}_{1,s} \Im G_{2,s} \reg{A}_{2,s}\rangle\frac{(\dd \Brwn_s)_{ab}}{\sqrt{N}}\right| \\
	&\le N^{\xi/2} \left(\int_{\tin}^t\frac{1}{N^2}\left(\frac{1}{\eta_{1,s}^2}+\frac{1}{\eta_{2,s}^2}\right)\langle \Im G_{1,s}\reg{A}_{1,s}\Im G_{2,s}\reg{A}_{2,s}\Im G_{1,s}\reg{A}_{2,s}^*\Im G_{2,s} \reg{A}_{1,s}^*\rangle\, \dd s\right)^{1/2} \\
	&\le  N^{\xi/2}  \left(\int_{\tin}^t\frac{1}{N}\left(\frac{1}{\eta_{1,s}^2}+\frac{1}{\eta_{2,s}^2}\right)\langle \Im G_{1,s}\reg{A}_{1,s}\Im G_{2,s}\reg{A}_{1,s}^*\rangle \langle \Im G_{1,s}\reg{A}_{2,s}^*\Im G_{2,s}\reg{A}_{2,s}\rangle\, \dd s\right)^{1/2} \\
		&\lesssim  N^{\xi/2} \frac{|\rho_{1,t}\rho_{2,t}|}{\sqrt{N\ell_t}} \vertiii{A_1}_{t,z_1, z_2} \vertiii{A_2}_{t,z_2, z_1},
\end{split}
\end{equation}
with very high probability, where in the second inequality we employed a \emph{reduction inequality} 
\[
\langle \Im G_{1,s}\reg{A}_{1,s}\Im G_{2,s}\reg{A}_{2,s}\Im G_{1,s}\reg{A}_{2,s}^*\Im G_{2,s}\reg{A}_{1,s}^* \rangle\le N\langle \Im G_{1,s}\reg{A}_{1,s}\Im G_{2,s}\reg{A}_{1,s}^*\rangle \langle \Im G_{1,s}\reg{A}_{2,s}^*\Im G_{2,s}\reg{A}_{2,s}\rangle \,. 
\]
This relies on the elementary trace inequality $\langle X Y \rangle \le N \langle X \rangle \langle Y \rangle$ for positive $X, Y \in \C^{N \times N}$, using operator positivity of $\sqrt{|\Im G_{2,s}|} B |\Im G_{1,s}| B^* \sqrt{|\Im G_{2,s}|}$ for $B \in \{\reg{A}_{2,s}, \reg{A}_{1,s}^*\}$.  The second term in the second line shall henceforth be neglected as it can be handled by a simple rescaling with the factor $\ee^{-t} \sim 1$. In particular, no Gronwall estimate will be applied to \eqref{eq:imGAimGA}. 

Next, involving the single resolvent law, the third line of \eqref{eq:imGAimGA} can easily be bounded as
\begin{equation*}
\int_{\tin}^{t} \dd s \, N^\xi \left(\frac{1}{N \eta_{1,s}^2} + \frac{1}{N \eta_{2,s}^2}\right) |\rho_{1,s} \rho_{2,s}| \vertiii{A_1}_{s,z_1, z_2} \vertiii{A_2}_{s,z_2, z_1} \lesssim \frac{N^{2 \xi}}{\sqrt{N \ell_t}} \cdot \frac{|\rho_{1,t} \rho_{2,t}|}{\sqrt{N \ell_t}} \vertiii{A_1}_{t,z_1, z_2} \vertiii{A_2}_{t,z_2, z_1} \,. 
\end{equation*}
Moreover, with the aid of a single resolvent law, we estimate the last two lines of \eqref{eq:imGAimGA} as
\begin{equation*}
\begin{split}
		\int_{\tin}^t \text{last two lines of \eqref{eq:imGAimGA}} \,\,\dd s &\le \int_{\tin}^t N^{\xi}\left(\frac{1}{N\eta_{1,s}^2}+\frac{1}{N\eta_{2,s}^2}\right)|\rho_{1,s}\rho_{2,s}| \,  \vertiii{A_1}_{s,z_1, z_2} \vertiii{A_2}_{s, z_2, z_1} \, \dd s\\
		&\lesssim \frac{N^{\xi}|\rho_{1,t}\rho_{2,t}|}{N\ell_t}\vertiii{A_1}_{t,z_1, z_2} \vertiii{A_2}_{t,z_2, z_1},
\end{split}
\end{equation*}
where we used that by a Schwarz inequality we have (writing $A_i \equiv \reg{A}_{i,t}$ for brevity)
\[
\begin{split}
\big| \langle \Im G_{1,s} A_1\Im G_{2,s} A_2 G_{1,s}\rangle\big|&\le\frac{1}{\eta_{1,s}}\big(\langle \Im G_{1,s} A_1\Im G_{2,s} A_1^*\rangle \langle \Im G_{1,s} A_2^*\Im G_{2,s} A_2\rangle\big)^{1/2}, \\
\big| \langle \Im G_{2,s} A_2\Im G_{1,s} A_1 G_{2,s}\rangle\big|&\le\frac{1}{\eta_{2,s}}\big(\langle \Im G_{1,s} A_1\Im G_{2,s} A_1^*\rangle \langle \Im G_{1,s} A_2^*\Im G_{2,s} A_2\rangle\big)^{1/2},
\end{split}
\]
similarly to \eqref{eq:normbound}. 

Similarly to \eqref{eq:algebraicimgimg} and \eqref{eq:algebraicimgaimg}, we rewrite the fourth, fifth, sixth, and seventh line of \eqref{eq:imGAimGA} as
\begin{equation} \label{eq:algebraic}
\begin{split}
	&- \langle G_{1,t}\reg{A}_{1,t}G_{2,t}\rangle\langle G_{2,t}\reg{A}_{2,t}G_{1,t}\rangle  
	+ \langle G_{1,t}^*\reg{A}_{1,t}G_{2,t}\rangle\langle G_{2,t}\reg{A}_{2,t}G_{1,t}^*\rangle \\
	&+ \langle G_{1,t}\reg{A}_{1,t}G^*_{2,t}\rangle\langle G^*_{2,t}\reg{A}_{2,t}G_{1,t}\rangle  - \langle G_{1,t}^*\reg{A}_{1,t}G^*_{2,t}\rangle\langle G^*_{2,t}\reg{A}_{2,t}G_{1,t}^*\rangle \\
	= \ &4 \big(  \langle \Im G_{1,t}\reg{A}_{1,t}\Im G_{2,t}\rangle\langle G^*_{2,t}\reg{A}_{2,t}G_{1,t}\rangle +  \langle G_{1,t}^*\reg{A}_{1,t}G_{2,t}\rangle\langle \Im G_{2,t}\reg{A}_{2,t}\Im G_{1,t}\rangle \\
	&+  \langle \Im G_{1,t}\reg{A}_{1,t}G_{2,t}\rangle\langle \Im G_{2,t}\reg{A}_{2,t}G_{1,t}\rangle  +  \langle G_{1,t}^*\reg{A}_{1,t}\Im G_{2,t}\rangle\langle G_{2,t}^*\reg{A}_{2,t}\Im G_{1,t}\rangle     \big)
\end{split}
\end{equation}
and analogously for the $M$-terms. As in the previous sections, we add and subtract the corresponding $G \times M$-type terms and control the resulting contributions as follows: First, all the resulting $(G- M) \times (G-M)$-type terms can easily be bounded as
\begin{equation} \label{eq:toobad3}
\int_{\tin}^{t} \dd s \, N^{4 \xi} \frac{|\rho_{1,s}\rho_{2,s}|}{N^2\eta_{1,s}\eta_{2,s}\ell_s} \vertiii{A_1}_{s,z_1, z_2} \vertiii{A_2}_{s,z_2, z_1} \lesssim \frac{N^{5 \xi}}{(N \ell_t)^{3/2}} \frac{|\rho_{1,t} \rho_{2,t}|}{\sqrt{N \ell_t}} \vertiii{A_1}_{t,z_1, z_2} \vertiii{A_2}_{t,z_2, z_1} \,, 
\end{equation}
where we did not exploit the presence of $\Im$'s, i.e.~simply used local laws for $\langle GAG \rangle$ as encoded in the stopping times \eqref{eq:stop1}--\eqref{eq:stop2}. 

Next, for the $M \times (G-M)$-type terms we distinguish, how many $\Im$'s are present in the $M$-term. If \emph{at least one} $\Im$ is present in the $M$-term (i.e., in particular for all the terms resulting from the third and fourth term on the right-hand side~of \eqref{eq:algebraic}), we have, with the aid of \eqref{eq:GImG_bounds} in Proposition \ref{prop:Mt_bound} and \eqref{eq:timerelations}, that
\begin{equation} \label{eq:G-MtimesM}
\begin{split}
	\int_{\tin}^{t} \dd s \, N^{2 \xi} \frac{|\rho_{1,s}| + |\rho_{2,s}| }{\other{\alpha}_s} \sqrt{\frac{|\rho_{1,s}\rho_{2,s}|}{N^2\eta_{1,s}\eta_{2,s}\ell_s}}   \vertiii{A_1}_{s,z_1, z_2}\, \vertiii{A_2}_{s,z_2, z_1} \lesssim \frac{N^{4 \xi}}{\sqrt{N \ell_t}} \frac{|\rho_{1,t} \rho_{2,t}|}{\sqrt{N \ell_t}} \vertiii{A_1}_{t,z_1, z_2} \vertiii{A_2}_{t,z_2, z_1} \,. 
\end{split}
\end{equation}

In case that there is \emph{no} $\Im$ in the $M$-term (as can only happen for the first and second term on the right-hand side~of \eqref{eq:algebraic}), we involve the eighth and ninth lines of \eqref{eq:imGAimGA} and find those contributions to be given by 
\begin{equation} \label{eq:UPscancel}
\begin{split}
	\int_{\tin}^t \dd s \, \Big(\langle M_s(z_1, I, \overline{z}_2) \reg{A}_{2,s}\rangle - \tfrac{\dd}{\dd s}{\Upsilon}_{s}(A_2)\Big)\Big(\langle \Im G_{1,s} \reg{A}_{1,s} \Im G_{2,s} \rangle - \langle \widehat{M}_s(z_1, I, z_2) \reg{A}_{1,s} \rangle\Big) \\
	+\int_{\tin}^t \dd s \, \Big(\langle M_s(\overline{z}_1, I, {z}_2) \reg{A}_{1,s}\rangle - \tfrac{\dd}{\dd s}{\Upsilon}_{s}(A_1)\Big)\Big(\langle \Im G_{1,s} \reg{A}_{2,s} \Im G_{2,s} \rangle - \langle \widehat{M}_s(z_1, I, z_2) \reg{A}_{2,s} \rangle\Big) \,. 
\end{split}
\end{equation}

We hence have that 
\begin{equation} \label{eq:UPscancelbound}
	|\eqref{eq:UPscancel}| \lesssim N^{2\xi}\int_{\tin}^t \dd s \, \mathcal{E}_{s}   \frac{|\rho_{1,s} \rho_{2,s}|}{N \ell_s} \gamma_{s}^{1/2}\vertiii{A_1}_{s, z_1, z_2} \vertiii{A_2}_{s, z_2, z_1}. 
\end{equation}
Then, following the estimates \eqref{eq:mixed1}--\eqref{eq:Upsest2} and \eqref{eq:Ups1}--\eqref{eq:Ups3}, we bound the two contributions to \eqref{eq:UPscancelbound} (after writing out the two terms of $\mathcal{E}_s$) as follows:\footnote{We shall henceforth, as above, ignore the observable norms that are always just bounded by monotonicity, as well as the global $N^{2\xi}/N$ factor and possible logarithmic corrections.} For the analog of \eqref{eq:mixed1} and \eqref{eq:Ups1} we have
\begin{equation} \label{eq:Upsest1fin}
\begin{split}
	&\int_{\tin}^{t} \dd s \, (|\rho_{1,s}| + |\rho_{2,s}|) \left(\frac{1}{|\rho_{1,s}|^2 \nu_{1,s}} + \frac{1}{|\rho_{2,s}|^2 \nu_{2,s}}\right) \big(\nu_{1,s} + \nu_{2,s}\big)^{1/2} \frac{ 1}{\other{\beta}_s \other{\beta}_{*,s}} \\
	\lesssim 	&\int_{\tin}^{t} \dd s \, (|\rho_{1,s}| + |\rho_{2,s}|)  \left(\frac{1}{|\rho_{1,s}|^2 \nu_{1,s}} + \frac{1}{|\rho_{2,s}|^2 \nu_{2,s}}\right)  \frac{ 1}{\other{\beta}_s \other{\beta}_{*,s}^{1/2}} 
	\lesssim \frac{\big(|\rho_{1,t}| + |\rho_{2,t}|\big)^3}{|\rho_{1,t} \rho_{2,t}|^2} \frac{1}{\other{\beta}_t \other{\beta}_{*,t}^{1/2}}  \lesssim \ell_t^{-1},
\end{split}
\end{equation}
where we additionally used \eqref{eq:lowerbeta}.
Finally, we estimate the second and last contribution to \eqref{eq:UPscancelbound}. Analogously to \eqref{eq:Upsest2} and \eqref{eq:Ups3}, this term admits the bound 
\begin{equation} \label{eq:Upsest3fin}
\begin{split}
	&	\int_{\tin}^t  \dd s \,    \frac{ 1}{ \other{\nu}_s  } \left(\frac{1}{|\rho_{1,s}|^2 \nu_{1,s}} + \frac{1}{|\rho_{2,s}|^2 \nu_{2,s}}\right) \lesssim \frac{1}{|\rho_{1,t}|^2 \nu_{1,t}} + \frac{1}{|\rho_{2,t}|^2 \nu_{2,t}} \lesssim \ell_t^{-1} \,. 
\end{split}
\end{equation}
Thus, the estimates \eqref{eq:Upsest1fin} and \eqref{eq:Upsest3fin} imply
\begin{equation*} 
|\eqref{eq:UPscancel}| \lesssim \frac{N^{3 \xi}}{\sqrt{N \ell_t}} \frac{|\rho_{1,t} \rho_{2,t}|}{\sqrt{N\ell_t}} \,. 
\end{equation*}

Therefore, by integrating \eqref{eq:imGAimGA} in time, collecting the above bounds, and using the estimate
\begin{equation*}
\begin{split}
	\big|\big\langle \big(\Im G_{1,\tin}\reg{A}_{1,\tin}\Im G_{2,\tin} - \widehat{M}_{\tin}(z_1, \reg{A}_{1,\tin}, z_2)\big)\reg{A}_{2,\tin}\big\rangle\big| \lesssim N^{\xi/2}\frac{|\rho_{1,t}\rho_{2,t}|}{\sqrt{N \ell_{t}}} \vertiii{A_1}_{t, z_1, z_2} \vertiii{A_2}_{t, z_2, z_1},
\end{split}
\end{equation*}
for the initial condition (which follows easily by monotonicity) we conclude that, uniformly in $t \in [\tin, \tfin \wedge \tau]$, 
\begin{equation} \label{eq:GAGAfinal}
\big|	\big\langle \big(\Im G_{1,t}\reg{A}_{1,t}\Im G_{2,t} - \widehat{M}_{t}(z_1, \reg{A}_{1,t} , z_2)\big)\reg{A}_{2,t}\big\rangle\big| \le N^{3\xi/4} \frac{|\rho_{1,t}\rho_{2,t}|}{\sqrt{N \ell_{t}}} \vertiii{A_1}_{t, z_1, z_2} \vertiii{A_2}_{t, z_2, z_1}\,, 
\end{equation}
where for the terms in \eqref{eq:toobad3}, \eqref{eq:G-MtimesM}, and \eqref{eq:UPscancel}, we additionally used that $\xi\le \epsilon/10$, to reduce the $N^\xi$ power by sacrificing an $\sqrt{N \ell_t} \ge N^{\epsilon/2} \ge N^{5 \xi}$ factor. That is, for any time $t \in [\tin, \tfin\wedge \tau]$, also the local law bound for $\langle \Im GA \Im GA \rangle$ is \emph{strictly better} than prescribed by the stopping time \eqref{eq:stoptime}. 

Hence, by combining \eqref{eq:GAGfinal}, \eqref{eq:GGfinal}, \eqref{eq:GAGfinal2}, and \eqref{eq:GAGAfinal}, we conclude that $\tau =  \tfin$, which finishes the proof of Proposition \ref{prop:zig}. \qed

 	\section{Green function comparison: Proof of Proposition \ref{prop:zag}} \label{sec:zagproof}
The goal of this section is to give the proof of Proposition \ref{prop:zag} and thereby conclude the argument for the \emph{zag} step of our proof. That is, we remove the Gaussian component introduced during the \emph{zig} step in Proposition \ref{prop:zig} by a \emph{Green function comparison} (GFT) approach. For simplicity {and in order to avoid unnecessary complications}, we will carry out the proof in the real symmetric case only; leaving the complex-Hermitian case to the reader, which can be dealt with minor modifications. In fact, the notation in the cumulant expansion is slightly more involved in the complex case, because the real and imaginary parts are treated separately; see~\cite[Appendix C]{slowcorr}. Finally, since throughout the argument, the time $t_k$ defined in \eqref{eq:tkdef} remains fixed, we shall henceforth drop the subscript/argument $t_k$ and, additionally,  abbreviate  $s_{\rm final} := s(\dift_k)$.  Moreover, we write $A_i \equiv \reg{A}_{i,t_k}$ for brevity, which violates the convention
started from Section~\ref{sec:Mboundsproof}, that
  $A$ denotes \emph{pre-regular} observable. In the entire Section~\ref{sec:zagproof}, except the proof of 
  Proposition \ref{prop:zag}, the
letter $A_i$ is used for this particular \emph{regular} observable. \nc
We also stress that the spectral parameters remain fixed in the zag step, and hence also the deterministic approximation $M$ as well as $\eta_1, \eta_2, \rho_1, \rho_2, \ell, \gamma$ are kept unchanged.

 	\subsection{Zag step: Proof of Proposition \ref{prop:zag}}

Our proof of Proposition \ref{prop:zag} relies on the following two Gronwall estimates, formulated in Propositions \ref{prop:gronwallImG}--\ref{prop:gronwallG}. Their proofs are given in Sections \ref{subsubsec:zagav2G} and \ref{subsubsec:zagav2G1A}, respectively. 

 	\begin{proposition}[Gronwall estimates for pure $ \Im G$ chains] \label{prop:gronwallImG}
 	Adopt the setting and notations from Proposition \ref{prop:zag}. For $s \in [0, s_{\rm final}]$ and any $B_1, B_2 \in \C^{N \times N}$, define
 		\begin{equation} \label{eq:Rsdef}
 			R_s^{B_1,B_2} := \langle \Im G^s(z_1) B_1 \Im G^s(z_2) B_2 \rangle - \langle \widehat{M}(z_1, B_1, z_2) B_2\rangle
 		\end{equation}
as well as
 		\begin{equation} \label{eq:Phidef}
 			\Phi_\delta^p(s; B_1, B_2) := \mathbb{E} \big|R_s^{B_1, B_2}\big|^p 
 			+ \left(N^{3\delta}\frac{ |\rho_1 \rho_2|}{\sqrt{N \ell}} \vertiii{B_1}_{z_1, z_2} \vertiii{B_2}_{z_2, z_1}\right)^p  \,. 
 		\end{equation}
 		Then, for any (large) even $p \in \mathbb{N}$ and arbitrarily small $\delta > 0$,we have the following: 
 		\begin{itemize}
\item[(i)] 	\textnormal{[Two regular observables]} It holds that
\begin{equation}
 			\frac{\mathrm{d}}{\mathrm{d}s}\mathbb{E} \big|R_s^{A_1, A_2}\big|^p \lesssim \left(1 + N^{-\delta}\max_i \left(\frac{|\rho_i|}{\eta_i}\right)\right) \, \left[\E \big|R_s^{A_1, A_2}\big|^p + \sqrt{\Phi_\delta^{p}(s; A_1, A_1^*) \Phi_\delta^{p}(s; A_2^*, A_2)} \,  \right] 
\end{equation}
uniformly in $s \in [0,s_{\rm final}]$ and $(z_1, z_2)$-regular $A_1, A_2\in \C^{N \times N}$.
\item[(ii)] \textnormal{[No regular observable]} It holds that, uniformly in $s \in [0,s_{\rm final}]$, 
\begin{equation}
	\frac{\mathrm{d}}{\mathrm{d}s}\mathbb{E} \big|R_s^{I, I}\big|^p \lesssim \left(1 + N^{-\delta}\max_i \left(\frac{|\rho_i|}{\eta_i}\right)\right) \, \left[\E \big|R_s^{I, I}\big|^p + \left(N^{3\delta}\frac{ |\rho_1 \rho_2|}{N \ell} \gamma(z_1, z_2)\right)^p  \right] \,. 
\end{equation}
\item[(iii)] \textnormal{[One regular observable]}	Assume that $$\left|R_s^{A_1, A_1^*}\right| \prec \frac{ |\rho_1 \rho_2|}{\sqrt{N \ell}} \vertiii{A_1}_{z_1, z_2}^2 \,, \qquad \left|R_s^{A_2^*, A_2}\right| \prec \frac{ |\rho_1 \rho_2|}{\sqrt{N \ell}} \vertiii{A_2}_{z_2, z_1}^2\,, \qquad \left|R_s^{I,I}\right| \prec \frac{ |\rho_1 \rho_2|}{N \ell} \gamma(z_1, z_2)$$
all hold uniformly in $s \in [0,s_{\rm final}]$ and $(z_1, z_2)$-regular $A_1 \in \C^{N \times N}$ as well as $(z_2, z_1)$-regular $A_2 \in \C^{N \times N}$. Then it holds that
\begin{equation}
	\frac{\mathrm{d}}{\mathrm{d}s}\mathbb{E} \big|R_s^{A_1, I}\big|^p \lesssim \left(1 + N^{-\delta}\max_i \left(\frac{|\rho_i|}{\eta_i}\right)\right) \, \left[\E \big|R_s^{A_1, I}\big|^p + \left(N^{3\delta}\frac{ |\rho_1 \rho_2|}{N \ell} \gamma(z_1, z_2)^{1/2} \vertiii{A_1}_{z_1, z_2}\right)^p  \right] 
\end{equation}
and similarly for $R_s^{I, A_2}$, both uniformly in $s \in [0,s_{\rm final}]$ and $(z_1, z_2)$-regular $A_1, A_2 \in \C^{N \times N}$. 
 		\end{itemize}

 	\end{proposition}

 	\begin{proposition}[Gronwall estimate for $\langle G A G \rangle$] \label{prop:gronwallG}
 	Adopt the setting and notations from Proposition \ref{prop:zag} and, for $s \in [0, s_{\rm final}]$, define
 		\begin{equation} \label{eq:Rdefinition}
 			S_s^{A_1} := \langle G^s(z_1) A_1 G^s(z_2) \rangle - \langle M(z_1, A_1, z_2) \rangle\,. 
 		\end{equation}
 		Using the same notations, assume that $R_s^{A_1, A_1^*}$ from \eqref{eq:Rsdef} satisfies $$\left|R_s^{A_1, A_1^*}\right| \prec \frac{ |\rho_1 \rho_2|}{\sqrt{N \ell}} \vertiii{A_1}_{z_1, z_2}^2$$
 			uniformly in $s \in [0,s_{\rm final}]$ and $(z_1, z_2)$-regular $A_1 \in \C^{N \times N}$. 
 			
 		Then, for any (large) even $p \in \mathbb{N}$ and arbitrarily small $\delta > 0$, it holds that
 		\begin{equation}
 			\frac{\mathrm{d}}{\mathrm{d}s}\mathbb{E} \big|S_s^{A_1}\big|^p \lesssim \left(1 + N^{-\delta}\max_i \left(\frac{|\rho_i|}{\eta_i}\right)\right) \, \left[\mathbb{E} \big|S_s^{A_1}\big|^p + \left(N^{3\delta}\sqrt{\frac{ |\rho_1 \rho_2|}{(N \ell) (N \eta_1 \eta_2)}} \vertiii{A_1}_{z_1, z_2} \right)^p \right] 
 		\end{equation}
 			uniformly in $s \in [0,s_{\rm final}]$ and $(z_1, z_2)$-regular $A_1 \in \C^{N \times N}$. 
 		
 			The same statement holds with $G^s_1$ and/or $G^s_2$ replaced by their adjoints $(G^s_1)^*$ and $(G^s_2)^*$, respectively. 
 	\end{proposition}
 	Armed with Propositions \ref{prop:gronwallImG}--\ref{prop:gronwallG}, we can easily prove Proposition \ref{prop:zag}. 
 	\begin{proof}[Proof of Proposition \ref{prop:zag}]  	
	Recall from the preamble of Section~\ref{sec:zagproof}, that $A_1, A_2$ in Propositions~\ref{prop:gronwallImG} and \ref{prop:gronwallG} are assumed to be regular, while in the setup of Proposition~\ref{prop:zag}, $A_i$ are only pre-regular. Throughout this proof, we adhere to the convention of  Proposition~\ref{prop:zag}, and apply  Propositions~\ref{prop:gronwallImG}--\ref{prop:gronwallG} with $A_i := \reg{A}_i$.
	\nc

 		We start with the case $A_1 = A_2^*$. Then, we simply apply Gronwall's lemma (using the notation $s_{\rm final} = s(\dift_k)$) to deduce from Proposition \ref{prop:gronwallImG}~(i), uniformly in $s \in [0, s_{\rm final}]$ and regular $A_1$, that
\begin{equation} \label{eq:integrate2G}
	\begin{split}
		\E \big|R_s^{A_1, A_1^*}\big|^p  \lesssim \E \big|R^{A_1, A_1^*}_{s_{\rm final}}\big|^p + \left(N^{3\delta} \frac{|\rho_1 \rho_2|}{\sqrt{N \ell}} \vertiii{A_1}_{z_1, z_2}^2\right)^p \lesssim  \left(N^{3\delta} \frac{|\rho_1 \rho_2|}{\sqrt{N \ell}} \vertiii{A_1}_{z_1, z_2}^2\right)^p, 
	\end{split}
\end{equation}
where we used that $ \max_i (|\rho_i| \eta_i^{-1}) \lesssim N^{k \delta}$ by \eqref{eq:tkdef} and $s_{\rm final} \lesssim N^{-(k-1)\delta}$ by \eqref{eq:dtkdef}, and that the terminal condition (spelled out in \eqref{eq:2G1Azag}--\eqref{eq:2G2Azag}) ensures the bound on $\E \big|R^{A_1, A_1^*}_{s_{\rm final}}\big|^p$. {Additionally we used that $\vertiii{A_1^*}_{z_2, z_1} \sim \vertiii{A_1}_{z_1, z_2}$ by \eqref{eq:symmetries}.} By arbitrariness of $p$ and $\delta$, this shows the local law for $R_s^{A_1, A_1^*}$ in $\prec$-sense. 
 	
This information (using uniformity in regular matrices) can then be plugged into Proposition \ref{prop:gronwallImG}~(i) to conclude the general $R_s^{A_1, A_2}$-case by a similar Gronwall argument. The same Gronwall argument can also be applied to the bounds in Proposition \ref{prop:gronwallImG}~(ii)+(iii), easily yielding the desired bounds in Proposition \ref{prop:zag}. 

Finally, having established the bounds for $R_s$, we can employ Proposition \ref{prop:gronwallG} and find, analogously to \eqref{eq:integrate2G}, 
\begin{equation*} 
	\E |S_s|^p  \lesssim \left(N^{3\delta} \sqrt{\frac{|\rho_1 \rho_2|}{(N\ell ) (N \eta_1 \eta_2)}} \vertiii{A_1}_{z_1, z_2} \right)^p 
\end{equation*}
uniformly in $s \in [0, s_{\rm final}]$ (and similarly with exchanged indices $1 \leftrightarrow 2$, as well as $G^s_1$ and/or $G^s_2$ replaced by their adjoints $(G^s_1)^*$ and $(G^s_2)^*$, respectively). Again, by arbitrariness of $p$ and $\delta$, we obtain the local law for $S_s$ in $\prec$-sense. This concludes the proof of Proposition \ref{prop:zag}
by noting that $\vertiii{A_i} = \vertiii{\reg{A}_i}$ by~\eqref{eq:tripl_I=0}. \nc
 	\end{proof}
 	
 	\subsection{Cumulant expansion: Proof of Propositions \ref{prop:gronwallImG}--\ref{prop:gronwallG}} \label{subsec:cumex}

 		In order to perform the GFT, i.e.,~compare initial and final $W$'s, given by $W^t = H^t- D = H^t - \E H^t$
 	with $H^t$ being the solution to \eqref{eq:zag_flow},   we employ Itô's formula: For a $C^2$-function $f(W^t)$, it holds that%
 	\begin{equation} \label{eq:cumexstart}
 		\frac{\rd }{\rd t} \E f(W^t) = - \frac{1}{2}\E \sum_\alpha w_\alpha(t) (\partial_\alpha f)(W^t) + \frac{1}{2N} \sum_{\alpha, \beta}\kappa_t(\alpha, \beta) \E (\partial_\alpha \partial_\beta f)(W^t) \,,
 	\end{equation}
 	where $\kappa_t(\alpha, \beta)$ denotes the second order cumulant of $w_\alpha(t)$ and $w_\beta(t)$, the matrix entries of $W^t$ (recall \eqref{eq:cumulants}).   
 	The first summand on the right-hand side~of \eqref{eq:cumexstart} can now be further treated by \emph{cumulant expansion}, which is the 	key tool for our proof. 
 	
 		\begin{proposition}[Multivariate cumulant expansion; cf.~Prop.~5.2 in \cite{cuspuniv}, Prop.~3.2 in \cite{slowcorr}, or Lemma 3.1 in \cite{HeKnowles}] \label{prop:cumex} Fix $L \in \N$. 
 		Assume that  %
 		$f : \R^{N \times N} \to \C$ is a $C^L$ function with bounded derivatives. Moreover, let $W$ be a random matrix, whose normalized cumulants satisfy Assumptions~\ref{ass:S_norms} and \ref{ass:cumulants}. Then, for any index $\alpha_0 \in \indset{N}^2$ it holds that (recall the definition of the neighborhood set $\mathcal{N}$ from Assumption \ref{ass:cumulants})
 		\begin{equation} \label{eq:cumulantexpansion}
 			\E w_{\alpha_0} f(W) = \sum_{k=0}^{L-1} \sum_{\bm \alpha \in \mathcal{N}(\alpha_0)^k} \frac{\kappa(\alpha_0 , \bm \alpha) }{N^{(k+1)/2} k!} \E (\partial_{\bm \alpha} f)(W)+ \Omega_L(f, \alpha_0),
 		\end{equation}
 		where $\bm \alpha = (\alpha_1, ... , \alpha_k)$ and $\partial_{\bm \alpha} = \partial_{w_{\alpha_1}} ... \partial_{w_{\alpha_k}}$ for $k \ge 1$, and for $k=0$, $\partial_{\bm \alpha}f$ is considered as the function $f$ itself.
 		Moreover, the error term in \eqref{eq:cumulantexpansion} satisfies
 		\begin{equation} \label{eq:errorbound}
 			\big|\Omega_L(f, \alpha_0)\big| \lesssim \frac{C_L}{N^{(L+1)/2}} \sum_{\bm \alpha \in \mathcal{N}(\alpha_0)^L} \sup_{ \lambda \in [0,1]} \left( \E \big| (\partial_{\bm \alpha}f)(\lambda W \vert_{\mathcal{N}(\alpha_0)} + W\vert_{\indset{N}^2 \setminus \mathcal{N}(\alpha_0)}) \big|^2\right)^{1/2},
 		\end{equation}
 		for some constant $C_L > 0$ depending only on $L$. The notation $W\vert_{\mathcal{N}}$ for $\mathcal{N} \subset \indset{N}^2$ in \eqref{eq:errorbound} refers to the matrix which equals $W$ at all entries $\alpha \in \mathcal{N}$ and is zero otherwise. 
 	\end{proposition}
 	We point out that the $k=1$  term   %
 	in the expansion of the first summand on the right-hand side~of \eqref{eq:cumexstart}   exactly cancels the second summand on the right-hand side~of \eqref{eq:cumexstart}. Finally, note that, for Proposition~\ref{prop:cumex} being practically applicable we need to control (i) every order of the expansion, and (ii) the truncation term $\Omega$. These will be guaranteed by Assumptions~\ref{ass:S_norms} and \ref{ass:cumulants} above. 
 	
In the rest of this section, we give the proofs of Propositions \ref{prop:gronwallImG}--\ref{prop:gronwallG}. Throughout the arguments, we will frequently use the entry-wise single-resolvent law (see, e.g., \cite[Theorem~2.8]{cuspuniv})
\begin{equation} \label{eq:singleGzag}
\max_{a,b \in \indset{N}}\big|(G^s(z) - M(z))_{ab}\big| \prec \sqrt{\frac{|\rho(z)|}{N \eta}}\,, 
\end{equation}
 	which implies that $\max_{a, b \in \indset{N}}\big|(G^s(z))_{ab}\big| \lesssim 1$ and $\max_{a, b \in \indset{N}}\big|(\Im G^s(z))_{ab}\big| \lesssim |\rho(z)|$ with very high probability.

 \subsubsection{Gronwall estimate for $\langle \Im GA \Im GA\rangle$: Proof of Proposition \ref{prop:gronwallImG}}	  \label{subsubsec:zagav2G}
 	
 	We start by giving the proof of Proposition \ref{prop:gronwallImG}~(i) and afterwards briefly discuss the necessary modifications for parts (ii) and (iii). 
 	
 We drop the $A_1, A_2$ superscript, whenever it does not lead to confusion. Moreover, we assume that $\Im z_1, \Im z_2 > 0$ so that $\rho(z_j) > 0$. By \eqref{eq:cumexstart} for $R_s$ we have 
 	\begin{equation} \label{eq:GFTstart}
 		\frac{\dif}{\dif s} \E |R_s|^{p} = - \frac{1}{2} \E \sum_{\alpha_1} w_{\alpha_1}(s) (\partial_{\alpha_1} |R_s|^{p}) + \frac{1}{2} \sum_{\alpha_1, \alpha_2} \kappa_s(\alpha_1, \alpha_2) \E \bigl[\partial_{\alpha_1} \partial_{\alpha_2} |R_s|^{p}\bigr],
 	\end{equation}
 	where $w_{\alpha_i}(s)$ is the $\alpha_i$-th entry of $W_s$, $\kappa_s(\alpha_1, \alpha_2, ...)$ is a joint normalized cumulant of $w_{\alpha_1}(s), w_{\alpha_2}(s), ...$ and $\partial_{\alpha_i} = \partial_{w_{\alpha_i}(s)}$ denotes the partial derivative in the direction of $w_{\alpha_i}(s)$.  
 	
 	The first term on the right-hand side~of \eqref{eq:GFTstart} can now be expanded by means of Proposition~\ref{prop:cumex}:
 	\begin{equation} \label{eq:cumexres}
 		\E  \bigl[ w_{\alpha_1}(s) (\partial_{\alpha_1} |R_s|^{p}) \bigr]  = \sum_{k=0}^{L-1} \sum_{\bm \alpha \in \mathcal{N}(\alpha_1)^k}\frac{\kappa_s(\alpha_1, \bm \alpha)}{N^{(k+1)/2} \, k!} \E \bigl[\partial_{\alpha_1} \partial_{\bm \alpha} |R_s|^{p}\bigr] + \Omega_L 
 	\end{equation}
 	where for $L$ large enough, by standard arguments using Assumption \ref{ass:cumulants} (see, e.g., \cite[Discussion around (5.17)]{cuspuniv}), the error term satisfies
 	\begin{equation*}
|\Omega_L| \lesssim  \left(\frac{ \rho_1 \rho_2}{\sqrt{N \ell}} \vertiii{A_1}_{z_1, z_2} \vertiii{A_2}_{z_2, z_1}\right)^p \,. 
 	\end{equation*}
 	
 		Hence, our goal is to estimate terms of the form (note that we changed $k$ in \eqref{eq:cumulantexpansion} to $k-1$)
 	\begin{equation}
 		 N^{-k/2} \sum_{\alpha_1, \dots, \alpha_k} \kappa(\alpha_1, \dots, \alpha_k) \E\Bigl[\partial_{\alpha_1}\dots\partial_{\alpha_{k}} \bigl\lvert R^s \bigr\rvert^p\Bigr], \qquad k \ge 3.
 	\end{equation}
 	 	Moreover, dropping the (sub-)superscript $s$,  
 	ignoring the difference between $R$ and $\overline{R}$, using the symmetry of $\kappa(\bm \alpha)$ for $\bm \alpha := (\alpha_j)_{j=1}^k$, the Leibniz rule, and ignoring the non-differentiated factors of $R$, we reduce the problem to estimating 
 	\begin{equation} \label{eq:GFT_structure}
 		N^{-k/2} \sum_{\bm \alpha} \kappa(\bm \alpha) \prod_{j=1}^{n}\bigl( \partial_{\bm \alpha^{(j)}} R\bigr),  \qquad  \ k \ge 3, 
 	\end{equation} 
 	where the concatenation of the vectors of index pairs $(\bm \alpha^{(j)})_{j=1}^n$ is equal to $\bm \alpha \in \indset{N}^{2k}$ with $k \ge 3$, and $\partial_{\bm \alpha^{(j)}}$ denotes the repeated derivative with respect to each component of the index-pair vector $\bm \alpha^{(j)}$.

 	The goal is to estimate \eqref{eq:GFT_structure} solely in terms of the $\vertiii{\kappa}_k$-norms and the non-negative tracial quantities $\langle \Im G_1 A_1 \Im G_2 A_1^*\rangle$, $\langle \Im G_1 A_2^* \Im G_2 A_2\rangle$.  
 	 Although $R$ itself is tracial, its derivatives involve individual entries of resolvent chains whose indices belong to the set
 	$\{a_i, b_i : i \in \indset{k}\}$, where $\alpha_i = (a_i,b_i) \in \indset{N}^2$.
 	Therefore, when estimating \eqref{eq:GFT_structure}, we need to systematically reconstruct traces from the summation over ${\bm \alpha}$. 
 	Since the norm $\vertiii{\kappa}_k$ already involves summation over $k-2$ index pairs, only two index pairs remain, each of which has to be summed separately in the $\ell^2$-sense. 
 	All remaining chain entries must be controlled pointwise, which is typically much cruder than averaged estimates; see, for example, \eqref{eq:X_bounds} below.
 	We devise a systematic procedure that carefully allocates the available summations and treats all contributions arising in \eqref{eq:GFT_structure} in a unified manner. 
 	Before presenting the full technical proof in the general setting, we first discuss several representative examples that illustrate the key mechanisms underlying our argument.	
 	 
 	 In the following, we  will frequently use the derivative rules for the resolvent $G := (H - z)^{-1}$: For all $i,j,a,b \in \indset{N}$, denoting $(E^{ab})_{k \ell} = \delta_{ak} \delta_{\ell b}$
 	 \begin{equation} \label{eq:partial_ab}
 	 	\begin{split}
 	 		-\delab_{ab} G_{ij} &= \bigl(GE^{ab}G\bigr)_{ij} = G_{ia} G_{bj},\\ 
 	 		- \delab_{ab} (\Im G)_{ij} &= \bigl(G E^{ab} \Im G + \Im G E^{ab} G^*\bigr)_{ij} = G_{ia} (\Im G)_{bj} + (\Im G)_{ia} (G^*)_{bj}.
 	 	\end{split}
 	 \end{equation}
 	 In particular, when differentiating $\Im G$, both terms on the right-hand side~contain one $\Im G$ each, i.e.~the number of $\Im G$'s is preserved.

 	\smallskip
 	\textbf{Example estimates}. 
 	For the purposes  of these examples, we ignore the different indices of the resolvents and observables and the difference between $A$ and $A^*$, as well as $G$ and $G^*$ (unless positivity of certain terms shall be stressed). 
 	If $k=3$, \eqref{eq:GFT_structure} reduces to terms of the form $(\partial^3 R)$, $(\partial^2 R)(\partial R)$ and $(\partial R)^3$, corresponding to $n=1$, $2$, and $3$, respectively. 
 	
 	We start with $k=3$, $n=1$,  estimate	one exemplary term  among the various possible ones arising from acting with three derivatives on one factor of $\langle \Im GA \Im GA\rangle$.  There are five types of such terms,   are given by (up to the symmetry of the chain, and ignoring the irrelevant $i$-indices of $a_i, b_i$)
 	\begin{equation} \label{eq:k3n1allterms_ex}
 		\begin{split}
 			&(\Im GA \Im G)_{ba} (GAG)_{ba} G_{ba}\,, \quad  (GA\Im G)_{ba} (GA \Im G)_{ba} G_{ba}\,, \quad (GAG)_{ba} (\Im GA  G)_{ba} (\Im G)_{ba} \,, \quad  \text{and} \\
 			&(\Im GA \Im GAG)_{ba} G_{ba} G_{ba}\,, \quad (GA\Im G A G)_{ba} (\Im G)_{ba} G_{ba} \,. 
 		\end{split}
 	\end{equation}
 	Note that  not only the number of $A$’s but also the number of $\Im G$’s  is preserved along differentiation, but their distribution among the various factors may change.
 	 
 	We estimate the contribution of the first term in \eqref{eq:k3n1allterms_ex} to \eqref{eq:GFT_structure} as follows,
 	\begin{equation} \label{eq:k3n1second_ex}
 		\begin{split}
 			N^{-5/2} &\sum_{\alpha_1, \alpha_2, \alpha_3} \left|\kappa(\alpha_1, \alpha_2, \alpha_3) (\Im G A \Im G)_{b_3 a_1} (GAG)_{b_1 a_2} G_{b_2 a_3}\right| \\
 			\lesssim  & ~N^{-5/2} \Bigl\Vert \sum_{\alpha_2} |\kappa(*, \alpha_2, *)|\Bigr\Vert \biggl( \sum_{\alpha_1, \alpha_3} \max_{\alpha_2} \left| (\Im G A \Im G)_{b_3 a_1} (GAG)_{b_1 a_2} G_{b_2 a_3}\right|^2 \biggr)^{1/2}\\
 			\lesssim &\frac{\Xi^{1/2}}{N^{5/2}\eta}  \vertiii{\kappa}_3  \biggl( \sum_{\alpha_1, \alpha_3} \eta\,(\Im G A^*\Im GA\Im G)_{a_1 a_1} \eta\, (GA\Im G A^*G^*)_{b_1b_1}\biggr)^{1/2}  \\
 			\lesssim &~\Xi \, \frac{\Psi}{\sqrt{N\ell}} \vertiii{\kappa}_3\,,
 		\end{split}
 	\end{equation}
 	where we recall $\ell := \rho\eta$, and  denote  $\Psi := \langle \Im GA\Im G A^* \rangle$, as well as  $\Xi := \eta^{-1} \rho \lesssim \ell^{-1}$. 
 	Here, in the second step,  we used the definition of the $\vertiii{\kappa}_3$-norm, 
 	employed a Schwarz inequality, together with $|G_{ab}|\lesssim 1$, $ (\Im G)_{b b}\lesssim \rho$
 	from the single resolvent entry-wise local law \eqref{eq:singleGzag} to bound
 	\begin{equation} \label{eq:imgaimg_ex}
 		|(\Im GA \Im G)_{b a}| \lesssim \big((\Im GA \Im G A^* \Im G)_{a a}\big)^{1/2} \big((\Im G)_{b b}\big)^{1/2} \lesssim  \Xi^{1/2}\big(\eta\, (\Im GA \Im G A^* \Im G)_{a a}\big)^{1/2}\,,
 	\end{equation}
 	\begin{equation} \label{eq:gag_ex}
 		|(GAG)_{b a} | \lesssim \eta^{-1}\big(\eta\,(GA \Im G A^* G^*)_{b b} \big)^{1/2}.
 	\end{equation}
 	In the final step of \eqref{eq:k3n1second_ex}  we performed the summation over $\alpha_1=(a_1, b_1)$, then used the Ward identity and the norm-bound $\norm{\Im G} \le \eta^{-1}$  to obtain
 	\begin{equation} \label{eq:X_tr}
 		\eta\, \langle \Im G A^*\Im GA\Im G\rangle \le \Psi,  \qquad  \eta \,\langle GA\Im G A^*G^* \rangle \le \Psi. 
 	\end{equation}
 	\nc 
 	
 	As our next example, we consider the case for $k=3$ and $n=3$ in \eqref{eq:GFT_structure}, where we estimate
 	\begin{equation} \label{eq:k3n3_ex}
 		\begin{split}
 			N^{-9/2} &\sum_{\alpha_1, \alpha_2, \alpha_3} \left|\kappa(\alpha_1, \alpha_2, \alpha_3) (\Im GA \Im G A G)_{b_1 a_1} (\Im GA \Im G A G)_{b_2 a_2} (\Im GA \Im G A G)_{b_3 a_3}\right| \\
			\lesssim &~ \frac{\Xi^{1/2} \Psi}{N^{7/2}\eta^2} \Bigl\Vert \sum_{\alpha_3} |\kappa(*, *, \alpha_3)|\Bigr\Vert \biggl( \sum_{\alpha_1} \eta\, (\Im GA \Im G A^* \Im G)_{b_1b_1} \eta\, (GA \Im G A^* G^*)_{a_1a_1}\biggr)^{1/2} \\ 
			&\qquad \times\biggl( \sum_{\alpha_2} \eta\, (\Im GA \Im G A^* \Im G)_{b_2b_2} \eta\, (GA \Im G A^* G^*)_{a_2a_2}\biggr)^{1/2}\\
			\lesssim &~ \ell^{1/2}\Xi^{3/2} \, \Bigl(\frac{\Psi}{\sqrt{N\ell}}\Bigr)^3  \vertiii{\kappa}_3  \lesssim \Xi \, \Bigl(\frac{\Psi}{\sqrt{N\ell}}\Bigr)^3  \vertiii{\kappa}_3\,. 
 		\end{split}
 	\end{equation}
 	Here, we first  bounded every $(\Im GA \Im G A G)_{ba}$ using the Schwarz inequality,
 	\begin{equation} \label{eq:onederiv_ex}
 			|(\Im GA \Im G A G)_{ba} | \lesssim \eta^{-1}\big( \eta\, (\Im GA \Im G A^* \Im G)_{bb}\,  \eta\, (GA \Im G A^* G^*)_{aa}\big)^{1/2},
 	\end{equation}
 	then estimated the two factors involving indices $\alpha_3 = (a_3, b_3)$ using
 	\begin{equation} \label{eq:gagag_max_ex}
 		\eta\,(\Im GA \Im G A^* \Im G)_{bb} \le N\eta\, (\Im G)_{bb} \Psi \lesssim N \rho\eta\, \Psi, \qquad \eta\,(GA \Im G A^* G^*)_{aa} \le N\eta\, \langle GA \Im G A^* G^* \rangle \le N \Psi. 
 	\end{equation}
 	The other two index pairs $\alpha_1 = (a_1, b_1)$ and $\alpha_2 = (a_2, b_2)$ in the second step of \eqref{eq:k3n3_ex} are summed  to produce traces.
 	
 	Finally, we consider the contribution coming from $(\Im G)_{ba}(GAG)_{ba}(\Im G)_{ba}(GAG)_{ba}$ in the case $k=4$ and $n=1$. Using the single-resolvent local law \eqref{eq:singleGzag} for $\Im G$ and \eqref{eq:gag_ex}, we obtain
 	\begin{equation} \label{eq:k4_worst}
 		\bigl\lvert (\Im G)_{b_4a_1}(GAG)_{b_1a_2}(\Im G)_{b_2a_3}(GAG)_{b_3a_4}\bigr\rvert \lesssim \Xi^2 \eta  \big((GA \Im G A^* G^*)_{b_1 b_1} (GA \Im G A^* G^*)_{b_3 b_3} \big)^{1/2}\,, 
 	\end{equation}
 	and hence, summing over $\alpha_2, \alpha_4$ within the definition of the $\vertiii{\kappa}_4$-norm,
 	while using the $\alpha_1, \alpha_3$ summations to recover traces, we obtain
 	\begin{equation} \label{eq:k4_worst_ex}
 		N^{-3} \sum_{\alpha_1, \dots, \alpha_4} \left|\kappa(\alpha_1, \dots, \alpha_4) (\Im G)_{b_4a_1}(GAG)_{b_1a_2}(\Im G)_{b_2a_3}(GAG)_{b_3a_4} \right| \lesssim N^{-1} \Xi^2 \Psi \vertiii{\kappa}_4 \lesssim   \Xi  \, \frac{\Psi}{N\ell} \vertiii{\kappa}_4 \,.
 	\end{equation}
 	In the sequel we will show that the $\Xi^2$ on the right-hand side of \eqref{eq:k4_worst} represents the worst power of $\Xi$ that can appear in a corresponding bound on a derivative of $\langle \Im GA\Im GA \rangle$ of any order (for comparison, the analogous factor is $\Xi^{1/2}$ in our first example and $\Xi^{3/2}$ in the second example). 
	This is because each $\Xi^{1/2}$-power arises from a severed connection between an $A$ and an $\Im G$, of which there are four in the initial tracial quantity $\langle \Im G A \Im G A \rangle$. 
 	
 	We also emphasize that,  although two index-pair summations are always available, traces cannot always be recovered from both. 
	In the second example, the $\alpha_1$ and $\alpha_2$ summations were both effective in this sense, only the $\alpha_3$ summation in the first example, and the $a_1, a_3$ summations in the last example, were ineffective.
 	Whether a summation contributes to a trace depends sensitively on the precise arrangement of the indices. The following procedure provides a systematic way to handle all cases simultaneously. 
 	
 	\smallskip
 	\textbf{General case}. We now present an estimate on \eqref{eq:GFT_structure} for a general partition $(\bm \alpha^{(1)}, \dots, \bm \alpha^{(n)})$ of $\bm \alpha \in \indset{N}^{2k}$.
 	Observe that the summation in \eqref{eq:GFT_structure} is of the form
 	\begin{equation} 
 		\sum_{\bm \alpha} \kappa(\bm \alpha) f(\bm \alpha), \qquad f(\bm \alpha) := \prod_{j=1}^n f_j\bigl(\bm \alpha^{(j)}\bigr).
 	\end{equation}
 	Applying the definition of higher-order cumulant norms $\vertiii{\kappa}_k$ from Assumption \ref{ass:cumulants} to the sum above leaves two index pairs $\alpha_{k-1}, \alpha_k$ that can be summed up in the $\ell^2$-sense after taking the supremum of $|f(\bm \alpha)|$ over the other variables $\alpha_1, \dots, \alpha_{k-2}$. In fact, since $\kappa$ is a totally symmetric function of the index-pairs, any pair of components of $\bm \alpha$ can be chosen for summation. 
 	The partition $(\bm \alpha^{(1)}, \dots, \bm \alpha^{(n)})$ induces two \textit{summation patterns}: both summation indices belong to the same block $\bm \alpha^{(i)}$ (see examples \eqref{eq:k3n1second_ex} and \eqref{eq:k4_worst_ex} above) or the summation indices belong to different blocks (as is the case in example \eqref{eq:k3n3_ex}). 
 	Hence, to estimate \eqref{eq:GFT_structure}, we define the following pair of quasi-norms on functions $f : \indset{N}^{2l} \to \mathbb{C}$,
 	\begin{equation} \label{eq:kappa_dual_norm}
 		\begin{split}
 			\norm{f}_{2,*} &:= \min_{\pi \in S_l} \inf\Biggl\{ \sum_i \biggl(\sum_{\alpha_1,\alpha_2} |g_i(\alpha_1) h_i(\alpha_2)|^2\biggr)^{1/2}\,:\, \max_{\alpha_3, \dots, \alpha_l} \bigl\lvert f\bigl(\pi (\bm \alpha)\bigr)\bigr\rvert \le \sum_i g_i(\alpha_1) h_i(\alpha_2) \Biggr\}, \qquad l \ge 2,\\
 			\norm{f}_{1,*} &:= \min_{\pi \in S_l} \biggl(\sum_{\alpha_1} \max_{\alpha_2, \dots, \alpha_l} \bigl\lvert f\bigl(\pi (\bm \alpha)\bigr)\bigr\rvert^2 \biggr)^{1/2}, \qquad l \ge 1, 			
 		\end{split}		
 	\end{equation}
 	where $S_l$ denotes the set of permutations on $l$ elements, $(\pi(\bm \alpha))_j := \alpha_{\pi(j)}$, and $\bm \alpha = (\alpha_j)_{j=1}^l \in \indset{N}^{2l}$. 
 	The subscript $p$ of the $\norm{\cdot}_{p,*}$ quasi-norm indicates the number of index-pairs that are summed.
 	The minimum over permutations simply expresses the freedom in selecting the summation index pairs. 
 	We also use $\norm{f}_{\infty} := \max_{\alpha_1, \dots, \alpha_l} |f(\bm \alpha)|$. The quasi-norm $\norm{\cdot}_{2,*}$ will be used in the first summation pattern to estimate the function $f_i$ if both summation indices belong to $\bm \alpha^{(i)}$, while the quasi-norm $\norm{\cdot}_{1,*}$ is used in the second summation pattern to estimate the two functions, each depending on a single summation index-pair. In both patterns all other $f_j$ estimated in $\norm{\cdot}_\infty\,$.
 	
 	Furthermore, since in our application the functions $f_j(\bm \alpha^{(j)}) = \partial_{\bm \alpha^{(j)}}R$ are given by sum of many terms, each of which might attain the minimum in \eqref{eq:kappa_dual_norm} with a different permutation $\pi$, we extend the quasi-norms  $\norm{\cdot}_{p,*}$ above to the following norms
 	\begin{equation}
 		\norm{f}_{p,\mathrm{s}} := \inf \Biggl\{ \sum_{j=1}^{C_{l,p}} \norm{f_j}_{p,*} \,:\, f = \sum_{j=1}^{C_{l,p}} f_j \Biggr\}, \qquad C_{l,p} := {l \choose p}, \qquad p \in \{1,2\}.
 	\end{equation} 
 	The $\norm{\cdot}_{p,\mathrm{s}}$-norms allow each term in $f_j$ to be summed over its own minimizing indices.  	
 	These norms give a natural upper bound on the dual to the higher-cumulant norms in the sense that 
 	\begin{equation} \label{eq:norm_is_dual}
 		\biggl\lvert \sum_{\bm \alpha \in \indset{N}^{2k}} \kappa(\bm\alpha) \prod_{j=1}^{n} f_j(\bm \alpha^{(j)}) \biggr\rvert \le \vertiii{\kappa}_k  
 		\biggl(\min_{i\in\indset{n}} \Bigl\{ \norm{f_i}_{2,\mathrm{s}} \prod_{j\neq i }\norm{f_j}_\infty \Bigr\} 
 		\wedge \min_{i_1 \neq i_2\in\indset{n}} \Bigl\{ \norm{f_{i_1}}_{1,\mathrm{s}} \norm{f_{i_2}}_{1,\mathrm{s}} \prod_{j\neq i_1,i_2}\norm{f_j}_\infty \Bigr\}\biggr),
 	\end{equation}
 	for any partition $(\bm\alpha^{(1)}, \dots\, \bm\alpha^{(n)}) \in \indset{N}^{2l_1}\times\dots\times \indset{N}^{2l_n}$ of $\bm \alpha \in \indset{N}^{2k}$ with $\sum_j l_j = k$, where an empty minimum is $+\infty$ by convention. 
 	The two minima to the left and to the right of the $\wedge$ sign on the right-hand side of \eqref{eq:norm_is_dual} correspond the first and second summation patterns, respectively.
 	For $n=1$, clearly, only the first pattern is available and for a general function $f(\bm \alpha)$.
 	However, for $n \ge 2$, the product $\| f_{i_1}\|_{1,s} \| f_{i_2}\|_{1,\mathrm{s}}$ would give our best estimate on $\| f_{i_1} f_{i_2}\|_{2,\mathrm{s}}$ in many cases.

	Indeed, examples \eqref{eq:k3n1second_ex}, \eqref{eq:k3n3_ex} and \eqref{eq:k4_worst_ex} above suggest that for $f_j := \partial_{\bm \alpha^{(j)}}R$, the first summation pattern (involving $\norm{\cdot}_{2,\mathrm{s}}$) in \eqref{eq:norm_is_dual} should be used only if $n=1$, while the second summation pattern (involving two $\norm{\cdot}_{1,\mathrm{s}}$ factors) is used for all $n\ge 2$. 
	Furthermore, for $n\ge 2$, when selecting the two factors $f_j = \delab_{\bm \alpha^{(j)}}R$ to be summed, priority should be given to first-order derivatives, that is, to $\bm \alpha^{(j)} = ((a,b))$. In this case, both summations over $a$ and $b$ can be ensured to be effective, as illustrated in \eqref{eq:k3n3_ex}. 		
	In the sequel, we show that restricting to this summation strategy is indeed sufficient for our purposes, so performing the full optimization in \eqref{eq:norm_is_dual} is not necessary for our proof.
	
 	We now address the structure of arbitrary-order derivatives $\partial_{\bm \alpha} \langle \Im G_1 A_1 \Im G_2 A_2 \rangle$ appearing in \eqref{eq:GFT_structure}. By inductively applying \eqref{eq:partial_ab} it is straightforward to check that for any $\bm{\alpha} = ((a_j,b_j))_{j=1}^{2l} \in \indset{N}^l$,
 	\begin{equation} \label{eq:deriv_struct}
 		(-1)^l\delab^l_{\bm \alpha} \Tr[ \Im G_1 A_1 \Im G_2 A_2] := (-1)^l\Bigl(\prod_{j=1}^{l}\delab_{a_jb_j}\Bigr)  \Tr\bigl[ \Im G_1 A_1 \Im G_2 A_2\bigr]
 	\end{equation}
 	is given by a sum of products of the form
 	\begin{equation} \label{eq:G_seq}
 		\bigl(\G^{(l+2)} A_1 \G^{(1)} \bigr)_{b'_{l}a'_1} \bigl(\G^{(2)}\bigr)_{b'_{1}a'_2}  \dots \bigl(\G^{(r)}\bigr)_{b'_{r-1}a'_r} \bigl(\G^{(r+1)} A_2 \G^{(r+2)} \bigr)_{b'_{r}a'_{r+1}} \bigl(\G^{(r+2)}\bigr)_{b'_{r+1}a'_{r+2}} \dots \bigl(\G^{(l+1)}\bigr)_{b'_{l-1}a'_{l}}~,
 	\end{equation}
 	where $0 \le r \le l$ is an integer, $\bm \alpha' := ((a'_j,b'_j))_{j=1}^l$ is some permutation of $\bm \alpha$, and there exist integers $1 \le l_1 \le r+1$ and $r+2 \le l_2 \le l+2$, such that sequence of resolvents $\{\mathcal{G}^{(i)}\}_{i=1}^{l+2}$ is given by
 	\begin{alignat*}{3} 
 		\G_i &:= G_1, \,\, \text{if} \,\, 1 \le i < l_1, \qquad &\G_i := \Im G_1\,\, \text{if} \,\, i = l_1, \qquad &\G_i := G_1^*, \,\, \text{if} \,\, l_1 < i \le r+1, \nonumber\\
 		\G_i &:= G_2, \,\, \text{if} \,\, r+2 \le i < l_2, \qquad &\G_i := \Im G_2\,\, \text{if} \,\, i = l_2, \qquad &\G_i := G_2^*, \,\, \text{if} \,\, l_2 < i \le l+2.
 	\end{alignat*}
 	To give meaning to $(\dots)$ in the expression \eqref{eq:G_seq} for the extreme cases $r=0$ and $r=l$, we replace, by convention,
 	\begin{equation} \label{eq:G_replace_conv}
 		\begin{split}
 			(\G^{(l+2)} A_1 \G^{(1)} )_{b'_{l}a'_1}\dots (\G^{(r+1)} A_2 \G^{(r+2)} )_{b'_{r}a'_{r+1}}\quad &\mapsto \quad (\G^{(l+2)} A_1 \Im G_1 A_2 \G^{(2)} )_{b'_{l}a'_{1}}, \quad \text{ if}\quad r = 0,\\
 			(\G^{(r+1)} A_2 \G^{(r+2)}  )_{b'_{r}a'_{r+1}} \dots  (\G^{(l+2)} A_1 \G^{(1)}  )_{b'_{l}a'_1}\quad &\mapsto \quad (\G^{(l+1)} A_2 \Im G_2 A_1 \G^{(1)}  )_{b'_{l}a'_1}, \quad \text{ if}\quad r = l.
 		\end{split}
 	\end{equation}
 	The formula \eqref{eq:G_seq} generalizes the terms in \eqref{eq:k3n1allterms_ex} to arbitrary chain lengths, while also encoding the precise index structure.
 	The key structural take-away are as follows: the two observables $A_1, A_2$ can appear either in the same chain,  $(\G_1 A_1 \G_2 A_2 \G_1)$ or $(\G_2 A_2 \G_1 A_1 \G_2)$, or in two separate chains, $(\G_1 A_1 \G_2)$ and $(\G_2 A_2 \G_1)$; the total number of $\G$'s is always $l+2$, with  exactly one $\Im G_1$ and one $\Im G_2$ among them (standalone or part of a longer chain). We remark that each $a_i, b_i$ appears exactly once among the indices of the resolvent chain entries. 
 	
 	We now estimate the elements of all chains which can appear in \eqref{eq:G_seq} including the replacement convention \eqref{eq:G_replace_conv} for the special cases $r\in\{0,l\}$.
 	For a fixed $\G_i \in \{G_i, G_i^*, \Im G_i\}$, let $\mathcal{X}^{\G_i}$ denote a matrix of the form
 	\begin{equation} \label{eq:X_form}
 		\mathcal{X}^{\G_i} \equiv\mathcal{X}^{\G_i}(A) 
 		= \mathcal{U} (\Im G_i)^{1/2} A \Im G_j A^* (\Im G_i)^{1/2} \mathcal{U}^*, \quad \mathcal{U} \equiv \mathcal{U}(\G_i) := \begin{cases}
 			\G_i |\G_i \G_i^*|^{-1/2}, \,\,  &\text{if}~ \G_i \in \{G_i, G_i^*\}\\
 			(\eta_i\Im G_i)^{1/2}, \,\, &\text{if}~ \G_i = \Im G
 		\end{cases},
 	\end{equation}
 	where $A \in \{A_i, A_j^*\}$, and $(i,j) \in \bigl\{(1,2), (2,1)\bigr\}$. We use the matrices $\mathcal{U}$ allow us to unify $\eta_i \Im G_i A \Im G_j A^*\Im G_i$ and $\eta_i G_i^* A \Im G_j A^* G_i$ (appearing in the examples \eqref{eq:k3n1second_ex} and \eqref{eq:k3n3_ex} above without indices), while the choices of the observable $A$ based on $(i,j)$ guarantee that the chain $\G_i A \G_j$ is consistent with the regularity of $A$. 
 	Note that for any choice of $\G$, the matrices $\mathcal{X}^{\G}$ and $\mathcal{U}(\G)$, defined in \eqref{eq:X_form}, satisfy
 	\begin{equation}
 		\mathcal{U}^*\mathcal{U} \le I, \qquad \mathcal{X}^{\G} \ge 0. 
 	\end{equation}
	The positive operators $\mathcal{X}^{\G}$ serve as an intermediate tool in the proof.
	
	First, we estimate every factor in~\eqref{eq:G_seq}
	in terms of the diagonal elements of $\mathcal{X}^{\G}$.  
 	Analogously to \eqref{eq:imgaimg_ex}, \eqref{eq:gag_ex} and \eqref{eq:onederiv_ex}, by using the Cauchy--Schwarz inequality, we have, with each $\G_i$ representing a (potentially different) matrix from the set $\{G_i, G_i^*, \Im G_i\}$,
 	\begin{equation} \label{eq:Schwarz_bounds}
 		\begin{split}
 			\bigl\lvert (\G_i A_i \Im G_j A_j G_i)_{ab} \bigr\rvert &\le \eta_i^{-1}  \bigl(\mathcal{X}_{aa}^{\G_i}(A_i) \bigr)^{1/2}  \bigl(\mathcal{X}_{bb}^{G_i^*}(A_j^*) \bigr)^{1/2}, 
 			\quad \bigl\lvert (G_i^* A_i G_j)_{ab} \bigr\rvert \le (\eta_i\eta_j)^{-1/2}  \bigl(\mathcal{X}^{G_i^*}_{aa}(A_i) \bigr)^{1/2},\\
 			\quad \bigl\lvert (\G_i A_i \Im G_j)_{ab} \bigr\rvert &\lesssim (\eta_i^{-1}\rho_j)^{1/2}  \bigl(\mathcal{X}^{\G_i}_{aa}(A_i) \bigr)^{1/2},
 			\quad \bigl\lvert (\Im G_j A_j G_i)_{ab} \bigr\rvert \lesssim (\eta_i^{-1}\rho_j)^{1/2}  \bigl(\mathcal{X}^{G_i^*}_{bb}(A_j^*) \bigr)^{1/2},
 		\end{split}
 	\end{equation}
 	where $\{(i,j)\} \in \{(1,2), (2,1)\}$, and $\mathcal{X}^{\G}(A)$ are defined in \eqref{eq:X_form}. 
 	A bound analogous to the first inequality in \eqref{eq:Schwarz_bounds} holds for $(G_i^* A_i \Im G_j A_j \G_i)_{ab}$ by symmetry.
 	Note that the estimates \eqref{eq:Schwarz_bounds} decouple the indices $a,b$, which facilitates summing them independently later. 
 	
 	Next, we estimate the diagonal elements of $\mathcal{X}^{\G}$ appearing in \eqref{eq:Schwarz_bounds} either in tracial sense (if their index  can be summed up effectively) or by its maximum-norm (if the corresponding index can not be summed up), using the following bounds,
 	Therefore, since $\mathcal{X}^{\G_i} \ge 0$, we have
 	\begin{equation} \label{eq:X_bounds}
 		\bigl\langle \mathcal{X}^{\G_i}(A) \bigr\rangle \le \langle \Im G_i A \Im G_j A^* \rangle, \qquad \norm{\mathcal{X}^{\G_i}(A) }_\mathrm{\max} \le N \langle \Im G_i A \Im G_j A^* \rangle \nc \cdot \begin{cases}
 			\rho_i\eta_i, \quad &\text{if}~  \G_i = \Im G_i,\\
 			1,\quad & \text{otherwise},
 		\end{cases}
 	\end{equation}
 	where $j=3-i$ is the complementary index to $i$ in the sense that $\{i,j\} = \{1,2\}$. \nc The tracial estimate in \eqref{eq:X_bounds} should be compared to \eqref{eq:X_tr}, and the $\norm{\cdot}_{\max}$-bound is the analog of \eqref{eq:gagag_max_ex}.

 	We define two deterministic control parameters $\ell, \Xi$, and a random control quantity $\Psi$,
 	\begin{equation} \label{eq:GFT_control}
 		\ell := \min_i\{\eta_i \rho_i \}, \quad \Xi := \max_i \bigl\{\eta_i^{-1} \rho_i\bigr\}	\quad \Psi := \langle \Im G_1 A_1 \Im G_2 A_1^* \rangle^{1/2}\langle \Im G_1 A_2^* \Im G_2 A_2 \rangle^{1/2}.
 	\end{equation}
 	Then, by plugging the first estimate in \eqref{eq:Schwarz_bounds} into \eqref{eq:G_seq} with the replacement convention \eqref{eq:G_replace_conv}, and using \eqref{eq:X_bounds}, we deduce that (c.f., \eqref{eq:k3n3_ex}), with very high probability\footnote{Note that $\infty$-norm here is taken over $\alpha$ and not the probability space.}, 
 	\begin{equation} \label{eq:d1_GFT}
 		\norm{\partial_{\alpha} \langle \Im G_1 A_1 \Im G_2 A_2 \rangle}_{1,\mathrm{s}} \lesssim \frac{1}{\sqrt{N\ell}} \sqrt{N} \Xi^{1/2} \Psi, \qquad \norm{\partial_{\alpha} \langle \Im G_1 A_1 \Im G_2 A_2 \rangle}_{\infty} \lesssim \Xi^{1/2}\Psi,
 	\end{equation}
 	where $\ell, \Xi, \Psi$ are defined in \eqref{eq:GFT_control}, and we used that $ \eta_i^{-2} \le \eta_i^{-1}\rho_i\,\Xi \le  \ell^{-1} \Xi$. 
 	The bounds \eqref{eq:d1_GFT} will be used to estimate \eqref{eq:deriv_struct} with $l=1$. Since 
 	$\partial_\alpha R$ depends only on one index pair $\alpha=(a,b)$, we estimate it only in the $\norm{\cdot}_{1,\mathrm{s}}$-norm.
 	
 	To estimate \eqref{eq:deriv_struct} for $l\ge 2$, after writing them it out in the form \eqref{eq:G_seq}--\eqref{eq:G_replace_conv},
 	we bound all the entries of single-resolvent objects using $\norm{G_i}_{\max} \lesssim 1$ and $\norm{\Im G_i}_{\max} \lesssim \rho_i$ by \eqref{eq:singleGzag}. 
 	What remains is either a single resolvent chain with two observables, or two chains with one observables each, which we estimate using \eqref{eq:Schwarz_bounds}. 
 	We find that, for any $l \ge 2$, every product of the form \eqref{eq:G_seq}--\eqref{eq:G_replace_conv} admits the bound
 	\begin{equation} \label{eq:GFT_generic_bound}
 		\begin{split}
 			&\Bigl(\sqrt{\frac{\rho_i}{\eta_i}} + \sqrt{\frac{\rho_1\rho_2}{\eta_1\eta_2}} \Bigr) \frac{1}{\sqrt{\rho_i\eta_i}}  \bigl(\mathcal{X}_{aa}^{\Im G_i}(A_i)  \mathcal{X}_{bb}^G(A_j) \bigr)^{1/2} \\
 			&~+  \Bigl(\sqrt{\frac{\rho_1\rho_2}{\eta_1\eta_2}} +  \frac{\rho_i}{\eta_i} + \mathbf{1}_{l\ge 3} \, \sqrt{\frac{\rho_1\rho_2}{\eta_1\eta_2}} \sqrt{\frac{\rho_i}{\eta_i}} + \mathbf{1}_{l\ge 4} \, \frac{\rho_1 \rho_2}{\eta_1\eta_2}   \Bigr) 
 			\bigl(\mathcal{X}_{aa}^G(A_1)  \mathcal{X}_{bb}^G(A_2) \bigr)^{1/2}\\
 			&~\lesssim \frac{\Xi}{\sqrt{\rho_i\eta_i}}  \bigl(\mathcal{X}_{aa}^{\Im G_i}(A_i)  \mathcal{X}_{bb}^G(A_j) \bigr)^{1/2} 
 			+   \Xi^{\min\{2, l/2\}}  \,
 			(\mathcal{X}_{aa}^G(A_1)  \mathcal{X}_{bb}^G(A_2) )^{1/2},
 		\end{split}
 	\end{equation}
 	where $a,b$ are two distinct indices from among the components of $\bm \alpha$, the matrices $\mathcal{X}^{\Im G}, \mathcal{X}^G$ are of the form \eqref{eq:X_form}, with the superscript $G \in \{G_1, G_1^*, G_2, G_2^*\}$, and $i$ is some index from $\{1,2\}$. Here in the second step we used the definition of~$\Xi$ from \eqref{eq:GFT_control}. 
 	Therefore, for derivatives of order $l \ge 2$, we obtain
 	\begin{equation} \label{eq:dl_GFT}
 		\begin{split}
 			\norm{\partial^l_{\bm \alpha} \langle \Im G_1 A_1 \Im G_2 A_2 \rangle}_{2,\mathrm{s}} &\lesssim  N\bigl(\ell^{-1/2}\Xi + \Xi^{\min\{2,l/2\}} \bigr) \Psi,\\
 			\norm{\partial^l_{\bm \alpha} \langle \Im G_1 A_1 \Im G_2 A_2 \rangle}_{1,\mathrm{s}} &\lesssim N^{1/2} \Xi^{\min \{2, l/2\}} \Psi, \quad  \norm{\partial^l_{\bm \alpha} \langle \Im G_1 A_1 \Im G_2 A_2 \rangle}_{\infty} \lesssim \Xi^{\min \{2, l/2\}} \Psi,
 		\end{split}
 	\end{equation}  
 	where we use the superscript $l$ in the notation $\partial^l_{\bm \alpha}$ to emphasize the order of the derivative, i.e., $\partial^l_{\bm \alpha} \equiv \partial_{\bm \alpha}$ for $\bm \alpha \in \indset{N}^{2l}$, and the $N^{p/2}$ factors in the $\norm{\cdot}_{p,\mathrm{s}}$-bounds with $p\in \{1,2\}$ arise from the remaining indices which cannot be summed effectively.  
 	Here, to obtain the $\norm{\cdot}_{1,\mathrm{s}}$-bounds, since we can not guarantee that $a,b$ belong to the same $\alpha$, we estimate $(\rho_i\eta_i)^{-1/2}(\mathcal{X}^{\Im G_i}_{aa})^{1/2}$ and $(\mathcal{X}^{G}_{aa})^{1/2}$ from~\eqref{eq:GFT_generic_bound} by their $\norm{\cdot}_{\max}$-norm, using the second bound in~\eqref{eq:X_bounds}, and sum the remaining $ \mathcal{X}^{G}_{bb}$. \nc

 	Finally, we estimate the desired \eqref{eq:GFT_structure}.
 	If $n=1$, using the first summation pattern in  \eqref{eq:norm_is_dual}, and the bounds \eqref{eq:dl_GFT}, we deduce that for any $k \ge 3$, with very high probability,
 	\begin{equation} \label{eq:order_k_bound_n1} 
		N^{-k/2}\Bigl\lvert \sum_{\bm \alpha \in \indset{N}^{2k}} \kappa(\bm\alpha) \partial^k_{\bm \alpha} \langle \Im G_1 A_1 \Im G_2 A_2 \rangle   \Bigr\rvert 
		\lesssim N^{1-k/2}\bigl(\ell^{-1/2}\Xi + \Xi^{\min\{2,k/2\}} \bigr) \Psi \lesssim \Xi \frac{\Psi}{\sqrt{N\ell}},
	\end{equation}
	where we used $N^{-1} \ll \ell \lesssim 1$, and  $\Xi \lesssim \ell^{-1}$, which follows from \eqref{eq:GFT_control} and $\max_i\rho_i \lesssim 1$. 	 

 	If $n\ge 2$, we use the second summation pattern in \eqref{eq:norm_is_dual},  and deduce from \eqref{eq:d1_GFT}--\eqref{eq:dl_GFT} that, for any $k \ge 2$, with very high probability,
 	\begin{equation} \label{eq:order_k_bound_nge2}
 		N^{-k/2}\Bigl\lvert \sum_{\bm \alpha \in \indset{N}^{2k}} \kappa(\bm\alpha) \prod_{j=1}^{n} \partial^{l_j}_{\bm \alpha^{(j)}} \langle \Im G_1 A_1 \Im G_2 A_2 \rangle   \Bigr\rvert \lesssim \Xi \biggl(\frac{\Psi}{\sqrt{N\ell}}\biggr)^n \times \ell^{n/2} N^{1-(k-n)/2}  \frac{\Xi^{J/2-1}}{(N\ell)^{n_1/2}} ,
 	\end{equation}
 	where $J := \sum_{j\in\indset{n}} \min\{4, l_j\}$, and  $n_1 := 2 \wedge \# \{j \in\indset{n} \, : \, l_j = 1 \}$.
 	The factor $(N\ell)^{-n_1/2}$ is collected from the additional $(N\ell)^{-1/2}$ factors appearing in the $\norm{\cdot}_{1,\mathrm{s}}$-bound on $\delab^1R$ in \eqref{eq:d1_GFT} compared to $\delab^lR$ for $l \ge 2$ in \eqref{eq:dl_GFT}. 
 	It is straightforward to verify that for all $k \ge 2$, the quantities $J$ and $n_1$ defined above satisfy
 	\begin{equation}
 		J \le k, \qquad n_1 \ge 2 - k + n.
 	\end{equation}
 	Therefore, the factor following the $\times$ sign on the right-hand side of \eqref{eq:order_k_bound_nge2} admits the bound
 	\begin{equation}
 		\ell^{\frac{n}{2}} N^{1-\frac{k-n}{2}}   \frac{\Xi^{\frac{J}{2}-1}}{(N\ell)^{n_1/2}}  
 		=  (N\ell)^{\frac{2-k+n - n_1}{2}}  \ell^{\frac{k-J}{2}}  \lesssim 1,\qquad k \ge 3,
 	\end{equation} 
 	where we again used $N^{-1} \ll \ell \lesssim 1$, and  $\Xi \lesssim \ell^{-1}$. 	\nc 

To summarize, we showed that, for any $k \ge 3$, 
	\begin{equation} \label{eq:zagfinalIm}
		\begin{split}
	&N^{-k/2} \left| \sum_{\alpha_1, \dots, \alpha_k} \kappa(\alpha_1, \dots, \alpha_k) \E\Bigl[\partial_{\alpha_1}\dots\partial_{\alpha_{k}} \bigl\lvert R_s^{A_1, A_2} \bigr\rvert^p\Bigr]\right| \\
	& \qquad \lesssim  \, \Xi \, \E \left[\sum_{n=1}^{k \wedge p}\biggl(\frac{\Psi}{\sqrt{N\ell}}\biggr)^n |R_s^{A_1, A_2}|^{p-n}\right] \lesssim N^{-\delta}\Xi \left(  \E\bigl[\big|R_s^{A_1, A_2}\big|^{p}\bigr]  + \sqrt{\Phi_\delta^p(s; A_1, A_1^*) \Phi_\delta^p(s; A_2^*, A_2)}\right)
		\end{split}
\end{equation}
by Young inequalities, additionally using (recall the definition of $\Psi$ in \eqref{eq:GFT_control} and $\Phi$ in \eqref{eq:Phidef})
\begin{equation*}
\E \left[\left|\frac{\Psi}{\sqrt{N \ell}}\right|^p\right] \lesssim \sqrt{\Phi_\delta^p(s; A_1, A_1^*) \Phi_\delta^p(s; A_2^*, A_2)} \,. 
\end{equation*}
This concludes the proof of Proposition \ref{prop:gronwallImG}~(i) and we are left with discussing the modifications needed for (ii)+(iii). 

\smallskip

\textbf{Modifications for Proposition \ref{prop:gronwallImG}~(ii)+(iii)} In both cases, the above proof is largely unchanged. The key difference is that an extra factor of $(N \ell)^{-1/2} \ll 1$ has to be gained compared to the proof of part (i). This is achieved, roughly speaking, since, due to the presence of the identity matrix as an observable, the operator norm of $\Im G_1 \Im G_2$ is a more effective estimate than the (unnormalized) trace norm, effectively gaining a factor $(N \ell)^{-1} \ll 1$. That is, we have
\begin{equation} \label{eq:improve}
	\begin{split}
\text{the "old" bound:} \quad \Vert \Im G_1 \Im G_2 \Vert_{\max} &\le N \langle \Im G_1 \Im G_2 \rangle \lesssim N |\rho_1 \rho_2| \, \gamma(z_1, z_2) \quad \text{and} \quad \\
\text{the "new" bound:} \quad \Vert \Im G_1 \Im G_2 \Vert_{\max} &\le 
\Vert \Im G_1 \Im G_2 \Vert \lesssim \ell^{-1} |\rho_1 \rho_2| \, \gamma(z_1, z_2) \,, 
	\end{split}
\end{equation}
where we employed the norm bound \eqref{eq:imgimgnorm} and \eqref{eq:gamma}. This makes, in particular, reduction-type estimates superfluous. 

In the general discussion above, this effect is manifest in the fact that second bound in \eqref{eq:X_bounds} can be improved  to  
\begin{equation}
\Vert \mathcal{X}^{\G_i}(I) \Vert_{\max}  \le \Vert \Im G_i \Im G_j \Vert \, \Vert \mathcal{U}^*(\G_i) \mathcal{U}(\G_i) \Vert_{\max} \lesssim \ell^{-1} |\rho_1 \rho_2| \, \gamma(z_1, z_2) \cdot \begin{cases}
\rho_i \eta_i \,, \quad &\text{if} \ \G_i = \Im G_i \\
1\,, \qquad &\text{otherwise} \,. 
\end{cases}
\end{equation}
We conclude this section by illustrating this effect in one exemplary low order term each for both parts (ii) and (iii), using freely the notation from above. 

For part (ii), we consider the analog of \eqref{eq:k3n1second_ex}, now carrying indices: 
	\begin{equation}
	\begin{split}
		N^{-5/2} &\sum_{\alpha_1, \alpha_2, \alpha_3} \left|\kappa(\alpha_1, \alpha_2, \alpha_3) (\Im G_1  \Im G_2)_{b_3 a_1} (G_2G_1)_{b_1 a_2} (G_1)_{b_2 a_3}\right| \\
		\lesssim  & ~N^{-5/2} \Bigl\Vert \sum_{\alpha_2} |\kappa(*, \alpha_2, *)|\Bigr\Vert \biggl( \sum_{\alpha_1, \alpha_3} \max_{\alpha_2} \left| (\Im G_1  \Im G_2)_{b_3 a_1} (G_2G_1)_{b_1 a_2} (G_1)_{b_2 a_3}\right|^2 \biggr)^{1/2}\\
		\lesssim &\frac{\Xi}{N}  \vertiii{\kappa}_3  \biggl(\frac{1}{N} \sum_{a_1, b_3} \left|(\Im G_1  \Im G_2)_{b_3 a_1}\right|^2 \biggr)^{1/2} \lesssim  \frac{\Xi}{N}   \langle \Im G_1 \Im G_2 \Im G_1 \Im G_2 \rangle^{1/2} \\
		\lesssim &~  \Xi \, \frac{(\rho_1 \rho_2)^{1/2}}{N\ell} \gamma^{1/2} \langle \Im G_1 \Im G_2 \rangle^{1/2}\,,
	\end{split}
\end{equation}
where we could even generously add an extra factor $\ell^{1/2}$ in the denominator in the last step and additionally used the second estimate in \eqref{eq:improve}. 

Finally, we consider the analog of \eqref{eq:k3n3_ex} for part (iii), also now carrying indices (but abbreviating $A \equiv A_1$): 
 	\begin{equation} 
	\begin{split}
		N^{-9/2} &\sum_{\alpha_1, \alpha_2, \alpha_3} \left|\kappa(\alpha_1, \alpha_2, \alpha_3) (\Im G_1 A \Im G_2 G_1)_{b_1 a_1} (\Im G_2 \Im G_1 A G_2)_{b_2 a_2} (\Im G_1 A \Im G_2 G_1)_{b_3 a_3}\right| \\
		\lesssim &~ \Xi^{1/2} \frac{ (\rho_1 \rho_2)^{1/2} \gamma^{1/2}}{N^{3}\ell^{1/2}} \langle \Im G_1 A \Im G_2 A^* \rangle^{1/2} \Bigl\Vert \sum_{\alpha_3} |\kappa(*, *, \alpha_3)|\Bigr\Vert \biggl( \frac{1}{N}\sum_{a_1, b_1} \left|(\Im G_1 A \Im G_2 G_1)_{b_1 a_1} \right|^2\biggr)^{1/2} \\ 
		&\qquad \times\biggl( \frac{1}{N}\sum_{a_2, b_2} \left|  (\Im G_2 \Im G_1 A G_2)_{b_2 a_2} \right|^2 \biggr)^{1/2}\\
		\lesssim &~ \Xi^{1/2} \frac{ (\rho_1 \rho_2)^{3/2} \gamma^{3/2}}{N^{3}\ell^{3/2} \eta_1 \eta_2 } \langle \Im G_1 A \Im G_2 A^* \rangle^{3/2} \vertiii{\kappa}_3  \lesssim \Xi \, \frac{ \big((\rho_1 \rho_2)^{3} \gamma^{3} \langle \Im G_1 A \Im G_2 A^* \rangle^{3}\big)^{1/2}}{(N \ell)^{3} } 
		\,, 
	\end{split}
\end{equation}
where we used the second bound in \eqref{eq:improve}. 
\begin{equation}
\big| (\Im G_1 A \Im G_2 G_1)_{b_3 a_3} \big| \lesssim \left(N \Xi  \langle \Im G_1 A \Im G_2 A^* \rangle \frac{\rho_1 \rho_2 \gamma}{\ell}\right)^{1/2} \,. 
\end{equation}
This concludes the proof of Proposition \ref{prop:gronwallImG}. 
\qed

 	\subsubsection{Gronwall estimate for $\langle GAG \rangle$: Proof of Proposition \ref{prop:gronwallG}} \label{subsubsec:zagav2G1A}
 	The goal of this section is to give the proof of Proposition~\ref{prop:gronwallG}, relying on a cumulant expansion exactly as in \eqref{eq:GFTstart}--\eqref{eq:cumexres}. In this section, we focus on presenting some representative low-order example terms with $k=3, n=1$, which require some new ingredient to the proof, namely the $\vertiii{\kappa}_{3}^{\rm av}$-norm and an isotropic two-resolvent bound, that was not discussed earlier (cf.~\eqref{eq:kappa3av} and \eqref{eq:G^2law12} below). The general abstract argument encompassing all other terms, which we presented in the previous section, can easily be adopted to the present setting, and we explain the minor required changes after completing the discussion of the low-order terms that involve the $\vertiii{\kappa}_{3}^{\rm av}$-norm.  For further notational simplicity, we abbreviate $A \equiv A_1$ and adopt the notations (cf.~\eqref{eq:GFT_control})
 	\begin{equation*}
	\ell := \min_i\{\eta_i |\rho_i| \}, \quad \Xi := \max_i \bigl\{\eta_i^{-1} |\rho_i|\bigr\}	\quad \Psi := |\langle \Im G_1 A \Im G_2 A^* \rangle| \,. 
 	\end{equation*}

 	As the first exemplary term for $k=3, n=1$, we estimate
 	\begin{equation} \label{eq:k3n1GAG1}
 		\begin{split}
 			N^{-5/2} \sum_{\alpha_1, \alpha_2 ,\alpha_3} \left| \kappa(\alpha_1, \alpha_2, \alpha_3)  \big(G_1 AG_2G_1\big)_{b_3 a_1} (G_1)_{b_1 a_2} (G_1)_{b_2 a_3}\right| 
 		\end{split}
 	\end{equation}
 	and decompose $(G_1)_{b_1 a_2}$ and  $(G_1)_{b_2 a_3}$ as $G = M + (G-M)$ and estimate the resulting four terms separately. In particular using $\Vert M_1 \Vert \lesssim 1$ and the entry-wise single resolvent law \eqref{eq:singleGzag}, we have that, for any $\delta > 0$, 
\begin{equation} \label{eq:k3n1firstGAG}
 		\begin{split}
		&N^{-5/2} \sum_{\alpha_1, \alpha_2 ,\alpha_3} \left| \kappa(\alpha_1, \alpha_2, \alpha_3)  \big(G_1 AG_2G_1\big)_{b_3 a_1} (G_1-M_1)_{b_1 a_2} (M_1)_{b_2 a_3}\right| \\
		\lesssim &N^{-3 + \delta} \frac{|\rho_1| }{\eta_1\eta_2^{1/2}} \sum_{\alpha_1, \alpha_2 ,\alpha_3} \left| \kappa(\alpha_1, \alpha_2, \alpha_3)  \big(\big(G_1A\Im G_2 A^* G_1^*\big)_{b_3 b_3}\big)^{1/2} \right| \\
		\lesssim & N^{-3 + \delta}  \frac{|\rho_1| }{\eta_1\eta_2^{1/2}}  \left\Vert \sum_{\alpha_1} |\kappa(\alpha_1, *,*)|\right\Vert \left(\sum_{a_3, b_3}\big| \big(G_1A\Im G_2 A^* G_1^*\big)_{b_3 b_3}\big|\right)^{1/2} \left(\sum_{a_2, b_2} 1\right)^{1/2} \\
		\lesssim & \vertiii{\kappa}_3 \Xi^{1/2} N^\delta\frac{ |\langle \Im G_1A \Im G_2A^* \rangle|^{1/2}}{\sqrt{N \ell} \sqrt{N\eta_1 \eta_2} } \lesssim \left(1 + N^{-\delta} \Xi \right)N^{3\delta}\sqrt{\frac{ \Psi}{(N \ell) (N\eta_1 \eta_2) }} \,. 
 		\end{split}
\end{equation}
 The terms with $M_{b_1 a_2} (G-M)_{b_2 a_3}$ and $(G-M)_{b_1 a_2} (G-M)_{b_2 a_3}$ are treated analogously and we are hence left with the $M_{b_1 a_2} M_{b_2 a_3}$ term. Here, as the main novelty compared to the previous section, we need to use the $\vertiii{\kappa}_3^{\rm av}$ norm, and estimate
 	\begin{equation} \label{eq:kappa3av}
 		\begin{split}
 			&N^{-5/2} \sum_{\alpha_1, \alpha_2 ,\alpha_3} \left| \kappa(\alpha_1, \alpha_2, \alpha_3)  \big(G_1AG_2G_1\big)_{b_3 a_1} (M_1)_{b_1 a_2} (M_1)_{b_2 a_3}\right| \\
 			\le \, & N^{-1} \vertiii{\kappa}_3^{\rm av} \Vert M \Vert^2 \Vert G_1AG_2G_1\Vert_{\rm hs} \lesssim \Xi^{1/2} \sqrt{\frac{ \Psi}{(N \ell) (N\eta_1 \eta_2) }} \,. 
 		\end{split}
 	\end{equation}
 	
 	We now turn to the second exemplary term for $k=3, n=1$: 
 	\begin{equation} \label{eq:k3n1GAG2}
 		\begin{split}
 			N^{-5/2} \sum_{\alpha_1, \alpha_2 ,\alpha_3} \left| \kappa(\alpha_1, \alpha_2, \alpha_3)  (G_1AG_2)_{b_3 a_1} (G_2G_1)_{b_1 a_2} (G_1)_{b_2 a_3}\right| \,. 
 		\end{split}
 	\end{equation}
 	To estimate it, we need another new ingredient not present in the previous section: In fact, we need the following local law for $(G_2 G_1)_{b_1 a_2}$, which easily follows by Cauchy's integral formula (see Footnote \ref{ftn:cauchy}) and the usual single resolvent law:
 	\begin{equation} \label{eq:G^2law12}
 		\left|\Big(G_2 G_1-\frac{M_2 - M_1}{z_2 - z_1}\Big)_{b_1 a_2}\right| \prec \frac{1}{\min_i \eta_i} \sqrt{\frac{\Xi}{N}}  \quad \text{with} \quad \left\Vert \frac{M_2 - M_1}{z_2 - z_1} \right\Vert \lesssim \Xi \,. 
 	\end{equation}
 	Then, performing a similar decomposition as in \eqref{eq:k3n1GAG1}, we find that, in particular
 	\begin{equation*}
 		\begin{split}
 			&N^{-5/2} \sum_{\alpha_1, \alpha_2 ,\alpha_3} \left| \kappa(\alpha_1, \alpha_2, \alpha_3)  \big(G_1AG_2\big)_{b_3 a_1} \Big(G_2 G_1-\frac{M_2 - M_1}{z_2 - z_1}\Big)_{b_1 a_2} (M_1)_{b_2 a_3}\right| \\
 			\lesssim &N^{-3 + \delta} \frac{1}{(\min_i \eta_i) \eta_2^{1/2}}\Xi^{1/2} \sum_{\alpha_1, \alpha_2 ,\alpha_3} \left| \kappa(\alpha_1, \alpha_2, \alpha_3)  \big|\big(G_1A\Im G_2 A^* G^*_1\big)_{b_3 b_3}\big|^{1/2} \right| \\
 			\lesssim & \vertiii{\kappa}_3\Xi^{1/2} \frac{N^\delta}{(\min_i \eta_i) \, \sqrt{\eta_1 \eta_2}}\frac{ \langle \Im GA \Im GA^* \rangle^{1/2}}{N} \lesssim \Xi N^\delta \sqrt{\frac{ \Psi}{(N \ell) (N\eta_1 \eta_2) }}  \,. 
 		\end{split}
 	\end{equation*}
 	
 	The other terms containing at least one $G-M$ can be handled similarly and we are hence left with the $\big(\tfrac{M_2 - M_1}{z_2 - z_1}\big)_{b_1 a_2} (M_1)_{b_2 a_3}$ term. Again using the $\vertiii{\kappa}_3^{\rm av}$ norm, we estimate 
 	\begin{equation}
 		\begin{split}
 			&N^{-5/2} \sum_{\alpha_1, \alpha_2 ,\alpha_3} \left| \kappa(\alpha_1, \alpha_2, \alpha_3)  \big(G_1AG_2\big)_{b_3 a_1} \Big(\frac{M_2 - M_1}{z_2 - z_1}\Big)_{b_1 a_2} (M_1)_{b_2 a_3}\right| \\
 			\le \, & N^{-1} \vertiii{\kappa}_3^{\rm av} \Vert M \Vert \big\Vert \tfrac{M_2 - M_1}{z_2 - z_1}\big\Vert  \Vert GAG \Vert_{\rm hs} \lesssim\sqrt{\frac{ \Psi}{(N \ell) (N\eta_1 \eta_2) }}  \,. 
 		\end{split}
 	\end{equation}

 	As discussed above, all other terms arising in the cumulant expansion can be handled similarly to the general treatment in Section \ref{subsubsec:zagav2G}. In fact, the previous proof works with only minor alterations, namely the bounds on resolvent chains from \eqref{eq:Schwarz_bounds} are replaced by
 	\begin{equation*} 
 			\bigl\lvert (G_i A G_j G_i)_{ab} \bigr\rvert \lesssim \sqrt{\frac{|\rho_i| }{\eta_i^2 \eta_j}} \bigl|\mathcal{X}_{aa}^{G_i}(A) \bigr|^{1/2}, 
 			\quad \bigl\lvert (G_i A G_j)_{ab} \bigr\rvert \le (\eta_i^2\eta_j)^{-1/2}  \bigl|\mathcal{X}^{G_i}_{aa}(A)\bigr|^{1/2},
 			\quad \bigl\lvert (G_i G_j)_{ab} \bigr\rvert \lesssim \sqrt{\frac{|\rho_i \rho_j|}{\eta_i\eta_j}} \,. 
 	\end{equation*}
Armed with these bounds and using \eqref{eq:X_bounds}, the proof can be concluded exactly as in the previous section; we leave the details to the reader.  
 	
To summarize the outcome of our argument, we have that, similarly to \eqref{eq:zagfinalIm}, and additionally using that $\Psi \lesssim |\rho_1 \rho_2| \vertiii{A}_{z_1, z_2}^2$ by assumption (see \eqref{eq:Rdefinition} and Proposition \ref{prop:Mt_bound}),  	
 		\begin{equation*} 
 		\begin{split}
 			&N^{-k/2} \left| \sum_{\alpha_1, \dots, \alpha_k} \kappa(\alpha_1, \dots, \alpha_k) \E\Bigl[\partial_{\alpha_1}\dots\partial_{\alpha_{k}} \bigl\lvert S_s^{A} \bigr\rvert^p\Bigr]\right| \\
 			 \lesssim  \, & \Xi \, \E \left[\sum_{n=1}^{k \wedge p}N^{2 \delta}\biggl(\sqrt{\frac{\Psi}{(N\ell) (N \eta_1 \eta_2)}}\biggr)^n |S_s^{A}|^{p-n}\right] \\
 			 \lesssim \, & N^{-\delta}\Xi \left(  \E\bigl[\big|S_s^{A}\big|^{p}\bigr]  + \left(N^{3 \delta} \sqrt{\frac{|\rho_1 \rho_2|}{(N \ell) (N \eta_1 \eta_2)}} \vertiii{A}_{z_1, z_2}\right)^p\right) \,. 
 		\end{split}
 	\end{equation*}
 	 	This concludes the proof of Proposition \ref{prop:gronwallG}. \qed
 	 	
 	\section{Characteristic Flow Analysis} \label{sec:M_flow}
	 In this section, we analyze the time dependence of  the quantity $z_{1,t} - z_{2,t}$ and derive the properties of the characteristic flow used in Section~\ref{sec:Mboundsproof}. 
	In particular, we prove Lemma~\ref{lemma:flow}. 
	The study of time-independent objects is deferred to Section~\ref{sec:pert}. 
	Before proceeding to the proofs, we first collect the necessary preliminaries, as established in \cite{AEK2020}. \nc 
	
	\subsection{Preliminaries: One-Body Analysis}
	The following lemma summarizes the basic properties of the solution to the matrix Dyson equation.  
	\begin{lemma} [Propositions~2.1 and 3.5 in \cite{AEK2020}] \label{lemma:M_structure} 
		Let $M(z)$ be the solution to \eqref{eq:MDE} with the data-pair $(D, \mathcal{S})$ satisfying $\lVert D\rVert \le C$ and the flatness estimates \eqref{eq:flatness} for some positive constants $c, C \sim 1$.
		Let $\rho(x):= \lim_{\eta \to +0} \langle M(x + \ii\eta) \rangle$ denote the corresponding self-consistent density of states. 
		Then, for all $z \in \mathbb{C}\backslash\mathbb{R}$, 
		\begin{equation} \label{eq:invM_bound}
			\bigl\lVert M(z)\bigr\rVert_{\mathrm{hs}}\lesssim 1, \quad \bigl\lVert  M(z)\bigr\rVert \le \frac{1}{\rho(z) + \kapd(z)},
			\quad \bigl\lVert \Im M(z)\bigr\rVert \le \frac{|\Im z|}{\kapd(z)^2}, 
			\quad \bigl\lVert  M(z)^{-1} \bigr\rVert \lesssim 1 + |z|, 
		\end{equation}
		where $\kapd(z) := \dist (z, \supp\,\rho\,  )$, and we recall that $\rho(z) := \pi^{-1}\langle \Im M(z) \rangle$.
		
		The density $\rho(x)$ satisfies
		\begin{equation} \label{eq:rho_props}
			\rho(x) \lesssim 1, \qquad \supp\, \rho \subset \Bigl[ -\bigl\lVert   D\bigr\rVert - 2\bigl\lVert \mathcal{S} \bigr\rVert^{1/2},\, \bigl\lVert  D\bigr\rVert + 2\bigl\lVert  \mathcal{S} \bigr\rVert^{1/2} \Bigr].
		\end{equation}
		The solution $M(z)$ admits the Stieltjes representation 
		\begin{equation} \label{eq:M_stieltjes}
			M(z) = \frac{1}{\pi} \int_\mathbb{R} \frac{1}{x- z} \Im M(x) \mathrm{d}x, \qquad z \in \mathbb{C}\backslash\mathbb{R},
		\end{equation}
		where the boundary value $\Im M(x)$ is given by $\lim_{\eta \to +0} \Im M(x+\ii\eta)$ in $\norm{\cdot}_\mathrm{hs}$. 
	\end{lemma}

	Under the additional assumption of boundedness, the solution $M(z)$ satisfies the following properties.
	\begin{lemma}[Lemmas~4.8~(ii) and~5.5 
		in \cite{AEK2020}] \label{lemma:MDE_props}
		Let $M(z)$ be the solution to \eqref{eq:MDE} with the data-pair $(D,\mathcal{S})$ satisfying Assumptions~\ref{ass:boundedexp},~\ref{ass:Mbdd} and the flatness estimates \eqref{eq:flatness}. 
		Let $\bddD$ be the domain defined in \eqref{eq:bdd_def}, and $\rho(z) := \pi^{-1}\langle \Im M(z) \rangle$. Then, 
		\begin{equation} \label{eq:stat_M_bound}
			\norm{M(z)} \lesssim 1, \qquad z \in \bddD,
		\end{equation}
		\begin{equation} \label{eq:imM_bound}
			  \Im M(z) \sim  \rho(z)\,I, \qquad z \in \bddD,
		\end{equation}
		in the sense of quadratic forms.
		Moreover, there exists a threshold $\rho_* \sim 1$, such that  \nc the maps\footnote{
			The function $\sigma(z)$, can be replaced by a regularized version $\other{\sigma}(z) := \sigma(z) \chi(|\rho(z)|)$, where $\chi$ is a $C^\infty$ bump functions taking values in $[0,1]$, which satisfies $\chi(x) = 1$ for $x \le \tfrac{1}{2}\rho_*$ and $\chi(x) = 0$ for $x \ge \rho_*$. The regularized function $\other{\sigma}(z)$ is $1/3$-H\"older in $\bddD\cap\mathbb{H}$ and satisfies $|\rho(z)| + |\other{\sigma}(z)| \sim |\rho(z)|+|\sigma(z)|$. 
			\label{foot:sigma}
		}
		\begin{equation} \label{eq:1/3Holder_rhoinveta}
			\begin{split}
				z &\mapsto \rho(z)^{-1} \Im z, \quad  z \mapsto \rho(z),  \quad \text{ are~ $1/3$-H\"older regular in } \bddD\cap\mathbb{H},\\
				z &\mapsto \sigma(z) \quad \text{ is ~$1/3$-H\"older regular in } \{z\in \bddD\cap\mathbb{H} \,:\, \rho(z) \le \rho_*\},
			\end{split}
		\end{equation}
		where $\sigma(z)$ defined in \eqref{eq:sigma_def}. In particular, $\rho(z)^{-1}\Im z$, $\rho(z)$ and $\sigma(z)$ admit $1/3$-H\"older regular extensions to the closure of the respective domains in \eqref{eq:1/3Holder_rhoinveta}. 
 		Finally, for any $z \in \bddD$,  the function $\kapd(z) := \dist(z, \supp\,\rho)$ \nc satisfies
		\begin{equation} \label{eq:kapd_rhoeta}
			\rho(z)^{-1} \Im z \gtrsim 1\quad   \Longleftrightarrow \quad \kapd(z) \gtrsim 1, \quad \text{and} \quad \rho(z)^{-1} \Im z = 0\quad   \Longleftrightarrow \quad \kapd(z) = 0. 
		\end{equation}
	\end{lemma}
	
	It is straightforward to check,   using \eqref{eq:adm_E} and \eqref{eq:bdd_def}, \nc that there exists a non-negative integer $L \le 3 c_M^{-1} (\norm{D} + \norm{\mathcal{S}}^{1/2})$, such that
	  the intersection of $\overline{\bddD}$ with the real line consists of~$L-1$ 
	intervals of order-one size and rays extending to~the cut-off of $\bddD$, $\pm C_{\rm bdd}$, which are mutually well-separated. More precisely, provided the $N$-independent constant $C_{\rm bdd}$ is sufficiently large,  
	\begin{equation} \label{eq:bddD_structure}
		\overline{\bddD} \cap \mathbb{R} = [-C_{\rm bdd}, b_1] \cup \bigcup_{j=2}^L [a_j, b_j] \cup [a_{L+1}, C_{\rm bdd}], \qquad b_j - a_j \ge c_M, \quad  a_{j+1} - b_{j} \ge c_M, \quad j \in \{1, \dots, L\},
	\end{equation}
	with $L=0$ corresponding to $\mathcal{I} = \mathbb{R}$, and hence $\overline{\bddD} \cap \mathbb{R} = [-C_{\rm bdd}, C_{\rm bdd}]$.

	For any   (small) \nc positive constant $c > 0$, we define the domain $\smallD_{c}$ as
	\begin{equation} \label{eq:shape_dom}
		\smallD_{c} := \bigl\{ z \in \bddD\,:\,   \rho(z)^{-1}\Im z  + \bigl\lvert \rho(z) \bigr\rvert \le c \bigr\}.
	\end{equation}
	  This is the set of spectral parameters $z$ which are admissible for non-trivial shape analysis, that is, it is a union of neighborhoods around all singularities and small local minima of $\rho$ in the set of admissible energies.
	We refer to it as the \emph{singular spectral regime}.
	The complementary set $\bddD\backslash \smallD_{c}$ consists of the bulk spectrum and points away from the support of~$\rho$. This regime is simpler, since the solution $M$ is uniformly Lipschitz continuous on $\bddD\cap\mathbb{H}\backslash \smallD_{c}$ and holomorphic on $\bddD\backslash (\smallD_{c}\cup\supp\,\rho)$. \nc
	
	\begin{definition}[Well-Structured Spectrum] \label{def:pert_cond}
		For a positive constant $\varepsilon \in (0, \tfrac{1}{2})$ and a norm  $\norm{\cdot}_\star$ on the space of linear maps $\mathbb{C}^{N\times N} \to \mathbb{C}^{N\times N}$, we 
		say that a super-operator $\mathcal{X}$ has an $\varepsilon$-\textit{well-structured spectrum} in the norm $\norm{\cdot}_\star$ if and only if 
		\begin{equation} \label{eq:pert_cond}
			\norm{(\zeta\,\mathrm{Id} - \mathcal{X})^{-1}}_\star \lesssim 1, 
		\end{equation}
		for all $\zeta \in\mathbb{C}$ satisfying $|\zeta| \ge \varepsilon$ and $|1 - \zeta| \ge 1 - 2\varepsilon$, and 
		\begin{equation} \label{eq:Pi_def}
			\mathrm{rank}\,\Pi_\mathcal{X}  = 1, \quad \text{where}
			\quad \Pi_\mathcal{X}  := \frac{1}{2\pi\ii} \oint_{|\zeta| = \varepsilon} (\zeta\, \mathrm{Id}- \mathcal{X})^{-1} \mathrm{d}\zeta.
		\end{equation}
	\end{definition}
	  Definition~\ref{def:pert_cond} expresses that the disk of radius $\varepsilon$ around the origin contains a single eigenvalue of $\mathcal{X}$ with algebraic multiplicity one with $\Pi_\mathcal{X} $ being the corresponding spectral projection, while the rest of the spectrum of $\mathcal{X}$ is confined to the disk of radius $1-2\varepsilon$ centered at $1$; in particular it is $\varepsilon$-separated from the smallest (in modulus) eigenvalue. \nc	
	For any operator $\mathcal{X}$ satisfying Definition~\ref{def:pert_cond} above, we have the bounds
	\begin{equation} \label{eq:PiQ_bounds}
		\norm{\Pi_\mathcal{X}}_\star  \lesssim 1, \qquad \norm{\mathcal{X}^{-1}(\mathrm{Id} - \Pi_\mathcal{X})}_\star \lesssim \varepsilon^{-1}.
	\end{equation}
	
	\begin{lemma}[Lemmas~4.3   and Eq~(5.9) \nc in \cite{AEK2020}] \label{lemma:satur}
		For all $z \in \mathbb{C}\backslash\mathbb{R}$, let $\mathcal{F} \equiv \mathcal{F}_z$ denote the \textit{saturated self-energy} super-operator 
		\begin{equation}
			\mathcal{F} \equiv \mathcal{F}_z := \mathcal{C}_{Q, Q^*} \mathcal{S}\, \mathcal{C}_{Q^*, Q}, \quad \text{where}\quad  \mathcal{C}_{X,Y}[R] := X R\, Y \quad \text{for}\quad X, Y, R\in\mathbb{C}^{N\times N},
		\end{equation}
		where $Q \equiv Q(z)$ is defined as 
		\begin{equation} \label{eq:polarQ_def}
			Q(z) := \bigl\lvert (\Im M(z))^{-1/2}(\Re M(z))(\Im M(z))^{-1/2} + \ii \, I \bigr\rvert^{1/2} (\Im M(z))^{1/2}.
		\end{equation}
		The super-operator $\mathcal{F}_z$ is self-adjoint,  positivity-preserving, and satisfies
		\begin{equation} \label{eq:satur_bound}
			1 - \norm{\mathcal{F}}_{\mathrm{hs}\to\mathrm{hs}} \sim \frac{\rho(z)}{\Im z} \le \other{\beta}(z,\bar{z})^{-1}, \qquad z \in \bddD.
		\end{equation}
		There exists a unique eigenvector $F \equiv F_z \in \mathbb{C}^{N\times N}$ with   $F>0$, 
		\nc  $\norm{F}_\mathrm{hs} = 1$, such that for all $z\in\bddD$,  $\mathcal{F}[F] = \norm{\mathcal{F}}_{\mathrm{hs}\to\mathrm{hs}} F$ and 
		\begin{equation} \label{eq:eigF}
			F \sim I, \qquad \norm{F - \norm{F_\mathscr{U}}_\mathrm{hs}^{-1}F_\mathscr{U}} \lesssim \rho(z)^{-1}\Im z, \qquad F_\mathscr{U} := \rho(z)^{-1}\Im \mathscr{U},
		\end{equation}
		where $\mathscr{U} \equiv \mathscr{U}(z)$ is defined in \eqref{eq:sigma_def}.  
		The vector $F_\mathscr{U}$ also satisfies
		\begin{equation} \label{eq:F_flat}
			F_\mathscr{U} \sim I. 
		\end{equation} 
		
		\nc Moreover, there exists a small constant $\varepsilon_* \sim 1$ such that for all $z \in \bddD$, the spectrum of $\mathcal{F}$ satisfies
		\begin{equation} \label{eq:satur_gap}
			\spec\bigl(\norm{\mathcal{F}}_{\mathrm{hs}\to\mathrm{hs}}^{-1}\mathcal{F}\bigr) \subset [-1+3\varepsilon_*, 1-3\varepsilon_*] \cup \{1\}.
		\end{equation} 
	\end{lemma}
	
	In the following lemma, we collect the key properties of the opposite-half plane stability operator 
	$\mathcal{B}_* \equiv \mathcal{B}_{z,\bar z}$, which has a well-structured in the regime when $\rho(z)^{-1}\Im z$ is small. 	
	
	\begin{lemma} \label{lemma:1stab_ohp} For all $z \in \bddD$, the stability operator $\mathcal{B}_* \equiv \mathcal{B}_{z,\bar{z}}$ satisfies
		\begin{equation} \label{eq:1stab_ohp}
			\norm{\mathcal{B}_*^{-1}} + \norm{\mathcal{B}_*^{-1}}_{\mathrm{hs}\to\mathrm{hs}} \lesssim 1 + \frac{\rho(z)}{ \Im z}. 
		\end{equation}
		
		Moreover, there exists a threshold $c_*$ for $z\in\bddD$ with $\rho(z)^{-1}\Im z \le c_*$, the super-operator $\mathcal{B}_* \equiv \mathcal{B}_{z,\bar z}$ has an $\varepsilon_*$-well-structured spectrum   in both $\norm{\cdot}$ and $\norm{\cdot}_{\mathrm{hs}\to\mathrm{hs}}$-norms, as in Definition~\ref{def:pert_cond}, where $\varepsilon_*$ is the constant from Lemma~\ref{lemma:satur}.
		
		Finally, for all $z\in\bddD$ with $\rho(z)^{-1}\Im z \le c_*$, the projector $\Pi_* \equiv \Pi_{\mathcal{B}_*}$, defined according to \eqref{eq:Pi_def}, and the eigenvectors $B_r \equiv B_r(z) := \Pi_* [\rho(z)^{-1}\Im M(z) ]$ and $B_\ell \equiv B_\ell(z) := (\Pi_*)^*\mathcal{C}_{M^*,M}^{-1}[\rho(z)^{-1}\Im M(z)]$ satisfy \nc 
		\begin{equation} \label{eq:ohp_Pi_ests}
			\Pi_*[\,\cdot\,] = \frac{\bigl\langle B_{\ell}^* (\,\cdot\,) \bigr\rangle}{\langle B_{\ell}^* B_{r} \rangle}B_{r}~, \qquad B_{r} = \frac{\Im M}{\rho} + \mathcal{O}(\rho^{-1}\Im z), \qquad B_{\ell} = \mathcal{C}_{M^*,M}^{-1}\Bigl[\frac{\Im M}{\rho} \Bigr] + \mathcal{O}(\rho^{-1}\Im z),
		\end{equation}
		where we abbreviate $M \equiv M(z)$, $\rho \equiv \rho(z)$.
	\end{lemma}
	We defer the proof of Lemma~\ref{lemma:1stab_ohp} to the end of the present subsection. 
	
	Next, we record the key properties of the  well-structured  one-body stability operator 
	$\mathcal{B} \equiv \mathcal{B}_{z,z}$, which was fully analyzed in \cite{AEK2020}. 
	In particular, we import asymptotic expansions for its isolated eigenvalue $\beta$ in the singular spectral regime.

	\begin{lemma} [Lemma 5.1 and Corollaries~5.2--5.3 in \cite{AEK2020}] \label{lemma:stab1} 
		For all $z \in \bddD$, the   one-body \nc stability operator $\mathcal{B} \equiv \mathcal{B}_{z,z}$ satisfies the bounds
		\begin{equation} \label{eq:1stab_bound}
			\norm{\mathcal{B}^{-1}} + \norm{\mathcal{B}^{-1}}_{\mathrm{hs}\to\mathrm{hs}}   \lesssim 1 + \frac{1}{\rho(z)^{-1}\Im z + \rho(z)^2 + |\rho(z)  \sigma(z)| }, 
		\end{equation}
		where $\sigma(z)$ is defined in \eqref{eq:sigma_def}. 
		
		There exist a (small) threshold $\rho_* \sim 1$ such that for all $z \in \smallD_{\rho_*}\backslash\mathbb{R}$, the super-operator $\mathcal{B} \equiv \mathcal{B}_{z,z}$ has an $\varepsilon_*$-well-structured spectrum in both $\norm{\cdot}$ and $\norm{\cdot}_{\mathrm{hs}\to\mathrm{hs}}$-norms in the sense of Definition~\ref{def:pert_cond}, where $\varepsilon_*$ is the constant from Lemma~\ref{lemma:satur}.

			For all $z \in \smallD_{\rho_*}\backslash\mathbb{R}$, let $\Pi \equiv \Pi_{\mathcal{B}_{z,z}}$ be the projector defined as in \eqref{eq:Pi_def}, and let 
			\begin{equation}
				V_r \equiv V_r(z) := \Pi[B_r], \qquad V_\ell \equiv V_\ell(z) := \Pi^*[B_\ell],
			\end{equation}			
			where $B_r \equiv B_r(z)$ and $B_\ell \equiv B_\ell(z)$ the right and left eigenvectors of $\mathcal{B}_* \equiv \mathcal{B}_{z,\bar z}$, respectively, defined above~\eqref{eq:ohp_Pi_ests}. Then, the eigenprojector $\Pi$ and the corresponding eigenvalue $\beta \equiv \beta_{z,z} := \Tr[\Pi\mathcal{B}]$ of $\mathcal{B}$ satisfy
			\begin{equation} \label{eq:Pi_beta}
				\Pi[\,\cdot\,] = \frac{\langle V_\ell^* (\,\cdot\,) \rangle}{\langle V_\ell^*V_r \rangle} V_r, \qquad 
				\beta\, \langle V_\ell^* V_r\rangle = \pi\rho^{-1} \Im z - 2\ii\rho \sigma +  2\rho^2 \Bigl(\psi + \frac{\sigma^2}{\langle F_\mathscr{U}^2 \rangle}\Bigr) + \mathcal{O}\bigl(|\rho|^3 + |\Im z| + (\rho^{-1}\Im z)^2\bigr),
			\end{equation}
			where $\rho\equiv\rho(z)$, $\sigma\equiv \sigma(z)$.
			Moreover, the $\psi$ and the matrices $V_\ell$, $V_r$ admit the estimates\footnote{
				In the original notation of \cite{AEK2020}, $V_r := b, V_\ell := l$. The comparison $\psi + \sigma^2 \sim 1$ follows from Eq. (5.8) and (5.35) in \cite{AEK2020}, while the properties of  $V_r$ and $V_\ell$ follow from Eq. (5.8), (5.14), (5.15), and (5.19) in \cite{AEK2020}.}
			\begin{equation} \label{eq:psi_LR_asymp}
				\psi + \sigma^2 \sim 1, \qquad V_r = \frac{\Im M}{\rho} + \mathcal{O}\bigl(|\rho| +\rho^{-1}\Im z\bigr),
				\qquad   V_\ell   = \mathcal{C}_{M^*,M}^{-1}\Bigl[\frac{\Im M}{\rho} \Bigr] +  \mathcal{O}\bigl(|\rho| +\rho^{-1}\Im z\bigr),
			\end{equation}
			where $M \equiv M(z)$. In particular, we have
			\begin{equation} \label{eq:LR_props}
				\langle V_r \rangle = \pi + \mathcal{O}\bigl(|\rho| +\rho^{-1}\Im z\bigr),
				\quad \langle V_\ell^* M^2  \rangle = \pi + \mathcal{O}\bigl(|\rho| +\rho^{-1}\Im z\bigr), \quad \bigl\lvert \langle V_\ell^*V_r \rangle \bigr\rvert \sim 1, \quad \norm{V_r} + \norm{V_\ell} \lesssim 1.
			\end{equation}
	\end{lemma}
	
Corollary 5.2 in \cite{AEK2020} provides more precise asymptotic expansions for the right and left eigenvectors $V_r, V_\ell$ of the one-body operator $\mathcal{B}_{z,z}$. However, for the purposes of the present work, the properties \eqref{eq:psi_LR_asymp}--\eqref{eq:LR_props} collected above are sufficient. \nc	

We conclude this section by proving Lemma~\ref{lemma:1stab_ohp}. 	 
	\begin{proof}[Proof of Lemma~\ref{lemma:1stab_ohp}]
		In \cite{AEK2020}, the operator $\mathcal{B}_*$ is analyzed in detail for $z \in \smallD_{\rho_*}\cap\mathbb{H}$. 		
		The proof for all $z \in \bddD$ satisfying $\rho(z)^{-1}\Im z \le c_*$ is analogous to that of Lemma~5.1 in \cite{AEK2020} for the operator $\mathrm{Id} - \mathcal{C}_{P, P}\mathcal{F}_z$ with $P \equiv P(z) := \sign\Re U(z)$, but we present it here for completeness.  
		
		  First we show the bound~\eqref{eq:1stab_ohp}. \nc Recall the \textit{balanced polar decomposition} from \cite{AEK2020},  
		\begin{equation} \label{eq:polar}
			M(z) = Q(z)^* U(z) Q(z), \qquad M(\bar z) = M(z)^* = Q(z)^* U(z)^*Q(z), \qquad z \in \mathbb{H},
		\end{equation}
		where $Q(z)$ is defined in \eqref{eq:polarQ_def}, and $U(z)$ is defined in \eqref{eq:sigma_def}. Then, the stability operator $\mathcal{B}_*$ admits the expression 
		\begin{equation} \label{eq:Bstar}
			\mathcal{B}_{z,\bar z}  = \mathcal{C}_{Q^*, Q}\bigl(\mathrm{Id} - \mathcal{C}_{U,U^*}\mathcal{F}\bigr)  \mathcal{C}_{Q^*, Q}^{-1},
		\end{equation}
		where $U \equiv U(z)$ and we define $U(z) := U(\bar z)^*$ if $\Im z < 0$. In particular,   the $\norm{\cdot}_{\mathrm{hs}\to\mathrm{hs}}$-bound in \nc \eqref{eq:1stab_ohp} follows immediately from~\eqref{eq:satur_bound}.
		  To conclude the bond in the norm induced by $\norm{\cdot}$, we \nc
		note that any symmetric super-operator $\mathcal{S}$ satisfying the upper flatness estimate \eqref{eq:flatness} admits the bound 
		\begin{equation} \label{eq:S_hs_to_op}
			\norm{\mathcal{S}}_{\mathrm{hs}\to\norm{\cdot}} \le 4C.
		\end{equation}
		Indeed, the upper bound in \eqref{eq:flatness} implies that $\lVert\mathcal{S}[R]\rVert \le C \norm{R}_\mathrm{hs}$ for any $N\times N$ matrix $R \ge 0$, and since any $X \in \mathbb{C}^{N\times N}$ can be expressed as the sum of four positive-definite matrices, \eqref{eq:S_hs_to_op} holds. Hence, it follows from \eqref{eq:stat_M_bound} that $\norm{\mathrm{Id}-\mathcal{B}_*}_{\mathrm{hs}\to\norm{\cdot}} \lesssim 1$, and therefore  
		\begin{equation} \label{eq:op_from_hs}
			\norm{\mathcal{B}_*^{-1}} \le 1 + \norm{\mathrm{Id}-\mathcal{B}_*}_{\mathrm{hs}\to\norm{\cdot}} \norm{\mathcal{B}_*^{-1}}_{\mathrm{hs}\to\mathrm{hs}} \lesssim 1 + (\Im z)^{-1}\rho(z). 
		\end{equation}
		
		Next, \nc to prove that the spectrum of $\mathcal{B}_{z,\bar z}$ is well-structured,   by~\eqref{eq:Bstar} \nc it suffices to establish the same property for the operator $\mathrm{Id} - \mathcal{C}_{U,U^*}\mathcal{F}$. To this end, we define 
		\begin{equation}
			\mathcal{U}_r := (1-r)\,\mathrm{Id} + r \,\mathcal{C}_{U,U^*}, \qquad r\in[0,1].
		\end{equation}
		It follows from \eqref{eq:eigF} that the Perron-Frobenius eigenvector $F$ of $\mathcal{F}$ satisfies  $\mathcal{U}_r[F] = F + \mathcal{O}_{\mathrm{hs}}(\rho^{-1}\Im z)$ for all $r\in[0,1]$. Hence, for any $X \in \mathbb{C}^{N\times N}$, and all $\zeta\in\mathbb{C}$ satisfying $|\zeta| \ge \varepsilon_*$ and $|1-\zeta|\ge 2 \varepsilon_*$,
		\begin{equation}
			\bigl\lVert ((\zeta-1)\,\mathrm{Id} + \mathcal{U}_r\mathcal{F})[X] \bigr\rVert_\mathrm{hs} \ge |\alpha|^2 \varepsilon_*^2 + \varepsilon_*^2 \norm{X^\perp}_\mathrm{hs}^2 = \varepsilon_*^2 \norm{X}_\mathrm{hs}^2, \qquad r \in [0,1],
		\end{equation}
		where $X = \alpha F + X^\perp$   for some $\alpha\in\C$ \nc and $\langle F^* X^\perp \rangle = 0$. Therefore, the operator $\mathcal{X} = \mathrm{I} - \mathcal{U}_r\mathcal{F}$ satisfies \eqref{eq:pert_cond} with $\norm{\cdot}_\star = \norm{\cdot}_{\mathrm{hs}\to\mathrm{hs}}$, for all $r\in [0,1]$. Hence, the map $r \mapsto \Tr[\Pi(\mathrm{I} - \mathcal{U}_r\mathcal{F})] = \mathrm{rank}\,\Pi(\mathrm{I} - \mathcal{U}_r\mathcal{F})$ is continuous on $r\in [0,1]$, and hence is identically equal to $\mathrm{rank}\,\Pi(\mathrm{I} - \mathcal{F}) = 1$ by \eqref{eq:satur_gap} provided $c_*$ is sufficiently small. 
		Therefore, $\mathcal{B}_*$ satisfies Definition~\ref{def:pert_cond} in the $\norm{\cdot}_{\mathrm{hs}\to\mathrm{hs}}$-norm with the constant $\varepsilon_*$. 
		  Analogously to \eqref{eq:op_from_hs}, \nc the bound \eqref{eq:pert_cond} for $\mathcal{X} = \mathcal{B}_*$ in the $\norm{\cdot}_{\mathrm{hs}\to \mathrm{hs}}$-norm implies the corresponding $\norm{\cdot}$-bound. 
		
		  Since $\mathcal{B}_*[\rho^{-1}\Im M] = (\rho^{-1}\Im z)\, MM^*$ by \eqref{eq:B12_identity}, and $\norm{MM^*} \lesssim 1$ by \eqref{eq:stat_M_bound}, \nc the estimates in \eqref{eq:ohp_Pi_ests} follow from \eqref{eq:B12_identity} via simple perturbation theory. 		
		This concludes the proof of Lemma~\ref{lemma:1stab_ohp}.
	\end{proof}

	\subsection{Properties of the Trajectories.   Proof of Lemma~\ref{lemma:flow}}	 \label{sec:flow_aux}
	\begin{lemma} [$u$-Lemma] \label{lemma:u}
		Let $M(z)$ be the solution to MDE \eqref{eq:MDE} with data-pair $(D, \mathcal{S})$ satisfying Assumption~\ref{ass:boundedexp} and the flatness bounds \eqref{eq:flatness}. 
		For all $z_1, z_2 \in \mathbb{C}\backslash\mathbb{R}$, define the function 
		\begin{equation} \label{eq:u_def}
			u(z_1,z_2) := \frac{m(z_1) - m(z_2)}{z_1-z_2},
		\end{equation}
		where we recall that $m(z) := \langle M(z) \rangle$. Then, the function $u(z_1,z_2)$ enjoys the following properties:
		\begin{enumerate}
			\item[(i)] For all $z_1, z_2\in\bddD$, either 
			\begin{subequations}
				\begin{equation} \label{eq:u_one}
					|u(z_1,z_2)| \lesssim 1 \quad \text{and}\quad   \other{\beta}(z_1,z_2) \sim 1.
				\end{equation}
				where $\other{\beta}(z_1,z_2)$ is defined in \eqref{eq:betaf_def}, or
				\begin{equation} \label{eq:u_sim}
					|u(z_1,z_2)| \sim \other{\beta}(z_1,z_2)^{-1},
				\end{equation}
			\end{subequations}
			
			\item[(ii)] There exists a positive constant $C \lesssim 1$, such that the real part of $u(z_1,z_2)$ satisfies
			\begin{equation} \label{eq:u_props}
				\Re u(z_1,z_2) \ge -C, \qquad z_1,z_2 \in \bddD.
			\end{equation}
		\end{enumerate} 
	\end{lemma}
	
	We prove Lemma~\ref{lemma:u} in Section~\ref{sec:pert}.

	\begin{proof}[Proof of Lemma~\ref{lemma:flow}]
		First, we prove $(i)$. The bound \eqref{eq:termT_bound} holds trivially for any $T \le \tfrac{1}{2}c_\mathrm{full}$, and the estimates in \eqref{eq:S_t_flatness} follows from \eqref{eq:termT_bound} and  \eqref{eq:flatness} since explicitly solving \eqref{eq:datapair_flow} yields $\mathcal{S}_t[\,\cdot\,] = \ee^{T-t}\mathcal{S}[\,\cdot\,] + (1-\ee^{T-t}) \langle \,\cdot \, \rangle$.
		
		Next, we prove $(ii)$. The identity \eqref{eq:dMt} follows by a direct computation using the invertibility  of the one-body stability operator $\mathrm{Id}[\,\cdot\,] - M_t(z_t) \mathcal{S}_t[\,\cdot\,] M_t(z_t)$. 
		
		The definition of $\mathcal{I}$ in \eqref{eq:adm_E} together with \eqref{eq:bdd_def} and \eqref{eq:dMt} implies that there exists a constant $\other{C}$, such that 
		\begin{equation}
			z_t \in \bddD + \{\zeta\in\mathbb{C}\,:\, |\zeta| \le    \other{C}\cdot(T-t) \nc \}, \qquad t \in [0,T], \qquad z_T\in\bddD.
		\end{equation}
		Therefore, it follows from Proposition 10.1 in \cite{AEK2020} that by suitably shrinking $T\sim 1$, we can guarantee that 
		\begin{equation} \label{eq:time_admis}
			\norm{M_t(w)} \lesssim 1, \qquad w \in \bddD + \{\zeta\in\mathbb{C}\,:\, |\zeta| \le   \other{C}\cdot(T-t) \nc \}, \qquad t\in[0,T].
		\end{equation}
		Hence, the flatness bounds \eqref{eq:S_t_flatness} together with Lemmas~4.8 and~5.4~(ii) in \cite{AEK2020} imply \eqref{eq:kapd_rhoeta_t}.

		Finally, we prove $(iii)$.
		Observe that it suffices to prove \eqref{eq:z_diff_comp} for $z_1, z_2 \in \bddD\cap\mathbb{H}$. 
		Otherwise, without loss of generality, we can assume that  $z_1, \overline{z}_2 \in \bddD\cap\mathbb{H}$ and \eqref{eq:z_diff_comp} holds for $z_{1,t} - \overline{z}_{2,t}$. Recall that $z_{j,T} = z_j$, then, denoting $\eta_{j,t} := \Im z_{j,t}$, $\eta_{j} := \Im z_j$,  $m_j := \langle M(z_j)\rangle$,	and using \eqref{eq:eta_t}, we obtain 
		\begin{equation} \label{eq:u_from_ubar}
			\begin{split}
				|z_{1,t} - z_{2,t}| &\sim |z_{1,t} - \overline{z}_{2,t}| +  \eta_{1,t}   +  \eta_{2,t} \sim \bigl(|z_{1} - \overline{z}_{2}| + |\eta_1| + |\eta_2|\bigr) + (T-t)\bigl(|m_{1} - \overline{m}_{2}| + |\Im m_1| + |\Im m_2|\bigr) \\
				&\sim |z_1 - z_2| + (T-t)|m_1 - m_2|,
			\end{split}
		\end{equation}
		where we used $|u -\overline{w}| \sim |u-w| + |\Im u| + |\Im w|$ for all $u, w \in \mathbb{H}$. The desired \eqref{eq:z_diff_comp} follows from \eqref{eq:u_from_ubar} after dividing by $|z_1 -z_2|$.
		
		We now consider $z_1, z_2 \in \bddD\cap\mathbb{H}$.		
		Using \eqref{eq:M_t} to solve the equation \eqref{eq:char_flow} for $z_t$ explicitly, we obtain
		\begin{equation} \label{eq:z_dif_sol}
			\ee^{(t-T)/2} \frac{z_{1,t} - z_{2,t}}{z_{1}-z_{2}} = 1 +  \bigl(1 - \ee^{t - T}\bigr)  u(z_1, z_2) , \qquad u(z_1, z_2) := \frac{m_{1} - m_{2}}{z_{1}-z_{2}}. 
		\end{equation}
		  We start with the simpler case assuming that \nc
		$\Re u(z_1, z_2) \ge 0$. Then it follows from \eqref{eq:z_dif_sol} that 
		\begin{equation}
			\Bigl\lvert\frac{z_{1,t} - z_{2,t}}{z_{1}-z_{2}}\Bigr\rvert \sim \bigl\lvert 1 + (T-t) \Re u(z_1, z_2) \bigr\rvert + \bigl\lvert \Im u(z_1, z_2) \bigr\rvert \sim 1 + \bigl\lvert u(z_1, z_2) \bigr\rvert, \qquad t\in[0,T], 
		\end{equation}
		which yields the desired \eqref{eq:z_diff_comp}. 
		
		  If $\Re u(z_1, z_2)  < 0$, then  we still have an effective lower bound, \nc  then $\Re u(z_1, z_2) \ge - C$ for some $C\sim1$ by \eqref{eq:u_props}, and there exists a threshold $t_* \equiv t_*(C) \sim 1$ such that $1 - \ee^{-t_*} \le 1/(2C)$; hence, 
		\begin{equation} \label{eq:z_dif_comp_short_t}
			\Bigl\lvert\frac{z_{1,t} - z_{2,t}}{z_{1}-z_{2}}\Bigr\rvert \sim 1 + \bigl\lvert \Im u(z_1, z_2)  \bigr\rvert \sim 1 + \bigl\lvert u(z_1, z_2)  \bigr\rvert, \qquad t\in[T-t_*, T].
		\end{equation}
		Next, we extend the bound \eqref{eq:z_dif_comp_short_t} to the remaining time range $t \in [0,T-t_*]$. Similarly to \eqref{eq:z_dif_sol}, we have
		\begin{equation} \label{eq:z_dif_st}
			\ee^{(t-s)/2} \frac{z_{1,t} - z_{2,t}}{z_{1,s}-z_{2,s}} = 1 +  \bigl(1 - \ee^{t - s}\bigr) \frac{m_{1,s} - m_{2,s}}{z_{1,s}-z_{2,s}},\qquad s,t \in [0,T].
		\end{equation}
		It follows from \eqref{eq:eta_t} and \eqref{eq:rho_t} that
		\begin{equation}
			\rho_{s}(z_{j,s})^{-1}\eta_{j,s} \gtrsim T-s \ge t_* \sim 1 \Longrightarrow \kapd_{j,s} := \dist(z_{j,s}, \supp\,\rho_s) \gtrsim 1, \qquad s \in [0,T-t_*], \quad j \in \{1,2\},
		\end{equation}
		where we used \eqref{eq:kapd_rhoeta_t} to obtain the implication. 
		Therefore, using the Stieltjes representation \eqref{eq:M_stieltjes}, we obtain
		\begin{equation} \label{eq:m_dif_bound1} 
			\Bigl\lvert \frac{m_{1,s}-m_{2,s}}{z_{1,s}-z_{2,s}}\Bigr\rvert \le \int_\mathbb{R} \frac{\rho_s(x)\mathrm{d}x}{|z_{1,s} - x| \,|z_{2,s} -x|} \le \frac{1}{\kapd_{1,s}\kapd_{2,s}} \lesssim 1, \qquad s \in [0, T-t_*].
		\end{equation} 
		Hence, plugging \eqref{eq:m_dif_bound1} into \eqref{eq:z_dif_st}, we conclude that 
		\begin{equation} \label{eq:z_dif_comp_long_t}
			\Bigl\lvert \frac{z_{1,t} - z_{2,t}}{z_{1,s}-z_{2,s}}  \Bigr\rvert \sim 1, \qquad s,t \in [0, T-t_*].
		\end{equation}
		Together with \eqref{eq:z_dif_comp_short_t}, \eqref{eq:z_dif_comp_long_t} yields the desired \eqref{eq:z_diff_comp}.
		This concludes the proof of Lemma~\ref{lemma:flow}.
	\end{proof}
	
	\section{Perturbative Analysis of the Two-Body Quantities} \label{sec:pert}
	In this section we give the proof of Lemmas~\ref{lemma:Gamma} and~\ref{lemma:u}, which contain the relevant properties of the key time-independent two-body quantities,
	$$
		\frac{M(z_1) - M(z_2)}{z_1 - z_2}  \quad\text{and}\quad u(z_1, z_2) := \frac{m(z_1) - m(z_2)}{z_1 - z_2}, \quad \text{respectively}.
	$$	
	In preparation for the proof, we record the following perturbative formulas for the solution $M(z)$ to the matrix Dyson equation \eqref{eq:MDE}, which is based on the one-body stability analysis in \cite{AEK2020}.
	\begin{lemma}[$M$-Perturbation] \label{lemma:xi}
		Assume that an MDE data-pair $(D,\mathcal{S})$ satisfies Assumption~\ref{ass:boundedexp} and the flatness bounds \eqref{eq:flatness}. 
		Then, for all $z,w\in\bddD \cap\mathbb{H}$,  the solution $M(z)$ to the MDE \eqref{eq:MDE} with data $(D, \mathcal{S})$ satisfies 
		\begin{equation} \label{eq:M_perturb}
			\norm{M(z) - M(w)} \lesssim  \other{\xi}(z,w),  
		\end{equation}
		\begin{equation} \label{eq:V_perturb}
			\norm{\rho(z)^{-1}\Im M(z) - \rho(w)^{-1} \Im M(w)} \lesssim \norm{M(z) - M(w)} \lesssim  \other{\xi}(z, w),   
		\end{equation}
		where, for all $z, w \in \mathbb{H}$, the control function $\other{\xi}(z,w)$ is defined as
		\begin{equation} \label{eq:xif_def}
			\other{\xi}(z,w) := |z-w|+\min\Bigl\{ \frac{|z-w|}{\rho(z)^{-1} \Im z + \rho(z)^2 + \rho(z) |\sigma(z)| },\, \frac{|z-w|^{1/2}}{\bigl(\rho(z) + |\sigma(z)|\bigr)^{1/2}},\, |z-w|^{1/3} \Bigr\} \sim \frac{|z-w|}{\other{\beta}(z,w)}.
		\end{equation}
		Furthermore, there exist thresholds $\rho_*, \delta_* \sim 1$ such that, for all $z\in \smallD_{\rho_*}$ and $w \in \bddD\cap\mathbb{H}$ satisfying $|z - w| \le \delta_*$,  
			\begin{equation} \label{eq:xi_bounds}
				\norm{M(w) - M(z) - \xi(z,w) V_r(z)} \lesssim \other{\xi}(z,w)^2 + |z-w|, \quad \xi(z,w) := \frac{\bigl\langle V_\ell(z)^* \bigl( M(w) - M(z)\bigr) \bigr\rangle}{\langle V_\ell(z)^*  V_r(z) \rangle}, \quad |\xi(z,w)| \sim \other{\xi}(z,w),
			\end{equation} 
			where $V_\ell(z)$ and $V_r(z)$ are the left and right eigenvectors of the one-body stability operator $\mathcal{B}_{z,z}$ from \eqref{eq:Pi_beta}.
	\end{lemma}
	We defer the proof of Lemma~\ref{lemma:xi} to the end of Section~\ref{sec:pert_aux}. 
	
	\subsection{Scalar Analysis. Proof of Lemma~\ref{lemma:u}}	
	Equipped with Lemma~\ref{lemma:xi}, we are ready to complete the analysis of the scalar quantity $u(z_1, z_2)$ and prove Lemma~\ref{lemma:u}. 	\nc
	\begin{proof}[Proof of Lemma~\ref{lemma:u}]
		Since $u(\overline{z}, \overline{w}) = \overline{u(z,w)}$, we can assume without loss of generality that $z_1 \in \bddD\cap\mathbb{H}$.
		First, we observe that if \eqref{eq:u_one} holds, then \eqref{eq:u_props} is satisfied automatically. 
		Furthermore, if $z_2 \in \bddD\cap\mathbb{H}$, then \eqref{eq:u_props} for $u(z_1,z_2)$ implies \eqref{eq:u_props} for $u(z_1,\overline{z}_2)$. Indeed, it follows from the Stieltjes representation \eqref{eq:M_stieltjes} that
		\begin{equation} \label{eq:u_stieltjes}
			u(z,w) = \int_\mathbb{R} \frac{\rho(x)\mathrm{d}x}{(x-z)(x-w)}, \qquad z,w \in \mathbb{C}\backslash\mathbb{R}.
		\end{equation}
		Hence, subtracting two copies of \eqref{eq:u_stieltjes} and taking the real part, we obtain
		\begin{equation} \label{eq:re_monot}
			\Re\bigl[ u(z,\overline{w}) - u(z,w)\bigr] = \int_\mathbb{R} \frac{2(\Im w)(\Im z)}{|x-z|^2  |x-w|^2}\rho(x)\mathrm{d}x \ge 0, \qquad z,w \in \mathbb{H}.
		\end{equation}
		
		 We proceed by considering six cases: In Cases 1--3, our goal is to establish \eqref{eq:u_one}. 
		   The simplest case, Case 1, corresponds to $z_1$ and $z_2$ being order-one distance apart. 
		 Cases 2 and 3 cover the situation where $z_1$ and $z_2$ are close to each other and lie in a region where the solution $M$ is uniformly Lipschitz continuous. 
		 In the remaining Cases 4--6 we will establish \eqref{eq:u_sim} and \eqref{eq:u_props}. 
		 The more straightforward Case 4 corresponds to $z_1$ and $z_2$ in the opposite half-planes in the neighborhood of the same bulk point. 
		 The final two cases require careful expansions and pertain to $z_1$ and $z_2$ in the singular spectral regime, close to each other, and either in the same half-plane (Case~5) or in the opposite half-planes (Case~6).  While these six cases have overlaps, they form an exhaustive partition of all configurations of $(z_1,z_2)\in \bddD\times \bddD$. \nc

		 \smallskip
		 \textbf{Case 1}. Assume that $z_1 \in \bddD\cap\mathbb{H}$ and $z_2 \in \bddD$ with  $|z_1 - z_2| \ge \delta_1$ for some $\delta_1\sim 1$. 
		Then, it follows from \eqref{eq:stat_M_bound} and $\delta_1 \sim 1$ that  
		\begin{equation} \label{eq:far_u_bound}
			|u(z_1, z_2)| \le \frac{|m_1| + |m_2|}{|z_1 - z_2|} \lesssim 1. 
		\end{equation}
		 Since $\other{\beta}(z_1, z_2) \gtrsim |z_1 - z_2|^{2/3}$, the desired \eqref{eq:u_one} holds.
		 
		 \smallskip
		 \textbf{Case 2}. Assume that $z_1 \in \bddD\cap\mathbb{H}$ with $\rho(z_1)^{-1}\Im z_1 \ge c_2$ and $|z_1 - z_2 | \le \delta_2$ for some thresholds $c_2, \delta_2 \sim 1$ to be chosen later. Then, it follows from \eqref{eq:kapd_rhoeta} that $\kapd_1 := \dist(z_1, \supp\,\rho) \gtrsim c_2^2$. By choosing $\delta_2 = \delta_2(c_2)$ small enough, we can guarantee that $\kapd_2 := \dist(z_2, \supp\,\rho) \gtrsim 1$. Hence, by \eqref{eq:m_dif_bound1}, we deduce that $\lvert u(z_1, z_2) \rvert \le (\kapd_{1}\kapd_{2})^{-1} \lesssim 1$.
		 On the other hand, since $\other{\beta}(z_1, z_2) \ge \rho(z_1)^{-1}\Im z_1   \gtrsim 1$,  
		 and $\other{\beta}\le 1$ by its definition \eqref{eq:betaf_def}, \nc we conclude that \eqref{eq:u_one} holds.	
		 
		 \smallskip
		 \textbf{Case 3}. Assume that $z_1 \in \bddD\cap\mathbb{H}$ with $\rho(z_1) \ge c_3$ and $z_2\in \bddD\cap\mathbb{H}$ with $|z_1 - z_2| \le \delta_3$, for some thresholds $c_3, \delta_3\sim 1$ to be chosen later. In this case, $1 \ge \other{\beta}(z_1, z_2) \ge \rho(z_1)^2 \gtrsim 1$. It follows from \eqref{eq:1/3Holder_rhoinveta} that $\rho(z) \ge c_3/2$ for all $z\in[z_1, z_2]$, provided $\delta_3 = \delta_3(c_3) \sim 1$ is sufficiently small. 
		 The quantity $u(z_1, z_2)$ admits the representation
		 \begin{equation} \label{eq:u_int}
		 	u(z_1, z_2) = \int_{0}^1 m'(w_r)  \mathrm{d}r, \qquad w_r := z_1 + (z_2 - z_1)r. 
		 \end{equation}
		 On the other hand, taking the   $z_1, z_2\to z$ limit \nc in \eqref{eq:B12_identity} at $t=T$, we obtain
		 \begin{equation} \label{eq:M'id}
		 	M'(z) = \mathcal{B}_{z,z}^{-1}\bigl[M(z)^2\bigr], \qquad z \in \mathbb{H},
		 \end{equation}
		 where $\mathcal{B}_{z,z}$ is the stability operator, defined in \eqref{eq:stab_def}. Therefore, since $w_r\in[z_1, z_2]$, it follows from \eqref{eq:1stab_bound} and \eqref{eq:stat_M_bound} that 
		 \begin{equation} \label{eq:case3_m'_bound}
		 	|m'(w_r)| \lesssim 1+\rho(w_r)^{-2} \lesssim 1, \qquad r \in [0,1].
		 \end{equation}
		 Integrating \eqref{eq:case3_m'_bound}, we conclude that 
		 \begin{equation} \label{eq:case3_u_bound}
		 	|u(z_1, z_2)| \lesssim 1,
		 \end{equation}
		 and, therefore, \eqref{eq:u_one} holds.
		 
		 \smallskip
		 In the remaining cases, we aim to establish \eqref{eq:u_sim} and \eqref{eq:u_props}.
		 
		 \smallskip
		 \textbf{Case 4}. Assume that $z_1 \in \bddD\cap\mathbb{H}$ with $\rho(z_1) \ge c_3$ and $\overline{z}_2\in \bddD\cap\mathbb{H}$ with $|z_1 - \overline{z}_2| \le 1$, where $c_3$ is the threshold from Case 3. Then, it follows from \eqref{eq:stat_M_bound} that 
		 \begin{equation} \label{eq:u_case3_est}
		 	|u(z_1, z_2)|= \frac{|m(z_1) - m(z_2)|}{|z_1 - z_2|} \sim \frac{\rho(z_1) + |m(z_1) - m(\overline{z}_2)|}{|z_1 - z_2|} \sim \frac{1}{|z_1 - z_2|}. 
		 \end{equation}
		 On the other hand, $\other{\beta}(z_1, z_2) \sim |z_1 - z_2|$ by \eqref{eq:betaf_def}. Therefore, \eqref{eq:u_sim} holds.  On the other hand, \eqref{eq:u_props} holds by \eqref{eq:re_monot}, 
		   and \eqref{eq:far_u_bound} or \eqref{eq:case3_u_bound} applied to $u(z_1, \overline{z}_2)$. \nc

		\smallskip		
		\textbf{Case 5.}
		Assume that $z_1 \in \smallD_{c_5}\cap\mathbb{H}$, defined in \eqref{eq:shape_dom}, $z_2\in \bddD\cap\mathbb{H}$, and $|z_1 - z_2| \le \delta_5$ for some positive thresholds  $c_5, \delta_5\sim 1$, satisfying $c_5 \le \rho_*/2$ and $\delta_5 \le \delta_*\wedge c_M$ to be chosen later. Here $\rho_*, \delta_* \sim 1$ are the constants from Lemma~\ref{lemma:xi}. 
		  It follows from \eqref{eq:xif_def} and \eqref{eq:betaf_def} that $|z_1 - z_2| \lesssim (\rho(z_1)^{-1}\Im z_1 + \rho(z_1) +  |z_1 - z_2|^{1/2}) \other{\xi}(z_1, z_2)$.
		Hence, by suitably shrinking $\delta_5 \sim 1$, the first relation in \nc 
		\eqref{eq:LR_props} and \eqref{eq:xi_bounds}   together with \eqref{eq:xif_def} \nc  imply that  
		\begin{equation} \label{eq:u_case1_est}
			|u(z_1, z_2)| \sim \frac{\other{\xi}(z_1, z_2)}{|z_1 - z_2|}  \sim \other{\beta}(z_1, z_2)^{-1}.
		\end{equation}
		Therefore, \eqref{eq:u_sim} holds. 
		Furthermore, since the map $z \mapsto \rho^{-1}(z)\Im z + \rho(z)$ is $1/3$-H\"older-regular on $\bddD\cap\mathbb{H}$ by \eqref{eq:1/3Holder_rhoinveta}, by shrinking $\delta_5 \sim 1$, we obtain
		\begin{equation} \label{eq:shpdom_incl}
			\{ \zeta \in \bddD\cap\mathbb{H} \, :\, |\zeta-z_1| \le \delta_5 \} \subset \smallD_{2c_5} \cap\mathbb{H}.
		\end{equation}
		Recall from \eqref{eq:bddD_structure} that if $z, w$ belong to different connected components of $\bddD$, then $|z - w| \ge c_M$.   
		Hence,   for small enough $\delta_5$, \nc the linear segment $[z_1, z_2] \subset \smallD_{2c_5}\cap\mathbb{H}$, and to prove \eqref{eq:u_props} it suffices to show that
		\begin{equation} \label{eq:deriv_goal}
			\Re m'(z) \ge -C, \qquad z \in \smallD_{2c_5}\cap\mathbb{H},
		\end{equation}
		where we used the identity \eqref{eq:u_int}.
		
		We now proceed to establish \eqref{eq:deriv_goal}. 
		Taking the trace of \eqref{eq:M'id}, decomposing $\mathcal{B}^{-1}_{z,z}$ according to $\mathrm{Id} = \Pi_{z,z} + (\mathrm{Id}-\Pi_{z,z})$, and using the first equality in \eqref{eq:Pi_beta}, we obtain
		\begin{equation} \label{eq:m'exp}
			m' = \frac{\langle V_\ell^* M^2 \rangle \langle V_r \rangle}{\beta \langle V_\ell^*V_r \rangle} + \bigl\langle \mathcal{B}_{z,z}^{-1} (\mathrm{Id}-\Pi_{z,z}) \bigl[M^2\bigr] \bigr\rangle,
		\end{equation}
		where we omit the argument $z$ of the functions $m', V_\ell, M, V_r$ and $\beta \equiv \beta_{z,z}$ for brevity. It follows immediately from \eqref{eq:stat_M_bound} and the second bound in \eqref{eq:PiQ_bounds}   for $\mathcal{X}=  \mathcal{B}_{z,z}$ \nc that the second term on the right-hand side of \eqref{eq:m'exp} is $\mathcal{O}(1)$. For the first term, we use \eqref{eq:Pi_beta}--\eqref{eq:LR_props} to obtain
		\begin{equation}
			\begin{split}
				\pi^{-2}\,\Re \Bigl[\frac{\langle V_\ell^* M^2 \rangle \langle V_r \rangle}{\beta \langle V_\ell^*V_r \rangle}\Bigr] &=  \frac{ \pi\rho^{-1}\Im z + 2\other{\psi}\rho^2 +  \mathcal{O}(\rho^2 |\sigma|) + \mathcal{O}(\rho^3 + \Im z + \rho^{-2}(\Im z)^2) }{\bigl\lvert \beta \langle V_\ell^*V_r \rangle \bigr\rvert^2}\\
				&\ge \frac{ \pi\rho^{-1}\Im z + \other{\psi}\rho^2 }{\bigl\lvert \beta \langle V_\ell^*V_r \rangle \bigr\rvert^2}\Bigl(1 + \mathcal{O}(\rho+ \rho^{-1}\Im z)\bigr) + \other{\psi}^{-1}\mathcal{O}\bigl(|\beta|^{-2}\rho^2 |\sigma|^2 \Bigr),
			\end{split}
		\end{equation}
		where $\other{\psi} := \psi + \langle F_\mathscr{U}^2 \rangle^{-1}\sigma^2$ in the second step we used the Cauchy--Schwarz inequality   $ C|\sigma| \le \tfrac{1}{2}\other{\psi} 
		+ \tfrac{1}{2}\other{\psi}^{-1}C^2|\sigma|^2$ for any $C > 0$, and  $|\langle V_\ell, V_r\rangle|\sim 1$ from \eqref{eq:LR_props}.  \nc
		Since $\other{\psi} \sim 1$ by \eqref{eq:psi_LR_asymp}, and $|\beta| \gtrsim \rho |\sigma|$ by \eqref{eq:Pi_beta},  we obtain the desired \eqref{eq:deriv_goal}. 
		
		\smallskip
		\textbf{Case 6}. Assume that $z_1 \in \smallD_{c_5}\cap\mathbb{H}$ and $\overline{z}_2 \in \bddD\cap\mathbb{H}$ with $|z_1 - \overline{z}_2| \le \delta_5$, where $c_5$ and $\delta_5$ are the thresholds from Case 1 above.  Then, 
		\begin{equation} \label{eq:u_case2_est}
			|u(z_1, z_2) | \sim \frac{\other{\xi}(z_1, \overline{z}_2) + \rho(z_1)}{|z_1 - z_2|} \sim \frac{|z_1 - \overline{z}_2|}{|z_1 - \overline{z}_2| + \Im z_1} \other{\beta}(z_1, \overline{z}_2)^{-1} + \frac{\rho(z_1)}{|z_1 - z_2|}.
		\end{equation}
		If $\Im z_1 \le |z_1 - \overline{z}_2|$, then $\eqref{eq:u_case2_est}$ and \eqref{eq:betaf_def} imply
		\begin{equation}
			|u(z_1, z_2)| \sim \frac{1}{\other{\beta}(z_1, \overline{z}_2)} + \frac{\rho(z_1)}{|z_1 - z_2|} \sim  \other{\beta}(z_1, z_2)^{-1}.
		\end{equation}
		On the other hand, if $\Im z_1 \ge |z_1 - \overline{z}_2|$, then   $\Im z_1 \ge |z_1 - z_2|$ as well, and  it follows from \eqref{eq:betaf_def} that \nc 
		\begin{equation} 
			\rho(z_1)^{-1}|z_1 - z_2|\lesssim\rho(z_1)^{-1}\Im z_1 \le \other{\beta}(z_1, \overline{z}_2) \quad \Longrightarrow \quad \other{\beta}(z_1, z_2) \sim \rho(z_1)^{-1}|z_1 - z_2|. 
		\end{equation}
		Hence, we deduce from \eqref{eq:u_case2_est} that $|u(z_1, z_2)| \sim |z_1 - z_2|^{-1}\rho(z_1) \sim  \other{\beta}(z_1, z_2)^{-1}$. 
		Therefore, \eqref{eq:u_sim} holds.
		The bound \eqref{eq:u_props} on $\Re u(z_1, z_2)$ holds by \eqref{eq:re_monot} and the corresponding bound on $\Re u(z_1, \overline{z}_2)$ established in Case 5.  
		
		It is straightforward to check that with appropriate choices of thresholds the six cases above are exhaustive. 
		This concludes the proof of Lemma~\ref{lemma:u}.
	\end{proof}
	
	\subsection{Matrix Analysis. Proof of Lemmas~\ref{lemma:Gamma} and Lemma~\ref{lemma:tripple_prop}} \label{sec:Gamma}
	Next, we perform the matrix-valued analysis and prove Lemma~\ref{lemma:Gamma}. 
	We begin by collecting the necessary perturbative estimates for the stability operator and related quantities. \nc 
	\begin{lemma} [Analytic Perturbation Theory] \label{lemma:perturb} 
		Fix a norm $\norm{\cdot}_\star \in \{ \norm{\cdot},\, \norm{\cdot}_{\mathrm{hs}\to \mathrm{hs}}\}$ on the space of (linear) super-operators on $\mathbb{C}^{N\times N}$.
		Let $\mathcal{B}_0$ and $\mathcal{B}$ be a pair of super-operator on $\mathbb{C}^{N\times N}$ with $\varepsilon$-well-structured spectrum in the $\norm{\cdot}_\star$, as in Definition~\ref{def:pert_cond}, for some $\varepsilon \sim 1$. Then, the following properties hold:
		\begin{itemize}
			\item[(i)] The eigenprojectors $\Pi_0 \equiv \Pi_{\mathcal{B}_0}$ and $\Pi \equiv \Pi_\mathcal{B}$, defined in \eqref{eq:Pi_def}, satisfy 
			\begin{equation} \label{eq:Pi_perturb}
				\Pi = \Pi_0 - \Pi_0 (\mathcal{B}-\mathcal{B}_0) \frac{\mathcal{Q}_0}{\beta_0\,\mathrm{Id} - \mathcal{B}_0} -  \frac{\mathcal{Q}_0}{\beta_0\,\mathrm{Id} - \mathcal{B}_0} (\mathcal{B}-\mathcal{B}_0) \Pi_0 + \Pi^{\mathrm{err}}, \qquad \norm{\Pi^{\mathrm{err}}}_\star   \lesssim \norm{\mathcal{B}-\mathcal{B}_0}_\star^2
			\end{equation}
			where $\beta_0 := \Tr[\mathcal{B}_0\Pi_0]$ is the smallest (in modulus) eigenvalue of $\mathcal{B}_0$, and $\mathcal{Q}_0 := \mathrm{Id} - \Pi_0$. 
			
			\item[(ii)] The smallest eigenvalue $\beta := \Tr[\mathcal{B}\Pi]$ of $\mathcal{B}$ satisfies
			\begin{equation} \label{eq:beta_perturb}
				\beta = \beta_0 + \Tr\bigl[(\mathcal{B}-\mathcal{B}_0)\Pi_0\bigr] + \beta^\mathrm{err}, \qquad |\beta^\mathrm{err}|   \lesssim \norm{\mathcal{B}-\mathcal{B}_0}_\star^2. 
			\end{equation}
			
			\item[(iii)] The projection of $\mathcal{B}^{-1}$ onto $\mathcal{Q} := \mathrm{Id} - \Pi$ satisfies
			\begin{equation} \label{eq:BinvQ_perturb}
				\norm{\mathcal{B}^{-1}\mathcal{Q} - \mathcal{B}_0^{-1}\mathcal{Q}_0 }_\star   \lesssim \norm{\mathcal{B}-\mathcal{B}_0}_\star. 
			\end{equation}
		\end{itemize}
	\end{lemma}

	\subsubsection{Second-Order Difference Cancellations. Proof of Lemma~\ref{lemma:Gamma}} \label{sec:Gamma_proof}

	\begin{definition}[Second-Order Difference Cancellation] \label{def:sodc}
		Let $F:\{1, \bar 1\} \times \{2, \bar 2\} \to \mathbb{X}$ be a matrix- or scalar-valued function, that is, with $\mathbb{X} = \mathbb{C}^{N\times N}$ or $\mathbb{X} = \mathbb{C}$, respectively. Let $\varepsilon : \{1, 2\} \to \mathbb{R}_{\ge 0}$ be a non-negative function. We say that $F$ satisfies a \emph{second-order difference cancellation with size} $\varepsilon$ if and only if the bounds
		\begin{equation}
			\bigl\lVert F_{\bar 1, \mathfrak{n}_2} - F_{1, \mathfrak{n}_2} \bigr\rVert_\mathrm{hs} \le \varepsilon_1,  \qquad  
			\bigl\lVert F_{\mathfrak{n}_1, \bar 2} - F_{\mathfrak{n}_1, 2} \bigr\rVert_\mathrm{hs} \le \varepsilon_2, \qquad 
			\bigl\lVert F_{1,2} + F_{\bar 1, \bar 2} - F_{1, \bar 2} - F_{\bar 1, 2} \bigr\rVert_\mathrm{hs} \le \varepsilon_1\varepsilon_2,
		\end{equation}
		hold for all $\mathfrak{n}_1 \in \{1, \bar 1\}$ and $\mathfrak{n}_2 \in \{2,\bar 2\}$. 
	\end{definition}
	
	\begin{claim} [Properties of second-order difference cancellations] \label{claim:sodc}
		Let $F :\{1, \bar 1\} \times \{2, \bar 2\} \to \mathbb{C}^{N \times N}$ be a matrix-valued function satisfying a second-order difference cancellation with size $\varepsilon$, and let $\langle F \rangle$ denote the scalar-valued function given by $\langle F \rangle : (\mathfrak{n}_1,\mathfrak{n}_2) \mapsto \langle F_{\mathfrak{n}_1,\mathfrak{n}_2} \rangle$ for all $\mathfrak{n}_1 \in \{1, \bar 1\}$ and $\mathfrak{n}_2 \in \{2,\bar 2\}$. Then, $F$ enjoys the following properties:
		\begin{itemize}
			\item[(i)] If $\langle F \rangle$ is bounded from below, that is, $|\langle F_{\mathfrak{n}_1, \mathfrak{n}_2} \rangle| \ge c$ for all $\mathfrak{n}_1 \in \{1, \bar 1\}$, $\mathfrak{n}_2 \in \{2,\bar 2\}$, and some small constant $c\in(0,1]$, then $\langle F\rangle^{-1}$ also satisfies a second-order difference cancellation with size~$c^{-2}\varepsilon$. 
			\item[(ii)] If we additionally assume that $F$ is bounded from above, that is, $\lVert F_{\mathfrak{n}_1,\mathfrak{n}_2} \rVert_\mathrm{hs} \le C$ for all $\mathfrak{n}_1 \in \{1, \bar 1\}$, $\mathfrak{n}_2 \in \{2,\bar 2\}$, and some constant $C \ge 1$, then $\langle F \rangle^{-1} F$ also satisfies a second-order difference cancellation with size~$Cc^{-2}\varepsilon$.
		\end{itemize}
	\end{claim}
	The proof of Claim~\ref{claim:sodc} is trivial and therefore omitted.  
	
	To make the presentation more concise, we introduce the following notation: for any object $f_{w_1, w_2}$ depending on a pair of spectral parameters $w_1, w_2 \in \{z_1, \overline{z}_1, z_2, \overline{z}_2\}$ with $z_1, z_2\in\mathbb{H}$, and for any  pair of labels $\mathfrak{n}_1, \mathfrak{n}_2 \in~\{1,\bar 1, 2, \bar 2\} $, we identify
	\begin{equation} \label{eq:ind_conv}
		f_{\mathfrak{n}_1 \mathfrak{n}_2} \equiv f_{w(\mathfrak{n}_1), w(\mathfrak{n}_2)}, \qquad w(\mathfrak{n}) := \begin{cases}
			z_j, \quad&\text{if } \mathfrak{n} = j \in \{1,2\},\\
			\overline{z}_j, \quad&\text{if } \mathfrak{n} = \bar j \in \{\bar 1,\bar 2\}. 
		\end{cases}
	\end{equation}
	For example, $\beta_{12} \equiv \beta_{z_1, z_2}$ and $\beta_{\bar 1 \bar 2} := \beta_{\bar z_1, \bar z_2}$. 
	
	For all $z,w \in \bddD$ satisfying $\min\{\other{\beta}(z,w), \other{\beta}(w,z)\} \le \beta_*$, it follows from \eqref{eq:M_bound}, \eqref{eq:B12_identity}, \eqref{eq:stab_beta_bound},  and \eqref{eq:beta_assymp} that
	\begin{equation}
		\bigl\lvert \bigl\langle M(z) - M(w) \bigr\rangle \bigr\rvert 
		\lesssim  \norm{M(z) - M(w)} 
		\lesssim \other{\beta}(z, w)^{-1} |z-w| \sim  \bigl\lvert \bigl\langle M(z) - M(w) \bigr\rangle \bigr\rvert.
	\end{equation}
	Therefore, for all $z,w \in \bddD\cap\mathbb{H}$ satisfying $\min\{\other{\beta}(z,w), \other{\beta}(w,z)\} \le \beta_*$,
	\begin{equation} \label{eq:Mdiff_norm}
		\norm{M(z) - M(w)} \sim \bigl\lvert \bigl\langle M(z) - M(w) \bigr\rangle \bigr\rvert \sim \other{\xi}(z,w), 
		\qquad \norm{M(z) - M(w)^*} \sim \bigl\lvert \bigl\langle M(z) - M(w)^* \bigr\rangle \bigr\rvert \sim \other{\xi}(z,w) + \rho(z),
	\end{equation} 
	where we additionally used  \eqref{eq:xif_def}--\eqref{eq:xi_bounds}.
	
	We now prove Lemma~\ref{lemma:Gamma} using the concept of second-order difference cancellations from Definition~\ref{def:sodc}.
	The key idea is that the divided difference quantities of the form $\frac{M_{1,t} - M_{2,t}}{\langle M_{1,t} - M_{2,t}\rangle}$ are time independent, while quantities of the form 
	$f_t := \frac{z_{1,t} - z_{2,t}}{\langle M_{1,t} - M_{2,t}\rangle}$ depend on time as $f_t = \ee^{-t}f_0 - (1-\ee^{-t})$. Therefore, they can be studied at time $t=0$ where the corresponding stability operators are bounded, which greatly simplifies the analysis (see, e.g., Claim~\ref{claim:stab_rho_cancel}).
	\begin{proof}[Proof of Lemma~\ref{lemma:Gamma}]
		We begin by proving \eqref{eq:Gamma_cancel} and \eqref{eq:Gamma_sizes}. To this end, we express the left-hand side of \eqref{eq:Gamma_cancel} as a second-order difference.
		We adopt the notation from \eqref{eq:U_def} and \eqref{eq:Vt_def}. Recall \eqref{eq:UVlin_dep}, and observe that it is equivalent to
		\begin{equation}
			(m_{1} - m_{2}) \biggl(\frac{U}{\langle U \rangle} - \frac{V}{\langle V \rangle}\biggr) + (\overline{m}_{1} - \overline{m}_{2}) \biggl(\frac{U^*}{\langle U^*\rangle} - \frac{V}{\langle V \rangle}\biggr) = (\overline{m}_{1} - m_{2})\biggl(\frac{ V^*}{\langle V^*\rangle} - \frac{V}{\langle V \rangle}\biggr),
		\end{equation}
		where $m_j := \langle M_j\rangle$. 
		In particular, we have the identity
		\begin{equation} \label{eq:M_dd_id}
			\frac{1}{\rho_2}\frac{U}{\langle U \rangle} + \frac{1}{\rho_1}\frac{U^*}{\langle U^* \rangle} - \frac{\rho_1 + \rho_2}{\rho_1\rho_2}\frac{V}{\langle V \rangle} = \ii\frac{\overline{m}_1 - m_2}{2\pi\rho_1\rho_2}\biggl(\frac{U}{\langle U \rangle} + \frac{U^*}{\langle U^*\rangle} - \frac{V}{\langle V \rangle}  - \frac{V^*}{\langle V^* \rangle} \biggr).
		\end{equation} 		
		Therefore, to establish \eqref{eq:Gamma_cancel}--\eqref{eq:Gamma_sizes}, it suffices to show that the map
		\begin{equation} \label{eq:M_dd}
			\{\mathfrak{n}_1,\mathfrak{n}_2\} \mapsto \frac{M_{\mathfrak{n}_1} - M_{\mathfrak{n}_2}}{\langle M_{\mathfrak{n}_1} - M_{\mathfrak{n}_2} \rangle}, \qquad \mathfrak{n}_1 \in \{1, \bar 1\}, \quad \mathfrak{n}_2 \in\{2,\bar 2\}
		\end{equation}
		satisfies a second-order difference cancellation with size $\rho$ (recall that  $\rho_j := \rho(z_j)$). Here we identify $M_j := M(z_j)$ and $M_{\bar j} := M(z_j)^*$  for $j \in \{1,2\}$.
		To show this, we first observe that by \eqref{eq:M_t}, we have
		\begin{equation}
			\frac{M(w_1) - M(w_2)}{\bigl\langle M(w_1) - M(w_2) \bigr\rangle} = \frac{M_t(w_{1,t}) - M_t(w_{2,t})}{\bigl\langle M_t(w_{1,t}) - M_t(w_{2,t}) \bigr\rangle}, \qquad t\in[0,T], \qquad w_1 \in \{z_1, \overline{z}_1\}, \quad w_2 \in \{z_2, \overline{z}_2\}.
		\end{equation} 
		In particular, setting $t=0$ and using \eqref{eq:B12_identity}, we have
		\begin{equation}\label{eq:Mid}
			\frac{M(w_1) - M(w_2)}{\bigl\langle M(w_1) - M(w_2) \bigr\rangle} = \frac{\mathcal{B}^{-1}_{0, w_1, w_2}[M_0(w_{1,0})M_0(w_{2,0})]}{\bigl\langle \mathcal{B}^{-1}_{0, w_1, w_2}[M_0(w_{1,0})M_0(w_{2,0})] \bigr\rangle},
		\end{equation}
		where we recall that $M_{j,0} = M_0(z_{j,0})$ is the value of $M_{j,t}$ at time $t=0$. Since at time $t=0$ we have $\other{\beta}_0(w_1, w_2) \sim 1$ by \eqref{eq:betaf_t_def}, it follows from Lemma~\ref{lemma:stab_t} that $\lVert \mathcal{B}^{-1}_{0, w_1, w_2} \rVert_{\mathrm{hs}\to\mathrm{hs}}  \lesssim 1$. Hence, \nc it follows from \eqref{eq:M_bound} and Claim~\ref{claim:stab_rho_cancel} that, adopting the convention \eqref{eq:ind_conv}, the function 
		\begin{equation} \label{eq:4foldF_def}
			F_{\mathfrak{n}_1,\mathfrak{n}_2} := \mathcal{B}^{-1}_{0, \mathfrak{n}_1,\mathfrak{n}_2}[M_{\mathfrak{n}_1,0}M_{\mathfrak{n}_2,0}], \qquad \mathfrak{n}_1 \in \{1, \bar1\},\quad \mathfrak{n}_2 \in \{2, \bar 2\},
		\end{equation}
		satisfies a second-order difference cancellation with size $C\rho$ for some $C \sim 1$.
		Furthermore, it follows from \eqref{eq:z_diff_comp}, \eqref{eq:M_t}, \eqref{eq:beta_assymp}, and $T\sim 1$, that 
		\begin{equation} \label{eq:F_tr_lower}
			\bigl\lvert \langle F_{\mathfrak{n}_1,\mathfrak{n}_2} \rangle \bigr\rvert = \frac{|m_{\mathfrak{n}_1,0} - m_{\mathfrak{n}_2,0}|}{|z_{\mathfrak{n}_1,0} - z_{\mathfrak{n}_2,0}|} 
			\sim \frac{|m_{\mathfrak{n}_1,T} - m_{\mathfrak{n}_2,T}|}{|z_{\mathfrak{n}_1,T} - z_{\mathfrak{n}_2,T}| + T |m_{\mathfrak{n}_1,T} - m_{\mathfrak{n}_2,T}|}  \sim 1. 
		\end{equation}
		Hence, by Claim~\ref{claim:sodc}, we conclude by~\eqref{eq:Mid} that the map \eqref{eq:M_dd} also satisfies a second-order difference cancellation of size $C\rho$ for some $C\sim1$. In particular, the bounds \eqref{eq:Gamma_sizes} hold. Furthermore,   
		\begin{equation}
			\biggl\lVert \frac{U}{\langle U \rangle} + \frac{U^*}{\langle U^*\rangle} - \frac{V}{\langle V \rangle}  - \frac{V^*}{\langle V^* \rangle} \biggr\rVert_\mathrm{hs} \lesssim \rho_1 \rho_2,
		\end{equation}
		and the desired bound \eqref{eq:Gamma_cancel} follows from \eqref{eq:M_dd_id}. 
		
		Next, we prove \eqref{eq:1/u-4fold}. To this end, we observe that, by \eqref{eq:M_t} and \eqref{eq:z_t}, we have, for all $t\in[0,T]$, $w_1 \in \{z_1, \overline{z}_1\}$, and $w_2 \in \{z_2, \overline{z}_2\}$, 
		\begin{equation}\label{eq:z-z/m-m}
			\frac{w_{1,t} - w_{2,t}}{\bigl\langle M_t(w_{1,t}) - M_t(w_{2,t})\bigr\rangle} = \ee^{-t}\frac{w_{1,0} - w_{2,0}}{\bigl\langle M_t(w_{1,0}) - M_0(w_{2,0})\bigr\rangle} - (1-\ee^{-t}), \qquad t\in[0,T].
		\end{equation}
		Therefore, establishing the desired \eqref{eq:1/u-4fold} at any time $t$ is equivalent to proving it at time $t=0$ since the additive $1-\ee^{-t}$ terms cancel and the multiplicative factors $\ee^{-t}$ are order-one.
		Furthermore, at time $t=0$, the left-hand side of \eqref{eq:1/u-4fold} can be expressed as
		\begin{equation}
			\Re\biggl[\frac{z_{1,0} - z_{2,0}}{\langle M_{1,0} - M_{2,0} \rangle} - \frac{z_{1,0} - \overline{z}_{2,0}}{\langle M_{1,0} - M_{2,0}^* \rangle}\biggr]  = \frac{1}{2}\biggl(\frac{1}{\langle F_{1,2}\rangle} + \frac{1}{\langle F_{\bar 1, \bar 2} \rangle} - \frac{1}{\langle F_{1, \bar 2} \rangle} - \frac{1}{\langle F_{\bar 1, 2}\rangle}\biggr), 
		\end{equation}
		where $F$ is defined in \eqref{eq:4foldF_def}. Since $F$ satisfies a second-order difference cancellation with size $\rho$, the desired bound \eqref{eq:1/u-4fold} follows immediately   by Claim~\ref{claim:sodc}(i).  
		
		Finally, we prove \eqref{eq:1/u-rho_eta}. As in the proof of \eqref{eq:4foldF_def} above, we reduce the problem to a second-order difference cancellation   at time $t=0$.  
		To this end, analogously to \eqref{eq:z-z/m-m}, we write, for all $t\in[0,T]$ and $j \in \{1,2\}$,
		\begin{equation}
			\frac{\Im z_{j,t}}{\pi\rho_t(z_{j,t})} = \frac{z_{j,t} - \overline{z}_{j,t}}{\bigl\langle M_t(z_{j,t}) - M_t(z_{j,t})^* \bigr\rangle} = \ee^{-t}\frac{z_{j,0} - \overline{z}_{j,0}}{\bigl\langle M_0(z_{j,0}) - M_0(z_{j,0})^* \bigr\rangle} - (1- \ee^{-t}). 
		\end{equation}
		Therefore, using \eqref{eq:z-z/m-m} once again, we deduce that it suffices to prove \eqref{eq:1/u-rho_eta} at time $t=0$, and the left-hand side of \eqref{eq:1/u-rho_eta} at $t=0$ is given by 
		\begin{equation} \label{eq:1/u-rho_eta_Kexpr}
			2\Re\biggl[\frac{z_{1,0} - z_{2,0}}{\langle M_{1,0} - M_{2,0} \rangle}\biggr] - \frac{\Im z_{1,0}}{\rho_t(z_{1,0})} - \frac{\Im z_{2,0}}{\rho_t(z_{2,0})} = \frac{1}{\langle K_{1, \bar 2}\rangle} + \frac{1}{\langle K_{\bar 1, 2}\rangle} - \frac{1}{\langle K_{1,2}\rangle} - \frac{1}{\langle K_{\bar 1, \bar 2}\rangle},
		\end{equation}
		where the function $K_{\mathfrak{n}_1,\mathfrak{n}_2}$ is given by
		\begin{equation}
			\begin{split}  				
				K_{1,2} &:= \mathcal{B}^{-1}_{0, z_1,\bar z_1}[M_{1,0}M_{1,0}^*], \quad K_{\bar 1,2} := \mathcal{B}^{-1}_{0, \bar z_2,\bar z_1}[M_{2,0}^*M_{1,0}^*],\\
				K_{1,\bar 2} &:= \mathcal{B}^{-1}_{0, z_1,z_2}[M_{1,0}M_{2,0}], \quad K_{\bar 1, \bar 2} := \mathcal{B}^{-1}_{0, \bar z_2,z_2}[M_{2,0}^*M_{2,0}].
			\end{split}
		\end{equation} 
		Completely analogously to the proof of Claim~\ref{claim:stab_rho_cancel}, it can be shown that $K$ satisfies a second-order difference cancellation with size $\varepsilon : \{1,2\} \mapsto C\norm{M_1 - M_2^*}$ for some $C\sim1 $. Furthermore, similarly to \eqref{eq:F_tr_lower}, we have $|\langle K \rangle| \sim 1$. Hence, by Claim~\ref{claim:sodc}, it follows that  
		\begin{equation} \label{eq:K_cancel}
			\biggl\lvert \frac{1}{\langle K_{1, \bar 2}\rangle} + \frac{1}{\langle K_{\bar 1, 2}\rangle} - \frac{1}{\langle K_{1,2}\rangle} - \frac{1}{\langle K_{\bar 1, \bar 2}\rangle} \biggr\rvert \lesssim \norm{M_1 - M_2^*}^2 \lesssim \rho_2^2 + \norm{M_1 - M_2}^2 \sim |m_1 - \overline{m}_2|^2,
		\end{equation}
		where in the last step we used \eqref{eq:Mdiff_norm}. Therefore, the desired bound \eqref{eq:1/u-rho_eta} follows from Claim~\ref{claim:sodc}, \eqref{eq:1/u-rho_eta_Kexpr}, and \eqref{eq:K_cancel}. 
		This concludes the proof of Lemma~\ref{lemma:Gamma}.   		
	\end{proof}

		\subsubsection{Quadratic Equation Analysis. Proof of Lemma~\ref{lemma:beta_reg}} \label{sec:quadratic}  
		\begin{proof}[Proof of Lemma~\ref{lemma:beta_reg}]
			Recall the definition of $u(z,w) := (z-w)^{-1}\langle M(z) - M(w)\rangle$ from \eqref{eq:u_def}. 
			Let  the spectral parameters $z \in \bddD\cap\mathbb{H}$ and $w \in \bddD$
			satisfy $\min\{\other{\beta}(z,w^+), \other{\beta}(w^+,z)\} \le \beta_*$, where $\beta_*$ is the threshold from Lemma~\ref{lemma:stab_bound}. Then, it follows from \eqref{eq:beta_assymp} that 
			\begin{equation} \label{eq:u_lower}
				|u(z,w)|^{-1} \sim \other{\beta}(z,w) \lesssim 1.
			\end{equation}
			Then, for all $w\in\{z_1, \overline{z}_1, z_2,\overline{z}_2\}$, the quantity on the left-hand side of \eqref{eq:u_reg} admit the expressions
			\begin{equation} \label{eq:u_exprs}
				\frac{z_{1,t} - z_{2,t}}{\langle M_t(z_{1,t}) - M_t(z_{2,t}) \rangle} - \frac{z_{1,t} - w_t}{\langle M_t(z_{1,t}) - M_t(w_t) \rangle}  = \ee^{T-t}\biggl(\frac{1}{u(z_1, z_2)} - \frac{1}{u(z_1, w)}\biggr),
			\end{equation}
			where we recall that $z_j \equiv z_{j,T}$ is the terminal condition of the characteristic $z_{j,t}$ solving \eqref{eq:char_flow}.
			
			Note that, by \eqref{eq:B12_identity},  $u(z,w)^{-1}$ admits the expression
			\begin{equation} \label{eq:u_expr}
				u(z,w)^{-1} = \bigl\langle\mathcal{B}_{z,w}^{-1}\bigl[M(z)M(w)\bigr]\bigr\rangle^{-1}.
			\end{equation}
			However, a straightforward application of the perturbative estimates from Lemma~\ref{lemma:perturb} to \eqref{eq:u_exprs} is not sufficient to deduce the desired \eqref{eq:beta_reg}, as it only yields
			\begin{equation} \label{eq:u_pert_trivial}
				\bigl\lvert u(z_1, z_2)^{-1} - u(z_1, w)^{-1} \bigr\rvert \lesssim \bigl\lvert \bigl\langle M(z_2) - M(w) \bigr\rangle \bigr\rvert,
			\end{equation}
			which is off by a (potentially small) factor of $\other{\alpha}(z_1, z_2)$. 		
			
			To amend this deficiency, we return to the quadratic equation for the difference $M(w) - M(z)$, studied in \cite[Section~6.1]{AEK2020}. Let $\Theta \equiv \Theta(z,w) := M(w) - M(z)$, then, subtracting two copies of \eqref{eq:MDE},  for all $z\in\mathbb{H}$ and $w \in \mathbb{C}\backslash\mathbb{R}$, we have (see \cite[Eq.~(6.9)]{AEK2020}) 
			\begin{equation} \label{eq:quad}
				\mathcal{B}_{z, z}[\Theta] = \mathcal{A}_{z}[\Theta, \Theta] + (w - z)\mathcal{K}_{z}[\Theta] + (w - z)M(z)^2,
			\end{equation} 
			where $\mathcal{B}_{z,z}$ is the stability operator defined in \eqref{eq:stab_def}, and the forms $\mathcal{A}_z$  and $\mathcal{K}_z$ are given by
			\begin{equation}
				\mathcal{A}_z[X,Y] := \tfrac{1}{2}\bigl(M(z)\mathcal{S}[X]Y + Y\mathcal{S}[X]M(z)\bigr), \qquad \mathcal{K}_z[X] := \frac{1}{2}\bigl(X M(z)   + M(z) X\bigr), \qquad X,Y \in \mathbb{C}^{N\times N}. 
			\end{equation}
			
			In \cite{AEK2020}, the quadratic equation \eqref{eq:quad} was used to study the difference $\Theta$, while in the present proof we use it to analyze the function $u(z,w)^{-1}$.  
			In the sequel we only consider spectral parameters $z\in\bddD\cap\mathbb{H}$ and $w\in\bddD$ satisfying 
			\begin{equation} \label{eq:pert_cond_u}
				|\rho(z)| + |\rho(w)| + \rho(z)^{-1}\Im z + \rho(w)^{-1}\Im w + |z-w| \le c,
			\end{equation}
			for some sufficiently small positive constant $c \sim 1$ to be chosen later. It is straightforward to check using \eqref{eq:betaf_def} that in the complementary case $\other{\alpha}(z^+, w^+) \sim 1$ and the desired bound \eqref{eq:u_reg} follows immediately from \eqref{eq:u_pert_trivial}. 
			Moreover, by suitably shrinking the threshold $c\sim1$, it follows from \eqref{eq:LR_props} that 
			\begin{equation} \label{eq:Vsim1}
				\bigl\lvert \langle \Pi_{z,z}[M(z)^2]\rangle \bigr\rvert \sim \bigl\lvert \bigl\langle V_\ell(z)^*M(z)^2 \bigr\rangle \bigr\rvert \sim \bigl\lvert \bigl\langle V_r(z) \bigr\rangle \bigr\rvert \sim 1, 
			\end{equation}
			where $\Pi_{z,z}$ is the spectral projector associated with the smallest eigenvalue of $\mathcal{B}_{z,z}$ from \eqref{eq:Pi_beta}.
			\nc
			
			To express $u(z,w)^{-1}$, we act on both sides of \eqref{eq:quad} by $\Pi_{z,z}$, take the trace, and divide by $\langle \Pi_{z,z}[M(z)^2]\rangle \langle \Theta \rangle$, obtaining
			\begin{equation} \label{eq:1/u_eq}
				u(z,w)^{-1} = b_z \bigl\langle  \Pi_{z,z}[\Xi] \bigr\rangle - \langle \Theta\rangle a_z[\Xi, \Xi] 
				- (w - z)k_z[\Xi],
			\end{equation}
			where we denote\footnote{
				For brevity, we omit the variables $z, w$ from $\Theta=\Theta(z,w)$ and $\Xi=\Xi(z, w)$ throughout the proof, while we carry them for all other quantities
			} $\Xi := \langle \Theta\rangle^{-1}\Theta$, and the coefficient $b_z$, and the forms  $a_z$ and $k_z$, are given by 
			\begin{equation} \label{eq:b_a_k_def}
				b_z := \frac{\beta_{z,z}}{\langle \Pi_{z,z}[M(z)^2]\rangle}, \qquad
				a_z[X,Y] := \frac{\bigl\langle  \Pi_{z,z}\bigl[\mathcal{A}_{z}[X, Y]\bigr]\bigr\rangle}{\langle \Pi_{z,z}[M(z)^2]\rangle}, \qquad
				k_z[X] := \frac{\bigl\langle  \Pi_{z,z}\bigl[\mathcal{K}_{z}[X]\bigr]\bigr\rangle}{\langle \Pi_{z,z}[M(z)^2]\rangle}, \qquad X,Y \in \mathbb{C}^{N\times N}. 
			\end{equation}
			Note that the denominators are of order unity by \eqref{eq:Vsim1}.

			We now estimate the terms on the right-hand side of \eqref{eq:1/u_eq}. To this end, we record the following properties of the matrix~$\Xi$. 
			The comparisons \eqref{eq:Mdiff_norm} imply that
			\begin{equation} \label{eq:Xi_norm}
				\norm{\Xi} \sim 1.
			\end{equation}
			Hence, multiplying both sides of the equation \eqref{eq:quad} by $\mathcal{Q}_{z,z}\mathcal{B}_{z,z}^{-1}$ with $\mathcal{Q}_{z,z} := \mathrm{Id} - \Pi_{z,z}$, and using  \eqref{eq:PiQ_bounds}, we deduce that 
			\begin{equation} \label{eq:Qbound}
				\bigl\lVert\mathcal{Q}_{z,z}[\Xi]\bigr\rVert  \lesssim  \lvert \langle\Theta\rangle \rvert + |u(z,w)|^{-1}, 
			\end{equation}
			where we used that $|w-z| \lesssim |u(z,w)|^{-1}$,   
			$\| \Theta\|\sim |\langle \Theta\rangle| = |u(z,w)| |z-w|$ by~\eqref{eq:Mdiff_norm},  
			and 
			\begin{equation} \label{eq:A_K_est}
				\norm{\mathcal{A}}_{\norm{\cdot},\norm{\cdot} \to \norm{\cdot}}  +  \norm{\mathcal{K}} \lesssim 1,
			\end{equation}
			which follows from \eqref{eq:M_bound} and \eqref{eq:S_hs_to_op} for all $z,w \in \bddD$. 
			
			Recall that $V_r(z)$ is the right eigenvector associated with the smallest eigenvalue $\beta_{z,z}$ of $\mathcal{B}_{z,z}$, see~\eqref{eq:Pi_beta}. 
			In particular $\Pi_{z,z}[ V_r(z)]=V_r(z)$, and $\| V_r\|\sim |\langle V_r \rangle| \sim 1$ by~\eqref{eq:LR_props} after suitably shrinking the threshold $c\sim1$ in~\eqref{eq:pert_cond_u}.
			Hence, it readily follows from~\eqref{eq:Qbound} and  $\langle \Xi \rangle = 1$ that   
			\begin{equation} \label{eq:Xi_est}
				\bigl\lVert \Xi - \langle V_r(z) \rangle^{-1} V_r(z)
				\bigr\rVert \lesssim \lvert \langle\Theta\rangle \rvert + |u(z,w)|^{-1}. 
			\end{equation}

			Plugging \eqref{eq:Xi_est} into $a_z[\Xi, \Xi]$, and using \eqref{eq:Vsim1} and \eqref{eq:A_K_est}, we conclude that 
			\begin{equation} \label{eq:a_formula}
				a_z[\Xi, \Xi] = \mu_z  + \mathcal{O}\bigl(  \lvert \langle\Theta\rangle \rvert  + |u(z,w)|^{-1} \bigr), \qquad \mu_z := \langle V_r(z) \rangle^{-2} a_z[V_r(z), V_r(z)].
			\end{equation}
			
			Therefore, it follows from \eqref{eq:b_a_k_def} and the estimate on $\langle V_\ell(z) \rangle$ \cite[Eq.~(6.24b) from Lemma~6.3]{AEK2020} that the coefficient $\mu_z$ from \eqref{eq:a_formula} admits the bound
			\begin{equation} \label{eq:mu_bound}
				|\mu_z| \lesssim  \lvert\sigma(z)\rvert + \rho(z) +  \rho(z)^{-1}\Im z \lesssim \other{\alpha}(z, w^+),
			\end{equation}
			where the second inequality follows from \eqref{eq:betaf_def} and  the definition of $\other{\alpha}$ in~\eqref{eq:alp_t_def}.  
			Furthermore, combining \eqref{eq:a_formula}, \eqref{eq:mu_bound}, 
			$\lvert \langle \Theta \rangle \rvert \lesssim    \other{\alpha}(z, w^+)$ by \eqref{eq:m-m_beta_bound},   
			and the bound $|u(z,w)|^{-1} \sim \other{\beta}(z,w)$ by \eqref{eq:beta_assymp}, we deduce that 
			\begin{equation} \label{eq:a_bound}
				\bigl\lvert a_z[\Xi, \Xi] \bigr\rvert \lesssim \other{\alpha}(z, w^+) + \other{\beta}(z,w) \lesssim \other{\alpha}(z, w^+),
			\end{equation}
			which is what allows us to recover the missing $\other{\alpha}(z_1, z_2)$ factor for the estimate \eqref{eq:beta_reg}, since the second term on the right-hand side of \eqref{eq:1/u_eq} is the most sensitive to perturbations in $w$.

			We proceed to prove \eqref{eq:u_reg}. Recalling \eqref{eq:u_exprs}, and subtracting \eqref{eq:1/u_eq} from its copy with $w := z_2$, we obtain
			\begin{equation}   \label{eq:imu_est}
				\begin{split}
					u(z,z_2)^{-1} - u(z,w)^{-1} =&~  \bigl\langle M(w)  - M(z_2) \bigr\rangle a_z[\Xi_2, \Xi_2] 
					+ (w-z_2)k_z[\Xi]  - b_z \bigl\langle  \Pi_{z,z}[\Xi-\Xi_2] \bigr\rangle\\
					& + \bigl\langle M(w)  - M(z) \bigr\rangle \bigl(a_z[\Xi-\Xi_2, \Xi] + a_z[\Xi_2, \Xi-\Xi_2] \bigr)+ (z_2 - z)k_z[\Xi-\Xi_2],   
				\end{split}
			\end{equation} 
			where we abbreviate $\Xi_2 := \Xi(z, z_2)$, while $\Xi := \Xi(z,w)$. 
			
			Observe that the identity \eqref{eq:B12_identity} implies that $\Xi(z,w)$ admits the expression
			\begin{equation} 
				\Xi(z,w) = \frac{\mathcal{B}_{z,w}^{-1}\bigl[M(z)M(w)\bigr]}{\bigl\langle \mathcal{B}_{z,w}^{-1}\bigl[M(z)M(w)\bigr] \bigr\rangle} 
				= \frac{\bigl(\Pi_{z,w}  + \beta_{z,w}\mathcal{Q}_{z,w}\mathcal{B}_{z,w}^{-1}\bigr)\bigl[M(z)M(w)\bigr]}{\bigl\langle \bigl(\Pi_{z,w} + \beta_{z,w}\mathcal{Q}_{z,w}\mathcal{B}_{z,w}^{-1}\bigr)\bigl[M(z)M(w)\bigr] \bigr\rangle}.
			\end{equation}
			It follows from \eqref{eq:stab_def},  \eqref{eq:M_bound}, \eqref{eq:S_hs_to_op}, and \eqref{eq:Mdiff_norm}, that for all $w\in\{z_1, \overline{z}_1, z_2,\overline{z}_2\}$, 
			\begin{equation}
				\norm{ \mathcal{B}_{z,w} - \mathcal{B}_{z,z_2}} \lesssim \norm{M(z)} \norm{\mathcal{S}}_{\mathrm{hs}\to\mathrm{op}} \norm{M(w) - M(z_2)} \lesssim \bigl\lvert\bigl\langle  M(w) - M(z_2) \bigr\rangle \bigr\rvert.
			\end{equation}
			Therefore, it follows from Lemma~\ref{lemma:perturb}, \eqref{eq:u_lower}, and \eqref{eq:u_expr},  that 
			\begin{equation} \label{eq:Xi_bound}
				\norm{\Xi(z,w) - \Xi(z,z_2)} \lesssim \bigl\lvert\bigl\langle  M(w) - M(z_2) \bigr\rangle \bigr\rvert. 
			\end{equation}
			
			Plugging \eqref{eq:Xi_bound} into \eqref{eq:imu_est}, and using \eqref{eq:a_bound} together with \eqref{eq:Xi_bound}, we obtain
			\begin{equation} 
				\Bigl\lvert \frac{1}{u(z,z_2)} - \frac{1}{u(z,w)} \Bigr\rvert \lesssim \bigl\lvert\bigl\langle  M(w) - M(z_2) \bigr\rangle \bigr\rvert \Bigl(|a_z[\Xi_2, \Xi_2]| + |b_z| + \bigl\lvert \langle M(w) - M(z)\rangle \bigr\rvert + |z_2 - z| )|\Bigr) +  |w-z_2|.  
			\end{equation}
			Using \eqref{eq:a_bound} to estimate $a_z[\Xi_2, \Xi_2]$, \eqref{eq:betaf_def}, \eqref{eq:b_a_k_def} and \eqref{eq:Vsim1} to estimate $b_z$, \eqref{eq:m-m_beta_bound} for $\langle M(w) - M(z)\rangle$, and $|z_2 - z| \lesssim \other{\beta}(z, z_2)^{3/2}$ by \eqref{eq:betaf_def}, we deduce that 		
			\begin{equation}
				|a_z[\Xi_2, \Xi_2]| + |b_z| + \bigl\lvert \langle M(w) - M(z)\rangle \bigr\rvert + |z_2 - z| + \frac{|w-z_2|}{\bigl\lvert\bigl\langle  M(w) - M(z_2) \bigr\rangle \bigr\rvert} \lesssim \other{\alpha}(z, z_2) + \other{\beta}(z, z_2) + \other{\beta}(z,w) + \other{\beta}(w,z_2).
			\end{equation}
			Setting $z := z_1$, it follows from \eqref{eq:betaf_def} and \eqref{eq:alp_t_def} that for all $w \in \{z_1, \overline{z}_1, z_2, \overline{z}_2\}$,  
			\begin{equation} 
				\Bigl\lvert \frac{1}{u(z_1,z_2)} - \frac{1}{u(z_1,w)} \Bigr\rvert \lesssim 		\other{\alpha}(z_1, z_2)\bigl\lvert\bigl\langle  M(w) - M(z_2) \bigr\rangle \bigr\rvert.
			\end{equation}
			This concludes the proof of Lemma~\ref{lemma:beta_reg}. \nc
		\end{proof}
		\nc 
		
		\subsection{Proof of Auxiliary Perturbative Results} \label{sec:pert_aux}
		We begin by proving Lemma~\ref{lemma:perturb}, which is a straightforward generalization of the non-Hermitian perturbation theory in \cite[Appendix C]{AEK2020}. \nc 
		\begin{proof}[Proof of Lemma~\ref{lemma:perturb}]
			Let $\mathcal{G}(\zeta) := (\zeta\,\mathrm{Id}-  \mathcal{B})^{-1}$ and $\mathcal{G}_0(\zeta) := (\zeta\,\mathrm{Id}-  \mathcal{B}_0)^{-1}$.		
			Note that 
			\begin{equation} \label{eq:PibetaQ_expr}
				\Pi = \frac{1}{2\pi\ii}\oint_{|\zeta|=\varepsilon} \mathcal{G}(\zeta)\mathrm{d}\zeta, \qquad \beta=\frac{1}{2\pi\ii}\oint_{|\zeta|=\varepsilon} \zeta\,\Tr\bigl[\mathcal{G}(\zeta)\bigr]\mathrm{d}\zeta, \qquad \mathcal{B}^{-1}\mathcal{Q} = \frac{1}{2\pi\ii}\oint_{|1-\zeta| = 1-2\varepsilon}  \frac{\mathcal{G}(\zeta)}{\zeta}\mathrm{d}\zeta.
			\end{equation}
			For all  $\zeta\in\mathbb{C}$ satisfying $|\zeta| \ge \varepsilon$ and $|1 - \zeta| \ge 1 - 2\varepsilon$, we have the identities 
			\begin{equation} \label{eq:zeta_expand}
				\begin{split}
					\mathcal{G}(\zeta) &= \mathcal{G}_0(\zeta) + \mathcal{G}_0(\zeta) (\mathcal{B} - \mathcal{B}_0) \mathcal{G}(\zeta)\\
					&= \mathcal{G}_0(\zeta) + \mathcal{G}_0(\zeta) (\mathcal{B} - \mathcal{B}_0) \mathcal{G}_0(\zeta) +  \mathcal{G}_0(\zeta) (\mathcal{B} - \mathcal{B}_0) \mathcal{G}_0(\zeta) (\mathcal{B} - \mathcal{B}_0) \mathcal{G}(\zeta). 
				\end{split}
			\end{equation}
			The remainder of the proof is analogous to Lemma~C.1 in \cite{AEK2020}, with the bounds \eqref{eq:Pi_perturb}--\eqref{eq:BinvQ_perturb} obtained by integrating \eqref{eq:zeta_expand} over $|\zeta| = \varepsilon$ or $|1-\zeta| = 1-2\varepsilon$,   
			and using the bounds $\norm{\mathcal{G}(\zeta)}_\star + \norm{\mathcal{G}_0(\zeta)}_\star \lesssim 1$ on these contours by \eqref{eq:pert_cond}. \nc	
			This concludes the proof of Lemma~\ref{lemma:perturb}.
		\end{proof}

		Next, we prove Lemma~\ref{lemma:xi}.	
		\begin{proof}[Proof of Lemma~\ref{lemma:xi}]
			First, we prove the bound \eqref{eq:M_perturb}. 
			Note that in the regime $|z - w| \gtrsim 1$, $\other{\xi}(z,w) \sim |z-w|$ and \eqref{eq:M_perturb} holds trivially by \eqref{eq:M_bound}. 
			If $\other{\beta}(z,z) \gtrsim 1$ and $|z-w| \le \delta$ for some sufficiently small constant $\delta\sim 1$, then  $\other{\xi}(z,w) \sim |z-w|$,  and it follows from \eqref{eq:1/3Holder_rhoinveta} and the definition~\eqref{eq:betaf_def} that 
			\begin{equation}
				\other{\beta}\bigl(\zeta(r),\zeta(r)\bigr) \gtrsim 1, \quad \text{for all} \quad \zeta \equiv
				\zeta(r) := z + r (w-z), \qquad r\in [0,1].
			\end{equation}
			Therefore, combining \eqref{eq:M_bound}, \eqref{eq:M'id} and Lemma~\ref{lemma:stab1}, we obtain
			\begin{equation}
				\norm{M(z) - M(w)} \lesssim |z-w| \sup\limits_{r\in[0,1]} \norm{M'\bigl(\zeta(r)\bigr)} \lesssim |z-w| \sup\limits_{r\in[0,1]} \norm{\mathcal{B}^{-1}_{\zeta(r),\zeta(r)}} \lesssim |z-w| \sim \other{\xi}(z,w). 
			\end{equation}
			Therefore, it remains to consider $z \in \smallD_{\delta}$ and $|z - w| \le \delta$ for some sufficiently small $c, \delta \sim 1$. In this regime, 
			Lemma~6.3 in \cite{AEK2020} guarantees the existence of a thresholds $\rho_*, \delta_* \sim 1$~, such that if $c\le \rho_*$ and $\delta \le \delta_*$, \nc then
			\begin{equation} \label{eq:M_decomp}
				M(z) - M(w) =  \xi(z,w) V_r(z)  + \mathcal{O}_{\norm{\cdot}}\bigl( |\xi(z,w)|^2 + |z-w|\bigr), 
			\end{equation} 
			where $V_\ell(z)$ and $V_r(z)$ are the left and right eigenvectors of $\mathcal{B}_{z,z}$ from \eqref{eq:Pi_beta}, and the function $\xi(z,w)$, defined in \eqref{eq:xi_bounds}, solves the approximate cubic equation
			\begin{equation} \label{eq:xi_cubic}
				\mu_3 \xi(z,w)^3 + \mu_2 \xi(z,w)^2 + \mu_1 \xi(z,w) +   (w-z) \mu_0= \mathcal{O}\bigl(|\xi(z,w)|^4 + |\xi(z,w)||w-z| + |w-z|^2 \bigr),
			\end{equation}
			with the coefficients $\mu_0 \equiv \mu_0(z,w)$ and $\mu_k \equiv \mu_k(z)$ for $k\in\{1, 2, 3\}$ satisfy
			\begin{equation} \label{eq:mu_coeff}
				\begin{split}
					\mu_3 &= \psi(z) + \mathcal{O}\bigl(|\rho(z)| + \rho(z)^{-1}\Im z\bigr), \qquad \mu_2 = \sigma(z) + \ii \rho(z) \Bigl(3\psi(z) + \langle F_\mathscr{U}(z)^2 \rangle^{-1}  \sigma(z)^2\Bigr) + \mathcal{O}\bigl(\rho(z)^2 + \rho(z)^{-1}\Im z\bigr), \\
					\mu_1 &= -\pi\rho(z)^{-1} \Im z + 2\ii\rho(z) \sigma(z) -  2\rho(z)^2 \Bigl(\psi(z) + \langle F_\mathscr{U}(z)^2\rangle^{-1}  \sigma(z)^2\Bigr) + \mathcal{O}\bigl(|\rho(z)|^3 + |\Im z| + \rho(z)^{-2}(\Im z)^2\bigr),\\
					\mu_0 &= \pi + \mathcal{O}\bigl(|\rho(z)| + \rho(z)^{-1}\Im z + |\xi(z,w)| \bigr)
				\end{split}
			\end{equation}
			where $\psi(z) \in \mathbb{R}$ satisfies $|\psi| \lesssim 1$ and $\psi(z) + \langle F_\mathscr{U}(z)^2\rangle \sigma(z)^2 \sim 1$. 
			The cubic inequality bootstrapping from Lemma~10.3 in \cite{AEK2020} therefore  implies 
			that $\xi(z,w)$ admits the bound
			\begin{equation} \label{eq:xi_boot}
				|\xi(z,w)| \lesssim \min\Bigl\{ |z-w|^{1/3},\, \frac{|z-w|^{1/2}}{\other{\xi}_2(z)^{1/2}},\, \frac{|z-w|}{\other{\beta}(z,z)} \Bigr\},
			\end{equation}
			where $\xi_2(z)$ is a positive function satisfying $\other{\xi}_2(z) \gtrsim \rho(z) + |\sigma(z)|$. Hence, using \eqref{eq:Pi_beta}--\eqref{eq:LR_props} and \eqref{eq:xif_def},  we deduce from \eqref{eq:xi_boot} that
			\begin{equation} \label{eq:xi_xi}
				|\xi(z,w)| \lesssim \other{\xi}(z,w). 
			\end{equation}
			Since $|z-w| \lesssim \other{\beta}(z, w)\other{\xi}(z,w)$ by \eqref{eq:xif_def},
			we conclude the desired \eqref{eq:M_perturb}
			from~\eqref{eq:M_decomp} and the last bound in \eqref{eq:LR_props}. \nc
			
			Next, we prove \eqref{eq:xi_bounds}. The first bound in \eqref{eq:xi_bounds} holds by \eqref{eq:M_decomp} and \eqref{eq:xi_xi}. 
			The upper bound on $|\xi(z,w)|$ was given in~\eqref{eq:xi_xi}. 
			The proof of \nc the remaining lower bound follows by contradiction from \eqref{eq:xi_cubic} given the estimates \eqref{eq:mu_coeff}. Indeed, assume that $|\xi(z,w)| \le a\, \other{\xi}(z,w)$, 
			for some small $a \sim 1$ \nc then
			\begin{equation} \label{eq:xi_cont}
				\begin{split}				
					|w-z|\Bigl\lvert \mu_0 + \mathcal{O}\bigl( \other{\xi}(z,w)  \bigr) \Bigr\rvert &= \Bigl\lvert \mu_3 \xi(z,w)^3 + \mu_2 \xi(z,w)^2 + \mu_1 \xi(z,w)  \Bigr\rvert
					\\&\lesssim a |z-w|\,\Bigl(a^2|\mu_3|  +  \frac{|\mu_1|}{\other{\beta}(z,z)} \Bigr)  + a^2|\mu_2| \other{\xi}(z,w)^2\\
					&\lesssim (a+a^2 + a^2\, \other{\xi}(z,w) + a^3) |w-z|,
				\end{split}
			\end{equation}
			where in the first inequality we used~\eqref{eq:xi_boot}, and in the last step \nc we used that 
			\begin{equation}
				|\mu_2| \other{\xi}(z, w)^{2} \lesssim \bigl(\rho(z)  + |\sigma(z) | + \rho(z)^{-1}\Im z\bigr)\other{\xi}(z,w)^{2}       \lesssim |w-z|\bigl(1+ \other{\xi}(z,w)\bigr),
			\end{equation}
			where the last inequality follows from the definition of $\other{\xi}$ in \eqref{eq:xif_def}. \nc
			The bound \eqref{eq:xi_cont} contradicts $|\mu_0| \sim 1$ for sufficiently small $a \sim 1$,
			since $\other{\xi}(z,w)\le|z-w|^{1/3}\lesssim 1$. \nc 
			
			Finally, we prove \eqref{eq:V_perturb}.
			Note that, similarly to \eqref{eq:M_perturb}, the bound \eqref{eq:V_perturb} holds trivially if $|z - w| \gtrsim 1$, since 
			\begin{equation} \label{eq:M-M_lowerbound}
				|z - w| = \bigl\lvert \bigl\langle M(z)^{-1} - M(w)^{-1} + \mathcal{S}\bigl[M(z) - M(w)\bigr] \bigr\rangle \bigr\rvert \lesssim \norm{M(z) - M(w)},
			\end{equation}
			where we used \eqref{eq:invM_bound}. 
			Furthermore, \eqref{eq:V_perturb} follows immediately  in the regime $\rho(z) \gtrsim 1$ provided  $|z - w| \lesssim 1$ is sufficiently small.
			Therefore, it remains to prove \eqref{eq:V_perturb} with $\rho(z) \le c_1$ and $|z - w| \le c_2$ for some sufficiently small $c_1, c_2 \sim 1$.
			To this end, we employ the expression 
			\begin{equation} \label{eq:imM_rep}
				\frac{\Im M(z)}{\pi\rho(z)} = \frac{\Im M(z)}{\langle \Im M(z) \rangle} = \frac{\mathcal{B}_{z,\bar z}^{-1}\bigl[M(z)M(z)^*\bigr] }{\bigl\langle \mathcal{B}_{z,\bar z}^{-1}\bigl[M(z)M(z)^*\bigr] \bigr\rangle}.  
			\end{equation} 
			If $\rho(z)^{-1} \Im z \le c_3$, for some sufficiently small $c_3\sim 1$, after suitably shrinking $c_1, c_2\sim 1$, we can assume that $\other{\beta}(z,\overline{z}) \vee \other{\beta}(w,\overline{w}) \le c_*$, where $c_*$ is the threshold from Lemma~\ref{lemma:1stab_ohp}, implying that both $\mathcal{B}_{z,\bar z}$ and $\mathcal{B}_{w,\bar w}$ have $\varepsilon_*$-well-structured spectrum in the $\norm{\cdot}$-norm, while \eqref{eq:M_perturb} established above together with \eqref{eq:stat_M_bound}, \eqref{eq:kapd_rhoeta}, \eqref{eq:S_hs_to_op} yields $\norm{\mathcal{B}_{z,\bar z} - \mathcal{B}_{w,\bar w}} \lesssim \norm{M(z) - M(w)}$. Therefore, it follows from Lemma~\ref{lemma:perturb} that
			\begin{equation} \label{eq:ImM_expand}
				\beta_{z,\bar z} \mathcal{B}_{z,\bar z}^{-1} \bigl[M(z)M(z)^*\bigr] = \beta_{w,\bar w} \mathcal{B}_{w,\bar w}^{-1} \bigl[M(w)M(w)^*\bigr] + \mathcal{O}_{\norm{\cdot}}\bigl(\norm{M(z) - M(w)}\bigr)
			\end{equation}
			The bounds \eqref{eq:stat_M_bound}, \eqref{eq:PiQ_bounds}, \eqref{eq:1stab_ohp}, and \eqref{eq:B12_identity} at $t=T$  imply that $|\beta_{z,\bar z} \langle\mathcal{B}_{z,\bar z}^{-1} [M(z)M(z)^*]\rangle| \sim 1$, and together with \eqref{eq:ImM_expand} yield the same bound with $z$ replaced by $w$, provided $c_1, c_2, c_3\sim 1$ are sufficiently small. Hence, plugging \eqref{eq:ImM_expand} into \eqref{eq:imM_rep}, we obtain the desired \eqref{eq:V_perturb}. \nc

			On the other hand, if $\other{\beta}(z,\overline{z}) = \rho(z)^{-1}\Im z > c_3$,  then \eqref{eq:kapd_rhoeta} implies that $\kapd(z) = \dist(z, \supp\,\rho) \gtrsim 1$, and after suitably shrinking $c_2\sim1$, we also have $\kapd(w) \sim \kapd(z) \sim 1$, since $|z|+|w|\lesssim 1$ by \eqref{eq:bdd_def}.  
			Hence, $\lVert\mathcal{B}_{z,\bar{z}}^{-1}\rVert + \lVert\mathcal{B}_{w,\bar{w}}^{-1}\rVert \lesssim 1$ by \eqref{eq:1stab_ohp} of Lemma~\ref{lemma:1stab_ohp}. 
			Therefore, it follows from \eqref{eq:stat_M_bound} 
			and the definition of $\mathcal{B}_{z,\bar z}$ in \eqref{eq:stab_def}, that 
			\begin{equation} \label{eq:stab_star_pert}
				\norm{\mathcal{B}_{z,\bar z}^{-1}\bigl[M(z)M(z)^*\bigr] - \mathcal{B}_{w,\bar w}^{-1}\bigl[M(w)M(w)^*\bigr]} \lesssim  \norm{M(z) - M(w)}, 
			\end{equation}
			where we used that $\kapd(z) \sim 1$. 
			Furthermore, it follows from \eqref{eq:B12_identity} at $t=T$ and the second bound in \eqref{eq:rho_props} \nc that 
			\begin{equation} \label{eq:stab_star_tr}
				\bigl\langle \mathcal{B}_{z,\bar z}^{-1}\bigl[M(z)M(z)^*\bigr] \bigr\rangle = \frac{\Im m(z)}{\Im z} = \int_\mathbb{R} \frac{\rho(x)\mathrm{d}x}{|x-z|^2} \sim  \kapd(z)^{-2} \sim 1.
			\end{equation}
			Combining \eqref{eq:stab_star_pert}--\eqref{eq:stab_star_tr}, and suitably shrinking the threshold $c_2\sim1 $, we also find $ \langle \mathcal{B}_{w,\bar w}^{-1} [M(w)M(w)* ]  \rangle   \sim 1$. 
			Hence, for sufficiently small $c_2\sim1$,  we deduce from \eqref{eq:stab_star_pert} and \eqref{eq:stab_star_tr} that  \nc
			\begin{equation}
				\biggl\lVert \frac{\mathcal{B}_{z,\bar z}^{-1}\bigl[M(z)M(z)^*\bigr] }{\bigl\langle \mathcal{B}_{z,\bar z}^{-1}\bigl[M(z)M(z)^*\bigr] \bigr\rangle} 
				- \frac{\mathcal{B}_{w,\bar w}^{-1}\bigl[M(w)M(w)^*\bigr] }{\bigl\langle \mathcal{B}_{w,\bar w}^{-1}\bigl[M(w)M(w)^*\bigr] \bigr\rangle} \biggr\rVert \lesssim   \other{\xi}(z,w),
			\end{equation}
			where we used that $\norm{M(z)} + \norm{M(w)} \lesssim 1$ by \eqref{eq:M_bound}.  
			Therefore, the desired bound \eqref{eq:V_perturb} follows from \eqref{eq:imM_rep} and the fact that $\lVert M(z)^{-1}\rVert \sim 1$ by \eqref{eq:M_bound} and \eqref{eq:invM_bound}.				
			This concludes the proof of Lemma~\ref{lemma:xi}. 
		\end{proof}

		\subsection{Regularity of the Two-Body Quantities as Functions of the Solution. Proof of Lemmas~\ref{lemma:tripple_prop} and~\ref{lemma:re-reg}} \label{sec:beta_reg}
		In the present section, we extend $\mathcal{R}_z$ from \eqref{eq:R_def} to the set of admissible energies on the real line via a regularity argument.
		We find that a particularly convenient framework for carrying out this analysis is to study the relevant two-body quantities as functions of the solution matrix $M(z)$ instead of $z$,  since it replaces the complicated H\"older regularity with varying exponents ($1$ in the bulk, $\tfrac{1}{2}$ at the edge, $\tfrac{1}{3}$ at the cusp) in $z$  with a uniformly Lipschitz behavior in $M(z)$. \nc 
		Subsequently, we use the same regularity properties of the two-body quantities to establish Lemma~\ref{lemma:re-reg}. 
		In preparation, we introduce the following notation. 
		
		Let $\mathfrak{M} \subset \mathbb{C}^{N\times N} \times \mathbb{C}^{N\times N}$ denote the set of ordered pairs $(M_1, M_2)$ obtained from a pair of spectral parameters $z_1, z_2 \in \bddD$ via $M_j = M(z_j)$, where $M(z)$ is the solution to the MDE \eqref{eq:MDE}, that is,
		\begin{equation}
			\mathfrak{M} = \Bigl\{\bigl(M(z_1), M(z_2)\bigr) \,:\, (z_1, z_2) \in (\bddD)^2 \Bigl\}. 
		\end{equation}
		Note that, by \eqref{eq:MDE},  for all $(M_1, M_2) \in \mathfrak{M}$, the corresponding pair of spectral parameters $(z_1, z_2)$ can be recovered via 
		\begin{equation} \label{eq:zM_correspond}
			z_j = z(M_j), \qquad z(M) := \bigl\langle D - M^{-1} - \mathcal{S}[M] \bigr\rangle, \qquad j \in \{ 1, 2 \}.
		\end{equation}
		
		Let $\mathfrak{M}_+$ denote the quadrant of $\mathfrak{M}$ in which $\Im M_1, \Im M_2$ are positive-definite in the sense of quadratic forms (or, equivalently, $\Im z(M_j) > 0$), that is
		\begin{equation} \label{eq:Mdomplus}
			\mathfrak{M}_+ := \Bigl\{ (M_1, M_2) \in \mathfrak{M} \,: \, \Im M_1 >0 ,\quad \Im M_2 >0 \Bigr\}.
		\end{equation}
		
		\subsubsection{Extension of $\mathcal{R}_z$}
		To construct the extension of the orthoprojector $\mathcal{R}_z$  with rank (at most) one, \nc defined in \eqref{eq:R_def}, to the set of admissible energies on the real line, we first establish the existence of an operator $\other{\mathcal{R}}_{z_1, z_2}$ with range in the span of $\vecL(z_1, z_2)$, defined in \eqref{eq:L_def}, for all $z_1, z_2 \in \bddD\cap\mathbb{H}$, which is uniformly Lipschitz as a function of $M(z_1)$ and $M(z_2)$. 
		Rather than considering just $z_1 = z_2 = z$, which would be sufficient to extend $\mathcal{R}_z$, we also perform the construction of the regular projector field $\other{\mathcal{R}}$ for general $z_1 \neq z_2$, in order to prove the second bound \eqref{eq:re-triple} from Lemma~\ref{lemma:re-reg} (see Section~\ref{sec:re-reg} below). 
		After setting $z_1 = z_2$, \nc this operator can subsequently be extended to the closure $\overline{\bddD\cap\mathbb{H}}$ and normalized to obtain $\mathcal{R}_E$.

		We remark that a na\"ive construction of $\mathcal{R}_E$ involving the extension of an appropriately normalized $M'$ to the real line first, followed by projecting it onto the orthogonal complement of $\rho^{-1} \Im M$ fails at the energies $E$ where \mbox{$\rho(E) = 0$}. Since (up to normalizations) $M'$ is an order $\rho$ perturbation of $\Im M$, this na\"ive procedure would yield a vanishing projection despite a non-trivial anomalous fluctuation mode existing. To counteract this issue, as described in the paragraph above, we first relate $\mathcal{R}_z$ to $\vecL(z,z)$ which correctly accounts for the blow-up of $M^\perp$ and the vanishing of $\rho$. 
		\nc
		\begin{lemma} [Extension of $\mathcal{R}$] \label{lemma:R_extend}
			For all $(M_1, M_2) \in \mathfrak{M}_+$, there exists a positive semi-definite super-operator $\other{\mathcal{R}}(M_1, M_2) : \mathbb{C}^{N\times N} \to \mathbb{C}^{N\times N}$ with rank at most one, satisfying the following properties: 
			\begin{itemize}
				\item[(i)] For any $z_1, z_2 \in \bddD\cap\mathbb{H}$, there exists a positive scalar multiplier $a(z_1, z_2) \in \mathbb{R}$ satisfying $a(z_1, z_2) \sim 1$, such that 
				\begin{equation} \label{eq:Rcolin}
					\other{\mathcal{R}}\bigl(M(z_1), M(z_2)\bigr)[X] = a(z_1, z_2) \bigl\langle \vecL(z_1, z_2)^*X\bigr\rangle \vecL(z_1, z_2), \qquad X \in \mathbb{C}^{N\times N}. 
				\end{equation}
				\item[(ii)] The map $(M_1, M_2) \mapsto \other{\mathcal{R}}(M_1, M_2)$ is bounded and uniformly Lipschitz regular on $\mathfrak{M}_+$ in $\norm{\cdot}_{\mathrm{hs}\to \mathrm{hs}}$-norm.
				In particular, $\other{\mathcal{R}}(M_1, M_2)$ admits a uniformly Lipschitz extension to
				\begin{equation} 
					\mathfrak{M}_+^{\mathrm{cl}} := \Bigl\{ \bigl(M(z_1), M(z_2)\bigr) \,:\, z_1, z_2 \in \overline{\bddD\cap\mathbb{H}} \Bigr\},
				\end{equation}
				where $M(E) := \lim_{\eta\to+0}M(E+\ii\eta)$ for all $E \in \overline{\bddD}\cap \mathbb{R}$. We 
				  continue to   denote this extension by $\other{\mathcal{R}}$. 
			\end{itemize}
			
			For all $E \in \overline{\bddD}\cap \mathbb{R}$, let $\other{\mathcal{R}}_E := \other{\mathcal{R}}(M(E), M(E))$. We define the extension $\mathcal{R}_E$ as 
			\begin{equation}
				\mathcal{R}_E := \bigl\lVert \other{\mathcal{R}}_E \bigr\rVert_{\mathrm{hs} \to \mathrm{hs}}^{-1}\other{\mathcal{R}}_E  \qquad \text{if}\,\,  \other{\mathcal{R}}_E \neq 0, \quad \text{and} \quad \mathcal{R}_E := 0 \,\,\text{otherwise}. 
			\end{equation}
			Then, for all $z_1, z_2 \in \bddD\cap\mathbb{H}$,  we have 
			\begin{equation} \label{eq:L_reg}
				\bigl\lvert \bigl\langle \vecL(z_1, z_2) A \bigr\rangle \bigr\rvert \lesssim \min\limits_{E \in \{\Re z_1, \Re z_2\}} \bigl\lVert\mathcal{R}_{E} [A ]\bigr\rVert_\mathrm{hs} +  \other{\alpha}(z_1, z_2)\norm{A}_\mathrm{hs}~,
			\end{equation}
			for all matrices $A \in \mathbb{C}^{N\times N}$. 
		\end{lemma}
		We defer the proof of Lemma~\ref{lemma:R_extend} to the end of the present section. 
		
		The  next ingredient we need to complete the proof of Lemma~\ref{lemma:re-reg} is the following general regularity property for $M_1 - M_2$. 
		Let $\Mdom \subset \mathfrak{M}$ denote the set
		\begin{equation} \label{eq:M_domain}
			\Mdom := \Bigl\{ \bigl( M(z_1), M(z_2) \bigr) \,:\, z_1, z_2 \in \bddD, \quad \other{\beta}(z_1, z_2) \wedge \other{\beta}(z_2, z_1) \le \beta_*  \Bigr\},
		\end{equation} 
		where $\beta_*$ is the threshold from Lemma~\ref{lemma:stab_bound}. 	
		Consider the symmetric function $F$ of $(M_1, M_2) \in \mathfrak{M}_\beta$~, given  by  
		\begin{equation} \label{eq:F_def}
			F(M_1, M_2) := \frac{M_1 - M_2}{\langle M_1 - M_2\rangle}.
		\end{equation}
		
		\begin{lemma}[Regularity of $F$] \label{lemma:F_reg}
			The function $F$, defined in \eqref{eq:F_def} is uniformly $C^{1,1}$-regular on the set $\Mdom$, defined in~\eqref{eq:M_domain}.	
			More precisely, for all $(X,Y) \in \Mdom$, there exists an $\mathcal{O}(1)$-bounded linear functional $\nabla F(X,Y) : \mathbb{C}^{N\times N} \to \mathbb{C}^{N\times N}$, such that
			\begin{equation}
				F(X, Y') - F(X, Y) = \bigl(\nabla F(X, Y)\bigr)[Y'-Y] +  \mathcal{O}_{\norm{\cdot}}\bigl(\norm{Y' - Y}^2\bigr),
			\end{equation}
			for all $(X,Y), (X,Y') \in \Mdom$, and the map $(X,Y) \mapsto \nabla F(X,Y)$ is uniformly Lipschitz continuous on $\Mdom$ in the norm induced by the operator norm $\norm{\cdot}$ on $\mathbb{C}^{N\times N}$, with an  $\mathcal{O}(1)$ Lipschitz constant. 
		\end{lemma} 
		We defer the proof of Lemma~\ref{lemma:F_reg} to Section~\ref{sec:F_reg}.
		\nc

		\subsubsection{Re-Regularization}  \label{sec:re-reg}
		Equipped with Lemmas~\ref{lemma:R_extend} and~\ref{lemma:F_reg}, we are ready to prove Lemma~\ref{lemma:re-reg}. 
		\begin{proof}[Proof of Lemma~\ref{lemma:re-reg}]
		Without loss of generality, we assume that the observable $A$ is pre-regular with respect to $(z,z)$ in the sense of Definition~\ref{def:prereg}. 
		We begin by establishing \eqref{eq:Exp_err}. 		
		Let $M(E) := \lim_{\eta\to+0} M(E+\ii\eta)$, then 
		\begin{equation} \label{eq:reg_diff1}
			\Bigl\langle \frac{\Im M(E)}{\pi \rho(E)} \reg{A}^{z,z}\Bigr\rangle = \biggl\langle \biggl(F\bigl(M(E),M(E)^*\bigr)-F\bigl(M(z),M(z)^*\bigr) - \frac{\widehat{M}(z,I,z)}{\langle \widehat{M}(z,I,z)\rangle} \biggr) A \biggr\rangle,
		\end{equation}
		where we used the definition of $\Upsilon(z,A,z)$ from \eqref{eq:Exp_err} and the pre-regularity of $A$. 
		
		Denoting  $\Theta := M(z) - M(E)$, it follows from Lemma~\ref{lemma:F_reg} and \eqref{eq:imM_bound}, that 
		\begin{equation}
			F\bigl(M(E),M(E)^*\bigr)-F\bigl(M(z),M(z)^*\bigr) = - 2\nabla F\bigl(M(z),M(z)^*\bigr)[\Re\Theta] + \mathcal{O}\bigl(\norm{\Theta}^2 + \rho(z)\norm{\Theta}\bigr). 
		\end{equation}
		Furthermore, with $\xi(E,z)$ defined as in \eqref{eq:xi_bounds}, it follows from \eqref{eq:psi_LR_asymp} and \eqref{eq:LR_props} that 
		\begin{equation}
			\Re \Theta = \pi \Re\bigl[\xi(E,z)\bigr] F\bigl(M(z), M(z)^*\bigr) + \mathcal{O}_{\norm{\cdot}}\Bigl(\bigl(\rho^{-1}\eta + \norm{\Theta}\bigr)\bigl(\rho+\norm{\Theta}\bigr)\Bigr),
		\end{equation}
		where we used that $F(M(E), M(E)^*) = F(M(z), M(z)^*) + \mathcal{O}(\norm{\Theta})$, and we use that $|\xi(E,z)| \lesssim \norm{\Theta}$.
		Next, we compute the leading order of the matrix
		\begin{equation}
			\nabla F\bigl(M(z),M(z)^*\bigr)[F(M(z), M(z)^*)].
		\end{equation}
		To this end, we consider $F(M(z), M(z)) := \lim_{w\to z} F(M(z), M(w)) = \langle \tfrac{\dd}{\dd z} M(z)\rangle^{-1}\tfrac{\dd}{\dd z} M(z)$, and, using Lemma~\ref{lemma:F_reg} together with $\norm{\Im M(z)} \lesssim \rho(z)$ by \eqref{eq:imM_bound}, we  deduce that 
		\begin{equation}
			2 \pi \nabla F\bigl(M(z),M(z)^*\bigr)[F(M(z), M(z)^*)] = \frac{\Im \bigl[F\bigl(M(z),M(z)\bigr)\bigr] }{\rho(z)} + \mathcal{O}\bigl(\rho(z)\bigr) = Y(z) + \mathcal{O}\bigl(\rho(z)\bigr), 
		\end{equation}
		where we recall that $Y(z) \equiv Y(z,z)$ from \eqref{eq:Ydef} is given by, with $M'(z) := \tfrac{\dd}{\dd z} M(z)$,
		\begin{equation} \label{eq:Yz_def}
			Y(z) := \rho(z)^{-1}\Im\Bigl[\bigl\langle M'(z) \bigr\rangle^{-1} M'(z)\Bigr].
		\end{equation}
		Next, we express $\widehat{M}(z,I,z)$, defined in \eqref{eq:ImM2_def}, in terms of $Y(z)$, as 
		\begin{equation} \label{eq:hatMz_expr}
			\widehat{M}(z,I,z) = \frac{\Im M(z)}{2\pi\rho(z)} - \frac{1}{2}\Re M'(z) = \frac{\Im M(z)}{2\pi\rho(z)} + \frac{\rho(z) \Im\langle M'(z)\rangle}{2} Y(z) - \frac{\ii\rho(z)\Re \langle M'(z)\rangle }{2}Z(z).
		\end{equation}
		where, analogously to \eqref{eq:Zdef}, we denote 
		\begin{equation}
			Z(z) := \frac{1}{\ii\rho(z)}\Re\biggl[\frac{M'(z)}{\bigl\langle M'(z) \bigr\rangle}\biggr]. 
		\end{equation}
		Finally, with $\other{\xi}(E,z) \equiv \other{\xi}(E+\ii0,z)$ defined as in \eqref{eq:xif_def}, we observe that 
		\begin{equation} \label{eq:theta_eta}
			\norm{\Theta} \lesssim \other{\xi}(E,z) \sim \other{\beta}(z,z)^{-1}\eta.
		\end{equation}
		by \eqref{eq:xi_bounds} and \eqref{eq:betaf_def}. In particular, we record the following consequences of \eqref{eq:betaf_def} and \eqref{eq:ImGImG_comp} at time $t=T$, that we use to simplify error terms in the sequel,
		\begin{equation} \label{eq:xi_eta_simp}
			\eta \lesssim \other{\beta}(z,z)^{-1}\eta \lesssim \rho(z), \qquad \bigl\langle\widehat{M}(z,I,z)\bigr\rangle \sim \eta^{-1}\rho(z),
		\end{equation}
		for all $z \in \bddD$ satisfying $\Re z \in \supp\,\rho$ and $\other{\beta}(z,z) \le \beta_*$. 
		Here, in the second bound we used that $\kapd(z) = \eta$ since $\Re z\in\supp\,\rho$ by assumption.
		
		Therefore, combining \eqref{eq:reg_diff1}--\eqref{eq:hatMz_expr}, using the pre-regularity of $A$ with respect to $(z,z)$, and $|\langle Z A\rangle| \lesssim \rho(z)\norm{A}_\mathrm{hs}$ by~\eqref{eq:Gamma_cancel}, we obtain
		\begin{equation} \label{eq:reg_diff_expr}
			\Bigl\langle \frac{\Im M(E)}{\pi \rho(E)} \reg{A}^{z,z}\Bigr\rangle = -\frac{2 \langle \widehat{M}(z,I,z)\rangle \Re\bigl[\xi(E,z)\bigr]  + \rho(z) \Im\langle M'\rangle }{2\langle \widehat{M}(z,I,z)\rangle} \bigl\langle Y(z)A \bigr\rangle  + \mathcal{O}\biggl(\frac{\eta}{\other{\alpha}(z,z)}\biggr)\norm{A}_\mathrm{hs},
		\end{equation}
		where we used \eqref{eq:alp_beta_in_supp}, \eqref{eq:theta_eta} and \eqref{eq:xi_eta_simp}, as well as  $\lvert \langle\Re M'(z) \rangle \rvert \lesssim \other{\beta}(z,z)^{-1}$ by \eqref{eq:beta_assymp}, to simplify the error term. 
		Analogously to \eqref{eq:YA_bounds}, we have the bound
		\begin{equation} \label{eq:YzA_bounds}
			\bigl\lvert \langle Y(z) A \rangle \bigr\rvert \lesssim \bigl\lvert \bigl\langle {\vecL}(z, z) A \bigr\rangle \bigr\rvert + \rho(z) \norm{A}_\mathrm{hs},
		\end{equation}
		where ${\vecL}(z,z)$ is defined in \eqref{eq:L_def}. Plugging \eqref{eq:YzA_bounds} into \eqref{eq:reg_diff_expr}, we deduce that 
		\begin{equation} \label{eq:reg_diff_bound}
			\biggl\lvert\Bigl\langle \frac{\Im M(E)}{\pi \rho(E)} \reg{A}^{z,z}\Bigr\rangle\biggr\rvert \lesssim \frac{\eta \bigl\lvert \theta(z) \bigr\rvert}{\rho(z)} \bigl\lvert \bigl\langle {\vecL}(z,z)A \bigr\rangle \bigr\rvert  +  \frac{\eta}{\other{\alpha}(z,z)} \norm{A}_\mathrm{hs},
		\end{equation}
		where we used \eqref{eq:Mhat_comp} to estimate $\langle \widehat{M}(z,I,z) \rangle$ and simplified the error term similarly to that of \eqref{eq:reg_diff_expr}. 
		Here, the scalar coefficient $\theta(z)$ is given by
		\begin{equation} \label{eq:theta_def}
			\theta(z) := 2 \langle \widehat{M}(z,I,z)\rangle \Re\bigl[\xi(E,z)\bigr]  + \rho(z) \Im\langle M'\rangle. 
		\end{equation}
		
		We now proceed to estimate $\theta(z)$. First, by \eqref{eq:M_bound}, \eqref{eq:PiQ_bounds},  \eqref{eq:Pi_beta}, \eqref{eq:LR_props}, and \eqref{eq:m'exp}, we have
		\begin{equation} \label{eq:imM'_expr}
			\Im \langle M'\rangle = \frac{2\sigma(z)\rho(z)}{|\beta(z,z)|^2} + \mathcal{O}\biggl(\frac{\bigl(\rho(z) + |\sigma(z)|\bigr)\rho(z)^2 + \eta + \rho(z)^{-2}\eta^2}{|\beta(z,z)|^2} + 1\biggr) = \frac{2\sigma(z)\rho(z)}{|\beta(z,z)|^2} + \mathcal{O}\biggl(\frac{1}{\other{\alpha}(z,z)}\biggr),
		\end{equation}
		where, in the second step, we used \eqref{eq:beta_assymp},  \eqref{eq:betaf_def}, and \eqref{eq:alp_beta_in_supp}. Similarly, we obtain
		\begin{equation}
			\Re \langle M'\rangle = \frac{\pi\rho(z)^{-1}\eta + 2 \psi (z)\rho(z)^2}{|\beta(z,z)|^2} + \mathcal{O}\biggl(\frac{1}{\other{\alpha}(z,z)}\biggr),
		\end{equation}
		where  $\psi(z)$ satisfies \eqref{eq:psi_LR_asymp}. 
		In particular, using \eqref{eq:ImM2_def}, \eqref{eq:Pi_beta} and~\eqref{eq:LR_props} once again, we deduce that 
		\begin{equation} \label{eq:Mhat_expr}
			2 \langle \widehat{M}(z,I,z)\rangle = \frac{2\rho(z)^3}{\pi|\beta(z,z)|^2\eta}\Bigl( \psi(z) \frac{\pi\eta}{\rho(z)} +  2\rho(z)^2  \psi(z)^2  + 2\sigma(z)^2 \Bigr) + \mathcal{O}\biggl(\frac{1}{\other{\alpha}(z,z)}+\frac{\rho(z)^2}{\eta}\biggr).
		\end{equation}
		Substituting \eqref{eq:imM'_expr} and \eqref{eq:Mhat_expr} into \eqref{eq:theta_def}, we find
		\begin{equation} \label{eq:theta_rep}
			\theta(z) = \frac{2\rho(z)^2}{\pi|\beta(z,z)|^2\eta}\Bigl(2\rho(z)\bigl(\rho(z)^2 \psi(z)^2  + \sigma(z)^2\bigr)\Re \xi + \pi\eta \bigl(\sigma(z) + \psi(z)\Re\xi\bigr)\Bigr)  + \mathcal{O}\biggl( \frac{\rho(z)}{\other{\alpha}(z,z)}\biggr),
		\end{equation}
		where we abbreviate $\xi := \xi(E,z)$.

		To proceed, we first, use \eqref{eq:xi_bounds} and \eqref{eq:LR_props}, deducing that 
		\begin{equation}
			\rho(z) -  \rho(E) = \pi^{-1}\langle \Im M(z) - \Im M(E) \rangle = \Im \xi  + \mathcal{O}(|\xi|^2 + \rho(E)|\xi| + \eta).
		\end{equation}
		In particular, using the fact that $|\rho(E) - \rho(z)|\lesssim|\xi|$ by \eqref{eq:xi_bounds}, we have 
		\begin{equation} \label{eq:imxi}
			\rho(E) + \Im \xi = \rho(z)\Bigl(1 + \mathcal{O}\bigl(\other{\beta}(z,z)^{-1}\eta + \rho(z)^{-1}\eta\bigr)\Bigr). 
		\end{equation}
		Moreover, since $|\rho(z) - \rho(E)| \lesssim |\xi| \lesssim \other{\beta}(z,z)^{-1}\eta \lesssim \rho(z)$, we also have 
		\begin{equation} \label{eq:vert_rho_monot}
			\rho(E) \lesssim \rho(z).
		\end{equation}
		
		Next, we recall from \eqref{eq:xi_cubic} that $\xi \equiv \xi(E,z)$ is the solution to the  cubic equation
		\begin{equation} \label{eq:eta_cubic}
			\mu_3(E)\xi^3 + \mu_2(E)\xi^2 + \mu_1(E)\xi + \ii\mu_0\eta = \mathcal{O}\bigl(|\xi|^4 + |\xi|\eta + \eta^2 \bigr),
		\end{equation}
		with the coefficients satisfying \eqref{eq:mu_coeff}. Recall that $\psi(E) \lesssim 1$ by \eqref{eq:psi_LR_asymp}.  Hence, taking the imaginary part of \eqref{eq:eta_cubic} and using \eqref{eq:imxi}, we obtain
		\begin{equation} \label{eq:eta1}
			\pi\eta  = - 2\sigma(E) \rho(z)   \Re \xi 
			+ \mathcal{O}\bigl( \bigl(\rho(z)^{-1}\eta + \other{\alpha}(z,z)^{-1}\rho(z)\bigr)\eta\bigr), 
		\end{equation} 
		where we used the fact that  $\rho(E) \lesssim \rho(z)$ by \eqref{eq:vert_rho_monot},   $|\sigma(z)| \lesssim 1$ by \eqref{eq:sigma_def} and \eqref{eq:F_flat},   $|\xi| \lesssim \other{\beta}(z,z)^{-1}\eta$ by \eqref{eq:xi_bounds}, and \eqref{eq:betaf_def} to simplify the error term. 
		
		Finally, we assert that  $\sigma(E)$ satisfies the estimate 
		\begin{equation} \label{eq:psisigma_reg}
			\sigma(E) = \sigma(z) + \mathcal{O}\bigl(\rho(z)\bigr).
		\end{equation} 
		Indeed, since $\langle \Im M \rangle^{-1} \Im M \sim 1$ by \eqref{eq:imM_bound}, $\norm{M^{-1}} \lesssim 1$ by \eqref{eq:invM_bound}, and the map $M \mapsto \langle \Im M \rangle^{-1} \Im M$	is uniformly Lipschitz on $\{M(z) \,: \, z \in \bddD\cap\mathbb{H} \}$, so is the map 
		\begin{equation}
			M \mapsto \langle \Im M \rangle^{-1}\Im\mathscr{U}(M) :=   \bigl\lvert(\langle \Im M \rangle^{-1}\Im M)^{-1/2} M (\langle \Im M \rangle^{-1}\Im M)^{-1/2}  \bigr\rvert^{-1},
		\end{equation} 
		where $\mathscr{U}(M(z)) := \mathscr{U}(z)$ we recall the definition of $\mathscr{U}(z)$ from \eqref{eq:sigma_def}. 
		Since $\sign \Re \mathscr{U}(z)$ is constant on each connected component of the set $\{z \in \bddD\cap\mathbb{H}\,:\, \rho(z) \le \rho_* \}$ for some sufficiently small $\rho_* \sim 1$, we conclude the desired estimate on $\sigma(E)$ in  \eqref{eq:psisigma_reg} from \eqref{eq:sigma_def}.

		Therefore, substituting \eqref{eq:psisigma_reg} into \eqref{eq:eta1}, we deduce that
		\begin{equation} \label{eq:eta_rep}
			\pi\eta  = - 2 \sigma(z)\rho(z)  \Re \xi
			+ \mathcal{O}\bigl( \bigl(\rho(z)^{-1}\eta + \other{\alpha}(z,z)^{-1}\rho(z)\bigr)\eta\bigr), 
		\end{equation} 
		In turn, plugging \eqref{eq:eta_rep} into \eqref{eq:theta_rep}, we obtain
		\begin{equation}  \label{eq:theta_bound}
			\bigl\lvert \theta(z) \bigr\rvert \lesssim \other{\alpha}(z,z)^{-2}\rho(z). 
		\end{equation}
		Hence, combining \eqref{eq:tripplenorm},  \eqref{eq:reg_diff_bound} and \eqref{eq:theta_bound}, we conclude the desired bound \eqref{eq:Exp_err}. 
		
		Next, we prove \eqref{eq:re-triple}. 
		We assume without loss of generality that $A\in\mathbb{C}^{N\times N}$
		is $(z_1, z_2)$-regular  in the sense of Definition~\ref{def:reg}, that is, $A=\reg{A}^{z_1,z_2}$.
		Note that to conclude that the bound \eqref{eq:re-triple} holds with the minimum over $E \in \{E_1, E_2\}$ on the right-hand side, it suffices to check the bound for both $E = E_1$ and $E = E_2$. We only treat the case $E = E_1$ in full detail, since $E=E_2$ is completely analogous. 
		To condense the presentation, we abbreviate $A_0 := \reg{A}^{E,E}$,  $\vecL_\eta := \vecL(z_1,z_2)$, 
		 and define $A_\eta$ to be the $(z_1, z_2)$-pre-regular version of $A$, given by 
		 (recall \eqref{eq:prereg_def}) 
		\begin{equation}
			A_\eta := A - \bigl\langle A F\bigl(M(z_1), M(z_2)\bigr)\bigr\rangle \,I,
		\end{equation} 
		where $F$ is defined in \eqref{eq:F_def}. 
		\nc Then, by \eqref{eq:tripplenorm}, we have 
		\begin{equation} \label{eq:trip_bound1}
			\begin{split}
				\vertiii{A}_{z_1,z_2} &= \frac{ \bigl\lvert \bigl\langle \vecL_\eta A  \bigr\rangle \bigr\rvert}{\other{\alpha}(z_1,z_2)} + \norm{A}_{\mathrm{hs}} 
				\le \frac{ \bigl\lvert \bigl\langle \vecL_\eta A_0 \bigr\rangle \bigr\rvert}{\other{\alpha}(z_1,z_2)} + \norm{A_0}_{\mathrm{hs}} 
				+\frac{ \bigl\lvert \bigl\langle \vecL_\eta (A-A_0)  \bigr\rangle \bigr\rvert}{\other{\alpha}(z_1,z_2)} + \norm{A - A_0}_{\mathrm{hs}} \\
				&\lesssim \frac{\bigl\lVert\mathcal{R}_{E} [A_0]\bigr\rVert_\mathrm{hs}}{\other{\alpha}(E_1, E_2) + N^{-1/4}} +  \norm{A_0}_\mathrm{hs} +\frac{\norm{A_\eta - A_0}_{\mathrm{hs}} + \norm{A - A_\eta }_{\mathrm{hs}}}{\other{\alpha}(z_1,z_2)},
			\end{split}
		\end{equation}
		where in the second line we used the upper bound from \eqref{eq:alpha_sym}, \eqref{eq:alpha_realline}, \eqref{eq:L_bound}, and \eqref{eq:L_reg}.

		Observe that for any matrices $P_1$, $P_2 \in \mathbb{C}^{N\times N}$, satisfying $\langle P_j\rangle = 1$ and $\norm{P_j}_\mathrm{hs} \lesssim 1$ the super-operators $$\mathcal{P}_j[\,\cdot\,] := \mathrm{Id}[\,\cdot\,] - \langle P_j (\,\cdot\,) \rangle\,I, \qquad  j\in\{1,2\}$$ satisfy $\norm{\mathcal{P}_j}_{\mathrm{hs}\to\mathrm{hs}} \lesssim 1$ and $\mathcal{P}_2 \mathcal{P}_1 = \mathcal{P}_2$. In particular, taking $P_1, P_2$ from among the matrices $F(M(z_1), M(z_2))$, $\frac{\widehat{M}(z_1, I, z_2)}{\langle \widehat{M}(z_1, I, z_2)\rangle}$, $\frac{\Im M(E)}{\pi \rho(E)}$, we deduce that 
		\begin{equation} \label{eq:regreg_rel}
			\reg{(A_\eta)}^{z_1, z_2} =   \reg{A}^{z_1, z_2}   = A ,   \qquad 
			A_\eta = A_0 - \bigl\langle A_0 F\bigl(M(z_1), M(z_2)\bigr)\bigr\rangle \,I, \qquad 
			\norm{A_\eta}_\mathrm{hs} \lesssim \norm{A_0}_\mathrm{hs},
		\end{equation}
		where we used  \eqref{eq:Mdiff_norm}, \eqref{eq:Mhat_flat}, \eqref{eq:imM_bound}, as well as the definitions \eqref{eq:Acirc_def} and \eqref{eq:regA}.  
		\nc

		It follows from the second relation in \eqref{eq:regreg_rel} and the definition of $A_0 = \reg{A}^{E,E}$ that 
		\begin{equation} 
			\begin{split}
				\norm{A_\eta - A_0}_{\mathrm{hs}} &= \bigl\lvert \bigl\langle \bigl(F\bigl(M(z_1), M(z_2)\bigr) - F\bigl(M(E), M(E)\bigr)\bigr) A_0 \bigr\rangle  \bigr\rvert
				\\&\lesssim \bigl(\norm{M(z_1) - M(E)} + \norm{M(z_2) - M(z_1)}\bigr)\norm{A_0}_\mathrm{hs} \lesssim \other{\alpha}(z_1, z_2)\norm{A_0}_\mathrm{hs}.
			\end{split}
		\end{equation}
		where $M(E) := \lim_{\eta \to +0} M(E+\ii\eta)$, $F$ is defined in \eqref{eq:F_def}, and in the second step we used the Lipschitz-continuity of~$F$ from Lemma~\ref{lemma:F_reg}, while in the third step we used \eqref{eq:M_perturb}, \eqref{eq:xif_def} and \eqref{eq:betaf_def} to estimate 
		\begin{equation} \label{eq:R_err_bound2}
			\norm{M(z_1) - M(E)} + \norm{M(z_2) - M(z_1)} \lesssim \other{\beta}(z_1, z_1)^{-1} \Im z_1 + \other{\xi}(z_1, z_2)  
			\lesssim |z_1 - z_2|^{1/3} + \rho(z_1) \lesssim \other{\alpha}(z_1, z_2). 
		\end{equation}
		To estimate $\norm{A_\eta - A}$, we use the first relation in \eqref{eq:regreg_rel}, obtaining	
		\begin{equation} \label{eq:trip_bound3}
			\norm{A_\eta - A} = \bigl\lvert\Upsilon(z_1, A_\eta, z_2)\bigr\rvert \lesssim \other{\alpha}(z_1, z_2) \norm{A_\eta}_\mathrm{hs} \lesssim \other{\alpha}(z_1, z_2) \norm{A_0}_\mathrm{hs}~,
		\end{equation}
		where we used \eqref{eq:Ups_bound} at $t=T$ and \eqref{eq:alpha_defin}, as well as the third relation in \eqref{eq:regreg_rel}. Combining \eqref{eq:trip_bound1}--\eqref{eq:trip_bound3}, we deduce the desired \eqref{eq:re-triple}.
		This concludes the proof of Lemma~\ref{lemma:re-reg}. 
		
		\nc 	
		\end{proof}
		
		\subsubsection{Regularity of $F(M_1, M_2)$} \label{sec:F_reg}
		
		For all $(M_1, M_2) \in \mathfrak{M}$, let $\mathcal{B}(M_1, M_2) : \mathbb{C}^{N\times N} \to \mathbb{C}^{N\times N}$ denote the corresponding stability operator (c.f. \eqref{eq:stab_def}), 
		\begin{equation} \label{eq:stabM_def}
			\mathcal{B}(M_1, M_2)[\,\cdot\,] := \mathrm{Id}[\,\cdot\,] - M_1 \mathcal{S}[\,\cdot\,] M_2. 
		\end{equation}		
		Note that $\mathcal{B}(M_1, M_2) = \mathcal{B}_{z_1, z_2}$ under the correspondence \eqref{eq:zM_correspond}. 
		
		\begin{proof}[Proof of Lemma~\ref{lemma:F_reg}]
			Note that for all $(M_1, M_2) \in \Mdom$, the identity \eqref{eq:B12_identity} implies that $F(M_1, M_2)$ admits the expression
			\begin{equation} \label{eq:F_expr}
				F(M_1, M_2) = \frac{\bigl(\mathcal{B}(M_1, M_2)\bigr)^{-1}[M_1M_2]}{\bigl\langle \bigl(\mathcal{B}(M_1, M_2)\bigr)^{-1}[M_1M_2] \bigr\rangle}, 
			\end{equation}
			where we recall the definition of $\mathcal{B}(M_1, M_2)$ from \eqref{eq:stabM_def}. 
			Hence, by definition of $\Mdom$ in \eqref{eq:M_domain} and Lemma~\ref{lemma:stab_bound},  $\mathcal{B}(M_1, M_2)$ satisfies the assumptions of Lemma~\ref{lemma:perturb}. In particular, 
			\begin{equation} \label{eq:B_decomp}
				\bigl(\mathcal{B}(M_1, M_2)\bigr)^{-1} = \beta(M_1, M_2)^{-1} \Pi(M_1, M_2) + \bigl(\mathcal{B}(M_1, M_2)\bigr)^{-1}\mathcal{Q}(M_1, M_2),
			\end{equation}
			where $\beta(M_1, M_2) := \beta_{\mathcal{B}(M_1, M_2)}$, $\Pi(M_1, M_2) := \Pi_{\mathcal{B}(M_1, M_2)}$ are defined in \eqref{eq:Pi_def} and \eqref{eq:beta_perturb}, respectively, while $\mathcal{Q}(M_1, M_2) := \mathrm{Id} - \Pi(M_1, M_2)$. Substituting \eqref{eq:B_decomp} into \eqref{eq:F_expr}, we deduce that 
			\begin{equation} \label{eq:F_decomp}
				F(M_1, M_2) = \frac{(\Pi + \beta\mathcal{B}^{-1}\mathcal{Q})[M_1M_2]}{\bigl\langle (\Pi + \beta\mathcal{B}^{-1}\mathcal{Q})[M_1M_2] \bigr\rangle},
			\end{equation}
			where $\beta$, $\Pi$, $\mathcal{Q}$, and $\mathcal{B}$ are evaluated at $(M_1, M_2) \in \Mdom$. Moreover, observe that by \eqref{eq:beta_assymp}, the denominator in \eqref{eq:F_decomp} satisfies
			\begin{equation} \label{eq:F_denom_bound}
				\bigl\lvert \bigl\langle (\Pi + \beta\mathcal{B}^{-1}\mathcal{Q})[M_1M_2] \bigr\rangle \bigr\rvert = \bigl\lvert \beta \bigl\langle \mathcal{B}^{-1}[M_1M_2] \bigr\rangle\bigr\rvert = \biggl\lvert \beta\frac{\langle M_1 - M_2\rangle}{z_1 - z_2}\biggr\rvert \sim 1.
			\end{equation}
			Therefore, to prove $C^{1,1}$-regularity of $F$, we first establish it for the maps $\mathcal{B}$, and then $\beta, \Pi, \mathcal{B}^{-1}\mathcal{Q}$ of $(M_1, M_2)\in \Mdom$. 
			
			First, we note that the map $\mathcal{B}(M_1, M_2)$ is affine in each of its arguments, and satisfies 
			\begin{equation}
				\mathcal{B}(M_1, M_2) - \mathcal{B}(M_1, M_3) =  \bigl(\nabla_2 \mathcal{B}(M_1, M_2)\bigr)[M_2-M_3], \qquad \Bigl(\bigl(\nabla_2 \mathcal{B}(M_1, M_2)\bigr)[X]\Bigr)[\,\cdot\,] := - M_1 \mathcal{S}[\,\cdot\,] X.
			\end{equation}
			Similarly, in its first argument, $\mathcal{B}$ satisfies 
			\begin{equation}
				\mathcal{B}(M_1, M_2) - \mathcal{B}(M_3, M_2) =  \bigl(\nabla_1 \mathcal{B}(M_1, M_2)\bigr)[M_1-M_3], \qquad \Bigl(\bigl(\nabla_1 \mathcal{B}(M_1, M_2)\bigr)[X]\Bigr)[\,\cdot\,] := - X \mathcal{S}[\,\cdot\,] M_2.
			\end{equation}
			By \eqref{eq:M_bound} and \eqref{eq:S_hs_to_op}, for all $(M_1, M_2) \in \Mdom$ and $j\in\{1,2\}$,  we have 
			\begin{equation}
				\bigl\lVert\nabla_j \mathcal{B}(M_1, M_2) [X]\bigr\rVert_{\mathrm{hs}\to\norm{\cdot}} \lesssim \norm{X}, \qquad X\in\mathbb{C}^{N\times N}
			\end{equation}
			And since $\nabla_j \mathcal{B}(M_1, M_2)$ depend linearly on $M_1$ and $M_2$, using a similar argument we conclude that they are uniformly Lipschitz on $\Mdom$.  Hence, uniform $C^{1,1}$-regularity of $(M_1, M_2)\mapsto\mathcal{B}(M_1, M_2)$ is established. 
			In particular,
			\begin{equation} \label{eq:trivial_B_boubd}
				\norm{\mathcal{B}(M_1,M_2) - \mathcal{B}(M_1, M_3)}_{\mathrm{hs}\to\norm{\cdot}} \lesssim \norm{M_2-M_3}, \qquad (M_1,M_2), (M_1, M_3)\in\mathcal{M},
			\end{equation}
			and an analogous bound holds for $\mathcal{B}(M_2,M_1) - \mathcal{B}(M_3, M_1)$.
			
			Next, it follows by \eqref{eq:Pi_perturb} of Lemma~\ref{lemma:perturb}, together with \eqref{eq:PiQ_bounds} and \eqref{eq:trivial_B_boubd}, that $\mathcal{B} \mapsto \Pi_\mathcal{B}$ is a uniformly $C^{1,1}$ function on the domain $\{\mathcal{B}(M_1, M_2) \,:\, (M_1,M_2)\in \Mdom\}$. Hence $(M_1, M_2) \mapsto \Pi(M_1, M_2)$ is a composition of two bounded uniformly $C^{1,1}$ maps on compatible domains, and so is itself $C^{1,1}$ on $\Mdom$. Analogously, we conclude uniform $C^{1,1}$-regularity of $\beta$, $\mathcal{Q}$, and $\mathcal{B}^{-1}\mathcal{Q}$ on $\Mdom$. 
			
			Finally, the map $(M_1, M_2) \mapsto M_1M_2$ is trivially uniformly $C^{1,1}$ in both of its arguments by \eqref{eq:M_bound}. Therefore, combining the uniform regularity and boundedness of each all the constituents on the right-hand side of \eqref{eq:F_decomp} with the bound \eqref{eq:F_denom_bound} on the denominator, we deduce the desired uniform $C^{1,1}$-regularity of $F$ on $\Mdom$. This concludes the proof of Lemma~\ref{lemma:F_reg}. 
		\end{proof}

	 	\subsubsection{Construction of the operator $\other{\mathcal{R}}$}
	 	We begin by preparing some notation. 
	 	Recall from Lemma~\ref{lemma:stab_bound} that for $(z_1, z_2)$ satisfying the condition in \eqref{eq:M_domain}, the stability operator $\mathcal{B}_{z_1, z_2}$, defined in \eqref{eq:stab_def}, has a well-structured spectrum in the sense of Definition~\ref{def:pert_cond}, and hence possesses a simple isolated small eigenvalue $\beta_{z_1, z_2}$ with the corresponding rank-one eigenprojector $\Pi_{z_1, z_2}$. Hence, for all $(M_1, M_2) \in \Mdom$, the same assertions hold for $\mathcal{B}(M_1, M_2)$. 
	 	In particular, recalling \eqref{eq:Pi_def}, we denote the corresponding eigenvalue and eigenprojector by $\beta(M_1, M_2)$ and $\Pi(M_1, M_2)$, respectively, 
	 	\begin{equation}
	 		\beta(M_1, M_2) := \Tr\bigl[\mathcal{B}(M_1, M_2)\Pi(M_1, M_2)\bigr], \qquad \Pi(M_1, M_2) := \Pi_{\mathcal{B}(M_1, M_2)}
	 	\end{equation}
		By analogy with \eqref{eq:V2_def}, for all $(M_1, M_2) \in \mathfrak{M}_+$, we denote 
		\begin{equation}
			U(M_1, M_2) := \frac{M_1 - M_2}{z(M_1) - z(M_2)}, \qquad V(M_1, M_2) := \frac{M_1 - M_2^*}{z(M_1) - z(M_2^*)}. 
		\end{equation}	
		Recall that the $\mathrm{Span}\{\vecL\} = \mathrm{Span}\{U^{\perp V}\}$, hence a correctly normalized version of $U^{\perp V}(M_1, M_2)$, defined analogously to \eqref{eq:UperpV}, is the natural candidate for the matrix generating the operator $\other{\mathcal{R}}$ from Lemma~\ref{lemma:R_extend}. 
		
		Since the regularity properties of $U(z_1, z_2)$ and $V(z_1, z_2)$ depend on $\other{\beta}(z_1, z_2)$ and  $\other{\beta}(z_1, \overline{z}_2)$, we need to consider separate cases. Define the covering $\{\mathfrak{M}_1, \mathfrak{M}_2, \mathfrak{M}_3\}$ of $\mathfrak{M}_+$ by  	
		\begin{equation} \label{eq:M123_domain}
			\begin{split}
				\mathfrak{M}_1 &:= \bigl\{ (M_1, M_2) \in \mathfrak{M}_+ \,:\,  \other{\beta}(z_1, 	z_2) \vee \other{\beta}(z_2, z_1) \le \beta_* \bigr\}, \\ 
				\mathfrak{M}_2 &:= \Bigl\{ (M_1, M_2) \in \mathfrak{M}_+ \,:\,        \other{\beta}(z_1, \overline{z}_2) \vee \other{\beta}(z_2, \overline{z}_1)  \le c_* < \tfrac{1}{2}\beta_* \le   \other{\beta}(z_1, z_2) \vee \other{\beta}(z_2, z_1) \Bigr\},\\
				\mathfrak{M}_3 &:= \bigl\{(M_1, M_2) \in \mathfrak{M}_+ \,:\,  \tfrac{1}{2}c_* \le \other{\beta}(z_1, \overline{z}_2) \vee \other{\beta}(z_2, \overline{z}_1)  \bigr\}.
			\end{split}
		\end{equation}
		where $c_* \sim 1$ is the threshold from Claim~\ref{claim:stab_diff_rho}, satisfying $c_* < \tfrac{1}{2}\beta_*$, and we identify $z_j \equiv z(M_j)$ as in \eqref{eq:zM_correspond}.
		 The set $\mathfrak{M}_1$ consists of pairs $(M_1, M_2)$ which come from $(z_1, z_2)$ close to the singularities of $\rho$, and hence requires the most delicate analysis. \nc

		\begin{proof}[Proof of Lemma~\ref{lemma:R_extend}]
			We begin by constructing a bounded,  uniformly  Lipschitz-continuous map$(M_1, M_2) \mapsto \tempR_j(M_1, M_1)$ on each of the subsets $\mathfrak{M}_j$ in the $\norm{\cdot} \to \norm{\cdot}_\mathrm{hs}$-norm. 
			For all $j \in \{1,2,3\}$, consider the functions 
			\begin{equation} \label{eq:Rj_def}
				\tempR_j(M_1, M_2) := \other{U}_j(M_1, M_2) - \frac{\bigl\langle \other{V}_j(M_1, M_2)^* \other{U}_j(M_1, M_2)\bigr\rangle \other{V}_j(M_1, M_2)}{\bigl\lVert \other{V}_j(M_1, M_2) \bigr\rVert_\mathrm{hs}^2}, \qquad (M_1, M_2) \in \mathfrak{M}_j~,
			\end{equation}
			where, for all $(M_1, M_2) \in \mathfrak{M}_j$, the matrices $\other{U}_j(M_1,M_2)$ and $\other{V}_j(M_1,M_2)$ are given by
			\begin{equation} \label{eq:UVj_def}
				\begin{aligned}
					\other{U}_1(M_1, M_2) &:= \frac{\beta(M_1, M_2) U(M_1, M_2) - \other{V}_1(M_1, M_2)}{\langle \Im M_2 \rangle}, \qquad &&\other{V}_1(M_1, M_2) := \beta(M_1, M_2^*) V(M_1, M_2),\\
					\other{U}_2(M_1, M_2) &:= U(M_1, M_2),  &&\other{V}_2(M_1, M_2) := \beta(M_1, M_2^*) V(M_1, M_2),\\
					\other{U}_3(M_1, M_2) &:= \frac{U(M_1, M_2) - V(M_1, M_2)}{\langle \Im M_2 \rangle},  &&\other{V}_3(M_1, M_2) := V(M_1, M_2). 		
				\end{aligned}
			\end{equation}
			 Here, we choose the definitions of $\other{U}_j$ and $\other{V}_j$ in such a way that they are uniformly Lipschitz and the resulting $\tempR_j(M_1, M_2)$ is an order one constant multiple of $\vecL(z_1, z_2)$.  Note that on the set $\mathfrak{M}_2$, $\langle \Im M_2\rangle \sim 1$, and hence we do not need to 
			 divide by $\langle \Im M_2\rangle$ in the definition of $\other{U}_2$ to 
			 account for the $\rho(w)$ factor in \eqref{eq:L_def}; similarly, $\other{\beta}(z,\overline{w}) \sim 1$ on $\mathfrak{M}_3$, so we do not need   to carry this factor in    $\other{U}_3$ and $\other{V}_3$. To allow the correct $1/\rho(w)$ normalization on $\mathfrak{M}_1$ and $\mathfrak{M}_3$, it is necessary to subtract an appropriately scaled $V$ from $U$ in the definition of $\other{U}$. \nc 
			
			Identifying $(z_1, z_2)$ with $(M_1, M_2) \in \mathfrak{M}$ via \eqref{eq:zM_correspond}, the matrices $\tempR_j(M_1, M_2)$ trivially satisfy
			\begin{equation} \label{eq:R_colin_proof}
				\tempR_j(M_1, M_2) = c_j(M_1, M_2) \vecL(z_1, z_2), \qquad (M_1, M_2) \in \mathfrak{M}_j,
			\end{equation}
			for scalars $c_j(M_1, M_2) \in \mathbb{C}$, given by 
			\begin{equation}
				c_1(M_1, M_2) := \frac{\beta(M_1, M_2)}{\pi \other{\beta}(z_1, z_2)}, \qquad c_2(M_1, M_2) := \frac{\rho(z_2)}{\other{\beta}(z_1, z_2)}, \qquad c_3(M_1, M_2) := \frac{1}{\pi\other{\beta}(z_1, z_2)}. 
			\end{equation}
			By \eqref{eq:rho_large} on $\mathfrak{M}_2$, the definition of $\mathfrak{M}_j$ in \eqref{eq:M123_domain}, and \eqref{eq:UVj_def}, we have $|c_j(M_1, M_2)| \sim 1$ for all $(M_1, M_2) \in \mathfrak{M}_j$ and $j \in \{1,2,3\}$.
			
			We first establish boundedness and uniform Lipschitz regularity of each $\tempR_j$ on the respective $\mathfrak{M}_j$ separately, and   then glue   them together using a Lipschitz partition of unity on the cover $\{\mathfrak{M}_j\}_{j=1}^3$ to define $\other{\mathcal{R}}$ and prove (i)--(ii) of Lemma~\ref{lemma:R_extend}. 
			
			\medskip
			\textbf{Case 1}. Consider $j=1$. To prove boundedness and Lipschitz-regularity of $\tempR_1$, it suffices to establish it for $\other{V}$ and $\other{U}$, since $\lVert \other{V} \rVert \sim \lVert \other{U} \rVert \sim 1$ by \eqref{eq:beta_assymp} and \eqref{eq:Mdiff_norm}. It follows from \eqref{eq:B12_identity} that 
			\begin{equation} \label{eq:otherV_rep}
				\other{V}_1(M_1, M_2) = \beta_{1\bar2} \mathcal{B}_{1\bar2}^{-1} [M_1M_2^*]
				= \bigl(\Pi_{1\bar2} + \beta_{1\bar2} \mathcal{B}_{1\bar2}^{-1}\mathcal{Q}_{1\bar2}\bigr)[M_1M_2^*],
			\end{equation}
			where we abbreviate $\mathcal{B}_{1\bar2} := \mathcal{B}(M_1, M_2^*)$, $\Pi_{1\bar2} := \Pi(M_1, M_2^*)$, $\beta_{1\bar2}:= \beta(M_1, M_2^*)$, and recall that $\mathcal{Q} := \mathrm{Id} - \Pi$. Hence, analogously to the proof of Lemma~\ref{lemma:F_reg} above, we conclude that $\other{V}_1$ is uniformly Lipschitz on $\mathfrak{M}_1$.
			Next, we observe that the matrix $\other{U}_1$ admits the expression
			\begin{equation}
				\begin{split}
					\other{U}_1(M_1, M_2) =&~ \frac{ \Pi_{12} - \Pi_{1\bar2}}{\langle \Im M_2\rangle} [M_1M_2]
					+\frac{ \beta_{12} - \beta_{1\bar2} }{\langle \Im M_2\rangle} \mathcal{B}_{12}^{-1}\mathcal{Q}_{12}[M_1M_2]
					+\beta_{1\bar2}  \frac{ \mathcal{B}_{12}^{-1}\mathcal{Q}_{12} - \mathcal{B}_{1\bar2}^{-1}\mathcal{Q}_{1\bar2}}{\langle \Im M_2\rangle} [M_1M_2]
					\\&+2\ii \bigl(\Pi_{1\bar2} + \beta_{1\bar2} \mathcal{B}_{1\bar2}^{-1}\mathcal{Q}_{1\bar2}\bigr)\Bigl[M_1\frac{\Im M_2}{\langle \Im M_2\rangle}\Bigr], 
				\end{split}
			\end{equation}
			where we abbreviate $\mathcal{B}_{12} := \mathcal{B}(M_1, M_2)$ and use analogous convention for  $\beta_{12}$, $\Pi_{12}$ for its smallest eigenvalue and the corresponding eigenprojector. 
			As we showed in the proof of Lemma~\ref{lemma:F_reg}, the maps 
			$$
			(M_1, M_2) \mapsto f(M_1,M_2), \qquad  
			f \in \bigl\{\mathcal{B},\,\Pi,\, \beta,\, \mathcal{B}^{-1}\mathcal{Q} \bigr\}
			$$
			are uniformly Lipschitz on $\mathfrak{M}_\beta$. Note that if $(M_1, M_2) \in \mathfrak{M}_1$, then $(M_1, M_2), (M_1, M_2^*) \in \mathfrak{M}_\beta$. 
			Moreover, the product maps
			$$
			(M_1,M_2) \mapsto M_1M_2, \qquad 
			(M_1, M_2) \mapsto M_1\frac{\Im M_2}{\langle \Im M_2 \rangle},
			$$
			are also uniformly Lipschitz on $\mathfrak{M}_1$, where the regularity of $\langle \Im M_2 \rangle^{-1}\Im M_2$ follows from \eqref{eq:V_perturb}. 
			Therefore, to conclude the regularity of $\other{U}_1$, it remains to establish Lipschitz continuity (on $\mathfrak{M}_1$) of the maps 
			\begin{equation} \label{eq:diff_maps}
					(M_1, M_2)   \mapsto \frac{ f(M_1, M_2) - f(M_1, M_2^*)}{\langle \Im M_2\rangle}, \qquad 
					f \in \bigl\{\Pi,\, \beta,\, \mathcal{B}^{-1}\mathcal{Q} \bigr\}.  
			\end{equation} 
			It follows from the integral representation \eqref{eq:PibetaQ_expr} and the bound \eqref{eq:pert_cond}, that the maps in \eqref{eq:diff_maps} inherit their Lipschitz regularity from the map
			\begin{equation} \label{eq:Gdiff_reg}
				(M_1, M_2) \mapsto \frac{\mathcal{G}_{M_1, M_2}(\zeta) - \mathcal{G}_{M_1, M_2^*}(\zeta)}{\langle \Im M_2\rangle} = - 2\ii  \mathcal{G}_{M_1, M_2}(\zeta) \circ \Bigl(M_1 \mathcal{S}[\,\cdot\,] \frac{\Im M_2}{\langle\Im M_2\rangle}\Bigr) \circ \mathcal{G}_{M_1, M_2^*}(\zeta),
			\end{equation}
			where $\mathcal{G}_{M_1, M_2}(\zeta)$ denotes the resolvent of the super-operator $\mathcal{B}(M_1, M_2)$, that is, 
			\begin{equation}
				\mathcal{G}_{M_1, M_2}(\zeta) := \bigl(\zeta \mathrm{Id} - \mathcal{B}(M_1, M_2)\bigr)^{-1}, 
			\end{equation}
			with the spectral parameter $\zeta$ satisfying $|\zeta| \ge \varepsilon_*$ and $|1 -\zeta| \ge 1 - 2\varepsilon_*$. Therefore, using \eqref{eq:V_perturb}, we deduce that the map \eqref{eq:Gdiff_reg} is indeed bounded and uniformly Lipschitz on $\mathfrak{M}_1$. Hence, we conclude the same about the maps in~\eqref{eq:diff_maps}, about~$\other{U}_1$, and, consequently, about~$\tempR_1$. 
			
			\medskip
			\textbf{Case 2}. Consider $j=2$. Recall from \eqref{eq:stab_beta_bound} that $\lVert \mathcal{B}(M_1, M_2)^{-1} \rVert_{\mathrm{hs}\to \norm{\cdot}} \lesssim 1$ for all $(M_1, M_2) \in \mathfrak{M}_2$; Hence, the boundedness and uniformly Lipschitz-continuity of the map $\other{U}_2$ on $\mathfrak{M}_2$ follows by subtracting two copies of \eqref{eq:B12_identity}. Analogously to \eqref{eq:otherV_rep} in the Case 1 above, the map $\other{V}_2$ is also uniformly Lipschitz on $\mathfrak{M}_2$, and we conclude the desired regularity for $\tempR_2$.
			
			\medskip
			\textbf{Case 3}. Consider $j=3$. Similarly to the discussion in the Case 2 above, $\other{V}_3$ is uniformly Lipschitz on  $\mathfrak{M}_3$ since the corresponding stability operator satisfies $\lVert \mathcal{B}(M_1, M_2^*)^{-1} \rVert_{\mathrm{hs}\to \norm{\cdot}} \lesssim 1$ for all $(M_1, M_2) \in \mathfrak{M}_3$. On the other hand, the matrix $\other{U}_3$ admits the expression
			\begin{equation}
				\other{U}_3(M_1, M_2) = 2\ii \mathcal{B}(M_1,M_2)^{-1}     \Bigl[ M_1 \mathcal{S} \bigl[\mathcal{B}(M_1,M_2^*)^{-1} [M_1M_2]\bigr]  \frac{\Im M_2 } {\langle \Im M_2 \rangle}\Bigr]  
				+ 2\ii \mathcal{B}(M_1,M_2^*)^{-1}\Bigl[M_1\frac{\Im M_2}{\langle \Im M_2 \rangle}\Bigr].
			\end{equation}
			Therefore, we deduce that $\other{U}_3$ is bounded and uniformly Lipschitz on $\mathfrak{M}_3$ from \eqref{eq:V_perturb} and the boundedness of the stability operators $\mathcal{B}(M_1, M_2^*)^{-1}$ and $\mathcal{B}(M_1, M_2)^{-1}$ by \eqref{eq:stab_beta_bound}. 
			Note that $\lVert \other{V}_3 \rVert \gtrsim 1$ by \eqref{eq:M-M_lowerbound}, and hence uniform Lipschitz continuity of $\tempR_3$ on $\mathfrak{M}_3$ follows.

			Therefore, $\tempR_j$ is bounded and uniformly Lipschitz on $\mathfrak{M}_j$ for all $j\in\{1,2,3\}$. 
			Let 
			\begin{equation}
				\other{\mathcal{R}}_j(M_1, M_2)[\,\cdot\,] := \bigl\langle \tempR_j(M_1, M_2)^* \, (\,\cdot\, )\bigr\rangle \tempR_j(M_1, M_2), \qquad j \in \{1,2,3\}
			\end{equation}			
			Moreover, it is straightforward to check by \eqref{eq:betaf_def} and \eqref{eq:1/3Holder_rhoinveta} that the covering  $\{\mathfrak{M}_i\}_{i=1}^3$ satisfies $\dist(\mathfrak{M}_i \backslash \mathfrak{M}_j, \mathfrak{M}_j \backslash \mathfrak{M}_i) \gtrsim 1$ for all $i \neq j \in \{1,2,3\}$ owing to the security factors $\tfrac{1}{2}$ in \eqref{eq:M123_domain}. Therefore, there exists a uniformly-Lipschitz partition of unity subordinated to the covering $\{\mathfrak{M}_i\}_{i=1}^3$, and the desired operator $\other{\mathcal{R}}$ exists by a standard Lipschitz gluing argument. The desired colinearity (i) with $\vecL$ in \eqref{eq:Rcolin} follows from \eqref{eq:R_colin_proof}, and regularity (ii) is satisfied by construction. 
			The scalar   $a(z_1,z_2)$   in \eqref{eq:Rcolin} arises from gluing the scalars $|c_j|^2 \sim 1$ on $\mathfrak{M}_j$ using the same partition unity on $\{\mathfrak{M}_j\}_{j=1}^3$. 
			
			Finally, we prove \eqref{eq:L_reg}.
			Note that to establish \eqref{eq:L_reg} with the minimum over $E \in \{\Re z_1, \Re z_2\}$ on the right-hand side, it suffices to check the bound both $E = \Re z_1$ and $E = \Re z_2$ separately.  
			We only present the case $E = \Re z_1$ in detail, since $E = \Re z_2$ is completely analogous. 
			To this end, we write, using \eqref{eq:Rcolin} and $|a(z_1, z_2)| \sim 1$, 
			\begin{equation} \label{eq:R_err_bound1}
				\bigl\lvert \bigl\langle \vecL(z_1, z_2) A \bigr\rangle\bigr\rvert^2 \sim \bigl\langle A^* \other{\mathcal{R}}(M_1,M_2)[A] \bigr\rangle \lesssim  \bigl\langle A^* \other{\mathcal{R}}\bigl(M(E),M(E)\bigr)[A] \bigr\rangle +  \bigl(\norm{M_1 - M(E)} + \norm{M_2 - M_1}\bigr) \norm{A}_\mathrm{hs}^2~, 
			\end{equation}
			where in the second step we used Lipschitz continuity of $\other{\mathcal{R}}$, and we denote $M(E) := \lim_{\eta\to+0} M(E+\ii\eta)$. 
			 Plugging \eqref{eq:R_err_bound2} into \eqref{eq:R_err_bound1}, we obtain the desired \eqref{eq:L_reg}. This concludes the proof of Lemma~\ref{lemma:R_extend}. 
		\end{proof}
		
		We conclude this section by proving Lemma~\ref{lemma:tripple_prop}. 
		\begin{proof}[Proof of Lemma~\ref{lemma:tripple_prop}]
			Without loss of generality, we assume that $z,w \in \regD\cap(\mathbb{H}^2)$. 
			First, the bound \eqref{eq:L_bound} follows immediately from \eqref{eq:R_colin_proof}, the comparison $|c_j| \sim 1$, and the boundedness of $\tempR_j$ on $\mathfrak{M}_j$ established in the proof of Lemma~\ref{lemma:R_extend} above. 
			
			Next, the vanishing of $\vertiii{\cdot}_{z,w}$ on the span of the identity is the trivial consequence of the definition of regularization \eqref{eq:regA_def}. Indeed, $B \mapsto \reg{B}^{z,w}$ is linear and $\reg{I}^{z,w} = 0$ for all $z,w \in \bddD$. 
			
			Finally, we prove \eqref{eq:symmetries}. For the first comparison in \eqref{eq:symmetries}, we observe that $\reg{B}^{z,w} = \reg{B}^{w,z}$ for all $z,w \in \bddD$ and $\reg{(B^*)}^{z,w} = (\reg{B}^{z,w})^*$; Hence, for every $z,w \in \bddD\cap\mathbb{H}$, it follows from \eqref{eq:alpha_sym} and \eqref{eq:Rcolin} that 
			\begin{equation} \label{eq:vertiistar}
				\vertiii{B^*}_{w,z} = \frac{ \bigl\lvert \bigl\langle \vecL(w, z) \reg{(B^*)}^{w, z}  \bigr\rangle \bigr\rvert}{\other{\alpha}(w, z)} + \norm{\reg{(B^*)}^{w, z}  }_{\mathrm{hs}} 
				\sim 
				\frac{ \bigl\lVert  \mathrm{Conj}\bigl(\other{\mathcal{R}}_{w,z}\bigr)\bigl[\reg{B}^{z, w}\bigr]   \bigr\rVert_\mathrm{hs}}{\other{\alpha}(z, w)} + \norm{\reg{B}^{z,w}  }_{\mathrm{hs}}~,
			\end{equation}
			where we abbreviate $\other{\mathcal{R}}_{w,z} := \other{\mathcal{R}}(M(w), M(z))$, and recall that for a super-operator $\mathcal{F}$, $\mathrm{Conj}(\mathcal{F)}\, :\, X \mapsto (\mathcal{F}[X^*])^*$. 
						
			Observe that the definitions of $U(M_1, M_2)$ and $V(M_1, M_2)$, and hence the construction of $\other{\mathcal{R}}(M_1, M_2)$ can be naturally extended to the negative quadrant   (recall \eqref{eq:Mdomplus})
			$$\mathfrak{M}_- := 
			\{(M_1^*, M_2^*)\,: \, (M_1, M_2) \in \mathfrak{M}_+\}, $$
			and satisfy, for all $(M_1, M_2) \in \mathfrak{M}_+$, 
			\begin{equation}
				U(M_1^*, M_2^*) = U(M_2^*, M_1^*) = U(M_1, M_2)^*, \qquad V(M_1^*, M_2^*) = V(M_2, M_1) = V(M_1, M_2)^*. 
			\end{equation}
			Hence, the construction of $\other{\mathcal{R}}(M_1, M_2)$ also extends to $\mathcal{M}_-$, and, since $\mathrm{Conj}\bigl(\langle R^* (\,\cdot\,)\rangle R\bigr) = \langle R (\,\cdot\,)\rangle  R^*$ for any $R\in\mathbb{C}^{N\times N}$, we have
			\begin{equation}
				\other{\mathcal{R}}(M_1^*, M_2^*) = \mathrm{Conj}\bigl( \other{\mathcal{R}}(M_2, M_1) \bigr), \qquad (M_1, M_2) \in \mathfrak{M}_+~. 
			\end{equation}
			Moreover, the argument for the Lipschitz-regularity of $\other{\mathcal{R}}(M_1, M_2)$ remains valid verbatim on the union~$\mathfrak{M}_+ \cup \mathfrak{M}_-$. 
			Therefore, for all $z,w \in \bddD\cap\mathbb{H}$, we deduce that 
			\begin{equation} \label{eq:conjR_est}
				\mathrm{Conj}\bigl(\other{\mathcal{R}}_{w,z}\bigr) = \other{\mathcal{R}}_{\bar z, \bar w} = \other{\mathcal{R}}_{z, w} + \mathcal{O}_{\mathrm{hs}\to\mathrm{hs}}\bigl(\rho(z) + \rho(w)\bigr),
			\end{equation}
			where we used that $\norm{M(\overline{z}) - M(z)} = 2\norm{\Im M(z)} \sim \rho(z)$ for all $z\in\bddD$ by \eqref{eq:imM_bound}. Since $\rho(z) + \rho(w) \lesssim \other{\alpha}(z,w)$, we conclude from \eqref{eq:vertiistar} and \eqref{eq:conjR_est} that
			\begin{equation}
				\vertiii{B^*}_{w,z} \sim \frac{ \bigl\lVert   \other{\mathcal{R}}_{z,w} \bigl[\reg{B}^{z, w}\bigr]   \bigr\rVert_\mathrm{hs}}{\other{\alpha}(z, w)} + \norm{\reg{B}^{z, w}  }_{\mathrm{hs}} = \vertiii{B}_{z,w}. 
			\end{equation}
			Hence, the first comparison in \eqref{eq:symmetries} holds. The second comparison is obtained analogously using the Lipschitz-continuity of $\other{\mathcal{R}}(M_1, M_2)$ to deduce that 
			\begin{equation} \label{eq:swapR_est}
				 \other{\mathcal{R}}_{w, z} = \other{\mathcal{R}}_{z, w} + \mathcal{O}_{\mathrm{hs}\to\mathrm{hs}}\bigl(\norm{M(z) - M(w)}\bigr),
			\end{equation}
			together with the fact that $\norm{M(z) - M(w)} \lesssim \other{\xi}(z,w) \lesssim |z-w|^{1/3} \lesssim \other{\beta}(z,w)^{1/2} \lesssim \other{\alpha}(z,w)$ by \eqref{eq:betaf_def}, \eqref{eq:alpha_defin},  \eqref{eq:M_perturb}, and~\eqref{eq:xif_def}. 
			This concludes the proof of Lemma~\ref{lemma:tripple_prop}. 
		\end{proof}

		\nc

	\section{Two-Body Stability Analysis} \label{sec:stab}	
	In this section, we analyze the behavior of the two-body stability operator $\mathcal{B}_{z_1, z_2}$, defined in \eqref{eq:stab_def}. 
	Our analysis is dynamical and relies on the time-dependent version of $\mathcal{B}_{z_1, z_2}$, namely $\mathcal{B}_{t,z_1, z_2}$, defined in \eqref{eq:stab_t_def}. 
	We recall that $z_{j,t}$ denote the trajectories of the characteristic flow 
	\eqref{eq:char_flow} with the final condition $z_{j,T} = z_j$ at the terminal time $T\sim 1$. Then, at the initial time $t=0$, the spectral parameter $z_{j,0}$ satisfy $\dist(z_{j,0}, \supp\,\rho_0) \sim 1$, and the operator $\mathcal{B}_{0,z_1, z_2}$ is controlled using the following general lemma. 
	\begin{lemma}[Stability away from the Spectrum] \label{lemma:stab_away}
		Let $(D_\gamma, \mathcal{S}_\gamma)$ be a data-pair with $D_\gamma = D_\gamma^*$ satisfying Assumption~\ref{ass:boundedexp}, and $\mathcal{S}_\gamma$ of the form
		\begin{equation} \label{eq:S_gamma}
			\mathcal{S}_\gamma[\, \cdot\,] = (1 - \gamma) \langle \,\cdot\,\rangle +  \gamma \, \mathcal{S}[\, \cdot\,], \qquad 1 + (c_\mathrm{full} - 1)\gamma \gtrsim 1,
		\end{equation}
		where $0 \le \gamma \lesssim 1$ and $\mathcal{S}$ is a self-energy operator satisfying Assumptions~\ref{ass:S_norms}--\ref{ass:full} with some constant $c_\mathrm{full} > 0$.
		
		Let $M_\gamma(z)$ be a solution to the MDE \eqref{eq:MDE} with data-pair $(D_\gamma, \mathcal{S}_\gamma)$, and $\rho_\gamma$ be the corresponding self-consistent density of states.
		Then, the stability operator $\mathcal{B}(\gamma,z_1, z_2) := \mathrm{Id}[\,\cdot\,] - M_\gamma(z_1)\mathcal{S}_\gamma[\,\cdot\,]M_\gamma(z_2)$
		admits the bound 
		\begin{equation} \label{eq:stab_apriori}
			\norm{\mathcal{B}(\gamma,z_1, z_2)^{-1}}_{\mathrm{hs}\to \mathrm{hs}} + \norm{\mathcal{B}(\gamma,z_1, z_2)^{-1}} 
			\lesssim 1,
		\end{equation}
		for all $z_1, z_2 \in \mathbb{C}$ satisfying $\dist(z_j, \supp\,\rho_\gamma) \gtrsim 1$, with $j \in \{1,2\}$.
	\end{lemma}
	We defer the proof of Lemma~\ref{lemma:stab_away} to Section \ref{sec:meta}. 
	In the sequel, we also use Lemma~\ref{lemma:stab_away} with $(D_\gamma, \mathcal{S}_\gamma) = (D_t, \mathcal{S}_t)$ and $\gamma := \ee^{T-t}$ to estimate $\mathcal{B}_{t, z_1, z_2} = \mathcal{B}(\gamma_t, z_{1,t}, z_{2,t})$.
	We are now ready to prove Lemma~\ref{lemma:stab_t}. 	 
	
	\subsection{Stability Operator Bounds. Proof of Lemmas~\ref{lemma:stab_t} and~\ref{lemma:stab_bound} }
	We begin by proving the bound on the time-dependent stability operator $\mathcal{B}_{t,z_1, z_2}$. 
	\begin{proof}[Proof of Lemma~\ref{lemma:stab_t}]
		First, we observe that it follows from  \eqref{eq:M_bound} and \eqref{eq:S_hs_to_op} that
		\begin{equation} \label{eq:1-B_bound}
			\norm{\mathrm{Id} - \mathcal{B}_{t,z_1, z_2}}_{\mathrm{hs}\to\norm{\cdot}} \lesssim \bigl(1+|z_1|\bigr)^{-1}\bigl(1+|z_2|\bigr)^{-1}, \qquad z_1, z_2 \in \bddD. 
		\end{equation}
		Hence, similarly to \eqref{eq:op_from_hs}, it suffices to establish the desired bound \eqref{eq:stab_t_bound} only in the $\norm{\,\cdot\,}_{\mathrm{hs}\to\mathrm{hs}}$-norm. 
		
		Recall that the bounds \eqref{eq:invM_bound}--\eqref{eq:rho_props} imply that
		\begin{equation} \label{eq:close_M_norms}
			\norm{M_1} + \norm{M_2} +\norm{M_1^{-1}} + \norm{M_2^{-1}} \lesssim 1, \qquad z_1, z_2 \in \bddD.
		\end{equation}
		It follows from \eqref{eq:dstab} that
		\begin{equation} \label{eq:stab_duhamel}
			\norm{\mathcal{B}_{t,z_1,z_2}^{-1}}_{\mathrm{hs}\to\mathrm{hs}} \le \norm{\mathcal{B}_{0,z_1,z_2}^{-1}}_{\mathrm{hs}\to\mathrm{hs}} + \biggl\lvert \int_{0}^t \frac{(z_{1} - z_{2})^2}{(z_{1,s} - z_{2,s})^2} \mathrm{d}s \biggr\rvert \times  \frac{\norm{M_1 - M_2}_{\mathrm{hs}}^2}{|z_1 - z_2|^2}.
		\end{equation}
		Furthermore, since $T\sim 1$, \eqref{eq:eta_t} and \eqref{eq:kapd_rhoeta_t} imply that $\dist(z_{j,0}, \supp\,\rho_0) \sim 1$. Therefore, by Lemma~\ref{lemma:stab_away}, the stability operator 
		$\mathcal{B}_{0,z_1,z_2} = \mathcal{B}(\ee^{T}, z_{1,0}, z_{2,0})$ \nc
		satisfies
		\begin{equation} \label{eq:init_stab}
			\norm{\mathcal{B}_{0,z_1,z_2}^{-1}}_{\mathrm{hs}\to\mathrm{hs}} \lesssim 1. 
		\end{equation} 
		
		We now estimate the second term in \eqref{eq:stab_duhamel}. 	
		The bound \eqref{eq:z_diff_comp}, together with \eqref{eq:close_M_norms} implies that
		\begin{equation}
			\biggl\lvert \int_{0}^t \frac{(z_{1} - z_{2})^2}{(z_{1,t} - z_{2,t})^2} \mathrm{d}s \biggr\rvert   \lesssim \frac{\other{\beta}(z_1, z_2)^2}{\other{\beta}_t(z_1, z_2)}. 
		\end{equation}
		
		Note that if $z_1, z_2$ are in the same half-plane (without loss of generality $z_1, z_2 \in \mathbb{H}$), then we use \eqref{eq:M_perturb}, \eqref{eq:xif_def}, and \eqref{eq:close_M_norms}, \nc to obtain
		\begin{equation} \label{eq:M_dd_shp}
			\frac{\norm{M_1 - M_2}_{\mathrm{hs}}^2}{|z_1 - z_2|^2} \lesssim \frac{\other{\xi}(z_1,z_2)^2}{|z_1 - z_2|^2} \lesssim \frac{1}{\other{\beta}(z_1, z_2)^2}.
		\end{equation}
		On the other hand, if $z_1, z_2$ are in opposite half-planes, then it follows from \eqref{eq:imM_bound} that
		\begin{equation} \label{eq:M_dd_ohp}
			\frac{\norm{M_1 - M_2}_{\mathrm{hs}}^2}{|z_1 - z_2|^2} \lesssim \frac{\norm{M_1 - M_2^*}_{\mathrm{hs}}^2}{|z_1 - \overline{z}_2|^2} + \frac{\rho(z_1)^2}{|z_1 - z_2|^2} \lesssim \frac{1}{\other{\beta}(z_1, z_2)^2},
		\end{equation}
		where in the second step we used \eqref{eq:M_dd_shp} above 
		and \eqref{eq:betaf_def}. 
		
		Plugging \eqref{eq:init_stab}--\eqref{eq:M_dd_ohp} into \eqref{eq:stab_duhamel}, we obtain the desired \eqref{eq:stab_t_bound} since $\other{\beta}_t(z_1, z_2) \lesssim 1$. This concludes the proof of Lemma~\ref{lemma:stab_t}.	
	\end{proof}
	
	Next, we prove the time-independent Lemma~\ref{lemma:stab_bound}. 
	
	\begin{proof}[Proof of Lemma~\ref{lemma:stab_bound}]
		The bound \eqref{eq:stab_beta_bound} follows from \eqref{eq:stab_t_bound} at time $t = T$.
		
		Next, we prove \eqref{eq:stab_well_struc}. Let $\other{z} := z$ if $(\Im z)(\Im w) >0$ and $\other{z} := \overline{z}$ if $(\Im z)(\Im w) < 0$.  Observe that, by \eqref{eq:betaf_def},  $\other{\beta}(z,\other{z}) \le 2 \other{\beta}(z,w)$. 
		Hence, there exists a threshold $\beta_*$ such that if $\other{\beta}(z,w) \le \beta_*$ for $z,w\in \bddD$, then  $\other{\beta}(z,\other{z}) \le 2\beta_*$, and
		\begin{equation} \label{eq:rho_cond}
			\begin{cases}
				|\rho(z)| + \rho(z)^{-1}\Im z \le \rho_*, \quad &\text{if}\quad (\Im z)(\Im w) > 0,\\
				\rho(z)^{-1}\Im z \le c_*, \quad &\text{if}\quad (\Im z)(\Im w) < 0,
			\end{cases}
		\end{equation}
		where $\rho_*$ and $c_*$ are the constants from Lemmas~\ref{lemma:stab1} and~\ref{lemma:1stab_ohp}. 
		Therefore, it follows from Lemmas~\ref{lemma:stab1} and~\ref{lemma:1stab_ohp} that the stability operator $\mathcal{B}_{z,\other{z}}$ has an $\varepsilon_*$-well-structured spectrum in both $\norm{\cdot}$ and $\norm{\cdot}_{\mathrm{hs} \to \mathrm{hs}}$-norms, in the sense of Definition~\ref{def:pert_cond}. On the other hand, it follows from \eqref{eq:M_bound}, \eqref{eq:S_hs_to_op}, and \eqref{eq:M_perturb} that 
		\begin{equation} \label{eq:beta_cond}
			\norm{\mathcal{B}_{z,w} - \mathcal{B}_{z,\other{z}}} \vee \norm{\mathcal{B}_{z,w} - \mathcal{B}_{z,\other{z}}}_{\mathrm{hs} \to \mathrm{hs}} \lesssim \other{\xi}(\other{z},w).
		\end{equation}
		Hence, by \eqref{eq:pert_cond} for $\mathcal{B}_{z,\other{z}}$ and a simple perturbation
		argument there exists a threshold $\delta_* \sim 1$ such that if $|w - \other{z}| \le \delta_*$, the super-operator $\mathcal{X}_r$, defined as
		\begin{equation} \label{eq:interp}
			\mathcal{X}_r := (1-r)\,\mathcal{B}_{z,\other{z}} + r \,\mathcal{B}_{z,w}, 
		\end{equation}
		satisfies \eqref{eq:pert_cond} with $\varepsilon = \varepsilon_*$ in both $\norm{\cdot}$ and $\norm{\cdot}_{\mathrm{hs} \to \mathrm{hs}}$-norms uniformly in $r\in[0,1]$; in particular we have \eqref{eq:pert_cond}
		for $\mathcal{X}_{r=1} = \mathcal{B}_{z,w}$.
		It follows from \eqref{eq:betaf_def} that by suitably shrinking the threshold $\beta_*$, we can guarantee that $\other{\beta}(z,w) \le \beta_*$ implies $|\other{z}-w| \le \delta_*$. Hence, the desired bound \eqref{eq:stab_well_struc} is established.
		
		Next, we prove \eqref{eq:Pi_rank1}. As we established above, the bound \eqref{eq:pert_cond} 
		holds for $\mathcal{X}_r$ defined in \eqref{eq:interp}. Hence, the map $r \mapsto \Tr[\Pi_{\mathcal{X}_r}\mathcal{X}_r] = \mathrm{rank}\, \Pi_{\mathcal{X}_r}$ is continuous on $r\in[0,1]$, and is therefore constant. Since $\mathcal{X}_0 = \mathcal{B}_{z,\other{z}}$ satisfies \eqref{eq:Pi_def}, so does $\mathcal{X}_1 = \mathcal{B}_{z,w}$, yielding the desired \eqref{eq:Pi_rank1}, provided $\other{\beta}(z,w) \le \beta_*$.
		
		Next, we prove \eqref{eq:beta_conj}. To this end, we observe that if $\mathcal{F}$ is a super-operator with eigenvalue $\beta$, then the operator $\mathrm{Conj}(\mathcal{F)}\, :\, X \mapsto (\mathcal{F}[X^*])^*$ has an eigenvalue $\overline{\beta}$, the complex conjugate of $\beta$. Computing $\mathrm{Conj}(\mathcal{B}_{z,w})$ explicitly, we obtain
		\begin{equation}
			\mathrm{Conj}(\mathcal{B}_{z,w})[X] = \bigl(X^* - M(z)X^*M(w)\bigr)^* = X - M(w)^* X M(z)^* = \mathcal{B}_{\bar w, \bar z}[X],
		\end{equation}  
		which implies \eqref{eq:beta_conj} immediately. 
		\nc

		We proceed by proving \eqref{eq:beta_assymp}, still under the
		condition $\other{\beta}(z,w) \le \beta_*$, which also implies  $|\other{z}-w| \le \delta_*$. 
		It follows from \eqref{eq:u_def}, \eqref{eq:B12_identity}, \eqref{eq:M_bound}, and \eqref{eq:PiQ_bounds}, that 
		\begin{equation} \label{eq:beta_identity}
			u(z,w) = \frac{\bigl\langle M(z) - M(w) \bigr\rangle}{z-w} = \bigl\langle \mathcal{B}_{z,w}^{-1}\bigl[M(z)M(w)\bigr]\bigr\rangle = \beta_{z,w}^{-1} \bigl\langle \Pi_{z,w}\bigl[M(z)M(w)\bigr]\bigr\rangle + \mathcal{O}(1).
		\end{equation}
		Using \eqref{eq:beta_cond}, \eqref{eq:M_perturb}, and Lemma~\ref{lemma:perturb}, we deduce that, by suitably shrinking $\delta_*, \beta_*\sim 1$, 
		\begin{equation} \label{eq:tr_Pi_large}
			\bigl\langle \Pi_{z,w}\bigl[M(z)M(w)\bigr]\bigr\rangle = \bigl\langle \Pi_{z,\other{z}}\bigl[M(z)M(\other{z})\bigr]\bigr\rangle + \mathcal{O}\bigl(\other{\xi}(\other{z},w)\bigr) \sim 1,
		\end{equation}
		where in the second step we used \eqref{eq:Pi_beta}--\eqref{eq:LR_props} if $(\Im z)(\Im w) > 0$ and \eqref{eq:ohp_Pi_ests} if $(\Im z)(\Im w) < 0$. Here we also used that $\other{\xi}(\other{z},w)\le |\other{z} - w| + |\other{z}-w|^{1/3}$.  
		On the other hand, the definitions $\beta_{z,w} = \Tr[\mathcal{B}_{z,w}\Pi_{z,w}]$ and \eqref{eq:Pi_rank1} imply that $|\beta_{z,w}| \le \varepsilon_* \sim 1$; Hence, multiplying both sides of \eqref{eq:beta_identity} by $\beta_{z,w} u(z,w)^{-1}$ and using \eqref{eq:tr_Pi_large}, we deduce that
		\begin{equation} \label{eq:beta_b_bound}
			|\beta_{z,w}| \lesssim \min \bigl\{1,\, \lvert u(z,w) \rvert^{-1}\bigl(1 + \mathcal{O}(|\beta_{z,w}|)\bigr) \bigr \} \lesssim 1 \wedge  \lvert u(z,w)  \rvert^{-1} \lesssim \other{\beta}(z,w). 
		\end{equation}
		Here, in the last step, we used the dichotomy between \eqref{eq:u_sim} and \eqref{eq:u_one} all $z, w \in \bddD$. 
		Plugging \eqref{eq:tr_Pi_large} and \eqref{eq:beta_b_bound} back into \eqref{eq:beta_identity}, we obtain
		\nc 
		\begin{equation} 
			|\beta_{z,w}| \sim  \lvert u(z,w) \rvert^{-1} \bigl(1 + \mathcal{O}\bigl(\other{\beta}(z,w)\bigr)\bigr),
		\end{equation}
		which implies \eqref{eq:beta_assymp} for sufficiently small $\beta_* \sim 1$. 		
	\end{proof}

	\subsection{Probabilistic Proof of Stability away from the Spectrum}	 \label{sec:meta} 	
	\begin{claim} \label{claim:S_bound}
		Let $\Lambda^\mathrm{av}$ and $\Lambda^\mathrm{iso}$ be a pair of deterministic control parameters, and let $X$ be a random matrix satisfying
		\begin{equation} \label{eq:radnX_assume}
			\bigl\lvert \bigl\langle X B \bigr\rangle  \bigr\rvert \prec \Lambda^\mathrm{av} \norm{B}_\mathrm{hs}, \qquad \bigl\lvert \langle \bm x, X \bm y\rangle \bigr\rvert  \prec \Lambda^\mathrm{iso} \norm{\bm x}\norm{\bm y}.
		\end{equation}
		for any deterministic matrix $B \in \mathbb{C}^{N\times N}$ and deterministic vectors $\bm x, \bm y \in \mathbb{C}^N$. Then,
		the self-energy operator $\mathcal{S}_\gamma$ of the form \eqref{eq:S_gamma} with $0 \le \gamma \lesssim 1$ and  $\mathcal{S}$ satisfying Assumptions~\ref{ass:S_norms}, admits the bounds
		\begin{equation} \label{eq:S_XYZ_bounds}
			\bigl\lvert \bigl\langle X \, Y \mathcal{S}_\gamma[Z] \bigr\rangle \bigr\rvert \prec \sqrt{N}\Lambda^\mathrm{av} \norm{Y}_\mathrm{hs} \norm{Z}_\mathrm{hs}, \qquad 
			\bigl\lvert \bigl\langle \mathcal{S}_\gamma[X] Z \, Y \bigr\rangle \bigr\rvert \prec \Lambda^\mathrm{iso} \norm{Y}_\mathrm{hs} \norm{Z}_\mathrm{hs},
		\end{equation}		
		for all deterministic matrices $Y \in \mathbb{C}^{N\times N}$, and all matrices  $Z \in \mathbb{C}^{N\times N}$ (deterministic or random). 
	\end{claim}
	
	\begin{proof} [Proof of Claim~\ref{claim:S_bound}]
		It follows from  from \eqref{eq:kappa2_norms} that 
		\begin{equation} \label{eq:S_max_op}
			\norm{\mathcal{S}}_{\max \to \norm{\cdot}} \le \vertiii{\mathcal{S}}_2.
		\end{equation}
		Hence, using $\norm{X}_{\max} \prec \Lambda^\mathrm{iso}$ by \eqref{eq:radnX_assume}, together with \eqref{eq:S_gamma} and \eqref{eq:S_max_op},  we obtain 
		\begin{equation}
			\bigl\lvert \bigl\langle \mathcal{S}_\gamma[X] Z \, Y \bigr\rangle \bigr\rvert 
			\le \bigl(|1 - \gamma| + \gamma \norm{\mathcal{S}}_{{\max}\to\norm{\cdot}}\bigr)  \norm{X}_{\max}\norm{Y}_\mathrm{hs} \norm{Z}_\mathrm{hs}  \prec \Lambda^\mathrm{iso}\norm{Y}_\mathrm{hs} \norm{Z}_\mathrm{hs}.
		\end{equation}
		Therefore, the second bound in \eqref{eq:S_XYZ_bounds} holds, and it remains to prove the first bound.

		It follows from \eqref{eq:S_gamma} and the decomposition $\mathcal{S} = \mathcal{S}_\mathrm{c} + \mathcal{S}_\mathrm{d}$ that 
		\begin{equation} \label{eq:XYZ_decomp}
			\bigl\lvert \bigl\langle X \, Y \mathcal{S}_\gamma[Z] \bigr\rangle \bigr\rvert \le |1-\gamma| \bigl\lvert \langle X \, Y\rangle \bigr\rvert \norm{Z}_\mathrm{hs} + \gamma \inf\limits_{\mathcal{S} = \mathcal{S}_\mathrm{c} + \mathcal{S}_\mathrm{d}}\Bigl(\bigl\lvert \bigl\langle X \, Y \mathcal{S}_\mathrm{c}[Z] \bigr\rangle \bigr\rvert + \bigl\lvert \bigl\langle X \, Y \mathcal{S}_\mathrm{d}[Z] \bigr\rangle \bigr\rvert\Bigr). 
		\end{equation}
		By \eqref{eq:radnX_assume}, the first term in \eqref{eq:XYZ_decomp} is stochastically  dominated by $\Lambda^\mathrm{av} \norm{Y}_\mathrm{hs}\norm{Z}_\mathrm{hs}$. We proceed to estimate the contribution of $\mathcal{S}_\mathrm{c}$. With $E^{ab}$ denoting the standard basis in the space of $N\times N$ matrices, that is $(E^{ab})_{ij} := \delta_{ia}\delta_{bj}$, we obtain
		\begin{equation} \label{eq:S_XYZ_cross_bound}
			\begin{split}
				\bigl\lvert \bigl\langle X \, Y \mathcal{S}_\mathrm{c}[Z] \bigr\rangle \bigr\rvert &= \Bigl\lvert \sum_{ab} Z_{ab} \bigl\langle X \, Y \mathcal{S}_\mathrm{c}[E^{ab}] \bigr\rangle \Bigr\rvert \le N^{1/2} \norm{Z}_\mathrm{hs}\biggl( \sum_{ab}  \bigl\lvert \bigl\langle X \, Y \mathcal{S}_\mathrm{c}[E^{ab}] \bigr\rangle \bigr\rvert^2 \bigr\rvert  \biggr)^{1/2}\\
				&\prec N^{1/2}\Lambda^\mathrm{av} \norm{Z}_\mathrm{hs}\biggl( \sum_{ab}  \norm{Y \mathcal{S}_\mathrm{c}[E^{ab}]}_\mathrm{hs}^2 \biggr)^{1/2},
			\end{split}
		\end{equation}
		where in the penultimate step we used Cauchy--Schwarz inequality, and in the last step we used \eqref{eq:radnX_assume} together with the fact that $Y$ is deterministic.  Computing the Hilbert-Schmidt norm squared in the last factor, we deduce that 
		\begin{equation}
			\sum_{ab}\norm{Y \mathcal{S}_\mathrm{c}[E^{ab}]}_\mathrm{hs}^2 = \sum_{ab} \bigl\langle \mathcal{S}_\mathrm{c}[E^{ab}]^*Y^* Y\mathcal{S}_{\mathrm{c}}[E^{ab}]  \bigr\rangle 
			\le \sum_{ab, ij} \Bigl\lvert \sum_p |Y_{ip}| |(\mathcal{S}_{\mathrm{c}})_{pj, ab}|  \Bigr\rvert^2 = \frac{1}{N}\sum_i \norm{R^i}_\mathrm{hs}^2,
		\end{equation}
		where $R^i$ are $N\times N$ matrices with entries
		\begin{equation}
			R_{ab}^i := N\biggl(\sum_j \Bigl\lvert \sum_p |Y_{ip}| |(\mathcal{S}_{\mathrm{c}})_{pj, ab}|  \Bigr\rvert^2\biggr)^{1/2}, \qquad a,b,i \in \indset{N}. 
		\end{equation}
		It follows from~\eqref{eq:cross_dir_norm_defs} that $\lVert R^i\rVert \le \vertiii{\mathcal{S}_{\mathrm{c}}}_\mathrm{c} |(YY^*)_{ii}|^{1/2}$. Therefore,
		\begin{equation} \label{eq:S_XYZ_cross}
			\sum_{ab}\norm{Y \mathcal{S}_\mathrm{c}[E^{ab}]}_\mathrm{hs}^2  \le \vertiii{\mathcal{S}_{\mathrm{c}}}_\mathrm{c}^2 \norm{Y}_\mathrm{hs}^2
			\quad \Longrightarrow \quad \bigl\lvert \bigl\langle X \, Y \mathcal{S}_\mathrm{c}[Z] \bigr\rangle \bigr\rvert \prec \Lambda^\mathrm{av} \vertiii{\mathcal{S}_{\mathrm{c}}}_\mathrm{c}  \norm{Y}_\mathrm{hs} \norm{Z}_\mathrm{hs},
		\end{equation}
		where the implication follows from \eqref{eq:S_XYZ_cross_bound}.  
		The desired first estimate in \eqref{eq:S_XYZ_bounds} follows form \eqref{eq:kappa2_norms}, \eqref{eq:XYZ_decomp},  \eqref{eq:S_XYZ_cross} and the analogous bound on $\mathcal{S}_\mathrm{d}$. This concludes the proof of Claim~\ref{claim:S_bound}.
	\end{proof}

	\begin{proof}[Proof of Lemma~\ref{lemma:stab_away}]
		Similarly to the reasoning below \eqref{eq:1-B_bound} in the proof of Lemma~\ref{lemma:stab_t}, it suffices to consider $z_1, z_2 \in \mathbb{C}\backslash\mathbb{R}$ satisfying $c \le \dist(z_j, \supp\,\rho_\gamma) \le C$ for some sufficiently large constant $C\sim 1$. Furthermore, we can assume that $|\Im z_j| \ge N^{-K}$ for some large constant $K \sim 1$, since the result can be easily extended into the complementary regime using the Lipschitz-continuity of $M_\gamma$ away from the support of $\rho_\gamma$ and a simple perturbative argument, the details of which we leave to the reader.
		
		Therefore, for the remainder of the proof we assume that $z_1, z_2 \in \mathbb{C}$ satisfy $c \le \dist(z_j, \supp\,\rho_\gamma) \le C$ and $|\Im z_j| \ge N^{-K}$. \nc
		To show that the operator $\mathcal{B} \equiv \mathcal{B}(\gamma, z_1, z_2)$ is invertible and satisfies the bound \eqref{eq:stab_apriori}, we use a minimalistic version of the meta-argument \cite{helton2007operator}, relating the action of $\mathcal{B}$ to an expected value of a two-resolvent chain. 
		To this end, we  define the covariance tensor $\other{\Sigma}_\gamma$ by
		\begin{equation}
			\other{\Sigma}_\gamma := (1-\gamma)\Sigma_{\mathrm{G}} + \gamma \Sigma,
		\end{equation}
		where $\Sigma_{\mathrm{G}}$ is the covariance tensor of a $N\times N$ Gaussian invariant ensemble (GUE or GOE) in the same symmetry class as $H$, and $\Sigma$ is the covariance tensor corresponding to $\mathcal{S}$ via \eqref{eq:SigmaSentries}. It follows from \eqref{eq:S_gamma} and Assumption~\ref{ass:full} that $\other{\Sigma}_\gamma \gtrsim \Sigma_{\mathrm{G}}$. 
		Consider a random matrix $\other{H} \equiv \other{H}(N)$, given by 
		\begin{equation}
			\other{H} := D_\gamma + \other{\Sigma}_\gamma^{1/2} \bigl[\mathcal{H}\bigr],
		\end{equation}
		where $\mathcal{H} \equiv \mathcal{H}(N)$ is an $N\times N$ self-adjoint matrix with standard Gaussian entries in the same symmetry class as $H$.
		Let $\mathcal{S}_\mathrm{G}$ be the self-energy operator corresponding to $\Sigma_{\mathrm{G}}$ via \eqref{eq:SigmaSentries}, and  $\mathcal{T}_{\mathrm{G}}[\,\cdot\,] := \mathcal{S}_\mathrm{G}[\,\cdot\,] - \langle \,\cdot\, \rangle$, then
		\begin{equation} \label{eq:torsion_def}
			\mathcal{T}_{\mathrm{G}}[X] = \begin{cases}
				0, \quad& H \text{ -- complex Hermitian},\\
				N^{-1}X^\mathfrak{t}, \qquad & H \text{ -- real symmetric},
			\end{cases}
		\end{equation}
		where $X^\mathfrak{t}$ denotes the transpose of $X \in \mathbb{C}^{N\times N}$. 
		Observe that the self-energy operator $\other{\mathcal{S}}$ corresponding to $\other{H}$ is given by
		\begin{equation} \label{eq:meta_DS}
			\other{\mathcal{S}} = \mathcal{S}_\gamma + (1-\gamma) \mathcal{T}_{\mathrm{G}}. 
		\end{equation}
		Since $\norm{\mathcal{T}_\mathrm{G}} \le N^{-1}$, the data-pair perturbation argument from Proposition 10.1 in \cite{AEK2020} implies that  the solution $\other{M}(z)$ to the MDE with data-pair $(D_\gamma, \other{S})$   satisfies 
		\begin{equation}
			\bigl\lVert\other{M}(z) - M_\gamma(z)\bigr\rVert \lesssim N^{-1}, \qquad z \in \mathbb{C} \,\, \text{with}\,\, \dist(z, \supp\, \rho_\gamma) \ge c.
		\end{equation}
		In summary, $M_\gamma$ is well approximated by the self-consistent resolvent $\other{M}$
		of a Gaussian ensemble $\other{H}$. The discrepancy of order $N^{-1}$ arises in the real symmetric case, where the $(1-\gamma)\mathcal{S}_{\mathrm{G}}$
		component of the self-energy operator of $\other{H}$ does not exactly coincide with the corresponding $(1-\gamma)\langle \cdot \rangle$ component of  $\mathcal{S}_\gamma$. 
		We now relate $\other{M}$ to the random resolvent $\other{G}(z) := (\other{H}-z)^{-1}$.  
		Indeed, the single-resolvent local laws proved in \cite{cuspuniv} imply that $\other{G}(z) := (\other{H}-z)^{-1}$ satisfies, for all $z  \in \mathbb{C}$ with $c \le \dist(z, \supp \,\rho_\gamma) \le C $, 
		\begin{equation} \label{eq:meta_ll}
			\Bigl\lvert \bigl(\other{G}(z) - M_\gamma(z)\bigr)_{\bm x \bm y}  \Bigr\rvert \prec
			\frac{1}{\sqrt{N}}, \qquad 
			\Bigl\lvert \bigl\langle \bigl(\other{G}(z) - M_\gamma(z)\bigr) B \bigr\rangle \Bigr\rvert \prec
			\frac{1}{N}, \qquad \bigl\lVert\Im \other{G}(z) \bigr\rVert \lesssim |\Im z|, \,\, \text{w.v.h.p.},
		\end{equation}
		for any deterministic vectors $\bm x, \bm y \in\mathbb{C}^{N}$ with $\norm{\bm x} = \norm{\bm y} = 1$, deterministic matrices $ B\in\mathbb{C}^{N\times N}$ with $\norm{B}_{\mathrm{hs}} = 1$. Note that for the estimates \eqref{eq:meta_ll} to hold for $z$ away from the support of $\rho_\gamma$, we need not require that $\norm{M_\gamma(w)}$ is uniformly bounded on a narrow vertical strip around $z$.  
		
		We are now ready to prove \eqref{eq:stab_apriori}. 
		Fix  a pair of spectral parameters $z_1, z_2 \in \mathbb{C}\backslash\mathbb{R}$ satisfying $c \le \dist(z_j,\supp \,\rho_\gamma) \le C$ for $j\in\{1,2\}$.
		Let $\other{G}_j := \other{G}(z_j)$ for $j \in \{1, 2\}$. It follows by Gaussian integration by parts that
		\begin{equation} \label{eq:2Gexpv}
			\E\Bigl[\mathcal{B}\bigl[ \other{G}_1 X \other{G}_2\bigr] \Bigr] = M_1 X M_2 + \mathcal{E}(X),\qquad X \in \mathbb{C}^{N\times N},
		\end{equation}
		where  $M_j := M_\gamma(z_j)$, $\mathcal{B}[\,\cdot\,] := \mathrm{Id}_{N}[\,\cdot\,] - M_1 \other{\mathcal{S}_\gamma}[\,\cdot\,]M_2$, and   $\mathcal{E}(X)$ is given by
		\begin{equation} \label{eq:meta_err}
			\begin{split}
				\mathcal{E}(X) := M_1 \E\Bigl[ & X \bigl(\other{G}_2 - M_2\bigr)  + (1-\gamma)  \mathcal{T}_\mathrm{G}[\other{G}_1 X \other{G}_2] \other{G}_2  +(1-\gamma)\mathcal{T}_\mathrm{G}[\other{G}_1]\other{G}_1X \other{G}_2 
				\\
				&+  \mathcal{S}_\gamma\bigl[\other{G}_1 - M_1\bigr] \other{G}_1 X \other{G}_2  
				+ \mathcal{S}_\gamma\bigl[\other{G}_1 X\other{G}_2\bigr]  \bigl(\other{G}_2 - M_2\bigr) \Bigr]. 
			\end{split}
		\end{equation}
		We now show that, for any fixed $\xi > 0$,
		\begin{equation} \label{eq:meta_err_bound}
			\norm{\mathcal{E}(X)}_\mathrm{hs} \lesssim N^{-1/2+\xi} \norm{X}_\mathrm{hs}.
		\end{equation}
		
		We compute $\norm{\mathcal{E}(X)}_\mathrm{hs}$ by testing $\mathcal{E}(X)$ against an arbitrary deterministic matrix $Y\in \mathbb{C}^{N\times N}$ and then taking the supremum over all such $Y$.
		For any fixed $Y$, it follows from the first inequality in \eqref{eq:invM_bound} and  the second bound in \eqref{eq:meta_ll} that
		\begin{equation} \label{eq:meta1}
			\Bigl\lvert \bigl\langle Y^*M_1 X \bigl(\other{G}_2 - M_2\bigr)  \bigr\rangle \Bigr\rvert \le N^{-1+\xi} \norm{Y^*M_1 X}_\mathrm{hs} 
			\lesssim N^{-1/2 + \xi} \norm{X}_\mathrm{hs}\norm{Y}_\mathrm{hs}. 
		\end{equation}
		Similarly, it follows from \eqref{eq:invM_bound} and the third bound in \eqref{eq:meta_ll} that 
		\begin{equation} \label{eq:meta2}
			\Bigl\lvert \bigl\langle Y^*M_1 \mathcal{T}_\mathrm{G}[\other{G}_1 X \other{G}_2] \other{G}_2  \bigr\rangle \Bigr\rvert \le \frac{1}{N} \sqrt{\frac{\bigl\langle M_1^*Y \Im \other{G}_2\, Y^*M_1  \bigr\rangle}{\Im z_2}\frac{\bigl\langle \Im \other{G}_1 X \,\Im \other{G}_2 X^* \bigr\rangle}{(\Im z_1)(\Im z_2)}} \lesssim N^{-1} \norm{X}_\mathrm{hs}\norm{Y}_\mathrm{hs},
		\end{equation}
		\begin{equation} \label{eq:meta3}
			\Bigl\lvert \bigl\langle Y^*M_1 \mathcal{T}_\mathrm{G}[\other{G}_1]\other{G}_1X \other{G}_2  \bigr\rangle \Bigr\rvert \le \frac{1}{N} \sqrt{\frac{\bigl\langle Y^*M_1 (\Im \other{G}_1)^\mathfrak{t} M_1^*Y  \bigr\rangle}{\Im z_1}\frac{\bigl\langle \Im \other{G}_1 X \,\Im \other{G}_2 X^* \bigr\rangle}{(\Im z_1)(\Im z_2)}} \lesssim N^{-1} \norm{X}_\mathrm{hs}\norm{Y}_\mathrm{hs}.
		\end{equation}
		
		Next, we estimate the terms in the second line of \eqref{eq:meta_err}. It follows from the second bound in \eqref{eq:S_XYZ_bounds} that 
		\begin{equation} \label{eq:meta4} 
			\Bigl\lvert \bigl\langle Y^*M_1 \mathcal{S}_\gamma\bigl[\other{G}_1 - M_1\bigr] \other{G}_1 X \other{G}_2 \bigr\rangle \Bigr\rvert 
			\lesssim N^{-1/2+\xi} \sqrt{\frac{\bigl\langle \Im \other{G}_1 X\Im \other{G}_2 X^*\bigr\rangle}{(\Im z_1)(\Im z_2)}} \bigr\rvert \norm{Y}_\mathrm{hs}
			\lesssim N^{-1/2+\xi}\norm{X}_\mathrm{hs} \norm{Y}_\mathrm{hs}, 
		\end{equation}
		where we used isotropic law from \eqref{eq:meta_ll} and the third bound in \eqref{eq:meta_ll}.	Similarly, using the first estimate in \eqref{eq:S_XYZ_bounds} together with the averaged bound in \eqref{eq:meta_ll}, we obtain
		\begin{equation} \label{eq:meta5}
			\Bigl\lvert \bigl\langle Y^* M_1 \mathcal{S}_{\gamma}\bigl[\other{G}_1 X \other{G}_2\bigr] (\other{G}_2-M_2) \bigr\rangle \Bigr\rvert  
			\lesssim N^{-1/2 +\xi} \norm{X}_\mathrm{hs}\norm{Y}_\mathrm{hs}.
		\end{equation}		
		Combining \eqref{eq:meta1}--\eqref{eq:meta5}, we conclude \eqref{eq:meta_err_bound}. 
		\nc 
		
		Next, we show that the kernel of $\mathcal{B}$ (or, equivalently, $\mathcal{B}^*$) is trivial. Assume the that $\mathrm{ker} \, \mathcal{B}^* \neq \{0\}$, and let $R \in \mathrm{ker} \,\mathcal{B}^*$ with $\norm{R}_\mathrm{hs} = 1$. We set $X := M_1^{-1} R\, M_2^{-1}$. By \eqref{eq:invM_bound}, we have $\norm{X}_{\mathrm{hs}} \lesssim \norm{R}_{\mathrm{hs}} = 1$.  
		Then, multiplying \eqref{eq:2Gexpv} by $R^*$ and taking the trace, we obtain
		\begin{equation}
			0 = \E\Bigl[\bigl\langle \bigl(\mathcal{B}^*[R]\bigr)^*  \other{G}_1 X \other{G}_2 \bigr\rangle \Bigr] 
			= \E\Bigl[\bigl\langle R^* \mathcal{B}\bigl[ \other{G}_1 X \other{G}_2\bigr] \bigr\rangle \Bigr] = \norm{R}_\mathrm{hs}^2 + \bigl\langle R^* \mathcal{E}(X) \bigr\rangle = 1 + \bigl\langle R^* \mathcal{E}(X) \bigr\rangle,
		\end{equation}
		which contradicts \eqref{eq:meta_err_bound} for sufficiently large $N$. Therefore, $\mathrm{ker}\, \mathcal{B}^* = \{0\} = \mathrm{ker}\, \mathcal{B}$. 
		
		Next, we quantify the same argument to obtain the desired estimate on the norm of $\mathcal{B}^{-1}$. Let $\mathfrak{b}\equiv \mathfrak{b}(z_1, z_2) := \lVert \mathcal{B}^{-1}\rVert_{\mathrm{hs}\to\mathrm{hs}}$.  Then, there exists  $Y \in \mathbb{C}^{N\times N}$ with $\norm{Y}_\mathrm{hs} = 1$, satisfying $\bigl\lVert \mathcal{B}^{-1}[Y] \bigr\rVert_\mathrm{hs} = \mathfrak{b}$.  Applying $\mathcal{B}^{-1}$ to both sides of  \eqref{eq:2Gexpv} with $X := M_1^{-1} Y M_2^{-1}$ yields
		\begin{equation} \label{eq:meta_beta}
			\E\Bigl[ \other{G}_1 M_1^{-1} Y M_2^{-1} \other{G}_2 \Bigr] = \other{\mathcal{B}}^{-1}\bigl[Y\bigr] + \other{\mathcal{B}}^{-1}\bigl[\mathcal{E}\bigl(M_1^{-1} Y M_2^{-1}\bigr)\bigr].
		\end{equation}
		Since $\lVert \mathcal{B}^{-1}[Y] \rVert_\mathrm{hs} = \mathfrak{b}$ by construction and 
		\begin{equation}
			\norm{\mathcal{B}^{-1}\bigl[\mathcal{E}_n\bigl(M_1^{-1} Y M_2^{-1}\bigr)\bigr]}_\mathrm{hs} 
			\le \mathfrak{b} \norm{\mathcal{E}_n\bigl(M_1^{-1} Y M_2^{-1}\bigr)}_\mathrm{hs} \lesssim N^{-1/4}\mathfrak{b} \norm{M_1^{-1}}\norm{M_2^{-1}},
		\end{equation}
		by \eqref{eq:invM_bound}  and \eqref{eq:meta_err_bound}, we conclude from \eqref{eq:meta_beta} that, with $X := M_1^{-1} Y M_2^{-1}$, 
		\begin{equation} \label{eq:meta_beta_bound}
			\mathfrak{b}\bigl(1 + \mathcal{O}(N^{-1/4})\bigr) = \norm{\E\Bigl[ \other{G}_1 X \other{G}_2 \Bigr]}_\mathrm{hs} \le  \E\norm{ \other{G}_1 X \other{G}_2  }_\mathrm{hs} 
			\lesssim \E\sqrt{\frac{\bigl\langle \Im\other{G}_1 X \Im \other{G}_2 X^* \bigr\rangle}{(\Im z_1)(\Im z_2)}} \lesssim 1.
		\end{equation}
		Here in the last line we used the last bound form \eqref{eq:meta_ll} and  $\norm{X}_{\mathrm{hs}} \lesssim \norm{Y}_{\mathrm{hs}} = 1$.  This concludes the proof  of \eqref{eq:stab_apriori}.
	\end{proof}

	\section{Global law: Proof of Proposition \ref{prop:global}} \label{sec:glob}
	
The goal of this section is to prove Proposition \ref{prop:global}. Note that for the global law, i.e.~when the spectral spectral parameter are at order one distance to
$\mbox{supp}\;\rho$, there is no need to single out regular observables, every observable can be handled 
in the same way. We continue to use the letter $A$ for observables, but in this section 
this does not indicate regularity. 
Moreover, for brevity of notation, we use  $z_i$, $G_i$ and $\rho$  instead of $z_{i,0}$, $G_{i,0}$ and $\rho_0$
 in the entire Section~\ref{sec:glob} (except in Remark~\ref{rem:torsion}),
  i.e. we drop any reference to the fact that the spectral parameter
 is a time-zero initial condition of the characteristic flow. Note that this differs from the convention in
 all other sections where $z_i$ denoted $z_{i, T}$ at the final time. \nc

Proposition \ref{prop:global} is obtained as a special case of  Lemma \ref{lemma:2G_glob_underline}, whose proof is given in Section \ref{subsec:globalproof} below.
	\begin{lemma} \label{lemma:2G_glob_underline}
		Using the notations and assumptions of Proposition \ref{prop:global}, we have that  
		\begin{equation} \label{eq:2G_glob_im_bound}
			\bigl\lvert \bigl\langle \mathcal{G}_{1} A_1 \mathcal{G}_{2} A_2 - \mathcal{M}[A_1]A_2 \bigr\rangle \bigr\rvert  \prec  \frac{\prod_{\mathcal{G}_j = \Im G_{j}} |\rho(z_j)|  }{N} \left(\norm{A_1}\norm{A_2}_\mathrm{hs} \wedge \norm{A_1}_\mathrm{hs}\norm{A_2}\right) ,
		\end{equation} 
		for any choice of $\mathcal{G}_{j} \in \{G_{j}, G_{j}^*,\Im G_{j}\}$, where $\mathcal{M}[A_1]$ is the deterministic approximation to the chain $\mathcal{G}_{1} A_1 \mathcal{G}_{2}$.  The product is
		understood for all $j\in \{ 1, 2\}$ for which $\mathcal{G}_{j}$ is chosen to be $\Im G_{j}$. 
	\end{lemma}

	\begin{proof}[Proof of Proposition \ref{prop:global}]
		First, Lemma \ref{lemma:2G_glob_underline} for $\mathcal{G}_j = \Im G_{j}$ for $j \in \indset{2}$ immediately implies \eqref{eq:2G2Aglobal}. Second, \eqref{eq:2G1Aglobal} easily follows from Lemma \ref{lemma:2G_glob_underline} with $\mathcal{G}_j = G_{j}^{(*)}$ for $j \in \indset{2}$ and $A_2 = I$, for which $\Vert A_2 \Vert = 1$. 
	\end{proof}

	It thus remains to give the proof of Lemma \ref{lemma:2G_glob_underline}. However, before doing that, we record a few necessary preliminary statements. Throughout the rest of this section, whenever it does not lead to confusion, we omit the subscript $0$ from Proposition \ref{prop:global} for brevity.  
	
	\subsection{Preliminaries} 	First, we make use of the following standard technical lemma, which shows how prove bounds in $\prec$-sense using high moment bounds.  Its straightforward proof is omitted. 
	\begin{lemma}[Underline bound] \label{lemma:high-moment}
		Let $\chi$ be a random variable satisfying $|\chi| \le N^C$ for some large constant $C > 0$, and let $\varphi$ be a deterministic control parameter satisfying $\varphi \ge N^{-C}$. Assume that $\chi$ admits the decomposition
		\begin{equation}
			\chi = \underline{\chi} + \mathcal{E},
		\end{equation}
		where $\underline{\chi}$ and $\mathcal{E}$ are random variables, which satisfy $\lvert \mathcal{E} \rvert \prec \varphi$ and, denoting the complex conjugate of $\chi$ by $\overline{\chi}$, 
		\begin{equation} \label{eq:und_require}
			\bigl\lvert \E  \bigl[ \underline{\chi}\cdot  \overline{\chi} \lvert \chi\rvert^{2p-2} \bigr]  \bigr\rvert \le C_{\nu_1,\nu_2, p}\left(N^{\nu_1} \varphi^{2p} + N^{-\nu_2} \E \bigl[\lvert \chi\rvert^{2p}\bigr]\right), \qquad p \in \mathbb{N}
		\end{equation}
		for any arbitrarily small $\nu_1 > 0$ and some $\nu_2 > 0$. Then, $\lvert \chi \rvert \prec \varphi$.
	\end{lemma}

  	As the next ingredient, as shown in \cite[Theorem~2.9]{cuspuniv}, for any spectral parameter $z \in \globD$, the operator norm bounds
	\begin{equation} \label{eq:glob_norm_bounds}
		\norm{G(z)} \lesssim 1, \qquad \norm{\Im G(z)} \lesssim |\rho(z)|,
	\end{equation}
	hold with very high probability. They come from the fact that the eigenvalues of $H$ are very close 
	to the support of $\rho$ with very high probability, so $z \in \globD$ also means that $\mbox{dist}( \mbox{Spec}(H),
z ) \gtrsim 1$ with very high probability.

	Moreover, we employ the method of \emph{self energy renormalization}, denoted by \emph{underline}, in agreement with the notation used in Lemma~\ref{lemma:high-moment}: For any given function $f$ of $W = H - \E H$, we define
	\begin{equation} \label{eq:underlinedef}
		\underline{W f(W)} := W f(W) - \widetilde{\E} \widetilde{W} (\partial_{\widetilde{W}} f)(W) 
	\end{equation}
	where $\partial_{\widetilde{W}} f(W)$ denotes the directional derivative of the function $f$ in the direction $\widetilde{W}$ at the point $W$, and $\widetilde{W}$ is an independent copy of $W$. 
	The definition is chosen in such a way, that it subtracts the second order term in the cumulant expansion, in particular, for all entries of $W$ being Gaussian, we would have $\E \underline{W f(W)} = 0$.

\begin{remark}[Torsion component of the self-energy operator]\label{rem:torsion}
 In this remark, we recall that we establish the global law for any random matrix $H_0$ with data $(D_0, \mathcal{S}_0)$, constructed as the initial condition of the flow \eqref{eq:datapair_flow}. In the real symmetric case, a small discrepancy arises from the torsion term~\eqref{eq:torsion_def}, due to the fact that we use the flow associated with the \emph{complex} Ornstein–Uhlenbeck process even in the real case. This simplifies the overall argument but produces an irrelevant error, which we now explain. \nc
		In the sequel, we use \eqref{eq:underlinedef} for $W := W_0 =  H_0 -\E H_0$ and 
		$f(W) := G_{1,0}AG_{2,0}$, where  $A$ is some deterministic matrix and 
		 we recall that $G_{j,0} := G_0(z_{j,0})= (H_0-z_{j,0})^{-1}$ (the index 0 will again be dropped
		after this remark). \nc 
		For such $f$, the underline can be computed explicitly, yielding
		\begin{equation}
			\underline{W G_{1,0}AG_{2,0}} = WG_{1,0}AG_{2,0} + \mathcal{S}_{H_0} [G_{1,0} ]G_{1,0}AG_{2,0} + \mathcal{S}_{H_0} [G_{1,0}AG_{2,0} ]G_{2,0},	
		\end{equation}
		where $\mathcal{S}_{H_0}$ is the self-energy operator associated with the random matrix $H_0$ via \eqref{eq:S_def}.
		We remark that $\mathcal{S}_{H_0}$ differs from the super-operator $\mathcal{S}_0$, solving the data-pair flow \eqref{eq:datapair_flow} at time $t=0$, in the real-symmetric case. More precisely, we have 
		\begin{equation}
			\mathcal{S}_{H_0} = \mathcal{S}_0 + (1-\ee^{T})  \mathcal{T}_\mathrm{G},
		\end{equation} 
		where $\mathcal{T}_\mathrm{G}$ is defined in \eqref{eq:torsion_def}. However, it is straightforward to check that the contribution of the super-operator $\mathcal{T}_\mathrm{G}[\,\cdot\,]$ is negligible for the purposes of the global laws established below. Indeed, for any (random) matrices $X_1, X_2, X_3$, using Cauchy--Schwarz inequality for the trace, we obtain
		\begin{equation}
				\bigl\lvert \bigl\langle \mathcal{T}_\mathrm{G}[ X_1 ]X_2A_1X_3 A_2\bigr\rangle \bigr\rvert \vee \bigl\lvert \bigl\langle \mathcal{T}_\mathrm{G}[ X_1A_1X_2]X_3 A_2\bigr\rangle \bigr\rvert  \le N^{-1} \norm{X_1}\norm{X_2} \norm{X_3}\norm{A_1}_\mathrm{hs}\norm{A_2}_\mathrm{hs},
		\end{equation}
		which is sufficient in view of the norm bounds \eqref{eq:glob_norm_bounds}.
		
		A similar discrepancy arises between the solutions to the MDE with data-pairs $(\E H_0, \mathcal{S}_{H_0})$ with $\E H_0 = D_0$ and $(D_0, \mathcal{S}_0)$. Claim~\ref{claim:glob_torsion} below establishes that this difference is also negligible. Therefore, we ignore it in the sequel. 
	\end{remark}
	We defer the proof of the following Claim~\ref{claim:glob_torsion} to Section \ref{subsec:globaux}. 
	
	\begin{claim} \label{claim:glob_torsion}
		Let $\mathcal{S}_\gamma$ be a self-energy operator of the form \eqref{eq:S_gamma} with $0 \le \gamma \lesssim 1$ and $\mathcal{S}$ satisfying Assumptions~\ref{ass:S_norms}--\ref{ass:full} with some constant $c_\mathrm{full} > 0$. Let $\other{\mathcal{S}}_\gamma$ be a self-energy operator of the form
		\begin{equation}
			\other{\mathcal{S}}_\gamma := (1 - \gamma) \mathcal{S}_\mathrm{G} +  \gamma \, \mathcal{S} = \mathcal{S}_\gamma + (1-\gamma)\mathcal{T}_\mathrm{G} , 
		\end{equation}
		where $\mathcal{T}_\mathrm{G}$ is defined in \eqref{eq:torsion_def}. Let $M_\gamma(z)$ and $\other{M}_\gamma(z)$ be the solutions to the MDEs with data-pairs $(D, \mathcal{S}_\gamma)$ and $(D, \other{\mathcal{S}}_\gamma)$, respectively. 
		Then, for any $z \in \mathbb{C}$ satisfying $\dist(z,\supp\,\rho_\gamma) \sim 1$, we have 
		\begin{equation} \label{eq:Theta_bounds}
			\norm{M_\gamma(z) - \other{M}_\gamma(z)} \lesssim N^{-1}, \qquad \norm{\Im M_\gamma(z) - \Im \other{M}_\gamma(z)} \lesssim N^{-1}|\rho_\gamma(z)|. 
		\end{equation}
	\end{claim}

	Finally, by \cite[Theorem~2.8]{cuspuniv}, for any $z \in \globD$, we have the following single-resolvent global laws, 
	\begin{equation} \label{eq:glob_1G_ll}
		\bigl\lvert (G(z) -  M(z))_{\bm x\bm y}  \bigr\rvert \prec \frac{\norm{\bm x}\norm{\bm y}}{\sqrt{N}}, \qquad \bigl\lvert \bigl\langle(G(z) -  M(z)) B \bigr\rangle  \bigr\rvert \prec \frac{\norm{B}_\mathrm{hs}}{N},
	\end{equation}
	for any deterministic vectors $\bm x, \bm y \in \mathbb{C}^N$ and deterministic matrices $B \in \mathbb{C}^{N\times N}$. 
	In the proof of Lemma \ref{lemma:2G_glob_underline}, we also need the following improved global laws for $\Im G$, which we establish in Section~\ref{subsec:globaux} below. 
	
	\begin{lemma}[Global law for a single $\Im G$] \label{lemma:glob_ImG_laws}
		Using the above notations and conventions, we have that for $z \in \globD$, 
		\begin{equation} \label{eq:global_ImG_ll}
			\bigl\lvert (\Im G(z) -  \Im M(z))_{\bm x\bm y}  \bigr\rvert \prec \frac{|\rho(z)|}{\sqrt{N}}  \norm{\bm x}\norm{\bm y}, \qquad 
			\bigl\lvert \bigl\langle(\Im G(z) -  \Im M(z)) B \bigr\rangle  \bigr\rvert \prec \frac{|\rho(z)|}{N} \norm{B}_\mathrm{hs} ,
		\end{equation}
		uniformly in deterministic vectors $\bm x, \bm y \in \mathbb{C}^N$ and deterministic matrices $B \in \mathbb{C}^{N\times N}$. 
	\end{lemma}
	
	\subsection{Proof of Lemma~\ref{lemma:2G_glob_underline}} \label{subsec:globalproof}
As mentioned above, we omit the subscript $0$ throughout this proof for brevity. 
We begin by proving the slightly weaker (due to $\Vert A \Vert_{\rm hs} \le \Vert A \Vert \le N^{1/2} \Vert A \Vert_{\rm hs}$) bound 
		\begin{equation} \label{eq:2G_glob_im_bound_weak}
	\bigl\lvert \bigl\langle \mathcal{G}_{1} A_1 \mathcal{G}_{2} A_2 - \mathcal{M}[A_1]A_2 \bigr\rangle \bigr\rvert  \prec  \frac{\prod_{\mathcal{G}_j = \Im G_{j}} |\rho_0(z_j)|  }{\sqrt{N}} \norm{A_1}_{\mathrm{hs}}\norm{A_2}_\mathrm{hs} 
\end{equation} 
and argue for the strengthened version in \eqref{eq:2G_glob_im_bound} (for which we need to gain a factor $1/\sqrt{N}$ at the expense of replacing one Hilbert-Schmidt norm by an operator norm) after identifying the main mechanisms leading to \eqref{eq:2G_glob_im_bound_weak}.  We point out that we present the weakened  \eqref{eq:2G_glob_im_bound_weak} merely as a toy example to showcase the $\rho$-gaining effect. In particular, the underline terms in \eqref{eq:2G_glob_case1_und}, \eqref{eq:2G_glob_case2_und}, and \eqref{eq:2G_glob_case3_und} would allow us to immediately determine the $1/N$ bound. Some other terms, however, require a decomposition of the form $G = M+ (G
-M)$ and an \emph{isotropic law}, as we discuss around \eqref{eq:2G_glob_im_bound_iso}.

	Now, for all $w_1 \in \{z_{1}, \overline{z}_{1}\}, w_2 \in \{z_{2}, \overline{z}_{2}\}$, and any deterministic observable $A\in\mathbb{C}^{N\times N}$,
	\begin{equation} \label{eq:2G_underline}
		G(w_1) A G (w_2) - M(w_1, A, w_2)  =  \mathcal{B}_{w_1,w_2}^{-1}\bigl[ -M(w_1) \underline{WG(w_1) A G(w_2)} + M(w_1)\mathcal{E}(w_1, A, w_2) \bigr], 
	\end{equation}
	where $\mathcal{B}_{w_1, w_2}$ is defined in \eqref{eq:stab_t_def} with $t = 0$, and
	\begin{equation} \label{eq:Eerr1}
		\begin{split}					
			\mathcal{E}(w_1, A, w_2) :=&~ A \bigl(G(w_2) - M(w_2)\bigr) + \mathcal{S}\bigl[G(w_1) A G(w_2)\bigr]\bigl(G(w_2)-M(w_2)\bigr)\\
			&+ \mathcal{S}\bigl[G(w_1) - M(w_1)\bigr] G(w_1) A G(w_2).
		\end{split}
	\end{equation}
	
	For a fixed choice of $\mathcal{G}_j \in \{G_j, G_j^*, \Im G_j\}$ with $j \in \{1, 2\}$, let $	\mathfrak{i} \equiv 	\mathfrak{i}(\mathcal{G}_1, \mathcal{G}_2)$ denote the number of $\Im G$'s among $\mathcal{G}_j$'s, that is
	\begin{equation} \label{eq:idef}
		\mathfrak{i} := \sum_{\mathcal{G}_j = \Im G_{j,0}} 1.
	\end{equation}  
	We prove the desired bound \eqref{eq:2G_glob_im_bound_weak} by gradually increasing $	\mathfrak{i} \in \{0, 1, 2\}$. 
	
	\smallskip
	\textbf{Step 1}. Assume first that $	\mathfrak{i}=0$. Then, taking the normalized trace of \eqref{eq:2G_underline} with $A := A_1$ against another deterministic observable $A_2 \in\mathbb{C}^{N\times N}$, we obtain 
	\begin{equation} \label{eq:2G_glob_case1}
		\bigl\langle G(w_1) A_1 G (w_2) A_2 - M(w_1, A_1, w_2)A_2\bigr\rangle  =  -\Bigl\langle \underline{WG(w_1) A_1 G(w_2)} A_2'\Bigr\rangle + \bigl\langle\mathcal{E}(w_1, A_1, w_2)A_2' \bigr\rangle,
	\end{equation}		
	where $A_2' \equiv A_2'(w_1, w_2) := (\mathcal{B}_{w_1,w_2}^{-1})^{*}[A_2^*]\bigr)^*M(w_1)$.
	
	Using \eqref{eq:M_bound} and \eqref{eq:stab_t_bound}--\eqref{eq:betaf_t_def} at time $t=0$ with $\other{\beta}_0 \sim 1$, we deduce that
	\begin{equation} \label{eq:A_2'_bound}
		\norm{A_2'}_\mathrm{hs} \lesssim \norm{A_2}_\mathrm{hs}~. 
	\end{equation}
	Denoting $\chi := \bigl\langle G(w_1) A_1 G (w_2) A_2 - M(w_1, A_1, w_2)A_2\bigr\rangle$, as we show in Section \ref{subsec:globalcumex} (see \eqref{eq:globalcum}), the first term on the right-hand side of \eqref{eq:2G_glob_case1} satisfies
	\begin{equation} \label{eq:2G_glob_case1_und}
		\left|\E \left[ \Bigl\langle\underline{WG(w_1) A_1 G(w_2)} A_2'\Bigr\rangle  \overline{\chi} |\chi|^{2p-2}\right]\right| \lesssim  N^{\nu_1}\left(\frac{1 	}{\sqrt{N}} \norm{A_1}_{\mathrm{hs}}\norm{A_2}_\mathrm{hs} \right)^{2p} + N^{-\nu_2} \E \big[|\chi|^{2p}\big] 
	\end{equation}
	for arbitrarily small $\nu_1 >0 $ and some $\nu_2 > 0$.
	Moreover, it follows from \eqref{eq:glob_norm_bounds} that, with very high probability,
	\begin{equation} \label{eq:GAG_hs}
		\norm{G(w_1) A G(w_2)}_\mathrm{hs} \lesssim \norm{G(w_1)}\norm{G(w_2)}\norm{A}_\mathrm{hs} \lesssim \norm{A}_\mathrm{hs}.
	\end{equation}
	Hence, it follows from Claim~\ref{claim:S_bound} and \eqref{eq:glob_1G_ll} that the second term on the right-hand side of \eqref{eq:2G_glob_case1} admits the bound
	\begin{equation} \label{eq:2G_glob_case1_quad}
		\bigl\lvert \bigl\langle\mathcal{E}(w_1, A_1, w_2)A_2' \bigr\rangle\bigr\rvert \prec N^{-1}\norm{A_2'A_1}_\mathrm{hs} + N^{-1/2} \norm{A_1}_\mathrm{hs} \norm{A_2'}_\mathrm{hs} \lesssim N^{-1/2} \norm{A_1}_\mathrm{hs} \norm{A_2'}_\mathrm{hs} \,. 
	\end{equation}
	Combining \eqref{eq:2G_glob_case1}--\eqref{eq:2G_glob_case1_quad}, we conclude the desired \eqref{eq:2G_glob_im_bound_weak} for $	\mathfrak{i}=0$ by Lemma~\ref{lemma:high-moment}.			 
	
	\smallskip
	\textbf{Step 2}. Assume that $	\mathfrak{i}=1$. We only analyze the case $(\mathcal{G}_1, \mathcal{G}_2) = (\Im G_1, G_2)$ in full detail, since the other cases are treated completely analogously. 
	
	Let $G_j := G(z_j)$, and let $M_{12}$, $M_{\bar 12}$, and $M_{\widehat{1}2}$ denote the deterministic approximations to resolvent chains $G_1A_1G_2$, $G_1^*A_1G_2$, and $\Im G_1 A_1 G_2$, respectively. Then, subtracting two copies of \eqref{eq:2G_underline} with $(w_1,w_2) = (z_1, z_2)$ and $(w_1,w_2) = (\overline{z}_1, z_2)$, we obtain
	\begin{equation} \label{eq:2G_glob_case2_exp}
		\Im G_1 A_1 G_2 - M_{\widehat{1}2}  = -  \mathcal{B}_{12}^{-1}\bigl[ M_1 \underline{W\Im G_1 A_1 G_2}\bigr] 
		+ \mathcal{B}_{12}^{-1}\bigl[\Im M_1 (M_1^*)^{-1}(G_1^*A_1 G_2 - M_{\bar{1}2}) \bigr] 
		+ \mathcal{B}_{12}^{-1}\bigl[M_1\mathcal{E}_{\widehat{1}2} \bigr]  , 
	\end{equation}
	where $M_j := M(z_{j})$,  $\mathcal{B}_{12} := \mathcal{B}_{z_1, z_2}$ as in \eqref{eq:ind_conv}, and  the error therm  $\mathcal{E}_{\widehat{1}2}$ is given by
	\begin{equation} \label{eq:E1hat_def}
		\mathcal{E}_{\widehat{1}2}
		= \mathcal{S}[\Im G_1 A_1 G_2](G_2-M_2) + \mathcal{S}[G_1 - M_1] \Im G_1 A_1 G_2 + \mathcal{S}[\Im G_1 - \Im M_1] G_1^* A_1 G_2.
	\end{equation}
	Here we  used the identity
	\begin{equation}
		(2\ii)^{-1}(\mathcal{B}_{\bar{1}2} - \mathcal{B}_{12})[\,\cdot\,] = \Im M_1 \mathcal{S}[\,\cdot\,] M_2 = \Im M_1 (M_1^*)^{-1} (\mathrm{Id} - \mathcal{B}_{\bar{1}2})[\,\cdot\,].
	\end{equation}
	Therefore, taking the normalized trace of \eqref{eq:2G_glob_case2_exp} against $A_2$, we deduce that
	\begin{equation} \label{eq:2G_glob_case2}
		\bigl\langle \Im G_1 A_1 G_2A_2 - M_{\widehat{1}2}A_2 \bigr\rangle = - \Bigl\langle \underline{W\Im G_1 A_1 G_2}A_2' \Bigr\rangle 
		+ \bigl\langle (G_1^*A_1 G_2 - M_{\bar{1}2})A_2''  \bigr\rangle + \bigl\langle  \mathcal{E}_{\widehat{1}2} A_2' \bigr\rangle ,
	\end{equation}
	where $A_2' \equiv A_2(z_1, z_2)$ is defined below \eqref{eq:2G_glob_case1}, and $A_2'' := A_2' M_1^{-1} \Im M_1 (M_1^*)^{-1}$. 
	
	Denoting $\chi := \bigl\langle \Im G_1 A_1 G_2A_2 - M_{\widehat{1}2}A_2 \bigr\rangle$, as we show in \eqref{eq:globalcum} in Section \ref{subsec:globalcumex}, the underline term in \eqref{eq:2G_glob_case2} satisfies 
	\begin{equation} \label{eq:2G_glob_case2_und}
		\left|\E \left[ \Bigl\langle \underline{W\Im G_1 A_1 G_2}A_2' \Bigr\rangle  \overline{\chi} |\chi|^{2p-2}\right]\right| \lesssim  N^{\nu_1}\left(\frac{|\rho(z_1)|	}{\sqrt{N}  } \norm{A_1}_{\mathrm{hs}}\norm{A_2}_\mathrm{hs} \right)^{2p} + N^{-\nu_2} \E \big[|\chi|^{2p}\big] 
	\end{equation}
	for arbitrarily small $\nu_1 >0 $ and some $\nu_2 > 0$. 
	Furthermore, it follows from \eqref{eq:2G_glob_im_bound} with $	\mathfrak{i}=0$ established in Step 1 that the second term on the right-hand side of \eqref{eq:2G_glob_case2} admits the bound
	\begin{equation} \label{eq:2G_glob_case2_bound2}
		\bigl\lvert \bigl\langle (G_1^*A_1 G_2 - M_{\bar{1}2})A_2''  \bigr\rangle  \bigr\rvert \prec   N^{-1/2}\rho(z_1)  \norm{A_1}_\mathrm{hs} \norm{A_2}_\mathrm{hs} , 
	\end{equation}
	where we used \eqref{eq:invM_bound} and \eqref{eq:imM_bound} to estimate $\norm{A_2''}_\mathrm{hs} \lesssim \rho(z_1)\norm{A_2}_\mathrm{hs}$.  
	Therefore, it remains to estimate the contribution from the term containing $\mathcal{E}_{\widehat{1}2}$ on the right-hand side of \eqref{eq:2G_glob_case2}.  To this end, we use Claim~\ref{claim:S_bound} and the laws \eqref{eq:glob_1G_ll}--\eqref{eq:global_ImG_ll}, to deduce from \eqref{eq:E1hat_def}, similarly to \eqref{eq:2G_glob_case1_quad}, that
	\begin{equation}\label{eq:2G_glob_case2_bound3}
		\bigl\lvert\bigl\langle  \mathcal{E}_{\widehat{1}2} A_2' \bigr\rangle \bigr\rvert 
		\prec N^{-1/2}\norm{A_2'}_\mathrm{hs} \bigl(\norm{\Im G_1 A_1 G_2}_\mathrm{hs} + \rho(z_1) \norm{G_1 A_1 G_2}_\mathrm{hs} \bigr) \le 
		N^{-1/2}\rho(z_1) \norm{A_1}_\mathrm{hs} \norm{A_2}_\mathrm{hs} ,
	\end{equation}
	where, with a slight abuse of notation, we abbreviate $\rho(z)$ instead of $\rho(z) + N^{-C}$.\footnote{This irrelevant technical modification
		is necessary to satisfy the $\varphi\ge N^{-C}$ condition of Lemma~\ref{lemma:high-moment}. }  Here, in the second step we used \eqref{eq:A_2'_bound}, \eqref{eq:GAG_hs}, and the bound 
	\begin{equation} \label{eq:ImGAG_hs}
		\norm{\Im G_1 A G_2}_\mathrm{hs} \lesssim \norm{\Im G_1}\norm{G_2}\norm{A}_\mathrm{hs} \lesssim \rho(z_1) \norm{A}_\mathrm{hs}.
	\end{equation}
	Hence, the desired \eqref{eq:2G_glob_im_bound_weak} with $(\mathcal{G}_1, \mathcal{G}_2) = (\Im G_1, G_2)$ follows from \eqref{eq:2G_glob_case2}--\eqref{eq:2G_glob_case2_bound3} by Lemma~\ref{lemma:high-moment}. The other choices of $\mathcal{G}_j$ with $	\mathfrak{i}=1$ are treated analogously. 
	
	\smallskip
	\textbf{Step 3}. Finally, assume that $	\mathfrak{i}=2$, that is, $\mathcal{G}_1 = \Im G_{1}$ and $\mathcal{G}_2 = \Im G_2$. Note that an expansion analogous to \eqref{eq:2G_glob_case2_exp} holds with $z_2$ replaced by $\overline{z}_2$; Hence, by subtracting these expansions, we obtain 
	\begin{equation}
		\begin{split}
			\Im G_1 A_1 \Im G_2 - \widehat{M}_{12}  = &-  \mathcal{B}_{12}^{-1}\bigl[ M_1 \underline{W\Im G_1 A_1 \Im G_2}\bigr] + \mathcal{B}_{12}^{-1}\bigl[M_1 \widehat{\mathcal{E}}_{12}  \bigr] \\
			&+ \mathcal{B}_{12}^{-1}\bigl[\Im M_1 (M_1^*)^{-1}(G_1^* A_1 \Im G_2 - M_{\bar{1}\widehat{2}})  \bigr]
			+ \mathcal{B}_{12}^{-1}\bigl[M_1\mathcal{S}[\Im G_1 A_1 G_2^* - M_{\widehat{1}\bar{2}}]\Im M_2\bigr], 
		\end{split}
	\end{equation}
	where $M_{\bar{1}\widehat{2}}$, $M_{\widehat{1}\bar{2}}$, and $\widehat{M}_{12}$ are the deterministic approximations to $G_1^* A_1 \Im G_2$, $\Im G_1 A_1 G_2^*$, and $\Im G_1 A_1 \Im G_2$, respectively, and the error term $\widehat{\mathcal{E}}_{12}$ is given by
	\begin{equation} \label{eq:widehatE_def}
		\begin{split}
			\widehat{\mathcal{E}}_{12} := (2\ii)^{-1}(\mathcal{E}_{\widehat{1}2} - \mathcal{E}_{\widehat{1}\bar 2}) =  &~
			\mathcal{S}[\Im G_1 A_1 \Im G_2](G_2-M_2) + \mathcal{S}[\Im G_1 A_1 G_2^*](\Im G_2 -\Im M_2) \\
			& + \mathcal{S}[G_1 - M_1] \Im G_1 A_1 \Im G_2 + \mathcal{S}[\Im G_1 - \Im M_1] G_1^* A_1 \Im G_2.
		\end{split}
	\end{equation}
	Therefore, multiplying \eqref{eq:widehatE_def} by $A_2$ and taking the normalized trace, we obtain
	\begin{equation} \label{eq:2G_glob_case3}
		\begin{split}
			\bigl\langle \Im G_1 A_1 \Im G_2 A_2 - \widehat{M}_{12} A_2 \bigr\rangle  = &-  \Bigl\langle \underline{W\Im G_1 A_1 \Im G_2} A_2'\Bigr\rangle 
			+ \bigl\langle \widehat{\mathcal{E}}_{12}  A_2' \bigr\rangle \\
			&+ \bigl\langle (G_1^* A_1 \Im G_2 - M_{\bar{1}\widehat{2}})  A_2'' \bigr\rangle
			+ \bigl\langle (\Im G_1 A_1 G_2^* - M_{\widehat{1}\bar{2}})A_2''' \bigr\rangle,
		\end{split}
	\end{equation}
	where $A_2'$ and $A_2''$ are defined below \eqref{eq:2G_glob_case1} and \eqref{eq:2G_glob_case2}, respectively, and $A_2''' := \mathcal{S}[ (\Im M_2) A_2']$. 
	
	Denoting $\chi := \bigl\langle \Im G_1 A_1 \Im G_2 A_2 - \widehat{M}_{12} A_2 \bigr\rangle$, as we show in \eqref{eq:globalcum} in Section \ref{subsec:globalcumex}, the underline term in \eqref{eq:2G_glob_case3} satisfies 
	\begin{equation} \label{eq:2G_glob_case3_und}
		\left|\E \left[ \Bigl\langle \underline{W\Im G_1 A_1 \Im G_2} A_2'\Bigr\rangle   \overline{\chi} |\chi|^{2p-2}\right]\right| \lesssim  N^{\nu_1} \left(\frac{|\rho(z_1) \rho(z_2)|	}{\sqrt{N}} \norm{A_1}_{\mathrm{hs}}\norm{A_2}_\mathrm{hs} \right)^{2p} + N^{-\nu_2} \E \big[|\chi|^{2p}\big] 
	\end{equation}
	for arbitrarily small $\nu_1 >0 $ and some $\nu_2 > 0$. 
	
	We proceed to bound the term involving $\widehat{\mathcal{E}}_{12}$ on the right-hand side of \eqref{eq:2G_glob_case3}. Similarly to \eqref{eq:2G_glob_case1_quad} and \eqref{eq:2G_glob_case2_bound3}, using Claim~\ref{claim:S_bound} and \eqref{eq:glob_1G_ll}--\eqref{eq:global_ImG_ll}, we deduce from \eqref{eq:widehatE_def} that 
	\begin{equation} \label{eq:2G_glob_case3_bound1}
		\begin{split}
			\bigl\lvert \bigl\langle \widehat{\mathcal{E}}_{12}  A_2' \bigr\rangle \bigr\rvert &\prec N^{-1/2}\norm{A_2'}_\mathrm{hs}\bigl( \norm{\Im G_1 A_1 \Im G_2}_\mathrm{hs} + \norm{\Im G_1 A_1 G_2^*}_\mathrm{hs} |\rho(z_2)| + |\rho(z_1)| \, \norm{G_1^* A_1 \Im G_2}_\mathrm{hs}  \bigr)\\
			&\prec N^{-1/2}|\rho(z_1)\rho(z_2)| \,  \norm{A_1}_\mathrm{hs} \norm{A_2}_\mathrm{hs},
		\end{split}
	\end{equation}
	where we used \eqref{eq:glob_norm_bounds} to bound $\norm{\Im G_1 A_1 \Im G_2}_\mathrm{hs} \lesssim \rho(z_1) \rho(z_2) \norm{A_1}_\mathrm{hs}$ with very high probability. 
	We estimate both terms in the second line of \eqref{eq:2G_glob_case3} using \eqref{eq:2G_glob_im_bound} with $	\mathfrak{i} = 1$ established in Step 2 above, which yields
	\begin{equation} \label{eq:2G_glob_case3_bound2}
		\begin{split}
			\bigl\lvert \bigl\langle & (G_1^* A_1 \Im G_2 - M_{\bar{1}\widehat{2}})  A_2'' \bigr\rangle \bigr\rvert
			+ \bigl\lvert\bigl\langle (\Im G_1 A_1 G_2^* - M_{\widehat{1}\bar{2}})A_2''' \bigr\rangle \bigr\rvert \\
			&\prec  N^{-1/2}  \norm{A_1}_\mathrm{hs} \bigl(|\rho(z_2)| \, \norm{A_2''}_\mathrm{hs} + |\rho(z_1)| \, \norm{A_2'''}_\mathrm{hs}\bigr)  
			\prec N^{-1/2} |\rho(z_1)\rho(z_2)| \norm{A_1}_{\mathrm{hs}}\norm{A_2}_\mathrm{hs} , 
		\end{split}
	\end{equation}
	where  we used \eqref{eq:A_2'_bound}  and the bounds $\norm{A_2'''}_\mathrm{hs} \lesssim |\rho(z_2)| \, \norm{A_2}_\mathrm{hs}$ and $\norm{A_2'''} \lesssim |\rho(z_2)| \, \norm{A_2}$, which are obtained analogously. Thus, combining \eqref{eq:2G_glob_case3}--\eqref{eq:2G_glob_case3_bound2}, we conclude the desired \eqref{eq:2G_glob_im_bound_weak} for $	\mathfrak{i}=2$ by Lemma~\ref{lemma:high-moment}. 
	
	We now briefly argue, how \eqref{eq:2G_glob_im_bound_weak} can be improved to \eqref{eq:2G_glob_im_bound}. First, following the same basic steps as above, one can show the following isotropic law
		\begin{equation} \label{eq:2G_glob_im_bound_iso}
	\bigl\lvert \bigl\langle \bm x, \big(\mathcal{G}_{1} A_1 \mathcal{G}_{2}  - \mathcal{M}[A_1]\big) \bm y\bigr\rangle \bigr\rvert  \prec  \frac{\prod_{\mathcal{G}_j = \Im G_{j}} |\rho(z_j)|  }{\sqrt{N}} \norm{A_1} \Vert \bm x \Vert \, \Vert \bm y \Vert 
\end{equation} 
uniformly in deterministic vectors $\bm x, \bm y \in \C^N$. Note that the norm of $A_1$ got changed from the Hilbert-Schmidt norm to the operator norm. Moreover, also the deterministic approximation $\mathcal{M}[A_1]$ admits a corresponding bound
\begin{equation} \label{eq:2G_glob_im_M}
|\langle \bm x, \mathcal{M}[A_1] \bm y \rangle | \lesssim \left(\prod_{\mathcal{G}_j = \Im G_{j}} |\rho(z_j)|  \right) \norm{A_1} \Vert \bm x \Vert \, \Vert \bm y \Vert \,, 
\end{equation}
which easily follows by realizing that Claim \ref{claim:stab_rho_cancel} remains valid when replacing $\Vert \cdot \Vert _{\mathrm{hs} \to \mathrm{hs}}$ by $\Vert \cdot \Vert _{\Vert \cdot \Vert \to \Vert \cdot \Vert}$ (see \eqref{eq:S_hs_to_op}). To control the analogs of the error terms in \eqref{eq:2G_glob_case1_quad}, \eqref{eq:2G_glob_case2_bound3}, \eqref{eq:2G_glob_case3_bound1}, one has to introduce an \emph{isotropic norm}, see, e.g., \cite[Section 7.1]{cuspuniv}) on an inductively generated set of vectors: This norm, defined in \cite[Eq.~(7.6)]{cuspuniv} gradually takes into larger sets of vectors and thus allows to treat the $\mathcal{E}$-error terms in \eqref{eq:2G_glob_case1}, \eqref{eq:2G_glob_case2}, and \eqref{eq:2G_glob_case3}, similarly to the treatment in \cite[Eqs.~(7.17)--(7.38)]{cuspuniv}.  Since, however, the main mechanisms allowing to gain a $\rho$-factor associated with every $\Im G$ is the same as in the proof of \eqref{eq:2G_glob_im_bound_weak}, we leave the details to the reader for brevity of the paper. The analogs of \eqref{eq:2G_glob_case1_und}, \eqref{eq:2G_glob_case2_und}, and \eqref{eq:2G_glob_case3_und} with $\Vert A_1 \Vert_{\rm hs} \Vert A_2 \Vert_{\rm hs}$ replaced by $\norm{A_1} \Vert \bm x \Vert \, \Vert \bm y \Vert $ are shown to hold in \eqref{eq:globalcum} in Section \ref{subsec:globalcumex} below (note that $\Vert N \ket{\bm x} \bra{\bm y} \Vert_{\rm hs} \lesssim N^{1/2}$). 

After having established \eqref{eq:2G_glob_im_bound_iso}--\eqref{eq:2G_glob_im_M}, we turn to proving \eqref{eq:2G_glob_im_bound}: Assume w.l.o.g. that $\Vert A_1 \Vert \Vert A_2 \Vert_{\rm hs} \le \Vert A_1 \Vert_{\rm hs} \Vert A_2 \Vert $. Now we run the whole argument once again, but now decompose every resolvent product $\G_1 A_1 \G_2$ in the error terms \eqref{eq:Eerr1}, \eqref{eq:E1hat_def}, and \eqref{eq:widehatE_def} as
\begin{equation} \label{eq:decomp}
\G_1 A_1 \G_2 = \mathcal{M}[A_1] + \big(\G_1 A_1 \G_2 - \mathcal{M}[A_1]\big) \,. 
\end{equation}
This allows us to gain a full $1/N$-factor in the global law, by treating the two terms in \eqref{eq:decomp} separately. For example, the second term on the right-hand side~of \eqref{eq:widehatE_def} admits the bound
\begin{equation*}
	\begin{split}
&\big| \langle \mathcal{S}[\Im G_1 A_1 G_2^*](\Im G_2 -\Im M_2) A_2' \rangle \big| \\[1mm]
\le &\big| \langle \mathcal{S}[\mathcal{M}[A_1]](\Im G_2 -\Im M_2) A_2' \rangle \big| + \big| \langle \mathcal{S}[\Im G_1 A_1 G_2^* - \mathcal{M}[A_1]](\Im G_2 -\Im M_2) A_2' \rangle \big| \\[1mm]
\prec  &\frac{|\rho(z_2)|}{N} \Vert A_2' \mathcal{S}[\mathcal{M}[A_1]] \Vert_{\rm hs} + \frac{1}{N}\sum_j |\sigma_j| \big| \big(\mathcal{S}[\Im G_1 A_1 G_2^* - \mathcal{M}[A_1]](\Im G_2 -\Im M_2)\big)_{\bm u_j \bm v_j} \big| \\
\prec  & \frac{| \rho(z_2)|}{N} \Vert A_2' \Vert_{\rm hs} \Vert \mathcal{S}[\mathcal{M}[A_1]] \Vert + \frac{|\rho(z_1) \rho(z_2)|}{N} \Vert A_1 \Vert \Vert A_2' \Vert_{\rm hs} \lesssim \frac{|\rho(z_1) \rho(z_2)|}{N} \Vert A_1 \Vert \Vert A_2 \Vert_{\rm hs} \,. 
	\end{split}
\end{equation*}
Here, to go to the third line, we employed a polar decomposition $A_2' = \sum_j \sigma_j \ket{\bm v_j} \bra{\bm u_j}$ of $A_2'$, where $\sigma_j = \sigma_j(A_2')$ and $\bm u_j = \bm u_j(A_2')$, $\bm v_j = \bm v_j(A_2')$ are the singular values and corresponding left and right, respectively, normalized singular vectors of $A_2'$. To go to the last line, we used 
\begin{equation}
\big| \big(\mathcal{S}[\Im G_1 A_1 G_2^* - \mathcal{M}[A_1]](\Im G_2 -\Im M_2)\big)_{\bm u_j \bm v_j} \big| \prec \frac{|\rho(z_1) \rho(z_2)|}{N} \Vert A_1 \Vert
\end{equation}
for the second term. This follows from \eqref{eq:global_ImG_ll} and \eqref{eq:2G_glob_im_bound_iso} together with an estimate similar to \eqref{eq:iso_cross} below (see also \cite[Eq.~(7.44)]{cuspuniv}). In the final step, we then employed \eqref{eq:2G_glob_im_M} and \eqref{eq:A_2'_bound}.

	This concludes the proof of Lemma~\ref{lemma:2G_glob_underline}. \qed

	\subsection{Cumulant expansions: Proofs of \eqref{eq:2G_glob_case1_und}, \eqref{eq:2G_glob_case2_und}, and \eqref{eq:2G_glob_case3_und}}
	 \label{subsec:globalcumex}
	 Throughout the argument, we will frequently use the norm bounds \eqref{eq:glob_norm_bounds}. 
	 Moreover, to unify the proofs of \eqref{eq:2G_glob_case1_und}, \eqref{eq:2G_glob_case2_und}, and \eqref{eq:2G_glob_case3_und}, we adopt the notation 
	\begin{equation}
		\mathcal{G}_i \in \{G_i, G_i^*, \Im G_i\}
	\end{equation}
	 and recall from \eqref{eq:partial_ab} that, when differentiating $\G_1A\G_2$ the total number $\mathfrak{i} \in \{0,1,2\}$ of $\Im G$'s is preserved (in particular also the index). Moreover,  for any fixed configuration $(\G_1, \G_2)$,  we abbreviate (with a mild abuse of notation)
		\begin{equation*}
			|\rho|^\mathfrak{i} \equiv \prod_{\mathcal{G}_j = \Im G_{j}} |\rho(z_j)|. 
		\end{equation*}

We now employ a cumulant expansion as in Section \ref{sec:zagproof} (see, in particular, Section \ref{subsec:cumex}) to give the proofs of the bounds \eqref{eq:2G_glob_case1_und}, \eqref{eq:2G_glob_case2_und}, and \eqref{eq:2G_glob_case3_und}. In fact, we immediately show the following stronger estimate: For any arbitrarily small $\nu_1 > 0$ there exists $\nu_2 > 0$ such that
	\begin{equation} \label{eq:globalcum}
		\left| \E \left[ \Bigl\langle \underline{W \G_1 A_1 \G_2 A_2'}\Bigr\rangle \right] \overline{\chi} |\chi|^{2p-2} \right| \lesssim N^{\nu_1} \left(\frac{|\rho|^{\mathfrak{i}}}{N} \big(\Vert A_1 \Vert \, \Vert A_2 \Vert_{\rm hs} \wedge \Vert A_1 \Vert_{\rm hs} \, \Vert A_2 \Vert\big)\right)^{2p} + N^{-\nu_2} \E \big[ |\chi|^{2p} \big] \,. 
	\end{equation}

	In fact, by cumulant expansion, denoting $\chi  = \chi^{(\mathfrak{i})} := \langle \G_1 A_1 \G_2 A_2 \rangle - \langle \mathcal{M}[A_1]A_2 \rangle$ (recall the notation from Lemma \ref{lemma:2G_glob_underline}), we have that (recall the definition of $A_2'$ from below \eqref{eq:2G_glob_case1})
	\begin{equation} \label{eq:cumexpglob}
		\begin{split}
			\left| \E \left[ \Bigl\langle \underline{W \G_1 A_1 \G_2 A_2'}\Bigr\rangle  \overline{\chi} |\chi|^{2p-2} \right] \right| \le &\Big| \E \Big[ \frac{1}{N^2} \sum_{ab} \sum_{\alpha_1} \kappa(ab, \alpha_1) (\G_1 A_1 \G_2 A_2')_{ba} \partial_{\alpha_1} \left\{ \overline{\chi} |\chi|^{2p-2}\right\} \Big] \Big| \\
			& + \sum_{k=2}^L \Bigg| \E \Bigg[ \frac{1}{N} \sum_{ab} \sum_{\bm \alpha \in (\indset{N}^2)^k} \frac{\kappa(ab, \bm \alpha)}{N^{(k+1)/2} k!}   \partial_{\bm \alpha} \left\{ (\G_1 A_1 \G_2A_2')_{ba} \overline{\chi} |\chi|^{2p-2}\right\}\Bigg] \Bigg| + \big|\Omega_L^{(\mathfrak{i})}\big|
		\end{split}
	\end{equation}
	Here, as in the proof of the zag step (cf.~the discussion below \eqref{eq:cumexres}), $\Omega_L^{(\mathfrak{i})}$ is an error term satisfying $\big|\Omega_L^{(\mathfrak{i})}\big| \lesssim \big(|\rho|^{\mathfrak{i}}/{N}\big)^{2p}$ for large enough $L$.

	We now discuss the new term involving second-order cumulants, for which we estimate (ignoring the common $|\chi|^{2p-2}$ factor)
	\begin{equation} \label{eq:imgaimgaglobal}
		\begin{split}
			&N^{-3} \left|\sum_{\alpha_1, \alpha_2} \kappa(\alpha_1, \alpha_2) \big(\G_1 A_1 \G_2  A_2'\big)_{b_1 a_1} \big(G_1 A_1\G_2 A_2 \G_1\big)_{b_2 a_2}\right| \\
			\lesssim &N^{-3} \vertiii{\kappa}_2\left(\sum_{a_1, b_1}  \big|\big(\G_1 A_1 \G_2 A_2'\big)_{b_1 a_1}\big|^2 \right)^{1/2}  \left(\sum_{a_1, b_1}  \big|\big(G_1A_1\G_2 A_2 \G_1\big)_{b_2 a_2}\big|^2 \right)^{1/2} \\
			\lesssim &N^{-2} |\langle \G_1 A_1 \G_2 A_2'(A_2')^* \G_2^* A_1^* \G_1^* \rangle|^{1/2} \, |\langle G_1A_1 \G_2 A_2 \G_1 \G_1^* A_2^* \G_2^* A_1^* G_1^* \rangle|^{1/2} \\
			\lesssim &\frac{|\rho|^{2 \mathfrak{i}}}{N^2} \big(\Vert A_1 \Vert \, \Vert A_2 \Vert_{\rm hs} \wedge \Vert A_1 \Vert_{\rm hs} \, \Vert A_2 \Vert\big)^2\,. 
		\end{split}
	\end{equation}
	In the last step, we employed the norm bound $\Vert \G_i \Vert \lesssim |\rho_i|^{I(\G_i = \Im G_i)}$ and the fact that
	$\Vert A_2'\Vert_{\rm hs} \lesssim \Vert A_2 \Vert_{\rm hs} $ as well as $\Vert A_2'\Vert \lesssim \Vert A_2 \Vert $ (which is obtained exactly as \eqref{eq:A_2'_bound}). 
	
	All the higher-order cumulant terms can be handled analogously, following similar arguments as presented in Section~\ref{subsec:cumex} and additionally involving the estimate 
	\begin{equation} \label{eq:generalderivboundglob}
		\left|\left(\prod_{i=1}^{m} \partial_{a_i b_i}\right) \langle \G_1 A_1 \G_2 A_2 \rangle\right| \lesssim  \frac{|\rho|^{\mathrm{i}}}{N} \Vert A_1 \Vert \, \Vert A_2 \Vert \le \frac{|\rho|^{\mathrm{i}}}{\sqrt{N}} \big(\Vert A_1 \Vert \, \Vert A_2 \Vert_{\rm hs} \wedge \Vert A_1 \Vert_{\rm hs} \, \Vert A_2 \Vert\big) 
	\end{equation}
	for any $m \in \N_0$ and $(a_i, b_i) \in \indset{N}^2$, which easily follows by norm bounds on $\G$. We leave the details to the reader. 
	
	To summarize, by application of weighted Young inequalities, to separate additively  the 
	common $|\chi|^{2p-2}$ factors,  we obtain \eqref{eq:globalcum}. 
	This concludes the proofs of  \eqref{eq:2G_glob_case1_und}, \eqref{eq:2G_glob_case2_und}, and \eqref{eq:2G_glob_case3_und}, for $\mathfrak{i} = 0,1,2$, respectively. \qed

	\subsection{Proofs of Auxiliary Statements} \label{subsec:globaux}
	In this section, we collect the proofs of Lemma \ref{lemma:glob_ImG_laws} and Claim~\ref{claim:glob_torsion}. 
	
	\begin{proof} [Proof of Lemma~\ref{lemma:glob_ImG_laws}]
	We assume that $\Im z \ge  0$ to avoid having to take absolute values of $\rho$; the complementary case simply follows by taking the adjoint.	Recall that the resolvent $G$ satisfies the quadratic equation
		\begin{equation} \label{eq:1G_underline}
			G - M = -\mathcal{B}_z^{-1}\bigl[M\underline{WG}\bigr] + \mathcal{B}_z^{-1}\bigl[M\mathcal{S}[G-M](G-M)\bigr],
		\end{equation}
		where $\mathcal{B}_z := \mathcal{B}_{0, z_0, z_0}$ is defined in \eqref{eq:stab_t_def}. 
		Rewriting \eqref{eq:1G_underline} as
		\begin{equation} \label{eq:1G_underline2}
G - M = - M \underline{WG} + M \mathcal{S}[G-M] G
		\end{equation}
		where we used that $\mathcal{B}_z [\cdot] = \mathrm{Id}[\, \cdot \, ] - M \mathcal{S}[\, \cdot \, ] M$ and subtracting two copies of \eqref{eq:1G_underline2}, one with $z$ replaced by $\overline{z}$, we deduce that
		\begin{equation} \label{eq:1ImG_expand} 
			\Im G  - \Im M =  -\mathcal{B}_z^{-1}\bigl[M\underline{W\Im G}\bigr] + \mathcal{E}^\mathrm{quad} + \mathcal{E}^{\mathrm{Im}} \,. 
		\end{equation}
		Here, the error terms $\mathcal{E}^\mathrm{quad}$ and $\mathcal{E}^{\mathrm{Im}}$ are given by
		\begin{align}
			\label{eq:QuadErr}
			\mathcal{E}^\mathrm{quad} &:= \mathcal{B}_z^{-1}\bigl[M\mathcal{S}[\Im G- \Im M](G-M) + M\mathcal{S}[G^*-M^*](\Im G - \Im M)\bigr],\\
			\label{eq:ImErr}
			\mathcal{E}^{\mathrm{Im}} &:= \mathcal{B}_{z}^{-1}\bigl[\Im M(M^*)^{-1}(G^* - M^*) + M\mathcal{S}[G^* - M^*]\Im M \bigr].
		\end{align}

		For any matrix $Z \in\mathbb{C}^{N\times N}$, it follows from \eqref{eq:stab_beta_bound}, \eqref{eq:M_bound}, \eqref{eq:S_max_op}, and \eqref{eq:kappa2_norms}, that, for all $\bm x, \bm y \in \mathbb{C}^N$,  
		\begin{equation} \label{eq:B_iso_bound}
			\bigl\lvert \bigl(\mathcal{B}_z^{-1}[Z]\bigr)_{\bm x \bm y} \bigr\rvert \le \lvert Z_{\bm x \bm y} \rvert + \norm{\mathcal{B}_z^{-1}}\norm{M}^2 \norm{\mathcal{S}}_{\max \to \norm{\cdot}} \norm{Z}_{\max} \norm{\bm x}\norm{\bm y} \lesssim \lvert Z_{\bm x \bm y} \rvert + \norm{Z}_{\max} \norm{\bm x}\norm{\bm y}. 
		\end{equation}
		Hence, by \eqref{eq:glob_1G_ll} and using $\Vert \Im M \Vert \lesssim \rho$, the last term in \eqref{eq:1ImG_expand} admits the bounds
		\begin{equation} \label{eq:imErr_bounds}
			\bigl\lvert \bigl(\mathcal{E}^{\mathrm{Im}}\bigr)_{\bm x\bm y} \bigr\rvert \prec \frac{\rho }{\sqrt{N}}\norm{\bm x}\norm{\bm y}, \qquad \bigl\lvert \bigl\langle \mathcal{E}^{\mathrm{Im}} B\bigr\rangle \bigr\rvert \prec \frac{\rho }{N}\norm{B}_\mathrm{hs},
		\end{equation}
		for any deterministic vectors $\bm x, \bm y\in \mathbb{C}^N$ and deterministic matrix $B\in \mathbb{C}^{N\times N}$.
		
		Next, we estimate $\mathcal{E}^\mathrm{quad}$, defined in \eqref{eq:QuadErr}.   Assume that $X$ and $Z$ are random matrices satisfying
		\begin{equation} \label{eq:iso_norm}
			\bigl\lvert X_{\bm x\bm y}  \bigr\rvert \prec \Psi_X^\mathrm{iso} \norm{\bm x} \norm{\bm y}, \qquad \bigl\lvert Z_{\bm x\bm y}  \bigr\rvert \prec \Psi_Z^\mathrm{iso} \norm{\bm x} \norm{\bm y}, 
		\end{equation}		
		for any deterministic vectors $\bm x, \bm y \in \mathbb{C}^N$ and with some deterministic control parameters $\Psi_X^\mathrm{iso} $ and $\Psi_Z^\mathrm{iso} $. Then, it follows from \eqref{eq:cross_dir_norm_defs} that the cross and direct parts of $\mathcal{S}$ satisfy
		\begin{equation} \label{eq:iso_cross}
			\begin{split}
				\bigl\lvert \bigl(M\mathcal{S}_\mathrm{c}[X] Z\bigr)_{\bm x\bm y} \bigr\rvert 
				&\le \norm{X}_{\max} \sum_{ab} \bigl\lvert \bigl\langle \mathcal{S}_\mathrm{c}[\bm e_a \bm e_b^*]M^*\bm x,  Z\bm y\bigr\rangle \bigr\rvert \prec \Psi_X^\mathrm{iso} \Psi_Z^\mathrm{iso} \sum_{ab} \norm{\mathcal{S}_\mathrm{c}[\bm e_a \bm e_b^*]M^*\bm x}   \prec  \vertiii{\mathcal{S}_\mathrm{c}}_\mathrm{c} \Psi_X^\mathrm{iso} \Psi_Z^\mathrm{iso}, \\
				\bigl\lvert \bigl(M\mathcal{S}_\mathrm{d}[X] Z\bigr)_{\bm x\bm y} \bigr\rvert 
				&\le \sum_{aj} \Bigl\lvert\sum_{b}  X_{ab} \bigl(M\mathcal{S}[E^{ab}]\bigr)_{\bm xj} \Bigr\lvert   \bigl\lvert Z_{j\bm y} \bigr\rvert
				\prec \Psi_X^\mathrm{iso} \Psi_Z^\mathrm{iso}\sum_{aj}  \norm{\mathcal{S}\bigl[\bm e_j (M^*\bm x)^*\bigr]\bm e_a } \prec  \vertiii{\mathcal{S}_\mathrm{d}}_\mathrm{d} \Psi_X^\mathrm{iso} \Psi_Z^\mathrm{iso},
			\end{split}
		\end{equation}
		for any deterministic vectors $\bm x, \bm y \in \mathbb{C}^N$ with $\norm{\bm x} = \norm{\bm y} = 1$. 
		Therefore, by optimizing over the decomposition $\mathcal{S} = \mathcal{S}_\mathrm{c} + \mathcal{S}_\mathrm{d}$, we deduce from \eqref{eq:kappa2_norms} and \eqref{eq:iso_cross} that		
		\begin{equation} \label{eq:S_quad_bound}
			\bigl\lvert \bigl(M\mathcal{S}[X] Z\bigr)_{\bm x\bm y} \bigr\rvert
			\prec  \Psi_X^\mathrm{iso} \Psi_Z^\mathrm{iso} \norm{\bm x} \norm{\bm y}, 
		\end{equation}
		for all random matrices $X, Z$ satisfying \eqref{eq:iso_norm}, and all all deterministic $\bm x, \bm y \in \mathbb{C}^N$. 
		
		Assume that, for some deterministic control parameter $\psi \ge N^{-C}$ with some large constant $C >0$,  
		\begin{equation} \label{eq:imG_assume}
			\bigl\lvert \bigl(\Im G - \Im M\bigr)_{\bm x \bm y} \bigr\rvert \prec \psi \norm{\bm x} \norm{\bm y},
		\end{equation}
		for any deterministic  vectors $\bm x, \bm y \in \mathbb{C}^N$. Then, together with \eqref{eq:B_iso_bound}, \eqref{eq:glob_1G_ll} and \eqref{eq:S_quad_bound} imply that the error term $\mathcal{E}^\mathrm{quad}$ admits the bounds
		\begin{equation} \label{eq:quad_bound}
			\bigl\lvert \bigl(\mathcal{E}^\mathrm{quad} \bigr)_{\bm x \bm y} \bigr\rvert \prec N^{-1/2} \psi\norm{\bm x}\norm{\bm y}, \qquad \bigl\lvert \bigl\langle\mathcal{E}^\mathrm{quad} B \bigr\rangle \bigr\rvert \prec N^{-1/2} \psi \norm{B}_\mathrm{hs},
		\end{equation}
		for any deterministic vectors $\bm x, \bm y\in\mathbb{C}^N$ and deterministic matrix $B\in \mathbb{C}^N$.
		Here, to obtain the averaged bound, we employed the spectral decomposition $R = \sum_i \lambda_i(R) \bm v_i^R (\bm v_i^R)^*$ for any $R=R^* \in \mathbb{C}^{N\times N}$, to deduce that
		\begin{equation} \label{eq:pseudo_av_bound}
			\bigl\lvert \bigl\langle \mathcal{E}^\mathrm{quad} R\bigr\rangle \bigr\rvert \le  N^{-1} \sum_{i} |\lambda_i(R)| \bigl\lvert \bigl(\mathcal{E}^\mathrm{quad}\bigr)_{\bm v_i^R \bm v_i^R} \bigr\rvert \prec   N^{-1/2} \psi \langle |R| \rangle \prec N^{-1/2} \psi \norm{R}_\mathrm{hs},
		\end{equation}		
		from the isotropic bound in \eqref{eq:quad_bound}, and subsequently applied \eqref{eq:pseudo_av_bound} to self-adjoint matrices $R = \Re B$ and $R = \Im B$. 
		
		Therefore, it suffices to bound the underline term in \eqref{eq:1ImG_expand}. We proceed by minimalist cumulant expansion.
		First, we address the isotropic bound. In view of \eqref{eq:B_iso_bound}, it suffices to estimate 
		\begin{equation}
			\E\Bigl[ \bigl(M\underline{W\Im G}\bigr)_{\bm u\bm v} \overline{S} |S|^{2p-1}\Bigr], \qquad S \equiv S^{\bm x\bm y} := (\Im G - \Im M)_{\bm x \bm y}, \qquad \bm u, \bm v \in \{\bm x, \bm y \} \cup \{\bm e_j\}_{j=1}^N,
		\end{equation}
		for any fixed deterministic $\bm x, \bm y \in \mathbb{C}^N$ with $\norm{\bm x} = \norm{\bm y} = 1$, and all $p\in\mathbb{N}$. Crucially, for any $k \in \mathbb{N}\cup\{0\}$ and $\bm\alpha = (a_jb_j)_{j=1}^k \in \indset{N}^{2k}$, we have 
		\begin{equation}
			\partial_{\bm\alpha} (\Im G)_{ba} = (-1)^k\rho\times \sum_{i=1}^{k+1} \prod_{j=1}^{k+1} \mathcal{G}^{(j;i)}_{b_{j-1}a_j}, \qquad \mathcal{G}^{(j;i)} := \begin{cases}
				G, \quad &j < i,\\
				\rho^{-1}\Im G, \quad &j=i,\\
				G^*, \quad &j > i. 
			\end{cases}
		\end{equation}
		where we identify the indices $b_0 := b$ and $a_{k+1} := a$. It follows from \eqref{eq:glob_norm_bounds} that $\lVert\mathcal{G}^{(j;i)}\rVert \lesssim 1$ for all possible $i,j$. By inspecting the proof of Lemma~7.3 in  \cite{cuspuniv}, this suffices to conclude that, for all $p\in\mathbb{N}$ and any $\nu > 0$,
		\begin{equation} \label{eq:ImG_iso_und}
			\Bigl\lvert \E\Bigl[ \bigl(M\underline{W\Im G}\bigr)_{\bm u\bm v} \overline{S} |S|^{2p-1}\Bigr] \Bigr\rvert \le C_p\Bigl(N^{2\nu}\frac{\rho}{\sqrt{N}} \Bigr)^{2p} + N^{-\nu} \E \bigl[|S|^{2p}\bigr], \qquad\bm u, \bm v \in \{\bm x, \bm y \} \cup \{\bm e_j\}_{j=1}^N.
		\end{equation}
		
		Similarly, by inspecting the proof in Section 7.3 of \cite{cuspuniv} (we leave the details to the reader),  we deduce that 
		\begin{equation} \label{eq:ImG_av_und}
			\Bigl\lvert \E\Bigl[  \bigl\langle B'\underline{W\Im G}\bigr\rangle \overline{R} |R|^{2p-1}\Bigr] \Bigr\rvert \le C_p\Bigl(N^{2\nu}\frac{\rho + \sqrt{N}\psi}{N}\norm{B}_\mathrm{hs}\Bigr)^{2p} + N^{-\nu} \E \bigl[|R|^{2p}\bigr], \qquad R := \bigl\langle (\Im G- \Im M) B \bigr\rangle,
		\end{equation}
		where $B$ is a deterministic matrix, and $B' := (\mathcal{B}_{z}^{-1})^{*}[B]\bigr)^*M$.

		Therefore, combining \eqref{eq:imErr_bounds}, \eqref{eq:quad_bound}, and \eqref{eq:ImG_iso_und}--\eqref{eq:ImG_av_und} assuming \eqref{eq:imG_assume}, we obtain 
		\begin{equation} \label{eq:imG_iter}
			\bigl\lvert \bigl(\Im G - \Im M\bigr)_{\bm x\bm y}\bigr\rvert \prec \frac{\rho + \psi}{\sqrt{N}}\norm{\bm x}\norm{\bm y}, \qquad \bigl\lvert  \bigl\langle (\Im G- \Im M) B \bigr\rangle\bigr\rvert \prec \frac{\rho + \sqrt{N}\psi}{N}\norm{B}_\mathrm{hs},
		\end{equation}
		for any deterministic $\bm x, \bm y\in\mathbb{C}^N$ and deterministic matrix $B\in\mathbb{C}^{N\times N}$. Since \eqref{eq:imG_assume} is satisfies with $\psi := \rho$ by \eqref{eq:glob_norm_bounds}, the desired isotropic bound in \eqref{eq:global_ImG_ll} holds. Plugging $\psi := N^{-1/2}\rho$ into the second bound in \eqref{eq:imG_iter} yields the desired averaged bound in \eqref{eq:global_ImG_ll}. This concludes the proof of Lemma~\ref{lemma:glob_ImG_laws}.
	\end{proof}
	
	\begin{proof} [Proof of Claim~\ref{claim:glob_torsion}] 
		For brevity, we drop the subscript $\gamma$ from $\rho_\gamma$, $M_\gamma$, and $\mathcal{S}_\gamma$.
		Fix a spectral parameter $z \in \mathbb{C}$ satisfying $\dist(z, \supp\,\rho_\gamma) \gtrsim 1$ and $\Im z \ge  0$ to avoid having to take absolute values of $\rho$ (the complementary case simply follows by taking the adjoint), and denote
		\begin{equation}
			\Theta \equiv \Theta(z) := \other{M}(z) - M_\gamma(z), \qquad \Im \Theta = \Im \other{M}(z) - \Im M_\gamma(z).
		\end{equation}
		Since $\mathcal{T} := \other{\mathcal{S}}_\gamma - \mathcal{S}_\gamma = (1-\gamma) \mathcal{T}_\mathrm{G}$, with $\mathcal{T}_\mathrm{G}$ defined in \eqref{eq:torsion_def}, satisfies $\norm{\mathcal{T}} \le N^{-1}$, Proposition~10.1 of~\cite{AEK2020} implies that 
		\begin{equation} \label{eq:Theta_bound}
			\norm{\Theta} \lesssim N^{-1}.
		\end{equation}
		Hence, the first bound in the desired \eqref{eq:Theta_bounds} is established. 
		
		Furthermore, subtracting the matrix Dyson equations \eqref{eq:MDE} with data-pairs $(D, \mathcal{S}_\gamma)$ and $(D, \other{\mathcal{S}}_\gamma)$, we deduce that $\Theta$ satisfies the quadratic equation
		\begin{equation} \label{eq:Theta_eq}
			\mathcal{B}_z[\Theta] = M \mathcal{S}[\Theta]\Theta + M\mathcal{T}[M+\Theta](M+\Theta),
		\end{equation}
		where $\mathcal{B}_z[\,\cdot\,] := \mathrm{Id}[\,\cdot\,] - M \mathcal{S}[\,\cdot\,]M$, and $M \equiv M(z)$. Subtracting two copies of \eqref{eq:Theta_eq}, similarly to the argument around \eqref{eq:1G_underline}--\eqref{eq:1G_underline2}, one with $z$ replaced by $\overline{z}$, we obtain 
		\begin{equation}
			\begin{split}
				\Im \Theta =&~  \mathcal{B}_z^{-1}\bigl[ \Im M (M^*)^{-1}\Theta^* + M \mathcal{S}[\Theta^*] \Im M   +  \mathcal{R}_\mathcal{\mathcal{S}}[\Im \Theta, \Theta] 
				+ \mathcal{R}_\mathcal{T}[\Im M+\Im\Theta, M + \Theta]\bigr]
			\end{split}
		\end{equation}
		where for any super-operator $\mathcal{X}$, we define $\mathcal{R}_\mathcal{X}[X,Y]$ as
		\begin{equation}
			\mathcal{R}_\mathcal{X}[X,Y] := M \mathcal{X}[X]Y + M \mathcal{X}[Y^*]X, \qquad X, Y \in \mathbb{C}^{N\times N}. 
		\end{equation}
		Therefore, it follows from \eqref{eq:stat_M_bound}, \eqref{eq:invM_bound}, \eqref{eq:imM_bound}, \eqref{eq:Theta_bound}, that
		\begin{equation}
			\norm{\Im \Theta} \lesssim  N^{-1} \rho + N^{-1} \norm{\Im \Theta}, 
		\end{equation}
		where we used $\norm{\mathcal{B}_z^{-1}} \lesssim 1$  and $\norm{\mathcal{T}} \le N^{-1}$. We conclude that $\norm{\Im \Theta} \lesssim N^{-1} \rho$, and hence the second bound in \eqref{eq:Theta_bounds} is established.
		This completes the proof of Claim~\ref{claim:glob_torsion}. 
	\end{proof}

	\section{Local laws without regularity: Proof of Theorem \ref{thm:locallawnonreg}} \label{sec:proofnonreg}
	The goal of this section is to provide the proof of Theorem \ref{thm:locallawnonreg}, i.e.~prove the two-resolvent local laws for spectral parameters $z_1, z_2 \in \C\setminus \R$ in the regime 
	\begin{equation} \label{eq:regime}
		\min\{ \other{\beta}(z_1^+, z_2^-), \other{\beta}(z_2^+, z_1^-)\} \gtrsim 1
	\end{equation}
	i.e.~for $(z_1, z_2) \notin \regD$, and additionally satisfying $N \ell \ge N^\epsilon$ and $|z_j| \le C$ for $j = 1,2$ for some constants $\epsilon, C > 0$, as assumed in Theorem \ref{thm:locallawnonreg}. Recall that in this regime, the observable norm from \eqref{eq:tripplenorm} is replaced by the Hilbert-Schmidt norm. 
	
	In order to prove Theorem \ref{thm:locallawnonreg}, we again follow the zigzag strategy from Section \ref{sec:zigzag} and freely use the notation from there.
	 Additionally, we again suppose that $\mathcal{I} = \R$, i.e.~the set of admissible energies from Assumption \ref{ass:Mbdd} exhausts the entire real line; the extensions to general $\mathcal{I}$ are discussed in Remark \ref{rmk:Igeneral}.
	
	As the input of the zigzag proof, the global law can simply be imported from Proposition \ref{prop:global}. Next, recall that, as shown in Theorem \ref{prop:M2_bounds}, in the regime \eqref{eq:regime}, the $M$-bounds \eqref{eq:ImM2_nobeta}--\eqref{eq:M2_nobeta} hold for arbitrary observables, instead of only regular ones. 
Also the local law will be shown to hold for arbitrary observables.

	Throughout the zigzag steps, we fix two target spectral parameters $z_1, z_2 \in \abvD(\epsilon)$ with $(z_1, z_2) \notin \regD$ and arbitrary deterministic matrices $B_1, B_2 \in \C^{N \times N}$. The zigzag induction is conducted along the following two propositions.
	
	\begin{proposition}[Zig step; cf.~Proposition \ref{prop:zig}] \label{prop:zignoreg}
		Upon replacing $\reg{A}_{i,t} \to B_i$ and $\vertiii{A_i}_{t,\cdot, \cdot} \to \Vert B_i \Vert_{\rm hs}$     for $i \in \indset{2}$, Proposition \ref{prop:zig} holds verbatim. 
	\end{proposition}
	
	\begin{proposition}[Zag step; cf.~Proposition \ref{prop:zag}] \label{prop:zagnoreg}
		Upon replacing $\reg{A}_{i,t} \to B_i$ and $\vertiii{A_i}_{t_k,\cdot , \cdot} \to \Vert B_i \Vert_{\rm hs}$ for $i \in \indset{2}$, Proposition \ref{prop:zag} holds verbatim. 
	\end{proposition}
	
	We can now easily conclude the proof of Theorem \ref{thm:locallawnonreg}. 
	
	\begin{proof}[Proof of Theorem \ref{thm:locallawnonreg}]
		Based on Propositions \ref{prop:zignoreg}--\ref{prop:zagnoreg}, the proof is identical to that of Theorem \ref{thm:locallawreg} in Section \ref{sec:zigzag}.
	\end{proof}
	
	We conclude this section by briefly sketching the main differences in the proofs of Propositions \ref{prop:zignoreg} and \ref{prop:zagnoreg}, in comparison with Propositions \ref{prop:zig} and \ref{prop:zag}, respectively. 
	
	\begin{proof}[Proof of Proposition \ref{prop:zignoreg}]
		The argument is largely unchanged compared to the proof of Proposition \ref{prop:zig} in Section \ref{sec:zig_proof}.
		
		The key difference, and major simplification, is that due to the condition \eqref{eq:regime}, we have that, for all times 
		\begin{equation}
			\other{\alpha}_t \sim \other{\beta}_t \sim \other{\beta}_{*,t} \sim 1
		\end{equation}
		which, in particular, implies that all propagator estimates in Lemma \ref{lem:prop} are  trivial, i.e.~$f_r + \Phi_{s,t} \lesssim 1$, and thus the Gronwall arguments are trivial as well. As the second difference, there are no time-dependent corrections to the observables, i.e.~all $\tfrac{\dd}{\dd t} \Upsilon_t$ terms in Section \ref{sec:zig_proof} are set to zero. Modulo these small simplifying adoptions, the proof of Proposition \ref{prop:zignoreg} follows along the same lines as the one of Proposition \ref{prop:zig} and is so omitted. 
	\end{proof}
	
	\begin{proof}[Proof of Proposition \ref{prop:zagnoreg}]
		The proof of Proposition \ref{prop:zagnoreg} is identical to that of Proposition \ref{prop:zag} and so omitted. 
	\end{proof}

	\section{Single resolvent local law: Proof of Theorem \ref{thm:singleG}} \label{sec:proof1G}
	The goal of this section is to provide the proof of the single resolvent local law with regular observable in Theorem \ref{thm:singleG}. To do so, we follow the zigzag strategy summarized in Section \ref{sec:zigzag} and we freely use the notation from there. In particular, we again suppose that $\mathcal{I} = \R$, i.e.~the set of admissible energies from Assumption \ref{ass:Mbdd} exhausts the entire real line; the extensions to general $\mathcal{I}$ are discussed in Remark \ref{rmk:Igeneral}. Moreover, we point out that the argument presented here will be rather short, since we have the already established two-resolvent local laws from Theorem \ref{thm:locallawreg} as powerful inputs at our disposal. In particular, the zig step (Proposition \ref{prop:zig1}) can be conducted without introducing a stopping time.

	For the zigzag proof, we start with the global law, which directly follows from \eqref{eq:glob_1G_ll}. 
	
	\begin{proposition}[Global law; cf.~Proposition \ref{prop:global}]
Fix $\epsilon, \kappa > 0$ and take $z \in \globD(\epsilon, \kappa)$ (with $\rho \to \rho_0$ in \eqref{eq:Dglobdef}). Then the resolvent $G_0 := (H_{k=0} - z)^{-1}$ satisfies the bound
\begin{equation*}
\big| \big\langle \big( G_0(z) - M_0(z)\big) A \big\rangle  \big| \prec \frac{\Vert A \Vert_{\rm hs}}{N}
\end{equation*}
uniformly in deterministic $A \in \C^{N \times N}$. 
	\end{proposition}
	
	Next, we formulate the zig and zag step, fixing a target spectral parameter $z \in \abvD(\epsilon)$ satisfying $(z,z) \in \regD_{\beta_*}$ with $\beta_* \sim 1$ from Lemma \ref{lemma:stab_bound}. Recall also the time-dependent norm \eqref{eq:size_t_def}. 
	 	\begin{proposition}[Zig step; cf.~Proposition \ref{prop:zig}] \label{prop:zig1}
		Fix $k \in \{1, ... , K\}$ and denote 
		\begin{equation*}
			G_{t}(z_t) := \big(\mathfrak{F}_{\rm zig}^{t-t_{k-1}}\big[H_{k-1}\big] - z_{t}\big)^{-1}\,, \quad t_{k-1} \le t \le t_k \,. 
		\end{equation*}
		Assume that  $G_{t}$ satisfies the local law
		\begin{equation} \label{eq:1Gzig}
			\left| \big\langle \big(G_{t}(z_t) - {M}_t(z_{t})\big) A \big\rangle \right| \prec \frac{|\rho_t|^{1/2}}{N \eta_t^{1/2}} \vertiii{A}_{t,z, z}
		\end{equation}
	uniformly in $(z, z)$-regular matrices $A  \in \C^{N \times N}$ at time $t = t_{k-1}$. Then $G_{t}(z_t)$ satisfies the local law \eqref{eq:1Gzig} uniformly in $(z,z)$-regular matrices $A  \in \C^{N \times N}$ and uniformly in all times $t\in [ t_{k-1}, t_k]$. 
	\end{proposition}
	
	\begin{proposition}[Zag step; cf.~Proposition \ref{prop:zag}] \label{prop:zag1}
		Fix $k \in \{1, ... , K\}$ and denote $s_k := s(\dift_k)$ (cf.~Lemma \ref{lemma:OUsurj}). Let $H_k$ be the random matrix defined in \eqref{eq:H_k_defs}, and denote 
		\begin{equation*}
			G^s(z_{t_k}) := \big(\mathfrak{F}_{\rm zag}^{s}\big[H_{k}\big] - z_{t_k}\big)^{-1}\,, \quad 0 \le s \le s_k \,. 
		\end{equation*}
		Assume that $G^s$ satisfies the local law
		\begin{equation} \label{eq:1Gzag}
			\left| \big\langle \big(G^s(z_{t_k})  - {M}_{t_k}(z_{t_k})\big) A \big\rangle \right| \prec \frac{|\rho_{t_k}|^{1/2}}{N \eta_{t_k}^{1/2}} \vertiii{A}_{t_k,z, z}
		\end{equation}
 uniformly in $(z, z)$-regular matrices $A  \in \C^{N \times N}$ at time $s= s_k$. Then $G^s$ satisfies the local law \eqref{eq:1Gzag}	uniformly in $(z, z)$-regular matrices $A  \in \C^{N \times N}$ and uniformly in all times $s\in [0, s_k]$. 
	\end{proposition}
	One can now easily conclude the proof of Theorem \ref{thm:singleG}. 
\begin{proof}[Proof of Theorem \ref{thm:singleG}]
First, by Lemma \ref{lem:Dtrel}, we have that $	\abvD_0 (\epsilon, c, C) \subset \globD(\epsilon, \kappa)$. Hence, using that $|\rho_{0}|/\eta_0 \gtrsim 1$ for $z\in \abvD(\epsilon)$, we have that the assumptions of Proposition \ref{prop:zig1} are satisfied. We can thus apply Propositions \ref{prop:zig1} and~\ref{prop:zag1} in tandem $K$ times, to deduce Theorem \ref{thm:singleG}. 
\end{proof}	
	
It remains to give the proofs of Propositions \ref{prop:zig1}--\ref{prop:zag1}. Due to them being largely similar to the arguments given in Sections \ref{sec:zig_proof} and \ref{sec:zagproof}, respectively, we will be rather brief. 

\begin{proof}[Proof of Proposition \ref{prop:zig1}]
We follow the argument given in Section \ref{sec:zig_proof}, employing the same simplifying notations and assumptions, in particular writing $\reg{A}_t := A - \Upsilon_t I$ with 
\begin{equation*}
\Upsilon_t := \frac{\langle \widehat{M}_t(z, I, z)A \rangle}{\langle \widehat{M}_t(z, I, z)\rangle} \,. 
\end{equation*}
The flow equation, analogously to \eqref{eq:GAG}--\eqref{eq:imGAimGA}, takes the form
\begin{equation}
		\label{eq:GA}
		\dd \langle (G_{t} - M_t)\reg{A}_t \rangle 
		=\frac{1}{\sqrt{N}}\sum_{ab=1}^N \partial_{ab}\langle G_{t}\reg{A}_t\rangle (\dd \Brwn_t)_{ab}+\frac{1}{2}\langle (G_{t} -M_t)\reg{A}_t\rangle \dd t + \langle G_t - M_t \rangle  \big(\langle G_t^2 \reg{A}_t \rangle -  \tfrac{\dd}{\dd t}{\Upsilon}_t\big) \dd t \,. 
\end{equation}
We now realize that the second term on the right-hand side~can easily be absorbed by a simple multiplication by $\ee^{-t/2} = O(1)$ and shall thus henceforth be ignored. 

Next, we fix an arbitrarily small $\xi \in (0, \epsilon/10)$, recalling $N\ell_t\ge N^\epsilon$. Also, recall that, by \cite[Theorem~2.9]{cuspuniv} and from Theorem \ref{thm:locallawreg} and Lemma \ref{lem:Upsderv}, together with \eqref{eq:timerelations}, \eqref{eq:betaf_t_def}, and \eqref{eq:alp_t_def}, we have that 
\begin{equation} \label{eq:2Ginputs}  
\big| \langle G_t - M_t \rangle\big| \le \frac{N^{\xi/2}}{N \varkappa_t}\,, \quad \big| \langle G_t^2 \reg{A}_t \rangle - \tfrac{\dd}{\dd t}{\Upsilon}_t\big| \lesssim  \frac{|\rho_t|^{1/2}}{\eta_t^{1/2}} \frac{\other{\beta}_t}{\other{\alpha}_t}\vertiii{A}_{t,z,z}\,,  \quad \big| \langle \Im G_t \reg{A}_t \Im G_t \reg{A}_t^* \rangle \big| \lesssim  |\rho_t|^2 \vertiii{A}_{t,z,z}^2
\end{equation}
with very high probability, uniformly in $t \in [\tin, \tfin]$. 
Armed with \eqref{eq:2Ginputs} and bearing in mind the monotonicity of the observable norm \eqref{eq:tripmonotone}, we now control the first and third term in~\eqref{eq:GA}. 

For the first term,  by the BDG inequality (see Appendix B.6, Eq.~(18) in \cite{martingale}) and application of \eqref{eq:intrule1}--\eqref{eq:tripmonotone} together with \eqref{eq:2Ginputs}, we have, with very high probability,
\begin{equation*} 
	\begin{split}
		\sup_{u \in [\tin, t]}  \left|\sum_{a,b=1}^N\int_{\tin}^{u}\partial_{ab} \langle  G_{s}\reg{A}_s \rangle\frac{(\dd \Brwn_s)_{ab}}{\sqrt{N}}\right| 
		\le \, N^{\xi/2} \left(\int_{\tin}^{t} \frac{1}{N^2\eta_{s}^2} |\langle \Im G_{s}\reg{A}_s\Im G_{s}\reg{A}_s^*\rangle| \,\dd s\right)^{1/2} 
		\lesssim \, 	 N^{\xi/2} \frac{|\rho_{t}|^{1/2}}{N\eta_{t}^{1/2}} \vertiii{A}_{t, z_1, z_2} \,.
	\end{split}
\end{equation*}
Next, we turn to the third term in \eqref{eq:GA}, for which, again by application of \eqref{eq:intrule1}--\eqref{eq:tripmonotone} together with \eqref{eq:2Ginputs}, we have the bound 
\begin{equation*}
\left|\int_{\tin}^t \langle G_s - M_s \rangle \big(\langle G_s^2 \reg{A}_s \rangle - \tfrac{\dd}{\dd s}\Upsilon_s\big) \dd s \right| \lesssim N^{\xi/2} \int_{\tin}^t \frac{1}{N \varkappa_s} \frac{|\rho_s|^{1/2}}{\eta_s^{1/2}} \frac{\other{\beta}_s}{\other{\alpha}_s}\vertiii{A}_{s, z, z}\dd s \lesssim N^{\xi/2}  \frac{|\rho_t|^{1/2}}{N \eta_t^{1/2}} \vertiii{A}_{t, z,z} \,, 
\end{equation*}
where in the last step we used that $\varkappa_s \sim \other{\beta}_s \eta_s/(\other{\alpha}_s |\rho_s|)$, as a consequence of \eqref{eq:gamma} for $z_1 = z_2$, and \eqref{eq:intrule1}. 

Collecting the above bounds and arguing analogously to the proof of Proposition \ref{prop:zig} in Section \ref{sec:zig_proof}, we conclude that, uniformly in $t \in [\tin, \tfin]$,
$$
\big| \langle (G_t - M_t) \reg{A}_t \rangle \big| \le N^{\xi} \frac{|\rho_t|^{1/2}}{N \eta_t^{1/2}} \vertiii{A}_{t, z,z}
$$
 and thus, by arbitrariness of $\xi$, the proof of Proposition \ref{prop:zig1}. 
\end{proof}
	
We are now left with proving Proposition \ref{prop:zag1}. 
	
\begin{proof}[Proof of Proposition \ref{prop:zag1}] We follow the argument presented in Section \ref{sec:zagproof} and conduct the proof of Proposition \ref{prop:zag1} by establishing the Gronwall estimate (neglecting the $t_k$-subscript as it remains fixed and writing $A \equiv \mathring{A}_{t_k}$ for short)
	\begin{equation} \label{eq:1gGron}
		\begin{split}
&\frac{\dd }{\dd s} \E \big| \langle (G^s(z) - M(z)) A \rangle \big|^{p} \lesssim \left(1 + N^{-\delta} \frac{|\rho|}{\eta}\right) \left[ \E \big| \langle (G^s(z) - M(z)) A \rangle \big|^{p} + \left(N^{3 \delta} \frac{|\rho|}{N \ell^{1/2}} \vertiii{A}_{z,z}\right)^p  \right]  \\
&\text{for} \quad s \in [0, s_{\rm final}] \quad \text{with} \quad s_{\rm final} := s(\dift_k)\,, 
		\end{split}
	\end{equation}
	for any large (even) $p \in \N$ and arbitrarily small $\delta > 0$, 	exactly as in Propositions \ref{prop:gronwallImG}--\ref{prop:gronwallG}. The proof is then concluded using Gronwall's lemma and arbitrariness of $p, \delta$ (cf.~the argument in Section \ref{sec:zagproof}).

	To establish \eqref{eq:1gGron}, we rely on a cumulant expansion exactly as in the proofs of Propositions \ref{prop:gronwallImG}--\ref{prop:gronwallG}, see in particular \eqref{eq:GFTstart}--\eqref{eq:cumexres}. Here, for brevity, we focus on presenting the estimates for one representative low-order example term with $k=3, n=1$ (using the nomenclature of Section \ref{subsec:cumex}: third order cumulant term, derivatives acting on one copy of $G$), as this term requires  the use of the $\vertiii{\kappa}_{3}^{\rm av}$-norm from \eqref{eq:kappa_3_av_norm}. A general abstract argument encompassing all other terms, which we presented in Section \ref{subsubsec:zagav2G}, can easily be adopted to the present setting; we leave the details to the reader.  
	
	For the $k=3, n=1$ term, we consider
	\begin{equation} \label{eq:k3n1GAG11}
		\begin{split}
			N^{-5/2} \sum_{\alpha_1, \alpha_2 ,\alpha_3} \left| \kappa(\alpha_1, \alpha_2, \alpha_3)  \big(G AG\big)_{b_3 a_1} G_{b_1 a_2} G_{b_2 a_3}\right| 
		\end{split}
	\end{equation}
	and, analogously to \eqref{eq:k3n1GAG11} and the estimate below, decompose $G_{b_1 a_2}$ and  $G_{b_2 a_3}$ as $G = M + (G-M)$. In particular using $\Vert M \Vert \lesssim 1$ and the isotropic single resolvent law \eqref{eq:singleGzag}, we have that, for any $\delta > 0$, 
	\begin{equation*} 
		\begin{split}
			&N^{-5/2} \sum_{\alpha_1, \alpha_2 ,\alpha_3} \left| \kappa(\alpha_1, \alpha_2, \alpha_3)  \big(GAG\big)_{b_3 a_1} (G-M)_{b_1 a_2} M_{b_2 a_3}\right| \\
			\lesssim &N^{-3 + \delta} \frac{|\rho|^{1/2} }{\eta} \sum_{\alpha_1, \alpha_2 ,\alpha_3} \left| \kappa(\alpha_1, \alpha_2, \alpha_3)  \big(\big(GA\Im G A^* G^*\big)_{b_3 b_3}\big)^{1/2} \right| \\
			\lesssim & \frac{|\rho|}{\eta} N^\delta\frac{ |\langle \Im GA \Im GA^* \rangle|^{1/2}}{N (\eta|\rho|)^{1/2} } \lesssim \left(1 + N^{-\delta} \frac{|\rho|}{\eta} \right)N^{3\delta} \frac{ |\rho|^{1/2}}{N \eta^{1/2} }  \vertiii{A}_{z,z}\,, 
		\end{split}
	\end{equation*}
	where, in the last step, we used the already established two resolvent bound $\langle \Im GA \Im GA^* \rangle \lesssim |\rho|^2 \vertiii{A}_{z,z}^2$. 
	The terms with $M_{b_1 a_2} (G-M)_{b_2 a_3}$ and $(G-M)_{b_1 a_2} (G-M)_{b_2 a_3}$ are treated analogously and we are hence left with the $M_{b_1 a_2} M_{b_2 a_3}$ term. Here, involving the $\vertiii{\kappa}_3^{\rm av}$ norm, we estimate
	\begin{equation*} 
		\begin{split}
			N^{-5/2} \sum_{\alpha_1, \alpha_2 ,\alpha_3} \left| \kappa(\alpha_1, \alpha_2, \alpha_3)  \big(GAG\big)_{b_3 a_1} M_{b_1 a_2} M_{b_2 a_3}\right| 
			\le \,  N^{-1} \vertiii{\kappa}_3^{\rm av} \Vert M \Vert^2 \Vert GAG\Vert_{\rm hs} \lesssim \left(1 + N^{-\delta} \frac{|\rho|}{\eta} \right)N^{3\delta} \frac{ |\rho|^{1/2}}{N \eta^{1/2} } \,, 
		\end{split}
	\end{equation*}
	again using the two resolvent bound $\langle \Im GA \Im GA^* \rangle \lesssim |\rho|^2 \vertiii{A}_{z,z}^2$. As mentioned above, all other terms arising in the cumulant expansion can be handled similarly, following the general power counting argument presented in Section \ref{subsubsec:zagav2G} and eventually leading to \eqref{eq:1gGron}. 
	
	This concludes the proof of Proposition \ref{prop:zag1}. 
\end{proof}

\appendix 

\section{Improved local laws outside of the support} \label{app:outside}

The goal of this section is to (briefly) explain how the zigzag proofs of the local laws have to be adjusted in order to obtain the improved bounds outside of the spectrum, as discussed in Section \ref{subsub:outside}. Like in Section \ref{sec:zigzag}, we shall henceforth assume that $\mathcal{I} = \R$ for simplicity of the presentation.

The first key ingredient to deducing the improved local laws with spectral parameters outside of the support of the scDoS is that the resolvent $G(z)$  and its imaginary part $\Im G(z)$ obey the following norm bounds with very high probability
\begin{equation} \label{eq:ImGnorm}
\Vert G(z) \Vert \lesssim \frac{1}{\kappa}\,, \qquad \Vert \Im G(z) \Vert \lesssim \frac{\eta}{\varkappa^2} \quad \text{with} \quad \eta := |\Im z|\,, \quad \varkappa := \dist(z, \supp \rho) 
\end{equation}
whenever $\widetilde{\ell} := |\rho(z)| \varkappa^2 /\eta$ satisfies $N \widetilde{\ell} \ge N^\epsilon$ for a fixed $\epsilon > 0$. These can easily be obtained from exclusion of spectrum outside of the support of the scDoS (see \cite[Theorem 2.9]{cuspuniv}) and spectral decomposition. 

The second key ingredient to the proof is the following integration rule \eqref{eq:kappaint}, replacing the ones \eqref{eq:intrule1}.  The proof of Lemma \ref{lem:kappaint} is given at the end of this section. 
\begin{lemma}[$\varkappa$-integration rule] \label{lem:kappaint}
 Consider a (target) spectral parameter $z= z_T \in \C \setminus \R$ and denote $\eta_t := |\Im z_t|$ and $\varkappa_t := \dist(z_t, \supp \rho_t)$ for any $t \in[0,T] $. Then, in the regime where $\varkappa_t \ge 10 \eta_t$ for all $t \in [0,T] $, we have that for any $s,t \in [0,T]$ with $s \le t$ it holds that
\begin{equation} \label{eq:kappaint}
	\int_s^t \dd r \, \frac{\eta_r^q}{\varkappa_r^p} \lesssim  \frac{\eta_t}{|\rho_t|}\frac{\eta_t^q}{\varkappa_t^p}  \quad \text{for any} \quad p,q \ge 0 \quad \text{with} \quad p > \frac{2q+2}{3} \,. 
\end{equation}
\end{lemma}
Note that the condition $\varkappa_t \ge 10 \eta_t$ ensures that $\Re z_t \notin \supp \rho_t$, i.e.~the spectral parameter is outside of the support. This can be guaranteed, e.g., by requiring that $\varkappa_T \ge C \eta_T$ for some constant $C > 10$. 

Armed with \eqref{eq:ImGnorm}--\eqref{eq:kappaint}, we now explain how the local laws in Section \ref{subsub:outside} can be obtained. In the discussion, we first focus on controlling the quadratic variation of the martingale terms arising in the zig step of the proof, as these terms "dictate" the best possible error terms. 

We start by illustrating the mechanism for the $\langle \Im G \rangle$ local law as a special case of \eqref{eq:generalImG}: For simplicity of the presentation, here and in the following, we just integrate from time $t=0$ as the resulting bounds on the quadratic variations are unaffected. Following the argument in \cite[Section~4]{cuspuniv}, for $\widetilde{\ell}_t := |\rho_t| \varkappa_t^2 /\eta_t \ge N^{-1+\epsilon}$ for some fixed $\epsilon > 0$ and all $t \in [0,T]$, we define the stopping time 
\begin{equation} \label{eq:stoptime1}
\tau:= \inf\left\{ t \in [0,T] \, : \,  |\langle \Im G_t - \Im M_t\rangle| \ge N^{\xi} \frac{|\rho_t|}{N \widetilde{\ell}_t}\right\} \qquad \text{for some} \quad \xi \le \epsilon/10 \,. 
\end{equation}
Then, by means of a Ward identity, \eqref{eq:ImGnorm}--\eqref{eq:kappaint}, and the definition of the stopping time \eqref{eq:stoptime1}, the quadratic variation of the martingale term (cf.~\cite[Eq.~(4.5)]{cuspuniv}) admits the bound
\begin{equation*}
\int_0^{t \wedge \tau} \dd s \, \frac{\langle |\Im G_s|^3 \rangle}{N^2 \eta_s} \lesssim \int_0^{t \wedge \tau} \dd s \, \frac{\eta_s}{N^2 \varkappa_s^4} |\langle \Im G_s \rangle|\lesssim \int_0^{t \wedge \tau} \dd s \, \frac{|\rho_s|\eta_s}{N^2 \varkappa_s^4}  \lesssim \left(\frac{|\rho_{t \wedge \tau}|}{N \widetilde{\ell}_{t \wedge \tau}}\right)^2 \,. 
\end{equation*}
By means of the BDG inequality, this allows to conclude the proof with the improved bound. 

Next, we consider the quadratic variation of the martingale term in the flow for $\langle \Im GA \Im GA \rangle$ in  \eqref{eq:imGAimGA}. Defining a suitably modified stopping time $\tau$ which encapsulates the desired improved local law and denoting $\varkappa_{i,s} := \dist(z_{i,s}, \supp \rho_s)$ and $\varkappa_s := \min_{i \in \indset{2}} \varkappa_{i,s}$, we find it to be bounded by (cf.~\eqref{eq:QVimgaimga} for the analogous term without $\widetilde{\ell}$-improvement)
\begin{equation*}  
	\begin{split}
&\int_0^{t \wedge \tau} \dd s \, \left(\frac{1}{N^2\varkappa_{1,s}^2} + \frac{1}{N^2\varkappa_{2,s}^2}\right)\langle \Im G_{1,s} \reg{A}_{1,s} \Im G_{2,s} \reg{A}_{2,s} \Im G_{1,s} \reg{A}_{2,s}^* \Im G_{2,s} \reg{A}_{1,s}^* \rangle \\
 \le &\int_0^{t \wedge \tau} \dd s \, \left(\frac{1}{N\varkappa_{1,s}^2} + \frac{1}{N\varkappa_{2,s}^2}\right)\langle \Im G_{1,s} \reg{A}_{1,s} \Im G_{2,s} \reg{A}_{1,s}^* \rangle \langle  \Im G_{1,s} \reg{A}_{2,s}^* \Im G_{2,s} \reg{A}_{2,s} \rangle \\
 \lesssim & \int_0^{t \wedge \tau} \dd s \, \left(\frac{|\rho_{1,s} \rho_{2,s}|^2}{N\varkappa_{1,s}^2} + \frac{|\rho_{1,s} \rho_{2,s}|^2}{N\varkappa_{2,s}^2}\right) \vertiii{A_1}_{s, z_1, z_2}^2 \, \vertiii{A_2}_{s, z_2, z_1}^2 \\
 \lesssim &\frac{|\rho_{1,t\wedge \tau} \rho_{2,t \wedge \tau}|^2}{N \widetilde{\ell}_{t \wedge \tau}} \vertiii{A_1}_{t\wedge \tau, z_1, z_2}^2 \, \vertiii{A_2}_{t \wedge \tau, z_2, z_1}^2
	\end{split}
\end{equation*}
additionally using \eqref{eq:tripmonotone}. Again, by means of the BDG inequality, this lies the basis for concluding the argument with the improved bound. The quadratic variations of $\langle GAG \rangle$ in \eqref{eq:GAG} and $\langle GA \rangle$ as well as $\langle \Im GA \rangle$ can be handled in a similar way, solely relying on the two ingredients \eqref{eq:ImGnorm} and \eqref{eq:kappaint}. We leave the details to the reader. 

Besides improving the bounds on the quadratic variations, there are certain terms in the evolutions of \eqref{eq:imGimG} and \eqref{eq:imGAimGA}, which require more care in their analysis. In fact, the $\Im$-ignoring overestimates  employed in \eqref{eq:rho/ell},  \eqref{eq:toobad}, \eqref{eq:toobad2}, \eqref{eq:toobad3}, and \eqref{eq:G-MtimesM}  are not affordable for the desired improved bound with $\ell \to \widetilde{\ell}$. To overcome this, we need to additionally track the local laws 
\begin{equation} \label{eq:newout}
\big| \big\langle \Im G_{1,t} G_{2,t} \big\rangle - (2 \ii)^{-1}\big\langle M_t(z_1, I, z_2) - M_t(\overline{z}_1, I, z_2) \big\rangle \big| \prec \frac{|\rho_1|}{N \widetilde{\ell}_t} \left( \frac{|\rho_{1,t} |}{\eta_{1,t} } + \frac{|\rho_{2,t}|}{\eta_{2,t}} \right)^{1/2}
\end{equation}
and
\begin{equation} \label{eq:newout2}
	\big| \big\langle \Im G_{1,t} \reg{A}_{1,t} G_{2,t} \big\rangle - (2 \ii)^{-1}\big\langle M_t(z_1, \reg{A}_{1,t}, z_2) - M_t(\overline{z}_1, \reg{A}_{1,t}, z_2) \big\rangle \big| \prec \frac{|\rho_1|}{N \widetilde{\ell}_t } \left( \frac{|\rho_{1,t} \rho_{2,t}|}{\eta_{1,t} \eta_{2,t}} \right)^{1/4} \vertiii{A_1}_{t, z_1, z_2}
\end{equation}
and similarly for $1 \leftrightarrow 2$ and/or $G_2 \to G_2^*$, by appropriate stopping times in the zig step of the proof. This follows along the same lines as our main characteristic flow proof in Section \ref{sec:zig_proof} with straightforward modifications. 

Finally, the zag step of our proof now has to track also \eqref{eq:newout}--\eqref{eq:newout2}. This is done along the same lines as in our main proof in Section \ref{sec:zagproof}; we leave the details to the reader. 

\medskip

It thus remains to give the proof of Lemma \ref{lem:kappaint}. 

\begin{proof}[Proof of Lemma \ref{lem:kappaint}]
We begin with recalling some notations and definitions from \cite[Section 6.1]{cuspuniv}; see also \cite{AEK2020}. We say that $\mathfrak{e}_{t}^\pm \in \R$ are right (left) endpoints of a gap in the of $\rho_t$ if and only if $\mathfrak{e}_{t}^{\pm} \in \partial \{x \in \R : \rho_t(x) > 0\} $ and $\rho_t(x) = 0$ for all $x \in [\mathfrak{e}_{t}^-, \mathfrak{e}_{t}^+]$. The \emph{gap size} is defined as $\Delta_t := \mathfrak{e}_{t}^+ - \mathfrak{e}_{t}^-$. In this proof, we focus on considering internal gaps and thus internal edges; the case of extremal edges is much simpler and so omitted. 

In the case of internal gaps, we have that 
\begin{equation}
|\rho_t| = |\rho_t(z_t)| \sim \frac{\eta_t}{\varkappa_t^{1/2} (\Delta_t + \varkappa_t)^{1/6}}\,, \quad \text{with} \quad \varkappa_t := \dist(z_t, \mathfrak{e}_t^\pm)
\end{equation}
as well as the time asymptotics (see \cite[Lemma 6.2]{cuspuniv})
\begin{equation*}
\Delta_s \sim \Delta_t + (t-s)^{3/2}\,, \qquad 
\eta_s \sim \eta_t + \rho_t (t-s) \,, \qquad \rho_s \sim \rho_t \quad \text{for all} \quad s , t \in [0,T] \quad \text{with} \quad s \le t \,. 
\end{equation*}
This implies that, since $\varkappa_s \lesssim \Delta_s$ for all $s \in [0,T]$, we have
\begin{equation} \label{eq:asymp}
\varkappa_s^{1/2} \Delta_s^{1/6} \sim \frac{\eta_s}{|\rho_s|} \qquad \text{and thus} \qquad \varkappa_s \sim \varkappa_t + \frac{(t-s)^2}{\Delta_t^{1/3} + (t-s)^{1/2}} \quad \text{for all} \quad  s , t \in [0,T] \quad \text{with} \quad s \le t \,. 
\end{equation}
Then we have that 
\begin{equation*}
	\int_s^t \dd r \, \frac{\eta_r^q}{\varkappa_r^p} \le 	\int_0^t \dd r \, \frac{\eta_r^q}{\varkappa_r^p} \sim 	\int_0^t \dd s \, \frac{\eta_t^q + \rho_t^q s^q}{\varkappa_t^p + \frac{s^{2p}}{\Delta_t^{p/3} + s^{p/2}}} \,. 
\end{equation*}
Splitting the integration regime in two parts, $[0,t] = [0, t \wedge \Delta_t^{2/3}] \cup [t \wedge \Delta_t^{2/3}, t]$ and using that $p > (2q+2)/3$ by assumption together with the first asymptotics in \eqref{eq:asymp}, we can easily deduce that 
\begin{equation*}
\int_0^t \dd s \, \frac{\eta_t^q + \rho_t^q s^q}{\varkappa_t^p + \frac{s^{2p}}{\Delta_t^{p/3} + s^{p/2}}} \sim \frac{\eta_t}{|\rho_t|}\frac{\eta_t^q}{\varkappa_t^p} \,. 
\end{equation*}
This concludes the proof of Lemma \ref{lem:kappaint}. 
\end{proof}

		\section{Optimality of the $M$-bounds} \label{app:Moptimal}
	The estimate \eqref{eq:ImM2_bounds} is optimal in the sense that it saturates for observables~$A_2^* = A_1 = A$, where $A$ is regular. Note that in this case $\langle \Im G_1 A_1 \Im G_2 A_2 \rangle$ is explicitly positive. 
	
	We illustrate this saturation in a simplified setting of \emph{deformed Wigner matrices} $H := D + \mathcal{W}$, where $\mathcal{W}$ is an $N\times N$ standard Wigner and $D = D^* \in \mathbb{C}^{N\times N}$ is a deterministic deformation satisfying $\norm{D} \lesssim 1$. 
	We choose this model since the corresponding $\widehat{M}(z_1, A_1, z_2)$ can be computed directly, and by appropriately choosing the deformation $D$ it can be made to exhibit any kind of singularity in the spectrum.
	To further simplify the presentation, we analyze the case $z_1 = z_2$ in detail, but analogous considerations apply for a general arrangement of spectral parameters. 
	Therefore, out goal is to calculate 
	\begin{equation}
		\widehat{m}(A) := \Bigl\langle \widehat{M}(z, \reg{A}, z) \reg{A}^* \Bigr\rangle, \qquad \reg{A} := A - \Upsilon(A), \qquad \Upsilon(A) := \frac{\bigl\langle \widehat{M}(z, I, z) A  \bigr\rangle}{\bigl\langle \widehat{M}(z, I, z) \bigr\rangle}, 
	\end{equation}
	for all  $(z,\overline{z})$-pre-regular $A\in\mathbb{C}^{N\times N}$, that is, $A$ satisfying $(\Im z + \langle \Im M(z) \rangle)^{-1}\langle A \Im M(z) \rangle = \langle A |M(z)|^2 \rangle = 0$. 	
	The explicit computation below also serves as a blueprint for our proof of the $M$-bounds in the general correlated setting, presented in Section~\ref{sec:Mboundsproof}.  
	
	For all $z \in \mathbb{H}$, $M(z)$ is in the spectral resolution of $D = D^*$ and hence $T := \Re M(z)$ and $S := \rho(z)^{-1}\Im M(z)$ commute.
	Furthermore, with $M^{(*)}$ denoting either $M \equiv M(z)$ or $M^*$, the stability operators $\mathcal{B}_{z,z^{(*)}}[\cdot] = \mathrm{Id}[\,\cdot\,] - MM^{(*)}\langle\,\cdot\,\rangle$ can be inverted explicitly,
	$$
	\bigl(\mathrm{Id}[\,\cdot\,] - MM^{(*)}\langle\,\cdot\,\rangle\bigr)^{-1}[X] = X + \frac{\langle X\rangle MM^{(*)}}{1 - \langle MM^{(*)}\rangle}. 
	$$
	Let $\beta\equiv\beta_{z,z} = 1 -\langle M^2 \rangle$ and $\theta := \beta_{z,\bar z} = 1 -\langle |M|^2 \rangle$ denote the smallest (in modulus) eigenvalues of  $\mathrm{Id}  - M^2\langle\,\cdot\,\rangle$ and $\mathrm{Id}[\,\cdot\,] - MM^{*}\langle\,\cdot\,\rangle$, respectively. Then, we have
	\begin{equation}
		\Re \beta = \theta + 2 \rho^2 \langle S^2  \rangle, \qquad \Im \beta = -2\rho \langle TS  \rangle,
	\end{equation}
	where we abbreviate $\rho := \rho(z)$. 
	Crucially, near a regular edge the typical size of $\Re \beta$ is of order $\rho^{-1}\eta + \rho$, where $\eta:= \Im z$, while $\Re \beta - \theta$ is of the order $\rho^2$,  which is much smaller. This cancellation is directly responsible for the size of the overlap fluctuations near the regular edge. 
	To recover this effect, we use \eqref{eq:M2_def}, \eqref{eq:ImM2_def}, the linearity of $A \mapsto \widehat{M}(z,A,z)$, the pre-regularity of $A$, and complete the square, arriving at the explicit expression
	\begin{equation} \label{m_lambda}
		\widehat{m}(A) =  
		\frac{\rho^2}{\alpha^2}\Bigl\lvert 
		\Bigl\langle  S^{1/2} \mathcal{P}\bigl[T \bigr] S^{1/2} A \Bigr\rangle \Bigr\rvert^2 
		+ \rho^2 \norm{\mathcal{P}\bigl[S^{1/2} A S^{1/2}\bigr]}_\mathrm{hs}^2 \,. 
	\end{equation} 
	Here, $\mathcal{P}$ is the orthoprojector onto the orthogonal complement of the span of $S$, that is, $\mathcal{P}[X] := X - \langle S^2 \rangle^{-1}\langle S X \rangle S$, and the quantity $\alpha^2$, with the square emphasizing its explicit positivity, is given by
	\begin{equation}
		\alpha^2 := \frac{|\beta|^2 - \theta\Re\beta}{4\rho^2\langle S^2  \rangle}   = \frac{\theta}{2} + \rho^2 \langle S^2\rangle + \frac{\langle TS  \rangle^2}{\langle S^2  \rangle}.
	\end{equation}
	Recall that $(1-\theta)S = \pi |M|^2$; hence, that in the regime where $\rho \le \rho_*$ for some small $\rho_*\sim 1$, we have, by \eqref{eq:sigma_def}, 
	\begin{equation} \label{eq:TS}
		\pi^2\langle TS \rangle = (1-\theta)^2 \sigma + \mathcal{O}(\rho^2).
	\end{equation}
	\nc 
	Therefore, it is straightforward to check, using $S\sim 1$, and \eqref{eq:TS} above, \nc that $\alpha^2$ satisfies
	\begin{equation}
		\alpha^2 \sim \rho^{-1}\eta + \rho^2 + |\sigma|^2 \sim |\sigma|^2 + |\beta| \sim \other{\alpha}(z,z)^2\,, 
	\end{equation}
	where we recall the definitions of $\sigma := \sigma(z)$ and $\other{\alpha}(z,z)$ from \eqref{eq:sigma_def} and \eqref{eq:alpha_defin}, respectively. 
	Moreover, since $\langle AS \rangle = 0$ by pre-regularity of $A$, and $S \sim I$ in the sense of quadratic forms, we have 
	\begin{equation}
		\norm{A}_\mathrm{hs}^2 \gtrsim \norm{\mathcal{P}\bigl[S^{1/2} A S^{1/2}\bigr]}_\mathrm{hs}^2 \ge \frac{\langle S \rangle^2}{\langle S \rangle^2 + \langle S^2 \rangle}\norm{S^{1/2} A S^{1/2}}_\mathrm{hs}^2  \gtrsim \norm{A}_\mathrm{hs}^2~.
	\end{equation}
	Since $|\langle TS\rangle| \lesssim |\sigma| + \rho^2 \lesssim \alpha$ by \eqref{eq:TS}, we conclude that 
	\begin{equation}
		\widehat{m}(A) \sim \frac{\rho^2}{\alpha^2}\bigl\lvert 
		\bigl\langle  T S A \bigr\rangle \bigr\rvert^2 
		+ \rho^2 \norm{A}_\mathrm{hs}^2~.
	\end{equation}
	In particular, this yields the optimality of the $M$-bound \eqref{eq:ImM2_bounds}, since 
	\begin{equation}
		\bigl\langle{\vecL} A\bigr\rangle = \frac{\other{\beta}(z,z)}{\rho}\bigl\langle M' A\bigr\rangle = \frac{C}{\rho}\bigl\langle M^2 A\bigr\rangle  = 2\ii C \langle TS A \rangle  + \mathcal{O}(\rho)\norm{A}_\mathrm{hs}, \qquad C := \frac{\other{\beta}(z,z)\langle M' \rangle}{\langle M^2\rangle}
	\end{equation}
	where the coefficient $C$ satisfies $|C|\sim 1$. \nc

\section{Proofs of Technical Statements} \label{app:technicals}

\subsection{Shape Analysis. Proof of Lemmas~\ref{lemma:m_hat}, and Claim~\ref{claim:alpha_kappa}} \label{app:shape}

In the following lemma we collect various estimates on the
self-consistent density of states $\rho$ from~\cite{AEK2020} that we will use
in the proof of Lemma~\ref{lemma:m_hat}.

\begin{lemma}[Shape Analysis for the Self-Consistent Density of States \cite{AEK2020}] \label{lemma:shape}
	Let $M$ be the solution to the matrix Dyson equation \eqref{eq:MDE} with data-pair $(D,\mathcal{S})$, where the self-energy $\mathcal{S}$ satisfies the flatness bounds \eqref{eq:flatness}. 
	
	Let $\mathbb{I} \subset \mathbb{R}$ be a compact interval, and 
	$\widehat{C}, \widehat{c} \sim 1$ a positive constants. Assume that the solution $M$ satisfies
	\begin{equation} \label{eq:Mbdd_cond}
		\norm{M(E+\ii\eta)} \le \widehat{C}, 
		\qquad \text{for all}\qquad  \dist(E,\mathbb{I}) \le \widehat{c}, \quad \eta > 0. 
	\end{equation} 
	Then, the self-consistent density of states $\rho := \pi^{-1}\langle \Im M \rangle$ and its boundary value $\rho(E) := \lim\limits_{\eta\to +0} \rho(E+\ii\eta)$ satisfy the following properties:
	\begin{itemize}
		\item[(a)] \emph{Structure of singularities and small local minima} \cite[Theorem~7.2(ii)]{AEK2020}. 
		For any constant $c > 0$, let $\mathbb{M}_c$ denote the set of small local minima of $\rho$ in the vicinity of $\mathbb{I}$,  that is, 
		\begin{equation} \label{eq:sing_set}
			\mathbb{M}_c \equiv \mathbb{M}_c(\rho) := \{ E \in \supp\,\rho\,:\, \dist(E, \mathbb{I}) \le \tfrac{3}{4}\widehat{c},\,\, \rho(E) \le c,\,\, E \text{ is a local minimum of }\rho\rvert_{\supp\,\rho} \},
		\end{equation} 
		Note that the boundary of the support $(\del\supp\,\rho)\cap\bddD \subset \mathbb{M}_c$ for all $c \ge 0$. Then, we have
		\begin{itemize}
			\item[(i)] There exists a threshold $\rho_*\sim 1$ such that the set $\mathbb{M}_{c}$ consists of $\mathcal{O}(1)$ points for any $c \le \rho_*$.
			\item[(ii)] Distinct points $E_1, E_2 \in \mathbb{M}_{\rho_*}$, $E_1 < E_2$, which belong to the boundary of the support $E_1, E_2 \in \del \supp\, \rho$, satisfy
			\begin{equation} \label{eq:boundary_isol}
				E_2 - E_1 \sim 1, \quad \text{or}\quad (E_1, E_2)\cap\supp\,\rho = \emptyset.
			\end{equation}
			That is, edge-points in $\mathbb{M}_{\rho_*}$ are isolated unless they are the end-points of the same gap in $\supp\,\rho$.
			
			\item[(iii)] Distinct points in $\mathbb{M}_{\rho_*}$, which belong to the interior of $\supp\, \rho$, are isolated from each-other and from the edge-points, that is, 
			for all $E_1 \in \mathbb{M}_{\rho_*} \backslash (\mathbb{M}_{\rho_*})$ and $E_2 \in \mathbb{M}_{\rho_*}$ satisfying $E_2 \neq E_1$, we have
			\begin{equation} \label{eq:smallmin_isol}
				|E_2 - E_1| \sim 1. 
			\end{equation}
		\end{itemize}
		
		\item[(b)] \emph{Scaling relations for the density} (\cite[Remark~7.3]{AEK2020}). 
		For all $E \in \mathbb{R}$, let $\mathfrak{e}(E)$  and $\delta(E)$ denote the closest point in $\mathbb{M}_{\rho_*}$ to $E$, and the corresponding distance, that is 
		\begin{equation} \label{eq:closest_sing}
			\mathfrak{e}(E) := \min\bigl\{ E' \in \mathbb{M}_{\rho_*} \,:\, \dist(E,\mathbb{M}_{\rho_*}) = |E-E'| \bigr\}, \qquad \delta(E) := \dist(E,\mathbb{M}_{\rho_*}) = |E-\mathfrak{e}(E)|.
		\end{equation}
		For any $E\in \mathbb{R}$, the local gap size $\Delta(E)$ around the energy $E$ is given by\footnote{Note that in \cite{AEK2020} the local gap size   may be   unbounded, but for the purposes of our analysis it is convenient to introduce an order one cut-off by adding $\wedge 1$ to the definition.}
		\begin{equation} \label{eq:gap_size}
			\Delta(E) := \sup\bigl\{ b-a \,:\, a \le E \le b,\, \text{ and }\, (a,b) \cap \supp\,\rho = \emptyset  \bigr\} \wedge 1. 
		\end{equation}
		In particular, $\Delta(E) = 0$ for all $E$ satisfying $\rho(E) > 0$. Then, for all $E\in \mathbb{I}$ and $0 \le \eta \lesssim 1$, the density $\rho(E+\ii\eta)$ satisfies
		\begin{equation} \label{eq:rho_comp}
			\rho(E+\ii\eta) \sim \begin{cases}
				\rho (\mathfrak{e} ) +  \bigl(\delta(E)+\eta\bigr)^{1/2}\bigl(\Delta(\mathfrak{e})+\delta(E) + \eta\bigr)^{-1/6},\quad&E \in \supp\,\rho,\\
				\eta\bigl(\delta(E)+\eta\bigr)^{-1/2}\bigl(\Delta(\mathfrak{e})+\delta(E) + \eta\bigr)^{-1/6},\quad&E \notin \supp\,\rho,
			\end{cases}
		\end{equation}
		where $\mathfrak{e} \equiv \mathfrak{e}(E)$.
		
		\item[(c)] \emph{Scaling relations for the stability coefficients} (\cite[Remark~10.4 and Eq.~(10.5a--d)]{AEK2020}). 
		For  all $z := E +\ii\eta$ satisfying $E \in \overline{\bddD}\cap\mathbb{R}$ and $0 \le \eta \lesssim 1$: 
		\begin{itemize}
			\item[(i)] The stability factor $\other{\beta}(z,z)$ (essentially equivalent to the coefficient $\other{\xi}_1$ from \cite[Remark~10.4]{AEK2020}) satisfies
			\begin{equation} \label{eq:beta1_shape}
				\other{\beta}(E+\ii\eta,E+\ii\eta)   \sim \rho(\mathfrak{e})^2 +  \bigl(\delta(E)+\eta\bigr)^{1/2}\bigl(\Delta(\mathfrak{e})+\delta(E) + \eta\bigr)^{1/6}.
			\end{equation}
			
			\item[(ii)] If we assume additionally that $E \in \supp\,\rho$ or $\delta(E) \le c_*\Delta(E)$ for some sufficiently small constant $c_*\sim1$, then (by scaling of the coefficient $\other{\xi}_2$ from \cite[Remark~10.4]{AEK2020})
			\begin{equation} \label{eq:rhosigma_shape_in}
				\rho(E+\ii\eta) + |\sigma(E+\ii\eta)| \sim \rho(\mathfrak{e}) + \bigl(\Delta(\mathfrak{e})+\delta(E) + \eta\bigr)^{1/3},
			\end{equation}
			while in the complementary regime $E \notin \supp\,\rho$ and $c_*\Delta(E) \le \delta(E) \lesssim 1$, we only have the upper bound 
			\begin{equation} \label{eq:rhosigma_shape_out}
				\rho(E+\ii\eta) + |\sigma(E+\ii\eta)| \lesssim \bigl(\Delta(\mathfrak{e})+\delta(E) + \eta\bigr)^{1/3} \lesssim \other{\beta}(E+\ii\eta, E+\ii\eta)^{1/2}.  
			\end{equation} 
			
		\end{itemize} 
	\end{itemize} 
\end{lemma}

Throughout the proof of Lemma~\ref{lemma:m_hat} we use the following technical claim to estimate the integral of Cauchy kernels.
\begin{claim} \label{claim:Cauchy_int}
	Let $0 < \eta_1 \le \eta_2 \le \dots \le \eta_n$ be an increasing sequence of scales, and let $x_* > 0$ be a positive number. 
	Assume that $g : (0, x_*] \to \mathbb{R}_{> 0}$ is a positive non-decreasing function satisfying
	\begin{equation} \label{eq:g_cond}
		\frac{g(x)}{g(y)} \le C \Bigl(\frac{x}{y}\Bigr)^{\gamma}, \qquad 0 < y < x < x_*,
	\end{equation}
	for some positive constant $0 \le \gamma < 1$ and a (large) constant $C > 1$. Then, for any constant $g_0 \ge 0$, 
	\begin{equation} \label{eq:g_int}
		c\, \eta_* \frac{g_0 + g(\eta_*) }{\prod_{j=1}^{n}  \eta_j^2 }   \le 
		\int_0^{x_*} \bigl(g_0 + g(x)\bigr)\prod_{j=1}^{n}\frac{1}{x^2 + \eta_j^2}  \mathrm{d}x \le  \frac{1}{c}\eta_* \frac{g_0 + g(\eta_*) }{\prod_{j=1}^{n}  \eta_j^2 },
	\end{equation}
	where $\eta_* := \min \{x_*, \eta_1, \dots, \eta_n\}$, and $c = c(\gamma, C, n) > 0$ is a small positive constant. 
\end{claim} 
\begin{proof}[Proof of Claim~\ref{claim:Cauchy_int}]
	It follows from the second bound in \eqref{eq:g_cond} and the monotonicity of $g$  that 
	\begin{equation}
		\frac{\eta_* g(\eta_*)}{C (1+\gamma)} \le  \int_0^{\eta_*}g(x)\mathrm{d}x \le   \eta_*   g(\eta_*).
	\end{equation}
	In particular, 
	\begin{equation}
		\frac{1}{2^n(1 + \gamma)C} \, \eta_* \frac{g_0 + g(\eta_*) }{\prod_{j=1}^{n}  \eta_j^2 } \le \int_0^{\eta_*}\bigl(g_0 + g(x)\bigr) \prod_{j=1}^{n}\frac{1}{x^2 + \eta_j^2} \mathrm{d}x \le   \eta_*   \frac{g_0 + g(\eta_*) }{\prod_{j=1}^{n}  \eta_j^2 }.
	\end{equation}
	Since the integrand in \eqref{eq:g_int} is positive, it remains to bound the contribution from the regime $\eta_*\le x \le x_*$ from above. Note that this regime is only relevant when $\eta_* < x_*$, that is, when $\eta_* = \eta_1$. Hence we estimating $x^2+\eta_1^2\ge x^2$ and $x^2 + \eta_j^2\ge \eta_j^2$ for $j \ge 2$,  and using the upper  bound in \eqref{eq:g_cond}, we obtain
	\begin{equation}
		\int_{\eta_*}^{x_*} \prod_{j=1}^{n}\frac{1}{x^2 + \eta_j^2} \bigl(g_0 + g(x)\bigr)\mathrm{d}x 
		\le \frac{g_0}{\prod_{j=2}^{n} \eta_j^2 } + C  \,\frac{g(\eta_*)}{\prod_{j=2}^{n} \eta_j^2 } \int_{\eta_*}^{x_*} \frac{1}{x^{2-\gamma}\eta_*^{\gamma}} \mathrm{d}x
		\le \frac{C}{1-\gamma}\, \eta_* \frac{g_0 + g(\eta_*)}{\prod_{j=1}^{n} \eta_j^2 }.
	\end{equation} 
	Here, in the last step we used the assumptions $C \ge 1$ and  $\gamma < 1$. This concludes the proof of Claim~\ref{claim:Cauchy_int}.  
\end{proof}

\begin{proof}[Proof of Lemma~\ref{lemma:m_hat}]
	We begin by plugging the integral representation \eqref{eq:M_stieltjes} into \eqref{eq:ImM2t_def}, using \eqref{eq:Mt_def} and \eqref{eq:B12_identity}, to obtain
	\begin{equation} \label{eq:M_hat_stieltjes}
		\widehat{M}_t(z_1, I, z_2) =  \int_\mathbb{R} \frac{\eta_{1,t}}{|z_{1,t} - x|^2 }\frac{\eta_{2,t}}{|z_{2,t} - x|^2 } \Im M_t(x)\mathrm{d}x. 
	\end{equation}
	In particular, since the integrand in \eqref{eq:M_hat_stieltjes} is positive semi-definite in the sense of quadratic forms, we have 
	\begin{equation}  
		\norm{\widehat{M}_t(z_1, I, z_2)} =  \sup_{X\ge 0} \langle X\rangle^{-1}\bigl\langle \widehat{M}_t(z_1, I, z_2) X \bigr\rangle, 
	\end{equation}
	where the supremum is taken over positive semi-definite matrices $X \in \mathbb{C}^{N\times N}$. 
	
	For the remainder of the proof, we fix $X \in \mathbb{C}^{N\times N}$ satisfying $X\ge0$, and analyze
	\begin{equation} \label{eq:m_hat_stieltjes}
		\bigl\langle  \widehat{M}_t(z_1, I, z_2) X\bigr\rangle =  \int_\mathbb{R} \frac{\eta_{1,t}}{|z_{1,t} - x|^2 }\frac{\eta_{2,t}}{|z_{2,t} - x|^2 } \bigl\langle \Im M_t(x) X\bigr\rangle \mathrm{d}x. 
	\end{equation}
	Note that to establish \eqref{eq:Mhat_flat}, it suffices to show that $\langle  \widehat{M}_t(z_1, I, z_2) X \rangle \lesssim \langle  \widehat{M}_t(z_1, I, z_2)   \rangle\langle    X \rangle$, since the opposite bound $\langle \widehat{M}_t(z_1, I, z_2) \rangle \lesssim \lVert \widehat{M}_t(z_1, I, z_2) \rVert$ is trivial. 
	
	The key regime to consider is when both $z_1, z_2$ are close to the same local minimum or singularity $\mathfrak{e}$ in the spectrum and the time $t$ is close to its terminal value $T$.
	Before proceeding with the analysis, we cover all the simpler complementary regimes.
	
	Recall that $\pi^{-1}\int_\mathbb{R} \Im M_t(x)\mathrm{d}x = I$.  First, we observe that \eqref{eq:m_hat_stieltjes} implies a trivial  lower and upper bounds
	\begin{equation} \label{eq:hatM_trivial}
		\eta_{1,t}\eta_{2,t} \langle X \rangle  \lesssim \bigl\langle  \widehat{M}_t(z_1, I, z_2) X\bigr\rangle \lesssim \frac{\eta_{1,t}}{\kapd_{1,t}^2}\frac{\eta_{2,t}}{\kapd_{2,t}^2} \langle X \rangle ,
	\end{equation}
	where we used the fact that for any $t\in[0,t]$, the matrix-valued probability measure $\pi^{-1}\Im M_t(x)\mathrm{d}x$ is supported on an order one interval by Lemma~\ref{lemma:M_structure}. 
	Hence, for any positive constant $c_1 \sim 1$, we have 
	\begin{equation}
		\bigl\langle \widehat{M}_t(z_1, I, z_2) X\bigr\rangle \sim \rho(z_1)\rho(z_2) \langle X \rangle, \qquad t \in [0, T-c_1], 
	\end{equation}
	for all $z_1, z_2 \in \bddD\cap\mathbb{H}$ satisfying $|z_1| + |z_2| \lesssim 1$. 
	In particular, the desired flatness \eqref{eq:Mhat_flat} and the  comparison \eqref{eq:ImGImG_comp} follows for all times $t \in [0, T-c_1]$, since in this regime $\kapd_{j,t} \sim 1$ and $\eta_{j,t}\sim \rho_{j,t}$ by \eqref{eq:kapd_rhoeta_t}, \eqref{eq:rho_t}--\eqref{eq:eta_t}.  
	
	Furthermore, since the Cauchy kernels $\Im z / |z-x|^2$ are localizing, the main contribution to the integral \eqref{eq:m_hat_stieltjes} comes from energies $x$ that are close to $\Re z_{1,t}$ or $\Re z_{2,t}$. More precisely, for any $r>0$, let $\mathbb{I}_r$ denote the set of energies
	\begin{equation}
		\mathbb{I}_r \equiv \mathbb{I}_r(z_{1,t}, z_{2,t}) := \bigl\{ x\in \mathbb{R} \,:\, |x-\Re z_{1,t}|\wedge|x-\Re z_{2,t}| \le r \bigr\}. 
	\end{equation}
	Then, by trivially estimating the denominators in the integrand of \eqref{eq:m_hat_stieltjes} for all $x \notin \mathbb{I}_r$, we deduce that
	\begin{equation} \label{eq:Mhat_r}
		\bigl\langle\widehat{M}_t(z_1, I, z_2)X\bigr\rangle = \frac{1}{\pi} \int_{\mathbb{I}_r} \frac{\eta_{1,t}}{|z_{1,t} - x|^2 }\frac{\eta_{2,t}}{|z_{2,t} - x|^2 } \bigl\langle \Im M_t(x) X\bigr\rangle\mathrm{d}x + \mathcal{O}\bigl(r^{-4}\eta_{1,t}\eta_{2,t}\bigr) \langle X \rangle.
	\end{equation}
	Moreover, if $|z_{1,t} - z_{2,t}| \ge c$ for some $c\sim1$, it is straightforward to check using \eqref{eq:Mhat_r} that 
	\begin{equation} \label{eq:hatM_far}
		\begin{split}
			\bigl\langle  \widehat{M}_t(z_1, I, z_2)X\bigr\rangle &\sim \bigl\langle \Im M_t(z_{1,t}) X\bigr\rangle \eta_{2,t} + \bigl\langle \Im M_t(z_{2,t}) X\bigr\rangle \eta_{1,t} + \mathcal{O}(\eta_{1,t}\eta_{2,t}) \langle X \rangle 
			\\&\sim \bigl(\rho_t(z_{1,t}) \eta_{2,t} + \rho_t(z_{2,t})  \eta_{1,t} + \mathcal{O}(\eta_{1,t}\eta_{2,t})\bigr) \langle X \rangle,
		\end{split}		
	\end{equation}
	where in the second step we used $z_1, z_2 \in \bddD$ together with  \eqref{eq:M_t} and \eqref{eq:imM_bound} to deduce that $ \langle \Im M_t(z_{j,t}) X \rangle \sim \rho_t(z_{j,t})  \langle X \rangle$.
	Comparison \eqref{eq:hatM_far}, together with the lower bound in \eqref{eq:hatM_trivial}, implies the desired \eqref{eq:Mhat_flat}--\eqref{eq:ImGImG_comp} provided $|z_{1,t} - z_{2,t}| \gtrsim 1$.

	Therefore, it remains to establish \eqref{eq:Mhat_flat}--\eqref{eq:ImGImG_comp} for times $t\in [0, T]$   and spectral parameters $z_1, z_2 \in \bddD\cap\mathbb{H}$ satisfying 
	\begin{equation} \label{eq:hatM_regime}
		|z_{1,t} - z_{2,t}|^2 \le c, \qquad  (T-t) \le c,
	\end{equation}
	for some sufficiently small positive constant $c \sim 1$, since in the complementary regime the desired \eqref{eq:Mhat_flat}--\eqref{eq:ImGImG_comp} follow from~\eqref{eq:hatM_trivial}--\eqref{eq:hatM_far}.
	
	For the remainder of the derivation, we fix a time $t \in [T-c, T]$. 
	By suitably shrinking the threshold $c \sim 1$ in \eqref{eq:hatM_regime} and reasoning as in \eqref{eq:time_admis}, we conclude that there exists a constant $r := r(z_{1}, z_{2}, t, c) \sim 1$, such that  the solution $M_t$ satisfies the assumption \eqref{eq:Mbdd_cond} of Lemma~\ref{lemma:shape} on the interval $\mathbb{I} = \mathbb{I}_r(z_{1,t}, z_{2,t})$ with $\widehat{c} = 2r$.
	Therefore, it follows from \eqref{eq:boundary_isol}--\eqref{eq:smallmin_isol} of Lemma~\ref{lemma:shape}(a) that by suitably shrinking $c\sim1$ and $r \sim 1$, we can guarantee that $\mathbb{I}_{2r}$ contains at most two points from the set $\mathbb{M} \equiv \mathbb{M}_{\rho_*}(\rho_t)$, defined in \eqref{eq:sing_set} with $\rho_* \sim 1$ the constant form item (i) of Lemma~\ref{lemma:shape}(a). Moreover, if $\mathbb{I}_{2r} \cap \mathbb{M} = \{\mathfrak{e}\}$, then either $\rho(\mathfrak{e}) > 0$ or $\Delta(\mathfrak{e}) \sim 1$, while if $\mathbb{I}_{2r} \cap \mathbb{M} = \{\mathfrak{e}_1, \mathfrak{e}_2\}$ with $\mathfrak{e}_1 < \mathfrak{e}_2$, then $\mathfrak{e}_1$ and $\mathfrak{e}_2$ are the left and right end-points of a small gap in the support of $\rho_t$.

	If $\mathbb{I}_{2r}\cap\supp\,\rho_t = \emptyset$, then $\kapd_{1,t} \sim \kapd_{2,t} \sim 1$ and the desired \eqref{eq:Mhat_flat}--\eqref{eq:ImGImG_comp} follow  trivially form \eqref{eq:hatM_trivial}. Therefore, in the sequel we assume that $\mathbb{I}_{2r}\cap\supp\,\rho_t \neq \emptyset$. 
	
	It follows from \eqref{eq:imM_bound} that $\langle \Im M_t(x) X \rangle \sim \rho_t(x) \langle X \rangle$ for all $x \in \mathbb{I}_{r}$, and hence
	\begin{equation} \label{eq:Mhat_normvstrace}
		\int_{\mathbb{I}_r} \frac{\eta_{1,t}}{|z_{1,t} - x|^2 }\frac{\eta_{2,t}}{|z_{2,t} - x|^2 } \bigl\langle \Im M_t(x) X \bigr\rangle\mathrm{d}x 
		\sim \langle X \rangle \int_{\mathbb{I}_r} \frac{\eta_{1,t}}{|z_{1,t} - x|^2 }\frac{\eta_{2,t}}{|z_{2,t} - x|^2 } \rho_t(x)\mathrm{d}x.    
	\end{equation}
	The comparison \eqref{eq:Mhat_normvstrace}, together with \eqref{eq:Mhat_r}  and the lower bound in \eqref{eq:hatM_trivial}, implies the desired \eqref{eq:Mhat_flat}. Therefore, it remains to establish \eqref{eq:ImGImG_comp}. 
	
	To proceed, we consider three cases, depending on the number of points in $\mathbb{I}_{2r} \cap \mathbb{M}$. 
	
	\medskip
	\noindent
	\textbf{Case 1.} \emph{Bulk spectrum}. If $\mathbb{I}_{2r} \cap \mathbb{M} = \emptyset$, then $\Re z_{1,t}, \Re z_{2,t} \in \supp\,\rho_t$, as well as  $\rho_t(x) \sim 1$ for all $x \in \mathbb{I}_r$ and, using the convolution formula for Cauchy kernels, we immediately obtain
	\begin{equation} \label{eq:Mhat_bulk_comp}
		\int_{\mathbb{I}_r} \frac{\eta_{1,t}}{|z_{1,t} - x|^2 }\frac{\eta_{2,t}}{|z_{2,t} - x|^2 } \rho_t(x)\mathrm{d}x \sim  \frac{\eta_{1,t} + \eta_{2,t}}{|z_1 - z_2|^2 + \eta_{1,t}^2 + \eta_{2,t}^2} + \mathcal{O}(\eta_{1,t}\eta_{2,t}).
	\end{equation}
	Therefore, the desired comparison \eqref{eq:ImGImG_comp} holds by combining \eqref{eq:Mhat_bulk_comp} with the lower bound from \eqref{eq:hatM_trivial}, since $\kapd_{j,t} = \eta_{j,t}$ if $\Re z_{j,t} \in \supp\,\rho_t$. 
	
	\medskip
	\noindent
	\textbf{Case 2.} \emph{Regular edge, sharp cusp or a small local minimum}.  If $\mathbb{I}_{2r} \cap \mathbb{M} = \{\mathfrak{e}\}$, then it follows from \eqref{eq:rho_comp} that 
	\begin{equation} \label{eq:case2_int}
		\int\limits_{\mathbb{I}_r} \frac{\eta_{1,t}}{|z_{1,t} - x|^2 }\frac{\eta_{2,t}}{|z_{2,t} - x|^2 } \rho_t(x)\mathrm{d}x \sim \int_{\mathbb{I}_r\cap\supp\rho_t} \frac{\eta_{1,t}}{|z_{1,t} - x|^2 }\frac{\eta_{2,t}}{|z_{2,t} - x|^2 } \biggl(\rho_t (\mathfrak{e} ) +  \frac{\delta(x)^{1/2}}{(\delta(x) + \Delta_t )^{1/6}}\biggr)\mathrm{d}x,
	\end{equation}
	where $\Delta_t := \Delta(\mathfrak{e})$ is the size of the local gap in the support of $\rho_t$ around $\mathfrak{e}$, defined in \eqref{eq:gap_size}. 
	Analogously to \eqref{eq:Mhat_bulk_comp}, the contribution from the term $\rho_t (\mathfrak{e} ) $ is given by 
	\begin{equation} \label{eq:rho_cont}
		\rho_t (\mathfrak{e}) \frac{\eta_{1,t} + \eta_{2,t}}{|z_1 - z_2|^2 + \eta_{1,t}^2 + \eta_{2,t}^2} + \mathcal{O}\bigl(\rho_t (\mathfrak{e})\eta_{1,t}\eta_{2,t}\bigr).
	\end{equation}
	Therefore, it remains to compute, 
	\begin{equation} \label{eq:case2_I}
		\widehat{I} :=  \int_{\mathbb{I}_r\cap\supp\rho_t} \frac{\eta_{1,t}}{|z_{1,t} - x|^2 }\frac{\eta_{2,t}}{|z_{2,t} - x|^2 } \frac{|x-\mathfrak{e}|^{1/2}}{(|x-\mathfrak{e}| + \Delta_t)^{1/6}}\mathrm{d}x,
	\end{equation}
	Recall that, under the assumption of Case 2, we have $\Delta_t \sim 1$ if $\rho_t(\mathfrak{e}) = 0$ and $\Delta_t =0$ if $\rho_t(\mathfrak{e}) > 0$.
	We defer the analysis of $\widehat{I}$ to the end of the proof, where we reduce it to a closely related integral from Case 3 below. 
	
	\medskip
	\noindent
	\textbf{Case 3.} \emph{Edge adjacent to a small gap}.  If $\mathbb{I}_{2r} \cap \mathbb{M} = \{\mathfrak{e}_1, \mathfrak{e}_2\}$ with $\mathfrak{e}_1 < \mathfrak{e}_2$, then $(\mathfrak{e}_1, \mathfrak{e}_2) \cap \supp\,\rho_t = \emptyset$, and we can assume without loss of generality that $\mathfrak{e}_2 - \mathfrak{e}_1 < 1$. Hence, $\mathfrak{e}_2 - \mathfrak{e}_1 = \Delta_t$, and \eqref{eq:rho_comp}  implies  that 
	\begin{equation} \label{eq:I_t_12}
		\int\limits_{\mathbb{I}_r} \frac{\eta_{1,t}}{|z_{1,t} - x|^2 }\frac{\eta_{2,t}}{|z_{2,t} - x|^2 } \rho_t(x)\mathrm{d}x  \sim \widehat{I}_1 + \widehat{I}_2,
	\end{equation}
	where the integrals $\widehat{I}_1$ and $\widehat{I}_2$ are given by
	\begin{equation}
		\begin{split}
			\widehat{I}_1 &:= \int_{\mathbb{I}_r\cap (-\infty, \mathfrak{e}_1]} \frac{\eta_{1,t}}{|z_{1,t} - x|^2 }\frac{\eta_{2,t}}{|z_{2,t} - x|^2 } \frac{(\mathfrak{e}_1-x)^{1/2}}{(\mathfrak{e} - x)^{1/6}} \mathrm{d}x,  
			\\
			\widehat{I}_2 &:= \int_{\mathbb{I}_r\cap [\mathfrak{e}_2, +\infty)} \frac{\eta_{1,t}}{|z_{1,t} - x|^2 }\frac{\eta_{2,t}}{|z_{2,t} - x|^2 } \frac{(x-\mathfrak{e}_2)^{1/2}}{(x-\mathfrak{e})^{1/6}} \mathrm{d}x,
		\end{split}
	\end{equation}
	where we define $\mathfrak{e} := \tfrac{1}{2}(\mathfrak{e}_1 + \mathfrak{e}_2)$ to be the mid-point of the gap and note that $|x-\mathfrak{e}| \gtrsim \Delta_t$ for all $x\in\supp\,\rho_t$.
	
	We now turn to the analysis of $\widehat{I}_1$ and $\widehat{I}_2$.	We partition the interval $\mathbb{I}_r$ into four disjoint parts, generated by intersecting it with
	\begin{equation} \label{eq:partition}
		(\infty, \mathfrak{e}_1],\quad (\mathfrak{e}_1, \mathfrak{e}], \quad (\mathfrak{e}, \mathfrak{e}_2], \quad (\mathfrak{e}_2, +\infty). 
	\end{equation}	
	Without loss of generality, we assume that $\Re z_{1,t} < \Re z_{2,t}$ and $\mathfrak{e} < \Re z_{2,t}$; hence, modulo symmetry\footnote{Arrangement 6 below has a symmetric counterpart $\Re z_{1,t} \le \mathfrak{e}_1 \le \mathfrak{e} \le \Re z_{2,t} \le \mathfrak{e}_2$. }, the partition \eqref{eq:partition} generates six distinct possible arrangements of $z_{1,t}, z_{2,t}$ depending on which intervals $\Re z_{j,t}$ fall into, which we now consider one-by-one.
	To shorten the presentation, we omit the trivial details and analogous steps.
	
	\medskip
	\noindent
	\textbf{Arrangement 1.} \emph{Same bump}. Assume that $\mathfrak{e}_2 < \Re z_{1,t} < \Re z_{2,t}$. We parameterize $z_{1,t} := \mathfrak{e}_2 + \delta + \ii\eta_1$ and $z_{2,t} := \mathfrak{e}_2 + \delta + \omega + \ii\eta_2$ with $\delta >0$ and $\omega > 0$. Here and in the sequel, we omit the subscript $t$ from $\eta_{j,t}$  and $\Delta_t$ for brevity. 
	Note that the parameterization  \eqref{eq:rho_comp} implies, abbreviating $\rho_j := \rho_t(z_{j,t})$,
	\begin{equation} \label{eq:arr1_params}
		\rho_1 \sim \frac{(\delta + \eta_1)^{1/2}}{(\Delta + \delta + \eta_1)^{1/6}}, \qquad 
		\rho_2 \sim \frac{(\delta +\omega + \eta_2)^{1/2}}{(\Delta + \delta + \omega + \eta_2)^{1/6}},
		\qquad |z_{1,t} - z_{2,t}| + \kapd_{1,t} + \kapd_{2,t} \sim \omega + \eta_1 + \eta_2.
	\end{equation}
	In this case, $\widehat{I}_j$ admit the equivalences
	\begin{equation} \label{eq:arr1}
		\begin{split}
			\widehat{I}_1 &\sim \int_0^1 \frac{\eta_{1}}{y^2 + (\Delta + \delta + \eta_1)^2}\frac{\eta_{2}}{y^2 + (\Delta + \delta + \omega + \eta_2)^2} g_\Delta(y) \mathrm{d}y, \qquad y := \mathfrak{e}_1-x, \\
			\widehat{I}_2 &\sim \int_0^1 \frac{\eta_{1}}{(\delta - y)^2 +\eta_1^2}\frac{\eta_{2}}{(\delta+\omega-y)^2 +\eta_2^2} g_\Delta(y)\mathrm{d}y, \qquad  y := x-\mathfrak{e}_2.		
		\end{split}
	\end{equation}
	Here, the function $g_\Delta(y)$ is defined as 
	\begin{equation} \label{eq:gDelta_def}
		g_\Delta(x) :=  y^{1/2}(\Delta + y)^{-1/6}, \qquad y > 0.
	\end{equation}
	It is straightforward to check that the integrand of $\widehat{I}_1$ in \eqref{eq:arr1} is smaller than the integrand of $\widehat{I}_2$ point-wise for all $y\in[0,1]$. Hence, it remains to estimate $\widehat{I}_2$. 
	
	By suitably shrinking $r\sim1$, we can assume without loss of generality that the $\delta + \omega < \tfrac{1}{2}$. We partition the integration domain $\widehat{I}_2$ into five intervals separated by the points $\{\tfrac{1}{2}\delta, \delta, \delta + \tfrac{1}{2}\omega, \delta+\omega\}$, and estimate the contribution coming from each interval separately using Claim~\ref{claim:Cauchy_int} with $g(x) := g_\Delta(x)$, defined in \eqref{eq:gDelta_def}, after shifting the integration variable. 
	Observe that $g_\Delta(x)$ is non-decreasing in $x > 0$ and satisfies \eqref{eq:g_cond} with $\gamma = \tfrac{1}{2}$ and $C = 2$, for all $\Delta \ge 0$. Moreover, for some absolute implicit constants,
	\begin{equation} \label{eq:g_max}
		g_\Delta(x+y) \sim g_\Delta(x) +  g_\Delta(y), \qquad g_\Delta(x \wedge y) \sim \frac{y\, g_\Delta(x) + x\,  g_\Delta(y)}{x+y}  \qquad x,y > 0.
	\end{equation}
	
	Let $f_1(y)$ denote the integrand of $\widehat{I}_2$ in \eqref{eq:arr1}. Then, it follows from Claim~\ref{claim:Cauchy_int} with $g_0 := 0$ and $g(x) := g_\Delta(x)$  that 
	\begin{equation} \label{eq:1regime1}
		\int_{0}^{\delta/2} f_1(y)\dd y \sim  \frac{\eta_{1}}{(\delta +\eta_1)^2}\frac{\eta_{2}}{(\delta + \omega +\eta_2)^2} \delta g_\Delta(\delta).
	\end{equation}
	On the interval $y \in [\tfrac{1}{2}\delta, \delta]$, we perform the change of variable $x:= \delta - y$, and use the first comparison in \eqref{eq:g_max} to obtain
	\begin{equation} \label{eq:1regime2}
		\int_{\delta/2}^\delta f_1(y)\dd y 
		\sim  g_\Delta(\delta) \int_{0}^{\delta/2} \frac{\eta_{1}}{x^2 +\eta_1^2}\frac{\eta_{2}}{x^2 + (\omega+\eta_2)^2} \dd x
		\sim  \frac{\eta_1\eta_2 \min\{\delta, \eta_1, \omega + \eta_2\} g_\Delta(\delta)}{ (\omega+\eta_1+\eta_2)^2 \min\{\eta_1, \omega+\eta_2\}^2},
	\end{equation}
	where in the first step we used that $g_\Delta(\delta - x) \sim g_\Delta(\delta)$ for all $x \in [0, \tfrac{1}{2}\delta]$, while in the second step we applied Claim~\ref{claim:Cauchy_int} with $g(x) = 1$, and $ab \sim (a + b)\min\{a,b\}$ for all $a,b \ge 0$. 
	On the interval $y \in [\delta, \delta+\tfrac{1}{2}\omega]$, after changing the variable $x := y - \delta$, we have
	\begin{equation} \label{eq:1regime3}
		\int_{\delta}^{\delta+\omega/2} f_1(y)\dd y \sim \int_{0}^{\omega/2} \frac{\eta_{1}}{x^2 +\eta_1^2}\frac{\eta_{2}}{(\omega+\eta_2)^2} \bigl( g_\Delta(\delta) + g_\Delta(x) \bigr)\dd x
		\sim  \frac{\eta_2 \omega (\omega + \eta_1) \bigl(g_\Delta(\delta) + g_\Delta(\min\{\omega, \eta_1\})\bigr)}{ (\omega + \eta_1 + \eta_2)^2(\omega+\min\{\eta_1, \eta_2\})^2}~,
	\end{equation}
	where in the first step we used that $g_\Delta(\delta + x) \sim g_\Delta(\delta) + g_\Delta(x)$ by \eqref{eq:g_max}, and in the second step we applied Claim~\ref{claim:Cauchy_int}.
	Proceeding similarly, in the regime $y \in [\delta+\omega/2, \delta+\omega]$, making the change of variable $x := \delta + \omega - y$, and using $g_\Delta(\delta + \omega - y) \sim g_\Delta(\delta + \omega)$ for all $y \in [0, \omega/2]$,  we deduce that 
	\begin{equation} \label{eq:1regime4}
		\int_{\delta+\omega/2}^{\delta+\omega} f_1(y)\dd y \sim  \int_{0}^{\omega/2} \frac{\eta_{1}}{(\omega +\eta_1)^2}\frac{\eta_{2}}{x^2 +\eta_2^2} g_\Delta(\delta + \omega)\mathrm{d}x 
		\sim  \frac{
			\eta_1 \eta_2\min\{\omega, \eta_2\} g_\Delta(\delta + \omega)  }{(\omega +\eta_1+\eta_2)^2 \min\{\omega+\eta_1, \eta_2\}^2}~. 
	\end{equation}
	Finally, for $y \in [\delta+\omega, 1]$, the change of variable $x := y - (\delta + \omega)$ and Claim~\ref{claim:Cauchy_int} yield
	\begin{equation} \label{eq:1regime5}
		\int_{\delta+\omega}^{1} f_1(y)\dd y \sim  \int_{0}^{1} \frac{\eta_{1}}{x^2 + (\omega+\eta_1)^2}\frac{\eta_{2}}{x^2 +\eta_2^2} \bigl(g_\Delta(\delta + \omega) + g_\Delta(x)\bigr)\mathrm{d}x 
		\sim \eta_1 \eta_2 \frac{g_\Delta(\delta + \omega) + g_\Delta( \min\{\omega+\eta_1, \eta_2\})}{(\omega + \eta_1+\eta_2)^2 \min\{\omega+\eta_1, \eta_2\}}.
	\end{equation}
	
	We combine \eqref{eq:1regime1}--\eqref{eq:1regime5}, and use the comparisons \eqref{eq:g_max} to express the result as a linear combination of the values of $g_\Delta$ at $\delta, \omega, \eta_1, \eta_2$, obtaining 
	\begin{equation} \label{eq:1I2_pre_comp}
		\widehat{I}_2 \sim \frac{1}{(\omega+\eta_1+\eta_2)^2} \biggl((\eta_1 + \eta_2)g_\Delta(\delta) + \eta_2 g_\Delta(\eta_1) +  \eta_1 g_\Delta(\eta_2) + \frac{\eta_1 \eta_2  g_\Delta(\omega)}{ \min\{\omega+\eta_1, \eta_2\} }    \biggr). 
	\end{equation}
	To identify $\rho_1$ and $\rho_2$ in the expression \eqref{eq:1I2_pre_comp} above and simplify the last term, we use the monotonicity of $g_\Delta$ and the second comparison in \eqref{eq:g_max} once again to deduce that
	\begin{equation} \label{eq:1last_term}
		\eta_1 g_\Delta(\omega) \lesssim \frac{\eta_1 \eta_2  g_\Delta(\omega)}{ \min\{\omega+\eta_1, \eta_2\} } \sim  \eta_1 g_\Delta(\omega) + \eta_2 (\omega\wedge\eta_1) \frac{ g_\Delta(\omega)}{ \omega } \lesssim \eta_1 g_\Delta(\omega) + \eta_2   g_\Delta(\omega\wedge\eta_1) \lesssim \eta_1 g_\Delta(\omega) + \eta_2   g_\Delta(\eta_1).
	\end{equation}
	Hence, plugging \eqref{eq:1last_term} back into \eqref{eq:1I2_pre_comp}, and using the first comparison in \eqref{eq:g_max}, we obtain 
	\begin{equation} \label{eq:arr1_I2_final}
		\widehat{I}_2 \sim \frac{ \eta_2 g_\Delta(\delta+\eta_1)  +  \eta_1 g_\Delta(\delta+\omega + \eta_2)}{(\omega+\eta_1+\eta_2)^2}
		\sim \frac{\rho_t(z_{1,t}) \eta_{2,t} + \rho_t (z_{2,t})\eta_{1,t}}{|z_{1,t}-z_{2,t}|^2 + \kapd_{1,t}^2 + \kapd_{2,t}^2},
	\end{equation}
	from which the desired \eqref{eq:ImGImG_comp} in this arrangement follows immediately by \eqref{eq:Mhat_r}, \eqref{eq:I_t_12}, and the lower bound in \eqref{eq:hatM_trivial}. Here, in the second step,  we used \eqref{eq:arr1_params}, the definition of $g_\Delta$ in \eqref{eq:gDelta_def},  and \eqref{eq:rho_t} for $\rho_j$. \nc 
	
	\medskip
	\noindent
	\textbf{Arrangement 2.} \emph{Opposite bumps}.
	Assume that $\Re z_{1,t} < \mathfrak{e}_1 < \mathfrak{e}_2 < \Re z_{2,t}$. We parameterize $z_{1,t} := \mathfrak{e}_1 - \delta_1 + \ii\eta_1$ and $z_{2,t} := \mathfrak{e}_2 + \delta_2 + \ii\eta_2$ with $\delta_1 >0$ and $\delta_2 > 0$. Then,  we have the comparisons
	\begin{equation} \label{eq:arr2_params}
		\rho_j \sim g_\Delta(\delta_j + \eta_j), 
		\qquad |z_{1,t} - z_{2,t}| + \kapd_{1,t} + \kapd_{2,t} \sim \Delta + \delta_1 + \delta_2 + \eta_1 + \eta_2,
	\end{equation}
	\begin{equation} \label{eq:arr2}
		\widehat{I}_1 \sim \int_0^1 \frac{\eta_{1}}{(\delta_1 - y)^2 +\eta_1^2}\frac{\eta_{2}}{y^2 + \Delta^2 + \delta_2^2 +\eta_2^2} g_\Delta(y)\mathrm{d}y,
	\end{equation}
	where $g_\Delta$ is defined in \eqref{eq:gDelta_def}. 
	The integral $\widehat{I}_2$ admits the same expression with the subscripts $1$ and $2$ interchanged in $\eta$ and $\delta$. We now compute $\widehat{I}_1$, applying Claim~\ref{claim:Cauchy_int} on intervals $y \in (0, \delta_1/2)$, $(\delta_1/2, \delta_1)$ and $(\delta_1, 1)$, obtaining
	\begin{equation} \label{eq:arr2_I1}
		\widehat{I}_1 \sim \frac{1}{(\Delta + \delta_1 + \delta_2+\eta_1+\eta_2)^2}\biggl(\eta_2 g_\Delta(\delta_1+\eta_1)  + \eta_1\frac{\eta_2 g_\Delta(\Delta + \delta_2 + \eta_2)}{\Delta + \delta_2 + \eta_2}\biggr). 
	\end{equation}
	It is straightforward to check, using the monotonicity of the map $x \mapsto x^{-1} g_\Delta(x)$ for $x > 0$, that $(\Delta + \delta_2 + \eta_2)^{-1} g_\Delta(\Delta + \delta_2 + \eta_2) \lesssim \eta_2^{-1}g_\Delta(\delta_2 + \eta_2)$. Hence, combining \eqref{eq:arr2_I1} with the symmetric expression for $\widehat{I}_2$, we obtain
	\begin{equation} \label{eq:arr2_final}
		\widehat{I}_1 + \widehat{I}_2 \sim \frac{\eta_2 g_\Delta(\delta_1 + \eta_1) + \eta_1 g_\Delta(\delta_2 + \eta_2)}{(\Delta + \delta_1 + \delta_2 + \eta_1 + \eta_2)^2},
	\end{equation}
	from which the desired \eqref{eq:ImGImG_comp} follows, analogously to the second step of \eqref{eq:arr1_I2_final}.
	
	\medskip
	\noindent
	\textbf{Arrangement 3.} \emph{Same half of the gap}.
	Assume that $\mathfrak{e} < \Re z_{1,t}  < \Re z_{2,t} < \mathfrak{e}_2$. We parameterize $z_{1,t} := \mathfrak{e}_2 - \delta - \omega + \ii\eta_1$ and $z_{2,t} := \mathfrak{e}_2 - \delta + \ii\eta_2$ with $\delta \ge 0, \omega \ge$ and $\delta + \omega  \le \tfrac{1}{2}\Delta$. Then, we have the comparisons
	\begin{equation} \label{eq:arr3_params} 
		\rho_1 \sim \frac{\eta_1 g_\Delta( \omega+ \delta +\eta_1)}{\omega+ \delta +\eta_1},  \qquad
		\rho_2 \sim \frac{\eta_2g_\Delta(\delta + \eta_2)}{\delta + \eta_2}, 
		\qquad |z_{1,t} - z_{2,t}| + \kapd_{1,t} + \kapd_{2,t}  \sim \delta+\omega +\eta_1 + \eta_2, 
	\end{equation}
	\begin{equation} \label{eq:arr3}
		\begin{split}
			\widehat{I}_1 &\sim \int_0^1 \frac{\eta_{1}}{y^2 + \Delta^2 +\eta_1^2}\frac{\eta_{2}}{y^2 + \Delta^2 +\eta_2^2}  g_\Delta(y) \mathrm{d}y,\qquad  \widehat{I}_2  \sim \int_0^1 \frac{\eta_{1}}{y^2 + \delta^2 + \omega^2 +\eta_1^2}\frac{\eta_{2}}{y^2 + \delta^2 +\eta_2^2}  g_\Delta(y) \mathrm{d}y.		
		\end{split}
	\end{equation}
	Similarly to the Arrangement 1~above, the integrand of $\widehat{I}_1$ in \eqref{eq:arr3} is point-wise smaller than that of $\widehat{I}_2$, hence we only need to analyze the latter. By a direct computation using Claim~\ref{claim:Cauchy_int}, \eqref{eq:g_max}, and \eqref{eq:arr3_params}, we obtain 
	\begin{equation} \label{eq:arr3_I2}
		\widehat{I}_2  \sim \frac{1}{(\delta + \omega +\eta_1+\eta_2)^2}\biggl(\eta_1\frac{\eta_2 g_\Delta(\delta + \eta_2)}{\delta + \eta_2} + \eta_2\frac{\eta_1  g_\Delta(\delta+\omega+\eta_1)}{\delta+\omega+\eta_1}\biggr),
	\end{equation} 
	which concludes the computation for this arrangement by \eqref{eq:arr3_params}.

	\medskip
	\noindent
	\textbf{Arrangement 4.} \emph{Opposite halves of the gap}.
	Assume that $\mathfrak{e}_1 <  \Re z_{1,t}  < \mathfrak{e} < \Re z_{2,t} < \mathfrak{e}_2$. We parameterize $z_{1,t} := \mathfrak{e}_1 + \delta_1 + \ii\eta_1$ and $z_{2,t} := \mathfrak{e}_2 - \delta_2 + \ii\eta_2$ with $\delta_1, \delta_2 \in [0, \tfrac{1}{2}\Delta]$. Then, we have the comparisons
	\begin{equation} \label{eq:arr4_params} 
		\rho_j  \sim \frac{\eta_j g_\Delta(\delta_j + \eta_j)}{ \delta_j + \eta_j }, \nc 
		\qquad |z_{1,t} - \overline{z}_{2,t}| + \kapd_{1,t} + \kapd_{2,t} \sim \Delta +\eta_1 + \eta_2, 
	\end{equation}
	\begin{equation} \label{eq:arr4}
		\widehat{I}_1 \sim \int_0^1 \frac{\eta_1}{y^2 + \delta_1^2 +\eta_1^2} \frac{\eta_2}{y^2 + \Delta^2 +\eta_2^2}  g_\Delta(y) \mathrm{d}y, \qquad
		\widehat{I}_2 \sim \int_0^1 \frac{\eta_1}{y^2 + \Delta^2 +\eta_1^2} \frac{\eta_2}{y^2 + \delta_2^2 +\eta_2^2}  g_\Delta(y) \mathrm{d}y.
	\end{equation}
	Proceeding as in Arrangement~3, we obtain
	\begin{equation}
		\widehat{I}_1 \sim \frac{1}{(\Delta + \eta_1 + \eta_2)^2 }\biggl(\eta_2\frac{\eta_1 g_\Delta(\delta_1 + \eta_1) }{\delta_1 + \eta_1} + \eta_1 \frac{ \eta_2 g_\Delta(\Delta + \eta_2)}{\Delta + \eta_2}\biggr) .
	\end{equation}
	Hence, by the symmetry between $\widehat{I}_1$ and $\widehat{I}_2$, we conclude, as desired, that
	\begin{equation}
		\widehat{I}_1 + \widehat{I}_2 \sim \frac{\rho_1\eta_2 + \rho_2\eta_1}{(\Delta + \eta_1 + \eta_2)^2},
	\end{equation}
	where we used \eqref{eq:arr4_params}, and $(\Delta + \eta_j)^{-1}g_\Delta(\Delta + \eta_j) \lesssim (\delta_j+ \eta_j)^{-1}g_\Delta(\delta_j + \eta_j)$ by monotonicity of the map $x^{-1}g_\Delta(x)$ and $\delta_j \le \Delta/2$.
	
	\medskip
	\noindent
	\textbf{Arrangement 5.} \emph{Bump and adjacent half of the gap}.
	Assume that $\mathfrak{e}  <  \Re z_{1,t} < \mathfrak{e}_2 < \Re z_{2,t} $, and parameterize $z_{1,t} := \mathfrak{e}_2 - \delta_1 + \ii\eta_1$, $z_{2,t} := \mathfrak{e}_2 + \delta_2 +\ii\eta_2$. In this arrangement, we have $|z_1 - z_2| + \kapd_{1,t} + \kapd_{2,t} \sim \delta_1 + \delta_2 + \eta_1 + \eta_2$, and it's easy to check that the main contribution is coming form $\widehat{I}_2$, given by
	\begin{equation}
		\widehat{I}_2 \sim \int_0^1 \frac{\eta_{1}}{y^2 + \delta_1^2 +\eta_1^2}\frac{\eta_{2}}{(\delta_2-y)^2 +\eta_2^2}  g_\Delta(y) \mathrm{d}y.
	\end{equation}
	Analogously to Arrangement~2, using Claim~\ref{claim:Cauchy_int}, \eqref{eq:rho_comp}, and \eqref{eq:g_max}, we obtain the desired comparison,
	\begin{equation} \label{eq:arr5_I2}
		\widehat{I}_2 
		\sim \frac{1}{(\delta_1 + \delta_2+\eta_1+\eta_2)^2}\biggl( \eta_2 \frac{\eta_1 g_\Delta(\delta_1 + \eta_1)}{\delta_1 + \eta_1} + \eta_1 g_\Delta(\delta_2+\eta_2) \biggr)
		\nc \sim \frac{\rho_1 \eta_2 + \rho_2\eta_1}{(\delta_1 + \delta_2 + \eta_1 + \eta_2)^2}.
	\end{equation} 
	
	\medskip
	\noindent
	\textbf{Arrangement 6.} \emph{Bump and opposite half of the gap}.
	Assume that $\mathfrak{e}_1  <  \Re z_{1,t} < \mathfrak{e}  < \mathfrak{e}_2 < \Re z_{2,t} $, and parameterize $z_{1,t} := \mathfrak{e}_1 + \delta_1 + \ii\eta_1$, $z_{2,t} := \mathfrak{e}_2 + \delta_2 +\ii\eta_2$. In this arrangement, $|z_1 - z_2| + \kapd_{1,t} + \kapd_{2,t} \sim \Delta + \delta_2 + \eta_1 + \eta_2$, and 
	\begin{equation}
		\widehat{I}_1 \sim \int_0^1 \frac{\eta_1}{y^2 + \delta_1^2 +\eta_1^2} \frac{\eta_2}{y^2 + \Delta^2 + \delta_2^2 +\eta_2^2}  g_\Delta(y) \mathrm{d}y, \qquad 
		\widehat{I}_2 \sim \int_0^1 \frac{\eta_{2}}{(\delta_2 - y)^2 +\eta_2^2}\frac{\eta_{1}}{y^2 + \Delta^2 +\eta_1^2} g_\Delta(y)\mathrm{d}y
	\end{equation}
	Computing $\widehat{I}_1$ and $\widehat{I}_2$ analogously to Arrangements~4 and~2, respectively, we obtain the estimates 
	\begin{equation}
		\begin{split}
			\widehat{I}_1 &\sim \frac{1}{(\Delta + \delta_2 + \eta_1 + \eta_2)^2}\biggl(\eta_2\frac{\eta_1g_\Delta(\delta_1 + \eta_1)}{\delta_1 + \eta_1} +  \eta_1\eta_2\frac{g_\Delta(\Delta + \delta_2 + \eta_2)}{\Delta + \delta_2 + \eta_2}\biggr)	
			\\	\widehat{I}_2 &\sim  \frac{1}{(\Delta + \delta_2+\eta_1+\eta_2)^2}\biggl(\eta_2 \frac{\eta_1 g_\Delta(\Delta + \eta_1)}{\Delta + \eta_1} +  \eta_1 g_\Delta(\delta_2+\eta_2) \biggr),	
		\end{split}			
	\end{equation}
	from which the desired \eqref{eq:ImGImG_comp} then follows similarly to previous arrangements. 
	\nc 
	Therefore, the desired comparison \eqref{eq:ImGImG_comp} is established in Case 3.
	
	It remains to complete the proof in Case 2. Without loss of generality, we can assume that $\Re z_{1,t} < \Re z_{2,t}$ and $\rho_t$ is positive on $(\mathfrak{e}, \mathfrak{e} + \delta)$ for some $\delta > 0$. 
	Recall that in this case $\Delta_t \sim 1$ or $\Delta_t =0$. 
	On the one hand, if $\Delta_t \sim 1$, then  $\widehat{I} \sim \widehat{I}_2$ and the desired comparison follows from \eqref{eq:arr1_I2_final}, \eqref{eq:arr3_I2} and \eqref{eq:arr5_I2}. 
	
	On the other hand, if $\Delta_t = 0$, then  $\mathbb{I}_r \subset \supp\,\rho_t$ and $\widehat{I} = \widehat{I}_1 + \widehat{I}_2$, where $\widehat{I}_j$ are defined in \eqref{eq:arr1} or \eqref{eq:arr2}. Hence, combining \eqref{eq:rho_comp},  \eqref{eq:case2_int}--\eqref{eq:case2_I}, and using estimates \eqref{eq:arr1_I2_final} and \eqref{eq:arr2_final}, we obtain 
	\begin{equation}
		\int\limits_{\mathbb{I}_r} \frac{\eta_{1,t}}{|z_{1,t} - x|^2 }\frac{\eta_{2,t}}{|z_{2,t} - x|^2 } \rho_t(x)\mathrm{d}x \sim \frac{\rho_t(z_{1,t}) \eta_{2,t} + \rho_t (z_{2,t})\eta_{1,t}}{|z_{1,t}-z_{2,t}|^2 + \kapd_{1,t}^2 + \kapd_{2,t}^2} + \mathcal{O}\bigl(\rho_t (\mathfrak{e})\eta_{1,t}\eta_{2,t}\bigr).
	\end{equation}
	This concludes the proof of Lemma~\ref{lemma:m_hat}. 
\end{proof}

\begin{proof}[Proof of Claim~\ref{claim:alpha_kappa}]
	For all $z_T, w_T \in \bddD\cap\mathbb{H}$ and $t \in [0,T]$, we define an auxiliary quantity $b_t(z_T, w_T)$ as 
	\begin{equation} \label{eq:bf_def}
		b_t(z_T, w_T) := \Bigl(\rho_t(z_t)^{-1} \Im z_t + \rho_t(z_t)^2+ \rho_t(z_t)|\sigma_t(z_t)| +  \bigl(\rho_t(z_t)+|\sigma_t(z_t)|\bigr)^{1/2}|z_t - w_t|^{1/2} + |z_t - w_t|^{2/3}\Bigr)\wedge 1,
	\end{equation}
	where we recall that $z_t$ and $w_t$ are the trajectories of \eqref{eq:char_flow} satisfying the final conditions $z_T$ and $w_T$, respectively.
	Here, the quantity $\sigma_t$ is defined analogously to $\sigma$ in \eqref{eq:sigma_def} but with $M$ replaced by $M_t$.
	Note that $\sigma_t(z_t) = \sigma_T(z_T)$ for all $t \in [t, T]$ by \eqref{eq:M_t}.
	Then, it follows from \eqref{eq:z_diff_comp} and \eqref{eq:eta_t}  that 
	\begin{equation} \label{eq:beta_b_comp}
		b_t(z_T, w_T) \sim b_T(z_T, w_T) + (T-t) \sim \other{\beta}_t(z_T, w_T).
	\end{equation}
	Hence, by definition of $\other{\alpha}_t$ in \eqref{eq:alp_t_def},  to derive \eqref{eq:alpha_comp}, it suffices to check that 
	\begin{equation} \label{eq:alpha_goal}
		|\sigma_t(z_t)|^2 + b_t(z_T,w_T) \sim \frac{ b_t(z_T, w_T)^2 b_t(z_T, \overline{w}_T)^2}{|z_t - w_t|^2 + \kapd_t(z_t)^2},
	\end{equation}
	where we recall that $\kapd_t(z_t) := \dist(z_t, \supp\,\rho_t)$, and we define
	\begin{equation} \label{eq:bf_ohp}
		b_t(z_T, \overline{w}_T) := b_t(z_T, w_T) \wedge \frac{|z_t - \overline{w}_t|}{\rho_t(z_t)} \sim \other{\beta}_t(z_T, \overline{w}_T). 
	\end{equation}
	Here we used that $\kapd_t(w_t) \lesssim |z_t - w_t| + \kapd_t(z_t)$ by triangle inequality. Furthermore, it suffices to prove \eqref{eq:alpha_goal} for $t \in [T-c, T]$ and $|z-w| + \kapd(z) \le c$ for some sufficiently small positive constant $c\sim1$, since in the complementary regime both the left- and the right-hand sides of \eqref{eq:alpha_goal} are trivially order unity. 
	Therefore, we can use the shape analysis of Lemma~\ref{lemma:shape} as in the proof of Lemma~\ref{lemma:m_hat} above. For the remainder of the proof the time $t$ remains fixed, and we drop it from the subscripts of $b, z, w, \rho, \sigma$, and $\kapd$. To prove \eqref{eq:alpha_goal}, we proceed by considering two cases. 
	
	\medskip
	\noindent
	\textbf{Case 1}. First, we consider the case $\rho(z) \le c$ for some sufficiently small constant $c\sim1$. Then, it follows immediately from \eqref{eq:bf_def} and  \eqref{eq:beta1_shape}--\eqref{eq:rhosigma_shape_out} that 
	\begin{equation} \label{eq:beta2_shape}
		b(z_T, w_T) \sim \bigl(\rho(\mathfrak{e})^3 + \delta + \eta + |z - w|\bigr)^{1/2}\bigl(\rho(\mathfrak{e})^3  + \Delta + \delta + \eta + |z-w|\bigr)^{1/6},
	\end{equation}
	where we abbreviate $\eta := \Im z$, $\mathfrak{e} := \mathfrak{e}(\Re z)$, $\delta := \delta(\Re z)$, and $\Delta := \Delta(\mathfrak{e})$ (see \eqref{eq:closest_sing}--\eqref{eq:gap_size}) recall that $\rho(\mathfrak{e}) \Delta = 0$). Here, in the regime where only the upper bound  \eqref{eq:rhosigma_shape_out} holds for $\rho(z) + |\sigma(z)|$, we combine it with Young's inequality and \eqref{eq:beta_b_comp} to obtain
	\begin{equation} 
		\bigl(\rho(z) + |\sigma(z)|\bigr)^{1/2} |z-w|^{1/2} \lesssim b(z_T,z_T) + |z-w|^{2/3}. 
	\end{equation}
	Moreover, by  \eqref{eq:bf_ohp}, we have
	\begin{equation}	\label{eq:beta2_shape_ohp}
		b(z_T,\overline{w}_T) \sim \frac{ |z-\overline{w}|}{\rho(z)b(z_T,w_T) + |z-\overline{w}|}b(z_T,w_T) \sim 
		\frac{|z-w| + \eta}{\rho(z)b(z_T,z_T) + |z-w|}b(z_T,w_T),
	\end{equation}
	where $\eta := \Im z$, and in the second step we used \eqref{eq:bf_def} together with Young's inequality. 
	Combining~\eqref{eq:rho_comp} and~\eqref{eq:beta1_shape}, we deduce that 
	\begin{equation} \label{eq:rho_beta_shape}
		\rho(z)b(z_T,z_T) \sim \begin{cases}
			\rho(\mathfrak{e})^3 + \delta + \eta, \qquad &\Re z \in \supp\,\rho,\\
			\eta, \qquad &\Re z \notin\supp\,\rho.
		\end{cases}
	\end{equation}
	
	Therefore, on the one hand,  using the fact that $\kapd(z) := \dist(z, \supp\,\rho)$ satisfies $\kapd(z) \sim \eta$ for  $\Re z \in \supp\,\rho$ and $\kapd(z) \sim \delta + \eta$ for  $\Re z \notin \supp\,\rho$, we deduce that the right-hand side of \eqref{eq:alpha_goal} satisfies
	\begin{equation}
		\frac{b(z_T,w_T)^2b(z_T,\overline{w}_T)^2}{|z - w|^2+\kapd(z)^2} \sim \rho(\mathfrak{e})^2 + 
		\bigl(\Delta + \delta + \eta \bigr)^{2/3} + |z-w|^{2/3}.
	\end{equation}
	On the other hand, it follows from \eqref{eq:bf_def} and the Young's inequality that the left-hand side of \eqref{eq:alpha_goal} satisfies
	\begin{equation} \label{eq:alp_b_comp}
		|\sigma(z)|^2 + b(z_T,w_T) \sim \bigl(\rho(z) + |\sigma(z)|\bigr)^2 + b(z_T,z_T) + |z-w|^{2/3} \sim \rho(\mathfrak{e})^2 + 
		\bigl(\Delta + \delta + \eta \bigr)^{2/3} + |z-w|^{2/3},
	\end{equation}
	where in the second step we used \eqref{eq:beta1_shape}--\eqref{eq:rhosigma_shape_out}.
	
	\medskip
	\noindent
	\textbf{Case 2}. Finally, we consider the case $\rho(z) \sim 1$. In this case, we have
	\begin{equation}
		b(z_T, w_T) \sim 1, \qquad b(z_T, \overline{w}_T) \sim |z-\overline{w}| \sim |z-w| + \eta.
	\end{equation}
	Moreover, since $\rho(z)\sim1$, we have $\kapd(z) \sim \eta$.
	Indeed, on the one hand, if $\Re z \in \supp\,\rho$, then $\kapd(z) = \eta$. On the other hand, if $\Re z \notin\supp\,\rho$, then \eqref{eq:rho_comp} implies that $1 \sim \rho(z) \lesssim \eta^{1/3}$, and, since $\kapd \lesssim 1$ by assumption, we have $\kapd(z) \sim 1 \sim \eta$. Therefore, we conclude that 
	\begin{equation}
		\frac{b(z_T, w_T)^2b(z_T, \overline{w}_T)^2}{|z - w|^2+\kapd(z)^2} \sim 1 \sim |\sigma(z)|^2 + b(z,w) \sim \other{\alpha}_t(z_T,w_T)^2,
	\end{equation}
	where in the second comparison we used the fact that $b(z_T, w_T) \gtrsim \rho(z)^2 \sim 1$ by \eqref{eq:bf_def}.
	This concludes the proof of Claim~\ref{claim:alpha_kappa}.	
\end{proof}

\subsection{Properties of the Two-Body Stability Factors. Proof of Lemma~\ref{lemma:alphabeta} and Claim~\ref{claim:stab_diff_rho}} \label{app:beta_prop} 
\begin{proof}[Proof of Lemma~\ref{lemma:alphabeta}]
	First, we prove \eqref{eq:beta_sym}. The equality in \eqref{eq:beta_sym} is a direct consequence of the fact that $M(\overline{z}) = M(z)^*$. 	
	To prove the comparison for $z,w$  lying in the same half-plane, recall from \eqref{eq:beta_b_comp} and \eqref{eq:beta2_shape} that, for all $z,w \in \bddD$, we have 
	\begin{equation}
		\other{\beta}(z,w) \sim \bigl(\rho(\mathfrak{e}(\Re z))^3 + \delta(\Re z) + \Im z + |z - w|\bigr)^{1/2}\bigl(\rho(\mathfrak{e}(\Re z))^3  + \Delta(\mathfrak{e}(\Re z)) + \delta(\Re z) + \Im w + |z-w|\bigr)^{1/6},
	\end{equation} 
	Since $|\Im w - \Im z|\lesssim|z-w|$ and $\mathfrak{e}(\Re z)$ is a locally constant function which only jumps when $\delta(\Re z) \gtrsim 1$, we conclude that 
	\begin{equation} \label{eq:comp_z_w}
		\begin{split}
			\rho(\mathfrak{e}(z))^3 + \delta(\Re z) + \Im z + |z - w| &\sim \rho(\mathfrak{e}(\Re z))^3 + \rho(\mathfrak{e}(\Re w))^3 + \delta(\Re z) + \delta(\Re w)  + \Im z + \Im w + |z - w|,\\
			\Delta(\mathfrak{e}(\Re z)) + \delta(\Re z) + |z-w| &\sim \Delta(\mathfrak{e}(\Re z)) + \Delta(\mathfrak{e}(\Re w)) + \delta(\Re z) + \delta(\Re w) + |z-w|
		\end{split}
	\end{equation}
	Therefore, $\other{\beta}(z,w) \sim \other{\beta}(w,z)$ for all $z,w \in \bddD$. To prove the comparison in \eqref{eq:beta_sym} in the opposite half-planes, recall from \eqref{eq:bf_ohp} and \eqref{eq:beta2_shape_ohp} that 
	\begin{equation}
		\other{\beta}(z,\overline{w}) \sim \other{\beta}(z,w) \frac{|z-w| + \Im z}{\rho(z)\other{\beta}(z,z) + |z-w|} \sim \other{\beta}(z,w) \frac{|z-w| + \Im z + \Im w}{\rho(z)\other{\beta}(z,z) + \rho(w)\other{\beta}(w,w) + |z-w|},
	\end{equation}
	where in the second comparison we used the fact that the right-hand sides of  \eqref{eq:rho_beta_shape} at $z$ and $w$ are comparable up to $|z-w|$ similarly to \eqref{eq:comp_z_w}. Hence, $\other{\beta}(z, \overline{w}) \sim \other{\beta}(w, \overline{z})$ for all $z,w \in \bddD$, and hence the desired \eqref{eq:beta_sym} is established. 
	
	The symmetry of $\other{\alpha}$ in \eqref{eq:alpha_sym} is an immediate consequence of \eqref{eq:alpha_defin}, \eqref{eq:beta_sym}, and the fact that $\sigma$ is $1/3$-H\"older regular on the set where $\rho$ is small by \eqref{eq:1/3Holder_rhoinveta}.  The upper bound in \eqref{eq:alpha_sym} follows from the fact that $\other{\beta}(z,w) \lesssim 1$ and $|\sigma(z)| \lesssim 1$ for all $z \in \bddD\cap\mathbb{H}$ by \eqref{eq:sigma_def} and \eqref{eq:F_flat}. 
	
	Next, to prove \eqref{eq:alp_beta_in_supp}, we recall that by \eqref{eq:alpha_defin},  \eqref{eq:beta_b_comp}, \eqref{eq:alp_b_comp}, \eqref{eq:rho_comp}, and \eqref{eq:beta1_shape}, we have
	\begin{equation}
		\other{\alpha}(z,w) \sim \rho (\mathfrak{e}(\Re z) )  + 
		\bigl(\Delta(\mathfrak{e}(\Re z)) + \delta(\Re z) + \Im z \bigr)^{1/3} \sim \rho(z)^{-1}\other{\beta}(z,z), 
	\end{equation}
	where we used the fact that $\Delta(\mathfrak{e})\rho(\mathfrak{e}) = 0$. 
	
	Similarly, \eqref{eq:alp_eta_comp} follows from \eqref{eq:alpha_defin},   \eqref{eq:beta_b_comp}, \eqref{eq:alp_b_comp}, and \eqref{eq:alpha_sym}. 
	
	Finally, to obtain \eqref{eq:alpha_realline}, we use \eqref{eq:etaf_def}, \eqref{eq:rho_comp}, and $1/3$-H\"older regularity of $\rho$ from \eqref{eq:1/3Holder_rhoinveta}, to deduce that for all $E\in \supp\,\rho$, we have 
	\begin{equation}
		\frac{1}{N} \sim \rho\bigl(E+\ii\etaf(E)\bigr)\etaf(E) \lesssim \rho\bigl(E+\ii\etaf(E)\bigr)^{4} + \etaf(E)^{4/3} \lesssim \rho\bigl(E\bigr)^{4} + \etaf(E)^{4/3}.
	\end{equation}
	Hence, it follows from \eqref{eq:ETH_alpha_jk} and \eqref{eq:alp_eta_comp} that 
	\begin{equation}
		\other{\alpha}(z_1, z_2) \sim \other{\alpha}(E_1, E_2) + \rho(E_1) + \rho(E_2) + \etaf(E_1)^{1/3} + \etaf(E_2)^{1/3} \gtrsim  \other{\alpha}(E_1, E_2) + N^{-1/4}. 
	\end{equation}
	This concludes the proof of Lemma~\ref{lemma:alphabeta}. 
\end{proof} \nc 

\begin{proof}[Proof of Claim~\ref{claim:stab_diff_rho}]
	\nc We only treat the case when $\other{\beta}(z, w) \ge \other{\beta}(w, z)$ in detail since the other case is analogous.
	In particular, $\other{\beta}(z, w) \ge \beta_*$ and $\other{\beta}(z, \overline{w}) < \other{\beta}(z, w)$.    
	By definition \eqref{eq:betaf_def} of $\other{\beta}(z, \overline{w})$, the first inequality in    \eqref{eq:2M_bound_case2} implies that   $\other{\beta}(z, \overline{w}) < 1 \wedge \other{\beta}(z, w) $, hence $\other{\beta}(z, \overline{w}) = \rho(z)^{-1} |z - \overline{w}|$, and  \nc 
	\begin{equation} \label{eq:max_young_prep}
		\max\{\Im z, |z - w|\} \le |z - \overline{w}| = \rho(z)\other{\beta}(z, \overline{w})  \le c_*\rho(z).
	\end{equation}
	Hence,   plugging \eqref{eq:max_young_prep} into the definition of $\other{\beta}(z, w)$ 
	in \eqref{eq:betaf_def},  \nc
	and using Young's inequality, we obtain
	\begin{equation} \label{eq:young1}
		\begin{split}
			\tfrac{1}{2}\beta_* &\le \other{\beta}(z, w) \le c_*  + \rho(z)\bigl(\rho(z) + |\sigma(z)|\bigr)   + \rho(z)^{1/2}\bigl(\rho(z) + |\sigma(z)|\bigr)^{1/2} c_*^{1/2} + c_*^{2/3} \rho(z)^{2/3}\\
			&\le 3c_* + 3\rho(z)\bigl(\rho(z) + |\sigma(z)|\bigr). 
		\end{split}
	\end{equation}
	By choosing $c_* \le \tfrac{1}{8}\beta_*$, \eqref{eq:young1} yields $\rho(z)^2 + \rho(z)|\sigma(z)| \ge \tfrac{1}{24}\beta_*$. 

	Since $|\sigma(z)| \lesssim 1$ by \eqref{eq:sigma_def} and \eqref{eq:F_flat},  and \nc we conclude that  $\rho(z) \gtrsim 1$. The second comparison in \eqref{eq:beta_sym} implies that there exists a constant $C\sim1$ such that $\other{\beta}(z,w) \le C \other{\beta}(w,z)$ for all $z,w\in \bddD$. Therefore, analogously to \eqref{eq:young1}, we obtain
	\begin{equation} \label{eq:impl_proof_last}
		\max\{\Im w, |z - w|\} \le |z - \overline{w}| \le C c_*\rho(w), \quad \text{and} \quad \tfrac{1}{2}C^{-1}\beta_* \le 3 C c_* + 3 \rho(w)\bigl(\rho(w) + |\sigma(w)|\bigr).
	\end{equation}
	Hence, by choosing $c_* \le \frac{1}{8}C^{-2}\beta_*$, we also have $\rho(w) \gtrsim 1$.
\end{proof}

\subsection{$\rho$-Cancellations in the Stable Domain} \label{app:stab_rho_cancel}

\begin{proof}[Proof of Claim~\ref{claim:stab_rho_cancel}]
	Let $\mathcal{C}_{w_1, w_2}$ denote the super-operators
	\begin{equation}
		\mathcal{C}_{w_1, w_2}[\,\cdot\,] \equiv \mathcal{C}_{t,w_1, w_2}[\,\cdot\,] := \ee^{t-T} M(w_1)^{-1} \mathrm{Id}[\,\cdot\,] M(w_2)^{-1}, \qquad w_j \in \{z_j, \overline{z}_j\}, \quad j\in \{1,2\}.
	\end{equation}
	It follows from \eqref{eq:stat_M_bound} and $T\sim 1$ that 
	\begin{equation}
		\norm{\mathcal{C}_{w_1, w_2}}_{\mathrm{hs}\to\mathrm{hs}} \lesssim \bigl(1+|z_1|\bigr)\bigl(1+|z_2|\bigr), \qquad \norm{\mathcal{C}_{w_1, w_2}^{-1}}_{\mathrm{hs}\to\mathrm{hs}} \lesssim \bigl(1+|z_1|\bigr)^{-1}\bigl(1+|z_2|\bigr)^{-1}.
	\end{equation} 
	Since $\mathcal{M}_{w_1, w_2}  = \mathcal{B}_{t,w_1, w_2}^{-1}\mathcal{C}_{w_1, w_2}^{-1}$ by \nc \eqref{eq:M_t} and \eqref{eq:M2op_def}, the bound \eqref{eq:stab_bound_assume} implies  $$\norm{\mathcal{M}_{w_1, w_2}}_{\mathrm{hs}\to\mathrm{hs}}  \lesssim (1+|z_1|)^{-1}(1+|z_2|)^{-1}.$$
	Observe that 
	\begin{equation} \label{eq:C_diff}
		\bigl( \mathcal{C}_{\overline{z}_1, w_2} - \mathcal{C}_{z_1, w_2}\bigr)[\,\cdot\,] = 2\ii M(\overline{z}_1)^{-1} \bigl(\Im M(z_1)\bigr) \mathcal{C}_{z_1, w_2}[\,\cdot\,].
	\end{equation}
	Therefore, $\lVert \mathcal{C}_{\overline{z}_1, w_2} - \mathcal{C}_{z_1, w_2} \rVert_{\mathrm{hs}\to\mathrm{hs}} \lesssim \rho(z_1) (1+|z_1|)^2(1+|z_2|)$ by \eqref{eq:invM_bound}, \eqref{eq:imM_bound}. 
	
	It follows from \eqref{eq:stab_def} and \eqref{eq:M2op_def} that
	$\mathcal{M}_{w_1, w_2} = (\mathcal{C}_{w_1, w_2} - \mathcal{S}_t)^{-1}$, hence we obtain the identity
	\begin{equation}
		\mathcal{M}_{z_1, w_2} - \mathcal{M}_{\overline{z}_1, w_2} = \mathcal{M}_{z_1, w_2} \bigl(\mathcal{C}_{\overline{z}_1, w_2} - \mathcal{C}_{z_1, w_2}\bigr) \mathcal{M}_{\overline{z}_1, w_2},
	\end{equation}
	from which the first bound in \eqref{eq:M2op_rho_cancel} follows immediately.  The second bound in \eqref{eq:M2op_rho_cancel} is completely analogous. 
	
	Finally, to obtain  the desired \eqref{eq:M2op_rhorho_cancel}, we use the straightforward identity
	\begin{equation}
		\begin{split}
			\mathcal{M}_{z_1, \overline{z}_2} + \mathcal{M}_{\overline{z}_1, z_2}  -\mathcal{M}_{z_1, z_2}  - \mathcal{M}_{\overline{z}_1, \overline{z}_2} 
			=&~ \mathcal{M}_{z_1, z_2} \bigl(\mathcal{C}_{\overline{z}_1, \overline{z}_2} - \mathcal{C}_{z_1, \overline{z}_2} - \mathcal{C}_{\overline{z}_1, z_2} + \mathcal{C}_{z_1, z_2}\bigr) \mathcal{M}_{\overline{z}_1, z_2} \\
			&+\mathcal{M}_{z_1, \overline{z}_2}\bigl(\mathcal{C}_{z_1, z_2} - \mathcal{C}_{z_1, \overline{z}_2}\bigr)\mathcal{M}_{z_1, z_2} \bigl(\mathcal{C}_{\overline{z}_1, \overline{z}_2} - \mathcal{C}_{z_1, \overline{z}_2}\bigr) \mathcal{M}_{\overline{z}_1, \overline{z}_2} \\
			&+\mathcal{M}_{z_1, z_2} \bigl(\mathcal{C}_{\overline{z}_1, \overline{z}_2}^{-1} - \mathcal{C}_{z_1, \overline{z}_2}^{-1}\bigr) \mathcal{M}_{\overline{z}_1, \overline{z}_2}\bigl(\mathcal{C}_{\overline{z}_1, z_2} - \mathcal{C}_{\overline{z}_1, \overline{z}_2}\bigr)\mathcal{M}_{\overline{z}_1, z_2},
		\end{split}
	\end{equation}
	together with \eqref{eq:C_diff}, $\norm{\Im M(z_j)} \lesssim \rho(z_j)$ by \eqref{eq:imM_bound},  and 
	\begin{equation}
		\bigl(\mathcal{C}_{\overline{z}_1, \overline{z}_2} - \mathcal{C}_{z_1, \overline{z}_2} - \mathcal{C}_{\overline{z}_1, z_2} + \mathcal{C}_{z_1, z_2}\bigr)[\,\cdot\,]  = - 4\,
		M(\overline{z}_1)^{-1} \bigl(\Im M(z_1)\bigr)  \mathcal{C}_{z_1, z_2}[\,\cdot\,] \bigl(\Im M(z_2)\bigr) M(\overline{z}_2)^{-1}.
	\end{equation}
	This concludes the proof of Claim \ref{claim:stab_rho_cancel}.		 
\end{proof}

\subsection{Equation of Motion for the Regularity Correction} \label{app:Upsderv}
\begin{proof}[Proof of Lemma \ref{lem:Upsderv}] We suppose that $z_{1,t}, z_{2,t} \in \mathbb{H}$ w.l.o.g.~and write $\eta_{i,t} := \Im z_{i,t}$. 
Moreover, before entering the proof, we introduce some convenient shorthand notation. First, as usual, we simply write $M_{i,t} := M_t(z_{i,t})$ for $i \in \indset{2}$, and also abbreviate
\begin{equation}
\widehat{M}_t := \widehat{M}_t(z_{1,t}, I, z_{2,t})\,, \quad U_t := \frac{M_{1,t} - M_{2,t}}{z_{1,t} - z_{2,t}}\,, \quad V_t := \frac{M_{1,t} - M_{2,t}^*}{z_{1,t} - \overline{z}_{2,t}}\,. 
\end{equation}
Moreover, we write $u_t := \langle U_t \rangle$ and $v_t := \langle V_t \rangle$ and drop the index $i$
 of $A$. 
 
We now differentiate the defining equation for $\Upsilon_t$, i.e.~$\langle \widehat{M}_t (A - \Upsilon_t I) \rangle = 0$, in time to get that
\begin{equation} \label{eq:Upsdot}
\frac{\dd}{\dd t}{\Upsilon}_t = \frac{\langle (u_tU_t + \overline{u}_t U_t^* - v_tV_t - \overline{v}_tV^*_t) \reg{A}_t\rangle}{u_t+\overline{u}_t - v_t - \overline{v}_t}
\end{equation}
where we used that 
\begin{equation} \label{eq:relations}
 \widehat{M}_t  = - \frac{1}{4} \big(U_t + U^*_t - V_t - V^*_t\big) \qquad \text{and} \qquad\langle (U_t +  U_t^* - V_t - V^*_t) \reg{A}_t\rangle = 0 \,. 
\end{equation}
The task thus reduced  to controlling the right-hand side~of \eqref{eq:Upsdot}. To do so, we first record two more identities: First
	\begin{equation}
				\widecheck{M}_t := \frac{1}{4} \Big(M_t(z_1, I, z_2) + M_t(\overline{z}_1,I , z_2) + M_t(z_1, , \overline{z}_2) + M_t(\overline{z}_1, I, \overline{z}_2)\Big)
				\end{equation}
				satisfies
	\begin{equation} \label{eq:relations2}
	\widecheck{M} = \frac{1}{4} \left( U + U^* + V + V^* \right) 
	\end{equation} 
	and we also have that 
	\begin{equation} \label{eq:relations3}
	(z_1 - z_2)U + (\overline{z}_1 - \overline{z_2}) U^* - (z_1 - \overline{z}_2) V - (\overline{z}_1 - z_2) V^* = 0 \,. 
	\end{equation} 
	Note that we  dropped the subscript $t$ and we continue to omit it in the rest of the proof. \nc
	Combining the second relation in \eqref{eq:relations} with \eqref{eq:relations3}, we infer
	\begin{equation}
	\langle (V - V^*) \reg{A} \rangle = \frac{\eta_1 - \eta_2}{\eta_1 + \eta_2} \langle (U - U^*) \reg{A} \rangle
	\end{equation}
	which allows us to obtain
	\begin{equation}
	\eqref{eq:Upsdot} = \frac{\langle (\omega U + \overline{\omega} U^*) \reg{A} \rangle}{2 \Re[u-v]} \quad \text{with} \quad \omega = u - \frac{\eta_1}{\eta_1 + \eta_2} v - \frac{\eta_2}{\eta_1 + \eta_2} \overline{v} \,. 
	\end{equation}
	Further, using again the second relation in \eqref{eq:relations}, we deduce
	\begin{equation*}
	\eqref{eq:Upsdot} = \langle \widecheck{M} \reg{A} \rangle + \frac{\langle ((\omega- \Re[u-v]) U + (\overline{\omega} - \Re [u-v]) U^*) \reg{A} \rangle}{2 \Re[u-v]} =  \langle \widecheck{M} \reg{A} \rangle - \frac{\Im[u] - \frac{\eta_1 - \eta_2}{\eta_1 + \eta_2} \Im[v]}{\Re [u-v]} \left\langle \frac{U - U^*}{2 \ii} \reg{A}\right\rangle \,. 
	\end{equation*}
	Note that 
	\begin{equation} \label{eq:Mdiffer}
	\big| \langle \widecheck{M} \reg{A} \rangle - \langle M(z_1, I, z_2) \reg{A} \rangle \big| \lesssim \frac{\rho_1 + \rho_2}{\other{\alpha}}. 
	\end{equation}
	as a consequence of \eqref{eq:GImG_bounds}. Moreover, as shown in \eqref{eq:GImG_bounds} as well, we have that 
	\begin{equation} \label{eq:imU}
	\left| \left\langle \frac{U - U^*}{2 \ii} \reg{A}\right\rangle \right| \lesssim \frac{|\rho_1| + |\rho_2| }{\other{\alpha}(z_1, z_2)}\vertiii{A}_{z_1,z_2}~,
	\end{equation}
	and it thus remains to control the prefactor, which we rewrite using \eqref{eq:ImV_id} as
	\begin{equation} \label{eq:pref}
	\frac{4\Im[u] \eta_1 \eta_2}{\Re [u-v] (\eta_1 + \eta_2)^2} - 2 \frac{(\eta_1 - \eta_2) \Re [z_1 - z_2]}{(\eta_1 + \eta_2)^2} \,.
	\end{equation}
	
	Therefore, combining \eqref{eq:Mdiffer}, \eqref{eq:imU}, \eqref{eq:pref}, as well as $|\Re[u-v]| \sim \rho_1 \rho_2 (\nu_1 + \nu_2) \big(\other{\alpha}(\other{\beta} \other{\beta}_*)^{-1}\big)^2$, as a consequence of \eqref{eq:m_hat_alpha}, and \eqref{eq:beta_reg} with \eqref{eq:beta_assymp}, we obtain
	\begin{equation} \label{eq:calEfirst}
	\mathcal{E} \lesssim \frac{\rho_1 + \rho_2}{\other{\alpha}} + \left(\frac{\other{\beta}_*}{\other{\alpha}}\right)^2 \frac{\nu_1 \nu_2}{(\nu_1 + \nu_2) \other{\nu}^2} + \frac{1}{\other{\nu} \other{\alpha}} \frac{|z_1 - z_2|^2}{|z_1 - \overline{z}_2|}, 
	\end{equation}
	where we additionally used that 
	\begin{equation*}
	|\eta_1 - \eta_2| \sim \frac{|z_1 - z_2| (\eta_1 + \eta_2)}{|z_1 - \overline{z}_2|} \quad \text{and} \quad |\Re[z_1 - z_2]| \le |z_1 - z_2| \,. 
	\end{equation*}
	Finally, we estimate the last two terms in \eqref{eq:calEfirst} as 
	\begin{equation} \label{eq:calEsecond}
	\left(\frac{\other{\beta}_*}{\other{\alpha}}\right)^2 \frac{\nu_1 \nu_2}{(\nu_1 + \nu_2) \other{\nu}^2} + \frac{1}{\other{\nu} \other{\alpha}} \frac{|z_1 - z_2|^2}{|z_1 - \overline{z}_2|} \lesssim \frac{\other{\beta} \other{\beta}_*}{(\nu_1 + \nu_2)^2 \other{\nu} \other{\alpha}} \,. 
	\end{equation}
	Here, we used that, by employing \eqref{eq:timerelations} and \eqref{eq:betaf_def}, 
	\begin{equation*}
	\frac{|z_1 - z_2|^2}{|z_1 - \overline{z}_2|} \lesssim \frac{\other{\beta} \other{\beta}_*}{(\nu_1 + \nu_2)^{1/2} } \quad \text{and} \quad \frac{\nu_1 \nu_2}{(\nu_1 + \nu_2)^{1/2} \other{\nu} \other{\alpha}} \lesssim 1. 
	\end{equation*}
	Hence, plugging \eqref{eq:calEsecond} into \eqref{eq:calEfirst}, we conclude the desired. 
\end{proof}

\subsection*{Glossary of Key Quantities}  In the following table, we collect a list of 
		 specialized quantities  related to the two-body stability operator, which depend on an ordered pair of spectral parameters $z_1, z_2 \in \mathbb{C}$.

		\captionof{table}{Glossary of Two-Body Quantities}
		\label{glossary} 
		
		\begin{tabularx}{\textwidth}{@{}p{0.12\textwidth}X@{}}  
			
			$\mathcal{B}_{z_1,z_2}$ &
			Two-body stability (super-)operator
			\(
			\mathcal{B}_{z_1,z_2}:\C^{N\times N}\to\C^{N\times N},
			\)
			defined in~\eqref{eq:stab_def} for \(z_1,z_2\in\C\setminus\R\). It controls the size of the deterministic approximation to the two-resolvent chain $G(z_1) A G(z_2)$. \\ 
			
			$\other{\beta}(z_1,z_2)$ &
			Two-body stability factor, defined in~\eqref{eq:betaf_def} for $z_1, z_2 \in \mathbb{C}\backslash\mathbb{R}$. It controls the norm of the inverse stability operator,
			\(\|\mathcal{B}_{z_1,z_2}^{-1}\|\); see~\eqref{eq:stab_beta_bound}. \\ 
			
			$\beta_{z_1,z_2}$ &
			Smallest (in modulus) eigenvalue of \(\mathcal{B}_{z_1,z_2}\). It is uniquely defined and isolated whenever
			\(\other{\beta}(z_1,z_2)\) is sufficiently small; see~\eqref{eq:stab_well_struc}--\eqref{eq:Pi_rank1}. In this regime,
			\(|\beta_{z_1,z_2}|\sim\other{\beta}(z_1,z_2)\); see~\eqref{eq:beta_assymp}. The corresponding eigenvector is referred to as the \emph{unstable direction}. \\ 
			
			$\other{\alpha}(z_1,z_2)$ &
			Control quantity, defined in~\eqref{eq:alpha_defin} for \(z_1,z_2\in\mathbb{H}\). It determines the size of the anomalous fluctuation mode in ETH and may be much larger than \(|\beta_{z_1,z_2}|\sim\other{\beta}(z_1,z_2)\). It is a key component of the optimal estimate on the deterministic approximation to $\langle \Im G(z_1) A_1 \Im G(z_2) A_2  \rangle$. \\ 
			
			$\widehat{M}(z_1,I,z_2)$ &
			Deterministic approximation to
			\(\Im G(z_1)\Im G(z_2)\), defined in~\eqref{eq:ImM2_def}. It generates the codimension-one subspace of \emph{regular} observables. \\ 
			
			$\reg{A}^{z_1,z_2}$ &
			Regularization of \(A\), obtained by projecting \(A\) onto the subspace orthogonal to
			\(\widehat{M}(z_1,I,z_2)\); see Definition~\ref{def:reg}. Equivalently,
			\(A\) is \((z_1,z_2)\)-regular iff
			\(\langle A\widehat{M}(z_1,I,z_2)\rangle=0\). \\ 
			
			$V(z_1,z_2)$ &
			Matrix defined in~\eqref{eq:V2_def}. It provides an effective approximation to the unstable direction of
			\(\mathcal{B}_{z_1,\bar z_2}\) and spans the one-dimensional complement of the space of pre-regular observables. \\ 
			
			$\ring{A}^{z_1,z_2}$ &
			Pre-regularization of \(A\), obtained by projecting onto the subspace orthogonal to \(V(z_1,z_2)\); see Definition~\ref{def:prereg}. Equivalently,
			\(A\) is \((z_1,z_2)\)-pre-regular iff
			\(\langle AV(z_1,z_2)\rangle=0\).
			 \\ 
			
			$U(z_1,z_2)$ &
			Matrix defined in~\eqref{eq:V2_def}. It provides an effective approximation to the unstable direction of the two-body stability operator
			\(\mathcal{B}_{z_1,z_2}\). \\ 
			
			$U^{\perp V}(z_1,z_2)$ &
			Orthogonal projection of \(U(z_1,z_2)\) onto the space of pre-regular observables; see~\eqref{eq:UperpV}. It measures the discrepancy between the unstable directions of
			\(\mathcal{B}_{z_1,z_2}\) and \(\mathcal{B}_{z_1,\bar z_2}\), and generates the anomalous fluctuation mode. \\ 
			
			$\vecL(z_1,z_2)$ &
			Rescaled version of \(U^{\perp V}(z_1,z_2)\), defined in~\eqref{eq:L_def}. The normalization is chosen so that
			\(\vecL\) enjoys convenient regularity properties as a function of \((z_1,z_2)\); see~\eqref{eq:L_reg}. \\
			
		\end{tabularx}

	\printbibliography
\end{document}